\documentclass[apjl,ip,twocolumn]{aastex61}
\usepackage{comment}
\usepackage{ifthen}

\newcommand{\forloop}[5][1]%
{%
\setcounter{#2}{#3}%
\ifthenelse{#4}%
	{%
	#5%
	\addtocounter{#2}{#1}%
	\forloop[#1]{#2}{\value{#2}}{#4}{#5}%
	}%
	{%
	}%
}%

\newcommand{\ctbd}[1]{}

\newcommand{\Lc}{Light curve}

\newcommand{\masy}{\ensuremath{\rm mas\,yr^{-1}}}
\newcommand{\kms}{\ensuremath{\rm km\,s^{-1}}}
\newcommand{\ms}{\ensuremath{\rm m\,s^{-1}}}

\newcommand{\gcmc}{\ensuremath{\rm g\,cm^{-3}}}

\newcommand{\logg}{\ensuremath{\log{g}}}
\newcommand{\vsini}{\ensuremath{v \sin{i}}}
\newcommand{\feh}{\ensuremath{\rm [Fe/H]}}

\newcommand{\vmac}{\ensuremath{v_{\rm mac}}}
\newcommand{\vmic}{\ensuremath{v_{\rm mic}}}

\newcommand{\rsun}{\ensuremath{R_\sun}}
\newcommand{\msun}{\ensuremath{M_\sun}}
\newcommand{\lsun}{\ensuremath{L_\sun}}

\newcommand{\rstar}{\ensuremath{R_\star}}
\newcommand{\mstar}{\ensuremath{M_\star}}
\newcommand{\lstar}{\ensuremath{L_\star}}

\newcommand{\teffstar}{\ensuremath{T_{\rm eff\star}}}
\newcommand{\rhostar}{\ensuremath{\rho_\star}}
\newcommand{\loggstar}{\ensuremath{\log{g_{\star}}}}

\newcommand{\rpl}{\ensuremath{R_{p}}}
\newcommand{\mpl}{\ensuremath{M_{p}}}

\newcommand{\rhopl}{\ensuremath{\rho_{p}}}

\newcommand{\arstar}{\ensuremath{a/\rstar}}
\newcommand{\zrstar}{\ensuremath{\zeta/\rstar}}

\newcommand{\rjup}{\ensuremath{R_{\rm J}}}
\newcommand{\mjup}{\ensuremath{M_{\rm J}}}

\newcommand{\reftabl}[1]{Table~\ref{tab:#1}}

\newcommand{\loopand}{\ifnum\value{planetcounter}=2 and \else\fi}
\newcommand{\loopcomma}{\ifnum\value{planetcounter}<2 ,\else. \fi}
\newcommand{\loopcommanoperiod}{\ifnum\value{planetcounter}<2 ,\else \space\fi}
\newcommand{\loopcommanospace}{\ifnum\value{planetcounter}<2 ,\else \fi}

\newcommand{\hatcurhtrxxxxxA}{HATS747-014}                      
\newcommand{\hatcurfieldxxxxxA}{\ensuremath{string}}            
\newcommand{\hatcurCCraxxxxxA}{\ensuremath{19^{\mathrm h}09^{\mathrm m}56.2504{\mathrm s}}}                   
\newcommand{\hatcurCCdecxxxxxA}{\ensuremath{-49{\arcdeg}39{\arcmin}53.868{\arcsec}}}                  
\newcommand{\hatcurCCmagxxxxxA}{14.829}                         
\newcommand{\hatcurCCtwomassxxxxxA}{2MASS~19095625-4939538}     
\newcommand{\hatcurCCgscxxxxxA}{GSC~}                           
\newcommand{\hatcurCCgaiaxxxxxA}{GAIA~6658373007402886400}      
\newcommand{\hatcurCCgaiadrtwoxxxxxA}{6658373007402886400} 
\newcommand{\hatcurCCtassmvxxxxxA}{\ensuremath{14.829\pm0.010}} 
\newcommand{\hatcurCCtassmvshortxxxxxA}{\ensuremath{14.8}}      
\newcommand{\hatcurCCtassmBxxxxxA}{\ensuremath{16.101\pm0.040}} 
\newcommand{\hatcurCCtassmBshortxxxxxA}{\ensuremath{16.1}}      
\newcommand{\hatcurCCtassmIxxxxxA}{\ensuremath{nff\pmnff}}      
\newcommand{\hatcurCCtassmIshortxxxxxA}{\ensuremath{0.0}}       
\newcommand{\hatcurCCtassmgxxxxxA}{\ensuremath{15.480\pm0.010}} 
\newcommand{\hatcurCCtassmgshortxxxxxA}{\ensuremath{15.5}}      
\newcommand{\hatcurCCtassmrxxxxxA}{\ensuremath{14.398\pm0.010}} 
\newcommand{\hatcurCCtassmrshortxxxxxA}{\ensuremath{14.4}}      
\newcommand{\hatcurCCtassmixxxxxA}{\ensuremath{14.009\pm0.010}} 
\newcommand{\hatcurCCtassmishortxxxxxA}{\ensuremath{14.0}}      
\newcommand{\hatcurCCparallaxxxxxxA}{\ensuremath{3.298\pm0.042}} 
\newcommand{\hatcurCCgaiamGxxxxxA}{\ensuremath{14.39980\pm0.00040}} 
\newcommand{\hatcurCCgaiamBPxxxxxA}{\ensuremath{15.0858\pm0.0021}} 
\newcommand{\hatcurCCgaiamRPxxxxxA}{\ensuremath{13.61140\pm0.00090}} 
\newcommand{\hatcurCCtwomassJmagxxxxxA}{\ensuremath{12.653\pm0.023}} 
\newcommand{\hatcurCCtwomassHmagxxxxxA}{\ensuremath{12.026\pm0.023}} 
\newcommand{\hatcurCCtwomassKmagxxxxxA}{\ensuremath{11.926\pm0.025}} 
\newcommand{\hatcurCCcitJmagxxxxxA}{\ensuremath{12.651\pm0.024}} 
\newcommand{\hatcurCCcitHmagxxxxxA}{\ensuremath{12.020\pm0.024}} 
\newcommand{\hatcurCCcitKmagxxxxxA}{\ensuremath{11.950\pm0.025}} 
\newcommand{\hatcurCCbbJmagxxxxxA}{\ensuremath{12.729\pm0.026}} 
\newcommand{\hatcurCCbbHmagxxxxxA}{\ensuremath{12.042\pm0.025}} 
\newcommand{\hatcurCCbbKmagxxxxxA}{\ensuremath{11.970\pm0.025}} 
\newcommand{\hatcurCCesoJmagxxxxxA}{\ensuremath{12.736\pm0.030}} 
\newcommand{\hatcurCCesoHmagxxxxxA}{\ensuremath{12.036\pm0.029}} 
\newcommand{\hatcurCCesoKmagxxxxxA}{\ensuremath{11.967\pm0.026}} 
\newcommand{\hatcurCCesoJHmagxxxxxA}{\ensuremath{0.699\pm0.018}} 
\newcommand{\hatcurCCesoJKmagxxxxxA}{\ensuremath{0.769\pm0.038}} 
\newcommand{\hatcurCCesoHKmagxxxxxA}{\ensuremath{0.069\pm0.038}} 
\newcommand{\hatcurCCWonemagxxxxxA}{\ensuremath{11.867\pm0.023}} 
\newcommand{\hatcurCCWtwomagxxxxxA}{\ensuremath{11.947\pm0.024}} 
\newcommand{\hatcurCCWthreemagxxxxxA}{\ensuremath{0\pm0}}       
\newcommand{\hatcurCCWfourmagxxxxxA}{\ensuremath{0\pm0}}        
\newcommand{\hatcurLCdipxxxxxA}{\ensuremath{31.1}}              
\newcommand{\hatcurLCrprstarxxxxxA}{\ensuremath{0.1746\pm0.0014}} 
\newcommand{\hatcurLCbsqxxxxxA}{\ensuremath{0.508_{-0.014}^{+0.014}}} 
\newcommand{\hatcurLCimpxxxxxA}{\ensuremath{0.7127_{-0.0096}^{+0.0096}}} 
\newcommand{\hatcurLCzetaxxxxxA}{\ensuremath{31.93\pm0.46}}     
\newcommand{\hatcurLCdurxxxxxA}{\ensuremath{0.08343\pm0.00089}} 
\newcommand{\hatcurLCdurshortxxxxxA}{\ensuremath{0.0834}}       
\newcommand{\hatcurLCdurhrxxxxxA}{\ensuremath{2.002\pm0.021}}   
\newcommand{\hatcurLCdurhrshortxxxxxA}{\ensuremath{2.002}}      
\newcommand{\hatcurLCqxxxxxA}{\ensuremath{0.02130\pm0.00024}}   
\newcommand{\hatcurLCqshortxxxxxA}{\ensuremath{0.021}}          
\newcommand{\hatcurLCingdurxxxxxA}{\ensuremath{0.02313\pm0.00067}} 
\newcommand{\hatcurLCPxxxxxA}{\ensuremath{3.9228038\pm0.0000022}} 
\newcommand{\hatcurLCPprecxxxxxA}{\ensuremath{3.9228038}}       
\newcommand{\hatcurLCPshortxxxxxA}{\ensuremath{3.9228}}         
\newcommand{\hatcurLCTxxxxxA}{\ensuremath{2457365.35804\pm0.00029}} 
\newcommand{\hatcurLCTAxxxxxA}{\ensuremath{2456368.96583\pm0.00066}} 
\newcommand{\hatcurLCTBxxxxxA}{\ensuremath{2457604.64907\pm0.00030}} 
\newcommand{\hatcurLChatnetmAxxxxxA}{\ensuremath{14.39324\pm0.00011}} 
\newcommand{\hatcurLCiblendAxxxxxA}{\ensuremath{0.920\pm0.037}} 
\newcommand{\hatcurLChatnetmBxxxxxA}{\ensuremath{-6.75171\pm0.00013}} 
\newcommand{\hatcurLCiblendBxxxxxA}{\ensuremath{0.869\pm0.012}} 
\newcommand{\hatcurLCrhoxxxxxA}{\ensuremath{3.360_{-0.099}^{+0.130}}} 
\newcommand{\hatcurSMEiteffxxxxxA}{\ensuremath{4479\pm51}}      
\newcommand{\hatcurSMEizfehxxxxxA}{\ensuremath{-0.140\pm0.066}} 
\newcommand{\hatcurSMEizfehshortxxxxxA}{\ensuremath{-0.14}}     
\newcommand{\hatcurSMEiloggxxxxxA}{\ensuremath{4.76\pm0.21}}    
\newcommand{\hatcurSMEivsinxxxxxA}{\ensuremath{2.47\pm0.70}}    
\newcommand{\hatcurSMEivmacxxxxxA}{\ensuremath{1.994\pm0.077}}      
\newcommand{\hatcurSMEivmicxxxxxA}{\ensuremath{0.326\pm0.048}}      
\newcommand{\hatcurSMEiiteffxxxxxA}{\ensuremath{4450\pm86}}     
\newcommand{\hatcurSMEiizfehxxxxxA}{\ensuremath{-0.14\pm0.10}}  
\newcommand{\hatcurSMEiizfehshortxxxxxA}{\ensuremath{-0.14}}    
\newcommand{\hatcurSMEiiloggxxxxxA}{\ensuremath{4.603\pm0.031}} 
\newcommand{\hatcurSMEiivsinxxxxxA}{\ensuremath{2.8\pm1.4}}     
\newcommand{\hatcurextraerrMgxxxxxA}{\ensuremath{0.0038\pm0.0031}} 
\newcommand{\hatcurextraerrMgtwosiglimxxxxxA}{\ensuremath{<0.0090}} 
\newcommand{\hatcurextraerrMrxxxxxA}{\ensuremath{0.0028\pm0.0022}} 
\newcommand{\hatcurextraerrMrtwosiglimxxxxxA}{\ensuremath{<0.0067}} 
\newcommand{\hatcurextraerrMixxxxxA}{\ensuremath{0.0037_{-0.0037}^{+0.0191}}} 
\newcommand{\hatcurextraerrMitwosiglimxxxxxA}{\ensuremath{<0.0461}} 
\newcommand{\hatcurextraerrMGxxxxxA}{\ensuremath{0.037_{-0.019}^{+0.064}}} 
\newcommand{\hatcurextraerrMGtwosiglimxxxxxA}{\ensuremath{<0.1548}} 
\newcommand{\hatcurextraerrMBPtwoxxxxxA}{\ensuremath{0.0039_{-0.0038}^{+0.0107}}} 
\newcommand{\hatcurextraerrMBPtwotwosiglimxxxxxA}{\ensuremath{<0.0246}} 
\newcommand{\hatcurextraerrMRPxxxxxA}{\ensuremath{0.0073_{-0.0068}^{+0.0404}}} 
\newcommand{\hatcurextraerrMRPtwosiglimxxxxxA}{\ensuremath{<0.1245}} 
\newcommand{\hatcurextraerrMJxxxxxA}{\ensuremath{0.00028_{-0.00027}^{+0.00247}}} 
\newcommand{\hatcurextraerrMJtwosiglimxxxxxA}{\ensuremath{<0.0070}} 
\newcommand{\hatcurextraerrMHxxxxxA}{\ensuremath{0.0013_{-0.0012}^{+0.0043}}} 
\newcommand{\hatcurextraerrMHtwosiglimxxxxxA}{\ensuremath{<0.0118}} 
\newcommand{\hatcurextraerrMKsxxxxxA}{\ensuremath{0.0018_{-0.0018}^{+0.0097}}} 
\newcommand{\hatcurextraerrMKstwosiglimxxxxxA}{\ensuremath{<0.0220}} 
\newcommand{\hatcurextraerrMWonexxxxxA}{\ensuremath{0.019_{-0.018}^{+0.132}}} 
\newcommand{\hatcurextraerrMWonetwosiglimxxxxxA}{\ensuremath{<0.2651}} 
\newcommand{\hatcurextraerrMWtwoxxxxxA}{\ensuremath{0.0028_{-0.0027}^{+0.0055}}} 
\newcommand{\hatcurextraerrMWtwotwosiglimxxxxxA}{\ensuremath{<0.0102}} 
\newcommand{\hatcurLBiBxxxxxA}{\ensuremath{0.9498}}             
\newcommand{\hatcurLBiiBxxxxxA}{\ensuremath{-0.0970}}           
\newcommand{\hatcurLBiVxxxxxA}{\ensuremath{0.7431}}             
\newcommand{\hatcurLBiiVxxxxxA}{\ensuremath{0.0539}}            
\newcommand{\hatcurLBiRxxxxxA}{\ensuremath{0.6121}}             
\newcommand{\hatcurLBiiRxxxxxA}{\ensuremath{0.1147}}            
\newcommand{\hatcurLBiIxxxxxA}{\ensuremath{0.4764}}             
\newcommand{\hatcurLBiiIxxxxxA}{\ensuremath{0.1543}}            
\newcommand{\hatcurLBiuxxxxxA}{\ensuremath{1.1009}}             
\newcommand{\hatcurLBiiuxxxxxA}{\ensuremath{-0.2662}}           
\newcommand{\hatcurLBigxxxxxA}{\ensuremath{0.8521}}             
\newcommand{\hatcurLBiigxxxxxA}{\ensuremath{-0.0323}}           
\newcommand{\hatcurLBirxxxxxA}{\ensuremath{0.49\pm0.13}}        
\newcommand{\hatcurLBiirxxxxxA}{\ensuremath{0.28_{-0.12}^{+0.15}}} 
\newcommand{\hatcurLBiixxxxxA}{\ensuremath{0.53\pm0.13}}        
\newcommand{\hatcurLBiiixxxxxA}{\ensuremath{0.38_{-0.16}^{+0.12}}} 
\newcommand{\hatcurLBizxxxxxA}{\ensuremath{0.4163}}             
\newcommand{\hatcurLBiizxxxxxA}{\ensuremath{0.1659}}            
\newcommand{\hatcurLBiJxxxxxA}{\ensuremath{0.2917}}             
\newcommand{\hatcurLBiiJxxxxxA}{\ensuremath{0.2132}}            
\newcommand{\hatcurLBiHxxxxxA}{\ensuremath{0.1585}}             
\newcommand{\hatcurLBiiHxxxxxA}{\ensuremath{0.2975}}            
\newcommand{\hatcurLBiKxxxxxA}{\ensuremath{0.1313}}             
\newcommand{\hatcurLBiiKxxxxxA}{\ensuremath{0.2488}}            
\newcommand{\hatcurLBiTxxxxxA}{\ensuremath{0.460\pm0.100}}      
\newcommand{\hatcurLBiiTxxxxxA}{\ensuremath{0.41\pm0.11}}       
\newcommand{\hatcurLBikepxxxxxA}{\ensuremath{0.6124}}           
\newcommand{\hatcurLBiikepxxxxxA}{\ensuremath{0.1247}}          
\newcommand{\hatcurLBiCxxxxxA}{\ensuremath{0.5860}}             
\newcommand{\hatcurLBiiCxxxxxA}{\ensuremath{0.1328}}            
\newcommand{\hatcurLBiMxxxxxA}{\ensuremath{0.7144}}             
\newcommand{\hatcurLBiiMxxxxxA}{\ensuremath{0.0734}}            
\newcommand{\hatcurLBiSonexxxxxA}{\ensuremath{0.1118}}            
\newcommand{\hatcurLBiiSonexxxxxA}{\ensuremath{0.1665}}           
\newcommand{\hatcurLBiStwoxxxxxA}{\ensuremath{0.0993}}            
\newcommand{\hatcurLBiiStwoxxxxxA}{\ensuremath{0.1321}}           
\newcommand{\hatcurLBiSthreexxxxxA}{\ensuremath{0.0826}}            
\newcommand{\hatcurLBiiSthreexxxxxA}{\ensuremath{0.1091}}           
\newcommand{\hatcurLBiSfourxxxxxA}{\ensuremath{0.0716}}            
\newcommand{\hatcurLBiiSfourxxxxxA}{\ensuremath{0.0934}}           
\newcommand{\hatcurISOmxxxxxA}{\ensuremath{0.674_{-0.012}^{+0.016}}} 
\newcommand{\hatcurISOmshortxxxxxA}{\ensuremath{0.67}}          
\newcommand{\hatcurISOmlongxxxxxA}{\ensuremath{0.674_{-0.012}^{+0.016}}} 
\newcommand{\hatcurISOrxxxxxA}{\ensuremath{0.6564\pm0.0055}}    
\newcommand{\hatcurISOrshortxxxxxA}{\ensuremath{0.66}}          
\newcommand{\hatcurISOrlongxxxxxA}{\ensuremath{0.6564\pm0.0055}} 
\newcommand{\hatcurISOrhoxxxxxA}{\ensuremath{3.360_{-0.099}^{+0.130}}} 
\newcommand{\hatcurISOrholongxxxxxA}{\ensuremath{3.360_{-0.099}^{+0.130}}} 
\newcommand{\hatcurISOloggxxxxxA}{\ensuremath{4.633\pm0.011}}   
\newcommand{\hatcurISOlumxxxxxA}{\ensuremath{0.1599\pm0.0031}}  
\newcommand{\hatcurISOlumshortxxxxxA}{\ensuremath{0.16}}        
\newcommand{\hatcurISOteffxxxxxA}{\ensuremath{4512\pm19}}       
\newcommand{\hatcurISOzfehxxxxxA}{\ensuremath{-0.113\pm0.035}}  
\newcommand{\hatcurISOagexxxxxA}{\ensuremath{8.1_{-4.3}^{+2.9}}} 
\newcommand{\hatcurISOspecxxxxxA}{K}                            
\newcommand{\hatcurRVKxxxxxA}{\ensuremath{61.7\pm4.3}}          
\newcommand{\hatcurRVKtwosiglimxxxxxA}{\ensuremath{<69.4}}      
\newcommand{\hatcurRVrkxxxxxA}{\ensuremath{0\pm0}}              
\newcommand{\hatcurRVrhxxxxxA}{\ensuremath{0\pm0}}              
\newcommand{\hatcurRVkxxxxxA}{\ensuremath{0\pm0}}               
\newcommand{\hatcurRVhxxxxxA}{\ensuremath{0\pm0}}               
\newcommand{\hatcurRVtronexxxxxA}{\ensuremath{0\pm0}}           
\newcommand{\hatcurRVtrtwoxxxxxA}{\ensuremath{0\pm0}}           
\newcommand{\hatcurRVgammaxxxxxA}{\ensuremath{13.0\pm3.0}}      
\newcommand{\hatcurRVjitterxxxxxA}{\ensuremath{40.4\pm8.9}}     
\newcommand{\hatcurRVjittertwosiglimxxxxxA}{\ensuremath{<57.6}} 
\newcommand{\hatcurRVfitrmsxxxxxA}{\ensuremath{.1fym}}          %
\newcommand{\hatcurRVeccenxxxxxA}{\ensuremath{0\pm0}}           
\newcommand{\hatcurRVeccentwosiglimxxxxxA}{\ensuremath{<0.000}} 
\newcommand{\hatcurRVomegaxxxxxA}{\ensuremath{0\pm0}}           
\newcommand{\hatcurPPixxxxxA}{\ensuremath{87.080\pm0.061}}      
\newcommand{\hatcurPPgxxxxxA}{\ensuremath{7.34_{-0.47}^{+0.61}}} 
\newcommand{\hatcurPPloggxxxxxA}{\ensuremath{2.866\pm0.032}}    
\newcommand{\hatcurPParxxxxxA}{\ensuremath{13.98\pm0.15}}       
\newcommand{\hatcurPParelxxxxxA}{\ensuremath{0.04269_{-0.00025}^{+0.00033}}} 
\newcommand{\hatcurPPrhoxxxxxA}{\ensuremath{0.331\pm0.027}}     
\newcommand{\hatcurPPmxxxxxA}{\ensuremath{0.369_{-0.021}^{+0.031}}} 
\newcommand{\hatcurPPmtwosiglimxxxxxA}{\ensuremath{<0.42}}      
\newcommand{\hatcurPPmshortxxxxxA}{\ensuremath{0.37}}           
\newcommand{\hatcurPPmlongxxxxxA}{\ensuremath{0.369_{-0.021}^{+0.031}}} 
\newcommand{\hatcurPPmexxxxxA}{\ensuremath{117.1_{-6.8}^{+10.0}}} 
\newcommand{\hatcurPPmeshortxxxxxA}{\ensuremath{117.1}}         
\newcommand{\hatcurPPmelongxxxxxA}{\ensuremath{117.1_{-6.8}^{+10.0}}} 
\newcommand{\hatcurPPrxxxxxA}{\ensuremath{1.117\pm0.014}}       
\newcommand{\hatcurPPrshortxxxxxA}{\ensuremath{1.12}}           
\newcommand{\hatcurPPrlongxxxxxA}{\ensuremath{1.117\pm0.014}}   
\newcommand{\hatcurPPrexxxxxA}{\ensuremath{12.51\pm0.16}}       
\newcommand{\hatcurPPreshortxxxxxA}{\ensuremath{12.5}}          
\newcommand{\hatcurPPrelongxxxxxA}{\ensuremath{12.51\pm0.16}}   
\newcommand{\hatcurPPmrcorrxxxxxA}{\ensuremath{-0.00}}          
\newcommand{\hatcurPPteffxxxxxA}{\ensuremath{852.9\pm4.7}}      
\newcommand{\hatcurPPthetaxxxxxA}{\ensuremath{0.0418_{-0.0025}^{+0.0034}}} 
\newcommand{\hatcurPPfluxperixxxxxA}{\ensuremath{1.194\pm0.026}} 
\newcommand{\hatcurPPfluxperidimxxxxxA}{\ensuremath{8}}         
\newcommand{\hatcurPPfluxapxxxxxA}{\ensuremath{1.194\pm0.026}}  
\newcommand{\hatcurPPfluxapdimxxxxxA}{\ensuremath{8}}           
\newcommand{\hatcurPPfluxavgxxxxxA}{\ensuremath{1.194\pm0.026}} 
\newcommand{\hatcurPPfluxavgdimxxxxxA}{\ensuremath{8}}          
\newcommand{\hatcurPPfluxavglogxxxxxA}{\ensuremath{8.0770\pm0.0095}} 
\newcommand{\hatcurXsecphasexxxxxA}{\ensuremath{0\pm0}}         
\newcommand{\hatcurXsecondaryxxxxxA}{\ensuremath{2457367.31944\pm0.00029}} 
\newcommand{\hatcurXsecdurxxxxxA}{\ensuremath{0.08343\pm0.00089}} 
\newcommand{\hatcurXsecingdurxxxxxA}{\ensuremath{0.02313\pm0.00067}} 
\newcommand{\hatcurPPphiconjxxxxxA}{\ensuremath{0\pm0}}         
\newcommand{\hatcurPPperixxxxxA}{\ensuremath{2457364.37733\pm0.00029}} 
\newcommand{\hatcurPPaequivxxxxxA}{\ensuremath{0.1068\pm0.0012}} 
\newcommand{\hatcurPPtcircxxxxxA}{\ensuremath{187\pm19}}        
\newcommand{\hatcurPPtinfallxxxxxA}{\ensuremath{36300\pm3200}}  
\newcommand{\hatcurXdistxxxxxA}{\ensuremath{301.7\pm1.9}}       
\newcommand{\hatcurXAvxxxxxA}{\ensuremath{0.108\pm0.032}}       
\newcommand{\hatcurXdistredxxxxxA}{\ensuremath{301.7\pm1.9}}    
\newcommand{\hatcurXEBVxxxxxA}{\ensuremath{0.035\pm0.010}}      
\newcommand{\hatcurCCpmraxxxxxA}{\ensuremath{3.827\pm0.058}}    
\newcommand{\hatcurCCpmdecxxxxxA}{\ensuremath{4.878\pm0.038}}   
\newcommand{\hatcurCCpmxxxxxA}{\ensuremath{6.200\pm0.069}}      
\newcommand{\hatcurhtrxxxxxB}{HATS778-005}                      
\newcommand{\hatcurfieldxxxxxB}{\ensuremath{string}}            
\newcommand{\hatcurCCraxxxxxB}{\ensuremath{19^{\mathrm h}14^{\mathrm m}41.2748{\mathrm s}}}                   
\newcommand{\hatcurCCdecxxxxxB}{\ensuremath{-59{\arcdeg}34{\arcmin}45.7571{\arcsec}}}                 
\newcommand{\hatcurCCmagxxxxxB}{14.352}                         
\newcommand{\hatcurCCtwomassxxxxxB}{2MASS~19144126-5934458}     
\newcommand{\hatcurCCgscxxxxxB}{GSC~}                           
\newcommand{\hatcurCCgaiaxxxxxB}{GAIA~6638412919991750912}      
\newcommand{\hatcurCCgaiadrtwoxxxxxB}{6638412919991750912} 
\newcommand{\hatcurCCtassmvxxxxxB}{\ensuremath{14.35\pm0.11}}   
\newcommand{\hatcurCCtassmvshortxxxxxB}{\ensuremath{14.4}}      
\newcommand{\hatcurCCtassmBxxxxxB}{\ensuremath{15.577\pm0.050}} 
\newcommand{\hatcurCCtassmBshortxxxxxB}{\ensuremath{15.6}}      
\newcommand{\hatcurCCtassmIxxxxxB}{\ensuremath{nff\pmnff}}      
\newcommand{\hatcurCCtassmIshortxxxxxB}{\ensuremath{0.0}}       
\newcommand{\hatcurCCtassmgxxxxxB}{\ensuremath{14.935\pm0.030}} 
\newcommand{\hatcurCCtassmgshortxxxxxB}{\ensuremath{14.9}}      
\newcommand{\hatcurCCtassmrxxxxxB}{\ensuremath{13.821\pm0.050}} 
\newcommand{\hatcurCCtassmrshortxxxxxB}{\ensuremath{13.8}}      
\newcommand{\hatcurCCtassmixxxxxB}{\ensuremath{13.69\pm0.17}}   
\newcommand{\hatcurCCtassmishortxxxxxB}{\ensuremath{13.7}}      
\newcommand{\hatcurCCparallaxxxxxxB}{\ensuremath{3.765\pm0.024}} 
\newcommand{\hatcurCCgaiamGxxxxxB}{\ensuremath{13.89510\pm0.00020}} 
\newcommand{\hatcurCCgaiamBPxxxxxB}{\ensuremath{14.5801\pm0.0016}} 
\newcommand{\hatcurCCgaiamRPxxxxxB}{\ensuremath{13.11260\pm0.00090}} 
\newcommand{\hatcurCCtwomassJmagxxxxxB}{\ensuremath{12.160\pm0.024}} 
\newcommand{\hatcurCCtwomassHmagxxxxxB}{\ensuremath{11.591\pm0.026}} 
\newcommand{\hatcurCCtwomassKmagxxxxxB}{\ensuremath{11.427\pm0.021}} 
\newcommand{\hatcurCCcitJmagxxxxxB}{\ensuremath{12.158\pm0.025}} 
\newcommand{\hatcurCCcitHmagxxxxxB}{\ensuremath{11.583\pm0.027}} 
\newcommand{\hatcurCCcitKmagxxxxxB}{\ensuremath{11.451\pm0.022}} 
\newcommand{\hatcurCCbbJmagxxxxxB}{\ensuremath{12.236\pm0.027}} 
\newcommand{\hatcurCCbbHmagxxxxxB}{\ensuremath{11.607\pm0.028}} 
\newcommand{\hatcurCCbbKmagxxxxxB}{\ensuremath{11.471\pm0.022}} 
\newcommand{\hatcurCCesoJmagxxxxxB}{\ensuremath{12.243\pm0.031}} 
\newcommand{\hatcurCCesoHmagxxxxxB}{\ensuremath{11.605\pm0.036}} 
\newcommand{\hatcurCCesoKmagxxxxxB}{\ensuremath{11.468\pm0.023}} 
\newcommand{\hatcurCCesoJHmagxxxxxB}{\ensuremath{0.638\pm0.044}} 
\newcommand{\hatcurCCesoJKmagxxxxxB}{\ensuremath{0.776\pm0.036}} 
\newcommand{\hatcurCCesoHKmagxxxxxB}{\ensuremath{0.137\pm0.042}} 
\newcommand{\hatcurCCWonemagxxxxxB}{\ensuremath{11.364\pm0.023}} 
\newcommand{\hatcurCCWtwomagxxxxxB}{\ensuremath{11.458\pm0.021}} 
\newcommand{\hatcurCCWthreemagxxxxxB}{\ensuremath{0\pm0}}       
\newcommand{\hatcurCCWfourmagxxxxxB}{\ensuremath{0\pm0}}        
\newcommand{\hatcurLCdipxxxxxB}{\ensuremath{16.5}}              
\newcommand{\hatcurLCrprstarxxxxxB}{\ensuremath{0.1148\pm0.0020}} 
\newcommand{\hatcurLCbsqxxxxxB}{\ensuremath{0.0069_{-0.0046}^{+0.0080}}} 
\newcommand{\hatcurLCimpxxxxxB}{\ensuremath{0.083_{-0.035}^{+0.039}}} 
\newcommand{\hatcurLCzetaxxxxxB}{\ensuremath{22.825_{-0.098}^{+0.130}}} 
\newcommand{\hatcurLCdurxxxxxB}{\ensuremath{0.09774\pm0.00050}} 
\newcommand{\hatcurLCdurshortxxxxxB}{\ensuremath{0.0977}}       
\newcommand{\hatcurLCdurhrxxxxxB}{\ensuremath{2.346\pm0.012}}   
\newcommand{\hatcurLCdurhrshortxxxxxB}{\ensuremath{2.346}}      
\newcommand{\hatcurLCqxxxxxB}{\ensuremath{0.03120\pm0.00016}}   
\newcommand{\hatcurLCqshortxxxxxB}{\ensuremath{0.031}}          
\newcommand{\hatcurLCingdurxxxxxB}{\ensuremath{0.01015\pm0.00020}} 
\newcommand{\hatcurLCPxxxxxB}{\ensuremath{3.1316666\pm0.0000037}} 
\newcommand{\hatcurLCPprecxxxxxB}{\ensuremath{3.1316666}}       
\newcommand{\hatcurLCPshortxxxxxB}{\ensuremath{3.1317}}         
\newcommand{\hatcurLCTxxxxxB}{\ensuremath{2457100.55022\pm0.00045}} 
\newcommand{\hatcurLCTAxxxxxB}{\ensuremath{2455678.7737\pm0.0017}} 
\newcommand{\hatcurLCTBxxxxxB}{\ensuremath{2457219.55354\pm0.00049}} 
\newcommand{\hatcurLChatnetmAxxxxxB}{\ensuremath{13.91272\pm0.00014}} 
\newcommand{\hatcurLCiblendAxxxxxB}{\ensuremath{0.998940\pm0.000094}} 
\newcommand{\hatcurLChatnetmBxxxxxB}{\ensuremath{-0.00017\pm0.00015}} 
\newcommand{\hatcurLCiblendBxxxxxB}{\ensuremath{0.597\pm0.024}} 
\newcommand{\hatcurLCrhoxxxxxB}{\ensuremath{2.804\pm0.036}}     
\newcommand{\hatcurSMEiteffxxxxxB}{\ensuremath{4190\pm100}}     
\newcommand{\hatcurSMEizfehxxxxxB}{\ensuremath{0.00\pm0.10}}    
\newcommand{\hatcurSMEizfehshortxxxxxB}{\ensuremath{0.00}}      
\newcommand{\hatcurSMEiloggxxxxxB}{\ensuremath{4.690\pm0.088}}  
\newcommand{\hatcurSMEivsinxxxxxB}{\ensuremath{0.73\pm0.55}}    
\newcommand{\hatcurSMEivmacxxxxxB}{\ensuremath{1.55\pm0.15}}      
\newcommand{\hatcurSMEivmicxxxxxB}{\ensuremath{0.000\pm0.085}}      
\newcommand{\hatcurextraerrMgxxxxxB}{\ensuremath{0.0098_{-0.0094}^{+0.0413}}} 
\newcommand{\hatcurextraerrMgtwosiglimxxxxxB}{\ensuremath{<0.0707}} 
\newcommand{\hatcurextraerrMrxxxxxB}{\ensuremath{0.0056_{-0.0055}^{+0.0426}}} 
\newcommand{\hatcurextraerrMrtwosiglimxxxxxB}{\ensuremath{<0.1586}} 
\newcommand{\hatcurextraerrMixxxxxB}{\ensuremath{0.35\pm0.20}}  
\newcommand{\hatcurextraerrMitwosiglimxxxxxB}{\ensuremath{<0.6740}} 
\newcommand{\hatcurextraerrMGxxxxxB}{\ensuremath{0.038_{-0.018}^{+0.026}}} 
\newcommand{\hatcurextraerrMGtwosiglimxxxxxB}{\ensuremath{<0.0914}} 
\newcommand{\hatcurextraerrMBPtwoxxxxxB}{\ensuremath{0.0060_{-0.0057}^{+0.0199}}} 
\newcommand{\hatcurextraerrMBPtwotwosiglimxxxxxB}{\ensuremath{<0.0361}} 
\newcommand{\hatcurextraerrMRPxxxxxB}{\ensuremath{0.0036_{-0.0035}^{+0.0145}}} 
\newcommand{\hatcurextraerrMRPtwosiglimxxxxxB}{\ensuremath{<0.0329}} 
\newcommand{\hatcurextraerrMJxxxxxB}{\ensuremath{0.0017_{-0.0017}^{+0.0101}}} 
\newcommand{\hatcurextraerrMJtwosiglimxxxxxB}{\ensuremath{<0.0232}} 
\newcommand{\hatcurextraerrMHxxxxxB}{\ensuremath{0.013_{-0.013}^{+0.034}}} 
\newcommand{\hatcurextraerrMHtwosiglimxxxxxB}{\ensuremath{<0.0674}} 
\newcommand{\hatcurextraerrMKsxxxxxB}{\ensuremath{0.0023_{-0.0022}^{+0.0153}}} 
\newcommand{\hatcurextraerrMKstwosiglimxxxxxB}{\ensuremath{<0.0297}} 
\newcommand{\hatcurextraerrMWonexxxxxB}{\ensuremath{0.023_{-0.021}^{+0.032}}} 
\newcommand{\hatcurextraerrMWonetwosiglimxxxxxB}{\ensuremath{<0.0792}} 
\newcommand{\hatcurextraerrMWtwoxxxxxB}{\ensuremath{0.032_{-0.031}^{+0.096}}} 
\newcommand{\hatcurextraerrMWtwotwosiglimxxxxxB}{\ensuremath{<0.1857}} 
\newcommand{\hatcurLBiBxxxxxB}{\ensuremath{0.8007}}             
\newcommand{\hatcurLBiiBxxxxxB}{\ensuremath{0.0222}}            
\newcommand{\hatcurLBiVxxxxxB}{\ensuremath{0.6354}}             
\newcommand{\hatcurLBiiVxxxxxB}{\ensuremath{0.1339}}            
\newcommand{\hatcurLBiRxxxxxB}{\ensuremath{0.5245}}             
\newcommand{\hatcurLBiiRxxxxxB}{\ensuremath{0.1825}}            
\newcommand{\hatcurLBiIxxxxxB}{\ensuremath{0.3905}}             
\newcommand{\hatcurLBiiIxxxxxB}{\ensuremath{0.2115}}            
\newcommand{\hatcurLBiuxxxxxB}{\ensuremath{0.8665}}             
\newcommand{\hatcurLBiiuxxxxxB}{\ensuremath{-0.0450}}           
\newcommand{\hatcurLBigxxxxxB}{\ensuremath{0.7223}}             
\newcommand{\hatcurLBiigxxxxxB}{\ensuremath{0.0663}}            
\newcommand{\hatcurLBirxxxxxB}{\ensuremath{0.36\pm0.14}}        
\newcommand{\hatcurLBiirxxxxxB}{\ensuremath{0.32\pm0.15}}       
\newcommand{\hatcurLBiixxxxxB}{\ensuremath{0.22\pm0.11}}        
\newcommand{\hatcurLBiiixxxxxB}{\ensuremath{0.17\pm0.16}}       
\newcommand{\hatcurLBizxxxxxB}{\ensuremath{0.3392}}             
\newcommand{\hatcurLBiizxxxxxB}{\ensuremath{0.2145}}            
\newcommand{\hatcurLBiJxxxxxB}{\ensuremath{0.2506}}             
\newcommand{\hatcurLBiiJxxxxxB}{\ensuremath{0.2171}}            
\newcommand{\hatcurLBiHxxxxxB}{\ensuremath{0.1694}}             
\newcommand{\hatcurLBiiHxxxxxB}{\ensuremath{0.2632}}            
\newcommand{\hatcurLBiKxxxxxB}{\ensuremath{0.1301}}             
\newcommand{\hatcurLBiiKxxxxxB}{\ensuremath{0.2308}}            
\newcommand{\hatcurLBiTxxxxxB}{\ensuremath{0.41\pm0.12}}        
\newcommand{\hatcurLBiiTxxxxxB}{\ensuremath{0.34\pm0.14}}       
\newcommand{\hatcurLBikepxxxxxB}{\ensuremath{0.5094}}           
\newcommand{\hatcurLBiikepxxxxxB}{\ensuremath{0.2164}}          
\newcommand{\hatcurLBiCxxxxxB}{\ensuremath{0.4819}}             
\newcommand{\hatcurLBiiCxxxxxB}{\ensuremath{0.2222}}            
\newcommand{\hatcurLBiMxxxxxB}{\ensuremath{0.5998}}             
\newcommand{\hatcurLBiiMxxxxxB}{\ensuremath{0.1782}}            
\newcommand{\hatcurLBiSonexxxxxB}{\ensuremath{0.0966}}            
\newcommand{\hatcurLBiiSonexxxxxB}{\ensuremath{0.1667}}           
\newcommand{\hatcurLBiStwoxxxxxB}{\ensuremath{0.0830}}            
\newcommand{\hatcurLBiiStwoxxxxxB}{\ensuremath{0.1376}}           
\newcommand{\hatcurLBiSthreexxxxxB}{\ensuremath{0.0699}}            
\newcommand{\hatcurLBiiSthreexxxxxB}{\ensuremath{0.1187}}           
\newcommand{\hatcurLBiSfourxxxxxB}{\ensuremath{0.0647}}            
\newcommand{\hatcurLBiiSfourxxxxxB}{\ensuremath{0.1036}}           
\newcommand{\hatcurISOmxxxxxB}{\ensuremath{0.7279\pm0.0066}}    
\newcommand{\hatcurISOmshortxxxxxB}{\ensuremath{0.73}}          
\newcommand{\hatcurISOmlongxxxxxB}{\ensuremath{0.7279\pm0.0066}} 
\newcommand{\hatcurISOrxxxxxB}{\ensuremath{0.7152\pm0.0038}}    
\newcommand{\hatcurISOrshortxxxxxB}{\ensuremath{0.72}}          
\newcommand{\hatcurISOrlongxxxxxB}{\ensuremath{0.7152\pm0.0038}} 
\newcommand{\hatcurISOrhoxxxxxB}{\ensuremath{2.804\pm0.036}}    
\newcommand{\hatcurISOrholongxxxxxB}{\ensuremath{2.804\pm0.036}} 
\newcommand{\hatcurISOloggxxxxxB}{\ensuremath{4.5909\pm0.0039}} 
\newcommand{\hatcurISOlumxxxxxB}{\ensuremath{0.1955\pm0.0042}}  
\newcommand{\hatcurISOlumshortxxxxxB}{\ensuremath{0.20}}        
\newcommand{\hatcurISOteffxxxxxB}{\ensuremath{4546_{-18}^{+23}}} 
\newcommand{\hatcurISOzfehxxxxxB}{\ensuremath{0.186\pm0.051}}   
\newcommand{\hatcurISOagexxxxxB}{\ensuremath{11.97_{-0.61}^{+0.42}}} 
\newcommand{\hatcurISOspecxxxxxB}{K}                            
\newcommand{\hatcurRVKxxxxxB}{\ensuremath{41.8\pm4.4}}          
\newcommand{\hatcurRVKtwosiglimxxxxxB}{\ensuremath{<47.7}}      
\newcommand{\hatcurRVrkxxxxxB}{\ensuremath{0\pm0}}              
\newcommand{\hatcurRVrhxxxxxB}{\ensuremath{0\pm0}}              
\newcommand{\hatcurRVkxxxxxB}{\ensuremath{0\pm0}}               
\newcommand{\hatcurRVhxxxxxB}{\ensuremath{0\pm0}}               
\newcommand{\hatcurRVtronexxxxxB}{\ensuremath{0\pm0}}           
\newcommand{\hatcurRVtrtwoxxxxxB}{\ensuremath{0\pm0}}           
\newcommand{\hatcurRVgammaxxxxxB}{\ensuremath{-15.6\pm5.3}}     
\newcommand{\hatcurRVjitterxxxxxB}{\ensuremath{25.2\pm6.2}}     
\newcommand{\hatcurRVjittertwosiglimxxxxxB}{\ensuremath{<37.0}} 
\newcommand{\hatcurRVfitrmsxxxxxB}{\ensuremath{.1fym}}          %
\newcommand{\hatcurRVeccenxxxxxB}{\ensuremath{0\pm0}}           
\newcommand{\hatcurRVeccentwosiglimxxxxxB}{\ensuremath{<0.000}} 
\newcommand{\hatcurRVomegaxxxxxB}{\ensuremath{0\pm0}}           
\newcommand{\hatcurPPixxxxxB}{\ensuremath{89.58\pm0.18}}        
\newcommand{\hatcurPPgxxxxxB}{\ensuremath{9.50_{-1.18}^{+0.84}}} 
\newcommand{\hatcurPPloggxxxxxB}{\ensuremath{2.978_{-0.058}^{+0.036}}} 
\newcommand{\hatcurPParxxxxxB}{\ensuremath{11.330\pm0.049}}     
\newcommand{\hatcurPParelxxxxxB}{\ensuremath{0.03769\pm0.00011}} 
\newcommand{\hatcurPPrhoxxxxxB}{\ensuremath{0.589\pm0.067}}     
\newcommand{\hatcurPPmxxxxxB}{\ensuremath{0.243_{-0.030}^{+0.022}}} 
\newcommand{\hatcurPPmtwosiglimxxxxxB}{\ensuremath{<0.28}}      
\newcommand{\hatcurPPmshortxxxxxB}{\ensuremath{0.24}}           
\newcommand{\hatcurPPmlongxxxxxB}{\ensuremath{0.243_{-0.030}^{+0.022}}} 
\newcommand{\hatcurPPmexxxxxB}{\ensuremath{77.3_{-9.6}^{+7.0}}} 
\newcommand{\hatcurPPmeshortxxxxxB}{\ensuremath{77.3}}          
\newcommand{\hatcurPPmelongxxxxxB}{\ensuremath{77.3_{-9.6}^{+7.0}}} 
\newcommand{\hatcurPPrxxxxxB}{\ensuremath{0.800\pm0.015}}       
\newcommand{\hatcurPPrshortxxxxxB}{\ensuremath{0.80}}           
\newcommand{\hatcurPPrlongxxxxxB}{\ensuremath{0.800\pm0.015}}   
\newcommand{\hatcurPPrexxxxxB}{\ensuremath{8.97\pm0.17}}        
\newcommand{\hatcurPPreshortxxxxxB}{\ensuremath{9.0}}           
\newcommand{\hatcurPPrelongxxxxxB}{\ensuremath{8.97\pm0.17}}    
\newcommand{\hatcurPPmrcorrxxxxxB}{\ensuremath{0.16}}           
\newcommand{\hatcurPPteffxxxxxB}{\ensuremath{954.6\pm4.8}}      
\newcommand{\hatcurPPthetaxxxxxB}{\ensuremath{0.0317_{-0.0041}^{+0.0027}}} 
\newcommand{\hatcurPPfluxperixxxxxB}{\ensuremath{1.873\pm0.038}} 
\newcommand{\hatcurPPfluxperidimxxxxxB}{\ensuremath{8}}         
\newcommand{\hatcurPPfluxapxxxxxB}{\ensuremath{1.873\pm0.038}}  
\newcommand{\hatcurPPfluxapdimxxxxxB}{\ensuremath{8}}           
\newcommand{\hatcurPPfluxavgxxxxxB}{\ensuremath{1.873\pm0.038}} 
\newcommand{\hatcurPPfluxavgdimxxxxxB}{\ensuremath{8}}          
\newcommand{\hatcurPPfluxavglogxxxxxB}{\ensuremath{8.2725\pm0.0086}} 
\newcommand{\hatcurXsecphasexxxxxB}{\ensuremath{0\pm0}}         
\newcommand{\hatcurXsecondaryxxxxxB}{\ensuremath{2457102.11605\pm0.00046}} 
\newcommand{\hatcurXsecdurxxxxxB}{\ensuremath{0.09774\pm0.00050}} 
\newcommand{\hatcurXsecingdurxxxxxB}{\ensuremath{0.01015\pm0.00020}} 
\newcommand{\hatcurPPphiconjxxxxxB}{\ensuremath{0\pm0}}         
\newcommand{\hatcurPPperixxxxxB}{\ensuremath{2457099.76730\pm0.00045}} 
\newcommand{\hatcurPPaequivxxxxxB}{\ensuremath{0.08520\pm0.00084}} 
\newcommand{\hatcurPPtcircxxxxxB}{\ensuremath{254\pm35}}        
\newcommand{\hatcurPPtinfallxxxxxB}{\ensuremath{16700_{-1500}^{+2200}}} 
\newcommand{\hatcurXdistxxxxxB}{\ensuremath{265.4\pm1.7}}       
\newcommand{\hatcurXAvxxxxxB}{\ensuremath{0.112\pm0.033}}       
\newcommand{\hatcurXdistredxxxxxB}{\ensuremath{265.4\pm1.7}}    
\newcommand{\hatcurXEBVxxxxxB}{\ensuremath{0.036\pm0.011}}      
\newcommand{\hatcurCCpmraxxxxxB}{\ensuremath{3.125\pm0.031}}    
\newcommand{\hatcurCCpmdecxxxxxB}{\ensuremath{6.146\pm0.029}}   
\newcommand{\hatcurCCpmxxxxxB}{\ensuremath{6.895\pm0.042}}      
\newcommand{\hatcurhtrxxxxxC}{HATS755-004}                      
\newcommand{\hatcurfieldxxxxxC}{\ensuremath{string}}            
\newcommand{\hatcurCCraxxxxxC}{\ensuremath{00^{\mathrm h}26^{\mathrm m}27.1829{\mathrm s}}}                   
\newcommand{\hatcurCCdecxxxxxC}{\ensuremath{-56{\arcdeg}20{\arcmin}39.5352{\arcsec}}}                 
\newcommand{\hatcurCCmagxxxxxC}{14.998}                         
\newcommand{\hatcurCCtwomassxxxxxC}{2MASS~00262717-5620395}     
\newcommand{\hatcurCCgscxxxxxC}{GSC~}                           
\newcommand{\hatcurCCgaiaxxxxxC}{GAIA~4919770108539385472}      
\newcommand{\hatcurCCgaiadrtwoxxxxxC}{4919770108539385472} 
\newcommand{\hatcurCCtassmvxxxxxC}{\ensuremath{14.998\pm0.040}} 
\newcommand{\hatcurCCtassmvshortxxxxxC}{\ensuremath{15.0}}      
\newcommand{\hatcurCCtassmBxxxxxC}{\ensuremath{16.378\pm0.040}} 
\newcommand{\hatcurCCtassmBshortxxxxxC}{\ensuremath{16.4}}      
\newcommand{\hatcurCCtassmIxxxxxC}{\ensuremath{nff\pmnff}}      
\newcommand{\hatcurCCtassmIshortxxxxxC}{\ensuremath{0.0}}       
\newcommand{\hatcurCCtassmgxxxxxC}{\ensuremath{15.668\pm0.040}} 
\newcommand{\hatcurCCtassmgshortxxxxxC}{\ensuremath{15.7}}      
\newcommand{\hatcurCCtassmrxxxxxC}{\ensuremath{14.496\pm0.010}} 
\newcommand{\hatcurCCtassmrshortxxxxxC}{\ensuremath{14.5}}      
\newcommand{\hatcurCCtassmixxxxxC}{\ensuremath{14.14\pm0.12}}   
\newcommand{\hatcurCCtassmishortxxxxxC}{\ensuremath{14.1}}      
\newcommand{\hatcurCCparallaxxxxxxC}{\ensuremath{3.054\pm0.022}} 
\newcommand{\hatcurCCgaiamGxxxxxC}{\ensuremath{14.54490\pm0.00030}} 
\newcommand{\hatcurCCgaiamBPxxxxxC}{\ensuremath{15.2886\pm0.0022}} 
\newcommand{\hatcurCCgaiamRPxxxxxC}{\ensuremath{13.7214\pm0.0013}} 
\newcommand{\hatcurCCtwomassJmagxxxxxC}{\ensuremath{12.692\pm0.024}} 
\newcommand{\hatcurCCtwomassHmagxxxxxC}{\ensuremath{12.105\pm0.024}} 
\newcommand{\hatcurCCtwomassKmagxxxxxC}{\ensuremath{11.938\pm0.023}} 
\newcommand{\hatcurCCcitJmagxxxxxC}{\ensuremath{12.688\pm0.025}} 
\newcommand{\hatcurCCcitHmagxxxxxC}{\ensuremath{12.097\pm0.025}} 
\newcommand{\hatcurCCcitKmagxxxxxC}{\ensuremath{11.962\pm0.023}} 
\newcommand{\hatcurCCbbJmagxxxxxC}{\ensuremath{12.769\pm0.027}} 
\newcommand{\hatcurCCbbHmagxxxxxC}{\ensuremath{12.121\pm0.026}} 
\newcommand{\hatcurCCbbKmagxxxxxC}{\ensuremath{11.982\pm0.023}} 
\newcommand{\hatcurCCesoJmagxxxxxC}{\ensuremath{12.776\pm0.031}} 
\newcommand{\hatcurCCesoHmagxxxxxC}{\ensuremath{12.118\pm0.035}} 
\newcommand{\hatcurCCesoKmagxxxxxC}{\ensuremath{11.979\pm0.025}} 
\newcommand{\hatcurCCesoJHmagxxxxxC}{\ensuremath{0.657\pm0.025}} 
\newcommand{\hatcurCCesoJKmagxxxxxC}{\ensuremath{0.797\pm0.038}} 
\newcommand{\hatcurCCesoHKmagxxxxxC}{\ensuremath{0.139\pm0.041}} 
\newcommand{\hatcurCCWonemagxxxxxC}{\ensuremath{11.903\pm0.023}} 
\newcommand{\hatcurCCWtwomagxxxxxC}{\ensuremath{11.990\pm0.022}} 
\newcommand{\hatcurCCWthreemagxxxxxC}{\ensuremath{0\pm0}}       
\newcommand{\hatcurCCWfourmagxxxxxC}{\ensuremath{0\pm0}}        
\newcommand{\hatcurLCdipxxxxxC}{\ensuremath{13.2}}              
\newcommand{\hatcurLCrprstarxxxxxC}{\ensuremath{0.1127\pm0.0015}} 
\newcommand{\hatcurLCbsqxxxxxC}{\ensuremath{0.176_{-0.028}^{+0.022}}} 
\newcommand{\hatcurLCimpxxxxxC}{\ensuremath{0.420_{-0.035}^{+0.025}}} 
\newcommand{\hatcurLCzetaxxxxxC}{\ensuremath{23.22\pm0.31}}     
\newcommand{\hatcurLCdurxxxxxC}{\ensuremath{0.09775\pm0.00095}} 
\newcommand{\hatcurLCdurshortxxxxxC}{\ensuremath{0.0977}}       
\newcommand{\hatcurLCdurhrxxxxxC}{\ensuremath{2.346\pm0.023}}   
\newcommand{\hatcurLCdurhrshortxxxxxC}{\ensuremath{2.346}}      
\newcommand{\hatcurLCqxxxxxC}{\ensuremath{0.02360\pm0.00023}}   
\newcommand{\hatcurLCqshortxxxxxC}{\ensuremath{0.024}}          
\newcommand{\hatcurLCingdurxxxxxC}{\ensuremath{0.01184\pm0.00036}} 
\newcommand{\hatcurLCPxxxxxC}{\ensuremath{4.1480467\pm0.0000037}} 
\newcommand{\hatcurLCPprecxxxxxC}{\ensuremath{4.1480467}}       
\newcommand{\hatcurLCPshortxxxxxC}{\ensuremath{4.1480}}         
\newcommand{\hatcurLCTxxxxxC}{\ensuremath{2457105.16480\pm0.00054}} 
\newcommand{\hatcurLCTAxxxxxC}{\ensuremath{2455761.1976\pm0.0015}} 
\newcommand{\hatcurLCTBxxxxxC}{\ensuremath{2457283.53081\pm0.00052}} 
\newcommand{\hatcurLChatnetmAxxxxxC}{\ensuremath{14.42634\pm0.00016}} 
\newcommand{\hatcurLCiblendAxxxxxC}{\ensuremath{0.811\pm0.049}} 
\newcommand{\hatcurLChatnetmBxxxxxC}{\ensuremath{14.42647\pm0.00021}} 
\newcommand{\hatcurLCiblendBxxxxxC}{\ensuremath{0.799\pm0.049}} 
\newcommand{\hatcurLChatnetmCxxxxxC}{\ensuremath{-0.000010\pm0.000090}} 
\newcommand{\hatcurLCiblendCxxxxxC}{\ensuremath{0.755\pm0.069}} 
\newcommand{\hatcurLCrhoxxxxxC}{\ensuremath{2.961\pm0.073}}     
\newcommand{\hatcurSMEiteffxxxxxC}{\ensuremath{4354\pm70}}      
\newcommand{\hatcurSMEizfehxxxxxC}{\ensuremath{0.080\pm0.084}}  
\newcommand{\hatcurSMEizfehshortxxxxxC}{\ensuremath{0.08}}      
\newcommand{\hatcurSMEiloggxxxxxC}{\ensuremath{4.55\pm0.19}}    
\newcommand{\hatcurSMEivsinxxxxxC}{\ensuremath{0.50\pm0.78}}    
\newcommand{\hatcurSMEivmacxxxxxC}{\ensuremath{1.80\pm0.11}}      
\newcommand{\hatcurSMEivmicxxxxxC}{\ensuremath{0.197\pm0.076}}      
\newcommand{\hatcurextraerrMgxxxxxC}{\ensuremath{0.0040_{-0.0038}^{+0.0209}}} 
\newcommand{\hatcurextraerrMgtwosiglimxxxxxC}{\ensuremath{<0.0528}} 
\newcommand{\hatcurextraerrMrxxxxxC}{\ensuremath{0.025\pm0.023}} 
\newcommand{\hatcurextraerrMrtwosiglimxxxxxC}{\ensuremath{<0.0788}} 
\newcommand{\hatcurextraerrMixxxxxC}{\ensuremath{0.068_{-0.065}^{+0.184}}} 
\newcommand{\hatcurextraerrMitwosiglimxxxxxC}{\ensuremath{<0.3698}} 
\newcommand{\hatcurextraerrMGxxxxxC}{\ensuremath{0.067_{-0.031}^{+0.054}}} 
\newcommand{\hatcurextraerrMGtwosiglimxxxxxC}{\ensuremath{<0.1491}} 
\newcommand{\hatcurextraerrMBPtwoxxxxxC}{\ensuremath{0.017\pm0.013}} 
\newcommand{\hatcurextraerrMBPtwotwosiglimxxxxxC}{\ensuremath{<0.0408}} 
\newcommand{\hatcurextraerrMRPxxxxxC}{\ensuremath{0.012\pm0.011}} 
\newcommand{\hatcurextraerrMRPtwosiglimxxxxxC}{\ensuremath{<0.0312}} 
\newcommand{\hatcurextraerrMJxxxxxC}{\ensuremath{0.0040_{-0.0039}^{+0.0171}}} 
\newcommand{\hatcurextraerrMJtwosiglimxxxxxC}{\ensuremath{<0.0337}} 
\newcommand{\hatcurextraerrMHxxxxxC}{\ensuremath{0.038_{-0.024}^{+0.033}}} 
\newcommand{\hatcurextraerrMHtwosiglimxxxxxC}{\ensuremath{<0.0972}} 
\newcommand{\hatcurextraerrMKsxxxxxC}{\ensuremath{0.028\pm0.021}} 
\newcommand{\hatcurextraerrMKstwosiglimxxxxxC}{\ensuremath{<0.0642}} 
\newcommand{\hatcurextraerrMWonexxxxxC}{\ensuremath{0.087_{-0.026}^{+0.035}}} 
\newcommand{\hatcurextraerrMWonetwosiglimxxxxxC}{\ensuremath{<0.1391}} 
\newcommand{\hatcurextraerrMWtwoxxxxxC}{\ensuremath{0.021_{-0.020}^{+0.055}}} 
\newcommand{\hatcurextraerrMWtwotwosiglimxxxxxC}{\ensuremath{<0.1182}} 
\newcommand{\hatcurLBiBxxxxxC}{\ensuremath{0.9623}}             
\newcommand{\hatcurLBiiBxxxxxC}{\ensuremath{-0.1066}}           
\newcommand{\hatcurLBiVxxxxxC}{\ensuremath{0.7587}}             
\newcommand{\hatcurLBiiVxxxxxC}{\ensuremath{0.0458}}            
\newcommand{\hatcurLBiRxxxxxC}{\ensuremath{0.6243}}             
\newcommand{\hatcurLBiiRxxxxxC}{\ensuremath{0.1118}}            
\newcommand{\hatcurLBiIxxxxxC}{\ensuremath{0.4811}}             
\newcommand{\hatcurLBiiIxxxxxC}{\ensuremath{0.1553}}            
\newcommand{\hatcurLBiuxxxxxC}{\ensuremath{1.0925}}             
\newcommand{\hatcurLBiiuxxxxxC}{\ensuremath{-0.2541}}           
\newcommand{\hatcurLBigxxxxxC}{\ensuremath{0.8657}}             
\newcommand{\hatcurLBiigxxxxxC}{\ensuremath{-0.0410}}           
\newcommand{\hatcurLBirxxxxxC}{\ensuremath{0.67\pm0.12}}        
\newcommand{\hatcurLBiirxxxxxC}{\ensuremath{0.16\pm0.17}}       
\newcommand{\hatcurLBiixxxxxC}{\ensuremath{0.34\pm0.11}}        
\newcommand{\hatcurLBiiixxxxxC}{\ensuremath{0.06\pm0.14}}       
\newcommand{\hatcurLBizxxxxxC}{\ensuremath{0.4191}}             
\newcommand{\hatcurLBiizxxxxxC}{\ensuremath{0.1681}}            
\newcommand{\hatcurLBiJxxxxxC}{\ensuremath{0.2967}}             
\newcommand{\hatcurLBiiJxxxxxC}{\ensuremath{0.2133}}            
\newcommand{\hatcurLBiHxxxxxC}{\ensuremath{0.1664}}             
\newcommand{\hatcurLBiiHxxxxxC}{\ensuremath{0.2985}}            
\newcommand{\hatcurLBiKxxxxxC}{\ensuremath{0.1330}}             
\newcommand{\hatcurLBiiKxxxxxC}{\ensuremath{0.2539}}            
\newcommand{\hatcurLBiTxxxxxC}{\ensuremath{0.56\pm0.17}}        
\newcommand{\hatcurLBiiTxxxxxC}{\ensuremath{0.24\pm0.15}}       
\newcommand{\hatcurLBikepxxxxxC}{\ensuremath{0.6246}}           
\newcommand{\hatcurLBiikepxxxxxC}{\ensuremath{0.1226}}          
\newcommand{\hatcurLBiCxxxxxC}{\ensuremath{0.5961}}             
\newcommand{\hatcurLBiiCxxxxxC}{\ensuremath{0.1316}}            
\newcommand{\hatcurLBiMxxxxxC}{\ensuremath{0.7290}}             
\newcommand{\hatcurLBiiMxxxxxC}{\ensuremath{0.0691}}            
\newcommand{\hatcurLBiSonexxxxxC}{\ensuremath{0.1108}}            
\newcommand{\hatcurLBiiSonexxxxxC}{\ensuremath{0.1733}}           
\newcommand{\hatcurLBiStwoxxxxxC}{\ensuremath{0.0986}}            
\newcommand{\hatcurLBiiStwoxxxxxC}{\ensuremath{0.1376}}           
\newcommand{\hatcurLBiSthreexxxxxC}{\ensuremath{0.0828}}            
\newcommand{\hatcurLBiiSthreexxxxxC}{\ensuremath{0.1135}}           
\newcommand{\hatcurLBiSfourxxxxxC}{\ensuremath{0.0731}}            
\newcommand{\hatcurLBiiSfourxxxxxC}{\ensuremath{0.0967}}           
\newcommand{\hatcurISOmxxxxxC}{\ensuremath{0.7133\pm0.0075}}    
\newcommand{\hatcurISOmshortxxxxxC}{\ensuremath{0.71}}          
\newcommand{\hatcurISOmlongxxxxxC}{\ensuremath{0.7133\pm0.0075}} 
\newcommand{\hatcurISOrxxxxxC}{\ensuremath{0.6977\pm0.0055}}    
\newcommand{\hatcurISOrshortxxxxxC}{\ensuremath{0.70}}          
\newcommand{\hatcurISOrlongxxxxxC}{\ensuremath{0.6977\pm0.0055}} 
\newcommand{\hatcurISOrhoxxxxxC}{\ensuremath{2.961\pm0.073}}    
\newcommand{\hatcurISOrholongxxxxxC}{\ensuremath{2.961\pm0.073}} 
\newcommand{\hatcurISOloggxxxxxC}{\ensuremath{4.6036\pm0.0077}} 
\newcommand{\hatcurISOlumxxxxxC}{\ensuremath{0.1641\pm0.0026}}  
\newcommand{\hatcurISOlumshortxxxxxC}{\ensuremath{0.16}}        
\newcommand{\hatcurISOteffxxxxxC}{\ensuremath{4405\pm15}}       
\newcommand{\hatcurISOzfehxxxxxC}{\ensuremath{0.208\pm0.053}}   
\newcommand{\hatcurISOagexxxxxC}{\ensuremath{10.5_{-2.0}^{+1.4}}} 
\newcommand{\hatcurISOspecxxxxxC}{K}                            
\newcommand{\hatcurRVKxxxxxC}{\ensuremath{55.9\pm5.5}}          
\newcommand{\hatcurRVKtwosiglimxxxxxC}{\ensuremath{<66.3}}      
\newcommand{\hatcurRVrkxxxxxC}{\ensuremath{0\pm0}}              
\newcommand{\hatcurRVrhxxxxxC}{\ensuremath{0\pm0}}              
\newcommand{\hatcurRVkxxxxxC}{\ensuremath{0\pm0}}               
\newcommand{\hatcurRVhxxxxxC}{\ensuremath{0\pm0}}               
\newcommand{\hatcurRVtronexxxxxC}{\ensuremath{0\pm0}}           
\newcommand{\hatcurRVtrtwoxxxxxC}{\ensuremath{0\pm0}}           
\newcommand{\hatcurRVgammaxxxxxC}{\ensuremath{-20.4\pm4.5}}     
\newcommand{\hatcurRVjitterxxxxxC}{\ensuremath{57\pm13}}        
\newcommand{\hatcurRVjittertwosiglimxxxxxC}{\ensuremath{<81.7}} 
\newcommand{\hatcurRVfitrmsxxxxxC}{\ensuremath{.1fym}}          %
\newcommand{\hatcurRVeccenxxxxxC}{\ensuremath{0\pm0}}           
\newcommand{\hatcurRVeccentwosiglimxxxxxC}{\ensuremath{<0.000}} 
\newcommand{\hatcurRVomegaxxxxxC}{\ensuremath{0\pm0}}           
\newcommand{\hatcurPPixxxxxC}{\ensuremath{88.27_{-0.11}^{+0.16}}} 
\newcommand{\hatcurPPgxxxxxC}{\ensuremath{15.0_{-1.2}^{+1.5}}}  
\newcommand{\hatcurPPloggxxxxxC}{\ensuremath{3.177\pm0.040}}    
\newcommand{\hatcurPParxxxxxC}{\ensuremath{13.91\pm0.11}}       
\newcommand{\hatcurPParelxxxxxC}{\ensuremath{0.04515\pm0.00016}} 
\newcommand{\hatcurPPrhoxxxxxC}{\ensuremath{0.986\pm0.094}}     
\newcommand{\hatcurPPmxxxxxC}{\ensuremath{0.353_{-0.027}^{+0.038}}} 
\newcommand{\hatcurPPmtwosiglimxxxxxC}{\ensuremath{<0.42}}      
\newcommand{\hatcurPPmshortxxxxxC}{\ensuremath{0.35}}           
\newcommand{\hatcurPPmlongxxxxxC}{\ensuremath{0.353_{-0.027}^{+0.038}}} 
\newcommand{\hatcurPPmexxxxxC}{\ensuremath{112.2_{-8.6}^{+12.2}}} 
\newcommand{\hatcurPPmeshortxxxxxC}{\ensuremath{112.2}}         
\newcommand{\hatcurPPmelongxxxxxC}{\ensuremath{112.2_{-8.6}^{+12.2}}} 
\newcommand{\hatcurPPrxxxxxC}{\ensuremath{0.765\pm0.013}}       
\newcommand{\hatcurPPrshortxxxxxC}{\ensuremath{0.77}}           
\newcommand{\hatcurPPrlongxxxxxC}{\ensuremath{0.765\pm0.013}}   
\newcommand{\hatcurPPrexxxxxC}{\ensuremath{8.58\pm0.14}}        
\newcommand{\hatcurPPreshortxxxxxC}{\ensuremath{8.6}}           
\newcommand{\hatcurPPrelongxxxxxC}{\ensuremath{8.58\pm0.14}}    
\newcommand{\hatcurPPmrcorrxxxxxC}{\ensuremath{0.31}}           
\newcommand{\hatcurPPteffxxxxxC}{\ensuremath{834.8\pm3.6}}      
\newcommand{\hatcurPPthetaxxxxxC}{\ensuremath{0.0585_{-0.0045}^{+0.0059}}} 
\newcommand{\hatcurPPfluxperixxxxxC}{\ensuremath{1.095\pm0.018}} 
\newcommand{\hatcurPPfluxperidimxxxxxC}{\ensuremath{8}}         
\newcommand{\hatcurPPfluxapxxxxxC}{\ensuremath{1.095\pm0.018}}  
\newcommand{\hatcurPPfluxapdimxxxxxC}{\ensuremath{8}}           
\newcommand{\hatcurPPfluxavgxxxxxC}{\ensuremath{1.095\pm0.018}} 
\newcommand{\hatcurPPfluxavgdimxxxxxC}{\ensuremath{8}}          
\newcommand{\hatcurPPfluxavglogxxxxxC}{\ensuremath{8.0394\pm0.0073}} 
\newcommand{\hatcurXsecphasexxxxxC}{\ensuremath{0\pm0}}         
\newcommand{\hatcurXsecondaryxxxxxC}{\ensuremath{2457107.23882\pm0.00054}} 
\newcommand{\hatcurXsecdurxxxxxC}{\ensuremath{0.09775\pm0.00095}} 
\newcommand{\hatcurXsecingdurxxxxxC}{\ensuremath{0.01184\pm0.00036}} 
\newcommand{\hatcurPPphiconjxxxxxC}{\ensuremath{0\pm0}}         
\newcommand{\hatcurPPperixxxxxC}{\ensuremath{2457104.12779\pm0.00054}} 
\newcommand{\hatcurPPaequivxxxxxC}{\ensuremath{0.11150\pm0.00095}} 
\newcommand{\hatcurPPtcircxxxxxC}{\ensuremath{1570\pm180}}      
\newcommand{\hatcurPPtinfallxxxxxC}{\ensuremath{41400\pm4200}}  
\newcommand{\hatcurXdistxxxxxC}{\ensuremath{324.6\pm2.2}}       
\newcommand{\hatcurXAvxxxxxC}{\ensuremath{0.046\pm0.014}}       
\newcommand{\hatcurXdistredxxxxxC}{\ensuremath{324.6\pm2.2}}    
\newcommand{\hatcurXEBVxxxxxC}{\ensuremath{0.0150\pm0.0045}}    
\newcommand{\hatcurCCpmraxxxxxC}{\ensuremath{42.581\pm0.035}}   
\newcommand{\hatcurCCpmdecxxxxxC}{\ensuremath{8.264\pm0.030}}   
\newcommand{\hatcurCCpmxxxxxC}{\ensuremath{43.376\pm0.046}}     
\newcommand{\hatcurhtrxxxxxD}{HATS537-014}                      
\newcommand{\hatcurfieldxxxxxD}{\ensuremath{string}}            
\newcommand{\hatcurCCraxxxxxD}{\ensuremath{22^{\mathrm h}36^{\mathrm m}06.3190{\mathrm s}}}                   
\newcommand{\hatcurCCdecxxxxxD}{\ensuremath{-16{\arcdeg}59{\arcmin}59.7882{\arcsec}}}                 
\newcommand{\hatcurCCmagxxxxxD}{12.469}                         
\newcommand{\hatcurCCtwomassxxxxxD}{2MASS~22360631-1659597}     
\newcommand{\hatcurCCgscxxxxxD}{GSC~6386-00784}                 
\newcommand{\hatcurCCgaiaxxxxxD}{GAIA~2594869603582993792}      
\newcommand{\hatcurCCgaiadrtwoxxxxxD}{2594869603582993792} 
\newcommand{\hatcurCCtassmvxxxxxD}{\ensuremath{12.469\pm0.010}} 
\newcommand{\hatcurCCtassmvshortxxxxxD}{\ensuremath{12.5}}      
\newcommand{\hatcurCCtassmBxxxxxD}{\ensuremath{13.572\pm0.030}} 
\newcommand{\hatcurCCtassmBshortxxxxxD}{\ensuremath{13.6}}      
\newcommand{\hatcurCCtassmIxxxxxD}{\ensuremath{nff\pmnff}}      
\newcommand{\hatcurCCtassmIshortxxxxxD}{\ensuremath{0.0}}       
\newcommand{\hatcurCCtassmgxxxxxD}{\ensuremath{12.995\pm0.010}} 
\newcommand{\hatcurCCtassmgshortxxxxxD}{\ensuremath{13.0}}      
\newcommand{\hatcurCCtassmrxxxxxD}{\ensuremath{11.998\pm0.010}} 
\newcommand{\hatcurCCtassmrshortxxxxxD}{\ensuremath{12.0}}      
\newcommand{\hatcurCCtassmixxxxxD}{\ensuremath{11.622\pm0.030}} 
\newcommand{\hatcurCCtassmishortxxxxxD}{\ensuremath{11.6}}      
\newcommand{\hatcurCCparallaxxxxxxD}{\ensuremath{7.809\pm0.037}} 
\newcommand{\hatcurCCgaiamGxxxxxD}{\ensuremath{12.07250\pm0.00030}} 
\newcommand{\hatcurCCgaiamBPxxxxxD}{\ensuremath{12.7084\pm0.0018}} 
\newcommand{\hatcurCCgaiamRPxxxxxD}{\ensuremath{11.3341\pm0.0010}} 
\newcommand{\hatcurCCtwomassJmagxxxxxD}{\ensuremath{10.424\pm0.023}} 
\newcommand{\hatcurCCtwomassHmagxxxxxD}{\ensuremath{9.907\pm0.026}} 
\newcommand{\hatcurCCtwomassKmagxxxxxD}{\ensuremath{9.764\pm0.021}} 
\newcommand{\hatcurCCcitJmagxxxxxD}{\ensuremath{10.425\pm0.024}} 
\newcommand{\hatcurCCcitHmagxxxxxD}{\ensuremath{9.900\pm0.027}} 
\newcommand{\hatcurCCcitKmagxxxxxD}{\ensuremath{9.788\pm0.022}} 
\newcommand{\hatcurCCbbJmagxxxxxD}{\ensuremath{10.498\pm0.026}} 
\newcommand{\hatcurCCbbHmagxxxxxD}{\ensuremath{9.923\pm0.027}}  
\newcommand{\hatcurCCbbKmagxxxxxD}{\ensuremath{9.808\pm0.022}}  
\newcommand{\hatcurCCesoJmagxxxxxD}{\ensuremath{10.504\pm0.029}} 
\newcommand{\hatcurCCesoHmagxxxxxD}{\ensuremath{9.920\pm0.034}} 
\newcommand{\hatcurCCesoKmagxxxxxD}{\ensuremath{9.806\pm0.023}} 
\newcommand{\hatcurCCesoJHmagxxxxxD}{\ensuremath{0.584\pm0.042}} 
\newcommand{\hatcurCCesoJKmagxxxxxD}{\ensuremath{0.699\pm0.035}} 
\newcommand{\hatcurCCesoHKmagxxxxxD}{\ensuremath{0.115\pm0.040}} 
\newcommand{\hatcurCCWonemagxxxxxD}{\ensuremath{9.687\pm0.024}} 
\newcommand{\hatcurCCWtwomagxxxxxD}{\ensuremath{9.772\pm0.020}} 
\newcommand{\hatcurCCWthreemagxxxxxD}{\ensuremath{9.675\pm0.043}} 
\newcommand{\hatcurCCWfourmagxxxxxD}{\ensuremath{0\pm0}}        
\newcommand{\hatcurLCdipxxxxxD}{\ensuremath{13.4}}              
\newcommand{\hatcurLCrprstarxxxxxD}{\ensuremath{0.10290\pm0.00034}} 
\newcommand{\hatcurLCbsqxxxxxD}{\ensuremath{0.0233_{-0.0066}^{+0.0054}}} 
\newcommand{\hatcurLCimpxxxxxD}{\ensuremath{0.153_{-0.023}^{+0.017}}} 
\newcommand{\hatcurLCzetaxxxxxD}{\ensuremath{17.195\pm0.049}}   
\newcommand{\hatcurLCdurxxxxxD}{\ensuremath{0.12853\pm0.00030}} 
\newcommand{\hatcurLCdurshortxxxxxD}{\ensuremath{0.1285}}       
\newcommand{\hatcurLCdurhrxxxxxD}{\ensuremath{3.0848\pm0.0072}} 
\newcommand{\hatcurLCdurhrshortxxxxxD}{\ensuremath{3.085}}      
\newcommand{\hatcurLCqxxxxxD}{\ensuremath{0.017500\pm0.000065}} 
\newcommand{\hatcurLCqshortxxxxxD}{\ensuremath{0.018}}          
\newcommand{\hatcurLCingdurxxxxxD}{\ensuremath{0.012258\pm0.000080}} 
\newcommand{\hatcurLCPxxxxxD}{\ensuremath{7.3279474\pm0.0000016}} 
\newcommand{\hatcurLCPprecxxxxxD}{\ensuremath{7.3279474}}       
\newcommand{\hatcurLCPshortxxxxxD}{\ensuremath{7.3279}}         
\newcommand{\hatcurLCTxxxxxD}{\ensuremath{2458087.647820\pm0.000075}} 
\newcommand{\hatcurLCTAxxxxxD}{\ensuremath{2457552.70766\pm0.00013}} 
\newcommand{\hatcurLCTBxxxxxD}{\ensuremath{2458380.765711\pm0.000099}} 
\newcommand{\hatcurLChatnetmAxxxxxD}{\ensuremath{12.049830\pm0.000052}} 
\newcommand{\hatcurLCiblendAxxxxxD}{\ensuremath{0.868\pm0.022}} 
\newcommand{\hatcurLChatnetmBxxxxxD}{\ensuremath{11.628240\pm0.000023}} 
\newcommand{\hatcurLCiblendBxxxxxD}{\ensuremath{0.97763\pm0.00052}} 
\newcommand{\hatcurLChatnetmCxxxxxD}{\ensuremath{0.000200\pm0.000095}} 
\newcommand{\hatcurLCiblendCxxxxxD}{\ensuremath{0.984\pm0.012}} 
\newcommand{\hatcurLCrhoxxxxxD}{\ensuremath{2.743\pm0.020}}     
\newcommand{\hatcurSMEiteffxxxxxD}{\ensuremath{4612\pm76}}      
\newcommand{\hatcurSMEizfehxxxxxD}{\ensuremath{-0.040\pm0.050}} 
\newcommand{\hatcurSMEizfehshortxxxxxD}{\ensuremath{-0.04}}     
\newcommand{\hatcurSMEiloggxxxxxD}{\ensuremath{4.55\pm0.16}}    
\newcommand{\hatcurSMEivsinxxxxxD}{\ensuremath{0.8\pm1.3}}      
\newcommand{\hatcurSMEivmacxxxxxD}{\ensuremath{2.20\pm0.12}}    
\newcommand{\hatcurSMEivmicxxxxxD}{\ensuremath{0.443\pm0.061}}  
\newcommand{\hatcurextraerrMBxxxxxD}{\ensuremath{0.00101_{-0.00098}^{+0.00384}}} 
\newcommand{\hatcurextraerrMBtwosiglimxxxxxD}{\ensuremath{<0.0084}} 
\newcommand{\hatcurextraerrMVxxxxxD}{\ensuremath{0.000082_{-0.000080}^{+0.000612}}} 
\newcommand{\hatcurextraerrMVtwosiglimxxxxxD}{\ensuremath{<0.0013}} 
\newcommand{\hatcurextraerrMgxxxxxD}{\ensuremath{0.046\pm0.020}} 
\newcommand{\hatcurextraerrMgtwosiglimxxxxxD}{\ensuremath{<0.0793}} 
\newcommand{\hatcurextraerrMrxxxxxD}{\ensuremath{0.0040_{-0.0040}^{+0.0136}}} 
\newcommand{\hatcurextraerrMrtwosiglimxxxxxD}{\ensuremath{<0.0288}} 
\newcommand{\hatcurextraerrMixxxxxD}{\ensuremath{0.063\pm0.043}} 
\newcommand{\hatcurextraerrMitwosiglimxxxxxD}{\ensuremath{<0.1424}} 
\newcommand{\hatcurextraerrMGxxxxxD}{\ensuremath{0.0172\pm0.0054}} 
\newcommand{\hatcurextraerrMGtwosiglimxxxxxD}{\ensuremath{<0.0278}} 
\newcommand{\hatcurextraerrMBPtwoxxxxxD}{\ensuremath{0.0199\pm0.0074}} 
\newcommand{\hatcurextraerrMBPtwotwosiglimxxxxxD}{\ensuremath{<0.0302}} 
\newcommand{\hatcurextraerrMRPxxxxxD}{\ensuremath{0.00012_{-0.00012}^{+0.00065}}} 
\newcommand{\hatcurextraerrMRPtwosiglimxxxxxD}{\ensuremath{<0.0015}} 
\newcommand{\hatcurextraerrMJxxxxxD}{\ensuremath{0.0042_{-0.0042}^{+0.0061}}} 
\newcommand{\hatcurextraerrMJtwosiglimxxxxxD}{\ensuremath{<0.0148}} 
\newcommand{\hatcurextraerrMHxxxxxD}{\ensuremath{0.00070_{-0.00069}^{+0.00866}}} 
\newcommand{\hatcurextraerrMHtwosiglimxxxxxD}{\ensuremath{<0.0160}} 
\newcommand{\hatcurextraerrMKsxxxxxD}{\ensuremath{0.00088_{-0.00087}^{+0.00618}}} 
\newcommand{\hatcurextraerrMKstwosiglimxxxxxD}{\ensuremath{<0.0143}} 
\newcommand{\hatcurextraerrMWonexxxxxD}{\ensuremath{0.057\pm0.024}} 
\newcommand{\hatcurextraerrMWonetwosiglimxxxxxD}{\ensuremath{<0.0978}} 
\newcommand{\hatcurextraerrMWtwoxxxxxD}{\ensuremath{0.079_{-0.026}^{+0.018}}} 
\newcommand{\hatcurextraerrMWtwotwosiglimxxxxxD}{\ensuremath{<0.1093}} 
\newcommand{\hatcurextraerrMWthreexxxxxD}{\ensuremath{0.123\pm0.068}} 
\newcommand{\hatcurextraerrMWthreetwosiglimxxxxxD}{\ensuremath{<0.2440}} 
\newcommand{\hatcurLBiBxxxxxD}{\ensuremath{0.9774}}             
\newcommand{\hatcurLBiiBxxxxxD}{\ensuremath{-0.1216}}           
\newcommand{\hatcurLBiVxxxxxD}{\ensuremath{0.7619}}             
\newcommand{\hatcurLBiiVxxxxxD}{\ensuremath{0.0363}}            
\newcommand{\hatcurLBiRxxxxxD}{\ensuremath{0.6278}}             
\newcommand{\hatcurLBiiRxxxxxD}{\ensuremath{0.0963}}            
\newcommand{\hatcurLBiIxxxxxD}{\ensuremath{0.262\pm0.060}}      
\newcommand{\hatcurLBiiIxxxxxD}{\ensuremath{0.305\pm0.078}}     
\newcommand{\hatcurLBiuxxxxxD}{\ensuremath{1.1854}}             
\newcommand{\hatcurLBiiuxxxxxD}{\ensuremath{-0.3416}}           
\newcommand{\hatcurLBigxxxxxD}{\ensuremath{0.8805}}             
\newcommand{\hatcurLBiigxxxxxD}{\ensuremath{-0.0557}}           
\newcommand{\hatcurLBirxxxxxD}{\ensuremath{0.423_{-0.080}^{+0.061}}} 
\newcommand{\hatcurLBiirxxxxxD}{\ensuremath{0.508\pm0.088}}     
\newcommand{\hatcurLBiixxxxxD}{\ensuremath{0.316\pm0.059}}      
\newcommand{\hatcurLBiiixxxxxD}{\ensuremath{0.444\pm0.097}}     
\newcommand{\hatcurLBizxxxxxD}{\ensuremath{0.4296}}             
\newcommand{\hatcurLBiizxxxxxD}{\ensuremath{0.1493}}            
\newcommand{\hatcurLBiJxxxxxD}{\ensuremath{0.2991}}             
\newcommand{\hatcurLBiiJxxxxxD}{\ensuremath{0.2072}}            
\newcommand{\hatcurLBiHxxxxxD}{\ensuremath{0.1534}}             
\newcommand{\hatcurLBiiHxxxxxD}{\ensuremath{0.3100}}            
\newcommand{\hatcurLBiKxxxxxD}{\ensuremath{0.1318}}             
\newcommand{\hatcurLBiiKxxxxxD}{\ensuremath{0.2539}}            
\newcommand{\hatcurLBiTxxxxxD}{\ensuremath{0.503\pm0.060}}      
\newcommand{\hatcurLBiiTxxxxxD}{\ensuremath{0.403\pm0.084}}     
\newcommand{\hatcurLBikepxxxxxD}{\ensuremath{0.524\pm0.077}}    
\newcommand{\hatcurLBiikepxxxxxD}{\ensuremath{0.349\pm0.074}}   
\newcommand{\hatcurLBiCxxxxxD}{\ensuremath{0.6027}}             
\newcommand{\hatcurLBiiCxxxxxD}{\ensuremath{0.1141}}            
\newcommand{\hatcurLBiMxxxxxD}{\ensuremath{0.7303}}             
\newcommand{\hatcurLBiiMxxxxxD}{\ensuremath{0.0552}}            
\newcommand{\hatcurLBiSonexxxxxD}{\ensuremath{0.1161}}            
\newcommand{\hatcurLBiiSonexxxxxD}{\ensuremath{0.1617}}           
\newcommand{\hatcurLBiStwoxxxxxD}{\ensuremath{0.1020}}            
\newcommand{\hatcurLBiiStwoxxxxxD}{\ensuremath{0.1270}}           
\newcommand{\hatcurLBiSthreexxxxxD}{\ensuremath{0.0831}}            
\newcommand{\hatcurLBiiSthreexxxxxD}{\ensuremath{0.1042}}           
\newcommand{\hatcurLBiSfourxxxxxD}{\ensuremath{0.0692}}            
\newcommand{\hatcurLBiiSfourxxxxxD}{\ensuremath{0.0886}}           
\newcommand{\hatcurISOmxxxxxD}{\ensuremath{0.7311\pm0.0028}}    
\newcommand{\hatcurISOmshortxxxxxD}{\ensuremath{0.73}}          
\newcommand{\hatcurISOmlongxxxxxD}{\ensuremath{0.7311\pm0.0028}} 
\newcommand{\hatcurISOrxxxxxD}{\ensuremath{0.7214\pm0.0021}}    
\newcommand{\hatcurISOrshortxxxxxD}{\ensuremath{0.72}}          
\newcommand{\hatcurISOrlongxxxxxD}{\ensuremath{0.7214\pm0.0021}} 
\newcommand{\hatcurISOrhoxxxxxD}{\ensuremath{2.743\pm0.020}}    
\newcommand{\hatcurISOrholongxxxxxD}{\ensuremath{2.743\pm0.020}} 
\newcommand{\hatcurISOloggxxxxxD}{\ensuremath{4.5853\pm0.0021}} 
\newcommand{\hatcurISOlumxxxxxD}{\ensuremath{0.2193\pm0.0021}}  
\newcommand{\hatcurISOlumshortxxxxxD}{\ensuremath{0.22}}        
\newcommand{\hatcurISOteffxxxxxD}{\ensuremath{4656.1\pm8.9}}    
\newcommand{\hatcurISOzfehxxxxxD}{\ensuremath{0.099\pm0.014}}   
\newcommand{\hatcurISOagexxxxxD}{\ensuremath{12.17_{-0.45}^{+0.24}}} 
\newcommand{\hatcurISOspecxxxxxD}{K}                            
\newcommand{\hatcurRVKxxxxxD}{\ensuremath{16.15\pm0.51}}        
\newcommand{\hatcurRVKtwosiglimxxxxxD}{\ensuremath{<17.1}}      
\newcommand{\hatcurRVrkxxxxxD}{\ensuremath{0\pm0}}              
\newcommand{\hatcurRVrhxxxxxD}{\ensuremath{0\pm0}}              
\newcommand{\hatcurRVkxxxxxD}{\ensuremath{0\pm0}}               
\newcommand{\hatcurRVhxxxxxD}{\ensuremath{0\pm0}}               
\newcommand{\hatcurRVtronexxxxxD}{\ensuremath{0\pm0}}           
\newcommand{\hatcurRVtrtwoxxxxxD}{\ensuremath{0\pm0}}           
\newcommand{\hatcurRVgammaAxxxxxD}{\ensuremath{15945.8\pm3.3}}  
\newcommand{\hatcurRVjitterAxxxxxD}{\ensuremath{18.7\pm6.3}}    
\newcommand{\hatcurRVjittertwosiglimAxxxxxD}{\ensuremath{<29.9}} 
\newcommand{\hatcurRVfitrmsAxxxxxD}{\ensuremath{0.0}}           
\newcommand{\hatcurRVgammaBxxxxxD}{\ensuremath{15954.8\pm1.8}}  
\newcommand{\hatcurRVjitterBxxxxxD}{\ensuremath{1.13\pm0.42}}   
\newcommand{\hatcurRVjittertwosiglimBxxxxxD}{\ensuremath{<1.8}} 
\newcommand{\hatcurRVfitrmsBxxxxxD}{\ensuremath{0.0}}           
\newcommand{\hatcurRVgammaCxxxxxD}{\ensuremath{-0.04\pm0.55}}   
\newcommand{\hatcurRVjitterCxxxxxD}{\ensuremath{6.4\pm1.7}}     
\newcommand{\hatcurRVjittertwosiglimCxxxxxD}{\ensuremath{<10.1}} 
\newcommand{\hatcurRVfitrmsCxxxxxD}{\ensuremath{0.0}}           
\newcommand{\hatcurRVgammaDxxxxxD}{\ensuremath{15996.51\pm0.18}} 
\newcommand{\hatcurRVjitterDxxxxxD}{\ensuremath{8.25\pm0.32}}   
\newcommand{\hatcurRVjittertwosiglimDxxxxxD}{\ensuremath{<8.9}} 
\newcommand{\hatcurRVfitrmsDxxxxxD}{\ensuremath{.1fym}}         %
\newcommand{\hatcurRVeccenxxxxxD}{\ensuremath{0\pm0}}           
\newcommand{\hatcurRVeccentwosiglimxxxxxD}{\ensuremath{<0.000}} 
\newcommand{\hatcurRVomegaxxxxxD}{\ensuremath{0\pm0}}           
\newcommand{\hatcurPPixxxxxD}{\ensuremath{89.560_{-0.050}^{+0.070}}} 
\newcommand{\hatcurPPgxxxxxD}{\ensuremath{5.95\pm0.18}}         
\newcommand{\hatcurPPloggxxxxxD}{\ensuremath{2.774\pm0.013}}    
\newcommand{\hatcurPParxxxxxD}{\ensuremath{19.821\pm0.048}}     
\newcommand{\hatcurPParelxxxxxD}{\ensuremath{0.066517\pm0.000085}} 
\newcommand{\hatcurPPrhoxxxxxD}{\ensuremath{0.4110_{-0.0100}^{+0.0150}}} 
\newcommand{\hatcurPPmxxxxxD}{\ensuremath{0.1254\pm0.0039}}     
\newcommand{\hatcurPPmtwosiglimxxxxxD}{\ensuremath{<0.13}}      
\newcommand{\hatcurPPmshortxxxxxD}{\ensuremath{0.13}}           
\newcommand{\hatcurPPmlongxxxxxD}{\ensuremath{0.1254\pm0.0039}} 
\newcommand{\hatcurPPmexxxxxD}{\ensuremath{39.9\pm1.3}}         
\newcommand{\hatcurPPmeshortxxxxxD}{\ensuremath{39.9}}          
\newcommand{\hatcurPPmelongxxxxxD}{\ensuremath{39.9\pm1.3}}     
\newcommand{\hatcurPPrxxxxxD}{\ensuremath{0.7224\pm0.0032}}     
\newcommand{\hatcurPPrshortxxxxxD}{\ensuremath{0.72}}           
\newcommand{\hatcurPPrlongxxxxxD}{\ensuremath{0.7224\pm0.0032}} 
\newcommand{\hatcurPPrexxxxxD}{\ensuremath{8.097\pm0.036}}      
\newcommand{\hatcurPPreshortxxxxxD}{\ensuremath{8.1}}           
\newcommand{\hatcurPPrelongxxxxxD}{\ensuremath{8.097\pm0.036}}  
\newcommand{\hatcurPPmrcorrxxxxxD}{\ensuremath{0.30}}           
\newcommand{\hatcurPPteffxxxxxD}{\ensuremath{739.3\pm1.6}}      
\newcommand{\hatcurPPthetaxxxxxD}{\ensuremath{0.03150\pm0.00097}} 
\newcommand{\hatcurPPfluxperixxxxxD}{\ensuremath{6.749\pm0.059}} 
\newcommand{\hatcurPPfluxperidimxxxxxD}{\ensuremath{7}}         
\newcommand{\hatcurPPfluxapxxxxxD}{\ensuremath{6.749\pm0.059}}  
\newcommand{\hatcurPPfluxapdimxxxxxD}{\ensuremath{7}}           
\newcommand{\hatcurPPfluxavgxxxxxD}{\ensuremath{6.749\pm0.059}} 
\newcommand{\hatcurPPfluxavgdimxxxxxD}{\ensuremath{7}}          
\newcommand{\hatcurPPfluxavglogxxxxxD}{\ensuremath{7.8292\pm0.0038}} 
\newcommand{\hatcurXsecphasexxxxxD}{\ensuremath{0\pm0}}         
\newcommand{\hatcurXsecondaryxxxxxD}{\ensuremath{2458091.311790\pm0.000075}} 
\newcommand{\hatcurXsecdurxxxxxD}{\ensuremath{0.12853\pm0.00030}} 
\newcommand{\hatcurXsecingdurxxxxxD}{\ensuremath{0.012258\pm0.000080}} 
\newcommand{\hatcurPPphiconjxxxxxD}{\ensuremath{0\pm0}}         
\newcommand{\hatcurPPperixxxxxD}{\ensuremath{2458085.815832\pm0.000075}} 
\newcommand{\hatcurPPaequivxxxxxD}{\ensuremath{0.14200\pm0.00062}} 
\newcommand{\hatcurPPtcircxxxxxD}{\ensuremath{8780_{-230}^{+330}}} 
\newcommand{\hatcurPPtinfallxxxxxD}{\ensuremath{1245000\pm41000}} 
\newcommand{\hatcurXdistxxxxxD}{\ensuremath{127.70\pm0.52}}     
\newcommand{\hatcurXAvxxxxxD}{\ensuremath{0.0270\pm0.0080}}     
\newcommand{\hatcurXdistredxxxxxD}{\ensuremath{127.66\pm0.52}}  
\newcommand{\hatcurXEBVxxxxxD}{\ensuremath{0.0090_{-0.0030}^{+0.0020}}} 
\newcommand{\hatcurCCpmraxxxxxD}{\ensuremath{-108.621\pm0.090}} 
\newcommand{\hatcurCCpmdecxxxxxD}{\ensuremath{-84.412\pm0.078}} 
\newcommand{\hatcurCCpmxxxxxD}{\ensuremath{137.56\pm0.12}}      
\newcommand{\hatcurCCbbHmag}[1]{\ifnum#1=47 %
\hatcurCCbbHmagxxxxxA
\else
\ifnum#1=48 %
\hatcurCCbbHmagxxxxxB
\else
\ifnum#1=49 %
\hatcurCCbbHmagxxxxxC
\else
\ifnum#1=72 %
\hatcurCCbbHmagxxxxxD
\else
??????\fi
\fi
\fi
\fi
}
\newcommand{\hatcurCCbbJmag}[1]{\ifnum#1=47 %
\hatcurCCbbJmagxxxxxA
\else
\ifnum#1=48 %
\hatcurCCbbJmagxxxxxB
\else
\ifnum#1=49 %
\hatcurCCbbJmagxxxxxC
\else
\ifnum#1=72 %
\hatcurCCbbJmagxxxxxD
\else
??????\fi
\fi
\fi
\fi
}
\newcommand{\hatcurCCbbKmag}[1]{\ifnum#1=47 %
\hatcurCCbbKmagxxxxxA
\else
\ifnum#1=48 %
\hatcurCCbbKmagxxxxxB
\else
\ifnum#1=49 %
\hatcurCCbbKmagxxxxxC
\else
\ifnum#1=72 %
\hatcurCCbbKmagxxxxxD
\else
??????\fi
\fi
\fi
\fi
}
\newcommand{\hatcurCCcitHmag}[1]{\ifnum#1=47 %
\hatcurCCcitHmagxxxxxA
\else
\ifnum#1=48 %
\hatcurCCcitHmagxxxxxB
\else
\ifnum#1=49 %
\hatcurCCcitHmagxxxxxC
\else
\ifnum#1=72 %
\hatcurCCcitHmagxxxxxD
\else
??????\fi
\fi
\fi
\fi
}
\newcommand{\hatcurCCcitJmag}[1]{\ifnum#1=47 %
\hatcurCCcitJmagxxxxxA
\else
\ifnum#1=48 %
\hatcurCCcitJmagxxxxxB
\else
\ifnum#1=49 %
\hatcurCCcitJmagxxxxxC
\else
\ifnum#1=72 %
\hatcurCCcitJmagxxxxxD
\else
??????\fi
\fi
\fi
\fi
}
\newcommand{\hatcurCCcitKmag}[1]{\ifnum#1=47 %
\hatcurCCcitKmagxxxxxA
\else
\ifnum#1=48 %
\hatcurCCcitKmagxxxxxB
\else
\ifnum#1=49 %
\hatcurCCcitKmagxxxxxC
\else
\ifnum#1=72 %
\hatcurCCcitKmagxxxxxD
\else
??????\fi
\fi
\fi
\fi
}
\newcommand{\hatcurCCdec}[1]{\ifnum#1=47 %
\hatcurCCdecxxxxxA
\else
\ifnum#1=48 %
\hatcurCCdecxxxxxB
\else
\ifnum#1=49 %
\hatcurCCdecxxxxxC
\else
\ifnum#1=72 %
\hatcurCCdecxxxxxD
\else
??????\fi
\fi
\fi
\fi
}
\newcommand{\hatcurCCesoHKmag}[1]{\ifnum#1=47 %
\hatcurCCesoHKmagxxxxxA
\else
\ifnum#1=48 %
\hatcurCCesoHKmagxxxxxB
\else
\ifnum#1=49 %
\hatcurCCesoHKmagxxxxxC
\else
\ifnum#1=72 %
\hatcurCCesoHKmagxxxxxD
\else
??????\fi
\fi
\fi
\fi
}
\newcommand{\hatcurCCesoHmag}[1]{\ifnum#1=47 %
\hatcurCCesoHmagxxxxxA
\else
\ifnum#1=48 %
\hatcurCCesoHmagxxxxxB
\else
\ifnum#1=49 %
\hatcurCCesoHmagxxxxxC
\else
\ifnum#1=72 %
\hatcurCCesoHmagxxxxxD
\else
??????\fi
\fi
\fi
\fi
}
\newcommand{\hatcurCCesoJHmag}[1]{\ifnum#1=47 %
\hatcurCCesoJHmagxxxxxA
\else
\ifnum#1=48 %
\hatcurCCesoJHmagxxxxxB
\else
\ifnum#1=49 %
\hatcurCCesoJHmagxxxxxC
\else
\ifnum#1=72 %
\hatcurCCesoJHmagxxxxxD
\else
??????\fi
\fi
\fi
\fi
}
\newcommand{\hatcurCCesoJKmag}[1]{\ifnum#1=47 %
\hatcurCCesoJKmagxxxxxA
\else
\ifnum#1=48 %
\hatcurCCesoJKmagxxxxxB
\else
\ifnum#1=49 %
\hatcurCCesoJKmagxxxxxC
\else
\ifnum#1=72 %
\hatcurCCesoJKmagxxxxxD
\else
??????\fi
\fi
\fi
\fi
}
\newcommand{\hatcurCCesoJmag}[1]{\ifnum#1=47 %
\hatcurCCesoJmagxxxxxA
\else
\ifnum#1=48 %
\hatcurCCesoJmagxxxxxB
\else
\ifnum#1=49 %
\hatcurCCesoJmagxxxxxC
\else
\ifnum#1=72 %
\hatcurCCesoJmagxxxxxD
\else
??????\fi
\fi
\fi
\fi
}
\newcommand{\hatcurCCesoKmag}[1]{\ifnum#1=47 %
\hatcurCCesoKmagxxxxxA
\else
\ifnum#1=48 %
\hatcurCCesoKmagxxxxxB
\else
\ifnum#1=49 %
\hatcurCCesoKmagxxxxxC
\else
\ifnum#1=72 %
\hatcurCCesoKmagxxxxxD
\else
??????\fi
\fi
\fi
\fi
}
\newcommand{\hatcurCCgaia}[1]{\ifnum#1=47 %
\hatcurCCgaiaxxxxxA
\else
\ifnum#1=48 %
\hatcurCCgaiaxxxxxB
\else
\ifnum#1=49 %
\hatcurCCgaiaxxxxxC
\else
\ifnum#1=72 %
\hatcurCCgaiaxxxxxD
\else
??????\fi
\fi
\fi
\fi
}
\newcommand{\hatcurCCgaiadrtwo}[1]{\ifnum#1=47 %
\hatcurCCgaiadrtwoxxxxxA
\else
\ifnum#1=48 %
\hatcurCCgaiadrtwoxxxxxB
\else
\ifnum#1=49 %
\hatcurCCgaiadrtwoxxxxxC
\else
\ifnum#1=72 %
\hatcurCCgaiadrtwoxxxxxD
\else
??????\fi
\fi
\fi
\fi
}
\newcommand{\hatcurCCgaiamBP}[1]{\ifnum#1=47 %
\hatcurCCgaiamBPxxxxxA
\else
\ifnum#1=48 %
\hatcurCCgaiamBPxxxxxB
\else
\ifnum#1=49 %
\hatcurCCgaiamBPxxxxxC
\else
\ifnum#1=72 %
\hatcurCCgaiamBPxxxxxD
\else
??????\fi
\fi
\fi
\fi
}
\newcommand{\hatcurCCgaiamG}[1]{\ifnum#1=47 %
\hatcurCCgaiamGxxxxxA
\else
\ifnum#1=48 %
\hatcurCCgaiamGxxxxxB
\else
\ifnum#1=49 %
\hatcurCCgaiamGxxxxxC
\else
\ifnum#1=72 %
\hatcurCCgaiamGxxxxxD
\else
??????\fi
\fi
\fi
\fi
}
\newcommand{\hatcurCCgaiamRP}[1]{\ifnum#1=47 %
\hatcurCCgaiamRPxxxxxA
\else
\ifnum#1=48 %
\hatcurCCgaiamRPxxxxxB
\else
\ifnum#1=49 %
\hatcurCCgaiamRPxxxxxC
\else
\ifnum#1=72 %
\hatcurCCgaiamRPxxxxxD
\else
??????\fi
\fi
\fi
\fi
}
\newcommand{\hatcurCCgsc}[1]{\ifnum#1=47 %
\hatcurCCgscxxxxxA
\else
\ifnum#1=48 %
\hatcurCCgscxxxxxB
\else
\ifnum#1=49 %
\hatcurCCgscxxxxxC
\else
\ifnum#1=72 %
\hatcurCCgscxxxxxD
\else
??????\fi
\fi
\fi
\fi
}
\newcommand{\hatcurCCmag}[1]{\ifnum#1=47 %
\hatcurCCmagxxxxxA
\else
\ifnum#1=48 %
\hatcurCCmagxxxxxB
\else
\ifnum#1=49 %
\hatcurCCmagxxxxxC
\else
\ifnum#1=72 %
\hatcurCCmagxxxxxD
\else
??????\fi
\fi
\fi
\fi
}
\newcommand{\hatcurCCparallax}[1]{\ifnum#1=47 %
\hatcurCCparallaxxxxxxA
\else
\ifnum#1=48 %
\hatcurCCparallaxxxxxxB
\else
\ifnum#1=49 %
\hatcurCCparallaxxxxxxC
\else
\ifnum#1=72 %
\hatcurCCparallaxxxxxxD
\else
??????\fi
\fi
\fi
\fi
}
\newcommand{\hatcurCCpm}[1]{\ifnum#1=47 %
\hatcurCCpmxxxxxA
\else
\ifnum#1=48 %
\hatcurCCpmxxxxxB
\else
\ifnum#1=49 %
\hatcurCCpmxxxxxC
\else
\ifnum#1=72 %
\hatcurCCpmxxxxxD
\else
??????\fi
\fi
\fi
\fi
}
\newcommand{\hatcurCCpmdec}[1]{\ifnum#1=47 %
\hatcurCCpmdecxxxxxA
\else
\ifnum#1=48 %
\hatcurCCpmdecxxxxxB
\else
\ifnum#1=49 %
\hatcurCCpmdecxxxxxC
\else
\ifnum#1=72 %
\hatcurCCpmdecxxxxxD
\else
??????\fi
\fi
\fi
\fi
}
\newcommand{\hatcurCCpmra}[1]{\ifnum#1=47 %
\hatcurCCpmraxxxxxA
\else
\ifnum#1=48 %
\hatcurCCpmraxxxxxB
\else
\ifnum#1=49 %
\hatcurCCpmraxxxxxC
\else
\ifnum#1=72 %
\hatcurCCpmraxxxxxD
\else
??????\fi
\fi
\fi
\fi
}
\newcommand{\hatcurCCra}[1]{\ifnum#1=47 %
\hatcurCCraxxxxxA
\else
\ifnum#1=48 %
\hatcurCCraxxxxxB
\else
\ifnum#1=49 %
\hatcurCCraxxxxxC
\else
\ifnum#1=72 %
\hatcurCCraxxxxxD
\else
??????\fi
\fi
\fi
\fi
}
\newcommand{\hatcurCCtassmB}[1]{\ifnum#1=47 %
\hatcurCCtassmBxxxxxA
\else
\ifnum#1=48 %
\hatcurCCtassmBxxxxxB
\else
\ifnum#1=49 %
\hatcurCCtassmBxxxxxC
\else
\ifnum#1=72 %
\hatcurCCtassmBxxxxxD
\else
??????\fi
\fi
\fi
\fi
}
\newcommand{\hatcurCCtassmBshort}[1]{\ifnum#1=47 %
\hatcurCCtassmBshortxxxxxA
\else
\ifnum#1=48 %
\hatcurCCtassmBshortxxxxxB
\else
\ifnum#1=49 %
\hatcurCCtassmBshortxxxxxC
\else
\ifnum#1=72 %
\hatcurCCtassmBshortxxxxxD
\else
??????\fi
\fi
\fi
\fi
}
\newcommand{\hatcurCCtassmg}[1]{\ifnum#1=47 %
\hatcurCCtassmgxxxxxA
\else
\ifnum#1=48 %
\hatcurCCtassmgxxxxxB
\else
\ifnum#1=49 %
\hatcurCCtassmgxxxxxC
\else
\ifnum#1=72 %
\hatcurCCtassmgxxxxxD
\else
??????\fi
\fi
\fi
\fi
}
\newcommand{\hatcurCCtassmgshort}[1]{\ifnum#1=47 %
\hatcurCCtassmgshortxxxxxA
\else
\ifnum#1=48 %
\hatcurCCtassmgshortxxxxxB
\else
\ifnum#1=49 %
\hatcurCCtassmgshortxxxxxC
\else
\ifnum#1=72 %
\hatcurCCtassmgshortxxxxxD
\else
??????\fi
\fi
\fi
\fi
}
\newcommand{\hatcurCCtassmi}[1]{\ifnum#1=47 %
\hatcurCCtassmixxxxxA
\else
\ifnum#1=48 %
\hatcurCCtassmixxxxxB
\else
\ifnum#1=49 %
\hatcurCCtassmixxxxxC
\else
\ifnum#1=72 %
\hatcurCCtassmixxxxxD
\else
??????\fi
\fi
\fi
\fi
}
\newcommand{\hatcurCCtassmI}[1]{\ifnum#1=47 %
\hatcurCCtassmIxxxxxA
\else
\ifnum#1=48 %
\hatcurCCtassmIxxxxxB
\else
\ifnum#1=49 %
\hatcurCCtassmIxxxxxC
\else
\ifnum#1=72 %
\hatcurCCtassmIxxxxxD
\else
??????\fi
\fi
\fi
\fi
}
\newcommand{\hatcurCCtassmishort}[1]{\ifnum#1=47 %
\hatcurCCtassmishortxxxxxA
\else
\ifnum#1=48 %
\hatcurCCtassmishortxxxxxB
\else
\ifnum#1=49 %
\hatcurCCtassmishortxxxxxC
\else
\ifnum#1=72 %
\hatcurCCtassmishortxxxxxD
\else
??????\fi
\fi
\fi
\fi
}
\newcommand{\hatcurCCtassmIshort}[1]{\ifnum#1=47 %
\hatcurCCtassmIshortxxxxxA
\else
\ifnum#1=48 %
\hatcurCCtassmIshortxxxxxB
\else
\ifnum#1=49 %
\hatcurCCtassmIshortxxxxxC
\else
\ifnum#1=72 %
\hatcurCCtassmIshortxxxxxD
\else
??????\fi
\fi
\fi
\fi
}
\newcommand{\hatcurCCtassmr}[1]{\ifnum#1=47 %
\hatcurCCtassmrxxxxxA
\else
\ifnum#1=48 %
\hatcurCCtassmrxxxxxB
\else
\ifnum#1=49 %
\hatcurCCtassmrxxxxxC
\else
\ifnum#1=72 %
\hatcurCCtassmrxxxxxD
\else
??????\fi
\fi
\fi
\fi
}
\newcommand{\hatcurCCtassmrshort}[1]{\ifnum#1=47 %
\hatcurCCtassmrshortxxxxxA
\else
\ifnum#1=48 %
\hatcurCCtassmrshortxxxxxB
\else
\ifnum#1=49 %
\hatcurCCtassmrshortxxxxxC
\else
\ifnum#1=72 %
\hatcurCCtassmrshortxxxxxD
\else
??????\fi
\fi
\fi
\fi
}
\newcommand{\hatcurCCtassmv}[1]{\ifnum#1=47 %
\hatcurCCtassmvxxxxxA
\else
\ifnum#1=48 %
\hatcurCCtassmvxxxxxB
\else
\ifnum#1=49 %
\hatcurCCtassmvxxxxxC
\else
\ifnum#1=72 %
\hatcurCCtassmvxxxxxD
\else
??????\fi
\fi
\fi
\fi
}
\newcommand{\hatcurCCtassmvshort}[1]{\ifnum#1=47 %
\hatcurCCtassmvshortxxxxxA
\else
\ifnum#1=48 %
\hatcurCCtassmvshortxxxxxB
\else
\ifnum#1=49 %
\hatcurCCtassmvshortxxxxxC
\else
\ifnum#1=72 %
\hatcurCCtassmvshortxxxxxD
\else
??????\fi
\fi
\fi
\fi
}
\newcommand{\hatcurCCtwomass}[1]{\ifnum#1=47 %
\hatcurCCtwomassxxxxxA
\else
\ifnum#1=48 %
\hatcurCCtwomassxxxxxB
\else
\ifnum#1=49 %
\hatcurCCtwomassxxxxxC
\else
\ifnum#1=72 %
\hatcurCCtwomassxxxxxD
\else
??????\fi
\fi
\fi
\fi
}
\newcommand{\hatcurCCtwomassHmag}[1]{\ifnum#1=47 %
\hatcurCCtwomassHmagxxxxxA
\else
\ifnum#1=48 %
\hatcurCCtwomassHmagxxxxxB
\else
\ifnum#1=49 %
\hatcurCCtwomassHmagxxxxxC
\else
\ifnum#1=72 %
\hatcurCCtwomassHmagxxxxxD
\else
??????\fi
\fi
\fi
\fi
}
\newcommand{\hatcurCCtwomassJmag}[1]{\ifnum#1=47 %
\hatcurCCtwomassJmagxxxxxA
\else
\ifnum#1=48 %
\hatcurCCtwomassJmagxxxxxB
\else
\ifnum#1=49 %
\hatcurCCtwomassJmagxxxxxC
\else
\ifnum#1=72 %
\hatcurCCtwomassJmagxxxxxD
\else
??????\fi
\fi
\fi
\fi
}
\newcommand{\hatcurCCtwomassKmag}[1]{\ifnum#1=47 %
\hatcurCCtwomassKmagxxxxxA
\else
\ifnum#1=48 %
\hatcurCCtwomassKmagxxxxxB
\else
\ifnum#1=49 %
\hatcurCCtwomassKmagxxxxxC
\else
\ifnum#1=72 %
\hatcurCCtwomassKmagxxxxxD
\else
??????\fi
\fi
\fi
\fi
}
\newcommand{\hatcurCCWfourmag}[1]{\ifnum#1=47 %
\hatcurCCWfourmagxxxxxA
\else
\ifnum#1=48 %
\hatcurCCWfourmagxxxxxB
\else
\ifnum#1=49 %
\hatcurCCWfourmagxxxxxC
\else
\ifnum#1=72 %
\hatcurCCWfourmagxxxxxD
\else
??????\fi
\fi
\fi
\fi
}
\newcommand{\hatcurCCWonemag}[1]{\ifnum#1=47 %
\hatcurCCWonemagxxxxxA
\else
\ifnum#1=48 %
\hatcurCCWonemagxxxxxB
\else
\ifnum#1=49 %
\hatcurCCWonemagxxxxxC
\else
\ifnum#1=72 %
\hatcurCCWonemagxxxxxD
\else
??????\fi
\fi
\fi
\fi
}
\newcommand{\hatcurCCWthreemag}[1]{\ifnum#1=47 %
\hatcurCCWthreemagxxxxxA
\else
\ifnum#1=48 %
\hatcurCCWthreemagxxxxxB
\else
\ifnum#1=49 %
\hatcurCCWthreemagxxxxxC
\else
\ifnum#1=72 %
\hatcurCCWthreemagxxxxxD
\else
??????\fi
\fi
\fi
\fi
}
\newcommand{\hatcurCCWtwomag}[1]{\ifnum#1=47 %
\hatcurCCWtwomagxxxxxA
\else
\ifnum#1=48 %
\hatcurCCWtwomagxxxxxB
\else
\ifnum#1=49 %
\hatcurCCWtwomagxxxxxC
\else
\ifnum#1=72 %
\hatcurCCWtwomagxxxxxD
\else
??????\fi
\fi
\fi
\fi
}
\newcommand{\hatcurextraerrMB}[1]{\ifnum#1=72 %
\hatcurextraerrMBxxxxxD
\else
??????\fi
}
\newcommand{\hatcurextraerrMBPtwo}[1]{\ifnum#1=47 %
\hatcurextraerrMBPtwoxxxxxA
\else
\ifnum#1=48 %
\hatcurextraerrMBPtwoxxxxxB
\else
\ifnum#1=49 %
\hatcurextraerrMBPtwoxxxxxC
\else
\ifnum#1=72 %
\hatcurextraerrMBPtwoxxxxxD
\else
??????\fi
\fi
\fi
\fi
}
\newcommand{\hatcurextraerrMBPtwotwosiglim}[1]{\ifnum#1=47 %
\hatcurextraerrMBPtwotwosiglimxxxxxA
\else
\ifnum#1=48 %
\hatcurextraerrMBPtwotwosiglimxxxxxB
\else
\ifnum#1=49 %
\hatcurextraerrMBPtwotwosiglimxxxxxC
\else
\ifnum#1=72 %
\hatcurextraerrMBPtwotwosiglimxxxxxD
\else
??????\fi
\fi
\fi
\fi
}
\newcommand{\hatcurextraerrMBtwosiglim}[1]{\ifnum#1=72 %
\hatcurextraerrMBtwosiglimxxxxxD
\else
??????\fi
}
\newcommand{\hatcurextraerrMg}[1]{\ifnum#1=47 %
\hatcurextraerrMgxxxxxA
\else
\ifnum#1=48 %
\hatcurextraerrMgxxxxxB
\else
\ifnum#1=49 %
\hatcurextraerrMgxxxxxC
\else
\ifnum#1=72 %
\hatcurextraerrMgxxxxxD
\else
??????\fi
\fi
\fi
\fi
}
\newcommand{\hatcurextraerrMG}[1]{\ifnum#1=47 %
\hatcurextraerrMGxxxxxA
\else
\ifnum#1=48 %
\hatcurextraerrMGxxxxxB
\else
\ifnum#1=49 %
\hatcurextraerrMGxxxxxC
\else
\ifnum#1=72 %
\hatcurextraerrMGxxxxxD
\else
??????\fi
\fi
\fi
\fi
}
\newcommand{\hatcurextraerrMgtwosiglim}[1]{\ifnum#1=47 %
\hatcurextraerrMgtwosiglimxxxxxA
\else
\ifnum#1=48 %
\hatcurextraerrMgtwosiglimxxxxxB
\else
\ifnum#1=49 %
\hatcurextraerrMgtwosiglimxxxxxC
\else
\ifnum#1=72 %
\hatcurextraerrMgtwosiglimxxxxxD
\else
??????\fi
\fi
\fi
\fi
}
\newcommand{\hatcurextraerrMGtwosiglim}[1]{\ifnum#1=47 %
\hatcurextraerrMGtwosiglimxxxxxA
\else
\ifnum#1=48 %
\hatcurextraerrMGtwosiglimxxxxxB
\else
\ifnum#1=49 %
\hatcurextraerrMGtwosiglimxxxxxC
\else
\ifnum#1=72 %
\hatcurextraerrMGtwosiglimxxxxxD
\else
??????\fi
\fi
\fi
\fi
}
\newcommand{\hatcurextraerrMH}[1]{\ifnum#1=47 %
\hatcurextraerrMHxxxxxA
\else
\ifnum#1=48 %
\hatcurextraerrMHxxxxxB
\else
\ifnum#1=49 %
\hatcurextraerrMHxxxxxC
\else
\ifnum#1=72 %
\hatcurextraerrMHxxxxxD
\else
??????\fi
\fi
\fi
\fi
}
\newcommand{\hatcurextraerrMHtwosiglim}[1]{\ifnum#1=47 %
\hatcurextraerrMHtwosiglimxxxxxA
\else
\ifnum#1=48 %
\hatcurextraerrMHtwosiglimxxxxxB
\else
\ifnum#1=49 %
\hatcurextraerrMHtwosiglimxxxxxC
\else
\ifnum#1=72 %
\hatcurextraerrMHtwosiglimxxxxxD
\else
??????\fi
\fi
\fi
\fi
}
\newcommand{\hatcurextraerrMi}[1]{\ifnum#1=47 %
\hatcurextraerrMixxxxxA
\else
\ifnum#1=48 %
\hatcurextraerrMixxxxxB
\else
\ifnum#1=49 %
\hatcurextraerrMixxxxxC
\else
\ifnum#1=72 %
\hatcurextraerrMixxxxxD
\else
??????\fi
\fi
\fi
\fi
}
\newcommand{\hatcurextraerrMitwosiglim}[1]{\ifnum#1=47 %
\hatcurextraerrMitwosiglimxxxxxA
\else
\ifnum#1=48 %
\hatcurextraerrMitwosiglimxxxxxB
\else
\ifnum#1=49 %
\hatcurextraerrMitwosiglimxxxxxC
\else
\ifnum#1=72 %
\hatcurextraerrMitwosiglimxxxxxD
\else
??????\fi
\fi
\fi
\fi
}
\newcommand{\hatcurextraerrMJ}[1]{\ifnum#1=47 %
\hatcurextraerrMJxxxxxA
\else
\ifnum#1=48 %
\hatcurextraerrMJxxxxxB
\else
\ifnum#1=49 %
\hatcurextraerrMJxxxxxC
\else
\ifnum#1=72 %
\hatcurextraerrMJxxxxxD
\else
??????\fi
\fi
\fi
\fi
}
\newcommand{\hatcurextraerrMJtwosiglim}[1]{\ifnum#1=47 %
\hatcurextraerrMJtwosiglimxxxxxA
\else
\ifnum#1=48 %
\hatcurextraerrMJtwosiglimxxxxxB
\else
\ifnum#1=49 %
\hatcurextraerrMJtwosiglimxxxxxC
\else
\ifnum#1=72 %
\hatcurextraerrMJtwosiglimxxxxxD
\else
??????\fi
\fi
\fi
\fi
}
\newcommand{\hatcurextraerrMKs}[1]{\ifnum#1=47 %
\hatcurextraerrMKsxxxxxA
\else
\ifnum#1=48 %
\hatcurextraerrMKsxxxxxB
\else
\ifnum#1=49 %
\hatcurextraerrMKsxxxxxC
\else
\ifnum#1=72 %
\hatcurextraerrMKsxxxxxD
\else
??????\fi
\fi
\fi
\fi
}
\newcommand{\hatcurextraerrMKstwosiglim}[1]{\ifnum#1=47 %
\hatcurextraerrMKstwosiglimxxxxxA
\else
\ifnum#1=48 %
\hatcurextraerrMKstwosiglimxxxxxB
\else
\ifnum#1=49 %
\hatcurextraerrMKstwosiglimxxxxxC
\else
\ifnum#1=72 %
\hatcurextraerrMKstwosiglimxxxxxD
\else
??????\fi
\fi
\fi
\fi
}
\newcommand{\hatcurextraerrMr}[1]{\ifnum#1=47 %
\hatcurextraerrMrxxxxxA
\else
\ifnum#1=48 %
\hatcurextraerrMrxxxxxB
\else
\ifnum#1=49 %
\hatcurextraerrMrxxxxxC
\else
\ifnum#1=72 %
\hatcurextraerrMrxxxxxD
\else
??????\fi
\fi
\fi
\fi
}
\newcommand{\hatcurextraerrMRP}[1]{\ifnum#1=47 %
\hatcurextraerrMRPxxxxxA
\else
\ifnum#1=48 %
\hatcurextraerrMRPxxxxxB
\else
\ifnum#1=49 %
\hatcurextraerrMRPxxxxxC
\else
\ifnum#1=72 %
\hatcurextraerrMRPxxxxxD
\else
??????\fi
\fi
\fi
\fi
}
\newcommand{\hatcurextraerrMRPtwosiglim}[1]{\ifnum#1=47 %
\hatcurextraerrMRPtwosiglimxxxxxA
\else
\ifnum#1=48 %
\hatcurextraerrMRPtwosiglimxxxxxB
\else
\ifnum#1=49 %
\hatcurextraerrMRPtwosiglimxxxxxC
\else
\ifnum#1=72 %
\hatcurextraerrMRPtwosiglimxxxxxD
\else
??????\fi
\fi
\fi
\fi
}
\newcommand{\hatcurextraerrMrtwosiglim}[1]{\ifnum#1=47 %
\hatcurextraerrMrtwosiglimxxxxxA
\else
\ifnum#1=48 %
\hatcurextraerrMrtwosiglimxxxxxB
\else
\ifnum#1=49 %
\hatcurextraerrMrtwosiglimxxxxxC
\else
\ifnum#1=72 %
\hatcurextraerrMrtwosiglimxxxxxD
\else
??????\fi
\fi
\fi
\fi
}
\newcommand{\hatcurextraerrMV}[1]{\ifnum#1=72 %
\hatcurextraerrMVxxxxxD
\else
??????\fi
}
\newcommand{\hatcurextraerrMVtwosiglim}[1]{\ifnum#1=72 %
\hatcurextraerrMVtwosiglimxxxxxD
\else
??????\fi
}
\newcommand{\hatcurextraerrMWone}[1]{\ifnum#1=47 %
\hatcurextraerrMWonexxxxxA
\else
\ifnum#1=48 %
\hatcurextraerrMWonexxxxxB
\else
\ifnum#1=49 %
\hatcurextraerrMWonexxxxxC
\else
\ifnum#1=72 %
\hatcurextraerrMWonexxxxxD
\else
??????\fi
\fi
\fi
\fi
}
\newcommand{\hatcurextraerrMWonetwosiglim}[1]{\ifnum#1=47 %
\hatcurextraerrMWonetwosiglimxxxxxA
\else
\ifnum#1=48 %
\hatcurextraerrMWonetwosiglimxxxxxB
\else
\ifnum#1=49 %
\hatcurextraerrMWonetwosiglimxxxxxC
\else
\ifnum#1=72 %
\hatcurextraerrMWonetwosiglimxxxxxD
\else
??????\fi
\fi
\fi
\fi
}
\newcommand{\hatcurextraerrMWthree}[1]{\ifnum#1=72 %
\hatcurextraerrMWthreexxxxxD
\else
??????\fi
}
\newcommand{\hatcurextraerrMWthreetwosiglim}[1]{\ifnum#1=72 %
\hatcurextraerrMWthreetwosiglimxxxxxD
\else
??????\fi
}
\newcommand{\hatcurextraerrMWtwo}[1]{\ifnum#1=47 %
\hatcurextraerrMWtwoxxxxxA
\else
\ifnum#1=48 %
\hatcurextraerrMWtwoxxxxxB
\else
\ifnum#1=49 %
\hatcurextraerrMWtwoxxxxxC
\else
\ifnum#1=72 %
\hatcurextraerrMWtwoxxxxxD
\else
??????\fi
\fi
\fi
\fi
}
\newcommand{\hatcurextraerrMWtwotwosiglim}[1]{\ifnum#1=47 %
\hatcurextraerrMWtwotwosiglimxxxxxA
\else
\ifnum#1=48 %
\hatcurextraerrMWtwotwosiglimxxxxxB
\else
\ifnum#1=49 %
\hatcurextraerrMWtwotwosiglimxxxxxC
\else
\ifnum#1=72 %
\hatcurextraerrMWtwotwosiglimxxxxxD
\else
??????\fi
\fi
\fi
\fi
}
\newcommand{\hatcurfield}[1]{\ifnum#1=47 %
\hatcurfieldxxxxxA
\else
\ifnum#1=48 %
\hatcurfieldxxxxxB
\else
\ifnum#1=49 %
\hatcurfieldxxxxxC
\else
\ifnum#1=72 %
\hatcurfieldxxxxxD
\else
??????\fi
\fi
\fi
\fi
}
\newcommand{\hatcurhtr}[1]{\ifnum#1=47 %
\hatcurhtrxxxxxA
\else
\ifnum#1=48 %
\hatcurhtrxxxxxB
\else
\ifnum#1=49 %
\hatcurhtrxxxxxC
\else
\ifnum#1=72 %
\hatcurhtrxxxxxD
\else
??????\fi
\fi
\fi
\fi
}
\newcommand{\hatcurISOage}[1]{\ifnum#1=47 %
\hatcurISOagexxxxxA
\else
\ifnum#1=48 %
\hatcurISOagexxxxxB
\else
\ifnum#1=49 %
\hatcurISOagexxxxxC
\else
\ifnum#1=72 %
\hatcurISOagexxxxxD
\else
??????\fi
\fi
\fi
\fi
}
\newcommand{\hatcurISOlogg}[1]{\ifnum#1=47 %
\hatcurISOloggxxxxxA
\else
\ifnum#1=48 %
\hatcurISOloggxxxxxB
\else
\ifnum#1=49 %
\hatcurISOloggxxxxxC
\else
\ifnum#1=72 %
\hatcurISOloggxxxxxD
\else
??????\fi
\fi
\fi
\fi
}
\newcommand{\hatcurISOlum}[1]{\ifnum#1=47 %
\hatcurISOlumxxxxxA
\else
\ifnum#1=48 %
\hatcurISOlumxxxxxB
\else
\ifnum#1=49 %
\hatcurISOlumxxxxxC
\else
\ifnum#1=72 %
\hatcurISOlumxxxxxD
\else
??????\fi
\fi
\fi
\fi
}
\newcommand{\hatcurISOlumshort}[1]{\ifnum#1=47 %
\hatcurISOlumshortxxxxxA
\else
\ifnum#1=48 %
\hatcurISOlumshortxxxxxB
\else
\ifnum#1=49 %
\hatcurISOlumshortxxxxxC
\else
\ifnum#1=72 %
\hatcurISOlumshortxxxxxD
\else
??????\fi
\fi
\fi
\fi
}
\newcommand{\hatcurISOm}[1]{\ifnum#1=47 %
\hatcurISOmxxxxxA
\else
\ifnum#1=48 %
\hatcurISOmxxxxxB
\else
\ifnum#1=49 %
\hatcurISOmxxxxxC
\else
\ifnum#1=72 %
\hatcurISOmxxxxxD
\else
??????\fi
\fi
\fi
\fi
}
\newcommand{\hatcurISOmlong}[1]{\ifnum#1=47 %
\hatcurISOmlongxxxxxA
\else
\ifnum#1=48 %
\hatcurISOmlongxxxxxB
\else
\ifnum#1=49 %
\hatcurISOmlongxxxxxC
\else
\ifnum#1=72 %
\hatcurISOmlongxxxxxD
\else
??????\fi
\fi
\fi
\fi
}
\newcommand{\hatcurISOmshort}[1]{\ifnum#1=47 %
\hatcurISOmshortxxxxxA
\else
\ifnum#1=48 %
\hatcurISOmshortxxxxxB
\else
\ifnum#1=49 %
\hatcurISOmshortxxxxxC
\else
\ifnum#1=72 %
\hatcurISOmshortxxxxxD
\else
??????\fi
\fi
\fi
\fi
}
\newcommand{\hatcurISOr}[1]{\ifnum#1=47 %
\hatcurISOrxxxxxA
\else
\ifnum#1=48 %
\hatcurISOrxxxxxB
\else
\ifnum#1=49 %
\hatcurISOrxxxxxC
\else
\ifnum#1=72 %
\hatcurISOrxxxxxD
\else
??????\fi
\fi
\fi
\fi
}
\newcommand{\hatcurISOrho}[1]{\ifnum#1=47 %
\hatcurISOrhoxxxxxA
\else
\ifnum#1=48 %
\hatcurISOrhoxxxxxB
\else
\ifnum#1=49 %
\hatcurISOrhoxxxxxC
\else
\ifnum#1=72 %
\hatcurISOrhoxxxxxD
\else
??????\fi
\fi
\fi
\fi
}
\newcommand{\hatcurISOrholong}[1]{\ifnum#1=47 %
\hatcurISOrholongxxxxxA
\else
\ifnum#1=48 %
\hatcurISOrholongxxxxxB
\else
\ifnum#1=49 %
\hatcurISOrholongxxxxxC
\else
\ifnum#1=72 %
\hatcurISOrholongxxxxxD
\else
??????\fi
\fi
\fi
\fi
}
\newcommand{\hatcurISOrlong}[1]{\ifnum#1=47 %
\hatcurISOrlongxxxxxA
\else
\ifnum#1=48 %
\hatcurISOrlongxxxxxB
\else
\ifnum#1=49 %
\hatcurISOrlongxxxxxC
\else
\ifnum#1=72 %
\hatcurISOrlongxxxxxD
\else
??????\fi
\fi
\fi
\fi
}
\newcommand{\hatcurISOrshort}[1]{\ifnum#1=47 %
\hatcurISOrshortxxxxxA
\else
\ifnum#1=48 %
\hatcurISOrshortxxxxxB
\else
\ifnum#1=49 %
\hatcurISOrshortxxxxxC
\else
\ifnum#1=72 %
\hatcurISOrshortxxxxxD
\else
??????\fi
\fi
\fi
\fi
}
\newcommand{\hatcurISOspec}[1]{\ifnum#1=47 %
\hatcurISOspecxxxxxA
\else
\ifnum#1=48 %
\hatcurISOspecxxxxxB
\else
\ifnum#1=49 %
\hatcurISOspecxxxxxC
\else
\ifnum#1=72 %
\hatcurISOspecxxxxxD
\else
??????\fi
\fi
\fi
\fi
}
\newcommand{\hatcurISOteff}[1]{\ifnum#1=47 %
\hatcurISOteffxxxxxA
\else
\ifnum#1=48 %
\hatcurISOteffxxxxxB
\else
\ifnum#1=49 %
\hatcurISOteffxxxxxC
\else
\ifnum#1=72 %
\hatcurISOteffxxxxxD
\else
??????\fi
\fi
\fi
\fi
}
\newcommand{\hatcurISOzfeh}[1]{\ifnum#1=47 %
\hatcurISOzfehxxxxxA
\else
\ifnum#1=48 %
\hatcurISOzfehxxxxxB
\else
\ifnum#1=49 %
\hatcurISOzfehxxxxxC
\else
\ifnum#1=72 %
\hatcurISOzfehxxxxxD
\else
??????\fi
\fi
\fi
\fi
}
\newcommand{\hatcurLBiB}[1]{\ifnum#1=47 %
\hatcurLBiBxxxxxA
\else
\ifnum#1=48 %
\hatcurLBiBxxxxxB
\else
\ifnum#1=49 %
\hatcurLBiBxxxxxC
\else
\ifnum#1=72 %
\hatcurLBiBxxxxxD
\else
??????\fi
\fi
\fi
\fi
}
\newcommand{\hatcurLBiC}[1]{\ifnum#1=47 %
\hatcurLBiCxxxxxA
\else
\ifnum#1=48 %
\hatcurLBiCxxxxxB
\else
\ifnum#1=49 %
\hatcurLBiCxxxxxC
\else
\ifnum#1=72 %
\hatcurLBiCxxxxxD
\else
??????\fi
\fi
\fi
\fi
}
\newcommand{\hatcurLBig}[1]{\ifnum#1=47 %
\hatcurLBigxxxxxA
\else
\ifnum#1=48 %
\hatcurLBigxxxxxB
\else
\ifnum#1=49 %
\hatcurLBigxxxxxC
\else
\ifnum#1=72 %
\hatcurLBigxxxxxD
\else
??????\fi
\fi
\fi
\fi
}
\newcommand{\hatcurLBiH}[1]{\ifnum#1=47 %
\hatcurLBiHxxxxxA
\else
\ifnum#1=48 %
\hatcurLBiHxxxxxB
\else
\ifnum#1=49 %
\hatcurLBiHxxxxxC
\else
\ifnum#1=72 %
\hatcurLBiHxxxxxD
\else
??????\fi
\fi
\fi
\fi
}
\newcommand{\hatcurLBii}[1]{\ifnum#1=47 %
\hatcurLBiixxxxxA
\else
\ifnum#1=48 %
\hatcurLBiixxxxxB
\else
\ifnum#1=49 %
\hatcurLBiixxxxxC
\else
\ifnum#1=72 %
\hatcurLBiixxxxxD
\else
??????\fi
\fi
\fi
\fi
}
\newcommand{\hatcurLBiI}[1]{\ifnum#1=47 %
\hatcurLBiIxxxxxA
\else
\ifnum#1=48 %
\hatcurLBiIxxxxxB
\else
\ifnum#1=49 %
\hatcurLBiIxxxxxC
\else
\ifnum#1=72 %
\hatcurLBiIxxxxxD
\else
??????\fi
\fi
\fi
\fi
}
\newcommand{\hatcurLBiiB}[1]{\ifnum#1=47 %
\hatcurLBiiBxxxxxA
\else
\ifnum#1=48 %
\hatcurLBiiBxxxxxB
\else
\ifnum#1=49 %
\hatcurLBiiBxxxxxC
\else
\ifnum#1=72 %
\hatcurLBiiBxxxxxD
\else
??????\fi
\fi
\fi
\fi
}
\newcommand{\hatcurLBiiC}[1]{\ifnum#1=47 %
\hatcurLBiiCxxxxxA
\else
\ifnum#1=48 %
\hatcurLBiiCxxxxxB
\else
\ifnum#1=49 %
\hatcurLBiiCxxxxxC
\else
\ifnum#1=72 %
\hatcurLBiiCxxxxxD
\else
??????\fi
\fi
\fi
\fi
}
\newcommand{\hatcurLBiig}[1]{\ifnum#1=47 %
\hatcurLBiigxxxxxA
\else
\ifnum#1=48 %
\hatcurLBiigxxxxxB
\else
\ifnum#1=49 %
\hatcurLBiigxxxxxC
\else
\ifnum#1=72 %
\hatcurLBiigxxxxxD
\else
??????\fi
\fi
\fi
\fi
}
\newcommand{\hatcurLBiiH}[1]{\ifnum#1=47 %
\hatcurLBiiHxxxxxA
\else
\ifnum#1=48 %
\hatcurLBiiHxxxxxB
\else
\ifnum#1=49 %
\hatcurLBiiHxxxxxC
\else
\ifnum#1=72 %
\hatcurLBiiHxxxxxD
\else
??????\fi
\fi
\fi
\fi
}
\newcommand{\hatcurLBiii}[1]{\ifnum#1=47 %
\hatcurLBiiixxxxxA
\else
\ifnum#1=48 %
\hatcurLBiiixxxxxB
\else
\ifnum#1=49 %
\hatcurLBiiixxxxxC
\else
\ifnum#1=72 %
\hatcurLBiiixxxxxD
\else
??????\fi
\fi
\fi
\fi
}
\newcommand{\hatcurLBiiI}[1]{\ifnum#1=47 %
\hatcurLBiiIxxxxxA
\else
\ifnum#1=48 %
\hatcurLBiiIxxxxxB
\else
\ifnum#1=49 %
\hatcurLBiiIxxxxxC
\else
\ifnum#1=72 %
\hatcurLBiiIxxxxxD
\else
??????\fi
\fi
\fi
\fi
}
\newcommand{\hatcurLBiiJ}[1]{\ifnum#1=47 %
\hatcurLBiiJxxxxxA
\else
\ifnum#1=48 %
\hatcurLBiiJxxxxxB
\else
\ifnum#1=49 %
\hatcurLBiiJxxxxxC
\else
\ifnum#1=72 %
\hatcurLBiiJxxxxxD
\else
??????\fi
\fi
\fi
\fi
}
\newcommand{\hatcurLBiiK}[1]{\ifnum#1=47 %
\hatcurLBiiKxxxxxA
\else
\ifnum#1=48 %
\hatcurLBiiKxxxxxB
\else
\ifnum#1=49 %
\hatcurLBiiKxxxxxC
\else
\ifnum#1=72 %
\hatcurLBiiKxxxxxD
\else
??????\fi
\fi
\fi
\fi
}
\newcommand{\hatcurLBiikep}[1]{\ifnum#1=47 %
\hatcurLBiikepxxxxxA
\else
\ifnum#1=48 %
\hatcurLBiikepxxxxxB
\else
\ifnum#1=49 %
\hatcurLBiikepxxxxxC
\else
\ifnum#1=72 %
\hatcurLBiikepxxxxxD
\else
??????\fi
\fi
\fi
\fi
}
\newcommand{\hatcurLBiiM}[1]{\ifnum#1=47 %
\hatcurLBiiMxxxxxA
\else
\ifnum#1=48 %
\hatcurLBiiMxxxxxB
\else
\ifnum#1=49 %
\hatcurLBiiMxxxxxC
\else
\ifnum#1=72 %
\hatcurLBiiMxxxxxD
\else
??????\fi
\fi
\fi
\fi
}
\newcommand{\hatcurLBiir}[1]{\ifnum#1=47 %
\hatcurLBiirxxxxxA
\else
\ifnum#1=48 %
\hatcurLBiirxxxxxB
\else
\ifnum#1=49 %
\hatcurLBiirxxxxxC
\else
\ifnum#1=72 %
\hatcurLBiirxxxxxD
\else
??????\fi
\fi
\fi
\fi
}
\newcommand{\hatcurLBiiR}[1]{\ifnum#1=47 %
\hatcurLBiiRxxxxxA
\else
\ifnum#1=48 %
\hatcurLBiiRxxxxxB
\else
\ifnum#1=49 %
\hatcurLBiiRxxxxxC
\else
\ifnum#1=72 %
\hatcurLBiiRxxxxxD
\else
??????\fi
\fi
\fi
\fi
}
\newcommand{\hatcurLBiiSfour}[1]{\ifnum#1=47 %
\hatcurLBiiSfourxxxxxA
\else
\ifnum#1=48 %
\hatcurLBiiSfourxxxxxB
\else
\ifnum#1=49 %
\hatcurLBiiSfourxxxxxC
\else
\ifnum#1=72 %
\hatcurLBiiSfourxxxxxD
\else
??????\fi
\fi
\fi
\fi
}
\newcommand{\hatcurLBiiSone}[1]{\ifnum#1=47 %
\hatcurLBiiSonexxxxxA
\else
\ifnum#1=48 %
\hatcurLBiiSonexxxxxB
\else
\ifnum#1=49 %
\hatcurLBiiSonexxxxxC
\else
\ifnum#1=72 %
\hatcurLBiiSonexxxxxD
\else
??????\fi
\fi
\fi
\fi
}
\newcommand{\hatcurLBiiSthree}[1]{\ifnum#1=47 %
\hatcurLBiiSthreexxxxxA
\else
\ifnum#1=48 %
\hatcurLBiiSthreexxxxxB
\else
\ifnum#1=49 %
\hatcurLBiiSthreexxxxxC
\else
\ifnum#1=72 %
\hatcurLBiiSthreexxxxxD
\else
??????\fi
\fi
\fi
\fi
}
\newcommand{\hatcurLBiiStwo}[1]{\ifnum#1=47 %
\hatcurLBiiStwoxxxxxA
\else
\ifnum#1=48 %
\hatcurLBiiStwoxxxxxB
\else
\ifnum#1=49 %
\hatcurLBiiStwoxxxxxC
\else
\ifnum#1=72 %
\hatcurLBiiStwoxxxxxD
\else
??????\fi
\fi
\fi
\fi
}
\newcommand{\hatcurLBiiT}[1]{\ifnum#1=47 %
\hatcurLBiiTxxxxxA
\else
\ifnum#1=48 %
\hatcurLBiiTxxxxxB
\else
\ifnum#1=49 %
\hatcurLBiiTxxxxxC
\else
\ifnum#1=72 %
\hatcurLBiiTxxxxxD
\else
??????\fi
\fi
\fi
\fi
}
\newcommand{\hatcurLBiiu}[1]{\ifnum#1=47 %
\hatcurLBiiuxxxxxA
\else
\ifnum#1=48 %
\hatcurLBiiuxxxxxB
\else
\ifnum#1=49 %
\hatcurLBiiuxxxxxC
\else
\ifnum#1=72 %
\hatcurLBiiuxxxxxD
\else
??????\fi
\fi
\fi
\fi
}
\newcommand{\hatcurLBiiV}[1]{\ifnum#1=47 %
\hatcurLBiiVxxxxxA
\else
\ifnum#1=48 %
\hatcurLBiiVxxxxxB
\else
\ifnum#1=49 %
\hatcurLBiiVxxxxxC
\else
\ifnum#1=72 %
\hatcurLBiiVxxxxxD
\else
??????\fi
\fi
\fi
\fi
}
\newcommand{\hatcurLBiiz}[1]{\ifnum#1=47 %
\hatcurLBiizxxxxxA
\else
\ifnum#1=48 %
\hatcurLBiizxxxxxB
\else
\ifnum#1=49 %
\hatcurLBiizxxxxxC
\else
\ifnum#1=72 %
\hatcurLBiizxxxxxD
\else
??????\fi
\fi
\fi
\fi
}
\newcommand{\hatcurLBiJ}[1]{\ifnum#1=47 %
\hatcurLBiJxxxxxA
\else
\ifnum#1=48 %
\hatcurLBiJxxxxxB
\else
\ifnum#1=49 %
\hatcurLBiJxxxxxC
\else
\ifnum#1=72 %
\hatcurLBiJxxxxxD
\else
??????\fi
\fi
\fi
\fi
}
\newcommand{\hatcurLBiK}[1]{\ifnum#1=47 %
\hatcurLBiKxxxxxA
\else
\ifnum#1=48 %
\hatcurLBiKxxxxxB
\else
\ifnum#1=49 %
\hatcurLBiKxxxxxC
\else
\ifnum#1=72 %
\hatcurLBiKxxxxxD
\else
??????\fi
\fi
\fi
\fi
}
\newcommand{\hatcurLBikep}[1]{\ifnum#1=47 %
\hatcurLBikepxxxxxA
\else
\ifnum#1=48 %
\hatcurLBikepxxxxxB
\else
\ifnum#1=49 %
\hatcurLBikepxxxxxC
\else
\ifnum#1=72 %
\hatcurLBikepxxxxxD
\else
??????\fi
\fi
\fi
\fi
}
\newcommand{\hatcurLBiM}[1]{\ifnum#1=47 %
\hatcurLBiMxxxxxA
\else
\ifnum#1=48 %
\hatcurLBiMxxxxxB
\else
\ifnum#1=49 %
\hatcurLBiMxxxxxC
\else
\ifnum#1=72 %
\hatcurLBiMxxxxxD
\else
??????\fi
\fi
\fi
\fi
}
\newcommand{\hatcurLBir}[1]{\ifnum#1=47 %
\hatcurLBirxxxxxA
\else
\ifnum#1=48 %
\hatcurLBirxxxxxB
\else
\ifnum#1=49 %
\hatcurLBirxxxxxC
\else
\ifnum#1=72 %
\hatcurLBirxxxxxD
\else
??????\fi
\fi
\fi
\fi
}
\newcommand{\hatcurLBiR}[1]{\ifnum#1=47 %
\hatcurLBiRxxxxxA
\else
\ifnum#1=48 %
\hatcurLBiRxxxxxB
\else
\ifnum#1=49 %
\hatcurLBiRxxxxxC
\else
\ifnum#1=72 %
\hatcurLBiRxxxxxD
\else
??????\fi
\fi
\fi
\fi
}
\newcommand{\hatcurLBiSfour}[1]{\ifnum#1=47 %
\hatcurLBiSfourxxxxxA
\else
\ifnum#1=48 %
\hatcurLBiSfourxxxxxB
\else
\ifnum#1=49 %
\hatcurLBiSfourxxxxxC
\else
\ifnum#1=72 %
\hatcurLBiSfourxxxxxD
\else
??????\fi
\fi
\fi
\fi
}
\newcommand{\hatcurLBiSone}[1]{\ifnum#1=47 %
\hatcurLBiSonexxxxxA
\else
\ifnum#1=48 %
\hatcurLBiSonexxxxxB
\else
\ifnum#1=49 %
\hatcurLBiSonexxxxxC
\else
\ifnum#1=72 %
\hatcurLBiSonexxxxxD
\else
??????\fi
\fi
\fi
\fi
}
\newcommand{\hatcurLBiSthree}[1]{\ifnum#1=47 %
\hatcurLBiSthreexxxxxA
\else
\ifnum#1=48 %
\hatcurLBiSthreexxxxxB
\else
\ifnum#1=49 %
\hatcurLBiSthreexxxxxC
\else
\ifnum#1=72 %
\hatcurLBiSthreexxxxxD
\else
??????\fi
\fi
\fi
\fi
}
\newcommand{\hatcurLBiStwo}[1]{\ifnum#1=47 %
\hatcurLBiStwoxxxxxA
\else
\ifnum#1=48 %
\hatcurLBiStwoxxxxxB
\else
\ifnum#1=49 %
\hatcurLBiStwoxxxxxC
\else
\ifnum#1=72 %
\hatcurLBiStwoxxxxxD
\else
??????\fi
\fi
\fi
\fi
}
\newcommand{\hatcurLBiT}[1]{\ifnum#1=47 %
\hatcurLBiTxxxxxA
\else
\ifnum#1=48 %
\hatcurLBiTxxxxxB
\else
\ifnum#1=49 %
\hatcurLBiTxxxxxC
\else
\ifnum#1=72 %
\hatcurLBiTxxxxxD
\else
??????\fi
\fi
\fi
\fi
}
\newcommand{\hatcurLBiu}[1]{\ifnum#1=47 %
\hatcurLBiuxxxxxA
\else
\ifnum#1=48 %
\hatcurLBiuxxxxxB
\else
\ifnum#1=49 %
\hatcurLBiuxxxxxC
\else
\ifnum#1=72 %
\hatcurLBiuxxxxxD
\else
??????\fi
\fi
\fi
\fi
}
\newcommand{\hatcurLBiV}[1]{\ifnum#1=47 %
\hatcurLBiVxxxxxA
\else
\ifnum#1=48 %
\hatcurLBiVxxxxxB
\else
\ifnum#1=49 %
\hatcurLBiVxxxxxC
\else
\ifnum#1=72 %
\hatcurLBiVxxxxxD
\else
??????\fi
\fi
\fi
\fi
}
\newcommand{\hatcurLBiz}[1]{\ifnum#1=47 %
\hatcurLBizxxxxxA
\else
\ifnum#1=48 %
\hatcurLBizxxxxxB
\else
\ifnum#1=49 %
\hatcurLBizxxxxxC
\else
\ifnum#1=72 %
\hatcurLBizxxxxxD
\else
??????\fi
\fi
\fi
\fi
}
\newcommand{\hatcurLCbsq}[1]{\ifnum#1=47 %
\hatcurLCbsqxxxxxA
\else
\ifnum#1=48 %
\hatcurLCbsqxxxxxB
\else
\ifnum#1=49 %
\hatcurLCbsqxxxxxC
\else
\ifnum#1=72 %
\hatcurLCbsqxxxxxD
\else
??????\fi
\fi
\fi
\fi
}
\newcommand{\hatcurLCdip}[1]{\ifnum#1=47 %
\hatcurLCdipxxxxxA
\else
\ifnum#1=48 %
\hatcurLCdipxxxxxB
\else
\ifnum#1=49 %
\hatcurLCdipxxxxxC
\else
\ifnum#1=72 %
\hatcurLCdipxxxxxD
\else
??????\fi
\fi
\fi
\fi
}
\newcommand{\hatcurLCdur}[1]{\ifnum#1=47 %
\hatcurLCdurxxxxxA
\else
\ifnum#1=48 %
\hatcurLCdurxxxxxB
\else
\ifnum#1=49 %
\hatcurLCdurxxxxxC
\else
\ifnum#1=72 %
\hatcurLCdurxxxxxD
\else
??????\fi
\fi
\fi
\fi
}
\newcommand{\hatcurLCdurhr}[1]{\ifnum#1=47 %
\hatcurLCdurhrxxxxxA
\else
\ifnum#1=48 %
\hatcurLCdurhrxxxxxB
\else
\ifnum#1=49 %
\hatcurLCdurhrxxxxxC
\else
\ifnum#1=72 %
\hatcurLCdurhrxxxxxD
\else
??????\fi
\fi
\fi
\fi
}
\newcommand{\hatcurLCdurhrshort}[1]{\ifnum#1=47 %
\hatcurLCdurhrshortxxxxxA
\else
\ifnum#1=48 %
\hatcurLCdurhrshortxxxxxB
\else
\ifnum#1=49 %
\hatcurLCdurhrshortxxxxxC
\else
\ifnum#1=72 %
\hatcurLCdurhrshortxxxxxD
\else
??????\fi
\fi
\fi
\fi
}
\newcommand{\hatcurLCdurshort}[1]{\ifnum#1=47 %
\hatcurLCdurshortxxxxxA
\else
\ifnum#1=48 %
\hatcurLCdurshortxxxxxB
\else
\ifnum#1=49 %
\hatcurLCdurshortxxxxxC
\else
\ifnum#1=72 %
\hatcurLCdurshortxxxxxD
\else
??????\fi
\fi
\fi
\fi
}
\newcommand{\hatcurLChatnetmA}[1]{\ifnum#1=47 %
\hatcurLChatnetmAxxxxxA
\else
\ifnum#1=48 %
\hatcurLChatnetmAxxxxxB
\else
\ifnum#1=49 %
\hatcurLChatnetmAxxxxxC
\else
\ifnum#1=72 %
\hatcurLChatnetmAxxxxxD
\else
??????\fi
\fi
\fi
\fi
}
\newcommand{\hatcurLChatnetmB}[1]{\ifnum#1=47 %
\hatcurLChatnetmBxxxxxA
\else
\ifnum#1=48 %
\hatcurLChatnetmBxxxxxB
\else
\ifnum#1=49 %
\hatcurLChatnetmBxxxxxC
\else
\ifnum#1=72 %
\hatcurLChatnetmBxxxxxD
\else
??????\fi
\fi
\fi
\fi
}
\newcommand{\hatcurLChatnetmC}[1]{\ifnum#1=49 %
\hatcurLChatnetmCxxxxxC
\else
\ifnum#1=72 %
\hatcurLChatnetmCxxxxxD
\else
??????\fi
\fi
}
\newcommand{\hatcurLCiblendA}[1]{\ifnum#1=47 %
\hatcurLCiblendAxxxxxA
\else
\ifnum#1=48 %
\hatcurLCiblendAxxxxxB
\else
\ifnum#1=49 %
\hatcurLCiblendAxxxxxC
\else
\ifnum#1=72 %
\hatcurLCiblendAxxxxxD
\else
??????\fi
\fi
\fi
\fi
}
\newcommand{\hatcurLCiblendB}[1]{\ifnum#1=47 %
\hatcurLCiblendBxxxxxA
\else
\ifnum#1=48 %
\hatcurLCiblendBxxxxxB
\else
\ifnum#1=49 %
\hatcurLCiblendBxxxxxC
\else
\ifnum#1=72 %
\hatcurLCiblendBxxxxxD
\else
??????\fi
\fi
\fi
\fi
}
\newcommand{\hatcurLCiblendC}[1]{\ifnum#1=49 %
\hatcurLCiblendCxxxxxC
\else
\ifnum#1=72 %
\hatcurLCiblendCxxxxxD
\else
??????\fi
\fi
}
\newcommand{\hatcurLCimp}[1]{\ifnum#1=47 %
\hatcurLCimpxxxxxA
\else
\ifnum#1=48 %
\hatcurLCimpxxxxxB
\else
\ifnum#1=49 %
\hatcurLCimpxxxxxC
\else
\ifnum#1=72 %
\hatcurLCimpxxxxxD
\else
??????\fi
\fi
\fi
\fi
}
\newcommand{\hatcurLCingdur}[1]{\ifnum#1=47 %
\hatcurLCingdurxxxxxA
\else
\ifnum#1=48 %
\hatcurLCingdurxxxxxB
\else
\ifnum#1=49 %
\hatcurLCingdurxxxxxC
\else
\ifnum#1=72 %
\hatcurLCingdurxxxxxD
\else
??????\fi
\fi
\fi
\fi
}
\newcommand{\hatcurLCP}[1]{\ifnum#1=47 %
\hatcurLCPxxxxxA
\else
\ifnum#1=48 %
\hatcurLCPxxxxxB
\else
\ifnum#1=49 %
\hatcurLCPxxxxxC
\else
\ifnum#1=72 %
\hatcurLCPxxxxxD
\else
??????\fi
\fi
\fi
\fi
}
\newcommand{\hatcurLCPprec}[1]{\ifnum#1=47 %
\hatcurLCPprecxxxxxA
\else
\ifnum#1=48 %
\hatcurLCPprecxxxxxB
\else
\ifnum#1=49 %
\hatcurLCPprecxxxxxC
\else
\ifnum#1=72 %
\hatcurLCPprecxxxxxD
\else
??????\fi
\fi
\fi
\fi
}
\newcommand{\hatcurLCPshort}[1]{\ifnum#1=47 %
\hatcurLCPshortxxxxxA
\else
\ifnum#1=48 %
\hatcurLCPshortxxxxxB
\else
\ifnum#1=49 %
\hatcurLCPshortxxxxxC
\else
\ifnum#1=72 %
\hatcurLCPshortxxxxxD
\else
??????\fi
\fi
\fi
\fi
}
\newcommand{\hatcurLCq}[1]{\ifnum#1=47 %
\hatcurLCqxxxxxA
\else
\ifnum#1=48 %
\hatcurLCqxxxxxB
\else
\ifnum#1=49 %
\hatcurLCqxxxxxC
\else
\ifnum#1=72 %
\hatcurLCqxxxxxD
\else
??????\fi
\fi
\fi
\fi
}
\newcommand{\hatcurLCqshort}[1]{\ifnum#1=47 %
\hatcurLCqshortxxxxxA
\else
\ifnum#1=48 %
\hatcurLCqshortxxxxxB
\else
\ifnum#1=49 %
\hatcurLCqshortxxxxxC
\else
\ifnum#1=72 %
\hatcurLCqshortxxxxxD
\else
??????\fi
\fi
\fi
\fi
}
\newcommand{\hatcurLCrho}[1]{\ifnum#1=47 %
\hatcurLCrhoxxxxxA
\else
\ifnum#1=48 %
\hatcurLCrhoxxxxxB
\else
\ifnum#1=49 %
\hatcurLCrhoxxxxxC
\else
\ifnum#1=72 %
\hatcurLCrhoxxxxxD
\else
??????\fi
\fi
\fi
\fi
}
\newcommand{\hatcurLCrprstar}[1]{\ifnum#1=47 %
\hatcurLCrprstarxxxxxA
\else
\ifnum#1=48 %
\hatcurLCrprstarxxxxxB
\else
\ifnum#1=49 %
\hatcurLCrprstarxxxxxC
\else
\ifnum#1=72 %
\hatcurLCrprstarxxxxxD
\else
??????\fi
\fi
\fi
\fi
}
\newcommand{\hatcurLCT}[1]{\ifnum#1=47 %
\hatcurLCTxxxxxA
\else
\ifnum#1=48 %
\hatcurLCTxxxxxB
\else
\ifnum#1=49 %
\hatcurLCTxxxxxC
\else
\ifnum#1=72 %
\hatcurLCTxxxxxD
\else
??????\fi
\fi
\fi
\fi
}
\newcommand{\hatcurLCTA}[1]{\ifnum#1=47 %
\hatcurLCTAxxxxxA
\else
\ifnum#1=48 %
\hatcurLCTAxxxxxB
\else
\ifnum#1=49 %
\hatcurLCTAxxxxxC
\else
\ifnum#1=72 %
\hatcurLCTAxxxxxD
\else
??????\fi
\fi
\fi
\fi
}
\newcommand{\hatcurLCTB}[1]{\ifnum#1=47 %
\hatcurLCTBxxxxxA
\else
\ifnum#1=48 %
\hatcurLCTBxxxxxB
\else
\ifnum#1=49 %
\hatcurLCTBxxxxxC
\else
\ifnum#1=72 %
\hatcurLCTBxxxxxD
\else
??????\fi
\fi
\fi
\fi
}
\newcommand{\hatcurLCzeta}[1]{\ifnum#1=47 %
\hatcurLCzetaxxxxxA
\else
\ifnum#1=48 %
\hatcurLCzetaxxxxxB
\else
\ifnum#1=49 %
\hatcurLCzetaxxxxxC
\else
\ifnum#1=72 %
\hatcurLCzetaxxxxxD
\else
??????\fi
\fi
\fi
\fi
}
\newcommand{\hatcurPPaequiv}[1]{\ifnum#1=47 %
\hatcurPPaequivxxxxxA
\else
\ifnum#1=48 %
\hatcurPPaequivxxxxxB
\else
\ifnum#1=49 %
\hatcurPPaequivxxxxxC
\else
\ifnum#1=72 %
\hatcurPPaequivxxxxxD
\else
??????\fi
\fi
\fi
\fi
}
\newcommand{\hatcurPPar}[1]{\ifnum#1=47 %
\hatcurPParxxxxxA
\else
\ifnum#1=48 %
\hatcurPParxxxxxB
\else
\ifnum#1=49 %
\hatcurPParxxxxxC
\else
\ifnum#1=72 %
\hatcurPParxxxxxD
\else
??????\fi
\fi
\fi
\fi
}
\newcommand{\hatcurPParel}[1]{\ifnum#1=47 %
\hatcurPParelxxxxxA
\else
\ifnum#1=48 %
\hatcurPParelxxxxxB
\else
\ifnum#1=49 %
\hatcurPParelxxxxxC
\else
\ifnum#1=72 %
\hatcurPParelxxxxxD
\else
??????\fi
\fi
\fi
\fi
}
\newcommand{\hatcurPPfluxap}[1]{\ifnum#1=47 %
\hatcurPPfluxapxxxxxA
\else
\ifnum#1=48 %
\hatcurPPfluxapxxxxxB
\else
\ifnum#1=49 %
\hatcurPPfluxapxxxxxC
\else
\ifnum#1=72 %
\hatcurPPfluxapxxxxxD
\else
??????\fi
\fi
\fi
\fi
}
\newcommand{\hatcurPPfluxapdim}[1]{\ifnum#1=47 %
\hatcurPPfluxapdimxxxxxA
\else
\ifnum#1=48 %
\hatcurPPfluxapdimxxxxxB
\else
\ifnum#1=49 %
\hatcurPPfluxapdimxxxxxC
\else
\ifnum#1=72 %
\hatcurPPfluxapdimxxxxxD
\else
??????\fi
\fi
\fi
\fi
}
\newcommand{\hatcurPPfluxavg}[1]{\ifnum#1=47 %
\hatcurPPfluxavgxxxxxA
\else
\ifnum#1=48 %
\hatcurPPfluxavgxxxxxB
\else
\ifnum#1=49 %
\hatcurPPfluxavgxxxxxC
\else
\ifnum#1=72 %
\hatcurPPfluxavgxxxxxD
\else
??????\fi
\fi
\fi
\fi
}
\newcommand{\hatcurPPfluxavgdim}[1]{\ifnum#1=47 %
\hatcurPPfluxavgdimxxxxxA
\else
\ifnum#1=48 %
\hatcurPPfluxavgdimxxxxxB
\else
\ifnum#1=49 %
\hatcurPPfluxavgdimxxxxxC
\else
\ifnum#1=72 %
\hatcurPPfluxavgdimxxxxxD
\else
??????\fi
\fi
\fi
\fi
}
\newcommand{\hatcurPPfluxavglog}[1]{\ifnum#1=47 %
\hatcurPPfluxavglogxxxxxA
\else
\ifnum#1=48 %
\hatcurPPfluxavglogxxxxxB
\else
\ifnum#1=49 %
\hatcurPPfluxavglogxxxxxC
\else
\ifnum#1=72 %
\hatcurPPfluxavglogxxxxxD
\else
??????\fi
\fi
\fi
\fi
}
\newcommand{\hatcurPPfluxperi}[1]{\ifnum#1=47 %
\hatcurPPfluxperixxxxxA
\else
\ifnum#1=48 %
\hatcurPPfluxperixxxxxB
\else
\ifnum#1=49 %
\hatcurPPfluxperixxxxxC
\else
\ifnum#1=72 %
\hatcurPPfluxperixxxxxD
\else
??????\fi
\fi
\fi
\fi
}
\newcommand{\hatcurPPfluxperidim}[1]{\ifnum#1=47 %
\hatcurPPfluxperidimxxxxxA
\else
\ifnum#1=48 %
\hatcurPPfluxperidimxxxxxB
\else
\ifnum#1=49 %
\hatcurPPfluxperidimxxxxxC
\else
\ifnum#1=72 %
\hatcurPPfluxperidimxxxxxD
\else
??????\fi
\fi
\fi
\fi
}
\newcommand{\hatcurPPg}[1]{\ifnum#1=47 %
\hatcurPPgxxxxxA
\else
\ifnum#1=48 %
\hatcurPPgxxxxxB
\else
\ifnum#1=49 %
\hatcurPPgxxxxxC
\else
\ifnum#1=72 %
\hatcurPPgxxxxxD
\else
??????\fi
\fi
\fi
\fi
}
\newcommand{\hatcurPPi}[1]{\ifnum#1=47 %
\hatcurPPixxxxxA
\else
\ifnum#1=48 %
\hatcurPPixxxxxB
\else
\ifnum#1=49 %
\hatcurPPixxxxxC
\else
\ifnum#1=72 %
\hatcurPPixxxxxD
\else
??????\fi
\fi
\fi
\fi
}
\newcommand{\hatcurPPlogg}[1]{\ifnum#1=47 %
\hatcurPPloggxxxxxA
\else
\ifnum#1=48 %
\hatcurPPloggxxxxxB
\else
\ifnum#1=49 %
\hatcurPPloggxxxxxC
\else
\ifnum#1=72 %
\hatcurPPloggxxxxxD
\else
??????\fi
\fi
\fi
\fi
}
\newcommand{\hatcurPPm}[1]{\ifnum#1=47 %
\hatcurPPmxxxxxA
\else
\ifnum#1=48 %
\hatcurPPmxxxxxB
\else
\ifnum#1=49 %
\hatcurPPmxxxxxC
\else
\ifnum#1=72 %
\hatcurPPmxxxxxD
\else
??????\fi
\fi
\fi
\fi
}
\newcommand{\hatcurPPme}[1]{\ifnum#1=47 %
\hatcurPPmexxxxxA
\else
\ifnum#1=48 %
\hatcurPPmexxxxxB
\else
\ifnum#1=49 %
\hatcurPPmexxxxxC
\else
\ifnum#1=72 %
\hatcurPPmexxxxxD
\else
??????\fi
\fi
\fi
\fi
}
\newcommand{\hatcurPPmelong}[1]{\ifnum#1=47 %
\hatcurPPmelongxxxxxA
\else
\ifnum#1=48 %
\hatcurPPmelongxxxxxB
\else
\ifnum#1=49 %
\hatcurPPmelongxxxxxC
\else
\ifnum#1=72 %
\hatcurPPmelongxxxxxD
\else
??????\fi
\fi
\fi
\fi
}
\newcommand{\hatcurPPmeshort}[1]{\ifnum#1=47 %
\hatcurPPmeshortxxxxxA
\else
\ifnum#1=48 %
\hatcurPPmeshortxxxxxB
\else
\ifnum#1=49 %
\hatcurPPmeshortxxxxxC
\else
\ifnum#1=72 %
\hatcurPPmeshortxxxxxD
\else
??????\fi
\fi
\fi
\fi
}
\newcommand{\hatcurPPmlong}[1]{\ifnum#1=47 %
\hatcurPPmlongxxxxxA
\else
\ifnum#1=48 %
\hatcurPPmlongxxxxxB
\else
\ifnum#1=49 %
\hatcurPPmlongxxxxxC
\else
\ifnum#1=72 %
\hatcurPPmlongxxxxxD
\else
??????\fi
\fi
\fi
\fi
}
\newcommand{\hatcurPPmrcorr}[1]{\ifnum#1=47 %
\hatcurPPmrcorrxxxxxA
\else
\ifnum#1=48 %
\hatcurPPmrcorrxxxxxB
\else
\ifnum#1=49 %
\hatcurPPmrcorrxxxxxC
\else
\ifnum#1=72 %
\hatcurPPmrcorrxxxxxD
\else
??????\fi
\fi
\fi
\fi
}
\newcommand{\hatcurPPmshort}[1]{\ifnum#1=47 %
\hatcurPPmshortxxxxxA
\else
\ifnum#1=48 %
\hatcurPPmshortxxxxxB
\else
\ifnum#1=49 %
\hatcurPPmshortxxxxxC
\else
\ifnum#1=72 %
\hatcurPPmshortxxxxxD
\else
??????\fi
\fi
\fi
\fi
}
\newcommand{\hatcurPPmtwosiglim}[1]{\ifnum#1=47 %
\hatcurPPmtwosiglimxxxxxA
\else
\ifnum#1=48 %
\hatcurPPmtwosiglimxxxxxB
\else
\ifnum#1=49 %
\hatcurPPmtwosiglimxxxxxC
\else
\ifnum#1=72 %
\hatcurPPmtwosiglimxxxxxD
\else
??????\fi
\fi
\fi
\fi
}
\newcommand{\hatcurPPperi}[1]{\ifnum#1=47 %
\hatcurPPperixxxxxA
\else
\ifnum#1=48 %
\hatcurPPperixxxxxB
\else
\ifnum#1=49 %
\hatcurPPperixxxxxC
\else
\ifnum#1=72 %
\hatcurPPperixxxxxD
\else
??????\fi
\fi
\fi
\fi
}
\newcommand{\hatcurPPphiconj}[1]{\ifnum#1=47 %
\hatcurPPphiconjxxxxxA
\else
\ifnum#1=48 %
\hatcurPPphiconjxxxxxB
\else
\ifnum#1=49 %
\hatcurPPphiconjxxxxxC
\else
\ifnum#1=72 %
\hatcurPPphiconjxxxxxD
\else
??????\fi
\fi
\fi
\fi
}
\newcommand{\hatcurPPr}[1]{\ifnum#1=47 %
\hatcurPPrxxxxxA
\else
\ifnum#1=48 %
\hatcurPPrxxxxxB
\else
\ifnum#1=49 %
\hatcurPPrxxxxxC
\else
\ifnum#1=72 %
\hatcurPPrxxxxxD
\else
??????\fi
\fi
\fi
\fi
}
\newcommand{\hatcurPPre}[1]{\ifnum#1=47 %
\hatcurPPrexxxxxA
\else
\ifnum#1=48 %
\hatcurPPrexxxxxB
\else
\ifnum#1=49 %
\hatcurPPrexxxxxC
\else
\ifnum#1=72 %
\hatcurPPrexxxxxD
\else
??????\fi
\fi
\fi
\fi
}
\newcommand{\hatcurPPrelong}[1]{\ifnum#1=47 %
\hatcurPPrelongxxxxxA
\else
\ifnum#1=48 %
\hatcurPPrelongxxxxxB
\else
\ifnum#1=49 %
\hatcurPPrelongxxxxxC
\else
\ifnum#1=72 %
\hatcurPPrelongxxxxxD
\else
??????\fi
\fi
\fi
\fi
}
\newcommand{\hatcurPPreshort}[1]{\ifnum#1=47 %
\hatcurPPreshortxxxxxA
\else
\ifnum#1=48 %
\hatcurPPreshortxxxxxB
\else
\ifnum#1=49 %
\hatcurPPreshortxxxxxC
\else
\ifnum#1=72 %
\hatcurPPreshortxxxxxD
\else
??????\fi
\fi
\fi
\fi
}
\newcommand{\hatcurPPrho}[1]{\ifnum#1=47 %
\hatcurPPrhoxxxxxA
\else
\ifnum#1=48 %
\hatcurPPrhoxxxxxB
\else
\ifnum#1=49 %
\hatcurPPrhoxxxxxC
\else
\ifnum#1=72 %
\hatcurPPrhoxxxxxD
\else
??????\fi
\fi
\fi
\fi
}
\newcommand{\hatcurPPrlong}[1]{\ifnum#1=47 %
\hatcurPPrlongxxxxxA
\else
\ifnum#1=48 %
\hatcurPPrlongxxxxxB
\else
\ifnum#1=49 %
\hatcurPPrlongxxxxxC
\else
\ifnum#1=72 %
\hatcurPPrlongxxxxxD
\else
??????\fi
\fi
\fi
\fi
}
\newcommand{\hatcurPPrshort}[1]{\ifnum#1=47 %
\hatcurPPrshortxxxxxA
\else
\ifnum#1=48 %
\hatcurPPrshortxxxxxB
\else
\ifnum#1=49 %
\hatcurPPrshortxxxxxC
\else
\ifnum#1=72 %
\hatcurPPrshortxxxxxD
\else
??????\fi
\fi
\fi
\fi
}
\newcommand{\hatcurPPtcirc}[1]{\ifnum#1=47 %
\hatcurPPtcircxxxxxA
\else
\ifnum#1=48 %
\hatcurPPtcircxxxxxB
\else
\ifnum#1=49 %
\hatcurPPtcircxxxxxC
\else
\ifnum#1=72 %
\hatcurPPtcircxxxxxD
\else
??????\fi
\fi
\fi
\fi
}
\newcommand{\hatcurPPteff}[1]{\ifnum#1=47 %
\hatcurPPteffxxxxxA
\else
\ifnum#1=48 %
\hatcurPPteffxxxxxB
\else
\ifnum#1=49 %
\hatcurPPteffxxxxxC
\else
\ifnum#1=72 %
\hatcurPPteffxxxxxD
\else
??????\fi
\fi
\fi
\fi
}
\newcommand{\hatcurPPtheta}[1]{\ifnum#1=47 %
\hatcurPPthetaxxxxxA
\else
\ifnum#1=48 %
\hatcurPPthetaxxxxxB
\else
\ifnum#1=49 %
\hatcurPPthetaxxxxxC
\else
\ifnum#1=72 %
\hatcurPPthetaxxxxxD
\else
??????\fi
\fi
\fi
\fi
}
\newcommand{\hatcurPPtinfall}[1]{\ifnum#1=47 %
\hatcurPPtinfallxxxxxA
\else
\ifnum#1=48 %
\hatcurPPtinfallxxxxxB
\else
\ifnum#1=49 %
\hatcurPPtinfallxxxxxC
\else
\ifnum#1=72 %
\hatcurPPtinfallxxxxxD
\else
??????\fi
\fi
\fi
\fi
}
\newcommand{\hatcurRVeccen}[1]{\ifnum#1=47 %
\hatcurRVeccenxxxxxA
\else
\ifnum#1=48 %
\hatcurRVeccenxxxxxB
\else
\ifnum#1=49 %
\hatcurRVeccenxxxxxC
\else
\ifnum#1=72 %
\hatcurRVeccenxxxxxD
\else
??????\fi
\fi
\fi
\fi
}
\newcommand{\hatcurRVeccentwosiglim}[1]{\ifnum#1=47 %
\hatcurRVeccentwosiglimxxxxxA
\else
\ifnum#1=48 %
\hatcurRVeccentwosiglimxxxxxB
\else
\ifnum#1=49 %
\hatcurRVeccentwosiglimxxxxxC
\else
\ifnum#1=72 %
\hatcurRVeccentwosiglimxxxxxD
\else
??????\fi
\fi
\fi
\fi
}
\newcommand{\hatcurRVfitrms}[1]{\ifnum#1=47 %
\hatcurRVfitrmsxxxxxA
\else
\ifnum#1=48 %
\hatcurRVfitrmsxxxxxB
\else
\ifnum#1=49 %
\hatcurRVfitrmsxxxxxC
\else
??????\fi
\fi
\fi
}
\newcommand{\hatcurRVfitrmsA}[1]{\ifnum#1=72 %
\hatcurRVfitrmsAxxxxxD
\else
??????\fi
}
\newcommand{\hatcurRVfitrmsB}[1]{\ifnum#1=72 %
\hatcurRVfitrmsBxxxxxD
\else
??????\fi
}
\newcommand{\hatcurRVfitrmsC}[1]{\ifnum#1=72 %
\hatcurRVfitrmsCxxxxxD
\else
??????\fi
}
\newcommand{\hatcurRVfitrmsD}[1]{\ifnum#1=72 %
\hatcurRVfitrmsDxxxxxD
\else
??????\fi
}
\newcommand{\hatcurRVgamma}[1]{\ifnum#1=47 %
\hatcurRVgammaxxxxxA
\else
\ifnum#1=48 %
\hatcurRVgammaxxxxxB
\else
\ifnum#1=49 %
\hatcurRVgammaxxxxxC
\else
??????\fi
\fi
\fi
}
\newcommand{\hatcurRVgammaA}[1]{\ifnum#1=72 %
\hatcurRVgammaAxxxxxD
\else
??????\fi
}
\newcommand{\hatcurRVgammaB}[1]{\ifnum#1=72 %
\hatcurRVgammaBxxxxxD
\else
??????\fi
}
\newcommand{\hatcurRVgammaC}[1]{\ifnum#1=72 %
\hatcurRVgammaCxxxxxD
\else
??????\fi
}
\newcommand{\hatcurRVgammaD}[1]{\ifnum#1=72 %
\hatcurRVgammaDxxxxxD
\else
??????\fi
}
\newcommand{\hatcurRVh}[1]{\ifnum#1=47 %
\hatcurRVhxxxxxA
\else
\ifnum#1=48 %
\hatcurRVhxxxxxB
\else
\ifnum#1=49 %
\hatcurRVhxxxxxC
\else
\ifnum#1=72 %
\hatcurRVhxxxxxD
\else
??????\fi
\fi
\fi
\fi
}
\newcommand{\hatcurRVjitter}[1]{\ifnum#1=47 %
\hatcurRVjitterxxxxxA
\else
\ifnum#1=48 %
\hatcurRVjitterxxxxxB
\else
\ifnum#1=49 %
\hatcurRVjitterxxxxxC
\else
??????\fi
\fi
\fi
}
\newcommand{\hatcurRVjitterA}[1]{\ifnum#1=72 %
\hatcurRVjitterAxxxxxD
\else
??????\fi
}
\newcommand{\hatcurRVjitterB}[1]{\ifnum#1=72 %
\hatcurRVjitterBxxxxxD
\else
??????\fi
}
\newcommand{\hatcurRVjitterC}[1]{\ifnum#1=72 %
\hatcurRVjitterCxxxxxD
\else
??????\fi
}
\newcommand{\hatcurRVjitterD}[1]{\ifnum#1=72 %
\hatcurRVjitterDxxxxxD
\else
??????\fi
}
\newcommand{\hatcurRVjittertwosiglim}[1]{\ifnum#1=47 %
\hatcurRVjittertwosiglimxxxxxA
\else
\ifnum#1=48 %
\hatcurRVjittertwosiglimxxxxxB
\else
\ifnum#1=49 %
\hatcurRVjittertwosiglimxxxxxC
\else
??????\fi
\fi
\fi
}
\newcommand{\hatcurRVjittertwosiglimA}[1]{\ifnum#1=72 %
\hatcurRVjittertwosiglimAxxxxxD
\else
??????\fi
}
\newcommand{\hatcurRVjittertwosiglimB}[1]{\ifnum#1=72 %
\hatcurRVjittertwosiglimBxxxxxD
\else
??????\fi
}
\newcommand{\hatcurRVjittertwosiglimC}[1]{\ifnum#1=72 %
\hatcurRVjittertwosiglimCxxxxxD
\else
??????\fi
}
\newcommand{\hatcurRVjittertwosiglimD}[1]{\ifnum#1=72 %
\hatcurRVjittertwosiglimDxxxxxD
\else
??????\fi
}
\newcommand{\hatcurRVk}[1]{\ifnum#1=47 %
\hatcurRVkxxxxxA
\else
\ifnum#1=48 %
\hatcurRVkxxxxxB
\else
\ifnum#1=49 %
\hatcurRVkxxxxxC
\else
\ifnum#1=72 %
\hatcurRVkxxxxxD
\else
??????\fi
\fi
\fi
\fi
}
\newcommand{\hatcurRVK}[1]{\ifnum#1=47 %
\hatcurRVKxxxxxA
\else
\ifnum#1=48 %
\hatcurRVKxxxxxB
\else
\ifnum#1=49 %
\hatcurRVKxxxxxC
\else
\ifnum#1=72 %
\hatcurRVKxxxxxD
\else
??????\fi
\fi
\fi
\fi
}
\newcommand{\hatcurRVKtwosiglim}[1]{\ifnum#1=47 %
\hatcurRVKtwosiglimxxxxxA
\else
\ifnum#1=48 %
\hatcurRVKtwosiglimxxxxxB
\else
\ifnum#1=49 %
\hatcurRVKtwosiglimxxxxxC
\else
\ifnum#1=72 %
\hatcurRVKtwosiglimxxxxxD
\else
??????\fi
\fi
\fi
\fi
}
\newcommand{\hatcurRVomega}[1]{\ifnum#1=47 %
\hatcurRVomegaxxxxxA
\else
\ifnum#1=48 %
\hatcurRVomegaxxxxxB
\else
\ifnum#1=49 %
\hatcurRVomegaxxxxxC
\else
\ifnum#1=72 %
\hatcurRVomegaxxxxxD
\else
??????\fi
\fi
\fi
\fi
}
\newcommand{\hatcurRVrh}[1]{\ifnum#1=47 %
\hatcurRVrhxxxxxA
\else
\ifnum#1=48 %
\hatcurRVrhxxxxxB
\else
\ifnum#1=49 %
\hatcurRVrhxxxxxC
\else
\ifnum#1=72 %
\hatcurRVrhxxxxxD
\else
??????\fi
\fi
\fi
\fi
}
\newcommand{\hatcurRVrk}[1]{\ifnum#1=47 %
\hatcurRVrkxxxxxA
\else
\ifnum#1=48 %
\hatcurRVrkxxxxxB
\else
\ifnum#1=49 %
\hatcurRVrkxxxxxC
\else
\ifnum#1=72 %
\hatcurRVrkxxxxxD
\else
??????\fi
\fi
\fi
\fi
}
\newcommand{\hatcurRVtrone}[1]{\ifnum#1=47 %
\hatcurRVtronexxxxxA
\else
\ifnum#1=48 %
\hatcurRVtronexxxxxB
\else
\ifnum#1=49 %
\hatcurRVtronexxxxxC
\else
\ifnum#1=72 %
\hatcurRVtronexxxxxD
\else
??????\fi
\fi
\fi
\fi
}
\newcommand{\hatcurRVtrtwo}[1]{\ifnum#1=47 %
\hatcurRVtrtwoxxxxxA
\else
\ifnum#1=48 %
\hatcurRVtrtwoxxxxxB
\else
\ifnum#1=49 %
\hatcurRVtrtwoxxxxxC
\else
\ifnum#1=72 %
\hatcurRVtrtwoxxxxxD
\else
??????\fi
\fi
\fi
\fi
}
\newcommand{\hatcurSMEiilogg}[1]{\ifnum#1=47 %
\hatcurSMEiiloggxxxxxA
\else
??????\fi
}
\newcommand{\hatcurSMEiiteff}[1]{\ifnum#1=47 %
\hatcurSMEiiteffxxxxxA
\else
??????\fi
}
\newcommand{\hatcurSMEiivsin}[1]{\ifnum#1=47 %
\hatcurSMEiivsinxxxxxA
\else
??????\fi
}
\newcommand{\hatcurSMEiizfeh}[1]{\ifnum#1=47 %
\hatcurSMEiizfehxxxxxA
\else
??????\fi
}
\newcommand{\hatcurSMEiizfehshort}[1]{\ifnum#1=47 %
\hatcurSMEiizfehshortxxxxxA
\else
??????\fi
}
\newcommand{\hatcurSMEilogg}[1]{\ifnum#1=47 %
\hatcurSMEiloggxxxxxA
\else
\ifnum#1=48 %
\hatcurSMEiloggxxxxxB
\else
\ifnum#1=49 %
\hatcurSMEiloggxxxxxC
\else
\ifnum#1=72 %
\hatcurSMEiloggxxxxxD
\else
??????\fi
\fi
\fi
\fi
}
\newcommand{\hatcurSMEiteff}[1]{\ifnum#1=47 %
\hatcurSMEiteffxxxxxA
\else
\ifnum#1=48 %
\hatcurSMEiteffxxxxxB
\else
\ifnum#1=49 %
\hatcurSMEiteffxxxxxC
\else
\ifnum#1=72 %
\hatcurSMEiteffxxxxxD
\else
??????\fi
\fi
\fi
\fi
}
\newcommand{\hatcurSMEivmac}[1]{\ifnum#1=47 %
\hatcurSMEivmacxxxxxA
\else
\ifnum#1=48 %
\hatcurSMEivmacxxxxxB
\else
\ifnum#1=49 %
\hatcurSMEivmacxxxxxC
\else
\ifnum#1=72 %
\hatcurSMEivmacxxxxxD
\else
??????\fi
\fi
\fi
\fi
}
\newcommand{\hatcurSMEivmic}[1]{\ifnum#1=47 %
\hatcurSMEivmicxxxxxA
\else
\ifnum#1=48 %
\hatcurSMEivmicxxxxxB
\else
\ifnum#1=49 %
\hatcurSMEivmicxxxxxC
\else
\ifnum#1=72 %
\hatcurSMEivmicxxxxxD
\else
??????\fi
\fi
\fi
\fi
}
\newcommand{\hatcurSMEivsin}[1]{\ifnum#1=47 %
\hatcurSMEivsinxxxxxA
\else
\ifnum#1=48 %
\hatcurSMEivsinxxxxxB
\else
\ifnum#1=49 %
\hatcurSMEivsinxxxxxC
\else
\ifnum#1=72 %
\hatcurSMEivsinxxxxxD
\else
??????\fi
\fi
\fi
\fi
}
\newcommand{\hatcurSMEizfeh}[1]{\ifnum#1=47 %
\hatcurSMEizfehxxxxxA
\else
\ifnum#1=48 %
\hatcurSMEizfehxxxxxB
\else
\ifnum#1=49 %
\hatcurSMEizfehxxxxxC
\else
\ifnum#1=72 %
\hatcurSMEizfehxxxxxD
\else
??????\fi
\fi
\fi
\fi
}
\newcommand{\hatcurSMEizfehshort}[1]{\ifnum#1=47 %
\hatcurSMEizfehshortxxxxxA
\else
\ifnum#1=48 %
\hatcurSMEizfehshortxxxxxB
\else
\ifnum#1=49 %
\hatcurSMEizfehshortxxxxxC
\else
\ifnum#1=72 %
\hatcurSMEizfehshortxxxxxD
\else
??????\fi
\fi
\fi
\fi
}
\newcommand{\hatcurXAv}[1]{\ifnum#1=47 %
\hatcurXAvxxxxxA
\else
\ifnum#1=48 %
\hatcurXAvxxxxxB
\else
\ifnum#1=49 %
\hatcurXAvxxxxxC
\else
\ifnum#1=72 %
\hatcurXAvxxxxxD
\else
??????\fi
\fi
\fi
\fi
}
\newcommand{\hatcurXdist}[1]{\ifnum#1=47 %
\hatcurXdistxxxxxA
\else
\ifnum#1=48 %
\hatcurXdistxxxxxB
\else
\ifnum#1=49 %
\hatcurXdistxxxxxC
\else
\ifnum#1=72 %
\hatcurXdistxxxxxD
\else
??????\fi
\fi
\fi
\fi
}
\newcommand{\hatcurXdistred}[1]{\ifnum#1=47 %
\hatcurXdistredxxxxxA
\else
\ifnum#1=48 %
\hatcurXdistredxxxxxB
\else
\ifnum#1=49 %
\hatcurXdistredxxxxxC
\else
\ifnum#1=72 %
\hatcurXdistredxxxxxD
\else
??????\fi
\fi
\fi
\fi
}
\newcommand{\hatcurXEBV}[1]{\ifnum#1=47 %
\hatcurXEBVxxxxxA
\else
\ifnum#1=48 %
\hatcurXEBVxxxxxB
\else
\ifnum#1=49 %
\hatcurXEBVxxxxxC
\else
\ifnum#1=72 %
\hatcurXEBVxxxxxD
\else
??????\fi
\fi
\fi
\fi
}
\newcommand{\hatcurXsecdur}[1]{\ifnum#1=47 %
\hatcurXsecdurxxxxxA
\else
\ifnum#1=48 %
\hatcurXsecdurxxxxxB
\else
\ifnum#1=49 %
\hatcurXsecdurxxxxxC
\else
\ifnum#1=72 %
\hatcurXsecdurxxxxxD
\else
??????\fi
\fi
\fi
\fi
}
\newcommand{\hatcurXsecingdur}[1]{\ifnum#1=47 %
\hatcurXsecingdurxxxxxA
\else
\ifnum#1=48 %
\hatcurXsecingdurxxxxxB
\else
\ifnum#1=49 %
\hatcurXsecingdurxxxxxC
\else
\ifnum#1=72 %
\hatcurXsecingdurxxxxxD
\else
??????\fi
\fi
\fi
\fi
}
\newcommand{\hatcurXsecondary}[1]{\ifnum#1=47 %
\hatcurXsecondaryxxxxxA
\else
\ifnum#1=48 %
\hatcurXsecondaryxxxxxB
\else
\ifnum#1=49 %
\hatcurXsecondaryxxxxxC
\else
\ifnum#1=72 %
\hatcurXsecondaryxxxxxD
\else
??????\fi
\fi
\fi
\fi
}
\newcommand{\hatcurXsecphase}[1]{\ifnum#1=47 %
\hatcurXsecphasexxxxxA
\else
\ifnum#1=48 %
\hatcurXsecphasexxxxxB
\else
\ifnum#1=49 %
\hatcurXsecphasexxxxxC
\else
\ifnum#1=72 %
\hatcurXsecphasexxxxxD
\else
??????\fi
\fi
\fi
\fi
}
\newcommand{\hatcurhtreccenxxxxxA}{HATS747-014}                      
\newcommand{\hatcurfieldeccenxxxxxA}{\ensuremath{string}}            
\newcommand{\hatcurCCraeccenxxxxxA}{\ensuremath{19^{\mathrm h}09^{\mathrm m}56.2504{\mathrm s}}}                   
\newcommand{\hatcurCCdececcenxxxxxA}{\ensuremath{-49{\arcdeg}39{\arcmin}53.868{\arcsec}}}                  
\newcommand{\hatcurCCmageccenxxxxxA}{14.829}                         
\newcommand{\hatcurCCtwomasseccenxxxxxA}{2MASS~19095625-4939538}     
\newcommand{\hatcurCCgsceccenxxxxxA}{GSC~}                           
\newcommand{\hatcurCCgaiaeccenxxxxxA}{GAIA~6658373007402886400}      
\newcommand{\hatcurCCgaiadrtwoeccenxxxxxA}{GAIA~DR2~6658373007402886400} 
\newcommand{\hatcurCCtassmveccenxxxxxA}{\ensuremath{14.829\pm0.010}} 
\newcommand{\hatcurCCtassmvshorteccenxxxxxA}{\ensuremath{14.8}}      
\newcommand{\hatcurCCtassmBeccenxxxxxA}{\ensuremath{16.101\pm0.040}} 
\newcommand{\hatcurCCtassmBshorteccenxxxxxA}{\ensuremath{16.1}}      
\newcommand{\hatcurCCtassmIeccenxxxxxA}{\ensuremath{nff\pmnff}}      
\newcommand{\hatcurCCtassmIshorteccenxxxxxA}{\ensuremath{0.0}}       
\newcommand{\hatcurCCtassmgeccenxxxxxA}{\ensuremath{15.480\pm0.010}} 
\newcommand{\hatcurCCtassmgshorteccenxxxxxA}{\ensuremath{15.5}}      
\newcommand{\hatcurCCtassmreccenxxxxxA}{\ensuremath{14.398\pm0.010}} 
\newcommand{\hatcurCCtassmrshorteccenxxxxxA}{\ensuremath{14.4}}      
\newcommand{\hatcurCCtassmieccenxxxxxA}{\ensuremath{14.009\pm0.010}} 
\newcommand{\hatcurCCtassmishorteccenxxxxxA}{\ensuremath{14.0}}      
\newcommand{\hatcurCCparallaxeccenxxxxxA}{\ensuremath{3.298\pm0.042}} 
\newcommand{\hatcurCCgaiamGeccenxxxxxA}{\ensuremath{14.39980\pm0.00040}} 
\newcommand{\hatcurCCgaiamBPeccenxxxxxA}{\ensuremath{15.0858\pm0.0021}} 
\newcommand{\hatcurCCgaiamRPeccenxxxxxA}{\ensuremath{13.61140\pm0.00090}} 
\newcommand{\hatcurCCtwomassJmageccenxxxxxA}{\ensuremath{12.653\pm0.023}} 
\newcommand{\hatcurCCtwomassHmageccenxxxxxA}{\ensuremath{12.026\pm0.023}} 
\newcommand{\hatcurCCtwomassKmageccenxxxxxA}{\ensuremath{11.926\pm0.025}} 
\newcommand{\hatcurCCcitJmageccenxxxxxA}{\ensuremath{12.651\pm0.024}} 
\newcommand{\hatcurCCcitHmageccenxxxxxA}{\ensuremath{12.020\pm0.024}} 
\newcommand{\hatcurCCcitKmageccenxxxxxA}{\ensuremath{11.950\pm0.025}} 
\newcommand{\hatcurCCbbJmageccenxxxxxA}{\ensuremath{12.729\pm0.026}} 
\newcommand{\hatcurCCbbHmageccenxxxxxA}{\ensuremath{12.042\pm0.025}} 
\newcommand{\hatcurCCbbKmageccenxxxxxA}{\ensuremath{11.970\pm0.025}} 
\newcommand{\hatcurCCesoJmageccenxxxxxA}{\ensuremath{12.736\pm0.030}} 
\newcommand{\hatcurCCesoHmageccenxxxxxA}{\ensuremath{12.036\pm0.029}} 
\newcommand{\hatcurCCesoKmageccenxxxxxA}{\ensuremath{11.967\pm0.026}} 
\newcommand{\hatcurCCesoJHmageccenxxxxxA}{\ensuremath{0.699\pm0.018}} 
\newcommand{\hatcurCCesoJKmageccenxxxxxA}{\ensuremath{0.769\pm0.038}} 
\newcommand{\hatcurCCesoHKmageccenxxxxxA}{\ensuremath{0.069\pm0.038}} 
\newcommand{\hatcurCCWonemageccenxxxxxA}{\ensuremath{11.867\pm0.023}} 
\newcommand{\hatcurCCWtwomageccenxxxxxA}{\ensuremath{11.947\pm0.024}} 
\newcommand{\hatcurCCWthreemageccenxxxxxA}{\ensuremath{0\pm0}}       
\newcommand{\hatcurCCWfourmageccenxxxxxA}{\ensuremath{0\pm0}}        
\newcommand{\hatcurLCdipeccenxxxxxA}{\ensuremath{31.1}}              
\newcommand{\hatcurLCrprstareccenxxxxxA}{\ensuremath{0.1772\pm0.0018}} 
\newcommand{\hatcurLCbsqeccenxxxxxA}{\ensuremath{0.564_{-0.024}^{+0.015}}} 
\newcommand{\hatcurLCimpeccenxxxxxA}{\ensuremath{0.751_{-0.016}^{+0.010}}} 
\newcommand{\hatcurLCzetaeccenxxxxxA}{\ensuremath{34.23\pm0.32}}     
\newcommand{\hatcurLCdureccenxxxxxA}{\ensuremath{0.0803\pm0.0011}}   
\newcommand{\hatcurLCdurshorteccenxxxxxA}{\ensuremath{0.0803}}       
\newcommand{\hatcurLCdurhreccenxxxxxA}{\ensuremath{1.927\pm0.027}}   
\newcommand{\hatcurLCdurhrshorteccenxxxxxA}{\ensuremath{1.927}}      
\newcommand{\hatcurLCqeccenxxxxxA}{\ensuremath{0.02050\pm0.00029}}   
\newcommand{\hatcurLCqshorteccenxxxxxA}{\ensuremath{0.021}}          
\newcommand{\hatcurLCingdureccenxxxxxA}{\ensuremath{0.0252\pm0.0017}} 
\newcommand{\hatcurLCPeccenxxxxxA}{\ensuremath{3.9228093\pm0.0000025}} 
\newcommand{\hatcurLCPprececcenxxxxxA}{\ensuremath{3.9228093}}       
\newcommand{\hatcurLCPshorteccenxxxxxA}{\ensuremath{3.9228}}         
\newcommand{\hatcurLCTeccenxxxxxA}{\ensuremath{2456988.76816\pm0.00027}} 
\newcommand{\hatcurLCTAeccenxxxxxA}{\ensuremath{2456368.96432\pm0.00051}} 
\newcommand{\hatcurLCTBeccenxxxxxA}{\ensuremath{2457604.64918\pm0.00046}} 
\newcommand{\hatcurLChatnetmAeccenxxxxxA}{\ensuremath{14.39327\pm0.00021}} 
\newcommand{\hatcurLCiblendAeccenxxxxxA}{\ensuremath{0.927\pm0.038}} 
\newcommand{\hatcurLChatnetmBeccenxxxxxA}{\ensuremath{-0.00002\pm0.00013}} 
\newcommand{\hatcurLCiblendBeccenxxxxxA}{\ensuremath{0.642\pm0.026}} 
\newcommand{\hatcurLCrhoeccenxxxxxA}{\ensuremath{3.31\pm0.13}}       
\newcommand{\hatcurSMEiteffeccenxxxxxA}{\ensuremath{4479\pm51}}      
\newcommand{\hatcurSMEizfeheccenxxxxxA}{\ensuremath{-0.140\pm0.066}} 
\newcommand{\hatcurSMEizfehshorteccenxxxxxA}{\ensuremath{-0.14}}     
\newcommand{\hatcurSMEiloggeccenxxxxxA}{\ensuremath{4.76\pm0.21}}    
\newcommand{\hatcurSMEivsineccenxxxxxA}{\ensuremath{2.47\pm0.70}}    
\newcommand{\hatcurSMEivmaceccenxxxxxA}{\ensuremath{nff\pmnff}}      
\newcommand{\hatcurSMEivmiceccenxxxxxA}{\ensuremath{nff\pmnff}}      
\newcommand{\hatcurSMEiiteffeccenxxxxxA}{\ensuremath{4450\pm86}}     
\newcommand{\hatcurSMEiizfeheccenxxxxxA}{\ensuremath{-0.14\pm0.10}}  
\newcommand{\hatcurSMEiizfehshorteccenxxxxxA}{\ensuremath{-0.14}}    
\newcommand{\hatcurSMEiiloggeccenxxxxxA}{\ensuremath{4.603\pm0.031}} 
\newcommand{\hatcurSMEiivsineccenxxxxxA}{\ensuremath{2.8\pm1.4}}     
\newcommand{\hatcurextraerrMgeccenxxxxxA}{\ensuremath{0.017_{-0.016}^{+0.033}}} 
\newcommand{\hatcurextraerrMgtwosiglimeccenxxxxxA}{\ensuremath{<0.0751}} 
\newcommand{\hatcurextraerrMreccenxxxxxA}{\ensuremath{0.098_{-0.070}^{+0.144}}} 
\newcommand{\hatcurextraerrMrtwosiglimeccenxxxxxA}{\ensuremath{<0.3298}} 
\newcommand{\hatcurextraerrMieccenxxxxxA}{\ensuremath{0.0053_{-0.0050}^{+0.0783}}} 
\newcommand{\hatcurextraerrMitwosiglimeccenxxxxxA}{\ensuremath{<0.2140}} 
\newcommand{\hatcurextraerrMGeccenxxxxxA}{\ensuremath{0.039_{-0.020}^{+0.028}}} 
\newcommand{\hatcurextraerrMGtwosiglimeccenxxxxxA}{\ensuremath{<0.0806}} 
\newcommand{\hatcurextraerrMBPtwoeccenxxxxxA}{\ensuremath{0.038\pm0.027}} 
\newcommand{\hatcurextraerrMBPtwotwosiglimeccenxxxxxA}{\ensuremath{<0.0855}} 
\newcommand{\hatcurextraerrMRPeccenxxxxxA}{\ensuremath{0.019_{-0.018}^{+0.051}}} 
\newcommand{\hatcurextraerrMRPtwosiglimeccenxxxxxA}{\ensuremath{<0.0992}} 
\newcommand{\hatcurextraerrMJeccenxxxxxA}{\ensuremath{0.023_{-0.021}^{+0.035}}} 
\newcommand{\hatcurextraerrMJtwosiglimeccenxxxxxA}{\ensuremath{<0.0776}} 
\newcommand{\hatcurextraerrMHeccenxxxxxA}{\ensuremath{0.025\pm0.022}} 
\newcommand{\hatcurextraerrMHtwosiglimeccenxxxxxA}{\ensuremath{<0.0682}} 
\newcommand{\hatcurextraerrMKseccenxxxxxA}{\ensuremath{0.037_{-0.028}^{+0.037}}} 
\newcommand{\hatcurextraerrMKstwosiglimeccenxxxxxA}{\ensuremath{<0.0938}} 
\newcommand{\hatcurextraerrMWoneeccenxxxxxA}{\ensuremath{0.014_{-0.013}^{+0.030}}} 
\newcommand{\hatcurextraerrMWonetwosiglimeccenxxxxxA}{\ensuremath{<0.0605}} 
\newcommand{\hatcurextraerrMWtwoeccenxxxxxA}{\ensuremath{0.012_{-0.012}^{+0.049}}} 
\newcommand{\hatcurextraerrMWtwotwosiglimeccenxxxxxA}{\ensuremath{<0.0975}} 
\newcommand{\hatcurLBiBeccenxxxxxA}{\ensuremath{0.9498}}             
\newcommand{\hatcurLBiiBeccenxxxxxA}{\ensuremath{-0.0970}}           
\newcommand{\hatcurLBiVeccenxxxxxA}{\ensuremath{0.7431}}             
\newcommand{\hatcurLBiiVeccenxxxxxA}{\ensuremath{0.0539}}            
\newcommand{\hatcurLBiReccenxxxxxA}{\ensuremath{0.6121}}             
\newcommand{\hatcurLBiiReccenxxxxxA}{\ensuremath{0.1147}}            
\newcommand{\hatcurLBiIeccenxxxxxA}{\ensuremath{0.4764}}             
\newcommand{\hatcurLBiiIeccenxxxxxA}{\ensuremath{0.1543}}            
\newcommand{\hatcurLBiueccenxxxxxA}{\ensuremath{1.1009}}             
\newcommand{\hatcurLBiiueccenxxxxxA}{\ensuremath{-0.2662}}           
\newcommand{\hatcurLBigeccenxxxxxA}{\ensuremath{0.8521}}             
\newcommand{\hatcurLBiigeccenxxxxxA}{\ensuremath{-0.0323}}           
\newcommand{\hatcurLBireccenxxxxxA}{\ensuremath{0.50\pm0.22}}        
\newcommand{\hatcurLBiireccenxxxxxA}{\ensuremath{-0.00\pm0.19}}      
\newcommand{\hatcurLBiieccenxxxxxA}{\ensuremath{0.41\pm0.10}}        
\newcommand{\hatcurLBiiieccenxxxxxA}{\ensuremath{0.24\pm0.13}}       
\newcommand{\hatcurLBizeccenxxxxxA}{\ensuremath{0.4163}}             
\newcommand{\hatcurLBiizeccenxxxxxA}{\ensuremath{0.1659}}            
\newcommand{\hatcurLBiJeccenxxxxxA}{\ensuremath{0.2917}}             
\newcommand{\hatcurLBiiJeccenxxxxxA}{\ensuremath{0.2132}}            
\newcommand{\hatcurLBiHeccenxxxxxA}{\ensuremath{0.1585}}             
\newcommand{\hatcurLBiiHeccenxxxxxA}{\ensuremath{0.2975}}            
\newcommand{\hatcurLBiKeccenxxxxxA}{\ensuremath{0.1313}}             
\newcommand{\hatcurLBiiKeccenxxxxxA}{\ensuremath{0.2488}}            
\newcommand{\hatcurLBiTeccenxxxxxA}{\ensuremath{0.52_{-0.14}^{+0.19}}} 
\newcommand{\hatcurLBiiTeccenxxxxxA}{\ensuremath{0.07_{-0.22}^{+0.30}}} 
\newcommand{\hatcurLBikepeccenxxxxxA}{\ensuremath{0.6124}}           
\newcommand{\hatcurLBiikepeccenxxxxxA}{\ensuremath{0.1247}}          
\newcommand{\hatcurLBiCeccenxxxxxA}{\ensuremath{0.5860}}             
\newcommand{\hatcurLBiiCeccenxxxxxA}{\ensuremath{0.1328}}            
\newcommand{\hatcurLBiMeccenxxxxxA}{\ensuremath{0.7144}}             
\newcommand{\hatcurLBiiMeccenxxxxxA}{\ensuremath{0.0734}}            
\newcommand{\hatcurLBiSoneeccenxxxxxA}{\ensuremath{0.1118}}            
\newcommand{\hatcurLBiiSoneeccenxxxxxA}{\ensuremath{0.1665}}           
\newcommand{\hatcurLBiStwoeccenxxxxxA}{\ensuremath{0.0993}}            
\newcommand{\hatcurLBiiStwoeccenxxxxxA}{\ensuremath{0.1321}}           
\newcommand{\hatcurLBiSthreeeccenxxxxxA}{\ensuremath{0.0826}}            
\newcommand{\hatcurLBiiSthreeeccenxxxxxA}{\ensuremath{0.1091}}           
\newcommand{\hatcurLBiSfoureccenxxxxxA}{\ensuremath{0.0716}}            
\newcommand{\hatcurLBiiSfoureccenxxxxxA}{\ensuremath{0.0934}}           
\newcommand{\hatcurISOmeccenxxxxxA}{\ensuremath{0.685_{-0.013}^{+0.020}}} 
\newcommand{\hatcurISOmshorteccenxxxxxA}{\ensuremath{0.68}}          
\newcommand{\hatcurISOmlongeccenxxxxxA}{\ensuremath{0.685_{-0.013}^{+0.020}}} 
\newcommand{\hatcurISOreccenxxxxxA}{\ensuremath{0.6645\pm0.0075}}    
\newcommand{\hatcurISOrshorteccenxxxxxA}{\ensuremath{0.66}}          
\newcommand{\hatcurISOrlongeccenxxxxxA}{\ensuremath{0.6645\pm0.0075}} 
\newcommand{\hatcurISOrhoeccenxxxxxA}{\ensuremath{3.31\pm0.13}}      
\newcommand{\hatcurISOrholongeccenxxxxxA}{\ensuremath{3.31\pm0.13}}  
\newcommand{\hatcurISOloggeccenxxxxxA}{\ensuremath{4.630\pm0.013}}   
\newcommand{\hatcurISOlumeccenxxxxxA}{\ensuremath{0.1673\pm0.0063}}  
\newcommand{\hatcurISOlumshorteccenxxxxxA}{\ensuremath{0.17}}        
\newcommand{\hatcurISOteffeccenxxxxxA}{\ensuremath{4534\pm47}}       
\newcommand{\hatcurISOzfeheccenxxxxxA}{\ensuremath{-0.082\pm0.059}}  
\newcommand{\hatcurISOageeccenxxxxxA}{\ensuremath{7.5\pm3.3}}        
\newcommand{\hatcurISOspececcenxxxxxA}{K}                            
\newcommand{\hatcurRVKeccenxxxxxA}{\ensuremath{42.4\pm3.4}}          
\newcommand{\hatcurRVKtwosiglimeccenxxxxxA}{\ensuremath{<47.5}}      
\newcommand{\hatcurRVrkeccenxxxxxA}{\ensuremath{0.08\pm0.11}}        
\newcommand{\hatcurRVrheccenxxxxxA}{\ensuremath{0.087\pm0.089}}      
\newcommand{\hatcurRVkeccenxxxxxA}{\ensuremath{0.009_{-0.013}^{+0.027}}} 
\newcommand{\hatcurRVheccenxxxxxA}{\ensuremath{0.011_{-0.011}^{+0.026}}} 
\newcommand{\hatcurRVtroneeccenxxxxxA}{\ensuremath{0\pm0}}           
\newcommand{\hatcurRVtrtwoeccenxxxxxA}{\ensuremath{0\pm0}}           
\newcommand{\hatcurRVgammaeccenxxxxxA}{\ensuremath{0.6\pm4.1}}       
\newcommand{\hatcurRVjittereccenxxxxxA}{\ensuremath{33.6\pm8.9}}     
\newcommand{\hatcurRVjittertwosiglimeccenxxxxxA}{\ensuremath{<49.1}} 
\newcommand{\hatcurRVfitrmseccenxxxxxA}{\ensuremath{.1fym}}          %
\newcommand{\hatcurRVecceneccenxxxxxA}{\ensuremath{0.022\pm0.028}}   
\newcommand{\hatcurRVeccentwosiglimeccenxxxxxA}{\ensuremath{<0.088}} 
\newcommand{\hatcurRVomegaeccenxxxxxA}{\ensuremath{61\pm91}}         
\newcommand{\hatcurPPieccenxxxxxA}{\ensuremath{86.860\pm0.078}}      
\newcommand{\hatcurPPgeccenxxxxxA}{\ensuremath{4.86\pm0.44}}         
\newcommand{\hatcurPPloggeccenxxxxxA}{\ensuremath{2.686_{-0.043}^{+0.031}}} 
\newcommand{\hatcurPPareccenxxxxxA}{\ensuremath{13.91\pm0.17}}       
\newcommand{\hatcurPPareleccenxxxxxA}{\ensuremath{0.04291_{-0.00027}^{+0.00041}}} 
\newcommand{\hatcurPPrhoeccenxxxxxA}{\ensuremath{0.212_{-0.023}^{+0.017}}} 
\newcommand{\hatcurPPmeccenxxxxxA}{\ensuremath{0.257_{-0.023}^{+0.018}}} 
\newcommand{\hatcurPPmtwosiglimeccenxxxxxA}{\ensuremath{<0.29}}      
\newcommand{\hatcurPPmshorteccenxxxxxA}{\ensuremath{0.26}}           
\newcommand{\hatcurPPmlongeccenxxxxxA}{\ensuremath{0.257_{-0.023}^{+0.018}}} 
\newcommand{\hatcurPPmeeccenxxxxxA}{\ensuremath{81.6_{-7.3}^{+5.6}}} 
\newcommand{\hatcurPPmeshorteccenxxxxxA}{\ensuremath{81.6}}          
\newcommand{\hatcurPPmelongeccenxxxxxA}{\ensuremath{81.6_{-7.3}^{+5.6}}} 
\newcommand{\hatcurPPreccenxxxxxA}{\ensuremath{1.146\pm0.016}}       
\newcommand{\hatcurPPrshorteccenxxxxxA}{\ensuremath{1.15}}           
\newcommand{\hatcurPPrlongeccenxxxxxA}{\ensuremath{1.146\pm0.016}}   
\newcommand{\hatcurPPreeccenxxxxxA}{\ensuremath{12.84\pm0.18}}       
\newcommand{\hatcurPPreshorteccenxxxxxA}{\ensuremath{12.8}}          
\newcommand{\hatcurPPrelongeccenxxxxxA}{\ensuremath{12.84\pm0.18}}   
\newcommand{\hatcurPPmrcorreccenxxxxxA}{\ensuremath{-0.15}}          
\newcommand{\hatcurPPteffeccenxxxxxA}{\ensuremath{859.9\pm8.0}}      
\newcommand{\hatcurPPthetaeccenxxxxxA}{\ensuremath{0.0279\pm0.0024}} 
\newcommand{\hatcurPPfluxperieccenxxxxxA}{\ensuremath{1.298_{-0.065}^{+0.114}}} 
\newcommand{\hatcurPPfluxperidimeccenxxxxxA}{\ensuremath{8}}         
\newcommand{\hatcurPPfluxapeccenxxxxxA}{\ensuremath{1.178_{-0.064}^{+0.045}}} 
\newcommand{\hatcurPPfluxapdimeccenxxxxxA}{\ensuremath{8}}           
\newcommand{\hatcurPPfluxavgeccenxxxxxA}{\ensuremath{1.233_{-0.033}^{+0.043}}} 
\newcommand{\hatcurPPfluxavgdimeccenxxxxxA}{\ensuremath{8}}          
\newcommand{\hatcurPPfluxavglogeccenxxxxxA}{\ensuremath{8.091\pm0.016}} 
\newcommand{\hatcurXsecphaseeccenxxxxxA}{\ensuremath{0.506\pm0.016}} 
\newcommand{\hatcurXsecondaryeccenxxxxxA}{\ensuremath{2456990.753\pm0.064}} 
\newcommand{\hatcurXsecdureccenxxxxxA}{\ensuremath{0.08085\pm0.00072}} 
\newcommand{\hatcurXsecingdureccenxxxxxA}{\ensuremath{0.0274\pm0.0037}} 
\newcommand{\hatcurPPphiconjeccenxxxxxA}{\ensuremath{0.091_{-0.223}^{+0.098}}} 
\newcommand{\hatcurPPperieccenxxxxxA}{\ensuremath{2456988.41\pm0.69}} 
\newcommand{\hatcurPPaequiveccenxxxxxA}{\ensuremath{0.1051\pm0.0019}} 
\newcommand{\hatcurPPtcirceccenxxxxxA}{\ensuremath{114\pm13}}        
\newcommand{\hatcurPPtinfalleccenxxxxxA}{\ensuremath{52400\pm5000}}  
\newcommand{\hatcurXdisteccenxxxxxA}{\ensuremath{305.9\pm2.8}}       
\newcommand{\hatcurXAveccenxxxxxA}{\ensuremath{0.135\pm0.061}}       
\newcommand{\hatcurXdistredeccenxxxxxA}{\ensuremath{305.9\pm2.8}}    
\newcommand{\hatcurXEBVeccenxxxxxA}{\ensuremath{0.043\pm0.020}}      
\newcommand{\hatcurCCpmraeccenxxxxxA}{\ensuremath{3.827\pm0.058}}    
\newcommand{\hatcurCCpmdececcenxxxxxA}{\ensuremath{4.878\pm0.038}}   
\newcommand{\hatcurCCpmeccenxxxxxA}{\ensuremath{6.200\pm0.069}}      
\newcommand{\hatcurhtreccenxxxxxB}{HATS778-005}                      
\newcommand{\hatcurfieldeccenxxxxxB}{\ensuremath{string}}            
\newcommand{\hatcurCCraeccenxxxxxB}{\ensuremath{19^{\mathrm h}14^{\mathrm m}41.2748{\mathrm s}}}                   
\newcommand{\hatcurCCdececcenxxxxxB}{\ensuremath{-59{\arcdeg}34{\arcmin}45.7571{\arcsec}}}                 
\newcommand{\hatcurCCmageccenxxxxxB}{14.352}                         
\newcommand{\hatcurCCtwomasseccenxxxxxB}{2MASS~19144126-5934458}     
\newcommand{\hatcurCCgsceccenxxxxxB}{GSC~}                           
\newcommand{\hatcurCCgaiaeccenxxxxxB}{GAIA~6638412919991750912}      
\newcommand{\hatcurCCgaiadrtwoeccenxxxxxB}{GAIA~DR2~6638412919991750912} 
\newcommand{\hatcurCCtassmveccenxxxxxB}{\ensuremath{14.35\pm0.11}}   
\newcommand{\hatcurCCtassmvshorteccenxxxxxB}{\ensuremath{14.4}}      
\newcommand{\hatcurCCtassmBeccenxxxxxB}{\ensuremath{15.577\pm0.050}} 
\newcommand{\hatcurCCtassmBshorteccenxxxxxB}{\ensuremath{15.6}}      
\newcommand{\hatcurCCtassmIeccenxxxxxB}{\ensuremath{nff\pmnff}}      
\newcommand{\hatcurCCtassmIshorteccenxxxxxB}{\ensuremath{0.0}}       
\newcommand{\hatcurCCtassmgeccenxxxxxB}{\ensuremath{14.935\pm0.030}} 
\newcommand{\hatcurCCtassmgshorteccenxxxxxB}{\ensuremath{14.9}}      
\newcommand{\hatcurCCtassmreccenxxxxxB}{\ensuremath{13.821\pm0.050}} 
\newcommand{\hatcurCCtassmrshorteccenxxxxxB}{\ensuremath{13.8}}      
\newcommand{\hatcurCCtassmieccenxxxxxB}{\ensuremath{13.69\pm0.17}}   
\newcommand{\hatcurCCtassmishorteccenxxxxxB}{\ensuremath{13.7}}      
\newcommand{\hatcurCCparallaxeccenxxxxxB}{\ensuremath{3.765\pm0.024}} 
\newcommand{\hatcurCCgaiamGeccenxxxxxB}{\ensuremath{13.89510\pm0.00020}} 
\newcommand{\hatcurCCgaiamBPeccenxxxxxB}{\ensuremath{14.5801\pm0.0016}} 
\newcommand{\hatcurCCgaiamRPeccenxxxxxB}{\ensuremath{13.11260\pm0.00090}} 
\newcommand{\hatcurCCtwomassJmageccenxxxxxB}{\ensuremath{12.160\pm0.024}} 
\newcommand{\hatcurCCtwomassHmageccenxxxxxB}{\ensuremath{11.591\pm0.026}} 
\newcommand{\hatcurCCtwomassKmageccenxxxxxB}{\ensuremath{11.427\pm0.021}} 
\newcommand{\hatcurCCcitJmageccenxxxxxB}{\ensuremath{12.158\pm0.025}} 
\newcommand{\hatcurCCcitHmageccenxxxxxB}{\ensuremath{11.583\pm0.027}} 
\newcommand{\hatcurCCcitKmageccenxxxxxB}{\ensuremath{11.451\pm0.022}} 
\newcommand{\hatcurCCbbJmageccenxxxxxB}{\ensuremath{12.236\pm0.027}} 
\newcommand{\hatcurCCbbHmageccenxxxxxB}{\ensuremath{11.607\pm0.028}} 
\newcommand{\hatcurCCbbKmageccenxxxxxB}{\ensuremath{11.471\pm0.022}} 
\newcommand{\hatcurCCesoJmageccenxxxxxB}{\ensuremath{12.243\pm0.031}} 
\newcommand{\hatcurCCesoHmageccenxxxxxB}{\ensuremath{11.605\pm0.036}} 
\newcommand{\hatcurCCesoKmageccenxxxxxB}{\ensuremath{11.468\pm0.023}} 
\newcommand{\hatcurCCesoJHmageccenxxxxxB}{\ensuremath{0.638\pm0.044}} 
\newcommand{\hatcurCCesoJKmageccenxxxxxB}{\ensuremath{0.776\pm0.036}} 
\newcommand{\hatcurCCesoHKmageccenxxxxxB}{\ensuremath{0.137\pm0.042}} 
\newcommand{\hatcurCCWonemageccenxxxxxB}{\ensuremath{11.364\pm0.023}} 
\newcommand{\hatcurCCWtwomageccenxxxxxB}{\ensuremath{11.458\pm0.021}} 
\newcommand{\hatcurCCWthreemageccenxxxxxB}{\ensuremath{0\pm0}}       
\newcommand{\hatcurCCWfourmageccenxxxxxB}{\ensuremath{0\pm0}}        
\newcommand{\hatcurLCdipeccenxxxxxB}{\ensuremath{16.5}}              
\newcommand{\hatcurLCrprstareccenxxxxxB}{\ensuremath{0.1135\pm0.0015}} 
\newcommand{\hatcurLCbsqeccenxxxxxB}{\ensuremath{0.024_{-0.018}^{+0.036}}} 
\newcommand{\hatcurLCimpeccenxxxxxB}{\ensuremath{0.156_{-0.076}^{+0.090}}} 
\newcommand{\hatcurLCzetaeccenxxxxxB}{\ensuremath{21.98\pm0.19}}     
\newcommand{\hatcurLCdureccenxxxxxB}{\ensuremath{0.1016\pm0.0010}}   
\newcommand{\hatcurLCdurshorteccenxxxxxB}{\ensuremath{0.1016}}       
\newcommand{\hatcurLCdurhreccenxxxxxB}{\ensuremath{2.438\pm0.024}}   
\newcommand{\hatcurLCdurhrshorteccenxxxxxB}{\ensuremath{2.438}}      
\newcommand{\hatcurLCqeccenxxxxxB}{\ensuremath{0.03240\pm0.00033}}   
\newcommand{\hatcurLCqshorteccenxxxxxB}{\ensuremath{0.032}}          
\newcommand{\hatcurLCingdureccenxxxxxB}{\ensuremath{0.01062\pm0.00039}} 
\newcommand{\hatcurLCPeccenxxxxxB}{\ensuremath{3.1316633\pm0.0000035}} 
\newcommand{\hatcurLCPprececcenxxxxxB}{\ensuremath{3.1316633}}       
\newcommand{\hatcurLCPshorteccenxxxxxB}{\ensuremath{3.1317}}         
\newcommand{\hatcurLCTeccenxxxxxB}{\ensuremath{2457160.05190\pm0.00030}} 
\newcommand{\hatcurLCTAeccenxxxxxB}{\ensuremath{2455678.7752\pm0.0016}} 
\newcommand{\hatcurLCTBeccenxxxxxB}{\ensuremath{2457219.55350\pm0.00033}} 
\newcommand{\hatcurLChatnetmAeccenxxxxxB}{\ensuremath{13.91271\pm0.00013}} 
\newcommand{\hatcurLCiblendAeccenxxxxxB}{\ensuremath{0.99905\pm0.00019}} 
\newcommand{\hatcurLChatnetmBeccenxxxxxB}{\ensuremath{-0.00021\pm0.00016}} 
\newcommand{\hatcurLCiblendBeccenxxxxxB}{\ensuremath{0.607\pm0.014}} 
\newcommand{\hatcurLCrhoeccenxxxxxB}{\ensuremath{2.840_{-0.064}^{+0.100}}} 
\newcommand{\hatcurSMEiteffeccenxxxxxB}{\ensuremath{4190\pm100}}     
\newcommand{\hatcurSMEizfeheccenxxxxxB}{\ensuremath{0.00\pm0.10}}    
\newcommand{\hatcurSMEizfehshorteccenxxxxxB}{\ensuremath{0.00}}      
\newcommand{\hatcurSMEiloggeccenxxxxxB}{\ensuremath{4.690\pm0.088}}  
\newcommand{\hatcurSMEivsineccenxxxxxB}{\ensuremath{0.73\pm0.55}}    
\newcommand{\hatcurSMEivmaceccenxxxxxB}{\ensuremath{nff\pmnff}}      
\newcommand{\hatcurSMEivmiceccenxxxxxB}{\ensuremath{nff\pmnff}}      
\newcommand{\hatcurextraerrMgeccenxxxxxB}{\ensuremath{0.051_{-0.038}^{+0.024}}} 
\newcommand{\hatcurextraerrMgtwosiglimeccenxxxxxB}{\ensuremath{<0.0988}} 
\newcommand{\hatcurextraerrMreccenxxxxxB}{\ensuremath{0.015_{-0.014}^{+0.101}}} 
\newcommand{\hatcurextraerrMrtwosiglimeccenxxxxxB}{\ensuremath{<0.2007}} 
\newcommand{\hatcurextraerrMieccenxxxxxB}{\ensuremath{0.293_{-0.107}^{+0.078}}} 
\newcommand{\hatcurextraerrMitwosiglimeccenxxxxxB}{\ensuremath{<0.4263}} 
\newcommand{\hatcurextraerrMGeccenxxxxxB}{\ensuremath{0.0225_{-0.0077}^{+0.0121}}} 
\newcommand{\hatcurextraerrMGtwosiglimeccenxxxxxB}{\ensuremath{<0.0433}} 
\newcommand{\hatcurextraerrMBPtwoeccenxxxxxB}{\ensuremath{0.00084_{-0.00082}^{+0.01035}}} 
\newcommand{\hatcurextraerrMBPtwotwosiglimeccenxxxxxB}{\ensuremath{<0.0232}} 
\newcommand{\hatcurextraerrMRPeccenxxxxxB}{\ensuremath{0.00096_{-0.00093}^{+0.00633}}} 
\newcommand{\hatcurextraerrMRPtwosiglimeccenxxxxxB}{\ensuremath{<0.0225}} 
\newcommand{\hatcurextraerrMJeccenxxxxxB}{\ensuremath{0.0049_{-0.0048}^{+0.0207}}} 
\newcommand{\hatcurextraerrMJtwosiglimeccenxxxxxB}{\ensuremath{<0.0414}} 
\newcommand{\hatcurextraerrMHeccenxxxxxB}{\ensuremath{0.018_{-0.018}^{+0.037}}} 
\newcommand{\hatcurextraerrMHtwosiglimeccenxxxxxB}{\ensuremath{<0.0748}} 
\newcommand{\hatcurextraerrMKseccenxxxxxB}{\ensuremath{0.012_{-0.011}^{+0.026}}} 
\newcommand{\hatcurextraerrMKstwosiglimeccenxxxxxB}{\ensuremath{<0.0511}} 
\newcommand{\hatcurextraerrMWoneeccenxxxxxB}{\ensuremath{0.0080_{-0.0077}^{+0.0243}}} 
\newcommand{\hatcurextraerrMWonetwosiglimeccenxxxxxB}{\ensuremath{<0.0435}} 
\newcommand{\hatcurextraerrMWtwoeccenxxxxxB}{\ensuremath{0.0033_{-0.0031}^{+0.0243}}} 
\newcommand{\hatcurextraerrMWtwotwosiglimeccenxxxxxB}{\ensuremath{<0.0846}} 
\newcommand{\hatcurLBiBeccenxxxxxB}{\ensuremath{0.8007}}             
\newcommand{\hatcurLBiiBeccenxxxxxB}{\ensuremath{0.0222}}            
\newcommand{\hatcurLBiVeccenxxxxxB}{\ensuremath{0.6354}}             
\newcommand{\hatcurLBiiVeccenxxxxxB}{\ensuremath{0.1339}}            
\newcommand{\hatcurLBiReccenxxxxxB}{\ensuremath{0.5245}}             
\newcommand{\hatcurLBiiReccenxxxxxB}{\ensuremath{0.1825}}            
\newcommand{\hatcurLBiIeccenxxxxxB}{\ensuremath{0.3905}}             
\newcommand{\hatcurLBiiIeccenxxxxxB}{\ensuremath{0.2115}}            
\newcommand{\hatcurLBiueccenxxxxxB}{\ensuremath{0.8665}}             
\newcommand{\hatcurLBiiueccenxxxxxB}{\ensuremath{-0.0450}}           
\newcommand{\hatcurLBigeccenxxxxxB}{\ensuremath{0.7223}}             
\newcommand{\hatcurLBiigeccenxxxxxB}{\ensuremath{0.0663}}            
\newcommand{\hatcurLBireccenxxxxxB}{\ensuremath{0.52\pm0.13}}        
\newcommand{\hatcurLBiireccenxxxxxB}{\ensuremath{0.30\pm0.13}}       
\newcommand{\hatcurLBiieccenxxxxxB}{\ensuremath{0.335\pm0.094}}      
\newcommand{\hatcurLBiiieccenxxxxxB}{\ensuremath{0.33\pm0.15}}       
\newcommand{\hatcurLBizeccenxxxxxB}{\ensuremath{0.3392}}             
\newcommand{\hatcurLBiizeccenxxxxxB}{\ensuremath{0.2145}}            
\newcommand{\hatcurLBiJeccenxxxxxB}{\ensuremath{0.2506}}             
\newcommand{\hatcurLBiiJeccenxxxxxB}{\ensuremath{0.2171}}            
\newcommand{\hatcurLBiHeccenxxxxxB}{\ensuremath{0.1694}}             
\newcommand{\hatcurLBiiHeccenxxxxxB}{\ensuremath{0.2632}}            
\newcommand{\hatcurLBiKeccenxxxxxB}{\ensuremath{0.1301}}             
\newcommand{\hatcurLBiiKeccenxxxxxB}{\ensuremath{0.2308}}            
\newcommand{\hatcurLBiTeccenxxxxxB}{\ensuremath{0.34_{-0.13}^{+0.18}}} 
\newcommand{\hatcurLBiiTeccenxxxxxB}{\ensuremath{0.45\pm0.16}}       
\newcommand{\hatcurLBikepeccenxxxxxB}{\ensuremath{0.5094}}           
\newcommand{\hatcurLBiikepeccenxxxxxB}{\ensuremath{0.2164}}          
\newcommand{\hatcurLBiCeccenxxxxxB}{\ensuremath{0.4819}}             
\newcommand{\hatcurLBiiCeccenxxxxxB}{\ensuremath{0.2222}}            
\newcommand{\hatcurLBiMeccenxxxxxB}{\ensuremath{0.5998}}             
\newcommand{\hatcurLBiiMeccenxxxxxB}{\ensuremath{0.1782}}            
\newcommand{\hatcurLBiSoneeccenxxxxxB}{\ensuremath{0.0966}}            
\newcommand{\hatcurLBiiSoneeccenxxxxxB}{\ensuremath{0.1667}}           
\newcommand{\hatcurLBiStwoeccenxxxxxB}{\ensuremath{0.0830}}            
\newcommand{\hatcurLBiiStwoeccenxxxxxB}{\ensuremath{0.1376}}           
\newcommand{\hatcurLBiSthreeeccenxxxxxB}{\ensuremath{0.0699}}            
\newcommand{\hatcurLBiiSthreeeccenxxxxxB}{\ensuremath{0.1187}}           
\newcommand{\hatcurLBiSfoureccenxxxxxB}{\ensuremath{0.0647}}            
\newcommand{\hatcurLBiiSfoureccenxxxxxB}{\ensuremath{0.1036}}           
\newcommand{\hatcurISOmeccenxxxxxB}{\ensuremath{0.739\pm0.012}}      
\newcommand{\hatcurISOmshorteccenxxxxxB}{\ensuremath{0.74}}          
\newcommand{\hatcurISOmlongeccenxxxxxB}{\ensuremath{0.739\pm0.012}}  
\newcommand{\hatcurISOreccenxxxxxB}{\ensuremath{0.7147\pm0.0060}}    
\newcommand{\hatcurISOrshorteccenxxxxxB}{\ensuremath{0.71}}          
\newcommand{\hatcurISOrlongeccenxxxxxB}{\ensuremath{0.7147\pm0.0060}} 
\newcommand{\hatcurISOrhoeccenxxxxxB}{\ensuremath{2.840_{-0.064}^{+0.100}}} 
\newcommand{\hatcurISOrholongeccenxxxxxB}{\ensuremath{2.840_{-0.064}^{+0.100}}} 
\newcommand{\hatcurISOloggeccenxxxxxB}{\ensuremath{4.596\pm0.012}}   
\newcommand{\hatcurISOlumeccenxxxxxB}{\ensuremath{0.1976\pm0.0052}}  
\newcommand{\hatcurISOlumshorteccenxxxxxB}{\ensuremath{0.20}}        
\newcommand{\hatcurISOteffeccenxxxxxB}{\ensuremath{4560\pm24}}       
\newcommand{\hatcurISOzfeheccenxxxxxB}{\ensuremath{0.257\pm0.069}}   
\newcommand{\hatcurISOageeccenxxxxxB}{\ensuremath{11.0_{-2.3}^{+1.1}}} 
\newcommand{\hatcurISOspececcenxxxxxB}{K}                            
\newcommand{\hatcurRVKeccenxxxxxB}{\ensuremath{42.1\pm3.7}}          
\newcommand{\hatcurRVKtwosiglimeccenxxxxxB}{\ensuremath{<48.4}}      
\newcommand{\hatcurRVrkeccenxxxxxB}{\ensuremath{-0.216_{-0.071}^{+0.052}}} 
\newcommand{\hatcurRVrheccenxxxxxB}{\ensuremath{-0.192\pm0.060}}     
\newcommand{\hatcurRVkeccenxxxxxB}{\ensuremath{-0.064_{-0.038}^{+0.023}}} 
\newcommand{\hatcurRVheccenxxxxxB}{\ensuremath{-0.057_{-0.025}^{+0.018}}} 
\newcommand{\hatcurRVtroneeccenxxxxxB}{\ensuremath{0\pm0}}           
\newcommand{\hatcurRVtrtwoeccenxxxxxB}{\ensuremath{0\pm0}}           
\newcommand{\hatcurRVgammaeccenxxxxxB}{\ensuremath{-17.8\pm2.7}}     
\newcommand{\hatcurRVjittereccenxxxxxB}{\ensuremath{19.9\pm5.2}}     
\newcommand{\hatcurRVjittertwosiglimeccenxxxxxB}{\ensuremath{<30.5}} 
\newcommand{\hatcurRVfitrmseccenxxxxxB}{\ensuremath{.1fym}}          %
\newcommand{\hatcurRVecceneccenxxxxxB}{\ensuremath{0.087\pm0.039}}   
\newcommand{\hatcurRVeccentwosiglimeccenxxxxxB}{\ensuremath{<0.162}} 
\newcommand{\hatcurRVomegaeccenxxxxxB}{\ensuremath{221\pm33}}        
\newcommand{\hatcurPPieccenxxxxxB}{\ensuremath{89.26\pm0.36}}        
\newcommand{\hatcurPPgeccenxxxxxB}{\ensuremath{9.84\pm0.76}}         
\newcommand{\hatcurPPloggeccenxxxxxB}{\ensuremath{2.993\pm0.034}}    
\newcommand{\hatcurPPareccenxxxxxB}{\ensuremath{11.378_{-0.086}^{+0.132}}} 
\newcommand{\hatcurPPareleccenxxxxxB}{\ensuremath{0.03787\pm0.00020}} 
\newcommand{\hatcurPPrhoeccenxxxxxB}{\ensuremath{0.624\pm0.049}}     
\newcommand{\hatcurPPmeccenxxxxxB}{\ensuremath{0.247\pm0.021}}       
\newcommand{\hatcurPPmtwosiglimeccenxxxxxB}{\ensuremath{<0.28}}      
\newcommand{\hatcurPPmshorteccenxxxxxB}{\ensuremath{0.25}}           
\newcommand{\hatcurPPmlongeccenxxxxxB}{\ensuremath{0.247\pm0.021}}   
\newcommand{\hatcurPPmeeccenxxxxxB}{\ensuremath{78.5\pm6.7}}         
\newcommand{\hatcurPPmeshorteccenxxxxxB}{\ensuremath{78.5}}          
\newcommand{\hatcurPPmelongeccenxxxxxB}{\ensuremath{78.5\pm6.7}}     
\newcommand{\hatcurPPreccenxxxxxB}{\ensuremath{0.788\pm0.012}}       
\newcommand{\hatcurPPrshorteccenxxxxxB}{\ensuremath{0.79}}           
\newcommand{\hatcurPPrlongeccenxxxxxB}{\ensuremath{0.788\pm0.012}}   
\newcommand{\hatcurPPreeccenxxxxxB}{\ensuremath{8.84\pm0.14}}        
\newcommand{\hatcurPPreshorteccenxxxxxB}{\ensuremath{8.8}}           
\newcommand{\hatcurPPrelongeccenxxxxxB}{\ensuremath{8.84\pm0.14}}    
\newcommand{\hatcurPPmrcorreccenxxxxxB}{\ensuremath{0.42}}           
\newcommand{\hatcurPPteffeccenxxxxxB}{\ensuremath{955.9\pm5.9}}      
\newcommand{\hatcurPPthetaeccenxxxxxB}{\ensuremath{0.0320\pm0.0026}} 
\newcommand{\hatcurPPfluxperieccenxxxxxB}{\ensuremath{2.25_{-0.12}^{+0.21}}} 
\newcommand{\hatcurPPfluxperidimeccenxxxxxB}{\ensuremath{8}}         
\newcommand{\hatcurPPfluxapeccenxxxxxB}{\ensuremath{1.585_{-0.128}^{+0.094}}} 
\newcommand{\hatcurPPfluxapdimeccenxxxxxB}{\ensuremath{8}}           
\newcommand{\hatcurPPfluxavgeccenxxxxxB}{\ensuremath{1.883\pm0.047}} 
\newcommand{\hatcurPPfluxavgdimeccenxxxxxB}{\ensuremath{8}}          
\newcommand{\hatcurPPfluxavglogeccenxxxxxB}{\ensuremath{8.275\pm0.011}} 
\newcommand{\hatcurXsecphaseeccenxxxxxB}{\ensuremath{0.459\pm0.021}} 
\newcommand{\hatcurXsecondaryeccenxxxxxB}{\ensuremath{2457161.490\pm0.067}} 
\newcommand{\hatcurXsecdureccenxxxxxB}{\ensuremath{0.0909\pm0.0035}} 
\newcommand{\hatcurXsecingdureccenxxxxxB}{\ensuremath{0.00946\pm0.00029}} 
\newcommand{\hatcurPPphiconjeccenxxxxxB}{\ensuremath{-0.344_{-0.027}^{+0.042}}} 
\newcommand{\hatcurPPperieccenxxxxxB}{\ensuremath{2457161.13\pm0.30}} 
\newcommand{\hatcurPPaequiveccenxxxxxB}{\ensuremath{0.0852\pm0.0011}} 
\newcommand{\hatcurPPtcirceccenxxxxxB}{\ensuremath{267\pm25}}        
\newcommand{\hatcurPPtinfalleccenxxxxxB}{\ensuremath{17200\pm1900}}  
\newcommand{\hatcurXdisteccenxxxxxB}{\ensuremath{265.6\pm1.4}}       
\newcommand{\hatcurXAveccenxxxxxB}{\ensuremath{0.132\pm0.041}}       
\newcommand{\hatcurXdistredeccenxxxxxB}{\ensuremath{265.6\pm1.4}}    
\newcommand{\hatcurXEBVeccenxxxxxB}{\ensuremath{0.043\pm0.013}}      
\newcommand{\hatcurCCpmraeccenxxxxxB}{\ensuremath{3.125\pm0.031}}    
\newcommand{\hatcurCCpmdececcenxxxxxB}{\ensuremath{6.146\pm0.029}}   
\newcommand{\hatcurCCpmeccenxxxxxB}{\ensuremath{6.895\pm0.042}}      
\newcommand{\hatcurhtreccenxxxxxC}{HATS755-004}                      
\newcommand{\hatcurfieldeccenxxxxxC}{\ensuremath{string}}            
\newcommand{\hatcurCCraeccenxxxxxC}{\ensuremath{00^{\mathrm h}26^{\mathrm m}27.1829{\mathrm s}}}                   
\newcommand{\hatcurCCdececcenxxxxxC}{\ensuremath{-56{\arcdeg}20{\arcmin}39.5352{\arcsec}}}                 
\newcommand{\hatcurCCmageccenxxxxxC}{14.998}                         
\newcommand{\hatcurCCtwomasseccenxxxxxC}{2MASS~00262717-5620395}     
\newcommand{\hatcurCCgsceccenxxxxxC}{GSC~}                           
\newcommand{\hatcurCCgaiaeccenxxxxxC}{GAIA~4919770108539385472}      
\newcommand{\hatcurCCgaiadrtwoeccenxxxxxC}{GAIA~DR2~4919770108539385472} 
\newcommand{\hatcurCCtassmveccenxxxxxC}{\ensuremath{14.998\pm0.040}} 
\newcommand{\hatcurCCtassmvshorteccenxxxxxC}{\ensuremath{15.0}}      
\newcommand{\hatcurCCtassmBeccenxxxxxC}{\ensuremath{16.378\pm0.040}} 
\newcommand{\hatcurCCtassmBshorteccenxxxxxC}{\ensuremath{16.4}}      
\newcommand{\hatcurCCtassmIeccenxxxxxC}{\ensuremath{nff\pmnff}}      
\newcommand{\hatcurCCtassmIshorteccenxxxxxC}{\ensuremath{0.0}}       
\newcommand{\hatcurCCtassmgeccenxxxxxC}{\ensuremath{15.668\pm0.040}} 
\newcommand{\hatcurCCtassmgshorteccenxxxxxC}{\ensuremath{15.7}}      
\newcommand{\hatcurCCtassmreccenxxxxxC}{\ensuremath{14.496\pm0.010}} 
\newcommand{\hatcurCCtassmrshorteccenxxxxxC}{\ensuremath{14.5}}      
\newcommand{\hatcurCCtassmieccenxxxxxC}{\ensuremath{14.14\pm0.12}}   
\newcommand{\hatcurCCtassmishorteccenxxxxxC}{\ensuremath{14.1}}      
\newcommand{\hatcurCCparallaxeccenxxxxxC}{\ensuremath{3.054\pm0.022}} 
\newcommand{\hatcurCCgaiamGeccenxxxxxC}{\ensuremath{14.54490\pm0.00030}} 
\newcommand{\hatcurCCgaiamBPeccenxxxxxC}{\ensuremath{15.2886\pm0.0022}} 
\newcommand{\hatcurCCgaiamRPeccenxxxxxC}{\ensuremath{13.7214\pm0.0013}} 
\newcommand{\hatcurCCtwomassJmageccenxxxxxC}{\ensuremath{12.692\pm0.024}} 
\newcommand{\hatcurCCtwomassHmageccenxxxxxC}{\ensuremath{12.105\pm0.024}} 
\newcommand{\hatcurCCtwomassKmageccenxxxxxC}{\ensuremath{11.938\pm0.023}} 
\newcommand{\hatcurCCcitJmageccenxxxxxC}{\ensuremath{12.688\pm0.025}} 
\newcommand{\hatcurCCcitHmageccenxxxxxC}{\ensuremath{12.097\pm0.025}} 
\newcommand{\hatcurCCcitKmageccenxxxxxC}{\ensuremath{11.962\pm0.023}} 
\newcommand{\hatcurCCbbJmageccenxxxxxC}{\ensuremath{12.769\pm0.027}} 
\newcommand{\hatcurCCbbHmageccenxxxxxC}{\ensuremath{12.121\pm0.026}} 
\newcommand{\hatcurCCbbKmageccenxxxxxC}{\ensuremath{11.982\pm0.023}} 
\newcommand{\hatcurCCesoJmageccenxxxxxC}{\ensuremath{12.776\pm0.031}} 
\newcommand{\hatcurCCesoHmageccenxxxxxC}{\ensuremath{12.118\pm0.035}} 
\newcommand{\hatcurCCesoKmageccenxxxxxC}{\ensuremath{11.979\pm0.025}} 
\newcommand{\hatcurCCesoJHmageccenxxxxxC}{\ensuremath{0.657\pm0.025}} 
\newcommand{\hatcurCCesoJKmageccenxxxxxC}{\ensuremath{0.797\pm0.038}} 
\newcommand{\hatcurCCesoHKmageccenxxxxxC}{\ensuremath{0.139\pm0.041}} 
\newcommand{\hatcurCCWonemageccenxxxxxC}{\ensuremath{11.903\pm0.023}} 
\newcommand{\hatcurCCWtwomageccenxxxxxC}{\ensuremath{11.990\pm0.022}} 
\newcommand{\hatcurCCWthreemageccenxxxxxC}{\ensuremath{0\pm0}}       
\newcommand{\hatcurCCWfourmageccenxxxxxC}{\ensuremath{0\pm0}}        
\newcommand{\hatcurLCdipeccenxxxxxC}{\ensuremath{13.2}}              
\newcommand{\hatcurLCrprstareccenxxxxxC}{\ensuremath{0.1156\pm0.0017}} 
\newcommand{\hatcurLCbsqeccenxxxxxC}{\ensuremath{0.188_{-0.032}^{+0.037}}} 
\newcommand{\hatcurLCimpeccenxxxxxC}{\ensuremath{0.434_{-0.038}^{+0.041}}} 
\newcommand{\hatcurLCzetaeccenxxxxxC}{\ensuremath{23.60\pm0.28}}     
\newcommand{\hatcurLCdureccenxxxxxC}{\ensuremath{0.0966\pm0.0011}}   
\newcommand{\hatcurLCdurshorteccenxxxxxC}{\ensuremath{0.0966}}       
\newcommand{\hatcurLCdurhreccenxxxxxC}{\ensuremath{2.318\pm0.027}}   
\newcommand{\hatcurLCdurhrshorteccenxxxxxC}{\ensuremath{2.318}}      
\newcommand{\hatcurLCqeccenxxxxxC}{\ensuremath{0.02330\pm0.00028}}   
\newcommand{\hatcurLCqshorteccenxxxxxC}{\ensuremath{0.023}}          
\newcommand{\hatcurLCingdureccenxxxxxC}{\ensuremath{0.01206\pm0.00049}} 
\newcommand{\hatcurLCPeccenxxxxxC}{\ensuremath{4.1480412\pm0.0000051}} 
\newcommand{\hatcurLCPprececcenxxxxxC}{\ensuremath{4.1480412}}       
\newcommand{\hatcurLCPshorteccenxxxxxC}{\ensuremath{4.1480}}         
\newcommand{\hatcurLCTeccenxxxxxC}{\ensuremath{2457171.53374\pm0.00057}} 
\newcommand{\hatcurLCTAeccenxxxxxC}{\ensuremath{2455761.1998\pm0.0020}} 
\newcommand{\hatcurLCTBeccenxxxxxC}{\ensuremath{2457283.53083\pm0.00056}} 
\newcommand{\hatcurLChatnetmAeccenxxxxxC}{\ensuremath{14.42626\pm0.00014}} 
\newcommand{\hatcurLCiblendAeccenxxxxxC}{\ensuremath{0.928\pm0.039}} 
\newcommand{\hatcurLChatnetmBeccenxxxxxC}{\ensuremath{14.42646\pm0.00019}} 
\newcommand{\hatcurLCiblendBeccenxxxxxC}{\ensuremath{0.877\pm0.043}} 
\newcommand{\hatcurLChatnetmCeccenxxxxxC}{\ensuremath{-0.000010\pm0.000097}} 
\newcommand{\hatcurLCiblendCeccenxxxxxC}{\ensuremath{0.738\pm0.070}} 
\newcommand{\hatcurLCrhoeccenxxxxxC}{\ensuremath{2.969_{-0.054}^{+0.086}}} 
\newcommand{\hatcurSMEiteffeccenxxxxxC}{\ensuremath{4354\pm70}}      
\newcommand{\hatcurSMEizfeheccenxxxxxC}{\ensuremath{0.080\pm0.084}}  
\newcommand{\hatcurSMEizfehshorteccenxxxxxC}{\ensuremath{0.08}}      
\newcommand{\hatcurSMEiloggeccenxxxxxC}{\ensuremath{4.55\pm0.19}}    
\newcommand{\hatcurSMEivsineccenxxxxxC}{\ensuremath{0.50\pm0.78}}    
\newcommand{\hatcurSMEivmaceccenxxxxxC}{\ensuremath{nff\pmnff}}      
\newcommand{\hatcurSMEivmiceccenxxxxxC}{\ensuremath{nff\pmnff}}      
\newcommand{\hatcurextraerrMgeccenxxxxxC}{\ensuremath{0.013_{-0.013}^{+0.026}}} 
\newcommand{\hatcurextraerrMgtwosiglimeccenxxxxxC}{\ensuremath{<0.0540}} 
\newcommand{\hatcurextraerrMreccenxxxxxC}{\ensuremath{0.112\pm0.076}} 
\newcommand{\hatcurextraerrMrtwosiglimeccenxxxxxC}{\ensuremath{<0.2519}} 
\newcommand{\hatcurextraerrMieccenxxxxxC}{\ensuremath{0.032_{-0.031}^{+0.088}}} 
\newcommand{\hatcurextraerrMitwosiglimeccenxxxxxC}{\ensuremath{<0.1942}} 
\newcommand{\hatcurextraerrMGeccenxxxxxC}{\ensuremath{0.022\pm0.011}} 
\newcommand{\hatcurextraerrMGtwosiglimeccenxxxxxC}{\ensuremath{<0.0403}} 
\newcommand{\hatcurextraerrMBPtwoeccenxxxxxC}{\ensuremath{0.054\pm0.011}} 
\newcommand{\hatcurextraerrMBPtwotwosiglimeccenxxxxxC}{\ensuremath{<0.0737}} 
\newcommand{\hatcurextraerrMRPeccenxxxxxC}{\ensuremath{0.030\pm0.018}} 
\newcommand{\hatcurextraerrMRPtwosiglimeccenxxxxxC}{\ensuremath{<0.0620}} 
\newcommand{\hatcurextraerrMJeccenxxxxxC}{\ensuremath{0.039\pm0.011}} 
\newcommand{\hatcurextraerrMJtwosiglimeccenxxxxxC}{\ensuremath{<0.0597}} 
\newcommand{\hatcurextraerrMHeccenxxxxxC}{\ensuremath{0.0040_{-0.0039}^{+0.0186}}} 
\newcommand{\hatcurextraerrMHtwosiglimeccenxxxxxC}{\ensuremath{<0.0319}} 
\newcommand{\hatcurextraerrMKseccenxxxxxC}{\ensuremath{0.017\pm0.017}} 
\newcommand{\hatcurextraerrMKstwosiglimeccenxxxxxC}{\ensuremath{<0.0520}} 
\newcommand{\hatcurextraerrMWoneeccenxxxxxC}{\ensuremath{0.050\pm0.014}} 
\newcommand{\hatcurextraerrMWonetwosiglimeccenxxxxxC}{\ensuremath{<0.0710}} 
\newcommand{\hatcurextraerrMWtwoeccenxxxxxC}{\ensuremath{0.049_{-0.048}^{+0.064}}} 
\newcommand{\hatcurextraerrMWtwotwosiglimeccenxxxxxC}{\ensuremath{<0.1510}} 
\newcommand{\hatcurLBiBeccenxxxxxC}{\ensuremath{0.9623}}             
\newcommand{\hatcurLBiiBeccenxxxxxC}{\ensuremath{-0.1066}}           
\newcommand{\hatcurLBiVeccenxxxxxC}{\ensuremath{0.7587}}             
\newcommand{\hatcurLBiiVeccenxxxxxC}{\ensuremath{0.0458}}            
\newcommand{\hatcurLBiReccenxxxxxC}{\ensuremath{0.6243}}             
\newcommand{\hatcurLBiiReccenxxxxxC}{\ensuremath{0.1118}}            
\newcommand{\hatcurLBiIeccenxxxxxC}{\ensuremath{0.4811}}             
\newcommand{\hatcurLBiiIeccenxxxxxC}{\ensuremath{0.1553}}            
\newcommand{\hatcurLBiueccenxxxxxC}{\ensuremath{1.0925}}             
\newcommand{\hatcurLBiiueccenxxxxxC}{\ensuremath{-0.2541}}           
\newcommand{\hatcurLBigeccenxxxxxC}{\ensuremath{0.8657}}             
\newcommand{\hatcurLBiigeccenxxxxxC}{\ensuremath{-0.0410}}           
\newcommand{\hatcurLBireccenxxxxxC}{\ensuremath{0.635_{-0.128}^{+0.092}}} 
\newcommand{\hatcurLBiireccenxxxxxC}{\ensuremath{0.08\pm0.16}}       
\newcommand{\hatcurLBiieccenxxxxxC}{\ensuremath{0.25\pm0.11}}        
\newcommand{\hatcurLBiiieccenxxxxxC}{\ensuremath{0.19\pm0.15}}       
\newcommand{\hatcurLBizeccenxxxxxC}{\ensuremath{0.4191}}             
\newcommand{\hatcurLBiizeccenxxxxxC}{\ensuremath{0.1681}}            
\newcommand{\hatcurLBiJeccenxxxxxC}{\ensuremath{0.2967}}             
\newcommand{\hatcurLBiiJeccenxxxxxC}{\ensuremath{0.2133}}            
\newcommand{\hatcurLBiHeccenxxxxxC}{\ensuremath{0.1664}}             
\newcommand{\hatcurLBiiHeccenxxxxxC}{\ensuremath{0.2985}}            
\newcommand{\hatcurLBiKeccenxxxxxC}{\ensuremath{0.1330}}             
\newcommand{\hatcurLBiiKeccenxxxxxC}{\ensuremath{0.2539}}            
\newcommand{\hatcurLBiTeccenxxxxxC}{\ensuremath{0.35\pm0.12}}        
\newcommand{\hatcurLBiiTeccenxxxxxC}{\ensuremath{0.55\pm0.12}}       
\newcommand{\hatcurLBikepeccenxxxxxC}{\ensuremath{0.6246}}           
\newcommand{\hatcurLBiikepeccenxxxxxC}{\ensuremath{0.1226}}          
\newcommand{\hatcurLBiCeccenxxxxxC}{\ensuremath{0.5961}}             
\newcommand{\hatcurLBiiCeccenxxxxxC}{\ensuremath{0.1316}}            
\newcommand{\hatcurLBiMeccenxxxxxC}{\ensuremath{0.7290}}             
\newcommand{\hatcurLBiiMeccenxxxxxC}{\ensuremath{0.0691}}            
\newcommand{\hatcurLBiSoneeccenxxxxxC}{\ensuremath{0.1108}}            
\newcommand{\hatcurLBiiSoneeccenxxxxxC}{\ensuremath{0.1733}}           
\newcommand{\hatcurLBiStwoeccenxxxxxC}{\ensuremath{0.0986}}            
\newcommand{\hatcurLBiiStwoeccenxxxxxC}{\ensuremath{0.1376}}           
\newcommand{\hatcurLBiSthreeeccenxxxxxC}{\ensuremath{0.0828}}            
\newcommand{\hatcurLBiiSthreeeccenxxxxxC}{\ensuremath{0.1135}}           
\newcommand{\hatcurLBiSfoureccenxxxxxC}{\ensuremath{0.0731}}            
\newcommand{\hatcurLBiiSfoureccenxxxxxC}{\ensuremath{0.0967}}           
\newcommand{\hatcurISOmeccenxxxxxC}{\ensuremath{0.7027\pm0.0077}}    
\newcommand{\hatcurISOmshorteccenxxxxxC}{\ensuremath{0.70}}          
\newcommand{\hatcurISOmlongeccenxxxxxC}{\ensuremath{0.7027\pm0.0077}} 
\newcommand{\hatcurISOreccenxxxxxC}{\ensuremath{0.6928\pm0.0065}}    
\newcommand{\hatcurISOrshorteccenxxxxxC}{\ensuremath{0.69}}          
\newcommand{\hatcurISOrlongeccenxxxxxC}{\ensuremath{0.6928\pm0.0065}} 
\newcommand{\hatcurISOrhoeccenxxxxxC}{\ensuremath{2.969_{-0.054}^{+0.086}}} 
\newcommand{\hatcurISOrholongeccenxxxxxC}{\ensuremath{2.969_{-0.054}^{+0.086}}} 
\newcommand{\hatcurISOloggeccenxxxxxC}{\ensuremath{4.6020\pm0.0078}} 
\newcommand{\hatcurISOlumeccenxxxxxC}{\ensuremath{0.1623\pm0.0027}}  
\newcommand{\hatcurISOlumshorteccenxxxxxC}{\ensuremath{0.16}}        
\newcommand{\hatcurISOteffeccenxxxxxC}{\ensuremath{4409\pm18}}       
\newcommand{\hatcurISOzfeheccenxxxxxC}{\ensuremath{0.136_{-0.043}^{+0.058}}} 
\newcommand{\hatcurISOageeccenxxxxxC}{\ensuremath{11.41_{-1.96}^{+0.81}}} 
\newcommand{\hatcurISOspececcenxxxxxC}{K}                            
\newcommand{\hatcurRVKeccenxxxxxC}{\ensuremath{58.3\pm4.8}}          
\newcommand{\hatcurRVKtwosiglimeccenxxxxxC}{\ensuremath{<67.0}}      
\newcommand{\hatcurRVrkeccenxxxxxC}{\ensuremath{-0.106\pm0.093}}     
\newcommand{\hatcurRVrheccenxxxxxC}{\ensuremath{0.052_{-0.118}^{+0.091}}} 
\newcommand{\hatcurRVkeccenxxxxxC}{\ensuremath{-0.015_{-0.024}^{+0.015}}} 
\newcommand{\hatcurRVheccenxxxxxC}{\ensuremath{0.005\pm0.021}}       
\newcommand{\hatcurRVtroneeccenxxxxxC}{\ensuremath{0\pm0}}           
\newcommand{\hatcurRVtrtwoeccenxxxxxC}{\ensuremath{0\pm0}}           
\newcommand{\hatcurRVgammaeccenxxxxxC}{\ensuremath{-28.0\pm2.2}}     
\newcommand{\hatcurRVjittereccenxxxxxC}{\ensuremath{26.9\pm8.7}}     
\newcommand{\hatcurRVjittertwosiglimeccenxxxxxC}{\ensuremath{<44.1}} 
\newcommand{\hatcurRVfitrmseccenxxxxxC}{\ensuremath{.1fym}}          %
\newcommand{\hatcurRVecceneccenxxxxxC}{\ensuremath{0.027\pm0.021}}   
\newcommand{\hatcurRVeccentwosiglimeccenxxxxxC}{\ensuremath{<0.071}} 
\newcommand{\hatcurRVomegaeccenxxxxxC}{\ensuremath{152\pm56}}        
\newcommand{\hatcurPPieccenxxxxxC}{\ensuremath{88.20\pm0.16}}        
\newcommand{\hatcurPPgeccenxxxxxC}{\ensuremath{14.9\pm1.5}}          
\newcommand{\hatcurPPloggeccenxxxxxC}{\ensuremath{3.172\pm0.042}}    
\newcommand{\hatcurPPareccenxxxxxC}{\ensuremath{13.928_{-0.086}^{+0.133}}} 
\newcommand{\hatcurPPareleccenxxxxxC}{\ensuremath{0.04492\pm0.00016}} 
\newcommand{\hatcurPPrhoeccenxxxxxC}{\ensuremath{0.949_{-0.086}^{+0.123}}} 
\newcommand{\hatcurPPmeccenxxxxxC}{\ensuremath{0.364\pm0.031}}       
\newcommand{\hatcurPPmtwosiglimeccenxxxxxC}{\ensuremath{<0.42}}      
\newcommand{\hatcurPPmshorteccenxxxxxC}{\ensuremath{0.36}}           
\newcommand{\hatcurPPmlongeccenxxxxxC}{\ensuremath{0.364\pm0.031}}   
\newcommand{\hatcurPPmeeccenxxxxxC}{\ensuremath{115.7\pm9.9}}        
\newcommand{\hatcurPPmeshorteccenxxxxxC}{\ensuremath{115.7}}         
\newcommand{\hatcurPPmelongeccenxxxxxC}{\ensuremath{115.7\pm9.9}}    
\newcommand{\hatcurPPreccenxxxxxC}{\ensuremath{0.778\pm0.011}}       
\newcommand{\hatcurPPrshorteccenxxxxxC}{\ensuremath{0.78}}           
\newcommand{\hatcurPPrlongeccenxxxxxC}{\ensuremath{0.778\pm0.011}}   
\newcommand{\hatcurPPreeccenxxxxxC}{\ensuremath{8.72\pm0.12}}        
\newcommand{\hatcurPPreshorteccenxxxxxC}{\ensuremath{8.7}}           
\newcommand{\hatcurPPrelongeccenxxxxxC}{\ensuremath{8.72\pm0.12}}    
\newcommand{\hatcurPPmrcorreccenxxxxxC}{\ensuremath{-0.29}}          
\newcommand{\hatcurPPteffeccenxxxxxC}{\ensuremath{835.2\pm3.0}}      
\newcommand{\hatcurPPthetaeccenxxxxxC}{\ensuremath{0.0597\pm0.0053}} 
\newcommand{\hatcurPPfluxperieccenxxxxxC}{\ensuremath{1.155_{-0.043}^{+0.056}}} 
\newcommand{\hatcurPPfluxperidimeccenxxxxxC}{\ensuremath{8}}         
\newcommand{\hatcurPPfluxapeccenxxxxxC}{\ensuremath{1.038\pm0.042}}  
\newcommand{\hatcurPPfluxapdimeccenxxxxxC}{\ensuremath{8}}           
\newcommand{\hatcurPPfluxavgeccenxxxxxC}{\ensuremath{1.096\pm0.016}} 
\newcommand{\hatcurPPfluxavgdimeccenxxxxxC}{\ensuremath{8}}          
\newcommand{\hatcurPPfluxavglogeccenxxxxxC}{\ensuremath{8.0398\pm0.0063}} 
\newcommand{\hatcurXsecphaseeccenxxxxxC}{\ensuremath{0.491\pm0.014}} 
\newcommand{\hatcurXsecondaryeccenxxxxxC}{\ensuremath{2457173.569\pm0.058}} 
\newcommand{\hatcurXsecdureccenxxxxxC}{\ensuremath{0.0975\pm0.0029}} 
\newcommand{\hatcurXsecingdureccenxxxxxC}{\ensuremath{0.01226\pm0.00037}} 
\newcommand{\hatcurPPphiconjeccenxxxxxC}{\ensuremath{-0.16\pm0.16}}  
\newcommand{\hatcurPPperieccenxxxxxC}{\ensuremath{2457172.19\pm0.65}} 
\newcommand{\hatcurPPaequiveccenxxxxxC}{\ensuremath{0.11150\pm0.00080}} 
\newcommand{\hatcurPPtcirceccenxxxxxC}{\ensuremath{1440_{-160}^{+220}}} 
\newcommand{\hatcurPPtinfalleccenxxxxxC}{\ensuremath{39900_{-3800}^{+5300}}} 
\newcommand{\hatcurXdisteccenxxxxxC}{\ensuremath{324.6\pm1.5}}       
\newcommand{\hatcurXAveccenxxxxxC}{\ensuremath{0.051\pm0.012}}       
\newcommand{\hatcurXdistredeccenxxxxxC}{\ensuremath{324.6\pm1.5}}    
\newcommand{\hatcurXEBVeccenxxxxxC}{\ensuremath{0.0160_{-0.0030}^{+0.0040}}} 
\newcommand{\hatcurCCpmraeccenxxxxxC}{\ensuremath{42.581\pm0.035}}   
\newcommand{\hatcurCCpmdececcenxxxxxC}{\ensuremath{8.264\pm0.030}}   
\newcommand{\hatcurCCpmeccenxxxxxC}{\ensuremath{43.376\pm0.046}}     
\newcommand{\hatcurhtreccenxxxxxD}{HATS537-014}                      
\newcommand{\hatcurfieldeccenxxxxxD}{\ensuremath{string}}            
\newcommand{\hatcurCCraeccenxxxxxD}{\ensuremath{22^{\mathrm h}36^{\mathrm m}06.3190{\mathrm s}}}                   
\newcommand{\hatcurCCdececcenxxxxxD}{\ensuremath{-16{\arcdeg}59{\arcmin}59.7882{\arcsec}}}                 
\newcommand{\hatcurCCmageccenxxxxxD}{12.469}                         
\newcommand{\hatcurCCtwomasseccenxxxxxD}{2MASS~22360631-1659597}     
\newcommand{\hatcurCCgsceccenxxxxxD}{GSC~6386-00784}                 
\newcommand{\hatcurCCgaiaeccenxxxxxD}{GAIA~2594869603582993792}      
\newcommand{\hatcurCCgaiadrtwoeccenxxxxxD}{GAIA~DR2~2594869603582993792} 
\newcommand{\hatcurCCtassmveccenxxxxxD}{\ensuremath{12.469\pm0.010}} 
\newcommand{\hatcurCCtassmvshorteccenxxxxxD}{\ensuremath{12.5}}      
\newcommand{\hatcurCCtassmBeccenxxxxxD}{\ensuremath{13.572\pm0.030}} 
\newcommand{\hatcurCCtassmBshorteccenxxxxxD}{\ensuremath{13.6}}      
\newcommand{\hatcurCCtassmIeccenxxxxxD}{\ensuremath{nff\pmnff}}      
\newcommand{\hatcurCCtassmIshorteccenxxxxxD}{\ensuremath{0.0}}       
\newcommand{\hatcurCCtassmgeccenxxxxxD}{\ensuremath{12.995\pm0.010}} 
\newcommand{\hatcurCCtassmgshorteccenxxxxxD}{\ensuremath{13.0}}      
\newcommand{\hatcurCCtassmreccenxxxxxD}{\ensuremath{11.998\pm0.010}} 
\newcommand{\hatcurCCtassmrshorteccenxxxxxD}{\ensuremath{12.0}}      
\newcommand{\hatcurCCtassmieccenxxxxxD}{\ensuremath{11.622\pm0.030}} 
\newcommand{\hatcurCCtassmishorteccenxxxxxD}{\ensuremath{11.6}}      
\newcommand{\hatcurCCparallaxeccenxxxxxD}{\ensuremath{7.809\pm0.037}} 
\newcommand{\hatcurCCgaiamGeccenxxxxxD}{\ensuremath{12.07250\pm0.00030}} 
\newcommand{\hatcurCCgaiamBPeccenxxxxxD}{\ensuremath{12.7084\pm0.0018}} 
\newcommand{\hatcurCCgaiamRPeccenxxxxxD}{\ensuremath{11.3341\pm0.0010}} 
\newcommand{\hatcurCCtwomassJmageccenxxxxxD}{\ensuremath{10.424\pm0.023}} 
\newcommand{\hatcurCCtwomassHmageccenxxxxxD}{\ensuremath{9.907\pm0.026}} 
\newcommand{\hatcurCCtwomassKmageccenxxxxxD}{\ensuremath{9.764\pm0.021}} 
\newcommand{\hatcurCCcitJmageccenxxxxxD}{\ensuremath{10.425\pm0.024}} 
\newcommand{\hatcurCCcitHmageccenxxxxxD}{\ensuremath{9.900\pm0.027}} 
\newcommand{\hatcurCCcitKmageccenxxxxxD}{\ensuremath{9.788\pm0.022}} 
\newcommand{\hatcurCCbbJmageccenxxxxxD}{\ensuremath{10.498\pm0.026}} 
\newcommand{\hatcurCCbbHmageccenxxxxxD}{\ensuremath{9.923\pm0.027}}  
\newcommand{\hatcurCCbbKmageccenxxxxxD}{\ensuremath{9.808\pm0.022}}  
\newcommand{\hatcurCCesoJmageccenxxxxxD}{\ensuremath{10.504\pm0.029}} 
\newcommand{\hatcurCCesoHmageccenxxxxxD}{\ensuremath{9.920\pm0.034}} 
\newcommand{\hatcurCCesoKmageccenxxxxxD}{\ensuremath{9.806\pm0.023}} 
\newcommand{\hatcurCCesoJHmageccenxxxxxD}{\ensuremath{0.584\pm0.042}} 
\newcommand{\hatcurCCesoJKmageccenxxxxxD}{\ensuremath{0.699\pm0.035}} 
\newcommand{\hatcurCCesoHKmageccenxxxxxD}{\ensuremath{0.115\pm0.040}} 
\newcommand{\hatcurCCWonemageccenxxxxxD}{\ensuremath{9.687\pm0.024}} 
\newcommand{\hatcurCCWtwomageccenxxxxxD}{\ensuremath{9.772\pm0.020}} 
\newcommand{\hatcurCCWthreemageccenxxxxxD}{\ensuremath{9.675\pm0.043}} 
\newcommand{\hatcurCCWfourmageccenxxxxxD}{\ensuremath{0\pm0}}        
\newcommand{\hatcurLCdipeccenxxxxxD}{\ensuremath{13.4}}              
\newcommand{\hatcurLCrprstareccenxxxxxD}{\ensuremath{0.10362\pm0.00024}} 
\newcommand{\hatcurLCbsqeccenxxxxxD}{\ensuremath{0.0543_{-0.0023}^{+0.0030}}} 
\newcommand{\hatcurLCimpeccenxxxxxD}{\ensuremath{0.2330_{-0.0050}^{+0.0064}}} 
\newcommand{\hatcurLCzetaeccenxxxxxD}{\ensuremath{17.532\pm0.051}}   
\newcommand{\hatcurLCdureccenxxxxxD}{\ensuremath{0.12654\pm0.00036}} 
\newcommand{\hatcurLCdurshorteccenxxxxxD}{\ensuremath{0.1265}}       
\newcommand{\hatcurLCdurhreccenxxxxxD}{\ensuremath{3.0370\pm0.0087}} 
\newcommand{\hatcurLCdurhrshorteccenxxxxxD}{\ensuremath{3.037}}      
\newcommand{\hatcurLCqeccenxxxxxD}{\ensuremath{0.017300\pm0.000066}} 
\newcommand{\hatcurLCqshorteccenxxxxxD}{\ensuremath{0.017}}          
\newcommand{\hatcurLCingdureccenxxxxxD}{\ensuremath{0.012501\pm0.000066}} 
\newcommand{\hatcurLCPeccenxxxxxD}{\ensuremath{7.3279491\pm0.0000013}} 
\newcommand{\hatcurLCPprececcenxxxxxD}{\ensuremath{7.3279491}}       
\newcommand{\hatcurLCPshorteccenxxxxxD}{\ensuremath{7.3279}}         
\newcommand{\hatcurLCTeccenxxxxxD}{\ensuremath{2458322.142290\pm0.000035}} 
\newcommand{\hatcurLCTAeccenxxxxxD}{\ensuremath{2457552.70764\pm0.00014}} 
\newcommand{\hatcurLCTBeccenxxxxxD}{\ensuremath{2458380.765881\pm0.000037}} 
\newcommand{\hatcurLChatnetmAeccenxxxxxD}{\ensuremath{12.049830\pm0.000051}} 
\newcommand{\hatcurLCiblendAeccenxxxxxD}{\ensuremath{0.862\pm0.023}} 
\newcommand{\hatcurLChatnetmBeccenxxxxxD}{\ensuremath{11.628240\pm0.000021}} 
\newcommand{\hatcurLCiblendBeccenxxxxxD}{\ensuremath{0.99352\pm0.00035}} 
\newcommand{\hatcurLChatnetmCeccenxxxxxD}{\ensuremath{0.000200\pm0.000097}} 
\newcommand{\hatcurLCiblendCeccenxxxxxD}{\ensuremath{0.967\pm0.017}} 
\newcommand{\hatcurLCrhoeccenxxxxxD}{\ensuremath{2.739\pm0.030}}     
\newcommand{\hatcurSMEiteffeccenxxxxxD}{\ensuremath{4612\pm76}}      
\newcommand{\hatcurSMEizfeheccenxxxxxD}{\ensuremath{-0.040\pm0.050}} 
\newcommand{\hatcurSMEizfehshorteccenxxxxxD}{\ensuremath{-0.04}}     
\newcommand{\hatcurSMEiloggeccenxxxxxD}{\ensuremath{4.55\pm0.16}}    
\newcommand{\hatcurSMEivsineccenxxxxxD}{\ensuremath{0.8\pm1.3}}      
\newcommand{\hatcurSMEivmaceccenxxxxxD}{\ensuremath{2.20\pm0.12}}    
\newcommand{\hatcurSMEivmiceccenxxxxxD}{\ensuremath{0.443\pm0.061}}  
\newcommand{\hatcurextraerrMBeccenxxxxxD}{\ensuremath{0.0119\pm0.0028}} 
\newcommand{\hatcurextraerrMBtwosiglimeccenxxxxxD}{\ensuremath{<0.0150}} 
\newcommand{\hatcurextraerrMVeccenxxxxxD}{\ensuremath{0.0070_{-0.0067}^{+0.0034}}} 
\newcommand{\hatcurextraerrMVtwosiglimeccenxxxxxD}{\ensuremath{<0.0127}} 
\newcommand{\hatcurextraerrMgeccenxxxxxD}{\ensuremath{0.037\pm0.013}} 
\newcommand{\hatcurextraerrMgtwosiglimeccenxxxxxD}{\ensuremath{<0.0647}} 
\newcommand{\hatcurextraerrMreccenxxxxxD}{\ensuremath{0.000057_{-0.000056}^{+0.000667}}} 
\newcommand{\hatcurextraerrMrtwosiglimeccenxxxxxD}{\ensuremath{<0.0018}} 
\newcommand{\hatcurextraerrMieccenxxxxxD}{\ensuremath{0.053\pm0.026}} 
\newcommand{\hatcurextraerrMitwosiglimeccenxxxxxD}{\ensuremath{<0.0961}} 
\newcommand{\hatcurextraerrMGeccenxxxxxD}{\ensuremath{0.0221_{-0.0059}^{+0.0083}}} 
\newcommand{\hatcurextraerrMGtwosiglimeccenxxxxxD}{\ensuremath{<0.0355}} 
\newcommand{\hatcurextraerrMBPtwoeccenxxxxxD}{\ensuremath{0.0120\pm0.0070}} 
\newcommand{\hatcurextraerrMBPtwotwosiglimeccenxxxxxD}{\ensuremath{<0.0250}} 
\newcommand{\hatcurextraerrMRPeccenxxxxxD}{\ensuremath{0.00013_{-0.00013}^{+0.00139}}} 
\newcommand{\hatcurextraerrMRPtwosiglimeccenxxxxxD}{\ensuremath{<0.0043}} 
\newcommand{\hatcurextraerrMJeccenxxxxxD}{\ensuremath{0.0014_{-0.0013}^{+0.0076}}} 
\newcommand{\hatcurextraerrMJtwosiglimeccenxxxxxD}{\ensuremath{<0.0144}} 
\newcommand{\hatcurextraerrMHeccenxxxxxD}{\ensuremath{0.0018_{-0.0017}^{+0.0076}}} 
\newcommand{\hatcurextraerrMHtwosiglimeccenxxxxxD}{\ensuremath{<0.0140}} 
\newcommand{\hatcurextraerrMKseccenxxxxxD}{\ensuremath{0.00017_{-0.00017}^{+0.00256}}} 
\newcommand{\hatcurextraerrMKstwosiglimeccenxxxxxD}{\ensuremath{<0.0064}} 
\newcommand{\hatcurextraerrMWoneeccenxxxxxD}{\ensuremath{0.071_{-0.011}^{+0.022}}} 
\newcommand{\hatcurextraerrMWonetwosiglimeccenxxxxxD}{\ensuremath{<0.1111}} 
\newcommand{\hatcurextraerrMWtwoeccenxxxxxD}{\ensuremath{0.096_{-0.018}^{+0.026}}} 
\newcommand{\hatcurextraerrMWtwotwosiglimeccenxxxxxD}{\ensuremath{<0.1377}} 
\newcommand{\hatcurextraerrMWthreeeccenxxxxxD}{\ensuremath{0.067\pm0.061}} 
\newcommand{\hatcurextraerrMWthreetwosiglimeccenxxxxxD}{\ensuremath{<0.1924}} 
\newcommand{\hatcurLBiBeccenxxxxxD}{\ensuremath{0.9774}}             
\newcommand{\hatcurLBiiBeccenxxxxxD}{\ensuremath{-0.1216}}           
\newcommand{\hatcurLBiVeccenxxxxxD}{\ensuremath{0.7619}}             
\newcommand{\hatcurLBiiVeccenxxxxxD}{\ensuremath{0.0363}}            
\newcommand{\hatcurLBiReccenxxxxxD}{\ensuremath{0.6278}}             
\newcommand{\hatcurLBiiReccenxxxxxD}{\ensuremath{0.0963}}            
\newcommand{\hatcurLBiIeccenxxxxxD}{\ensuremath{0.124\pm0.041}}      
\newcommand{\hatcurLBiiIeccenxxxxxD}{\ensuremath{0.365_{-0.060}^{+0.045}}} 
\newcommand{\hatcurLBiueccenxxxxxD}{\ensuremath{1.1854}}             
\newcommand{\hatcurLBiiueccenxxxxxD}{\ensuremath{-0.3416}}           
\newcommand{\hatcurLBigeccenxxxxxD}{\ensuremath{0.8805}}             
\newcommand{\hatcurLBiigeccenxxxxxD}{\ensuremath{-0.0557}}           
\newcommand{\hatcurLBireccenxxxxxD}{\ensuremath{0.453\pm0.064}}      
\newcommand{\hatcurLBiireccenxxxxxD}{\ensuremath{0.416_{-0.109}^{+0.049}}} 
\newcommand{\hatcurLBiieccenxxxxxD}{\ensuremath{0.339_{-0.046}^{+0.035}}} 
\newcommand{\hatcurLBiiieccenxxxxxD}{\ensuremath{0.263\pm0.073}}     
\newcommand{\hatcurLBizeccenxxxxxD}{\ensuremath{0.4296}}             
\newcommand{\hatcurLBiizeccenxxxxxD}{\ensuremath{0.1493}}            
\newcommand{\hatcurLBiJeccenxxxxxD}{\ensuremath{0.2991}}             
\newcommand{\hatcurLBiiJeccenxxxxxD}{\ensuremath{0.2072}}            
\newcommand{\hatcurLBiHeccenxxxxxD}{\ensuremath{0.1534}}             
\newcommand{\hatcurLBiiHeccenxxxxxD}{\ensuremath{0.3100}}            
\newcommand{\hatcurLBiKeccenxxxxxD}{\ensuremath{0.1318}}             
\newcommand{\hatcurLBiiKeccenxxxxxD}{\ensuremath{0.2539}}            
\newcommand{\hatcurLBiTeccenxxxxxD}{\ensuremath{0.282\pm0.063}}      
\newcommand{\hatcurLBiiTeccenxxxxxD}{\ensuremath{0.607\pm0.081}}     
\newcommand{\hatcurLBikepeccenxxxxxD}{\ensuremath{0.660_{-0.113}^{+0.053}}} 
\newcommand{\hatcurLBiikepeccenxxxxxD}{\ensuremath{0.176\pm0.078}}   
\newcommand{\hatcurLBiCeccenxxxxxD}{\ensuremath{0.6027}}             
\newcommand{\hatcurLBiiCeccenxxxxxD}{\ensuremath{0.1141}}            
\newcommand{\hatcurLBiMeccenxxxxxD}{\ensuremath{0.7303}}             
\newcommand{\hatcurLBiiMeccenxxxxxD}{\ensuremath{0.0552}}            
\newcommand{\hatcurLBiSoneeccenxxxxxD}{\ensuremath{0.1161}}            
\newcommand{\hatcurLBiiSoneeccenxxxxxD}{\ensuremath{0.1617}}           
\newcommand{\hatcurLBiStwoeccenxxxxxD}{\ensuremath{0.1020}}            
\newcommand{\hatcurLBiiStwoeccenxxxxxD}{\ensuremath{0.1270}}           
\newcommand{\hatcurLBiSthreeeccenxxxxxD}{\ensuremath{0.0831}}            
\newcommand{\hatcurLBiiSthreeeccenxxxxxD}{\ensuremath{0.1042}}           
\newcommand{\hatcurLBiSfoureccenxxxxxD}{\ensuremath{0.0692}}            
\newcommand{\hatcurLBiiSfoureccenxxxxxD}{\ensuremath{0.0886}}           
\newcommand{\hatcurISOmeccenxxxxxD}{\ensuremath{0.7381_{-0.0027}^{+0.0019}}} 
\newcommand{\hatcurISOmshorteccenxxxxxD}{\ensuremath{0.74}}          
\newcommand{\hatcurISOmlongeccenxxxxxD}{\ensuremath{0.7381_{-0.0027}^{+0.0019}}} 
\newcommand{\hatcurISOreccenxxxxxD}{\ensuremath{0.7241\pm0.0024}}    
\newcommand{\hatcurISOrshorteccenxxxxxD}{\ensuremath{0.72}}          
\newcommand{\hatcurISOrlongeccenxxxxxD}{\ensuremath{0.7241\pm0.0024}} 
\newcommand{\hatcurISOrhoeccenxxxxxD}{\ensuremath{2.739\pm0.030}}    
\newcommand{\hatcurISOrholongeccenxxxxxD}{\ensuremath{2.739\pm0.030}} 
\newcommand{\hatcurISOloggeccenxxxxxD}{\ensuremath{4.5862\pm0.0033}} 
\newcommand{\hatcurISOlumeccenxxxxxD}{\ensuremath{0.2219\pm0.0018}}  
\newcommand{\hatcurISOlumshorteccenxxxxxD}{\ensuremath{0.22}}        
\newcommand{\hatcurISOteffeccenxxxxxD}{\ensuremath{4660.8\pm7.6}}    
\newcommand{\hatcurISOzfeheccenxxxxxD}{\ensuremath{0.1269_{-0.0139}^{+0.0098}}} 
\newcommand{\hatcurISOageeccenxxxxxD}{\ensuremath{11.75\pm0.60}}     
\newcommand{\hatcurISOspececcenxxxxxD}{K}                            
\newcommand{\hatcurRVKeccenxxxxxD}{\ensuremath{17.92\pm0.34}}        
\newcommand{\hatcurRVKtwosiglimeccenxxxxxD}{\ensuremath{<18.4}}      
\newcommand{\hatcurRVrkeccenxxxxxD}{\ensuremath{0.045_{-0.044}^{+0.031}}} 
\newcommand{\hatcurRVrheccenxxxxxD}{\ensuremath{0.043_{-0.048}^{+0.034}}} 
\newcommand{\hatcurRVkeccenxxxxxD}{\ensuremath{0.0030_{-0.0030}^{+0.0045}}} 
\newcommand{\hatcurRVheccenxxxxxD}{\ensuremath{0.0026_{-0.0028}^{+0.0050}}} 
\newcommand{\hatcurRVtroneeccenxxxxxD}{\ensuremath{0\pm0}}           
\newcommand{\hatcurRVtrtwoeccenxxxxxD}{\ensuremath{0\pm0}}           
\newcommand{\hatcurRVgammaAeccenxxxxxD}{\ensuremath{15970.9\pm2.7}}  
\newcommand{\hatcurRVjitterAeccenxxxxxD}{\ensuremath{30.5\pm5.2}}    
\newcommand{\hatcurRVjittertwosiglimAeccenxxxxxD}{\ensuremath{<39.1}} 
\newcommand{\hatcurRVfitrmsAeccenxxxxxD}{\ensuremath{0.0}}           
\newcommand{\hatcurRVgammaBeccenxxxxxD}{\ensuremath{15955.0\pm1.1}}  
\newcommand{\hatcurRVjitterBeccenxxxxxD}{\ensuremath{1.11\pm0.25}}   
\newcommand{\hatcurRVjittertwosiglimBeccenxxxxxD}{\ensuremath{<1.6}} 
\newcommand{\hatcurRVfitrmsBeccenxxxxxD}{\ensuremath{0.0}}           
\newcommand{\hatcurRVgammaCeccenxxxxxD}{\ensuremath{-1.31\pm0.28}}   
\newcommand{\hatcurRVjitterCeccenxxxxxD}{\ensuremath{6.7\pm1.7}}     
\newcommand{\hatcurRVjittertwosiglimCeccenxxxxxD}{\ensuremath{<10.3}} 
\newcommand{\hatcurRVfitrmsCeccenxxxxxD}{\ensuremath{0.0}}           
\newcommand{\hatcurRVgammaDeccenxxxxxD}{\ensuremath{15995.63\pm0.14}} 
\newcommand{\hatcurRVjitterDeccenxxxxxD}{\ensuremath{4.27\pm0.19}}   
\newcommand{\hatcurRVjittertwosiglimDeccenxxxxxD}{\ensuremath{<4.6}} 
\newcommand{\hatcurRVfitrmsDeccenxxxxxD}{\ensuremath{.1fym}}         %
\newcommand{\hatcurRVecceneccenxxxxxD}{\ensuremath{0.0050\pm0.0046}} 
\newcommand{\hatcurRVeccentwosiglimeccenxxxxxD}{\ensuremath{<0.013}} 
\newcommand{\hatcurRVomegaeccenxxxxxD}{\ensuremath{50\pm110}}        
\newcommand{\hatcurPPieccenxxxxxD}{\ensuremath{89.3200_{-0.0100}^{+0.0200}}} 
\newcommand{\hatcurPPgeccenxxxxxD}{\ensuremath{6.50_{-0.14}^{+0.11}}} 
\newcommand{\hatcurPPloggeccenxxxxxD}{\ensuremath{2.8130\pm0.0093}}  
\newcommand{\hatcurPPareccenxxxxxD}{\ensuremath{19.812\pm0.071}}     
\newcommand{\hatcurPPareleccenxxxxxD}{\ensuremath{0.066729_{-0.000082}^{+0.000057}}} 
\newcommand{\hatcurPPrhoeccenxxxxxD}{\ensuremath{0.4460_{-0.0110}^{+0.0080}}} 
\newcommand{\hatcurPPmeccenxxxxxD}{\ensuremath{0.1398_{-0.0027}^{+0.0020}}} 
\newcommand{\hatcurPPmtwosiglimeccenxxxxxD}{\ensuremath{<0.14}}      
\newcommand{\hatcurPPmshorteccenxxxxxD}{\ensuremath{0.14}}           
\newcommand{\hatcurPPmlongeccenxxxxxD}{\ensuremath{0.1398_{-0.0027}^{+0.0020}}} 
\newcommand{\hatcurPPmeeccenxxxxxD}{\ensuremath{44.43_{-0.86}^{+0.64}}} 
\newcommand{\hatcurPPmeshorteccenxxxxxD}{\ensuremath{44.4}}          
\newcommand{\hatcurPPmelongeccenxxxxxD}{\ensuremath{44.43_{-0.86}^{+0.64}}} 
\newcommand{\hatcurPPreccenxxxxxD}{\ensuremath{0.7302\pm0.0033}}     
\newcommand{\hatcurPPrshorteccenxxxxxD}{\ensuremath{0.73}}           
\newcommand{\hatcurPPrlongeccenxxxxxD}{\ensuremath{0.7302\pm0.0033}} 
\newcommand{\hatcurPPreeccenxxxxxD}{\ensuremath{8.185\pm0.038}}      
\newcommand{\hatcurPPreshorteccenxxxxxD}{\ensuremath{8.2}}           
\newcommand{\hatcurPPrelongeccenxxxxxD}{\ensuremath{8.185\pm0.038}}  
\newcommand{\hatcurPPmrcorreccenxxxxxD}{\ensuremath{-0.02}}          
\newcommand{\hatcurPPteffeccenxxxxxD}{\ensuremath{740.5\pm1.3}}      
\newcommand{\hatcurPPthetaeccenxxxxxD}{\ensuremath{0.03460_{-0.00070}^{+0.00050}}} 
\newcommand{\hatcurPPfluxperieccenxxxxxD}{\ensuremath{6.852_{-0.063}^{+0.109}}} 
\newcommand{\hatcurPPfluxperidimeccenxxxxxD}{\ensuremath{7}}         
\newcommand{\hatcurPPfluxapeccenxxxxxD}{\ensuremath{6.704\pm0.071}}  
\newcommand{\hatcurPPfluxapdimeccenxxxxxD}{\ensuremath{7}}           
\newcommand{\hatcurPPfluxavgeccenxxxxxD}{\ensuremath{6.785\pm0.054}} 
\newcommand{\hatcurPPfluxavgdimeccenxxxxxD}{\ensuremath{7}}          
\newcommand{\hatcurPPfluxavglogeccenxxxxxD}{\ensuremath{7.8315\pm0.0034}} 
\newcommand{\hatcurXsecphaseeccenxxxxxD}{\ensuremath{0.5019\pm0.0030}} 
\newcommand{\hatcurXsecondaryeccenxxxxxD}{\ensuremath{2458325.820\pm0.022}} 
\newcommand{\hatcurXsecdureccenxxxxxD}{\ensuremath{0.12729\pm0.00084}} 
\newcommand{\hatcurXsecingdureccenxxxxxD}{\ensuremath{0.01260\pm0.00011}} 
\newcommand{\hatcurPPphiconjeccenxxxxxD}{\ensuremath{0.12\pm0.16}}   
\newcommand{\hatcurPPperieccenxxxxxD}{\ensuremath{2458321.3\pm1.2}}  
\newcommand{\hatcurPPaequiveccenxxxxxD}{\ensuremath{0.14160\pm0.00057}} 
\newcommand{\hatcurPPtcirceccenxxxxxD}{\ensuremath{9360\pm290}}      
\newcommand{\hatcurPPtinfalleccenxxxxxD}{\ensuremath{1125000\pm30000}} 
\newcommand{\hatcurXdisteccenxxxxxD}{\ensuremath{128.00\pm0.37}}     
\newcommand{\hatcurXAveccenxxxxxD}{\ensuremath{0.0380\pm0.0090}}     
\newcommand{\hatcurXdistredeccenxxxxxD}{\ensuremath{128.04\pm0.37}}  
\newcommand{\hatcurXEBVeccenxxxxxD}{\ensuremath{0.0120_{-0.0020}^{+0.0030}}} 
\newcommand{\hatcurCCpmraeccenxxxxxD}{\ensuremath{-108.621\pm0.090}} 
\newcommand{\hatcurCCpmdececcenxxxxxD}{\ensuremath{-84.412\pm0.078}} 
\newcommand{\hatcurCCpmeccenxxxxxD}{\ensuremath{137.56\pm0.12}}      
\newcommand{\hatcurCCbbHmageccen}[1]{\ifnum#1=47 %
\hatcurCCbbHmageccenxxxxxA
\else
\ifnum#1=48 %
\hatcurCCbbHmageccenxxxxxB
\else
\ifnum#1=49 %
\hatcurCCbbHmageccenxxxxxC
\else
\ifnum#1=72 %
\hatcurCCbbHmageccenxxxxxD
\else
??????\fi
\fi
\fi
\fi
}
\newcommand{\hatcurCCbbJmageccen}[1]{\ifnum#1=47 %
\hatcurCCbbJmageccenxxxxxA
\else
\ifnum#1=48 %
\hatcurCCbbJmageccenxxxxxB
\else
\ifnum#1=49 %
\hatcurCCbbJmageccenxxxxxC
\else
\ifnum#1=72 %
\hatcurCCbbJmageccenxxxxxD
\else
??????\fi
\fi
\fi
\fi
}
\newcommand{\hatcurCCbbKmageccen}[1]{\ifnum#1=47 %
\hatcurCCbbKmageccenxxxxxA
\else
\ifnum#1=48 %
\hatcurCCbbKmageccenxxxxxB
\else
\ifnum#1=49 %
\hatcurCCbbKmageccenxxxxxC
\else
\ifnum#1=72 %
\hatcurCCbbKmageccenxxxxxD
\else
??????\fi
\fi
\fi
\fi
}
\newcommand{\hatcurCCcitHmageccen}[1]{\ifnum#1=47 %
\hatcurCCcitHmageccenxxxxxA
\else
\ifnum#1=48 %
\hatcurCCcitHmageccenxxxxxB
\else
\ifnum#1=49 %
\hatcurCCcitHmageccenxxxxxC
\else
\ifnum#1=72 %
\hatcurCCcitHmageccenxxxxxD
\else
??????\fi
\fi
\fi
\fi
}
\newcommand{\hatcurCCcitJmageccen}[1]{\ifnum#1=47 %
\hatcurCCcitJmageccenxxxxxA
\else
\ifnum#1=48 %
\hatcurCCcitJmageccenxxxxxB
\else
\ifnum#1=49 %
\hatcurCCcitJmageccenxxxxxC
\else
\ifnum#1=72 %
\hatcurCCcitJmageccenxxxxxD
\else
??????\fi
\fi
\fi
\fi
}
\newcommand{\hatcurCCcitKmageccen}[1]{\ifnum#1=47 %
\hatcurCCcitKmageccenxxxxxA
\else
\ifnum#1=48 %
\hatcurCCcitKmageccenxxxxxB
\else
\ifnum#1=49 %
\hatcurCCcitKmageccenxxxxxC
\else
\ifnum#1=72 %
\hatcurCCcitKmageccenxxxxxD
\else
??????\fi
\fi
\fi
\fi
}
\newcommand{\hatcurCCdececcen}[1]{\ifnum#1=47 %
\hatcurCCdececcenxxxxxA
\else
\ifnum#1=48 %
\hatcurCCdececcenxxxxxB
\else
\ifnum#1=49 %
\hatcurCCdececcenxxxxxC
\else
\ifnum#1=72 %
\hatcurCCdececcenxxxxxD
\else
??????\fi
\fi
\fi
\fi
}
\newcommand{\hatcurCCesoHKmageccen}[1]{\ifnum#1=47 %
\hatcurCCesoHKmageccenxxxxxA
\else
\ifnum#1=48 %
\hatcurCCesoHKmageccenxxxxxB
\else
\ifnum#1=49 %
\hatcurCCesoHKmageccenxxxxxC
\else
\ifnum#1=72 %
\hatcurCCesoHKmageccenxxxxxD
\else
??????\fi
\fi
\fi
\fi
}
\newcommand{\hatcurCCesoHmageccen}[1]{\ifnum#1=47 %
\hatcurCCesoHmageccenxxxxxA
\else
\ifnum#1=48 %
\hatcurCCesoHmageccenxxxxxB
\else
\ifnum#1=49 %
\hatcurCCesoHmageccenxxxxxC
\else
\ifnum#1=72 %
\hatcurCCesoHmageccenxxxxxD
\else
??????\fi
\fi
\fi
\fi
}
\newcommand{\hatcurCCesoJHmageccen}[1]{\ifnum#1=47 %
\hatcurCCesoJHmageccenxxxxxA
\else
\ifnum#1=48 %
\hatcurCCesoJHmageccenxxxxxB
\else
\ifnum#1=49 %
\hatcurCCesoJHmageccenxxxxxC
\else
\ifnum#1=72 %
\hatcurCCesoJHmageccenxxxxxD
\else
??????\fi
\fi
\fi
\fi
}
\newcommand{\hatcurCCesoJKmageccen}[1]{\ifnum#1=47 %
\hatcurCCesoJKmageccenxxxxxA
\else
\ifnum#1=48 %
\hatcurCCesoJKmageccenxxxxxB
\else
\ifnum#1=49 %
\hatcurCCesoJKmageccenxxxxxC
\else
\ifnum#1=72 %
\hatcurCCesoJKmageccenxxxxxD
\else
??????\fi
\fi
\fi
\fi
}
\newcommand{\hatcurCCesoJmageccen}[1]{\ifnum#1=47 %
\hatcurCCesoJmageccenxxxxxA
\else
\ifnum#1=48 %
\hatcurCCesoJmageccenxxxxxB
\else
\ifnum#1=49 %
\hatcurCCesoJmageccenxxxxxC
\else
\ifnum#1=72 %
\hatcurCCesoJmageccenxxxxxD
\else
??????\fi
\fi
\fi
\fi
}
\newcommand{\hatcurCCesoKmageccen}[1]{\ifnum#1=47 %
\hatcurCCesoKmageccenxxxxxA
\else
\ifnum#1=48 %
\hatcurCCesoKmageccenxxxxxB
\else
\ifnum#1=49 %
\hatcurCCesoKmageccenxxxxxC
\else
\ifnum#1=72 %
\hatcurCCesoKmageccenxxxxxD
\else
??????\fi
\fi
\fi
\fi
}
\newcommand{\hatcurCCgaiadrtwoeccen}[1]{\ifnum#1=47 %
\hatcurCCgaiadrtwoeccenxxxxxA
\else
\ifnum#1=48 %
\hatcurCCgaiadrtwoeccenxxxxxB
\else
\ifnum#1=49 %
\hatcurCCgaiadrtwoeccenxxxxxC
\else
\ifnum#1=72 %
\hatcurCCgaiadrtwoeccenxxxxxD
\else
??????\fi
\fi
\fi
\fi
}
\newcommand{\hatcurCCgaiaeccen}[1]{\ifnum#1=47 %
\hatcurCCgaiaeccenxxxxxA
\else
\ifnum#1=48 %
\hatcurCCgaiaeccenxxxxxB
\else
\ifnum#1=49 %
\hatcurCCgaiaeccenxxxxxC
\else
\ifnum#1=72 %
\hatcurCCgaiaeccenxxxxxD
\else
??????\fi
\fi
\fi
\fi
}
\newcommand{\hatcurCCgaiamBPeccen}[1]{\ifnum#1=47 %
\hatcurCCgaiamBPeccenxxxxxA
\else
\ifnum#1=48 %
\hatcurCCgaiamBPeccenxxxxxB
\else
\ifnum#1=49 %
\hatcurCCgaiamBPeccenxxxxxC
\else
\ifnum#1=72 %
\hatcurCCgaiamBPeccenxxxxxD
\else
??????\fi
\fi
\fi
\fi
}
\newcommand{\hatcurCCgaiamGeccen}[1]{\ifnum#1=47 %
\hatcurCCgaiamGeccenxxxxxA
\else
\ifnum#1=48 %
\hatcurCCgaiamGeccenxxxxxB
\else
\ifnum#1=49 %
\hatcurCCgaiamGeccenxxxxxC
\else
\ifnum#1=72 %
\hatcurCCgaiamGeccenxxxxxD
\else
??????\fi
\fi
\fi
\fi
}
\newcommand{\hatcurCCgaiamRPeccen}[1]{\ifnum#1=47 %
\hatcurCCgaiamRPeccenxxxxxA
\else
\ifnum#1=48 %
\hatcurCCgaiamRPeccenxxxxxB
\else
\ifnum#1=49 %
\hatcurCCgaiamRPeccenxxxxxC
\else
\ifnum#1=72 %
\hatcurCCgaiamRPeccenxxxxxD
\else
??????\fi
\fi
\fi
\fi
}
\newcommand{\hatcurCCgsceccen}[1]{\ifnum#1=47 %
\hatcurCCgsceccenxxxxxA
\else
\ifnum#1=48 %
\hatcurCCgsceccenxxxxxB
\else
\ifnum#1=49 %
\hatcurCCgsceccenxxxxxC
\else
\ifnum#1=72 %
\hatcurCCgsceccenxxxxxD
\else
??????\fi
\fi
\fi
\fi
}
\newcommand{\hatcurCCmageccen}[1]{\ifnum#1=47 %
\hatcurCCmageccenxxxxxA
\else
\ifnum#1=48 %
\hatcurCCmageccenxxxxxB
\else
\ifnum#1=49 %
\hatcurCCmageccenxxxxxC
\else
\ifnum#1=72 %
\hatcurCCmageccenxxxxxD
\else
??????\fi
\fi
\fi
\fi
}
\newcommand{\hatcurCCparallaxeccen}[1]{\ifnum#1=47 %
\hatcurCCparallaxeccenxxxxxA
\else
\ifnum#1=48 %
\hatcurCCparallaxeccenxxxxxB
\else
\ifnum#1=49 %
\hatcurCCparallaxeccenxxxxxC
\else
\ifnum#1=72 %
\hatcurCCparallaxeccenxxxxxD
\else
??????\fi
\fi
\fi
\fi
}
\newcommand{\hatcurCCpmdececcen}[1]{\ifnum#1=47 %
\hatcurCCpmdececcenxxxxxA
\else
\ifnum#1=48 %
\hatcurCCpmdececcenxxxxxB
\else
\ifnum#1=49 %
\hatcurCCpmdececcenxxxxxC
\else
\ifnum#1=72 %
\hatcurCCpmdececcenxxxxxD
\else
??????\fi
\fi
\fi
\fi
}
\newcommand{\hatcurCCpmeccen}[1]{\ifnum#1=47 %
\hatcurCCpmeccenxxxxxA
\else
\ifnum#1=48 %
\hatcurCCpmeccenxxxxxB
\else
\ifnum#1=49 %
\hatcurCCpmeccenxxxxxC
\else
\ifnum#1=72 %
\hatcurCCpmeccenxxxxxD
\else
??????\fi
\fi
\fi
\fi
}
\newcommand{\hatcurCCpmraeccen}[1]{\ifnum#1=47 %
\hatcurCCpmraeccenxxxxxA
\else
\ifnum#1=48 %
\hatcurCCpmraeccenxxxxxB
\else
\ifnum#1=49 %
\hatcurCCpmraeccenxxxxxC
\else
\ifnum#1=72 %
\hatcurCCpmraeccenxxxxxD
\else
??????\fi
\fi
\fi
\fi
}
\newcommand{\hatcurCCraeccen}[1]{\ifnum#1=47 %
\hatcurCCraeccenxxxxxA
\else
\ifnum#1=48 %
\hatcurCCraeccenxxxxxB
\else
\ifnum#1=49 %
\hatcurCCraeccenxxxxxC
\else
\ifnum#1=72 %
\hatcurCCraeccenxxxxxD
\else
??????\fi
\fi
\fi
\fi
}
\newcommand{\hatcurCCtassmBeccen}[1]{\ifnum#1=47 %
\hatcurCCtassmBeccenxxxxxA
\else
\ifnum#1=48 %
\hatcurCCtassmBeccenxxxxxB
\else
\ifnum#1=49 %
\hatcurCCtassmBeccenxxxxxC
\else
\ifnum#1=72 %
\hatcurCCtassmBeccenxxxxxD
\else
??????\fi
\fi
\fi
\fi
}
\newcommand{\hatcurCCtassmBshorteccen}[1]{\ifnum#1=47 %
\hatcurCCtassmBshorteccenxxxxxA
\else
\ifnum#1=48 %
\hatcurCCtassmBshorteccenxxxxxB
\else
\ifnum#1=49 %
\hatcurCCtassmBshorteccenxxxxxC
\else
\ifnum#1=72 %
\hatcurCCtassmBshorteccenxxxxxD
\else
??????\fi
\fi
\fi
\fi
}
\newcommand{\hatcurCCtassmgeccen}[1]{\ifnum#1=47 %
\hatcurCCtassmgeccenxxxxxA
\else
\ifnum#1=48 %
\hatcurCCtassmgeccenxxxxxB
\else
\ifnum#1=49 %
\hatcurCCtassmgeccenxxxxxC
\else
\ifnum#1=72 %
\hatcurCCtassmgeccenxxxxxD
\else
??????\fi
\fi
\fi
\fi
}
\newcommand{\hatcurCCtassmgshorteccen}[1]{\ifnum#1=47 %
\hatcurCCtassmgshorteccenxxxxxA
\else
\ifnum#1=48 %
\hatcurCCtassmgshorteccenxxxxxB
\else
\ifnum#1=49 %
\hatcurCCtassmgshorteccenxxxxxC
\else
\ifnum#1=72 %
\hatcurCCtassmgshorteccenxxxxxD
\else
??????\fi
\fi
\fi
\fi
}
\newcommand{\hatcurCCtassmieccen}[1]{\ifnum#1=47 %
\hatcurCCtassmieccenxxxxxA
\else
\ifnum#1=48 %
\hatcurCCtassmieccenxxxxxB
\else
\ifnum#1=49 %
\hatcurCCtassmieccenxxxxxC
\else
\ifnum#1=72 %
\hatcurCCtassmieccenxxxxxD
\else
??????\fi
\fi
\fi
\fi
}
\newcommand{\hatcurCCtassmIeccen}[1]{\ifnum#1=47 %
\hatcurCCtassmIeccenxxxxxA
\else
\ifnum#1=48 %
\hatcurCCtassmIeccenxxxxxB
\else
\ifnum#1=49 %
\hatcurCCtassmIeccenxxxxxC
\else
\ifnum#1=72 %
\hatcurCCtassmIeccenxxxxxD
\else
??????\fi
\fi
\fi
\fi
}
\newcommand{\hatcurCCtassmishorteccen}[1]{\ifnum#1=47 %
\hatcurCCtassmishorteccenxxxxxA
\else
\ifnum#1=48 %
\hatcurCCtassmishorteccenxxxxxB
\else
\ifnum#1=49 %
\hatcurCCtassmishorteccenxxxxxC
\else
\ifnum#1=72 %
\hatcurCCtassmishorteccenxxxxxD
\else
??????\fi
\fi
\fi
\fi
}
\newcommand{\hatcurCCtassmIshorteccen}[1]{\ifnum#1=47 %
\hatcurCCtassmIshorteccenxxxxxA
\else
\ifnum#1=48 %
\hatcurCCtassmIshorteccenxxxxxB
\else
\ifnum#1=49 %
\hatcurCCtassmIshorteccenxxxxxC
\else
\ifnum#1=72 %
\hatcurCCtassmIshorteccenxxxxxD
\else
??????\fi
\fi
\fi
\fi
}
\newcommand{\hatcurCCtassmreccen}[1]{\ifnum#1=47 %
\hatcurCCtassmreccenxxxxxA
\else
\ifnum#1=48 %
\hatcurCCtassmreccenxxxxxB
\else
\ifnum#1=49 %
\hatcurCCtassmreccenxxxxxC
\else
\ifnum#1=72 %
\hatcurCCtassmreccenxxxxxD
\else
??????\fi
\fi
\fi
\fi
}
\newcommand{\hatcurCCtassmrshorteccen}[1]{\ifnum#1=47 %
\hatcurCCtassmrshorteccenxxxxxA
\else
\ifnum#1=48 %
\hatcurCCtassmrshorteccenxxxxxB
\else
\ifnum#1=49 %
\hatcurCCtassmrshorteccenxxxxxC
\else
\ifnum#1=72 %
\hatcurCCtassmrshorteccenxxxxxD
\else
??????\fi
\fi
\fi
\fi
}
\newcommand{\hatcurCCtassmveccen}[1]{\ifnum#1=47 %
\hatcurCCtassmveccenxxxxxA
\else
\ifnum#1=48 %
\hatcurCCtassmveccenxxxxxB
\else
\ifnum#1=49 %
\hatcurCCtassmveccenxxxxxC
\else
\ifnum#1=72 %
\hatcurCCtassmveccenxxxxxD
\else
??????\fi
\fi
\fi
\fi
}
\newcommand{\hatcurCCtassmvshorteccen}[1]{\ifnum#1=47 %
\hatcurCCtassmvshorteccenxxxxxA
\else
\ifnum#1=48 %
\hatcurCCtassmvshorteccenxxxxxB
\else
\ifnum#1=49 %
\hatcurCCtassmvshorteccenxxxxxC
\else
\ifnum#1=72 %
\hatcurCCtassmvshorteccenxxxxxD
\else
??????\fi
\fi
\fi
\fi
}
\newcommand{\hatcurCCtwomasseccen}[1]{\ifnum#1=47 %
\hatcurCCtwomasseccenxxxxxA
\else
\ifnum#1=48 %
\hatcurCCtwomasseccenxxxxxB
\else
\ifnum#1=49 %
\hatcurCCtwomasseccenxxxxxC
\else
\ifnum#1=72 %
\hatcurCCtwomasseccenxxxxxD
\else
??????\fi
\fi
\fi
\fi
}
\newcommand{\hatcurCCtwomassHmageccen}[1]{\ifnum#1=47 %
\hatcurCCtwomassHmageccenxxxxxA
\else
\ifnum#1=48 %
\hatcurCCtwomassHmageccenxxxxxB
\else
\ifnum#1=49 %
\hatcurCCtwomassHmageccenxxxxxC
\else
\ifnum#1=72 %
\hatcurCCtwomassHmageccenxxxxxD
\else
??????\fi
\fi
\fi
\fi
}
\newcommand{\hatcurCCtwomassJmageccen}[1]{\ifnum#1=47 %
\hatcurCCtwomassJmageccenxxxxxA
\else
\ifnum#1=48 %
\hatcurCCtwomassJmageccenxxxxxB
\else
\ifnum#1=49 %
\hatcurCCtwomassJmageccenxxxxxC
\else
\ifnum#1=72 %
\hatcurCCtwomassJmageccenxxxxxD
\else
??????\fi
\fi
\fi
\fi
}
\newcommand{\hatcurCCtwomassKmageccen}[1]{\ifnum#1=47 %
\hatcurCCtwomassKmageccenxxxxxA
\else
\ifnum#1=48 %
\hatcurCCtwomassKmageccenxxxxxB
\else
\ifnum#1=49 %
\hatcurCCtwomassKmageccenxxxxxC
\else
\ifnum#1=72 %
\hatcurCCtwomassKmageccenxxxxxD
\else
??????\fi
\fi
\fi
\fi
}
\newcommand{\hatcurCCWfourmageccen}[1]{\ifnum#1=47 %
\hatcurCCWfourmageccenxxxxxA
\else
\ifnum#1=48 %
\hatcurCCWfourmageccenxxxxxB
\else
\ifnum#1=49 %
\hatcurCCWfourmageccenxxxxxC
\else
\ifnum#1=72 %
\hatcurCCWfourmageccenxxxxxD
\else
??????\fi
\fi
\fi
\fi
}
\newcommand{\hatcurCCWonemageccen}[1]{\ifnum#1=47 %
\hatcurCCWonemageccenxxxxxA
\else
\ifnum#1=48 %
\hatcurCCWonemageccenxxxxxB
\else
\ifnum#1=49 %
\hatcurCCWonemageccenxxxxxC
\else
\ifnum#1=72 %
\hatcurCCWonemageccenxxxxxD
\else
??????\fi
\fi
\fi
\fi
}
\newcommand{\hatcurCCWthreemageccen}[1]{\ifnum#1=47 %
\hatcurCCWthreemageccenxxxxxA
\else
\ifnum#1=48 %
\hatcurCCWthreemageccenxxxxxB
\else
\ifnum#1=49 %
\hatcurCCWthreemageccenxxxxxC
\else
\ifnum#1=72 %
\hatcurCCWthreemageccenxxxxxD
\else
??????\fi
\fi
\fi
\fi
}
\newcommand{\hatcurCCWtwomageccen}[1]{\ifnum#1=47 %
\hatcurCCWtwomageccenxxxxxA
\else
\ifnum#1=48 %
\hatcurCCWtwomageccenxxxxxB
\else
\ifnum#1=49 %
\hatcurCCWtwomageccenxxxxxC
\else
\ifnum#1=72 %
\hatcurCCWtwomageccenxxxxxD
\else
??????\fi
\fi
\fi
\fi
}
\newcommand{\hatcurextraerrMBeccen}[1]{\ifnum#1=72 %
\hatcurextraerrMBeccenxxxxxD
\else
??????\fi
}
\newcommand{\hatcurextraerrMBPtwoeccen}[1]{\ifnum#1=47 %
\hatcurextraerrMBPtwoeccenxxxxxA
\else
\ifnum#1=48 %
\hatcurextraerrMBPtwoeccenxxxxxB
\else
\ifnum#1=49 %
\hatcurextraerrMBPtwoeccenxxxxxC
\else
\ifnum#1=72 %
\hatcurextraerrMBPtwoeccenxxxxxD
\else
??????\fi
\fi
\fi
\fi
}
\newcommand{\hatcurextraerrMBPtwotwosiglimeccen}[1]{\ifnum#1=47 %
\hatcurextraerrMBPtwotwosiglimeccenxxxxxA
\else
\ifnum#1=48 %
\hatcurextraerrMBPtwotwosiglimeccenxxxxxB
\else
\ifnum#1=49 %
\hatcurextraerrMBPtwotwosiglimeccenxxxxxC
\else
\ifnum#1=72 %
\hatcurextraerrMBPtwotwosiglimeccenxxxxxD
\else
??????\fi
\fi
\fi
\fi
}
\newcommand{\hatcurextraerrMBtwosiglimeccen}[1]{\ifnum#1=72 %
\hatcurextraerrMBtwosiglimeccenxxxxxD
\else
??????\fi
}
\newcommand{\hatcurextraerrMgeccen}[1]{\ifnum#1=47 %
\hatcurextraerrMgeccenxxxxxA
\else
\ifnum#1=48 %
\hatcurextraerrMgeccenxxxxxB
\else
\ifnum#1=49 %
\hatcurextraerrMgeccenxxxxxC
\else
\ifnum#1=72 %
\hatcurextraerrMgeccenxxxxxD
\else
??????\fi
\fi
\fi
\fi
}
\newcommand{\hatcurextraerrMGeccen}[1]{\ifnum#1=47 %
\hatcurextraerrMGeccenxxxxxA
\else
\ifnum#1=48 %
\hatcurextraerrMGeccenxxxxxB
\else
\ifnum#1=49 %
\hatcurextraerrMGeccenxxxxxC
\else
\ifnum#1=72 %
\hatcurextraerrMGeccenxxxxxD
\else
??????\fi
\fi
\fi
\fi
}
\newcommand{\hatcurextraerrMgtwosiglimeccen}[1]{\ifnum#1=47 %
\hatcurextraerrMgtwosiglimeccenxxxxxA
\else
\ifnum#1=48 %
\hatcurextraerrMgtwosiglimeccenxxxxxB
\else
\ifnum#1=49 %
\hatcurextraerrMgtwosiglimeccenxxxxxC
\else
\ifnum#1=72 %
\hatcurextraerrMgtwosiglimeccenxxxxxD
\else
??????\fi
\fi
\fi
\fi
}
\newcommand{\hatcurextraerrMGtwosiglimeccen}[1]{\ifnum#1=47 %
\hatcurextraerrMGtwosiglimeccenxxxxxA
\else
\ifnum#1=48 %
\hatcurextraerrMGtwosiglimeccenxxxxxB
\else
\ifnum#1=49 %
\hatcurextraerrMGtwosiglimeccenxxxxxC
\else
\ifnum#1=72 %
\hatcurextraerrMGtwosiglimeccenxxxxxD
\else
??????\fi
\fi
\fi
\fi
}
\newcommand{\hatcurextraerrMHeccen}[1]{\ifnum#1=47 %
\hatcurextraerrMHeccenxxxxxA
\else
\ifnum#1=48 %
\hatcurextraerrMHeccenxxxxxB
\else
\ifnum#1=49 %
\hatcurextraerrMHeccenxxxxxC
\else
\ifnum#1=72 %
\hatcurextraerrMHeccenxxxxxD
\else
??????\fi
\fi
\fi
\fi
}
\newcommand{\hatcurextraerrMHtwosiglimeccen}[1]{\ifnum#1=47 %
\hatcurextraerrMHtwosiglimeccenxxxxxA
\else
\ifnum#1=48 %
\hatcurextraerrMHtwosiglimeccenxxxxxB
\else
\ifnum#1=49 %
\hatcurextraerrMHtwosiglimeccenxxxxxC
\else
\ifnum#1=72 %
\hatcurextraerrMHtwosiglimeccenxxxxxD
\else
??????\fi
\fi
\fi
\fi
}
\newcommand{\hatcurextraerrMieccen}[1]{\ifnum#1=47 %
\hatcurextraerrMieccenxxxxxA
\else
\ifnum#1=48 %
\hatcurextraerrMieccenxxxxxB
\else
\ifnum#1=49 %
\hatcurextraerrMieccenxxxxxC
\else
\ifnum#1=72 %
\hatcurextraerrMieccenxxxxxD
\else
??????\fi
\fi
\fi
\fi
}
\newcommand{\hatcurextraerrMitwosiglimeccen}[1]{\ifnum#1=47 %
\hatcurextraerrMitwosiglimeccenxxxxxA
\else
\ifnum#1=48 %
\hatcurextraerrMitwosiglimeccenxxxxxB
\else
\ifnum#1=49 %
\hatcurextraerrMitwosiglimeccenxxxxxC
\else
\ifnum#1=72 %
\hatcurextraerrMitwosiglimeccenxxxxxD
\else
??????\fi
\fi
\fi
\fi
}
\newcommand{\hatcurextraerrMJeccen}[1]{\ifnum#1=47 %
\hatcurextraerrMJeccenxxxxxA
\else
\ifnum#1=48 %
\hatcurextraerrMJeccenxxxxxB
\else
\ifnum#1=49 %
\hatcurextraerrMJeccenxxxxxC
\else
\ifnum#1=72 %
\hatcurextraerrMJeccenxxxxxD
\else
??????\fi
\fi
\fi
\fi
}
\newcommand{\hatcurextraerrMJtwosiglimeccen}[1]{\ifnum#1=47 %
\hatcurextraerrMJtwosiglimeccenxxxxxA
\else
\ifnum#1=48 %
\hatcurextraerrMJtwosiglimeccenxxxxxB
\else
\ifnum#1=49 %
\hatcurextraerrMJtwosiglimeccenxxxxxC
\else
\ifnum#1=72 %
\hatcurextraerrMJtwosiglimeccenxxxxxD
\else
??????\fi
\fi
\fi
\fi
}
\newcommand{\hatcurextraerrMKseccen}[1]{\ifnum#1=47 %
\hatcurextraerrMKseccenxxxxxA
\else
\ifnum#1=48 %
\hatcurextraerrMKseccenxxxxxB
\else
\ifnum#1=49 %
\hatcurextraerrMKseccenxxxxxC
\else
\ifnum#1=72 %
\hatcurextraerrMKseccenxxxxxD
\else
??????\fi
\fi
\fi
\fi
}
\newcommand{\hatcurextraerrMKstwosiglimeccen}[1]{\ifnum#1=47 %
\hatcurextraerrMKstwosiglimeccenxxxxxA
\else
\ifnum#1=48 %
\hatcurextraerrMKstwosiglimeccenxxxxxB
\else
\ifnum#1=49 %
\hatcurextraerrMKstwosiglimeccenxxxxxC
\else
\ifnum#1=72 %
\hatcurextraerrMKstwosiglimeccenxxxxxD
\else
??????\fi
\fi
\fi
\fi
}
\newcommand{\hatcurextraerrMreccen}[1]{\ifnum#1=47 %
\hatcurextraerrMreccenxxxxxA
\else
\ifnum#1=48 %
\hatcurextraerrMreccenxxxxxB
\else
\ifnum#1=49 %
\hatcurextraerrMreccenxxxxxC
\else
\ifnum#1=72 %
\hatcurextraerrMreccenxxxxxD
\else
??????\fi
\fi
\fi
\fi
}
\newcommand{\hatcurextraerrMRPeccen}[1]{\ifnum#1=47 %
\hatcurextraerrMRPeccenxxxxxA
\else
\ifnum#1=48 %
\hatcurextraerrMRPeccenxxxxxB
\else
\ifnum#1=49 %
\hatcurextraerrMRPeccenxxxxxC
\else
\ifnum#1=72 %
\hatcurextraerrMRPeccenxxxxxD
\else
??????\fi
\fi
\fi
\fi
}
\newcommand{\hatcurextraerrMRPtwosiglimeccen}[1]{\ifnum#1=47 %
\hatcurextraerrMRPtwosiglimeccenxxxxxA
\else
\ifnum#1=48 %
\hatcurextraerrMRPtwosiglimeccenxxxxxB
\else
\ifnum#1=49 %
\hatcurextraerrMRPtwosiglimeccenxxxxxC
\else
\ifnum#1=72 %
\hatcurextraerrMRPtwosiglimeccenxxxxxD
\else
??????\fi
\fi
\fi
\fi
}
\newcommand{\hatcurextraerrMrtwosiglimeccen}[1]{\ifnum#1=47 %
\hatcurextraerrMrtwosiglimeccenxxxxxA
\else
\ifnum#1=48 %
\hatcurextraerrMrtwosiglimeccenxxxxxB
\else
\ifnum#1=49 %
\hatcurextraerrMrtwosiglimeccenxxxxxC
\else
\ifnum#1=72 %
\hatcurextraerrMrtwosiglimeccenxxxxxD
\else
??????\fi
\fi
\fi
\fi
}
\newcommand{\hatcurextraerrMVeccen}[1]{\ifnum#1=72 %
\hatcurextraerrMVeccenxxxxxD
\else
??????\fi
}
\newcommand{\hatcurextraerrMVtwosiglimeccen}[1]{\ifnum#1=72 %
\hatcurextraerrMVtwosiglimeccenxxxxxD
\else
??????\fi
}
\newcommand{\hatcurextraerrMWoneeccen}[1]{\ifnum#1=47 %
\hatcurextraerrMWoneeccenxxxxxA
\else
\ifnum#1=48 %
\hatcurextraerrMWoneeccenxxxxxB
\else
\ifnum#1=49 %
\hatcurextraerrMWoneeccenxxxxxC
\else
\ifnum#1=72 %
\hatcurextraerrMWoneeccenxxxxxD
\else
??????\fi
\fi
\fi
\fi
}
\newcommand{\hatcurextraerrMWonetwosiglimeccen}[1]{\ifnum#1=47 %
\hatcurextraerrMWonetwosiglimeccenxxxxxA
\else
\ifnum#1=48 %
\hatcurextraerrMWonetwosiglimeccenxxxxxB
\else
\ifnum#1=49 %
\hatcurextraerrMWonetwosiglimeccenxxxxxC
\else
\ifnum#1=72 %
\hatcurextraerrMWonetwosiglimeccenxxxxxD
\else
??????\fi
\fi
\fi
\fi
}
\newcommand{\hatcurextraerrMWthreeeccen}[1]{\ifnum#1=72 %
\hatcurextraerrMWthreeeccenxxxxxD
\else
??????\fi
}
\newcommand{\hatcurextraerrMWthreetwosiglimeccen}[1]{\ifnum#1=72 %
\hatcurextraerrMWthreetwosiglimeccenxxxxxD
\else
??????\fi
}
\newcommand{\hatcurextraerrMWtwoeccen}[1]{\ifnum#1=47 %
\hatcurextraerrMWtwoeccenxxxxxA
\else
\ifnum#1=48 %
\hatcurextraerrMWtwoeccenxxxxxB
\else
\ifnum#1=49 %
\hatcurextraerrMWtwoeccenxxxxxC
\else
\ifnum#1=72 %
\hatcurextraerrMWtwoeccenxxxxxD
\else
??????\fi
\fi
\fi
\fi
}
\newcommand{\hatcurextraerrMWtwotwosiglimeccen}[1]{\ifnum#1=47 %
\hatcurextraerrMWtwotwosiglimeccenxxxxxA
\else
\ifnum#1=48 %
\hatcurextraerrMWtwotwosiglimeccenxxxxxB
\else
\ifnum#1=49 %
\hatcurextraerrMWtwotwosiglimeccenxxxxxC
\else
\ifnum#1=72 %
\hatcurextraerrMWtwotwosiglimeccenxxxxxD
\else
??????\fi
\fi
\fi
\fi
}
\newcommand{\hatcurfieldeccen}[1]{\ifnum#1=47 %
\hatcurfieldeccenxxxxxA
\else
\ifnum#1=48 %
\hatcurfieldeccenxxxxxB
\else
\ifnum#1=49 %
\hatcurfieldeccenxxxxxC
\else
\ifnum#1=72 %
\hatcurfieldeccenxxxxxD
\else
??????\fi
\fi
\fi
\fi
}
\newcommand{\hatcurhtreccen}[1]{\ifnum#1=47 %
\hatcurhtreccenxxxxxA
\else
\ifnum#1=48 %
\hatcurhtreccenxxxxxB
\else
\ifnum#1=49 %
\hatcurhtreccenxxxxxC
\else
\ifnum#1=72 %
\hatcurhtreccenxxxxxD
\else
??????\fi
\fi
\fi
\fi
}
\newcommand{\hatcurISOageeccen}[1]{\ifnum#1=47 %
\hatcurISOageeccenxxxxxA
\else
\ifnum#1=48 %
\hatcurISOageeccenxxxxxB
\else
\ifnum#1=49 %
\hatcurISOageeccenxxxxxC
\else
\ifnum#1=72 %
\hatcurISOageeccenxxxxxD
\else
??????\fi
\fi
\fi
\fi
}
\newcommand{\hatcurISOloggeccen}[1]{\ifnum#1=47 %
\hatcurISOloggeccenxxxxxA
\else
\ifnum#1=48 %
\hatcurISOloggeccenxxxxxB
\else
\ifnum#1=49 %
\hatcurISOloggeccenxxxxxC
\else
\ifnum#1=72 %
\hatcurISOloggeccenxxxxxD
\else
??????\fi
\fi
\fi
\fi
}
\newcommand{\hatcurISOlumeccen}[1]{\ifnum#1=47 %
\hatcurISOlumeccenxxxxxA
\else
\ifnum#1=48 %
\hatcurISOlumeccenxxxxxB
\else
\ifnum#1=49 %
\hatcurISOlumeccenxxxxxC
\else
\ifnum#1=72 %
\hatcurISOlumeccenxxxxxD
\else
??????\fi
\fi
\fi
\fi
}
\newcommand{\hatcurISOlumshorteccen}[1]{\ifnum#1=47 %
\hatcurISOlumshorteccenxxxxxA
\else
\ifnum#1=48 %
\hatcurISOlumshorteccenxxxxxB
\else
\ifnum#1=49 %
\hatcurISOlumshorteccenxxxxxC
\else
\ifnum#1=72 %
\hatcurISOlumshorteccenxxxxxD
\else
??????\fi
\fi
\fi
\fi
}
\newcommand{\hatcurISOmeccen}[1]{\ifnum#1=47 %
\hatcurISOmeccenxxxxxA
\else
\ifnum#1=48 %
\hatcurISOmeccenxxxxxB
\else
\ifnum#1=49 %
\hatcurISOmeccenxxxxxC
\else
\ifnum#1=72 %
\hatcurISOmeccenxxxxxD
\else
??????\fi
\fi
\fi
\fi
}
\newcommand{\hatcurISOmlongeccen}[1]{\ifnum#1=47 %
\hatcurISOmlongeccenxxxxxA
\else
\ifnum#1=48 %
\hatcurISOmlongeccenxxxxxB
\else
\ifnum#1=49 %
\hatcurISOmlongeccenxxxxxC
\else
\ifnum#1=72 %
\hatcurISOmlongeccenxxxxxD
\else
??????\fi
\fi
\fi
\fi
}
\newcommand{\hatcurISOmshorteccen}[1]{\ifnum#1=47 %
\hatcurISOmshorteccenxxxxxA
\else
\ifnum#1=48 %
\hatcurISOmshorteccenxxxxxB
\else
\ifnum#1=49 %
\hatcurISOmshorteccenxxxxxC
\else
\ifnum#1=72 %
\hatcurISOmshorteccenxxxxxD
\else
??????\fi
\fi
\fi
\fi
}
\newcommand{\hatcurISOreccen}[1]{\ifnum#1=47 %
\hatcurISOreccenxxxxxA
\else
\ifnum#1=48 %
\hatcurISOreccenxxxxxB
\else
\ifnum#1=49 %
\hatcurISOreccenxxxxxC
\else
\ifnum#1=72 %
\hatcurISOreccenxxxxxD
\else
??????\fi
\fi
\fi
\fi
}
\newcommand{\hatcurISOrhoeccen}[1]{\ifnum#1=47 %
\hatcurISOrhoeccenxxxxxA
\else
\ifnum#1=48 %
\hatcurISOrhoeccenxxxxxB
\else
\ifnum#1=49 %
\hatcurISOrhoeccenxxxxxC
\else
\ifnum#1=72 %
\hatcurISOrhoeccenxxxxxD
\else
??????\fi
\fi
\fi
\fi
}
\newcommand{\hatcurISOrholongeccen}[1]{\ifnum#1=47 %
\hatcurISOrholongeccenxxxxxA
\else
\ifnum#1=48 %
\hatcurISOrholongeccenxxxxxB
\else
\ifnum#1=49 %
\hatcurISOrholongeccenxxxxxC
\else
\ifnum#1=72 %
\hatcurISOrholongeccenxxxxxD
\else
??????\fi
\fi
\fi
\fi
}
\newcommand{\hatcurISOrlongeccen}[1]{\ifnum#1=47 %
\hatcurISOrlongeccenxxxxxA
\else
\ifnum#1=48 %
\hatcurISOrlongeccenxxxxxB
\else
\ifnum#1=49 %
\hatcurISOrlongeccenxxxxxC
\else
\ifnum#1=72 %
\hatcurISOrlongeccenxxxxxD
\else
??????\fi
\fi
\fi
\fi
}
\newcommand{\hatcurISOrshorteccen}[1]{\ifnum#1=47 %
\hatcurISOrshorteccenxxxxxA
\else
\ifnum#1=48 %
\hatcurISOrshorteccenxxxxxB
\else
\ifnum#1=49 %
\hatcurISOrshorteccenxxxxxC
\else
\ifnum#1=72 %
\hatcurISOrshorteccenxxxxxD
\else
??????\fi
\fi
\fi
\fi
}
\newcommand{\hatcurISOspececcen}[1]{\ifnum#1=47 %
\hatcurISOspececcenxxxxxA
\else
\ifnum#1=48 %
\hatcurISOspececcenxxxxxB
\else
\ifnum#1=49 %
\hatcurISOspececcenxxxxxC
\else
\ifnum#1=72 %
\hatcurISOspececcenxxxxxD
\else
??????\fi
\fi
\fi
\fi
}
\newcommand{\hatcurISOteffeccen}[1]{\ifnum#1=47 %
\hatcurISOteffeccenxxxxxA
\else
\ifnum#1=48 %
\hatcurISOteffeccenxxxxxB
\else
\ifnum#1=49 %
\hatcurISOteffeccenxxxxxC
\else
\ifnum#1=72 %
\hatcurISOteffeccenxxxxxD
\else
??????\fi
\fi
\fi
\fi
}
\newcommand{\hatcurISOzfeheccen}[1]{\ifnum#1=47 %
\hatcurISOzfeheccenxxxxxA
\else
\ifnum#1=48 %
\hatcurISOzfeheccenxxxxxB
\else
\ifnum#1=49 %
\hatcurISOzfeheccenxxxxxC
\else
\ifnum#1=72 %
\hatcurISOzfeheccenxxxxxD
\else
??????\fi
\fi
\fi
\fi
}
\newcommand{\hatcurLBiBeccen}[1]{\ifnum#1=47 %
\hatcurLBiBeccenxxxxxA
\else
\ifnum#1=48 %
\hatcurLBiBeccenxxxxxB
\else
\ifnum#1=49 %
\hatcurLBiBeccenxxxxxC
\else
\ifnum#1=72 %
\hatcurLBiBeccenxxxxxD
\else
??????\fi
\fi
\fi
\fi
}
\newcommand{\hatcurLBiCeccen}[1]{\ifnum#1=47 %
\hatcurLBiCeccenxxxxxA
\else
\ifnum#1=48 %
\hatcurLBiCeccenxxxxxB
\else
\ifnum#1=49 %
\hatcurLBiCeccenxxxxxC
\else
\ifnum#1=72 %
\hatcurLBiCeccenxxxxxD
\else
??????\fi
\fi
\fi
\fi
}
\newcommand{\hatcurLBigeccen}[1]{\ifnum#1=47 %
\hatcurLBigeccenxxxxxA
\else
\ifnum#1=48 %
\hatcurLBigeccenxxxxxB
\else
\ifnum#1=49 %
\hatcurLBigeccenxxxxxC
\else
\ifnum#1=72 %
\hatcurLBigeccenxxxxxD
\else
??????\fi
\fi
\fi
\fi
}
\newcommand{\hatcurLBiHeccen}[1]{\ifnum#1=47 %
\hatcurLBiHeccenxxxxxA
\else
\ifnum#1=48 %
\hatcurLBiHeccenxxxxxB
\else
\ifnum#1=49 %
\hatcurLBiHeccenxxxxxC
\else
\ifnum#1=72 %
\hatcurLBiHeccenxxxxxD
\else
??????\fi
\fi
\fi
\fi
}
\newcommand{\hatcurLBiiBeccen}[1]{\ifnum#1=47 %
\hatcurLBiiBeccenxxxxxA
\else
\ifnum#1=48 %
\hatcurLBiiBeccenxxxxxB
\else
\ifnum#1=49 %
\hatcurLBiiBeccenxxxxxC
\else
\ifnum#1=72 %
\hatcurLBiiBeccenxxxxxD
\else
??????\fi
\fi
\fi
\fi
}
\newcommand{\hatcurLBiiCeccen}[1]{\ifnum#1=47 %
\hatcurLBiiCeccenxxxxxA
\else
\ifnum#1=48 %
\hatcurLBiiCeccenxxxxxB
\else
\ifnum#1=49 %
\hatcurLBiiCeccenxxxxxC
\else
\ifnum#1=72 %
\hatcurLBiiCeccenxxxxxD
\else
??????\fi
\fi
\fi
\fi
}
\newcommand{\hatcurLBiieccen}[1]{\ifnum#1=47 %
\hatcurLBiieccenxxxxxA
\else
\ifnum#1=48 %
\hatcurLBiieccenxxxxxB
\else
\ifnum#1=49 %
\hatcurLBiieccenxxxxxC
\else
\ifnum#1=72 %
\hatcurLBiieccenxxxxxD
\else
??????\fi
\fi
\fi
\fi
}
\newcommand{\hatcurLBiIeccen}[1]{\ifnum#1=47 %
\hatcurLBiIeccenxxxxxA
\else
\ifnum#1=48 %
\hatcurLBiIeccenxxxxxB
\else
\ifnum#1=49 %
\hatcurLBiIeccenxxxxxC
\else
\ifnum#1=72 %
\hatcurLBiIeccenxxxxxD
\else
??????\fi
\fi
\fi
\fi
}
\newcommand{\hatcurLBiigeccen}[1]{\ifnum#1=47 %
\hatcurLBiigeccenxxxxxA
\else
\ifnum#1=48 %
\hatcurLBiigeccenxxxxxB
\else
\ifnum#1=49 %
\hatcurLBiigeccenxxxxxC
\else
\ifnum#1=72 %
\hatcurLBiigeccenxxxxxD
\else
??????\fi
\fi
\fi
\fi
}
\newcommand{\hatcurLBiiHeccen}[1]{\ifnum#1=47 %
\hatcurLBiiHeccenxxxxxA
\else
\ifnum#1=48 %
\hatcurLBiiHeccenxxxxxB
\else
\ifnum#1=49 %
\hatcurLBiiHeccenxxxxxC
\else
\ifnum#1=72 %
\hatcurLBiiHeccenxxxxxD
\else
??????\fi
\fi
\fi
\fi
}
\newcommand{\hatcurLBiiieccen}[1]{\ifnum#1=47 %
\hatcurLBiiieccenxxxxxA
\else
\ifnum#1=48 %
\hatcurLBiiieccenxxxxxB
\else
\ifnum#1=49 %
\hatcurLBiiieccenxxxxxC
\else
\ifnum#1=72 %
\hatcurLBiiieccenxxxxxD
\else
??????\fi
\fi
\fi
\fi
}
\newcommand{\hatcurLBiiIeccen}[1]{\ifnum#1=47 %
\hatcurLBiiIeccenxxxxxA
\else
\ifnum#1=48 %
\hatcurLBiiIeccenxxxxxB
\else
\ifnum#1=49 %
\hatcurLBiiIeccenxxxxxC
\else
\ifnum#1=72 %
\hatcurLBiiIeccenxxxxxD
\else
??????\fi
\fi
\fi
\fi
}
\newcommand{\hatcurLBiiJeccen}[1]{\ifnum#1=47 %
\hatcurLBiiJeccenxxxxxA
\else
\ifnum#1=48 %
\hatcurLBiiJeccenxxxxxB
\else
\ifnum#1=49 %
\hatcurLBiiJeccenxxxxxC
\else
\ifnum#1=72 %
\hatcurLBiiJeccenxxxxxD
\else
??????\fi
\fi
\fi
\fi
}
\newcommand{\hatcurLBiiKeccen}[1]{\ifnum#1=47 %
\hatcurLBiiKeccenxxxxxA
\else
\ifnum#1=48 %
\hatcurLBiiKeccenxxxxxB
\else
\ifnum#1=49 %
\hatcurLBiiKeccenxxxxxC
\else
\ifnum#1=72 %
\hatcurLBiiKeccenxxxxxD
\else
??????\fi
\fi
\fi
\fi
}
\newcommand{\hatcurLBiikepeccen}[1]{\ifnum#1=47 %
\hatcurLBiikepeccenxxxxxA
\else
\ifnum#1=48 %
\hatcurLBiikepeccenxxxxxB
\else
\ifnum#1=49 %
\hatcurLBiikepeccenxxxxxC
\else
\ifnum#1=72 %
\hatcurLBiikepeccenxxxxxD
\else
??????\fi
\fi
\fi
\fi
}
\newcommand{\hatcurLBiiMeccen}[1]{\ifnum#1=47 %
\hatcurLBiiMeccenxxxxxA
\else
\ifnum#1=48 %
\hatcurLBiiMeccenxxxxxB
\else
\ifnum#1=49 %
\hatcurLBiiMeccenxxxxxC
\else
\ifnum#1=72 %
\hatcurLBiiMeccenxxxxxD
\else
??????\fi
\fi
\fi
\fi
}
\newcommand{\hatcurLBiireccen}[1]{\ifnum#1=47 %
\hatcurLBiireccenxxxxxA
\else
\ifnum#1=48 %
\hatcurLBiireccenxxxxxB
\else
\ifnum#1=49 %
\hatcurLBiireccenxxxxxC
\else
\ifnum#1=72 %
\hatcurLBiireccenxxxxxD
\else
??????\fi
\fi
\fi
\fi
}
\newcommand{\hatcurLBiiReccen}[1]{\ifnum#1=47 %
\hatcurLBiiReccenxxxxxA
\else
\ifnum#1=48 %
\hatcurLBiiReccenxxxxxB
\else
\ifnum#1=49 %
\hatcurLBiiReccenxxxxxC
\else
\ifnum#1=72 %
\hatcurLBiiReccenxxxxxD
\else
??????\fi
\fi
\fi
\fi
}
\newcommand{\hatcurLBiiSfoureccen}[1]{\ifnum#1=47 %
\hatcurLBiiSfoureccenxxxxxA
\else
\ifnum#1=48 %
\hatcurLBiiSfoureccenxxxxxB
\else
\ifnum#1=49 %
\hatcurLBiiSfoureccenxxxxxC
\else
\ifnum#1=72 %
\hatcurLBiiSfoureccenxxxxxD
\else
??????\fi
\fi
\fi
\fi
}
\newcommand{\hatcurLBiiSoneeccen}[1]{\ifnum#1=47 %
\hatcurLBiiSoneeccenxxxxxA
\else
\ifnum#1=48 %
\hatcurLBiiSoneeccenxxxxxB
\else
\ifnum#1=49 %
\hatcurLBiiSoneeccenxxxxxC
\else
\ifnum#1=72 %
\hatcurLBiiSoneeccenxxxxxD
\else
??????\fi
\fi
\fi
\fi
}
\newcommand{\hatcurLBiiSthreeeccen}[1]{\ifnum#1=47 %
\hatcurLBiiSthreeeccenxxxxxA
\else
\ifnum#1=48 %
\hatcurLBiiSthreeeccenxxxxxB
\else
\ifnum#1=49 %
\hatcurLBiiSthreeeccenxxxxxC
\else
\ifnum#1=72 %
\hatcurLBiiSthreeeccenxxxxxD
\else
??????\fi
\fi
\fi
\fi
}
\newcommand{\hatcurLBiiStwoeccen}[1]{\ifnum#1=47 %
\hatcurLBiiStwoeccenxxxxxA
\else
\ifnum#1=48 %
\hatcurLBiiStwoeccenxxxxxB
\else
\ifnum#1=49 %
\hatcurLBiiStwoeccenxxxxxC
\else
\ifnum#1=72 %
\hatcurLBiiStwoeccenxxxxxD
\else
??????\fi
\fi
\fi
\fi
}
\newcommand{\hatcurLBiiTeccen}[1]{\ifnum#1=47 %
\hatcurLBiiTeccenxxxxxA
\else
\ifnum#1=48 %
\hatcurLBiiTeccenxxxxxB
\else
\ifnum#1=49 %
\hatcurLBiiTeccenxxxxxC
\else
\ifnum#1=72 %
\hatcurLBiiTeccenxxxxxD
\else
??????\fi
\fi
\fi
\fi
}
\newcommand{\hatcurLBiiueccen}[1]{\ifnum#1=47 %
\hatcurLBiiueccenxxxxxA
\else
\ifnum#1=48 %
\hatcurLBiiueccenxxxxxB
\else
\ifnum#1=49 %
\hatcurLBiiueccenxxxxxC
\else
\ifnum#1=72 %
\hatcurLBiiueccenxxxxxD
\else
??????\fi
\fi
\fi
\fi
}
\newcommand{\hatcurLBiiVeccen}[1]{\ifnum#1=47 %
\hatcurLBiiVeccenxxxxxA
\else
\ifnum#1=48 %
\hatcurLBiiVeccenxxxxxB
\else
\ifnum#1=49 %
\hatcurLBiiVeccenxxxxxC
\else
\ifnum#1=72 %
\hatcurLBiiVeccenxxxxxD
\else
??????\fi
\fi
\fi
\fi
}
\newcommand{\hatcurLBiizeccen}[1]{\ifnum#1=47 %
\hatcurLBiizeccenxxxxxA
\else
\ifnum#1=48 %
\hatcurLBiizeccenxxxxxB
\else
\ifnum#1=49 %
\hatcurLBiizeccenxxxxxC
\else
\ifnum#1=72 %
\hatcurLBiizeccenxxxxxD
\else
??????\fi
\fi
\fi
\fi
}
\newcommand{\hatcurLBiJeccen}[1]{\ifnum#1=47 %
\hatcurLBiJeccenxxxxxA
\else
\ifnum#1=48 %
\hatcurLBiJeccenxxxxxB
\else
\ifnum#1=49 %
\hatcurLBiJeccenxxxxxC
\else
\ifnum#1=72 %
\hatcurLBiJeccenxxxxxD
\else
??????\fi
\fi
\fi
\fi
}
\newcommand{\hatcurLBiKeccen}[1]{\ifnum#1=47 %
\hatcurLBiKeccenxxxxxA
\else
\ifnum#1=48 %
\hatcurLBiKeccenxxxxxB
\else
\ifnum#1=49 %
\hatcurLBiKeccenxxxxxC
\else
\ifnum#1=72 %
\hatcurLBiKeccenxxxxxD
\else
??????\fi
\fi
\fi
\fi
}
\newcommand{\hatcurLBikepeccen}[1]{\ifnum#1=47 %
\hatcurLBikepeccenxxxxxA
\else
\ifnum#1=48 %
\hatcurLBikepeccenxxxxxB
\else
\ifnum#1=49 %
\hatcurLBikepeccenxxxxxC
\else
\ifnum#1=72 %
\hatcurLBikepeccenxxxxxD
\else
??????\fi
\fi
\fi
\fi
}
\newcommand{\hatcurLBiMeccen}[1]{\ifnum#1=47 %
\hatcurLBiMeccenxxxxxA
\else
\ifnum#1=48 %
\hatcurLBiMeccenxxxxxB
\else
\ifnum#1=49 %
\hatcurLBiMeccenxxxxxC
\else
\ifnum#1=72 %
\hatcurLBiMeccenxxxxxD
\else
??????\fi
\fi
\fi
\fi
}
\newcommand{\hatcurLBireccen}[1]{\ifnum#1=47 %
\hatcurLBireccenxxxxxA
\else
\ifnum#1=48 %
\hatcurLBireccenxxxxxB
\else
\ifnum#1=49 %
\hatcurLBireccenxxxxxC
\else
\ifnum#1=72 %
\hatcurLBireccenxxxxxD
\else
??????\fi
\fi
\fi
\fi
}
\newcommand{\hatcurLBiReccen}[1]{\ifnum#1=47 %
\hatcurLBiReccenxxxxxA
\else
\ifnum#1=48 %
\hatcurLBiReccenxxxxxB
\else
\ifnum#1=49 %
\hatcurLBiReccenxxxxxC
\else
\ifnum#1=72 %
\hatcurLBiReccenxxxxxD
\else
??????\fi
\fi
\fi
\fi
}
\newcommand{\hatcurLBiSfoureccen}[1]{\ifnum#1=47 %
\hatcurLBiSfoureccenxxxxxA
\else
\ifnum#1=48 %
\hatcurLBiSfoureccenxxxxxB
\else
\ifnum#1=49 %
\hatcurLBiSfoureccenxxxxxC
\else
\ifnum#1=72 %
\hatcurLBiSfoureccenxxxxxD
\else
??????\fi
\fi
\fi
\fi
}
\newcommand{\hatcurLBiSoneeccen}[1]{\ifnum#1=47 %
\hatcurLBiSoneeccenxxxxxA
\else
\ifnum#1=48 %
\hatcurLBiSoneeccenxxxxxB
\else
\ifnum#1=49 %
\hatcurLBiSoneeccenxxxxxC
\else
\ifnum#1=72 %
\hatcurLBiSoneeccenxxxxxD
\else
??????\fi
\fi
\fi
\fi
}
\newcommand{\hatcurLBiSthreeeccen}[1]{\ifnum#1=47 %
\hatcurLBiSthreeeccenxxxxxA
\else
\ifnum#1=48 %
\hatcurLBiSthreeeccenxxxxxB
\else
\ifnum#1=49 %
\hatcurLBiSthreeeccenxxxxxC
\else
\ifnum#1=72 %
\hatcurLBiSthreeeccenxxxxxD
\else
??????\fi
\fi
\fi
\fi
}
\newcommand{\hatcurLBiStwoeccen}[1]{\ifnum#1=47 %
\hatcurLBiStwoeccenxxxxxA
\else
\ifnum#1=48 %
\hatcurLBiStwoeccenxxxxxB
\else
\ifnum#1=49 %
\hatcurLBiStwoeccenxxxxxC
\else
\ifnum#1=72 %
\hatcurLBiStwoeccenxxxxxD
\else
??????\fi
\fi
\fi
\fi
}
\newcommand{\hatcurLBiTeccen}[1]{\ifnum#1=47 %
\hatcurLBiTeccenxxxxxA
\else
\ifnum#1=48 %
\hatcurLBiTeccenxxxxxB
\else
\ifnum#1=49 %
\hatcurLBiTeccenxxxxxC
\else
\ifnum#1=72 %
\hatcurLBiTeccenxxxxxD
\else
??????\fi
\fi
\fi
\fi
}
\newcommand{\hatcurLBiueccen}[1]{\ifnum#1=47 %
\hatcurLBiueccenxxxxxA
\else
\ifnum#1=48 %
\hatcurLBiueccenxxxxxB
\else
\ifnum#1=49 %
\hatcurLBiueccenxxxxxC
\else
\ifnum#1=72 %
\hatcurLBiueccenxxxxxD
\else
??????\fi
\fi
\fi
\fi
}
\newcommand{\hatcurLBiVeccen}[1]{\ifnum#1=47 %
\hatcurLBiVeccenxxxxxA
\else
\ifnum#1=48 %
\hatcurLBiVeccenxxxxxB
\else
\ifnum#1=49 %
\hatcurLBiVeccenxxxxxC
\else
\ifnum#1=72 %
\hatcurLBiVeccenxxxxxD
\else
??????\fi
\fi
\fi
\fi
}
\newcommand{\hatcurLBizeccen}[1]{\ifnum#1=47 %
\hatcurLBizeccenxxxxxA
\else
\ifnum#1=48 %
\hatcurLBizeccenxxxxxB
\else
\ifnum#1=49 %
\hatcurLBizeccenxxxxxC
\else
\ifnum#1=72 %
\hatcurLBizeccenxxxxxD
\else
??????\fi
\fi
\fi
\fi
}
\newcommand{\hatcurLCbsqeccen}[1]{\ifnum#1=47 %
\hatcurLCbsqeccenxxxxxA
\else
\ifnum#1=48 %
\hatcurLCbsqeccenxxxxxB
\else
\ifnum#1=49 %
\hatcurLCbsqeccenxxxxxC
\else
\ifnum#1=72 %
\hatcurLCbsqeccenxxxxxD
\else
??????\fi
\fi
\fi
\fi
}
\newcommand{\hatcurLCdipeccen}[1]{\ifnum#1=47 %
\hatcurLCdipeccenxxxxxA
\else
\ifnum#1=48 %
\hatcurLCdipeccenxxxxxB
\else
\ifnum#1=49 %
\hatcurLCdipeccenxxxxxC
\else
\ifnum#1=72 %
\hatcurLCdipeccenxxxxxD
\else
??????\fi
\fi
\fi
\fi
}
\newcommand{\hatcurLCdureccen}[1]{\ifnum#1=47 %
\hatcurLCdureccenxxxxxA
\else
\ifnum#1=48 %
\hatcurLCdureccenxxxxxB
\else
\ifnum#1=49 %
\hatcurLCdureccenxxxxxC
\else
\ifnum#1=72 %
\hatcurLCdureccenxxxxxD
\else
??????\fi
\fi
\fi
\fi
}
\newcommand{\hatcurLCdurhreccen}[1]{\ifnum#1=47 %
\hatcurLCdurhreccenxxxxxA
\else
\ifnum#1=48 %
\hatcurLCdurhreccenxxxxxB
\else
\ifnum#1=49 %
\hatcurLCdurhreccenxxxxxC
\else
\ifnum#1=72 %
\hatcurLCdurhreccenxxxxxD
\else
??????\fi
\fi
\fi
\fi
}
\newcommand{\hatcurLCdurhrshorteccen}[1]{\ifnum#1=47 %
\hatcurLCdurhrshorteccenxxxxxA
\else
\ifnum#1=48 %
\hatcurLCdurhrshorteccenxxxxxB
\else
\ifnum#1=49 %
\hatcurLCdurhrshorteccenxxxxxC
\else
\ifnum#1=72 %
\hatcurLCdurhrshorteccenxxxxxD
\else
??????\fi
\fi
\fi
\fi
}
\newcommand{\hatcurLCdurshorteccen}[1]{\ifnum#1=47 %
\hatcurLCdurshorteccenxxxxxA
\else
\ifnum#1=48 %
\hatcurLCdurshorteccenxxxxxB
\else
\ifnum#1=49 %
\hatcurLCdurshorteccenxxxxxC
\else
\ifnum#1=72 %
\hatcurLCdurshorteccenxxxxxD
\else
??????\fi
\fi
\fi
\fi
}
\newcommand{\hatcurLChatnetmAeccen}[1]{\ifnum#1=47 %
\hatcurLChatnetmAeccenxxxxxA
\else
\ifnum#1=48 %
\hatcurLChatnetmAeccenxxxxxB
\else
\ifnum#1=49 %
\hatcurLChatnetmAeccenxxxxxC
\else
\ifnum#1=72 %
\hatcurLChatnetmAeccenxxxxxD
\else
??????\fi
\fi
\fi
\fi
}
\newcommand{\hatcurLChatnetmBeccen}[1]{\ifnum#1=47 %
\hatcurLChatnetmBeccenxxxxxA
\else
\ifnum#1=48 %
\hatcurLChatnetmBeccenxxxxxB
\else
\ifnum#1=49 %
\hatcurLChatnetmBeccenxxxxxC
\else
\ifnum#1=72 %
\hatcurLChatnetmBeccenxxxxxD
\else
??????\fi
\fi
\fi
\fi
}
\newcommand{\hatcurLChatnetmCeccen}[1]{\ifnum#1=49 %
\hatcurLChatnetmCeccenxxxxxC
\else
\ifnum#1=72 %
\hatcurLChatnetmCeccenxxxxxD
\else
??????\fi
\fi
}
\newcommand{\hatcurLCiblendAeccen}[1]{\ifnum#1=47 %
\hatcurLCiblendAeccenxxxxxA
\else
\ifnum#1=48 %
\hatcurLCiblendAeccenxxxxxB
\else
\ifnum#1=49 %
\hatcurLCiblendAeccenxxxxxC
\else
\ifnum#1=72 %
\hatcurLCiblendAeccenxxxxxD
\else
??????\fi
\fi
\fi
\fi
}
\newcommand{\hatcurLCiblendBeccen}[1]{\ifnum#1=47 %
\hatcurLCiblendBeccenxxxxxA
\else
\ifnum#1=48 %
\hatcurLCiblendBeccenxxxxxB
\else
\ifnum#1=49 %
\hatcurLCiblendBeccenxxxxxC
\else
\ifnum#1=72 %
\hatcurLCiblendBeccenxxxxxD
\else
??????\fi
\fi
\fi
\fi
}
\newcommand{\hatcurLCiblendCeccen}[1]{\ifnum#1=49 %
\hatcurLCiblendCeccenxxxxxC
\else
\ifnum#1=72 %
\hatcurLCiblendCeccenxxxxxD
\else
??????\fi
\fi
}
\newcommand{\hatcurLCimpeccen}[1]{\ifnum#1=47 %
\hatcurLCimpeccenxxxxxA
\else
\ifnum#1=48 %
\hatcurLCimpeccenxxxxxB
\else
\ifnum#1=49 %
\hatcurLCimpeccenxxxxxC
\else
\ifnum#1=72 %
\hatcurLCimpeccenxxxxxD
\else
??????\fi
\fi
\fi
\fi
}
\newcommand{\hatcurLCingdureccen}[1]{\ifnum#1=47 %
\hatcurLCingdureccenxxxxxA
\else
\ifnum#1=48 %
\hatcurLCingdureccenxxxxxB
\else
\ifnum#1=49 %
\hatcurLCingdureccenxxxxxC
\else
\ifnum#1=72 %
\hatcurLCingdureccenxxxxxD
\else
??????\fi
\fi
\fi
\fi
}
\newcommand{\hatcurLCPeccen}[1]{\ifnum#1=47 %
\hatcurLCPeccenxxxxxA
\else
\ifnum#1=48 %
\hatcurLCPeccenxxxxxB
\else
\ifnum#1=49 %
\hatcurLCPeccenxxxxxC
\else
\ifnum#1=72 %
\hatcurLCPeccenxxxxxD
\else
??????\fi
\fi
\fi
\fi
}
\newcommand{\hatcurLCPprececcen}[1]{\ifnum#1=47 %
\hatcurLCPprececcenxxxxxA
\else
\ifnum#1=48 %
\hatcurLCPprececcenxxxxxB
\else
\ifnum#1=49 %
\hatcurLCPprececcenxxxxxC
\else
\ifnum#1=72 %
\hatcurLCPprececcenxxxxxD
\else
??????\fi
\fi
\fi
\fi
}
\newcommand{\hatcurLCPshorteccen}[1]{\ifnum#1=47 %
\hatcurLCPshorteccenxxxxxA
\else
\ifnum#1=48 %
\hatcurLCPshorteccenxxxxxB
\else
\ifnum#1=49 %
\hatcurLCPshorteccenxxxxxC
\else
\ifnum#1=72 %
\hatcurLCPshorteccenxxxxxD
\else
??????\fi
\fi
\fi
\fi
}
\newcommand{\hatcurLCqeccen}[1]{\ifnum#1=47 %
\hatcurLCqeccenxxxxxA
\else
\ifnum#1=48 %
\hatcurLCqeccenxxxxxB
\else
\ifnum#1=49 %
\hatcurLCqeccenxxxxxC
\else
\ifnum#1=72 %
\hatcurLCqeccenxxxxxD
\else
??????\fi
\fi
\fi
\fi
}
\newcommand{\hatcurLCqshorteccen}[1]{\ifnum#1=47 %
\hatcurLCqshorteccenxxxxxA
\else
\ifnum#1=48 %
\hatcurLCqshorteccenxxxxxB
\else
\ifnum#1=49 %
\hatcurLCqshorteccenxxxxxC
\else
\ifnum#1=72 %
\hatcurLCqshorteccenxxxxxD
\else
??????\fi
\fi
\fi
\fi
}
\newcommand{\hatcurLCrhoeccen}[1]{\ifnum#1=47 %
\hatcurLCrhoeccenxxxxxA
\else
\ifnum#1=48 %
\hatcurLCrhoeccenxxxxxB
\else
\ifnum#1=49 %
\hatcurLCrhoeccenxxxxxC
\else
\ifnum#1=72 %
\hatcurLCrhoeccenxxxxxD
\else
??????\fi
\fi
\fi
\fi
}
\newcommand{\hatcurLCrprstareccen}[1]{\ifnum#1=47 %
\hatcurLCrprstareccenxxxxxA
\else
\ifnum#1=48 %
\hatcurLCrprstareccenxxxxxB
\else
\ifnum#1=49 %
\hatcurLCrprstareccenxxxxxC
\else
\ifnum#1=72 %
\hatcurLCrprstareccenxxxxxD
\else
??????\fi
\fi
\fi
\fi
}
\newcommand{\hatcurLCTAeccen}[1]{\ifnum#1=47 %
\hatcurLCTAeccenxxxxxA
\else
\ifnum#1=48 %
\hatcurLCTAeccenxxxxxB
\else
\ifnum#1=49 %
\hatcurLCTAeccenxxxxxC
\else
\ifnum#1=72 %
\hatcurLCTAeccenxxxxxD
\else
??????\fi
\fi
\fi
\fi
}
\newcommand{\hatcurLCTBeccen}[1]{\ifnum#1=47 %
\hatcurLCTBeccenxxxxxA
\else
\ifnum#1=48 %
\hatcurLCTBeccenxxxxxB
\else
\ifnum#1=49 %
\hatcurLCTBeccenxxxxxC
\else
\ifnum#1=72 %
\hatcurLCTBeccenxxxxxD
\else
??????\fi
\fi
\fi
\fi
}
\newcommand{\hatcurLCTeccen}[1]{\ifnum#1=47 %
\hatcurLCTeccenxxxxxA
\else
\ifnum#1=48 %
\hatcurLCTeccenxxxxxB
\else
\ifnum#1=49 %
\hatcurLCTeccenxxxxxC
\else
\ifnum#1=72 %
\hatcurLCTeccenxxxxxD
\else
??????\fi
\fi
\fi
\fi
}
\newcommand{\hatcurLCzetaeccen}[1]{\ifnum#1=47 %
\hatcurLCzetaeccenxxxxxA
\else
\ifnum#1=48 %
\hatcurLCzetaeccenxxxxxB
\else
\ifnum#1=49 %
\hatcurLCzetaeccenxxxxxC
\else
\ifnum#1=72 %
\hatcurLCzetaeccenxxxxxD
\else
??????\fi
\fi
\fi
\fi
}
\newcommand{\hatcurPPaequiveccen}[1]{\ifnum#1=47 %
\hatcurPPaequiveccenxxxxxA
\else
\ifnum#1=48 %
\hatcurPPaequiveccenxxxxxB
\else
\ifnum#1=49 %
\hatcurPPaequiveccenxxxxxC
\else
\ifnum#1=72 %
\hatcurPPaequiveccenxxxxxD
\else
??????\fi
\fi
\fi
\fi
}
\newcommand{\hatcurPPareccen}[1]{\ifnum#1=47 %
\hatcurPPareccenxxxxxA
\else
\ifnum#1=48 %
\hatcurPPareccenxxxxxB
\else
\ifnum#1=49 %
\hatcurPPareccenxxxxxC
\else
\ifnum#1=72 %
\hatcurPPareccenxxxxxD
\else
??????\fi
\fi
\fi
\fi
}
\newcommand{\hatcurPPareleccen}[1]{\ifnum#1=47 %
\hatcurPPareleccenxxxxxA
\else
\ifnum#1=48 %
\hatcurPPareleccenxxxxxB
\else
\ifnum#1=49 %
\hatcurPPareleccenxxxxxC
\else
\ifnum#1=72 %
\hatcurPPareleccenxxxxxD
\else
??????\fi
\fi
\fi
\fi
}
\newcommand{\hatcurPPfluxapdimeccen}[1]{\ifnum#1=47 %
\hatcurPPfluxapdimeccenxxxxxA
\else
\ifnum#1=48 %
\hatcurPPfluxapdimeccenxxxxxB
\else
\ifnum#1=49 %
\hatcurPPfluxapdimeccenxxxxxC
\else
\ifnum#1=72 %
\hatcurPPfluxapdimeccenxxxxxD
\else
??????\fi
\fi
\fi
\fi
}
\newcommand{\hatcurPPfluxapeccen}[1]{\ifnum#1=47 %
\hatcurPPfluxapeccenxxxxxA
\else
\ifnum#1=48 %
\hatcurPPfluxapeccenxxxxxB
\else
\ifnum#1=49 %
\hatcurPPfluxapeccenxxxxxC
\else
\ifnum#1=72 %
\hatcurPPfluxapeccenxxxxxD
\else
??????\fi
\fi
\fi
\fi
}
\newcommand{\hatcurPPfluxavgdimeccen}[1]{\ifnum#1=47 %
\hatcurPPfluxavgdimeccenxxxxxA
\else
\ifnum#1=48 %
\hatcurPPfluxavgdimeccenxxxxxB
\else
\ifnum#1=49 %
\hatcurPPfluxavgdimeccenxxxxxC
\else
\ifnum#1=72 %
\hatcurPPfluxavgdimeccenxxxxxD
\else
??????\fi
\fi
\fi
\fi
}
\newcommand{\hatcurPPfluxavgeccen}[1]{\ifnum#1=47 %
\hatcurPPfluxavgeccenxxxxxA
\else
\ifnum#1=48 %
\hatcurPPfluxavgeccenxxxxxB
\else
\ifnum#1=49 %
\hatcurPPfluxavgeccenxxxxxC
\else
\ifnum#1=72 %
\hatcurPPfluxavgeccenxxxxxD
\else
??????\fi
\fi
\fi
\fi
}
\newcommand{\hatcurPPfluxavglogeccen}[1]{\ifnum#1=47 %
\hatcurPPfluxavglogeccenxxxxxA
\else
\ifnum#1=48 %
\hatcurPPfluxavglogeccenxxxxxB
\else
\ifnum#1=49 %
\hatcurPPfluxavglogeccenxxxxxC
\else
\ifnum#1=72 %
\hatcurPPfluxavglogeccenxxxxxD
\else
??????\fi
\fi
\fi
\fi
}
\newcommand{\hatcurPPfluxperidimeccen}[1]{\ifnum#1=47 %
\hatcurPPfluxperidimeccenxxxxxA
\else
\ifnum#1=48 %
\hatcurPPfluxperidimeccenxxxxxB
\else
\ifnum#1=49 %
\hatcurPPfluxperidimeccenxxxxxC
\else
\ifnum#1=72 %
\hatcurPPfluxperidimeccenxxxxxD
\else
??????\fi
\fi
\fi
\fi
}
\newcommand{\hatcurPPfluxperieccen}[1]{\ifnum#1=47 %
\hatcurPPfluxperieccenxxxxxA
\else
\ifnum#1=48 %
\hatcurPPfluxperieccenxxxxxB
\else
\ifnum#1=49 %
\hatcurPPfluxperieccenxxxxxC
\else
\ifnum#1=72 %
\hatcurPPfluxperieccenxxxxxD
\else
??????\fi
\fi
\fi
\fi
}
\newcommand{\hatcurPPgeccen}[1]{\ifnum#1=47 %
\hatcurPPgeccenxxxxxA
\else
\ifnum#1=48 %
\hatcurPPgeccenxxxxxB
\else
\ifnum#1=49 %
\hatcurPPgeccenxxxxxC
\else
\ifnum#1=72 %
\hatcurPPgeccenxxxxxD
\else
??????\fi
\fi
\fi
\fi
}
\newcommand{\hatcurPPieccen}[1]{\ifnum#1=47 %
\hatcurPPieccenxxxxxA
\else
\ifnum#1=48 %
\hatcurPPieccenxxxxxB
\else
\ifnum#1=49 %
\hatcurPPieccenxxxxxC
\else
\ifnum#1=72 %
\hatcurPPieccenxxxxxD
\else
??????\fi
\fi
\fi
\fi
}
\newcommand{\hatcurPPloggeccen}[1]{\ifnum#1=47 %
\hatcurPPloggeccenxxxxxA
\else
\ifnum#1=48 %
\hatcurPPloggeccenxxxxxB
\else
\ifnum#1=49 %
\hatcurPPloggeccenxxxxxC
\else
\ifnum#1=72 %
\hatcurPPloggeccenxxxxxD
\else
??????\fi
\fi
\fi
\fi
}
\newcommand{\hatcurPPmeccen}[1]{\ifnum#1=47 %
\hatcurPPmeccenxxxxxA
\else
\ifnum#1=48 %
\hatcurPPmeccenxxxxxB
\else
\ifnum#1=49 %
\hatcurPPmeccenxxxxxC
\else
\ifnum#1=72 %
\hatcurPPmeccenxxxxxD
\else
??????\fi
\fi
\fi
\fi
}
\newcommand{\hatcurPPmeeccen}[1]{\ifnum#1=47 %
\hatcurPPmeeccenxxxxxA
\else
\ifnum#1=48 %
\hatcurPPmeeccenxxxxxB
\else
\ifnum#1=49 %
\hatcurPPmeeccenxxxxxC
\else
\ifnum#1=72 %
\hatcurPPmeeccenxxxxxD
\else
??????\fi
\fi
\fi
\fi
}
\newcommand{\hatcurPPmelongeccen}[1]{\ifnum#1=47 %
\hatcurPPmelongeccenxxxxxA
\else
\ifnum#1=48 %
\hatcurPPmelongeccenxxxxxB
\else
\ifnum#1=49 %
\hatcurPPmelongeccenxxxxxC
\else
\ifnum#1=72 %
\hatcurPPmelongeccenxxxxxD
\else
??????\fi
\fi
\fi
\fi
}
\newcommand{\hatcurPPmeshorteccen}[1]{\ifnum#1=47 %
\hatcurPPmeshorteccenxxxxxA
\else
\ifnum#1=48 %
\hatcurPPmeshorteccenxxxxxB
\else
\ifnum#1=49 %
\hatcurPPmeshorteccenxxxxxC
\else
\ifnum#1=72 %
\hatcurPPmeshorteccenxxxxxD
\else
??????\fi
\fi
\fi
\fi
}
\newcommand{\hatcurPPmlongeccen}[1]{\ifnum#1=47 %
\hatcurPPmlongeccenxxxxxA
\else
\ifnum#1=48 %
\hatcurPPmlongeccenxxxxxB
\else
\ifnum#1=49 %
\hatcurPPmlongeccenxxxxxC
\else
\ifnum#1=72 %
\hatcurPPmlongeccenxxxxxD
\else
??????\fi
\fi
\fi
\fi
}
\newcommand{\hatcurPPmrcorreccen}[1]{\ifnum#1=47 %
\hatcurPPmrcorreccenxxxxxA
\else
\ifnum#1=48 %
\hatcurPPmrcorreccenxxxxxB
\else
\ifnum#1=49 %
\hatcurPPmrcorreccenxxxxxC
\else
\ifnum#1=72 %
\hatcurPPmrcorreccenxxxxxD
\else
??????\fi
\fi
\fi
\fi
}
\newcommand{\hatcurPPmshorteccen}[1]{\ifnum#1=47 %
\hatcurPPmshorteccenxxxxxA
\else
\ifnum#1=48 %
\hatcurPPmshorteccenxxxxxB
\else
\ifnum#1=49 %
\hatcurPPmshorteccenxxxxxC
\else
\ifnum#1=72 %
\hatcurPPmshorteccenxxxxxD
\else
??????\fi
\fi
\fi
\fi
}
\newcommand{\hatcurPPmtwosiglimeccen}[1]{\ifnum#1=47 %
\hatcurPPmtwosiglimeccenxxxxxA
\else
\ifnum#1=48 %
\hatcurPPmtwosiglimeccenxxxxxB
\else
\ifnum#1=49 %
\hatcurPPmtwosiglimeccenxxxxxC
\else
\ifnum#1=72 %
\hatcurPPmtwosiglimeccenxxxxxD
\else
??????\fi
\fi
\fi
\fi
}
\newcommand{\hatcurPPperieccen}[1]{\ifnum#1=47 %
\hatcurPPperieccenxxxxxA
\else
\ifnum#1=48 %
\hatcurPPperieccenxxxxxB
\else
\ifnum#1=49 %
\hatcurPPperieccenxxxxxC
\else
\ifnum#1=72 %
\hatcurPPperieccenxxxxxD
\else
??????\fi
\fi
\fi
\fi
}
\newcommand{\hatcurPPphiconjeccen}[1]{\ifnum#1=47 %
\hatcurPPphiconjeccenxxxxxA
\else
\ifnum#1=48 %
\hatcurPPphiconjeccenxxxxxB
\else
\ifnum#1=49 %
\hatcurPPphiconjeccenxxxxxC
\else
\ifnum#1=72 %
\hatcurPPphiconjeccenxxxxxD
\else
??????\fi
\fi
\fi
\fi
}
\newcommand{\hatcurPPreccen}[1]{\ifnum#1=47 %
\hatcurPPreccenxxxxxA
\else
\ifnum#1=48 %
\hatcurPPreccenxxxxxB
\else
\ifnum#1=49 %
\hatcurPPreccenxxxxxC
\else
\ifnum#1=72 %
\hatcurPPreccenxxxxxD
\else
??????\fi
\fi
\fi
\fi
}
\newcommand{\hatcurPPreeccen}[1]{\ifnum#1=47 %
\hatcurPPreeccenxxxxxA
\else
\ifnum#1=48 %
\hatcurPPreeccenxxxxxB
\else
\ifnum#1=49 %
\hatcurPPreeccenxxxxxC
\else
\ifnum#1=72 %
\hatcurPPreeccenxxxxxD
\else
??????\fi
\fi
\fi
\fi
}
\newcommand{\hatcurPPrelongeccen}[1]{\ifnum#1=47 %
\hatcurPPrelongeccenxxxxxA
\else
\ifnum#1=48 %
\hatcurPPrelongeccenxxxxxB
\else
\ifnum#1=49 %
\hatcurPPrelongeccenxxxxxC
\else
\ifnum#1=72 %
\hatcurPPrelongeccenxxxxxD
\else
??????\fi
\fi
\fi
\fi
}
\newcommand{\hatcurPPreshorteccen}[1]{\ifnum#1=47 %
\hatcurPPreshorteccenxxxxxA
\else
\ifnum#1=48 %
\hatcurPPreshorteccenxxxxxB
\else
\ifnum#1=49 %
\hatcurPPreshorteccenxxxxxC
\else
\ifnum#1=72 %
\hatcurPPreshorteccenxxxxxD
\else
??????\fi
\fi
\fi
\fi
}
\newcommand{\hatcurPPrhoeccen}[1]{\ifnum#1=47 %
\hatcurPPrhoeccenxxxxxA
\else
\ifnum#1=48 %
\hatcurPPrhoeccenxxxxxB
\else
\ifnum#1=49 %
\hatcurPPrhoeccenxxxxxC
\else
\ifnum#1=72 %
\hatcurPPrhoeccenxxxxxD
\else
??????\fi
\fi
\fi
\fi
}
\newcommand{\hatcurPPrlongeccen}[1]{\ifnum#1=47 %
\hatcurPPrlongeccenxxxxxA
\else
\ifnum#1=48 %
\hatcurPPrlongeccenxxxxxB
\else
\ifnum#1=49 %
\hatcurPPrlongeccenxxxxxC
\else
\ifnum#1=72 %
\hatcurPPrlongeccenxxxxxD
\else
??????\fi
\fi
\fi
\fi
}
\newcommand{\hatcurPPrshorteccen}[1]{\ifnum#1=47 %
\hatcurPPrshorteccenxxxxxA
\else
\ifnum#1=48 %
\hatcurPPrshorteccenxxxxxB
\else
\ifnum#1=49 %
\hatcurPPrshorteccenxxxxxC
\else
\ifnum#1=72 %
\hatcurPPrshorteccenxxxxxD
\else
??????\fi
\fi
\fi
\fi
}
\newcommand{\hatcurPPtcirceccen}[1]{\ifnum#1=47 %
\hatcurPPtcirceccenxxxxxA
\else
\ifnum#1=48 %
\hatcurPPtcirceccenxxxxxB
\else
\ifnum#1=49 %
\hatcurPPtcirceccenxxxxxC
\else
\ifnum#1=72 %
\hatcurPPtcirceccenxxxxxD
\else
??????\fi
\fi
\fi
\fi
}
\newcommand{\hatcurPPteffeccen}[1]{\ifnum#1=47 %
\hatcurPPteffeccenxxxxxA
\else
\ifnum#1=48 %
\hatcurPPteffeccenxxxxxB
\else
\ifnum#1=49 %
\hatcurPPteffeccenxxxxxC
\else
\ifnum#1=72 %
\hatcurPPteffeccenxxxxxD
\else
??????\fi
\fi
\fi
\fi
}
\newcommand{\hatcurPPthetaeccen}[1]{\ifnum#1=47 %
\hatcurPPthetaeccenxxxxxA
\else
\ifnum#1=48 %
\hatcurPPthetaeccenxxxxxB
\else
\ifnum#1=49 %
\hatcurPPthetaeccenxxxxxC
\else
\ifnum#1=72 %
\hatcurPPthetaeccenxxxxxD
\else
??????\fi
\fi
\fi
\fi
}
\newcommand{\hatcurPPtinfalleccen}[1]{\ifnum#1=47 %
\hatcurPPtinfalleccenxxxxxA
\else
\ifnum#1=48 %
\hatcurPPtinfalleccenxxxxxB
\else
\ifnum#1=49 %
\hatcurPPtinfalleccenxxxxxC
\else
\ifnum#1=72 %
\hatcurPPtinfalleccenxxxxxD
\else
??????\fi
\fi
\fi
\fi
}
\newcommand{\hatcurRVecceneccen}[1]{\ifnum#1=47 %
\hatcurRVecceneccenxxxxxA
\else
\ifnum#1=48 %
\hatcurRVecceneccenxxxxxB
\else
\ifnum#1=49 %
\hatcurRVecceneccenxxxxxC
\else
\ifnum#1=72 %
\hatcurRVecceneccenxxxxxD
\else
??????\fi
\fi
\fi
\fi
}
\newcommand{\hatcurRVeccentwosiglimeccen}[1]{\ifnum#1=47 %
\hatcurRVeccentwosiglimeccenxxxxxA
\else
\ifnum#1=48 %
\hatcurRVeccentwosiglimeccenxxxxxB
\else
\ifnum#1=49 %
\hatcurRVeccentwosiglimeccenxxxxxC
\else
\ifnum#1=72 %
\hatcurRVeccentwosiglimeccenxxxxxD
\else
??????\fi
\fi
\fi
\fi
}
\newcommand{\hatcurRVfitrmsAeccen}[1]{\ifnum#1=72 %
\hatcurRVfitrmsAeccenxxxxxD
\else
??????\fi
}
\newcommand{\hatcurRVfitrmsBeccen}[1]{\ifnum#1=72 %
\hatcurRVfitrmsBeccenxxxxxD
\else
??????\fi
}
\newcommand{\hatcurRVfitrmsCeccen}[1]{\ifnum#1=72 %
\hatcurRVfitrmsCeccenxxxxxD
\else
??????\fi
}
\newcommand{\hatcurRVfitrmsDeccen}[1]{\ifnum#1=72 %
\hatcurRVfitrmsDeccenxxxxxD
\else
??????\fi
}
\newcommand{\hatcurRVfitrmseccen}[1]{\ifnum#1=47 %
\hatcurRVfitrmseccenxxxxxA
\else
\ifnum#1=48 %
\hatcurRVfitrmseccenxxxxxB
\else
\ifnum#1=49 %
\hatcurRVfitrmseccenxxxxxC
\else
??????\fi
\fi
\fi
}
\newcommand{\hatcurRVgammaAeccen}[1]{\ifnum#1=72 %
\hatcurRVgammaAeccenxxxxxD
\else
??????\fi
}
\newcommand{\hatcurRVgammaBeccen}[1]{\ifnum#1=72 %
\hatcurRVgammaBeccenxxxxxD
\else
??????\fi
}
\newcommand{\hatcurRVgammaCeccen}[1]{\ifnum#1=72 %
\hatcurRVgammaCeccenxxxxxD
\else
??????\fi
}
\newcommand{\hatcurRVgammaDeccen}[1]{\ifnum#1=72 %
\hatcurRVgammaDeccenxxxxxD
\else
??????\fi
}
\newcommand{\hatcurRVgammaeccen}[1]{\ifnum#1=47 %
\hatcurRVgammaeccenxxxxxA
\else
\ifnum#1=48 %
\hatcurRVgammaeccenxxxxxB
\else
\ifnum#1=49 %
\hatcurRVgammaeccenxxxxxC
\else
??????\fi
\fi
\fi
}
\newcommand{\hatcurRVheccen}[1]{\ifnum#1=47 %
\hatcurRVheccenxxxxxA
\else
\ifnum#1=48 %
\hatcurRVheccenxxxxxB
\else
\ifnum#1=49 %
\hatcurRVheccenxxxxxC
\else
\ifnum#1=72 %
\hatcurRVheccenxxxxxD
\else
??????\fi
\fi
\fi
\fi
}
\newcommand{\hatcurRVjitterAeccen}[1]{\ifnum#1=72 %
\hatcurRVjitterAeccenxxxxxD
\else
??????\fi
}
\newcommand{\hatcurRVjitterBeccen}[1]{\ifnum#1=72 %
\hatcurRVjitterBeccenxxxxxD
\else
??????\fi
}
\newcommand{\hatcurRVjitterCeccen}[1]{\ifnum#1=72 %
\hatcurRVjitterCeccenxxxxxD
\else
??????\fi
}
\newcommand{\hatcurRVjitterDeccen}[1]{\ifnum#1=72 %
\hatcurRVjitterDeccenxxxxxD
\else
??????\fi
}
\newcommand{\hatcurRVjittereccen}[1]{\ifnum#1=47 %
\hatcurRVjittereccenxxxxxA
\else
\ifnum#1=48 %
\hatcurRVjittereccenxxxxxB
\else
\ifnum#1=49 %
\hatcurRVjittereccenxxxxxC
\else
??????\fi
\fi
\fi
}
\newcommand{\hatcurRVjittertwosiglimAeccen}[1]{\ifnum#1=72 %
\hatcurRVjittertwosiglimAeccenxxxxxD
\else
??????\fi
}
\newcommand{\hatcurRVjittertwosiglimBeccen}[1]{\ifnum#1=72 %
\hatcurRVjittertwosiglimBeccenxxxxxD
\else
??????\fi
}
\newcommand{\hatcurRVjittertwosiglimCeccen}[1]{\ifnum#1=72 %
\hatcurRVjittertwosiglimCeccenxxxxxD
\else
??????\fi
}
\newcommand{\hatcurRVjittertwosiglimDeccen}[1]{\ifnum#1=72 %
\hatcurRVjittertwosiglimDeccenxxxxxD
\else
??????\fi
}
\newcommand{\hatcurRVjittertwosiglimeccen}[1]{\ifnum#1=47 %
\hatcurRVjittertwosiglimeccenxxxxxA
\else
\ifnum#1=48 %
\hatcurRVjittertwosiglimeccenxxxxxB
\else
\ifnum#1=49 %
\hatcurRVjittertwosiglimeccenxxxxxC
\else
??????\fi
\fi
\fi
}
\newcommand{\hatcurRVkeccen}[1]{\ifnum#1=47 %
\hatcurRVkeccenxxxxxA
\else
\ifnum#1=48 %
\hatcurRVkeccenxxxxxB
\else
\ifnum#1=49 %
\hatcurRVkeccenxxxxxC
\else
\ifnum#1=72 %
\hatcurRVkeccenxxxxxD
\else
??????\fi
\fi
\fi
\fi
}
\newcommand{\hatcurRVKeccen}[1]{\ifnum#1=47 %
\hatcurRVKeccenxxxxxA
\else
\ifnum#1=48 %
\hatcurRVKeccenxxxxxB
\else
\ifnum#1=49 %
\hatcurRVKeccenxxxxxC
\else
\ifnum#1=72 %
\hatcurRVKeccenxxxxxD
\else
??????\fi
\fi
\fi
\fi
}
\newcommand{\hatcurRVKtwosiglimeccen}[1]{\ifnum#1=47 %
\hatcurRVKtwosiglimeccenxxxxxA
\else
\ifnum#1=48 %
\hatcurRVKtwosiglimeccenxxxxxB
\else
\ifnum#1=49 %
\hatcurRVKtwosiglimeccenxxxxxC
\else
\ifnum#1=72 %
\hatcurRVKtwosiglimeccenxxxxxD
\else
??????\fi
\fi
\fi
\fi
}
\newcommand{\hatcurRVomegaeccen}[1]{\ifnum#1=47 %
\hatcurRVomegaeccenxxxxxA
\else
\ifnum#1=48 %
\hatcurRVomegaeccenxxxxxB
\else
\ifnum#1=49 %
\hatcurRVomegaeccenxxxxxC
\else
\ifnum#1=72 %
\hatcurRVomegaeccenxxxxxD
\else
??????\fi
\fi
\fi
\fi
}
\newcommand{\hatcurRVrheccen}[1]{\ifnum#1=47 %
\hatcurRVrheccenxxxxxA
\else
\ifnum#1=48 %
\hatcurRVrheccenxxxxxB
\else
\ifnum#1=49 %
\hatcurRVrheccenxxxxxC
\else
\ifnum#1=72 %
\hatcurRVrheccenxxxxxD
\else
??????\fi
\fi
\fi
\fi
}
\newcommand{\hatcurRVrkeccen}[1]{\ifnum#1=47 %
\hatcurRVrkeccenxxxxxA
\else
\ifnum#1=48 %
\hatcurRVrkeccenxxxxxB
\else
\ifnum#1=49 %
\hatcurRVrkeccenxxxxxC
\else
\ifnum#1=72 %
\hatcurRVrkeccenxxxxxD
\else
??????\fi
\fi
\fi
\fi
}
\newcommand{\hatcurRVtroneeccen}[1]{\ifnum#1=47 %
\hatcurRVtroneeccenxxxxxA
\else
\ifnum#1=48 %
\hatcurRVtroneeccenxxxxxB
\else
\ifnum#1=49 %
\hatcurRVtroneeccenxxxxxC
\else
\ifnum#1=72 %
\hatcurRVtroneeccenxxxxxD
\else
??????\fi
\fi
\fi
\fi
}
\newcommand{\hatcurRVtrtwoeccen}[1]{\ifnum#1=47 %
\hatcurRVtrtwoeccenxxxxxA
\else
\ifnum#1=48 %
\hatcurRVtrtwoeccenxxxxxB
\else
\ifnum#1=49 %
\hatcurRVtrtwoeccenxxxxxC
\else
\ifnum#1=72 %
\hatcurRVtrtwoeccenxxxxxD
\else
??????\fi
\fi
\fi
\fi
}
\newcommand{\hatcurSMEiiloggeccen}[1]{\ifnum#1=47 %
\hatcurSMEiiloggeccenxxxxxA
\else
??????\fi
}
\newcommand{\hatcurSMEiiteffeccen}[1]{\ifnum#1=47 %
\hatcurSMEiiteffeccenxxxxxA
\else
??????\fi
}
\newcommand{\hatcurSMEiivsineccen}[1]{\ifnum#1=47 %
\hatcurSMEiivsineccenxxxxxA
\else
??????\fi
}
\newcommand{\hatcurSMEiizfeheccen}[1]{\ifnum#1=47 %
\hatcurSMEiizfeheccenxxxxxA
\else
??????\fi
}
\newcommand{\hatcurSMEiizfehshorteccen}[1]{\ifnum#1=47 %
\hatcurSMEiizfehshorteccenxxxxxA
\else
??????\fi
}
\newcommand{\hatcurSMEiloggeccen}[1]{\ifnum#1=47 %
\hatcurSMEiloggeccenxxxxxA
\else
\ifnum#1=48 %
\hatcurSMEiloggeccenxxxxxB
\else
\ifnum#1=49 %
\hatcurSMEiloggeccenxxxxxC
\else
\ifnum#1=72 %
\hatcurSMEiloggeccenxxxxxD
\else
??????\fi
\fi
\fi
\fi
}
\newcommand{\hatcurSMEiteffeccen}[1]{\ifnum#1=47 %
\hatcurSMEiteffeccenxxxxxA
\else
\ifnum#1=48 %
\hatcurSMEiteffeccenxxxxxB
\else
\ifnum#1=49 %
\hatcurSMEiteffeccenxxxxxC
\else
\ifnum#1=72 %
\hatcurSMEiteffeccenxxxxxD
\else
??????\fi
\fi
\fi
\fi
}
\newcommand{\hatcurSMEivmaceccen}[1]{\ifnum#1=47 %
\hatcurSMEivmaceccenxxxxxA
\else
\ifnum#1=48 %
\hatcurSMEivmaceccenxxxxxB
\else
\ifnum#1=49 %
\hatcurSMEivmaceccenxxxxxC
\else
\ifnum#1=72 %
\hatcurSMEivmaceccenxxxxxD
\else
??????\fi
\fi
\fi
\fi
}
\newcommand{\hatcurSMEivmiceccen}[1]{\ifnum#1=47 %
\hatcurSMEivmiceccenxxxxxA
\else
\ifnum#1=48 %
\hatcurSMEivmiceccenxxxxxB
\else
\ifnum#1=49 %
\hatcurSMEivmiceccenxxxxxC
\else
\ifnum#1=72 %
\hatcurSMEivmiceccenxxxxxD
\else
??????\fi
\fi
\fi
\fi
}
\newcommand{\hatcurSMEivsineccen}[1]{\ifnum#1=47 %
\hatcurSMEivsineccenxxxxxA
\else
\ifnum#1=48 %
\hatcurSMEivsineccenxxxxxB
\else
\ifnum#1=49 %
\hatcurSMEivsineccenxxxxxC
\else
\ifnum#1=72 %
\hatcurSMEivsineccenxxxxxD
\else
??????\fi
\fi
\fi
\fi
}
\newcommand{\hatcurSMEizfeheccen}[1]{\ifnum#1=47 %
\hatcurSMEizfeheccenxxxxxA
\else
\ifnum#1=48 %
\hatcurSMEizfeheccenxxxxxB
\else
\ifnum#1=49 %
\hatcurSMEizfeheccenxxxxxC
\else
\ifnum#1=72 %
\hatcurSMEizfeheccenxxxxxD
\else
??????\fi
\fi
\fi
\fi
}
\newcommand{\hatcurSMEizfehshorteccen}[1]{\ifnum#1=47 %
\hatcurSMEizfehshorteccenxxxxxA
\else
\ifnum#1=48 %
\hatcurSMEizfehshorteccenxxxxxB
\else
\ifnum#1=49 %
\hatcurSMEizfehshorteccenxxxxxC
\else
\ifnum#1=72 %
\hatcurSMEizfehshorteccenxxxxxD
\else
??????\fi
\fi
\fi
\fi
}
\newcommand{\hatcurXAveccen}[1]{\ifnum#1=47 %
\hatcurXAveccenxxxxxA
\else
\ifnum#1=48 %
\hatcurXAveccenxxxxxB
\else
\ifnum#1=49 %
\hatcurXAveccenxxxxxC
\else
\ifnum#1=72 %
\hatcurXAveccenxxxxxD
\else
??????\fi
\fi
\fi
\fi
}
\newcommand{\hatcurXdisteccen}[1]{\ifnum#1=47 %
\hatcurXdisteccenxxxxxA
\else
\ifnum#1=48 %
\hatcurXdisteccenxxxxxB
\else
\ifnum#1=49 %
\hatcurXdisteccenxxxxxC
\else
\ifnum#1=72 %
\hatcurXdisteccenxxxxxD
\else
??????\fi
\fi
\fi
\fi
}
\newcommand{\hatcurXdistredeccen}[1]{\ifnum#1=47 %
\hatcurXdistredeccenxxxxxA
\else
\ifnum#1=48 %
\hatcurXdistredeccenxxxxxB
\else
\ifnum#1=49 %
\hatcurXdistredeccenxxxxxC
\else
\ifnum#1=72 %
\hatcurXdistredeccenxxxxxD
\else
??????\fi
\fi
\fi
\fi
}
\newcommand{\hatcurXEBVeccen}[1]{\ifnum#1=47 %
\hatcurXEBVeccenxxxxxA
\else
\ifnum#1=48 %
\hatcurXEBVeccenxxxxxB
\else
\ifnum#1=49 %
\hatcurXEBVeccenxxxxxC
\else
\ifnum#1=72 %
\hatcurXEBVeccenxxxxxD
\else
??????\fi
\fi
\fi
\fi
}
\newcommand{\hatcurXsecdureccen}[1]{\ifnum#1=47 %
\hatcurXsecdureccenxxxxxA
\else
\ifnum#1=48 %
\hatcurXsecdureccenxxxxxB
\else
\ifnum#1=49 %
\hatcurXsecdureccenxxxxxC
\else
\ifnum#1=72 %
\hatcurXsecdureccenxxxxxD
\else
??????\fi
\fi
\fi
\fi
}
\newcommand{\hatcurXsecingdureccen}[1]{\ifnum#1=47 %
\hatcurXsecingdureccenxxxxxA
\else
\ifnum#1=48 %
\hatcurXsecingdureccenxxxxxB
\else
\ifnum#1=49 %
\hatcurXsecingdureccenxxxxxC
\else
\ifnum#1=72 %
\hatcurXsecingdureccenxxxxxD
\else
??????\fi
\fi
\fi
\fi
}
\newcommand{\hatcurXsecondaryeccen}[1]{\ifnum#1=47 %
\hatcurXsecondaryeccenxxxxxA
\else
\ifnum#1=48 %
\hatcurXsecondaryeccenxxxxxB
\else
\ifnum#1=49 %
\hatcurXsecondaryeccenxxxxxC
\else
\ifnum#1=72 %
\hatcurXsecondaryeccenxxxxxD
\else
??????\fi
\fi
\fi
\fi
}
\newcommand{\hatcurXsecphaseeccen}[1]{\ifnum#1=47 %
\hatcurXsecphaseeccenxxxxxA
\else
\ifnum#1=48 %
\hatcurXsecphaseeccenxxxxxB
\else
\ifnum#1=49 %
\hatcurXsecphaseeccenxxxxxC
\else
\ifnum#1=72 %
\hatcurXsecphaseeccenxxxxxD
\else
??????\fi
\fi
\fi
\fi
}
\newcommand{\hatcurhtrempiricalxxxxxA}{HATS747-014}                      
\newcommand{\hatcurfieldempiricalxxxxxA}{\ensuremath{string}}            
\newcommand{\hatcurCCraempiricalxxxxxA}{\ensuremath{19^{\mathrm h}09^{\mathrm m}56.2504{\mathrm s}}}                   
\newcommand{\hatcurCCdecempiricalxxxxxA}{\ensuremath{-49{\arcdeg}39{\arcmin}53.868{\arcsec}}}                  
\newcommand{\hatcurCCmagempiricalxxxxxA}{14.829}                         
\newcommand{\hatcurCCtwomassempiricalxxxxxA}{2MASS~19095625-4939538}     
\newcommand{\hatcurCCgscempiricalxxxxxA}{GSC~}                           
\newcommand{\hatcurCCgaiaempiricalxxxxxA}{GAIA~6658373007402886400}      
\newcommand{\hatcurCCgaiadrtwoempiricalxxxxxA}{GAIA~DR2~6658373007402886400} 
\newcommand{\hatcurCCtassmvempiricalxxxxxA}{\ensuremath{14.829\pm0.010}} 
\newcommand{\hatcurCCtassmvshortempiricalxxxxxA}{\ensuremath{14.8}}      
\newcommand{\hatcurCCtassmBempiricalxxxxxA}{\ensuremath{16.101\pm0.040}} 
\newcommand{\hatcurCCtassmBshortempiricalxxxxxA}{\ensuremath{16.1}}      
\newcommand{\hatcurCCtassmIempiricalxxxxxA}{\ensuremath{nff\pmnff}}      
\newcommand{\hatcurCCtassmIshortempiricalxxxxxA}{\ensuremath{0.0}}       
\newcommand{\hatcurCCtassmgempiricalxxxxxA}{\ensuremath{15.480\pm0.010}} 
\newcommand{\hatcurCCtassmgshortempiricalxxxxxA}{\ensuremath{15.5}}      
\newcommand{\hatcurCCtassmrempiricalxxxxxA}{\ensuremath{14.398\pm0.010}} 
\newcommand{\hatcurCCtassmrshortempiricalxxxxxA}{\ensuremath{14.4}}      
\newcommand{\hatcurCCtassmiempiricalxxxxxA}{\ensuremath{14.009\pm0.010}} 
\newcommand{\hatcurCCtassmishortempiricalxxxxxA}{\ensuremath{14.0}}      
\newcommand{\hatcurCCparallaxempiricalxxxxxA}{\ensuremath{3.298\pm0.042}} 
\newcommand{\hatcurCCgaiamGempiricalxxxxxA}{\ensuremath{14.39980\pm0.00040}} 
\newcommand{\hatcurCCgaiamBPempiricalxxxxxA}{\ensuremath{15.0858\pm0.0021}} 
\newcommand{\hatcurCCgaiamRPempiricalxxxxxA}{\ensuremath{13.61140\pm0.00090}} 
\newcommand{\hatcurCCtwomassJmagempiricalxxxxxA}{\ensuremath{12.653\pm0.023}} 
\newcommand{\hatcurCCtwomassHmagempiricalxxxxxA}{\ensuremath{12.026\pm0.023}} 
\newcommand{\hatcurCCtwomassKmagempiricalxxxxxA}{\ensuremath{11.926\pm0.025}} 
\newcommand{\hatcurCCcitJmagempiricalxxxxxA}{\ensuremath{12.651\pm0.024}} 
\newcommand{\hatcurCCcitHmagempiricalxxxxxA}{\ensuremath{12.020\pm0.024}} 
\newcommand{\hatcurCCcitKmagempiricalxxxxxA}{\ensuremath{11.950\pm0.025}} 
\newcommand{\hatcurCCbbJmagempiricalxxxxxA}{\ensuremath{12.729\pm0.026}} 
\newcommand{\hatcurCCbbHmagempiricalxxxxxA}{\ensuremath{12.042\pm0.025}} 
\newcommand{\hatcurCCbbKmagempiricalxxxxxA}{\ensuremath{11.970\pm0.025}} 
\newcommand{\hatcurCCesoJmagempiricalxxxxxA}{\ensuremath{12.736\pm0.030}} 
\newcommand{\hatcurCCesoHmagempiricalxxxxxA}{\ensuremath{12.036\pm0.029}} 
\newcommand{\hatcurCCesoKmagempiricalxxxxxA}{\ensuremath{11.967\pm0.026}} 
\newcommand{\hatcurCCesoJHmagempiricalxxxxxA}{\ensuremath{0.699\pm0.018}} 
\newcommand{\hatcurCCesoJKmagempiricalxxxxxA}{\ensuremath{0.769\pm0.038}} 
\newcommand{\hatcurCCesoHKmagempiricalxxxxxA}{\ensuremath{0.069\pm0.038}} 
\newcommand{\hatcurLCdipempiricalxxxxxA}{\ensuremath{31.1}}              
\newcommand{\hatcurLCrprstarempiricalxxxxxA}{\ensuremath{}}              
\newcommand{\hatcurLCrprstarnoisorestrictempiricalxxxxxA}{\ensuremath{0.1721\pm0.0031}} 
\newcommand{\hatcurLCbsqempiricalxxxxxA}{\ensuremath{}}                  
\newcommand{\hatcurLCbsqnoisorestrictempiricalxxxxxA}{\ensuremath{0.500_{-0.033}^{+0.037}}} 
\newcommand{\hatcurLCimpempiricalxxxxxA}{\ensuremath{}}                  
\newcommand{\hatcurLCimpnoisorestrictempiricalxxxxxA}{\ensuremath{0.707_{-0.023}^{+0.026}}} 
\newcommand{\hatcurLCzetaempiricalxxxxxA}{\ensuremath{}}                 
\newcommand{\hatcurLCzetanoisorestrictempiricalxxxxxA}{\ensuremath{32.88\pm0.37}} 
\newcommand{\hatcurLCdurempiricalxxxxxA}{\ensuremath{}}                  
\newcommand{\hatcurLCdurnoisorestrictempiricalxxxxxA}{\ensuremath{0.0806\pm0.0018}} 
\newcommand{\hatcurLCdurshortempiricalxxxxxA}{\ensuremath{}}             
\newcommand{\hatcurLCdurshortnoisorestrictempiricalxxxxxA}{\ensuremath{0.0806}} 
\newcommand{\hatcurLCdurhrempiricalxxxxxA}{\ensuremath{}}                
\newcommand{\hatcurLCdurhrnoisorestrictempiricalxxxxxA}{\ensuremath{1.934\pm0.044}} 
\newcommand{\hatcurLCdurhrshortempiricalxxxxxA}{\ensuremath{}}           
\newcommand{\hatcurLCdurhrshortnoisorestrictempiricalxxxxxA}{\ensuremath{1.934}} 
\newcommand{\hatcurLCqempiricalxxxxxA}{\ensuremath{}}                    
\newcommand{\hatcurLCqnoisorestrictempiricalxxxxxA}{\ensuremath{0.02050\pm0.00047}} 
\newcommand{\hatcurLCqshortempiricalxxxxxA}{\ensuremath{}}               
\newcommand{\hatcurLCqshortnoisorestrictempiricalxxxxxA}{\ensuremath{0.021}} 
\newcommand{\hatcurLCingdurempiricalxxxxxA}{\ensuremath{}}               
\newcommand{\hatcurLCingdurnoisorestrictempiricalxxxxxA}{\ensuremath{0.0217\pm0.0027}} 
\newcommand{\hatcurLCPempiricalxxxxxA}{\ensuremath{}}                    
\newcommand{\hatcurLCPnoisorestrictempiricalxxxxxA}{\ensuremath{3.9228146\pm0.0000037}} 
\newcommand{\hatcurLCPprecempiricalxxxxxA}{\ensuremath{}}                
\newcommand{\hatcurLCPprecnoisorestrictempiricalxxxxxA}{\ensuremath{3.9228146}} 
\newcommand{\hatcurLCPshortempiricalxxxxxA}{\ensuremath{}}               
\newcommand{\hatcurLCPshortnoisorestrictempiricalxxxxxA}{\ensuremath{3.9228}} 
\newcommand{\hatcurLCTempiricalxxxxxA}{\ensuremath{}}                    
\newcommand{\hatcurLCTnoisorestrictempiricalxxxxxA}{\ensuremath{2457306.51608\pm0.00031}} 
\newcommand{\hatcurLCTAempiricalxxxxxA}{\ensuremath{}}                   
\newcommand{\hatcurLCTAnoisorestrictempiricalxxxxxA}{\ensuremath{2456368.96337\pm0.00085}} 
\newcommand{\hatcurLCTBempiricalxxxxxA}{\ensuremath{}}                   
\newcommand{\hatcurLCTBnoisorestrictempiricalxxxxxA}{\ensuremath{2457604.65001\pm0.00048}} 
\newcommand{\hatcurLChatnetmAempiricalxxxxxA}{\ensuremath{}}             
\newcommand{\hatcurLChatnetmAnoisorestrictempiricalxxxxxA}{\ensuremath{14.39325\pm0.00017}} 
\newcommand{\hatcurLCiblendAempiricalxxxxxA}{\ensuremath{}}              
\newcommand{\hatcurLCiblendAnoisorestrictempiricalxxxxxA}{\ensuremath{0.924\pm0.039}} 
\newcommand{\hatcurLChatnetmBempiricalxxxxxA}{\ensuremath{}}             
\newcommand{\hatcurLChatnetmBnoisorestrictempiricalxxxxxA}{\ensuremath{0.00007\pm0.00025}} 
\newcommand{\hatcurLCiblendBempiricalxxxxxA}{\ensuremath{}}              
\newcommand{\hatcurLCiblendBnoisorestrictempiricalxxxxxA}{\ensuremath{0.602\pm0.015}} 
\newcommand{\hatcurLCrhoempiricalxxxxxA}{\ensuremath{}}                  
\newcommand{\hatcurLCrhonoisorestrictempiricalxxxxxA}{\ensuremath{3.73\pm0.44}} 
\newcommand{\hatcurSMEiteffempiricalxxxxxA}{\ensuremath{4479\pm51}}      
\newcommand{\hatcurSMEizfehempiricalxxxxxA}{\ensuremath{-0.140\pm0.066}} 
\newcommand{\hatcurSMEizfehshortempiricalxxxxxA}{\ensuremath{-0.14}}     
\newcommand{\hatcurSMEiloggempiricalxxxxxA}{\ensuremath{4.76\pm0.21}}    
\newcommand{\hatcurSMEivsinempiricalxxxxxA}{\ensuremath{2.47\pm0.70}}    
\newcommand{\hatcurSMEivmacempiricalxxxxxA}{\ensuremath{nff\pmnff}}      
\newcommand{\hatcurSMEivmicempiricalxxxxxA}{\ensuremath{nff\pmnff}}      
\newcommand{\hatcurSMEiiteffempiricalxxxxxA}{\ensuremath{4450\pm86}}     
\newcommand{\hatcurSMEiizfehempiricalxxxxxA}{\ensuremath{-0.14\pm0.10}}  
\newcommand{\hatcurSMEiizfehshortempiricalxxxxxA}{\ensuremath{-0.14}}    
\newcommand{\hatcurSMEiiloggempiricalxxxxxA}{\ensuremath{4.603\pm0.031}} 
\newcommand{\hatcurSMEiivsinempiricalxxxxxA}{\ensuremath{2.8\pm1.4}}     
\newcommand{\hatcurextraerrMgempiricalxxxxxA}{\ensuremath{}}             
\newcommand{\hatcurextraerrMgnoisorestrictempiricalxxxxxA}{\ensuremath{0.050_{-0.046}^{+0.070}}} 
\newcommand{\hatcurextraerrMgtwosiglimempiricalxxxxxA}{\ensuremath{}}    
\newcommand{\hatcurextraerrMgtwosiglimnoisorestrictempiricalxxxxxA}{\ensuremath{<0.1594}} 
\newcommand{\hatcurextraerrMrempiricalxxxxxA}{\ensuremath{}}             
\newcommand{\hatcurextraerrMrnoisorestrictempiricalxxxxxA}{\ensuremath{0.047_{-0.029}^{+0.053}}} 
\newcommand{\hatcurextraerrMrtwosiglimempiricalxxxxxA}{\ensuremath{}}    
\newcommand{\hatcurextraerrMrtwosiglimnoisorestrictempiricalxxxxxA}{\ensuremath{<0.1460}} 
\newcommand{\hatcurextraerrMiempiricalxxxxxA}{\ensuremath{}}             
\newcommand{\hatcurextraerrMinoisorestrictempiricalxxxxxA}{\ensuremath{0.045_{-0.044}^{+0.449}}} 
\newcommand{\hatcurextraerrMitwosiglimempiricalxxxxxA}{\ensuremath{}}    
\newcommand{\hatcurextraerrMitwosiglimnoisorestrictempiricalxxxxxA}{\ensuremath{<0.6763}} 
\newcommand{\hatcurextraerrMGempiricalxxxxxA}{\ensuremath{}}             
\newcommand{\hatcurextraerrMGnoisorestrictempiricalxxxxxA}{\ensuremath{0.061\pm0.031}} 
\newcommand{\hatcurextraerrMGtwosiglimempiricalxxxxxA}{\ensuremath{}}    
\newcommand{\hatcurextraerrMGtwosiglimnoisorestrictempiricalxxxxxA}{\ensuremath{<0.1196}} 
\newcommand{\hatcurextraerrMBPtwoempiricalxxxxxA}{\ensuremath{}}         
\newcommand{\hatcurextraerrMBPtwonoisorestrictempiricalxxxxxA}{\ensuremath{0.065\pm0.043}} 
\newcommand{\hatcurextraerrMBPtwotwosiglimempiricalxxxxxA}{\ensuremath{}} 
\newcommand{\hatcurextraerrMBPtwotwosiglimnoisorestrictempiricalxxxxxA}{\ensuremath{<0.1530}} 
\newcommand{\hatcurextraerrMRPempiricalxxxxxA}{\ensuremath{}}            
\newcommand{\hatcurextraerrMRPnoisorestrictempiricalxxxxxA}{\ensuremath{0.017\pm0.015}} 
\newcommand{\hatcurextraerrMRPtwosiglimempiricalxxxxxA}{\ensuremath{}}   
\newcommand{\hatcurextraerrMRPtwosiglimnoisorestrictempiricalxxxxxA}{\ensuremath{<0.0465}} 
\newcommand{\hatcurextraerrMJempiricalxxxxxA}{\ensuremath{}}             
\newcommand{\hatcurextraerrMJnoisorestrictempiricalxxxxxA}{\ensuremath{0.023_{-0.019}^{+0.027}}} 
\newcommand{\hatcurextraerrMJtwosiglimempiricalxxxxxA}{\ensuremath{}}    
\newcommand{\hatcurextraerrMJtwosiglimnoisorestrictempiricalxxxxxA}{\ensuremath{<0.0705}} 
\newcommand{\hatcurextraerrMHempiricalxxxxxA}{\ensuremath{}}             
\newcommand{\hatcurextraerrMHnoisorestrictempiricalxxxxxA}{\ensuremath{0.0044_{-0.0041}^{+0.0158}}} 
\newcommand{\hatcurextraerrMHtwosiglimempiricalxxxxxA}{\ensuremath{}}    
\newcommand{\hatcurextraerrMHtwosiglimnoisorestrictempiricalxxxxxA}{\ensuremath{<0.0348}} 
\newcommand{\hatcurextraerrMKsempiricalxxxxxA}{\ensuremath{}}            
\newcommand{\hatcurextraerrMKsnoisorestrictempiricalxxxxxA}{\ensuremath{0.013_{-0.012}^{+0.028}}} 
\newcommand{\hatcurextraerrMKstwosiglimempiricalxxxxxA}{\ensuremath{}}   
\newcommand{\hatcurextraerrMKstwosiglimnoisorestrictempiricalxxxxxA}{\ensuremath{<0.0719}} 
\newcommand{\hatcurextraerrMWoneempiricalxxxxxA}{\ensuremath{}}          
\newcommand{\hatcurextraerrMWonenoisorestrictempiricalxxxxxA}{\ensuremath{0.0076_{-0.0073}^{+0.0188}}} 
\newcommand{\hatcurextraerrMWonetwosiglimempiricalxxxxxA}{\ensuremath{}} 
\newcommand{\hatcurextraerrMWonetwosiglimnoisorestrictempiricalxxxxxA}{\ensuremath{<0.0387}} 
\newcommand{\hatcurextraerrMWtwoempiricalxxxxxA}{\ensuremath{}}          
\newcommand{\hatcurextraerrMWtwonoisorestrictempiricalxxxxxA}{\ensuremath{0.053_{-0.050}^{+0.084}}} 
\newcommand{\hatcurextraerrMWtwotwosiglimempiricalxxxxxA}{\ensuremath{}} 
\newcommand{\hatcurextraerrMWtwotwosiglimnoisorestrictempiricalxxxxxA}{\ensuremath{<0.2122}} 
\newcommand{\hatcurLBiBempiricalxxxxxA}{\ensuremath{0.9498}}             
\newcommand{\hatcurLBiiBempiricalxxxxxA}{\ensuremath{-0.0970}}           
\newcommand{\hatcurLBiVempiricalxxxxxA}{\ensuremath{0.7431}}             
\newcommand{\hatcurLBiiVempiricalxxxxxA}{\ensuremath{0.0539}}            
\newcommand{\hatcurLBiRempiricalxxxxxA}{\ensuremath{0.6121}}             
\newcommand{\hatcurLBiiRempiricalxxxxxA}{\ensuremath{0.1147}}            
\newcommand{\hatcurLBiIempiricalxxxxxA}{\ensuremath{0.4764}}             
\newcommand{\hatcurLBiiIempiricalxxxxxA}{\ensuremath{0.1543}}            
\newcommand{\hatcurLBiuempiricalxxxxxA}{\ensuremath{1.1009}}             
\newcommand{\hatcurLBiiuempiricalxxxxxA}{\ensuremath{-0.2662}}           
\newcommand{\hatcurLBigempiricalxxxxxA}{\ensuremath{0.8521}}             
\newcommand{\hatcurLBiigempiricalxxxxxA}{\ensuremath{-0.0323}}           
\newcommand{\hatcurLBirempiricalxxxxxA}{\ensuremath{}}                   
\newcommand{\hatcurLBirnoisorestrictempiricalxxxxxA}{\ensuremath{0.34\pm0.15}} 
\newcommand{\hatcurLBiirempiricalxxxxxA}{\ensuremath{}}                  
\newcommand{\hatcurLBiirnoisorestrictempiricalxxxxxA}{\ensuremath{0.50\pm0.15}} 
\newcommand{\hatcurLBiiempiricalxxxxxA}{\ensuremath{}}                   
\newcommand{\hatcurLBiinoisorestrictempiricalxxxxxA}{\ensuremath{0.52\pm0.18}} 
\newcommand{\hatcurLBiiiempiricalxxxxxA}{\ensuremath{}}                  
\newcommand{\hatcurLBiiinoisorestrictempiricalxxxxxA}{\ensuremath{0.22\pm0.16}} 
\newcommand{\hatcurLBizempiricalxxxxxA}{\ensuremath{0.4163}}             
\newcommand{\hatcurLBiizempiricalxxxxxA}{\ensuremath{0.1659}}            
\newcommand{\hatcurLBiJempiricalxxxxxA}{\ensuremath{0.2917}}             
\newcommand{\hatcurLBiiJempiricalxxxxxA}{\ensuremath{0.2132}}            
\newcommand{\hatcurLBiHempiricalxxxxxA}{\ensuremath{0.1585}}             
\newcommand{\hatcurLBiiHempiricalxxxxxA}{\ensuremath{0.2975}}            
\newcommand{\hatcurLBiKempiricalxxxxxA}{\ensuremath{0.1313}}             
\newcommand{\hatcurLBiiKempiricalxxxxxA}{\ensuremath{0.2488}}            
\newcommand{\hatcurLBiTempiricalxxxxxA}{\ensuremath{}}                   
\newcommand{\hatcurLBiTnoisorestrictempiricalxxxxxA}{\ensuremath{0.45\pm0.14}} 
\newcommand{\hatcurLBiiTempiricalxxxxxA}{\ensuremath{}}                  
\newcommand{\hatcurLBiiTnoisorestrictempiricalxxxxxA}{\ensuremath{0.44\pm0.15}} 
\newcommand{\hatcurLBikepempiricalxxxxxA}{\ensuremath{0.6124}}           
\newcommand{\hatcurLBiikepempiricalxxxxxA}{\ensuremath{0.1247}}          
\newcommand{\hatcurLBiCempiricalxxxxxA}{\ensuremath{0.5860}}             
\newcommand{\hatcurLBiiCempiricalxxxxxA}{\ensuremath{0.1328}}            
\newcommand{\hatcurLBiMempiricalxxxxxA}{\ensuremath{0.7144}}             
\newcommand{\hatcurLBiiMempiricalxxxxxA}{\ensuremath{0.0734}}            
\newcommand{\hatcurLBiSoneempiricalxxxxxA}{\ensuremath{0.1118}}            
\newcommand{\hatcurLBiiSoneempiricalxxxxxA}{\ensuremath{0.1665}}           
\newcommand{\hatcurLBiStwoempiricalxxxxxA}{\ensuremath{0.0993}}            
\newcommand{\hatcurLBiiStwoempiricalxxxxxA}{\ensuremath{0.1321}}           
\newcommand{\hatcurLBiSthreeempiricalxxxxxA}{\ensuremath{0.0826}}            
\newcommand{\hatcurLBiiSthreeempiricalxxxxxA}{\ensuremath{0.1091}}           
\newcommand{\hatcurLBiSfourempiricalxxxxxA}{\ensuremath{0.0716}}            
\newcommand{\hatcurLBiiSfourempiricalxxxxxA}{\ensuremath{0.0934}}           
\newcommand{\hatcurISOmempiricalxxxxxA}{\ensuremath{}}                   
\newcommand{\hatcurISOmnoisorestrictempiricalxxxxxA}{\ensuremath{0.76_{-0.10}^{+0.18}}} 
\newcommand{\hatcurISOmshortempiricalxxxxxA}{\ensuremath{}}              
\newcommand{\hatcurISOmshortnoisorestrictempiricalxxxxxA}{\ensuremath{0.76}} 
\newcommand{\hatcurISOmlongempiricalxxxxxA}{\ensuremath{}}               
\newcommand{\hatcurISOmlongnoisorestrictempiricalxxxxxA}{\ensuremath{0.76_{-0.10}^{+0.18}}} 
\newcommand{\hatcurISOrempiricalxxxxxA}{\ensuremath{}}                   
\newcommand{\hatcurISOrnoisorestrictempiricalxxxxxA}{\ensuremath{0.665_{-0.011}^{+0.022}}} 
\newcommand{\hatcurISOrshortempiricalxxxxxA}{\ensuremath{}}              
\newcommand{\hatcurISOrshortnoisorestrictempiricalxxxxxA}{\ensuremath{0.66}} 
\newcommand{\hatcurISOrlongempiricalxxxxxA}{\ensuremath{}}               
\newcommand{\hatcurISOrlongnoisorestrictempiricalxxxxxA}{\ensuremath{0.665_{-0.011}^{+0.022}}} 
\newcommand{\hatcurISOloggempiricalxxxxxA}{\ensuremath{}}                
\newcommand{\hatcurISOloggnoisorestrictempiricalxxxxxA}{\ensuremath{4.681\pm0.058}} 
\newcommand{\hatcurISOlumempiricalxxxxxA}{\ensuremath{}}                 
\newcommand{\hatcurISOlumnoisorestrictempiricalxxxxxA}{\ensuremath{0.1738_{-0.0053}^{+0.0156}}} 
\newcommand{\hatcurISOlumshortempiricalxxxxxA}{\ensuremath{}}            
\newcommand{\hatcurISOlumshortnoisorestrictempiricalxxxxxA}{\ensuremath{0.17}} 
\newcommand{\hatcurISOfehempiricalxxxxxA}{\ensuremath{}}                 
\newcommand{\hatcurISOfehnoisorestrictempiricalxxxxxA}{\ensuremath{-0.10\pm0.16}} 
\newcommand{\hatcurISOteffempiricalxxxxxA}{\ensuremath{}}                
\newcommand{\hatcurISOteffnoisorestrictempiricalxxxxxA}{\ensuremath{4570\pm150}} 
\newcommand{\hatcurISOageempiricalxxxxxA}{\ensuremath{}}                 
\newcommand{\hatcurISOagenoisorestrictempiricalxxxxxA}{\ensuremath{0\pm0}} 
\newcommand{\hatcurISOspecempiricalxxxxxA}{K}                            
\newcommand{\hatcurRVKempiricalxxxxxA}{\ensuremath{}}                    
\newcommand{\hatcurRVKnoisorestrictempiricalxxxxxA}{\ensuremath{48.4\pm4.8}} 
\newcommand{\hatcurRVrkempiricalxxxxxA}{\ensuremath{}}                   
\newcommand{\hatcurRVrknoisorestrictempiricalxxxxxA}{\ensuremath{0\pm0}} 
\newcommand{\hatcurRVrhempiricalxxxxxA}{\ensuremath{}}                   
\newcommand{\hatcurRVrhnoisorestrictempiricalxxxxxA}{\ensuremath{0\pm0}} 
\newcommand{\hatcurRVkempiricalxxxxxA}{\ensuremath{}}                    
\newcommand{\hatcurRVknoisorestrictempiricalxxxxxA}{\ensuremath{0\pm0}}  
\newcommand{\hatcurRVhempiricalxxxxxA}{\ensuremath{}}                    
\newcommand{\hatcurRVhnoisorestrictempiricalxxxxxA}{\ensuremath{0\pm0}}  
\newcommand{\hatcurRVtroneempiricalxxxxxA}{\ensuremath{}}                
\newcommand{\hatcurRVtronenoisorestrictempiricalxxxxxA}{\ensuremath{0\pm0}} 
\newcommand{\hatcurRVtrtwoempiricalxxxxxA}{\ensuremath{}}                
\newcommand{\hatcurRVtrtwonoisorestrictempiricalxxxxxA}{\ensuremath{0\pm0}} 
\newcommand{\hatcurRVgammaempiricalxxxxxA}{\ensuremath{}}                
\newcommand{\hatcurRVgammanoisorestrictempiricalxxxxxA}{\ensuremath{8.7\pm3.4}} 
\newcommand{\hatcurRVjitterempiricalxxxxxA}{\ensuremath{}}               
\newcommand{\hatcurRVjitternoisorestrictempiricalxxxxxA}{\ensuremath{42\pm19}} 
\newcommand{\hatcurRVjittertwosiglimempiricalxxxxxA}{\ensuremath{}}      
\newcommand{\hatcurRVjittertwosiglimnoisorestrictempiricalxxxxxA}{\ensuremath{<82.3}} 
\newcommand{\hatcurRVfitrmsempiricalxxxxxA}{\ensuremath{.1fym}}          %
\newcommand{\hatcurRVeccenempiricalxxxxxA}{\ensuremath{}}                
\newcommand{\hatcurRVeccennoisorestrictempiricalxxxxxA}{\ensuremath{0\pm0}} 
\newcommand{\hatcurRVeccentwosiglimempiricalxxxxxA}{\ensuremath{}}       
\newcommand{\hatcurRVeccentwosiglimnoisorestrictempiricalxxxxxA}{\ensuremath{<0.000}} 
\newcommand{\hatcurRVomegaempiricalxxxxxA}{\ensuremath{}}                
\newcommand{\hatcurRVomeganoisorestrictempiricalxxxxxA}{\ensuremath{0\pm0}} 
\newcommand{\hatcurPPiempiricalxxxxxA}{\ensuremath{}}                    
\newcommand{\hatcurPPinoisorestrictempiricalxxxxxA}{\ensuremath{87.20\pm0.23}} 
\newcommand{\hatcurPPgempiricalxxxxxA}{\ensuremath{}}                    
\newcommand{\hatcurPPgnoisorestrictempiricalxxxxxA}{\ensuremath{6.39_{-0.96}^{+0.71}}} 
\newcommand{\hatcurPPloggempiricalxxxxxA}{\ensuremath{}}                 
\newcommand{\hatcurPPloggnoisorestrictempiricalxxxxxA}{\ensuremath{2.805_{-0.070}^{+0.046}}} 
\newcommand{\hatcurPParempiricalxxxxxA}{\ensuremath{}}                   
\newcommand{\hatcurPParnoisorestrictempiricalxxxxxA}{\ensuremath{14.48\pm0.58}} 
\newcommand{\hatcurPParelempiricalxxxxxA}{\ensuremath{}}                 
\newcommand{\hatcurPParelnoisorestrictempiricalxxxxxA}{\ensuremath{0.0445_{-0.0021}^{+0.0033}}} 
\newcommand{\hatcurPPrhoempiricalxxxxxA}{\ensuremath{}}                  
\newcommand{\hatcurPPrhonoisorestrictempiricalxxxxxA}{\ensuremath{0.286_{-0.043}^{+0.032}}} 
\newcommand{\hatcurPPmempiricalxxxxxA}{\ensuremath{}}                    
\newcommand{\hatcurPPmnoisorestrictempiricalxxxxxA}{\ensuremath{0.318\pm0.040}} 
\newcommand{\hatcurPPmshortempiricalxxxxxA}{\ensuremath{}}               
\newcommand{\hatcurPPmshortnoisorestrictempiricalxxxxxA}{\ensuremath{0.32}} 
\newcommand{\hatcurPPmlongempiricalxxxxxA}{\ensuremath{}}                
\newcommand{\hatcurPPmlongnoisorestrictempiricalxxxxxA}{\ensuremath{0.318\pm0.040}} 
\newcommand{\hatcurPPmeempiricalxxxxxA}{\ensuremath{}}                   
\newcommand{\hatcurPPmenoisorestrictempiricalxxxxxA}{\ensuremath{101\pm13}} 
\newcommand{\hatcurPPmeshortempiricalxxxxxA}{\ensuremath{}}              
\newcommand{\hatcurPPmeshortnoisorestrictempiricalxxxxxA}{\ensuremath{101.2}} 
\newcommand{\hatcurPPmelongempiricalxxxxxA}{\ensuremath{}}               
\newcommand{\hatcurPPmelongnoisorestrictempiricalxxxxxA}{\ensuremath{101\pm13}} 
\newcommand{\hatcurPPrempiricalxxxxxA}{\ensuremath{}}                    
\newcommand{\hatcurPPrnoisorestrictempiricalxxxxxA}{\ensuremath{1.115_{-0.021}^{+0.029}}} 
\newcommand{\hatcurPPrshortempiricalxxxxxA}{\ensuremath{}}               
\newcommand{\hatcurPPrshortnoisorestrictempiricalxxxxxA}{\ensuremath{1.12}} 
\newcommand{\hatcurPPrlongempiricalxxxxxA}{\ensuremath{}}                
\newcommand{\hatcurPPrlongnoisorestrictempiricalxxxxxA}{\ensuremath{1.115_{-0.021}^{+0.029}}} 
\newcommand{\hatcurPPreempiricalxxxxxA}{\ensuremath{}}                   
\newcommand{\hatcurPPrenoisorestrictempiricalxxxxxA}{\ensuremath{12.50_{-0.24}^{+0.32}}} 
\newcommand{\hatcurPPreshortempiricalxxxxxA}{\ensuremath{}}              
\newcommand{\hatcurPPreshortnoisorestrictempiricalxxxxxA}{\ensuremath{12.5}} 
\newcommand{\hatcurPPrelongempiricalxxxxxA}{\ensuremath{}}               
\newcommand{\hatcurPPrelongnoisorestrictempiricalxxxxxA}{\ensuremath{12.50_{-0.24}^{+0.32}}} 
\newcommand{\hatcurPPmrcorrempiricalxxxxxA}{\ensuremath{0.00}}           
\newcommand{\hatcurPPteffempiricalxxxxxA}{\ensuremath{}}                 
\newcommand{\hatcurPPteffnoisorestrictempiricalxxxxxA}{\ensuremath{851\pm40}} 
\newcommand{\hatcurPPthetaempiricalxxxxxA}{\ensuremath{}}                
\newcommand{\hatcurPPthetanoisorestrictempiricalxxxxxA}{\ensuremath{0.0324\pm0.0035}} 
\newcommand{\hatcurPPfluxperiempiricalxxxxxA}{\ensuremath{}}             
\newcommand{\hatcurPPfluxperinoisorestrictempiricalxxxxxA}{\ensuremath{}} 
\newcommand{\hatcurPPfluxperidimempiricalxxxxxA}{\ensuremath{2d}}        
\newcommand{\hatcurPPfluxapempiricalxxxxxA}{\ensuremath{}}               
\newcommand{\hatcurPPfluxapnoisorestrictempiricalxxxxxA}{\ensuremath{}}  
\newcommand{\hatcurPPfluxapdimempiricalxxxxxA}{\ensuremath{2d}}          
\newcommand{\hatcurPPfluxavgempiricalxxxxxA}{\ensuremath{}}              
\newcommand{\hatcurPPfluxavgnoisorestrictempiricalxxxxxA}{\ensuremath{}} 
\newcommand{\hatcurPPfluxavgdimempiricalxxxxxA}{\ensuremath{2d}}         
\newcommand{\hatcurPPfluxavglogempiricalxxxxxA}{\ensuremath{}}           
\newcommand{\hatcurPPfluxavglognoisorestrictempiricalxxxxxA}{\ensuremath{8.076\pm0.077}} 
\newcommand{\hatcurXsecphaseempiricalxxxxxA}{\ensuremath{}}              
\newcommand{\hatcurXsecphasenoisorestrictempiricalxxxxxA}{\ensuremath{0\pm0}} 
\newcommand{\hatcurXsecondaryempiricalxxxxxA}{\ensuremath{}}             
\newcommand{\hatcurXsecondarynoisorestrictempiricalxxxxxA}{\ensuremath{2457308.47749\pm0.00031}} 
\newcommand{\hatcurXsecdurempiricalxxxxxA}{\ensuremath{}}                
\newcommand{\hatcurXsecdurnoisorestrictempiricalxxxxxA}{\ensuremath{0.0806\pm0.0018}} 
\newcommand{\hatcurXsecingdurempiricalxxxxxA}{\ensuremath{}}             
\newcommand{\hatcurXsecingdurnoisorestrictempiricalxxxxxA}{\ensuremath{0.0217\pm0.0027}} 
\newcommand{\hatcurPPphiconjempiricalxxxxxA}{\ensuremath{}}              
\newcommand{\hatcurPPphiconjnoisorestrictempiricalxxxxxA}{\ensuremath{0\pm0}} 
\newcommand{\hatcurPPperiempiricalxxxxxA}{\ensuremath{}}                 
\newcommand{\hatcurPPperinoisorestrictempiricalxxxxxA}{\ensuremath{2457305.53538\pm0.00031}} 
\newcommand{\hatcurPPaequivempiricalxxxxxA}{\ensuremath{}}               
\newcommand{\hatcurPPaequivnoisorestrictempiricalxxxxxA}{\ensuremath{0.1069\pm0.0084}} 
\newcommand{\hatcurPPtcircempiricalxxxxxA}{\ensuremath{}}                
\newcommand{\hatcurPPtcircnoisorestrictempiricalxxxxxA}{\ensuremath{182_{-40}^{+30}}} 
\newcommand{\hatcurPPtinfallempiricalxxxxxA}{\ensuremath{}}              
\newcommand{\hatcurPPtinfallnoisorestrictempiricalxxxxxA}{\ensuremath{56000_{-12000}^{+26000}}} 
\newcommand{\hatcurXdistempiricalxxxxxA}{\ensuremath{}}                  
\newcommand{\hatcurXdistnoisorestrictempiricalxxxxxA}{\ensuremath{306.1\pm2.6}} 
\newcommand{\hatcurXAvempiricalxxxxxA}{\ensuremath{}}                    
\newcommand{\hatcurXAvnoisorestrictempiricalxxxxxA}{\ensuremath{0.164_{-0.065}^{+0.043}}} 
\newcommand{\hatcurXdistredempiricalxxxxxA}{\ensuremath{}}               
\newcommand{\hatcurXdistrednoisorestrictempiricalxxxxxA}{\ensuremath{306.1\pm2.6}} 
\newcommand{\hatcurXEBVempiricalxxxxxA}{\ensuremath{}}                   
\newcommand{\hatcurXEBVnoisorestrictempiricalxxxxxA}{\ensuremath{0.053_{-0.021}^{+0.014}}} 
\newcommand{\hatcurCCpmraempiricalxxxxxA}{\ensuremath{3.827\pm0.058}}    
\newcommand{\hatcurCCpmdecempiricalxxxxxA}{\ensuremath{4.878\pm0.038}}   
\newcommand{\hatcurCCpmempiricalxxxxxA}{\ensuremath{6.200\pm0.069}}      
\newcommand{\hatcurhtrempiricalxxxxxB}{HATS778-005}                      
\newcommand{\hatcurfieldempiricalxxxxxB}{\ensuremath{string}}            
\newcommand{\hatcurCCraempiricalxxxxxB}{\ensuremath{19^{\mathrm h}14^{\mathrm m}41.2748{\mathrm s}}}                   
\newcommand{\hatcurCCdecempiricalxxxxxB}{\ensuremath{-59{\arcdeg}34{\arcmin}45.7571{\arcsec}}}                 
\newcommand{\hatcurCCmagempiricalxxxxxB}{14.352}                         
\newcommand{\hatcurCCtwomassempiricalxxxxxB}{2MASS~19144126-5934458}     
\newcommand{\hatcurCCgscempiricalxxxxxB}{GSC~}                           
\newcommand{\hatcurCCgaiaempiricalxxxxxB}{GAIA~6638412919991750912}      
\newcommand{\hatcurCCgaiadrtwoempiricalxxxxxB}{GAIA~DR2~6638412919991750912} 
\newcommand{\hatcurCCtassmvempiricalxxxxxB}{\ensuremath{14.35\pm0.11}}   
\newcommand{\hatcurCCtassmvshortempiricalxxxxxB}{\ensuremath{14.4}}      
\newcommand{\hatcurCCtassmBempiricalxxxxxB}{\ensuremath{15.577\pm0.050}} 
\newcommand{\hatcurCCtassmBshortempiricalxxxxxB}{\ensuremath{15.6}}      
\newcommand{\hatcurCCtassmIempiricalxxxxxB}{\ensuremath{nff\pmnff}}      
\newcommand{\hatcurCCtassmIshortempiricalxxxxxB}{\ensuremath{0.0}}       
\newcommand{\hatcurCCtassmgempiricalxxxxxB}{\ensuremath{14.935\pm0.030}} 
\newcommand{\hatcurCCtassmgshortempiricalxxxxxB}{\ensuremath{14.9}}      
\newcommand{\hatcurCCtassmrempiricalxxxxxB}{\ensuremath{13.821\pm0.050}} 
\newcommand{\hatcurCCtassmrshortempiricalxxxxxB}{\ensuremath{13.8}}      
\newcommand{\hatcurCCtassmiempiricalxxxxxB}{\ensuremath{13.69\pm0.17}}   
\newcommand{\hatcurCCtassmishortempiricalxxxxxB}{\ensuremath{13.7}}      
\newcommand{\hatcurCCparallaxempiricalxxxxxB}{\ensuremath{3.765\pm0.024}} 
\newcommand{\hatcurCCgaiamGempiricalxxxxxB}{\ensuremath{13.89510\pm0.00020}} 
\newcommand{\hatcurCCgaiamBPempiricalxxxxxB}{\ensuremath{14.5801\pm0.0016}} 
\newcommand{\hatcurCCgaiamRPempiricalxxxxxB}{\ensuremath{13.11260\pm0.00090}} 
\newcommand{\hatcurCCtwomassJmagempiricalxxxxxB}{\ensuremath{12.160\pm0.024}} 
\newcommand{\hatcurCCtwomassHmagempiricalxxxxxB}{\ensuremath{11.591\pm0.026}} 
\newcommand{\hatcurCCtwomassKmagempiricalxxxxxB}{\ensuremath{11.427\pm0.021}} 
\newcommand{\hatcurCCcitJmagempiricalxxxxxB}{\ensuremath{12.158\pm0.025}} 
\newcommand{\hatcurCCcitHmagempiricalxxxxxB}{\ensuremath{11.583\pm0.027}} 
\newcommand{\hatcurCCcitKmagempiricalxxxxxB}{\ensuremath{11.451\pm0.022}} 
\newcommand{\hatcurCCbbJmagempiricalxxxxxB}{\ensuremath{12.236\pm0.027}} 
\newcommand{\hatcurCCbbHmagempiricalxxxxxB}{\ensuremath{11.607\pm0.028}} 
\newcommand{\hatcurCCbbKmagempiricalxxxxxB}{\ensuremath{11.471\pm0.022}} 
\newcommand{\hatcurCCesoJmagempiricalxxxxxB}{\ensuremath{12.243\pm0.031}} 
\newcommand{\hatcurCCesoHmagempiricalxxxxxB}{\ensuremath{11.605\pm0.036}} 
\newcommand{\hatcurCCesoKmagempiricalxxxxxB}{\ensuremath{11.468\pm0.023}} 
\newcommand{\hatcurCCesoJHmagempiricalxxxxxB}{\ensuremath{0.638\pm0.044}} 
\newcommand{\hatcurCCesoJKmagempiricalxxxxxB}{\ensuremath{0.776\pm0.036}} 
\newcommand{\hatcurCCesoHKmagempiricalxxxxxB}{\ensuremath{0.137\pm0.042}} 
\newcommand{\hatcurLCdipempiricalxxxxxB}{\ensuremath{16.5}}              
\newcommand{\hatcurLCrprstarempiricalxxxxxB}{\ensuremath{}}              
\newcommand{\hatcurLCrprstarnoisorestrictempiricalxxxxxB}{\ensuremath{0.1118\pm0.0016}} 
\newcommand{\hatcurLCbsqempiricalxxxxxB}{\ensuremath{}}                  
\newcommand{\hatcurLCbsqnoisorestrictempiricalxxxxxB}{\ensuremath{0.052_{-0.034}^{+0.068}}} 
\newcommand{\hatcurLCimpempiricalxxxxxB}{\ensuremath{}}                  
\newcommand{\hatcurLCimpnoisorestrictempiricalxxxxxB}{\ensuremath{0.228_{-0.095}^{+0.118}}} 
\newcommand{\hatcurLCzetaempiricalxxxxxB}{\ensuremath{}}                 
\newcommand{\hatcurLCzetanoisorestrictempiricalxxxxxB}{\ensuremath{22.41\pm0.42}} 
\newcommand{\hatcurLCdurempiricalxxxxxB}{\ensuremath{}}                  
\newcommand{\hatcurLCdurnoisorestrictempiricalxxxxxB}{\ensuremath{0.0999\pm0.0019}} 
\newcommand{\hatcurLCdurshortempiricalxxxxxB}{\ensuremath{}}             
\newcommand{\hatcurLCdurshortnoisorestrictempiricalxxxxxB}{\ensuremath{0.0999}} 
\newcommand{\hatcurLCdurhrempiricalxxxxxB}{\ensuremath{}}                
\newcommand{\hatcurLCdurhrnoisorestrictempiricalxxxxxB}{\ensuremath{2.397\pm0.047}} 
\newcommand{\hatcurLCdurhrshortempiricalxxxxxB}{\ensuremath{}}           
\newcommand{\hatcurLCdurhrshortnoisorestrictempiricalxxxxxB}{\ensuremath{2.397}} 
\newcommand{\hatcurLCqempiricalxxxxxB}{\ensuremath{}}                    
\newcommand{\hatcurLCqnoisorestrictempiricalxxxxxB}{\ensuremath{0.03190\pm0.00062}} 
\newcommand{\hatcurLCqshortempiricalxxxxxB}{\ensuremath{}}               
\newcommand{\hatcurLCqshortnoisorestrictempiricalxxxxxB}{\ensuremath{0.032}} 
\newcommand{\hatcurLCingdurempiricalxxxxxB}{\ensuremath{}}               
\newcommand{\hatcurLCingdurnoisorestrictempiricalxxxxxB}{\ensuremath{0.01057\pm0.00069}} 
\newcommand{\hatcurLCPempiricalxxxxxB}{\ensuremath{}}                    
\newcommand{\hatcurLCPnoisorestrictempiricalxxxxxB}{\ensuremath{3.1316644\pm0.0000029}} 
\newcommand{\hatcurLCPprecempiricalxxxxxB}{\ensuremath{}}                
\newcommand{\hatcurLCPprecnoisorestrictempiricalxxxxxB}{\ensuremath{3.1316644}} 
\newcommand{\hatcurLCPshortempiricalxxxxxB}{\ensuremath{}}               
\newcommand{\hatcurLCPshortnoisorestrictempiricalxxxxxB}{\ensuremath{3.1317}} 
\newcommand{\hatcurLCTempiricalxxxxxB}{\ensuremath{}}                    
\newcommand{\hatcurLCTnoisorestrictempiricalxxxxxB}{\ensuremath{2456984.67939\pm0.00071}} 
\newcommand{\hatcurLCTAempiricalxxxxxB}{\ensuremath{}}                   
\newcommand{\hatcurLCTAnoisorestrictempiricalxxxxxB}{\ensuremath{2455678.7754\pm0.0016}} 
\newcommand{\hatcurLCTBempiricalxxxxxB}{\ensuremath{}}                   
\newcommand{\hatcurLCTBnoisorestrictempiricalxxxxxB}{\ensuremath{2457219.55421\pm0.00067}} 
\newcommand{\hatcurLChatnetmAempiricalxxxxxB}{\ensuremath{}}             
\newcommand{\hatcurLChatnetmAnoisorestrictempiricalxxxxxB}{\ensuremath{13.91277\pm0.00018}} 
\newcommand{\hatcurLCiblendAempiricalxxxxxB}{\ensuremath{}}              
\newcommand{\hatcurLCiblendAnoisorestrictempiricalxxxxxB}{\ensuremath{0.99890\pm0.00019}} 
\newcommand{\hatcurLChatnetmBempiricalxxxxxB}{\ensuremath{}}             
\newcommand{\hatcurLChatnetmBnoisorestrictempiricalxxxxxB}{\ensuremath{-0.00009\pm0.00019}} 
\newcommand{\hatcurLCiblendBempiricalxxxxxB}{\ensuremath{}}              
\newcommand{\hatcurLCiblendBnoisorestrictempiricalxxxxxB}{\ensuremath{0.590\pm0.016}} 
\newcommand{\hatcurLCrhoempiricalxxxxxB}{\ensuremath{}}                  
\newcommand{\hatcurLCrhonoisorestrictempiricalxxxxxB}{\ensuremath{2.45_{-0.29}^{+0.19}}} 
\newcommand{\hatcurSMEiteffempiricalxxxxxB}{\ensuremath{4190\pm100}}     
\newcommand{\hatcurSMEizfehempiricalxxxxxB}{\ensuremath{0.00\pm0.10}}    
\newcommand{\hatcurSMEizfehshortempiricalxxxxxB}{\ensuremath{0.00}}      
\newcommand{\hatcurSMEiloggempiricalxxxxxB}{\ensuremath{4.690\pm0.088}}  
\newcommand{\hatcurSMEivsinempiricalxxxxxB}{\ensuremath{0.73\pm0.55}}    
\newcommand{\hatcurSMEivmacempiricalxxxxxB}{\ensuremath{nff\pmnff}}      
\newcommand{\hatcurSMEivmicempiricalxxxxxB}{\ensuremath{nff\pmnff}}      
\newcommand{\hatcurextraerrMgempiricalxxxxxB}{\ensuremath{}}             
\newcommand{\hatcurextraerrMgnoisorestrictempiricalxxxxxB}{\ensuremath{0.095\pm0.052}} 
\newcommand{\hatcurextraerrMgtwosiglimempiricalxxxxxB}{\ensuremath{}}    
\newcommand{\hatcurextraerrMgtwosiglimnoisorestrictempiricalxxxxxB}{\ensuremath{<0.1783}} 
\newcommand{\hatcurextraerrMrempiricalxxxxxB}{\ensuremath{}}             
\newcommand{\hatcurextraerrMrnoisorestrictempiricalxxxxxB}{\ensuremath{0.037_{-0.036}^{+0.122}}} 
\newcommand{\hatcurextraerrMrtwosiglimempiricalxxxxxB}{\ensuremath{}}    
\newcommand{\hatcurextraerrMrtwosiglimnoisorestrictempiricalxxxxxB}{\ensuremath{<0.2553}} 
\newcommand{\hatcurextraerrMiempiricalxxxxxB}{\ensuremath{}}             
\newcommand{\hatcurextraerrMinoisorestrictempiricalxxxxxB}{\ensuremath{0.018_{-0.018}^{+0.066}}} 
\newcommand{\hatcurextraerrMitwosiglimempiricalxxxxxB}{\ensuremath{}}    
\newcommand{\hatcurextraerrMitwosiglimnoisorestrictempiricalxxxxxB}{\ensuremath{<0.1418}} 
\newcommand{\hatcurextraerrMGempiricalxxxxxB}{\ensuremath{}}             
\newcommand{\hatcurextraerrMGnoisorestrictempiricalxxxxxB}{\ensuremath{0.0249_{-0.0100}^{+0.0156}}} 
\newcommand{\hatcurextraerrMGtwosiglimempiricalxxxxxB}{\ensuremath{}}    
\newcommand{\hatcurextraerrMGtwosiglimnoisorestrictempiricalxxxxxB}{\ensuremath{<0.0494}} 
\newcommand{\hatcurextraerrMBPtwoempiricalxxxxxB}{\ensuremath{}}         
\newcommand{\hatcurextraerrMBPtwonoisorestrictempiricalxxxxxB}{\ensuremath{0.0013_{-0.0013}^{+0.0095}}} 
\newcommand{\hatcurextraerrMBPtwotwosiglimempiricalxxxxxB}{\ensuremath{}} 
\newcommand{\hatcurextraerrMBPtwotwosiglimnoisorestrictempiricalxxxxxB}{\ensuremath{<0.0225}} 
\newcommand{\hatcurextraerrMRPempiricalxxxxxB}{\ensuremath{}}            
\newcommand{\hatcurextraerrMRPnoisorestrictempiricalxxxxxB}{\ensuremath{0.0034_{-0.0033}^{+0.0170}}} 
\newcommand{\hatcurextraerrMRPtwosiglimempiricalxxxxxB}{\ensuremath{}}   
\newcommand{\hatcurextraerrMRPtwosiglimnoisorestrictempiricalxxxxxB}{\ensuremath{<0.0346}} 
\newcommand{\hatcurextraerrMJempiricalxxxxxB}{\ensuremath{}}             
\newcommand{\hatcurextraerrMJnoisorestrictempiricalxxxxxB}{\ensuremath{0.0026_{-0.0025}^{+0.0147}}} 
\newcommand{\hatcurextraerrMJtwosiglimempiricalxxxxxB}{\ensuremath{}}    
\newcommand{\hatcurextraerrMJtwosiglimnoisorestrictempiricalxxxxxB}{\ensuremath{<0.0340}} 
\newcommand{\hatcurextraerrMHempiricalxxxxxB}{\ensuremath{}}             
\newcommand{\hatcurextraerrMHnoisorestrictempiricalxxxxxB}{\ensuremath{0.0088_{-0.0084}^{+0.0180}}} 
\newcommand{\hatcurextraerrMHtwosiglimempiricalxxxxxB}{\ensuremath{}}    
\newcommand{\hatcurextraerrMHtwosiglimnoisorestrictempiricalxxxxxB}{\ensuremath{<0.0402}} 
\newcommand{\hatcurextraerrMKsempiricalxxxxxB}{\ensuremath{}}            
\newcommand{\hatcurextraerrMKsnoisorestrictempiricalxxxxxB}{\ensuremath{0.0050_{-0.0048}^{+0.0260}}} 
\newcommand{\hatcurextraerrMKstwosiglimempiricalxxxxxB}{\ensuremath{}}   
\newcommand{\hatcurextraerrMKstwosiglimnoisorestrictempiricalxxxxxB}{\ensuremath{<0.0476}} 
\newcommand{\hatcurextraerrMWoneempiricalxxxxxB}{\ensuremath{}}          
\newcommand{\hatcurextraerrMWonenoisorestrictempiricalxxxxxB}{\ensuremath{0.014_{-0.013}^{+0.029}}} 
\newcommand{\hatcurextraerrMWonetwosiglimempiricalxxxxxB}{\ensuremath{}} 
\newcommand{\hatcurextraerrMWonetwosiglimnoisorestrictempiricalxxxxxB}{\ensuremath{<0.0741}} 
\newcommand{\hatcurextraerrMWtwoempiricalxxxxxB}{\ensuremath{}}          
\newcommand{\hatcurextraerrMWtwonoisorestrictempiricalxxxxxB}{\ensuremath{0.0033_{-0.0032}^{+0.0231}}} 
\newcommand{\hatcurextraerrMWtwotwosiglimempiricalxxxxxB}{\ensuremath{}} 
\newcommand{\hatcurextraerrMWtwotwosiglimnoisorestrictempiricalxxxxxB}{\ensuremath{<0.0617}} 
\newcommand{\hatcurLBiBempiricalxxxxxB}{\ensuremath{0.8007}}             
\newcommand{\hatcurLBiiBempiricalxxxxxB}{\ensuremath{0.0222}}            
\newcommand{\hatcurLBiVempiricalxxxxxB}{\ensuremath{0.6354}}             
\newcommand{\hatcurLBiiVempiricalxxxxxB}{\ensuremath{0.1339}}            
\newcommand{\hatcurLBiRempiricalxxxxxB}{\ensuremath{0.5245}}             
\newcommand{\hatcurLBiiRempiricalxxxxxB}{\ensuremath{0.1825}}            
\newcommand{\hatcurLBiIempiricalxxxxxB}{\ensuremath{0.3905}}             
\newcommand{\hatcurLBiiIempiricalxxxxxB}{\ensuremath{0.2115}}            
\newcommand{\hatcurLBiuempiricalxxxxxB}{\ensuremath{0.8665}}             
\newcommand{\hatcurLBiiuempiricalxxxxxB}{\ensuremath{-0.0450}}           
\newcommand{\hatcurLBigempiricalxxxxxB}{\ensuremath{0.7223}}             
\newcommand{\hatcurLBiigempiricalxxxxxB}{\ensuremath{0.0663}}            
\newcommand{\hatcurLBirempiricalxxxxxB}{\ensuremath{}}                   
\newcommand{\hatcurLBirnoisorestrictempiricalxxxxxB}{\ensuremath{0.49\pm0.14}} 
\newcommand{\hatcurLBiirempiricalxxxxxB}{\ensuremath{}}                  
\newcommand{\hatcurLBiirnoisorestrictempiricalxxxxxB}{\ensuremath{0.28\pm0.15}} 
\newcommand{\hatcurLBiiempiricalxxxxxB}{\ensuremath{}}                   
\newcommand{\hatcurLBiinoisorestrictempiricalxxxxxB}{\ensuremath{0.35\pm0.13}} 
\newcommand{\hatcurLBiiiempiricalxxxxxB}{\ensuremath{}}                  
\newcommand{\hatcurLBiiinoisorestrictempiricalxxxxxB}{\ensuremath{0.24\pm0.15}} 
\newcommand{\hatcurLBizempiricalxxxxxB}{\ensuremath{0.3392}}             
\newcommand{\hatcurLBiizempiricalxxxxxB}{\ensuremath{0.2145}}            
\newcommand{\hatcurLBiJempiricalxxxxxB}{\ensuremath{0.2506}}             
\newcommand{\hatcurLBiiJempiricalxxxxxB}{\ensuremath{0.2171}}            
\newcommand{\hatcurLBiHempiricalxxxxxB}{\ensuremath{0.1694}}             
\newcommand{\hatcurLBiiHempiricalxxxxxB}{\ensuremath{0.2632}}            
\newcommand{\hatcurLBiKempiricalxxxxxB}{\ensuremath{0.1301}}             
\newcommand{\hatcurLBiiKempiricalxxxxxB}{\ensuremath{0.2308}}            
\newcommand{\hatcurLBiTempiricalxxxxxB}{\ensuremath{}}                   
\newcommand{\hatcurLBiTnoisorestrictempiricalxxxxxB}{\ensuremath{0.45\pm0.14}} 
\newcommand{\hatcurLBiiTempiricalxxxxxB}{\ensuremath{}}                  
\newcommand{\hatcurLBiiTnoisorestrictempiricalxxxxxB}{\ensuremath{0.26\pm0.17}} 
\newcommand{\hatcurLBikepempiricalxxxxxB}{\ensuremath{0.5094}}           
\newcommand{\hatcurLBiikepempiricalxxxxxB}{\ensuremath{0.2164}}          
\newcommand{\hatcurLBiCempiricalxxxxxB}{\ensuremath{0.4819}}             
\newcommand{\hatcurLBiiCempiricalxxxxxB}{\ensuremath{0.2222}}            
\newcommand{\hatcurLBiMempiricalxxxxxB}{\ensuremath{0.5998}}             
\newcommand{\hatcurLBiiMempiricalxxxxxB}{\ensuremath{0.1782}}            
\newcommand{\hatcurLBiSoneempiricalxxxxxB}{\ensuremath{0.0966}}            
\newcommand{\hatcurLBiiSoneempiricalxxxxxB}{\ensuremath{0.1667}}           
\newcommand{\hatcurLBiStwoempiricalxxxxxB}{\ensuremath{0.0830}}            
\newcommand{\hatcurLBiiStwoempiricalxxxxxB}{\ensuremath{0.1376}}           
\newcommand{\hatcurLBiSthreeempiricalxxxxxB}{\ensuremath{0.0699}}            
\newcommand{\hatcurLBiiSthreeempiricalxxxxxB}{\ensuremath{0.1187}}           
\newcommand{\hatcurLBiSfourempiricalxxxxxB}{\ensuremath{0.0647}}            
\newcommand{\hatcurLBiiSfourempiricalxxxxxB}{\ensuremath{0.1036}}           
\newcommand{\hatcurISOmempiricalxxxxxB}{\ensuremath{}}                   
\newcommand{\hatcurISOmnoisorestrictempiricalxxxxxB}{\ensuremath{0.637\pm0.066}} 
\newcommand{\hatcurISOmshortempiricalxxxxxB}{\ensuremath{}}              
\newcommand{\hatcurISOmshortnoisorestrictempiricalxxxxxB}{\ensuremath{0.64}} 
\newcommand{\hatcurISOmlongempiricalxxxxxB}{\ensuremath{}}               
\newcommand{\hatcurISOmlongnoisorestrictempiricalxxxxxB}{\ensuremath{0.637\pm0.066}} 
\newcommand{\hatcurISOrempiricalxxxxxB}{\ensuremath{}}                   
\newcommand{\hatcurISOrnoisorestrictempiricalxxxxxB}{\ensuremath{0.716\pm0.011}} 
\newcommand{\hatcurISOrshortempiricalxxxxxB}{\ensuremath{}}              
\newcommand{\hatcurISOrshortnoisorestrictempiricalxxxxxB}{\ensuremath{0.72}} 
\newcommand{\hatcurISOrlongempiricalxxxxxB}{\ensuremath{}}               
\newcommand{\hatcurISOrlongnoisorestrictempiricalxxxxxB}{\ensuremath{0.716\pm0.011}} 
\newcommand{\hatcurISOloggempiricalxxxxxB}{\ensuremath{}}                
\newcommand{\hatcurISOloggnoisorestrictempiricalxxxxxB}{\ensuremath{4.534\pm0.044}} 
\newcommand{\hatcurISOlumempiricalxxxxxB}{\ensuremath{}}                 
\newcommand{\hatcurISOlumnoisorestrictempiricalxxxxxB}{\ensuremath{0.200\pm0.011}} 
\newcommand{\hatcurISOlumshortempiricalxxxxxB}{\ensuremath{}}            
\newcommand{\hatcurISOlumshortnoisorestrictempiricalxxxxxB}{\ensuremath{0.20}} 
\newcommand{\hatcurISOfehempiricalxxxxxB}{\ensuremath{}}                 
\newcommand{\hatcurISOfehnoisorestrictempiricalxxxxxB}{\ensuremath{0.18\pm0.11}} 
\newcommand{\hatcurISOteffempiricalxxxxxB}{\ensuremath{}}                
\newcommand{\hatcurISOteffnoisorestrictempiricalxxxxxB}{\ensuremath{4562\pm49}} 
\newcommand{\hatcurISOageempiricalxxxxxB}{\ensuremath{}}                 
\newcommand{\hatcurISOagenoisorestrictempiricalxxxxxB}{\ensuremath{0\pm0}} 
\newcommand{\hatcurISOspecempiricalxxxxxB}{K}                            
\newcommand{\hatcurRVKempiricalxxxxxB}{\ensuremath{}}                    
\newcommand{\hatcurRVKnoisorestrictempiricalxxxxxB}{\ensuremath{41.4\pm4.0}} 
\newcommand{\hatcurRVrkempiricalxxxxxB}{\ensuremath{}}                   
\newcommand{\hatcurRVrknoisorestrictempiricalxxxxxB}{\ensuremath{0\pm0}} 
\newcommand{\hatcurRVrhempiricalxxxxxB}{\ensuremath{}}                   
\newcommand{\hatcurRVrhnoisorestrictempiricalxxxxxB}{\ensuremath{0\pm0}} 
\newcommand{\hatcurRVkempiricalxxxxxB}{\ensuremath{}}                    
\newcommand{\hatcurRVknoisorestrictempiricalxxxxxB}{\ensuremath{0\pm0}}  
\newcommand{\hatcurRVhempiricalxxxxxB}{\ensuremath{}}                    
\newcommand{\hatcurRVhnoisorestrictempiricalxxxxxB}{\ensuremath{0\pm0}}  
\newcommand{\hatcurRVtroneempiricalxxxxxB}{\ensuremath{}}                
\newcommand{\hatcurRVtronenoisorestrictempiricalxxxxxB}{\ensuremath{0\pm0}} 
\newcommand{\hatcurRVtrtwoempiricalxxxxxB}{\ensuremath{}}                
\newcommand{\hatcurRVtrtwonoisorestrictempiricalxxxxxB}{\ensuremath{0\pm0}} 
\newcommand{\hatcurRVgammaempiricalxxxxxB}{\ensuremath{}}                
\newcommand{\hatcurRVgammanoisorestrictempiricalxxxxxB}{\ensuremath{-14.6\pm3.7}} 
\newcommand{\hatcurRVjitterempiricalxxxxxB}{\ensuremath{}}               
\newcommand{\hatcurRVjitternoisorestrictempiricalxxxxxB}{\ensuremath{25.3\pm6.6}} 
\newcommand{\hatcurRVjittertwosiglimempiricalxxxxxB}{\ensuremath{}}      
\newcommand{\hatcurRVjittertwosiglimnoisorestrictempiricalxxxxxB}{\ensuremath{<39.1}} 
\newcommand{\hatcurRVfitrmsempiricalxxxxxB}{\ensuremath{.1fym}}          %
\newcommand{\hatcurRVeccenempiricalxxxxxB}{\ensuremath{}}                
\newcommand{\hatcurRVeccennoisorestrictempiricalxxxxxB}{\ensuremath{0\pm0}} 
\newcommand{\hatcurRVeccentwosiglimempiricalxxxxxB}{\ensuremath{}}       
\newcommand{\hatcurRVeccentwosiglimnoisorestrictempiricalxxxxxB}{\ensuremath{<0.000}} 
\newcommand{\hatcurRVomegaempiricalxxxxxB}{\ensuremath{}}                
\newcommand{\hatcurRVomeganoisorestrictempiricalxxxxxB}{\ensuremath{0\pm0}} 
\newcommand{\hatcurPPiempiricalxxxxxB}{\ensuremath{}}                    
\newcommand{\hatcurPPinoisorestrictempiricalxxxxxB}{\ensuremath{88.79_{-0.70}^{+0.53}}} 
\newcommand{\hatcurPPgempiricalxxxxxB}{\ensuremath{}}                    
\newcommand{\hatcurPPgnoisorestrictempiricalxxxxxB}{\ensuremath{8.9\pm1.1}} 
\newcommand{\hatcurPPloggempiricalxxxxxB}{\ensuremath{}}                 
\newcommand{\hatcurPPloggnoisorestrictempiricalxxxxxB}{\ensuremath{2.950\pm0.052}} 
\newcommand{\hatcurPParempiricalxxxxxB}{\ensuremath{}}                   
\newcommand{\hatcurPParnoisorestrictempiricalxxxxxB}{\ensuremath{10.83_{-0.44}^{+0.28}}} 
\newcommand{\hatcurPParelempiricalxxxxxB}{\ensuremath{}}                 
\newcommand{\hatcurPParelnoisorestrictempiricalxxxxxB}{\ensuremath{0.0360\pm0.0013}} 
\newcommand{\hatcurPPrhoempiricalxxxxxB}{\ensuremath{}}                  
\newcommand{\hatcurPPrhonoisorestrictempiricalxxxxxB}{\ensuremath{0.567_{-0.065}^{+0.090}}} 
\newcommand{\hatcurPPmempiricalxxxxxB}{\ensuremath{}}                    
\newcommand{\hatcurPPmnoisorestrictempiricalxxxxxB}{\ensuremath{0.219\pm0.023}} 
\newcommand{\hatcurPPmshortempiricalxxxxxB}{\ensuremath{}}               
\newcommand{\hatcurPPmshortnoisorestrictempiricalxxxxxB}{\ensuremath{0.22}} 
\newcommand{\hatcurPPmlongempiricalxxxxxB}{\ensuremath{}}                
\newcommand{\hatcurPPmlongnoisorestrictempiricalxxxxxB}{\ensuremath{0.219\pm0.023}} 
\newcommand{\hatcurPPmeempiricalxxxxxB}{\ensuremath{}}                   
\newcommand{\hatcurPPmenoisorestrictempiricalxxxxxB}{\ensuremath{69.7\pm7.3}} 
\newcommand{\hatcurPPmeshortempiricalxxxxxB}{\ensuremath{}}              
\newcommand{\hatcurPPmeshortnoisorestrictempiricalxxxxxB}{\ensuremath{69.7}} 
\newcommand{\hatcurPPmelongempiricalxxxxxB}{\ensuremath{}}               
\newcommand{\hatcurPPmelongnoisorestrictempiricalxxxxxB}{\ensuremath{69.7\pm7.3}} 
\newcommand{\hatcurPPrempiricalxxxxxB}{\ensuremath{}}                    
\newcommand{\hatcurPPrnoisorestrictempiricalxxxxxB}{\ensuremath{0.779\pm0.017}} 
\newcommand{\hatcurPPrshortempiricalxxxxxB}{\ensuremath{}}               
\newcommand{\hatcurPPrshortnoisorestrictempiricalxxxxxB}{\ensuremath{0.78}} 
\newcommand{\hatcurPPrlongempiricalxxxxxB}{\ensuremath{}}                
\newcommand{\hatcurPPrlongnoisorestrictempiricalxxxxxB}{\ensuremath{0.779\pm0.017}} 
\newcommand{\hatcurPPreempiricalxxxxxB}{\ensuremath{}}                   
\newcommand{\hatcurPPrenoisorestrictempiricalxxxxxB}{\ensuremath{8.73\pm0.19}} 
\newcommand{\hatcurPPreshortempiricalxxxxxB}{\ensuremath{}}              
\newcommand{\hatcurPPreshortnoisorestrictempiricalxxxxxB}{\ensuremath{8.7}} 
\newcommand{\hatcurPPrelongempiricalxxxxxB}{\ensuremath{}}               
\newcommand{\hatcurPPrelongnoisorestrictempiricalxxxxxB}{\ensuremath{8.73\pm0.19}} 
\newcommand{\hatcurPPmrcorrempiricalxxxxxB}{\ensuremath{0.00}}           
\newcommand{\hatcurPPteffempiricalxxxxxB}{\ensuremath{}}                 
\newcommand{\hatcurPPteffnoisorestrictempiricalxxxxxB}{\ensuremath{982\pm18}} 
\newcommand{\hatcurPPthetaempiricalxxxxxB}{\ensuremath{}}                
\newcommand{\hatcurPPthetanoisorestrictempiricalxxxxxB}{\ensuremath{0.0321\pm0.0034}} 
\newcommand{\hatcurPPfluxperiempiricalxxxxxB}{\ensuremath{}}             
\newcommand{\hatcurPPfluxperinoisorestrictempiricalxxxxxB}{\ensuremath{}} 
\newcommand{\hatcurPPfluxperidimempiricalxxxxxB}{\ensuremath{2d}}        
\newcommand{\hatcurPPfluxapempiricalxxxxxB}{\ensuremath{}}               
\newcommand{\hatcurPPfluxapnoisorestrictempiricalxxxxxB}{\ensuremath{}}  
\newcommand{\hatcurPPfluxapdimempiricalxxxxxB}{\ensuremath{2d}}          
\newcommand{\hatcurPPfluxavgempiricalxxxxxB}{\ensuremath{}}              
\newcommand{\hatcurPPfluxavgnoisorestrictempiricalxxxxxB}{\ensuremath{}} 
\newcommand{\hatcurPPfluxavgdimempiricalxxxxxB}{\ensuremath{2d}}         
\newcommand{\hatcurPPfluxavglogempiricalxxxxxB}{\ensuremath{}}           
\newcommand{\hatcurPPfluxavglognoisorestrictempiricalxxxxxB}{\ensuremath{8.324\pm0.032}} 
\newcommand{\hatcurXsecphaseempiricalxxxxxB}{\ensuremath{}}              
\newcommand{\hatcurXsecphasenoisorestrictempiricalxxxxxB}{\ensuremath{0\pm0}} 
\newcommand{\hatcurXsecondaryempiricalxxxxxB}{\ensuremath{}}             
\newcommand{\hatcurXsecondarynoisorestrictempiricalxxxxxB}{\ensuremath{2456986.24522\pm0.00071}} 
\newcommand{\hatcurXsecdurempiricalxxxxxB}{\ensuremath{}}                
\newcommand{\hatcurXsecdurnoisorestrictempiricalxxxxxB}{\ensuremath{0.0999\pm0.0019}} 
\newcommand{\hatcurXsecingdurempiricalxxxxxB}{\ensuremath{}}             
\newcommand{\hatcurXsecingdurnoisorestrictempiricalxxxxxB}{\ensuremath{0.01057\pm0.00069}} 
\newcommand{\hatcurPPphiconjempiricalxxxxxB}{\ensuremath{}}              
\newcommand{\hatcurPPphiconjnoisorestrictempiricalxxxxxB}{\ensuremath{0\pm0}} 
\newcommand{\hatcurPPperiempiricalxxxxxB}{\ensuremath{}}                 
\newcommand{\hatcurPPperinoisorestrictempiricalxxxxxB}{\ensuremath{2456983.89648\pm0.00071}} 
\newcommand{\hatcurPPaequivempiricalxxxxxB}{\ensuremath{}}               
\newcommand{\hatcurPPaequivnoisorestrictempiricalxxxxxB}{\ensuremath{0.0803\pm0.0029}} 
\newcommand{\hatcurPPtcircempiricalxxxxxB}{\ensuremath{}}                
\newcommand{\hatcurPPtcircnoisorestrictempiricalxxxxxB}{\ensuremath{237_{-40}^{+52}}} 
\newcommand{\hatcurPPtinfallempiricalxxxxxB}{\ensuremath{}}              
\newcommand{\hatcurPPtinfallnoisorestrictempiricalxxxxxB}{\ensuremath{12400\pm3100}} 
\newcommand{\hatcurXdistempiricalxxxxxB}{\ensuremath{}}                  
\newcommand{\hatcurXdistnoisorestrictempiricalxxxxxB}{\ensuremath{265.9\pm1.5}} 
\newcommand{\hatcurXAvempiricalxxxxxB}{\ensuremath{}}                    
\newcommand{\hatcurXAvnoisorestrictempiricalxxxxxB}{\ensuremath{0.127\pm0.052}} 
\newcommand{\hatcurXdistredempiricalxxxxxB}{\ensuremath{}}               
\newcommand{\hatcurXdistrednoisorestrictempiricalxxxxxB}{\ensuremath{265.9\pm1.5}} 
\newcommand{\hatcurXEBVempiricalxxxxxB}{\ensuremath{}}                   
\newcommand{\hatcurXEBVnoisorestrictempiricalxxxxxB}{\ensuremath{0.041\pm0.017}} 
\newcommand{\hatcurCCpmraempiricalxxxxxB}{\ensuremath{3.125\pm0.031}}    
\newcommand{\hatcurCCpmdecempiricalxxxxxB}{\ensuremath{6.146\pm0.029}}   
\newcommand{\hatcurCCpmempiricalxxxxxB}{\ensuremath{6.895\pm0.042}}      
\newcommand{\hatcurhtrempiricalxxxxxC}{HATS755-004}                      
\newcommand{\hatcurfieldempiricalxxxxxC}{\ensuremath{string}}            
\newcommand{\hatcurCCraempiricalxxxxxC}{\ensuremath{00^{\mathrm h}26^{\mathrm m}27.1829{\mathrm s}}}                   
\newcommand{\hatcurCCdecempiricalxxxxxC}{\ensuremath{-56{\arcdeg}20{\arcmin}39.5352{\arcsec}}}                 
\newcommand{\hatcurCCmagempiricalxxxxxC}{14.998}                         
\newcommand{\hatcurCCtwomassempiricalxxxxxC}{2MASS~00262717-5620395}     
\newcommand{\hatcurCCgscempiricalxxxxxC}{GSC~}                           
\newcommand{\hatcurCCgaiaempiricalxxxxxC}{GAIA~4919770108539385472}      
\newcommand{\hatcurCCgaiadrtwoempiricalxxxxxC}{GAIA~DR2~4919770108539385472} 
\newcommand{\hatcurCCtassmvempiricalxxxxxC}{\ensuremath{14.998\pm0.040}} 
\newcommand{\hatcurCCtassmvshortempiricalxxxxxC}{\ensuremath{15.0}}      
\newcommand{\hatcurCCtassmBempiricalxxxxxC}{\ensuremath{16.378\pm0.040}} 
\newcommand{\hatcurCCtassmBshortempiricalxxxxxC}{\ensuremath{16.4}}      
\newcommand{\hatcurCCtassmIempiricalxxxxxC}{\ensuremath{nff\pmnff}}      
\newcommand{\hatcurCCtassmIshortempiricalxxxxxC}{\ensuremath{0.0}}       
\newcommand{\hatcurCCtassmgempiricalxxxxxC}{\ensuremath{15.668\pm0.040}} 
\newcommand{\hatcurCCtassmgshortempiricalxxxxxC}{\ensuremath{15.7}}      
\newcommand{\hatcurCCtassmrempiricalxxxxxC}{\ensuremath{14.496\pm0.010}} 
\newcommand{\hatcurCCtassmrshortempiricalxxxxxC}{\ensuremath{14.5}}      
\newcommand{\hatcurCCtassmiempiricalxxxxxC}{\ensuremath{14.14\pm0.12}}   
\newcommand{\hatcurCCtassmishortempiricalxxxxxC}{\ensuremath{14.1}}      
\newcommand{\hatcurCCparallaxempiricalxxxxxC}{\ensuremath{3.054\pm0.022}} 
\newcommand{\hatcurCCgaiamGempiricalxxxxxC}{\ensuremath{14.54490\pm0.00030}} 
\newcommand{\hatcurCCgaiamBPempiricalxxxxxC}{\ensuremath{15.2886\pm0.0022}} 
\newcommand{\hatcurCCgaiamRPempiricalxxxxxC}{\ensuremath{13.7214\pm0.0013}} 
\newcommand{\hatcurCCtwomassJmagempiricalxxxxxC}{\ensuremath{12.692\pm0.024}} 
\newcommand{\hatcurCCtwomassHmagempiricalxxxxxC}{\ensuremath{12.105\pm0.024}} 
\newcommand{\hatcurCCtwomassKmagempiricalxxxxxC}{\ensuremath{11.938\pm0.023}} 
\newcommand{\hatcurCCcitJmagempiricalxxxxxC}{\ensuremath{12.688\pm0.025}} 
\newcommand{\hatcurCCcitHmagempiricalxxxxxC}{\ensuremath{12.097\pm0.025}} 
\newcommand{\hatcurCCcitKmagempiricalxxxxxC}{\ensuremath{11.962\pm0.023}} 
\newcommand{\hatcurCCbbJmagempiricalxxxxxC}{\ensuremath{12.769\pm0.027}} 
\newcommand{\hatcurCCbbHmagempiricalxxxxxC}{\ensuremath{12.121\pm0.026}} 
\newcommand{\hatcurCCbbKmagempiricalxxxxxC}{\ensuremath{11.982\pm0.023}} 
\newcommand{\hatcurCCesoJmagempiricalxxxxxC}{\ensuremath{12.776\pm0.031}} 
\newcommand{\hatcurCCesoHmagempiricalxxxxxC}{\ensuremath{12.118\pm0.035}} 
\newcommand{\hatcurCCesoKmagempiricalxxxxxC}{\ensuremath{11.979\pm0.025}} 
\newcommand{\hatcurCCesoJHmagempiricalxxxxxC}{\ensuremath{0.657\pm0.025}} 
\newcommand{\hatcurCCesoJKmagempiricalxxxxxC}{\ensuremath{0.797\pm0.038}} 
\newcommand{\hatcurCCesoHKmagempiricalxxxxxC}{\ensuremath{0.139\pm0.041}} 
\newcommand{\hatcurLCdipempiricalxxxxxC}{\ensuremath{13.2}}              
\newcommand{\hatcurLCrprstarempiricalxxxxxC}{\ensuremath{}}              
\newcommand{\hatcurLCrprstarnoisorestrictempiricalxxxxxC}{\ensuremath{0.1115\pm0.0019}} 
\newcommand{\hatcurLCbsqempiricalxxxxxC}{\ensuremath{}}                  
\newcommand{\hatcurLCbsqnoisorestrictempiricalxxxxxC}{\ensuremath{0.235_{-0.048}^{+0.025}}} 
\newcommand{\hatcurLCimpempiricalxxxxxC}{\ensuremath{}}                  
\newcommand{\hatcurLCimpnoisorestrictempiricalxxxxxC}{\ensuremath{0.485_{-0.053}^{+0.025}}} 
\newcommand{\hatcurLCzetaempiricalxxxxxC}{\ensuremath{}}                 
\newcommand{\hatcurLCzetanoisorestrictempiricalxxxxxC}{\ensuremath{22.76\pm0.28}} 
\newcommand{\hatcurLCdurempiricalxxxxxC}{\ensuremath{}}                  
\newcommand{\hatcurLCdurnoisorestrictempiricalxxxxxC}{\ensuremath{0.1005\pm0.0013}} 
\newcommand{\hatcurLCdurshortempiricalxxxxxC}{\ensuremath{}}             
\newcommand{\hatcurLCdurshortnoisorestrictempiricalxxxxxC}{\ensuremath{0.1005}} 
\newcommand{\hatcurLCdurhrempiricalxxxxxC}{\ensuremath{}}                
\newcommand{\hatcurLCdurhrnoisorestrictempiricalxxxxxC}{\ensuremath{2.411\pm0.030}} 
\newcommand{\hatcurLCdurhrshortempiricalxxxxxC}{\ensuremath{}}           
\newcommand{\hatcurLCdurhrshortnoisorestrictempiricalxxxxxC}{\ensuremath{2.411}} 
\newcommand{\hatcurLCqempiricalxxxxxC}{\ensuremath{}}                    
\newcommand{\hatcurLCqnoisorestrictempiricalxxxxxC}{\ensuremath{0.02420\pm0.00031}} 
\newcommand{\hatcurLCqshortempiricalxxxxxC}{\ensuremath{}}               
\newcommand{\hatcurLCqshortnoisorestrictempiricalxxxxxC}{\ensuremath{0.024}} 
\newcommand{\hatcurLCingdurempiricalxxxxxC}{\ensuremath{}}               
\newcommand{\hatcurLCingdurnoisorestrictempiricalxxxxxC}{\ensuremath{0.01280\pm0.00086}} 
\newcommand{\hatcurLCPempiricalxxxxxC}{\ensuremath{}}                    
\newcommand{\hatcurLCPnoisorestrictempiricalxxxxxC}{\ensuremath{4.1480419\pm0.0000043}} 
\newcommand{\hatcurLCPprecempiricalxxxxxC}{\ensuremath{}}                
\newcommand{\hatcurLCPprecnoisorestrictempiricalxxxxxC}{\ensuremath{4.1480419}} 
\newcommand{\hatcurLCPshortempiricalxxxxxC}{\ensuremath{}}               
\newcommand{\hatcurLCPshortnoisorestrictempiricalxxxxxC}{\ensuremath{4.1480}} 
\newcommand{\hatcurLCTempiricalxxxxxC}{\ensuremath{}}                    
\newcommand{\hatcurLCTnoisorestrictempiricalxxxxxC}{\ensuremath{2457080.27658\pm0.00064}} 
\newcommand{\hatcurLCTAempiricalxxxxxC}{\ensuremath{}}                   
\newcommand{\hatcurLCTAnoisorestrictempiricalxxxxxC}{\ensuremath{2455761.1991\pm0.0016}} 
\newcommand{\hatcurLCTBempiricalxxxxxC}{\ensuremath{}}                   
\newcommand{\hatcurLCTBnoisorestrictempiricalxxxxxC}{\ensuremath{2457283.53064\pm0.00064}} 
\newcommand{\hatcurLChatnetmAempiricalxxxxxC}{\ensuremath{}}             
\newcommand{\hatcurLChatnetmAnoisorestrictempiricalxxxxxC}{\ensuremath{14.42630\pm0.00018}} 
\newcommand{\hatcurLCiblendAempiricalxxxxxC}{\ensuremath{}}              
\newcommand{\hatcurLCiblendAnoisorestrictempiricalxxxxxC}{\ensuremath{0.823\pm0.050}} 
\newcommand{\hatcurLChatnetmBempiricalxxxxxC}{\ensuremath{}}             
\newcommand{\hatcurLChatnetmBnoisorestrictempiricalxxxxxC}{\ensuremath{14.42650\pm0.00018}} 
\newcommand{\hatcurLCiblendBempiricalxxxxxC}{\ensuremath{}}              
\newcommand{\hatcurLCiblendBnoisorestrictempiricalxxxxxC}{\ensuremath{0.888\pm0.034}} 
\newcommand{\hatcurLChatnetmCempiricalxxxxxC}{\ensuremath{}}             
\newcommand{\hatcurLChatnetmCnoisorestrictempiricalxxxxxC}{\ensuremath{0.00005\pm0.00011}} 
\newcommand{\hatcurLCiblendCempiricalxxxxxC}{\ensuremath{}}              
\newcommand{\hatcurLCiblendCnoisorestrictempiricalxxxxxC}{\ensuremath{0.685\pm0.064}} 
\newcommand{\hatcurLCrhoempiricalxxxxxC}{\ensuremath{}}                  
\newcommand{\hatcurLCrhonoisorestrictempiricalxxxxxC}{\ensuremath{2.49_{-0.10}^{+0.32}}} 
\newcommand{\hatcurSMEiteffempiricalxxxxxC}{\ensuremath{4354\pm70}}      
\newcommand{\hatcurSMEizfehempiricalxxxxxC}{\ensuremath{0.080\pm0.084}}  
\newcommand{\hatcurSMEizfehshortempiricalxxxxxC}{\ensuremath{0.08}}      
\newcommand{\hatcurSMEiloggempiricalxxxxxC}{\ensuremath{4.55\pm0.19}}    
\newcommand{\hatcurSMEivsinempiricalxxxxxC}{\ensuremath{0.50\pm0.78}}    
\newcommand{\hatcurSMEivmacempiricalxxxxxC}{\ensuremath{nff\pmnff}}      
\newcommand{\hatcurSMEivmicempiricalxxxxxC}{\ensuremath{nff\pmnff}}      
\newcommand{\hatcurextraerrMgempiricalxxxxxC}{\ensuremath{}}             
\newcommand{\hatcurextraerrMgnoisorestrictempiricalxxxxxC}{\ensuremath{0.018_{-0.017}^{+0.036}}} 
\newcommand{\hatcurextraerrMgtwosiglimempiricalxxxxxC}{\ensuremath{}}    
\newcommand{\hatcurextraerrMgtwosiglimnoisorestrictempiricalxxxxxC}{\ensuremath{<0.0742}} 
\newcommand{\hatcurextraerrMrempiricalxxxxxC}{\ensuremath{}}             
\newcommand{\hatcurextraerrMrnoisorestrictempiricalxxxxxC}{\ensuremath{0.237\pm0.074}} 
\newcommand{\hatcurextraerrMrtwosiglimempiricalxxxxxC}{\ensuremath{}}    
\newcommand{\hatcurextraerrMrtwosiglimnoisorestrictempiricalxxxxxC}{\ensuremath{<0.3740}} 
\newcommand{\hatcurextraerrMiempiricalxxxxxC}{\ensuremath{}}             
\newcommand{\hatcurextraerrMinoisorestrictempiricalxxxxxC}{\ensuremath{0.052_{-0.051}^{+0.142}}} 
\newcommand{\hatcurextraerrMitwosiglimempiricalxxxxxC}{\ensuremath{}}    
\newcommand{\hatcurextraerrMitwosiglimnoisorestrictempiricalxxxxxC}{\ensuremath{<0.2795}} 
\newcommand{\hatcurextraerrMGempiricalxxxxxC}{\ensuremath{}}             
\newcommand{\hatcurextraerrMGnoisorestrictempiricalxxxxxC}{\ensuremath{0.0256\pm0.0094}} 
\newcommand{\hatcurextraerrMGtwosiglimempiricalxxxxxC}{\ensuremath{}}    
\newcommand{\hatcurextraerrMGtwosiglimnoisorestrictempiricalxxxxxC}{\ensuremath{<0.0417}} 
\newcommand{\hatcurextraerrMBPtwoempiricalxxxxxC}{\ensuremath{}}         
\newcommand{\hatcurextraerrMBPtwonoisorestrictempiricalxxxxxC}{\ensuremath{0.016\pm0.012}} 
\newcommand{\hatcurextraerrMBPtwotwosiglimempiricalxxxxxC}{\ensuremath{}} 
\newcommand{\hatcurextraerrMBPtwotwosiglimnoisorestrictempiricalxxxxxC}{\ensuremath{<0.0372}} 
\newcommand{\hatcurextraerrMRPempiricalxxxxxC}{\ensuremath{}}            
\newcommand{\hatcurextraerrMRPnoisorestrictempiricalxxxxxC}{\ensuremath{0.0055_{-0.0052}^{+0.0119}}} 
\newcommand{\hatcurextraerrMRPtwosiglimempiricalxxxxxC}{\ensuremath{}}   
\newcommand{\hatcurextraerrMRPtwosiglimnoisorestrictempiricalxxxxxC}{\ensuremath{<0.0232}} 
\newcommand{\hatcurextraerrMJempiricalxxxxxC}{\ensuremath{}}             
\newcommand{\hatcurextraerrMJnoisorestrictempiricalxxxxxC}{\ensuremath{0.00079_{-0.00075}^{+0.00598}}} 
\newcommand{\hatcurextraerrMJtwosiglimempiricalxxxxxC}{\ensuremath{}}    
\newcommand{\hatcurextraerrMJtwosiglimnoisorestrictempiricalxxxxxC}{\ensuremath{<0.0177}} 
\newcommand{\hatcurextraerrMHempiricalxxxxxC}{\ensuremath{}}             
\newcommand{\hatcurextraerrMHnoisorestrictempiricalxxxxxC}{\ensuremath{0.0051_{-0.0049}^{+0.0141}}} 
\newcommand{\hatcurextraerrMHtwosiglimempiricalxxxxxC}{\ensuremath{}}    
\newcommand{\hatcurextraerrMHtwosiglimnoisorestrictempiricalxxxxxC}{\ensuremath{<0.0272}} 
\newcommand{\hatcurextraerrMKsempiricalxxxxxC}{\ensuremath{}}            
\newcommand{\hatcurextraerrMKsnoisorestrictempiricalxxxxxC}{\ensuremath{0.030\pm0.024}} 
\newcommand{\hatcurextraerrMKstwosiglimempiricalxxxxxC}{\ensuremath{}}   
\newcommand{\hatcurextraerrMKstwosiglimnoisorestrictempiricalxxxxxC}{\ensuremath{<0.0683}} 
\newcommand{\hatcurextraerrMWoneempiricalxxxxxC}{\ensuremath{}}          
\newcommand{\hatcurextraerrMWonenoisorestrictempiricalxxxxxC}{\ensuremath{0.012_{-0.011}^{+0.015}}} 
\newcommand{\hatcurextraerrMWonetwosiglimempiricalxxxxxC}{\ensuremath{}} 
\newcommand{\hatcurextraerrMWonetwosiglimnoisorestrictempiricalxxxxxC}{\ensuremath{<0.0402}} 
\newcommand{\hatcurextraerrMWtwoempiricalxxxxxC}{\ensuremath{}}          
\newcommand{\hatcurextraerrMWtwonoisorestrictempiricalxxxxxC}{\ensuremath{0.096\pm0.065}} 
\newcommand{\hatcurextraerrMWtwotwosiglimempiricalxxxxxC}{\ensuremath{}} 
\newcommand{\hatcurextraerrMWtwotwosiglimnoisorestrictempiricalxxxxxC}{\ensuremath{<0.2051}} 
\newcommand{\hatcurLBiBempiricalxxxxxC}{\ensuremath{0.9623}}             
\newcommand{\hatcurLBiiBempiricalxxxxxC}{\ensuremath{-0.1066}}           
\newcommand{\hatcurLBiVempiricalxxxxxC}{\ensuremath{0.7587}}             
\newcommand{\hatcurLBiiVempiricalxxxxxC}{\ensuremath{0.0458}}            
\newcommand{\hatcurLBiRempiricalxxxxxC}{\ensuremath{0.6243}}             
\newcommand{\hatcurLBiiRempiricalxxxxxC}{\ensuremath{0.1118}}            
\newcommand{\hatcurLBiIempiricalxxxxxC}{\ensuremath{0.4811}}             
\newcommand{\hatcurLBiiIempiricalxxxxxC}{\ensuremath{0.1553}}            
\newcommand{\hatcurLBiuempiricalxxxxxC}{\ensuremath{1.0925}}             
\newcommand{\hatcurLBiiuempiricalxxxxxC}{\ensuremath{-0.2541}}           
\newcommand{\hatcurLBigempiricalxxxxxC}{\ensuremath{0.8657}}             
\newcommand{\hatcurLBiigempiricalxxxxxC}{\ensuremath{-0.0410}}           
\newcommand{\hatcurLBirempiricalxxxxxC}{\ensuremath{}}                   
\newcommand{\hatcurLBirnoisorestrictempiricalxxxxxC}{\ensuremath{0.555\pm0.098}} 
\newcommand{\hatcurLBiirempiricalxxxxxC}{\ensuremath{}}                  
\newcommand{\hatcurLBiirnoisorestrictempiricalxxxxxC}{\ensuremath{0.37\pm0.11}} 
\newcommand{\hatcurLBiiempiricalxxxxxC}{\ensuremath{}}                   
\newcommand{\hatcurLBiinoisorestrictempiricalxxxxxC}{\ensuremath{0.53_{-0.13}^{+0.10}}} 
\newcommand{\hatcurLBiiiempiricalxxxxxC}{\ensuremath{}}                  
\newcommand{\hatcurLBiiinoisorestrictempiricalxxxxxC}{\ensuremath{0.30_{-0.12}^{+0.16}}} 
\newcommand{\hatcurLBizempiricalxxxxxC}{\ensuremath{0.4191}}             
\newcommand{\hatcurLBiizempiricalxxxxxC}{\ensuremath{0.1681}}            
\newcommand{\hatcurLBiJempiricalxxxxxC}{\ensuremath{0.2967}}             
\newcommand{\hatcurLBiiJempiricalxxxxxC}{\ensuremath{0.2133}}            
\newcommand{\hatcurLBiHempiricalxxxxxC}{\ensuremath{0.1664}}             
\newcommand{\hatcurLBiiHempiricalxxxxxC}{\ensuremath{0.2985}}            
\newcommand{\hatcurLBiKempiricalxxxxxC}{\ensuremath{0.1330}}             
\newcommand{\hatcurLBiiKempiricalxxxxxC}{\ensuremath{0.2539}}            
\newcommand{\hatcurLBiTempiricalxxxxxC}{\ensuremath{}}                   
\newcommand{\hatcurLBiTnoisorestrictempiricalxxxxxC}{\ensuremath{0.56\pm0.11}} 
\newcommand{\hatcurLBiiTempiricalxxxxxC}{\ensuremath{}}                  
\newcommand{\hatcurLBiiTnoisorestrictempiricalxxxxxC}{\ensuremath{0.28\pm0.12}} 
\newcommand{\hatcurLBikepempiricalxxxxxC}{\ensuremath{0.6246}}           
\newcommand{\hatcurLBiikepempiricalxxxxxC}{\ensuremath{0.1226}}          
\newcommand{\hatcurLBiCempiricalxxxxxC}{\ensuremath{0.5961}}             
\newcommand{\hatcurLBiiCempiricalxxxxxC}{\ensuremath{0.1316}}            
\newcommand{\hatcurLBiMempiricalxxxxxC}{\ensuremath{0.7290}}             
\newcommand{\hatcurLBiiMempiricalxxxxxC}{\ensuremath{0.0691}}            
\newcommand{\hatcurLBiSoneempiricalxxxxxC}{\ensuremath{0.1108}}            
\newcommand{\hatcurLBiiSoneempiricalxxxxxC}{\ensuremath{0.1733}}           
\newcommand{\hatcurLBiStwoempiricalxxxxxC}{\ensuremath{0.0986}}            
\newcommand{\hatcurLBiiStwoempiricalxxxxxC}{\ensuremath{0.1376}}           
\newcommand{\hatcurLBiSthreeempiricalxxxxxC}{\ensuremath{0.0828}}            
\newcommand{\hatcurLBiiSthreeempiricalxxxxxC}{\ensuremath{0.1135}}           
\newcommand{\hatcurLBiSfourempiricalxxxxxC}{\ensuremath{0.0731}}            
\newcommand{\hatcurLBiiSfourempiricalxxxxxC}{\ensuremath{0.0967}}           
\newcommand{\hatcurISOmempiricalxxxxxC}{\ensuremath{}}                   
\newcommand{\hatcurISOmnoisorestrictempiricalxxxxxC}{\ensuremath{0.587_{-0.026}^{+0.089}}} 
\newcommand{\hatcurISOmshortempiricalxxxxxC}{\ensuremath{}}              
\newcommand{\hatcurISOmshortnoisorestrictempiricalxxxxxC}{\ensuremath{0.59}} 
\newcommand{\hatcurISOmlongempiricalxxxxxC}{\ensuremath{}}               
\newcommand{\hatcurISOmlongnoisorestrictempiricalxxxxxC}{\ensuremath{0.587_{-0.026}^{+0.089}}} 
\newcommand{\hatcurISOrempiricalxxxxxC}{\ensuremath{}}                   
\newcommand{\hatcurISOrnoisorestrictempiricalxxxxxC}{\ensuremath{0.6935\pm0.0080}} 
\newcommand{\hatcurISOrshortempiricalxxxxxC}{\ensuremath{}}              
\newcommand{\hatcurISOrshortnoisorestrictempiricalxxxxxC}{\ensuremath{0.69}} 
\newcommand{\hatcurISOrlongempiricalxxxxxC}{\ensuremath{}}               
\newcommand{\hatcurISOrlongnoisorestrictempiricalxxxxxC}{\ensuremath{0.6935\pm0.0080}} 
\newcommand{\hatcurISOloggempiricalxxxxxC}{\ensuremath{}}                
\newcommand{\hatcurISOloggnoisorestrictempiricalxxxxxC}{\ensuremath{4.524\pm0.041}} 
\newcommand{\hatcurISOlumempiricalxxxxxC}{\ensuremath{}}                 
\newcommand{\hatcurISOlumnoisorestrictempiricalxxxxxC}{\ensuremath{0.1699\pm0.0075}} 
\newcommand{\hatcurISOlumshortempiricalxxxxxC}{\ensuremath{}}            
\newcommand{\hatcurISOlumshortnoisorestrictempiricalxxxxxC}{\ensuremath{0.17}} 
\newcommand{\hatcurISOfehempiricalxxxxxC}{\ensuremath{}}                 
\newcommand{\hatcurISOfehnoisorestrictempiricalxxxxxC}{\ensuremath{0.093\pm0.066}} 
\newcommand{\hatcurISOteffempiricalxxxxxC}{\ensuremath{}}                
\newcommand{\hatcurISOteffnoisorestrictempiricalxxxxxC}{\ensuremath{4453\pm39}} 
\newcommand{\hatcurISOageempiricalxxxxxC}{\ensuremath{}}                 
\newcommand{\hatcurISOagenoisorestrictempiricalxxxxxC}{\ensuremath{0\pm0}} 
\newcommand{\hatcurISOspecempiricalxxxxxC}{K}                            
\newcommand{\hatcurRVKempiricalxxxxxC}{\ensuremath{}}                    
\newcommand{\hatcurRVKnoisorestrictempiricalxxxxxC}{\ensuremath{51.2\pm2.6}} 
\newcommand{\hatcurRVrkempiricalxxxxxC}{\ensuremath{}}                   
\newcommand{\hatcurRVrknoisorestrictempiricalxxxxxC}{\ensuremath{0\pm0}} 
\newcommand{\hatcurRVrhempiricalxxxxxC}{\ensuremath{}}                   
\newcommand{\hatcurRVrhnoisorestrictempiricalxxxxxC}{\ensuremath{0\pm0}} 
\newcommand{\hatcurRVkempiricalxxxxxC}{\ensuremath{}}                    
\newcommand{\hatcurRVknoisorestrictempiricalxxxxxC}{\ensuremath{0\pm0}}  
\newcommand{\hatcurRVhempiricalxxxxxC}{\ensuremath{}}                    
\newcommand{\hatcurRVhnoisorestrictempiricalxxxxxC}{\ensuremath{0\pm0}}  
\newcommand{\hatcurRVtroneempiricalxxxxxC}{\ensuremath{}}                
\newcommand{\hatcurRVtronenoisorestrictempiricalxxxxxC}{\ensuremath{0\pm0}} 
\newcommand{\hatcurRVtrtwoempiricalxxxxxC}{\ensuremath{}}                
\newcommand{\hatcurRVtrtwonoisorestrictempiricalxxxxxC}{\ensuremath{0\pm0}} 
\newcommand{\hatcurRVgammaempiricalxxxxxC}{\ensuremath{}}                
\newcommand{\hatcurRVgammanoisorestrictempiricalxxxxxC}{\ensuremath{-20.0\pm5.4}} 
\newcommand{\hatcurRVjitterempiricalxxxxxC}{\ensuremath{}}               
\newcommand{\hatcurRVjitternoisorestrictempiricalxxxxxC}{\ensuremath{27\pm10}} 
\newcommand{\hatcurRVjittertwosiglimempiricalxxxxxC}{\ensuremath{}}      
\newcommand{\hatcurRVjittertwosiglimnoisorestrictempiricalxxxxxC}{\ensuremath{<46.6}} 
\newcommand{\hatcurRVfitrmsempiricalxxxxxC}{\ensuremath{.1fym}}          %
\newcommand{\hatcurRVeccenempiricalxxxxxC}{\ensuremath{}}                
\newcommand{\hatcurRVeccennoisorestrictempiricalxxxxxC}{\ensuremath{0\pm0}} 
\newcommand{\hatcurRVeccentwosiglimempiricalxxxxxC}{\ensuremath{}}       
\newcommand{\hatcurRVeccentwosiglimnoisorestrictempiricalxxxxxC}{\ensuremath{<0.000}} 
\newcommand{\hatcurRVomegaempiricalxxxxxC}{\ensuremath{}}                
\newcommand{\hatcurRVomeganoisorestrictempiricalxxxxxC}{\ensuremath{0\pm0}} 
\newcommand{\hatcurPPiempiricalxxxxxC}{\ensuremath{}}                    
\newcommand{\hatcurPPinoisorestrictempiricalxxxxxC}{\ensuremath{87.89_{-0.14}^{+0.29}}} 
\newcommand{\hatcurPPgempiricalxxxxxC}{\ensuremath{}}                    
\newcommand{\hatcurPPgnoisorestrictempiricalxxxxxC}{\ensuremath{12.54\pm0.91}} 
\newcommand{\hatcurPPloggempiricalxxxxxC}{\ensuremath{}}                 
\newcommand{\hatcurPPloggnoisorestrictempiricalxxxxxC}{\ensuremath{3.098\pm0.031}} 
\newcommand{\hatcurPParempiricalxxxxxC}{\ensuremath{}}                   
\newcommand{\hatcurPParnoisorestrictempiricalxxxxxC}{\ensuremath{13.13_{-0.18}^{+0.54}}} 
\newcommand{\hatcurPParelempiricalxxxxxC}{\ensuremath{}}                 
\newcommand{\hatcurPParelnoisorestrictempiricalxxxxxC}{\ensuremath{0.04232_{-0.00064}^{+0.00204}}} 
\newcommand{\hatcurPPrhoempiricalxxxxxC}{\ensuremath{}}                  
\newcommand{\hatcurPPrhonoisorestrictempiricalxxxxxC}{\ensuremath{0.835\pm0.070}} 
\newcommand{\hatcurPPmempiricalxxxxxC}{\ensuremath{}}                    
\newcommand{\hatcurPPmnoisorestrictempiricalxxxxxC}{\ensuremath{0.286\pm0.020}} 
\newcommand{\hatcurPPmshortempiricalxxxxxC}{\ensuremath{}}               
\newcommand{\hatcurPPmshortnoisorestrictempiricalxxxxxC}{\ensuremath{0.29}} 
\newcommand{\hatcurPPmlongempiricalxxxxxC}{\ensuremath{}}                
\newcommand{\hatcurPPmlongnoisorestrictempiricalxxxxxC}{\ensuremath{0.286\pm0.020}} 
\newcommand{\hatcurPPmeempiricalxxxxxC}{\ensuremath{}}                   
\newcommand{\hatcurPPmenoisorestrictempiricalxxxxxC}{\ensuremath{90.9\pm6.4}} 
\newcommand{\hatcurPPmeshortempiricalxxxxxC}{\ensuremath{}}              
\newcommand{\hatcurPPmeshortnoisorestrictempiricalxxxxxC}{\ensuremath{90.9}} 
\newcommand{\hatcurPPmelongempiricalxxxxxC}{\ensuremath{}}               
\newcommand{\hatcurPPmelongnoisorestrictempiricalxxxxxC}{\ensuremath{90.9\pm6.4}} 
\newcommand{\hatcurPPrempiricalxxxxxC}{\ensuremath{}}                    
\newcommand{\hatcurPPrnoisorestrictempiricalxxxxxC}{\ensuremath{0.753\pm0.016}} 
\newcommand{\hatcurPPrshortempiricalxxxxxC}{\ensuremath{}}               
\newcommand{\hatcurPPrshortnoisorestrictempiricalxxxxxC}{\ensuremath{0.75}} 
\newcommand{\hatcurPPrlongempiricalxxxxxC}{\ensuremath{}}                
\newcommand{\hatcurPPrlongnoisorestrictempiricalxxxxxC}{\ensuremath{0.753\pm0.016}} 
\newcommand{\hatcurPPreempiricalxxxxxC}{\ensuremath{}}                   
\newcommand{\hatcurPPrenoisorestrictempiricalxxxxxC}{\ensuremath{8.44\pm0.18}} 
\newcommand{\hatcurPPreshortempiricalxxxxxC}{\ensuremath{}}              
\newcommand{\hatcurPPreshortnoisorestrictempiricalxxxxxC}{\ensuremath{8.4}} 
\newcommand{\hatcurPPrelongempiricalxxxxxC}{\ensuremath{}}               
\newcommand{\hatcurPPrelongnoisorestrictempiricalxxxxxC}{\ensuremath{8.44\pm0.18}} 
\newcommand{\hatcurPPmrcorrempiricalxxxxxC}{\ensuremath{0.00}}           
\newcommand{\hatcurPPteffempiricalxxxxxC}{\ensuremath{}}                 
\newcommand{\hatcurPPteffnoisorestrictempiricalxxxxxC}{\ensuremath{869.5_{-14.8}^{+6.5}}} 
\newcommand{\hatcurPPthetaempiricalxxxxxC}{\ensuremath{}}                
\newcommand{\hatcurPPthetanoisorestrictempiricalxxxxxC}{\ensuremath{0.0540\pm0.0027}} 
\newcommand{\hatcurPPfluxperiempiricalxxxxxC}{\ensuremath{}}             
\newcommand{\hatcurPPfluxperinoisorestrictempiricalxxxxxC}{\ensuremath{}} 
\newcommand{\hatcurPPfluxperidimempiricalxxxxxC}{\ensuremath{2d}}        
\newcommand{\hatcurPPfluxapempiricalxxxxxC}{\ensuremath{}}               
\newcommand{\hatcurPPfluxapnoisorestrictempiricalxxxxxC}{\ensuremath{}}  
\newcommand{\hatcurPPfluxapdimempiricalxxxxxC}{\ensuremath{2d}}          
\newcommand{\hatcurPPfluxavgempiricalxxxxxC}{\ensuremath{}}              
\newcommand{\hatcurPPfluxavgnoisorestrictempiricalxxxxxC}{\ensuremath{}} 
\newcommand{\hatcurPPfluxavgdimempiricalxxxxxC}{\ensuremath{2d}}         
\newcommand{\hatcurPPfluxavglogempiricalxxxxxC}{\ensuremath{}}           
\newcommand{\hatcurPPfluxavglognoisorestrictempiricalxxxxxC}{\ensuremath{8.113_{-0.030}^{+0.013}}} 
\newcommand{\hatcurXsecphaseempiricalxxxxxC}{\ensuremath{}}              
\newcommand{\hatcurXsecphasenoisorestrictempiricalxxxxxC}{\ensuremath{0\pm0}} 
\newcommand{\hatcurXsecondaryempiricalxxxxxC}{\ensuremath{}}             
\newcommand{\hatcurXsecondarynoisorestrictempiricalxxxxxC}{\ensuremath{2457082.35060\pm0.00064}} 
\newcommand{\hatcurXsecdurempiricalxxxxxC}{\ensuremath{}}                
\newcommand{\hatcurXsecdurnoisorestrictempiricalxxxxxC}{\ensuremath{0.1005\pm0.0013}} 
\newcommand{\hatcurXsecingdurempiricalxxxxxC}{\ensuremath{}}             
\newcommand{\hatcurXsecingdurnoisorestrictempiricalxxxxxC}{\ensuremath{0.01280\pm0.00086}} 
\newcommand{\hatcurPPphiconjempiricalxxxxxC}{\ensuremath{}}              
\newcommand{\hatcurPPphiconjnoisorestrictempiricalxxxxxC}{\ensuremath{0\pm0}} 
\newcommand{\hatcurPPperiempiricalxxxxxC}{\ensuremath{}}                 
\newcommand{\hatcurPPperinoisorestrictempiricalxxxxxC}{\ensuremath{2457079.23957\pm0.00065}} 
\newcommand{\hatcurPPaequivempiricalxxxxxC}{\ensuremath{}}               
\newcommand{\hatcurPPaequivnoisorestrictempiricalxxxxxC}{\ensuremath{0.1025_{-0.0016}^{+0.0035}}} 
\newcommand{\hatcurPPtcircempiricalxxxxxC}{\ensuremath{}}                
\newcommand{\hatcurPPtcircnoisorestrictempiricalxxxxxC}{\ensuremath{1220_{-140}^{+200}}} 
\newcommand{\hatcurPPtinfallempiricalxxxxxC}{\ensuremath{}}              
\newcommand{\hatcurPPtinfallnoisorestrictempiricalxxxxxC}{\ensuremath{31600_{-2900}^{+10000}}} 
\newcommand{\hatcurXdistempiricalxxxxxC}{\ensuremath{}}                  
\newcommand{\hatcurXdistnoisorestrictempiricalxxxxxC}{\ensuremath{327.2\pm1.7}} 
\newcommand{\hatcurXAvempiricalxxxxxC}{\ensuremath{}}                    
\newcommand{\hatcurXAvnoisorestrictempiricalxxxxxC}{\ensuremath{0.057_{-0.011}^{+0.015}}} 
\newcommand{\hatcurXdistredempiricalxxxxxC}{\ensuremath{}}               
\newcommand{\hatcurXdistrednoisorestrictempiricalxxxxxC}{\ensuremath{327.2\pm1.7}} 
\newcommand{\hatcurXEBVempiricalxxxxxC}{\ensuremath{}}                   
\newcommand{\hatcurXEBVnoisorestrictempiricalxxxxxC}{\ensuremath{0.0180_{-0.0030}^{+0.0050}}} 
\newcommand{\hatcurCCpmraempiricalxxxxxC}{\ensuremath{42.581\pm0.035}}   
\newcommand{\hatcurCCpmdecempiricalxxxxxC}{\ensuremath{8.264\pm0.030}}   
\newcommand{\hatcurCCpmempiricalxxxxxC}{\ensuremath{43.376\pm0.046}}     
\newcommand{\hatcurhtrempiricalxxxxxD}{HATS537-014}                      
\newcommand{\hatcurfieldempiricalxxxxxD}{\ensuremath{string}}            
\newcommand{\hatcurCCraempiricalxxxxxD}{\ensuremath{22^{\mathrm h}36^{\mathrm m}06.3190{\mathrm s}}}                   
\newcommand{\hatcurCCdecempiricalxxxxxD}{\ensuremath{-16{\arcdeg}59{\arcmin}59.7882{\arcsec}}}                 
\newcommand{\hatcurCCmagempiricalxxxxxD}{12.469}                         
\newcommand{\hatcurCCtwomassempiricalxxxxxD}{2MASS~22360631-1659597}     
\newcommand{\hatcurCCgscempiricalxxxxxD}{GSC~6386-00784}                 
\newcommand{\hatcurCCgaiaempiricalxxxxxD}{GAIA~2594869603582993792}      
\newcommand{\hatcurCCgaiadrtwoempiricalxxxxxD}{GAIA~DR2~2594869603582993792} 
\newcommand{\hatcurCCtassmvempiricalxxxxxD}{\ensuremath{12.469\pm0.010}} 
\newcommand{\hatcurCCtassmvshortempiricalxxxxxD}{\ensuremath{12.5}}      
\newcommand{\hatcurCCtassmBempiricalxxxxxD}{\ensuremath{13.572\pm0.030}} 
\newcommand{\hatcurCCtassmBshortempiricalxxxxxD}{\ensuremath{13.6}}      
\newcommand{\hatcurCCtassmIempiricalxxxxxD}{\ensuremath{nff\pmnff}}      
\newcommand{\hatcurCCtassmIshortempiricalxxxxxD}{\ensuremath{0.0}}       
\newcommand{\hatcurCCtassmgempiricalxxxxxD}{\ensuremath{12.995\pm0.010}} 
\newcommand{\hatcurCCtassmgshortempiricalxxxxxD}{\ensuremath{13.0}}      
\newcommand{\hatcurCCtassmrempiricalxxxxxD}{\ensuremath{11.998\pm0.010}} 
\newcommand{\hatcurCCtassmrshortempiricalxxxxxD}{\ensuremath{12.0}}      
\newcommand{\hatcurCCtassmiempiricalxxxxxD}{\ensuremath{11.622\pm0.030}} 
\newcommand{\hatcurCCtassmishortempiricalxxxxxD}{\ensuremath{11.6}}      
\newcommand{\hatcurCCparallaxempiricalxxxxxD}{\ensuremath{7.809\pm0.037}} 
\newcommand{\hatcurCCgaiamGempiricalxxxxxD}{\ensuremath{12.07250\pm0.00030}} 
\newcommand{\hatcurCCgaiamBPempiricalxxxxxD}{\ensuremath{12.7084\pm0.0018}} 
\newcommand{\hatcurCCgaiamRPempiricalxxxxxD}{\ensuremath{11.3341\pm0.0010}} 
\newcommand{\hatcurCCtwomassJmagempiricalxxxxxD}{\ensuremath{10.424\pm0.023}} 
\newcommand{\hatcurCCtwomassHmagempiricalxxxxxD}{\ensuremath{9.907\pm0.026}} 
\newcommand{\hatcurCCtwomassKmagempiricalxxxxxD}{\ensuremath{9.764\pm0.021}} 
\newcommand{\hatcurCCcitJmagempiricalxxxxxD}{\ensuremath{10.425\pm0.024}} 
\newcommand{\hatcurCCcitHmagempiricalxxxxxD}{\ensuremath{9.900\pm0.027}} 
\newcommand{\hatcurCCcitKmagempiricalxxxxxD}{\ensuremath{9.788\pm0.022}} 
\newcommand{\hatcurCCbbJmagempiricalxxxxxD}{\ensuremath{10.498\pm0.026}} 
\newcommand{\hatcurCCbbHmagempiricalxxxxxD}{\ensuremath{9.923\pm0.027}}  
\newcommand{\hatcurCCbbKmagempiricalxxxxxD}{\ensuremath{9.808\pm0.022}}  
\newcommand{\hatcurCCesoJmagempiricalxxxxxD}{\ensuremath{10.504\pm0.029}} 
\newcommand{\hatcurCCesoHmagempiricalxxxxxD}{\ensuremath{9.920\pm0.034}} 
\newcommand{\hatcurCCesoKmagempiricalxxxxxD}{\ensuremath{9.806\pm0.023}} 
\newcommand{\hatcurCCesoJHmagempiricalxxxxxD}{\ensuremath{0.584\pm0.042}} 
\newcommand{\hatcurCCesoJKmagempiricalxxxxxD}{\ensuremath{0.699\pm0.035}} 
\newcommand{\hatcurCCesoHKmagempiricalxxxxxD}{\ensuremath{0.115\pm0.040}} 
\newcommand{\hatcurLCdipempiricalxxxxxD}{\ensuremath{13.4}}              
\newcommand{\hatcurLCrprstarempiricalxxxxxD}{\ensuremath{}}              
\newcommand{\hatcurLCrprstarnoisorestrictempiricalxxxxxD}{\ensuremath{0.10392\pm0.00050}} 
\newcommand{\hatcurLCbsqempiricalxxxxxD}{\ensuremath{}}                  
\newcommand{\hatcurLCbsqnoisorestrictempiricalxxxxxD}{\ensuremath{0.1206_{-0.0096}^{+0.0115}}} 
\newcommand{\hatcurLCimpempiricalxxxxxD}{\ensuremath{}}                  
\newcommand{\hatcurLCimpnoisorestrictempiricalxxxxxD}{\ensuremath{0.347_{-0.014}^{+0.016}}} 
\newcommand{\hatcurLCzetaempiricalxxxxxD}{\ensuremath{}}                 
\newcommand{\hatcurLCzetanoisorestrictempiricalxxxxxD}{\ensuremath{17.444\pm0.098}} 
\newcommand{\hatcurLCdurempiricalxxxxxD}{\ensuremath{}}                  
\newcommand{\hatcurLCdurnoisorestrictempiricalxxxxxD}{\ensuremath{0.12817\pm0.00076}} 
\newcommand{\hatcurLCdurshortempiricalxxxxxD}{\ensuremath{}}             
\newcommand{\hatcurLCdurshortnoisorestrictempiricalxxxxxD}{\ensuremath{0.1282}} 
\newcommand{\hatcurLCdurhrempiricalxxxxxD}{\ensuremath{}}                
\newcommand{\hatcurLCdurhrnoisorestrictempiricalxxxxxD}{\ensuremath{3.076\pm0.018}} 
\newcommand{\hatcurLCdurhrshortempiricalxxxxxD}{\ensuremath{}}           
\newcommand{\hatcurLCdurhrshortnoisorestrictempiricalxxxxxD}{\ensuremath{3.076}} 
\newcommand{\hatcurLCqempiricalxxxxxD}{\ensuremath{}}                    
\newcommand{\hatcurLCqnoisorestrictempiricalxxxxxD}{\ensuremath{0.01750\pm0.00011}} 
\newcommand{\hatcurLCqshortempiricalxxxxxD}{\ensuremath{}}               
\newcommand{\hatcurLCqshortnoisorestrictempiricalxxxxxD}{\ensuremath{0.018}} 
\newcommand{\hatcurLCingdurempiricalxxxxxD}{\ensuremath{}}               
\newcommand{\hatcurLCingdurnoisorestrictempiricalxxxxxD}{\ensuremath{0.01357\pm0.00020}} 
\newcommand{\hatcurLCPempiricalxxxxxD}{\ensuremath{}}                    
\newcommand{\hatcurLCPnoisorestrictempiricalxxxxxD}{\ensuremath{7.3279460\pm0.0000029}} 
\newcommand{\hatcurLCPprecempiricalxxxxxD}{\ensuremath{}}                
\newcommand{\hatcurLCPprecnoisorestrictempiricalxxxxxD}{\ensuremath{7.3279460}} 
\newcommand{\hatcurLCPshortempiricalxxxxxD}{\ensuremath{}}               
\newcommand{\hatcurLCPshortnoisorestrictempiricalxxxxxD}{\ensuremath{7.3279}} 
\newcommand{\hatcurLCTempiricalxxxxxD}{\ensuremath{}}                    
\newcommand{\hatcurLCTnoisorestrictempiricalxxxxxD}{\ensuremath{2458051.00814\pm0.00015}} 
\newcommand{\hatcurLCTAempiricalxxxxxD}{\ensuremath{}}                   
\newcommand{\hatcurLCTAnoisorestrictempiricalxxxxxD}{\ensuremath{2457552.70780\pm0.00024}} 
\newcommand{\hatcurLCTBempiricalxxxxxD}{\ensuremath{}}                   
\newcommand{\hatcurLCTBnoisorestrictempiricalxxxxxD}{\ensuremath{2458380.76570\pm0.00020}} 
\newcommand{\hatcurLChatnetmAempiricalxxxxxD}{\ensuremath{}}             
\newcommand{\hatcurLChatnetmAnoisorestrictempiricalxxxxxD}{\ensuremath{12.049830\pm0.000052}} 
\newcommand{\hatcurLCiblendAempiricalxxxxxD}{\ensuremath{}}              
\newcommand{\hatcurLCiblendAnoisorestrictempiricalxxxxxD}{\ensuremath{0.903\pm0.027}} 
\newcommand{\hatcurLChatnetmBempiricalxxxxxD}{\ensuremath{}}             
\newcommand{\hatcurLChatnetmBnoisorestrictempiricalxxxxxD}{\ensuremath{11.628240\pm0.000020}} 
\newcommand{\hatcurLCiblendBempiricalxxxxxD}{\ensuremath{}}              
\newcommand{\hatcurLCiblendBnoisorestrictempiricalxxxxxD}{\ensuremath{0.9958\pm0.0017}} 
\newcommand{\hatcurLChatnetmCempiricalxxxxxD}{\ensuremath{}}             
\newcommand{\hatcurLChatnetmCnoisorestrictempiricalxxxxxD}{\ensuremath{0.000200\pm0.000091}} 
\newcommand{\hatcurLCiblendCempiricalxxxxxD}{\ensuremath{}}              
\newcommand{\hatcurLCiblendCnoisorestrictempiricalxxxxxD}{\ensuremath{0.987\pm0.015}} 
\newcommand{\hatcurLCrhoempiricalxxxxxD}{\ensuremath{}}                  
\newcommand{\hatcurLCrhonoisorestrictempiricalxxxxxD}{\ensuremath{2.441\pm0.065}} 
\newcommand{\hatcurSMEiteffempiricalxxxxxD}{\ensuremath{4612\pm76}}      
\newcommand{\hatcurSMEizfehempiricalxxxxxD}{\ensuremath{-0.040\pm0.050}} 
\newcommand{\hatcurSMEizfehshortempiricalxxxxxD}{\ensuremath{-0.04}}     
\newcommand{\hatcurSMEiloggempiricalxxxxxD}{\ensuremath{4.55\pm0.16}}    
\newcommand{\hatcurSMEivsinempiricalxxxxxD}{\ensuremath{0.8\pm1.3}}      
\newcommand{\hatcurSMEivmacempiricalxxxxxD}{\ensuremath{2.20\pm0.12}}    
\newcommand{\hatcurSMEivmicempiricalxxxxxD}{\ensuremath{0.443\pm0.061}}  
\newcommand{\hatcurextraerrMBempiricalxxxxxD}{\ensuremath{}}             
\newcommand{\hatcurextraerrMBnoisorestrictempiricalxxxxxD}{\ensuremath{0.0163\pm0.0043}} 
\newcommand{\hatcurextraerrMBtwosiglimempiricalxxxxxD}{\ensuremath{}}    
\newcommand{\hatcurextraerrMBtwosiglimnoisorestrictempiricalxxxxxD}{\ensuremath{<0.0230}} 
\newcommand{\hatcurextraerrMVempiricalxxxxxD}{\ensuremath{}}             
\newcommand{\hatcurextraerrMVnoisorestrictempiricalxxxxxD}{\ensuremath{0.0056_{-0.0053}^{+0.0112}}} 
\newcommand{\hatcurextraerrMVtwosiglimempiricalxxxxxD}{\ensuremath{}}    
\newcommand{\hatcurextraerrMVtwosiglimnoisorestrictempiricalxxxxxD}{\ensuremath{<0.0257}} 
\newcommand{\hatcurextraerrMgempiricalxxxxxD}{\ensuremath{}}             
\newcommand{\hatcurextraerrMgnoisorestrictempiricalxxxxxD}{\ensuremath{0.058\pm0.020}} 
\newcommand{\hatcurextraerrMgtwosiglimempiricalxxxxxD}{\ensuremath{}}    
\newcommand{\hatcurextraerrMgtwosiglimnoisorestrictempiricalxxxxxD}{\ensuremath{<0.0946}} 
\newcommand{\hatcurextraerrMrempiricalxxxxxD}{\ensuremath{}}             
\newcommand{\hatcurextraerrMrnoisorestrictempiricalxxxxxD}{\ensuremath{0.047\pm0.025}} 
\newcommand{\hatcurextraerrMrtwosiglimempiricalxxxxxD}{\ensuremath{}}    
\newcommand{\hatcurextraerrMrtwosiglimnoisorestrictempiricalxxxxxD}{\ensuremath{<0.0930}} 
\newcommand{\hatcurextraerrMiempiricalxxxxxD}{\ensuremath{}}             
\newcommand{\hatcurextraerrMinoisorestrictempiricalxxxxxD}{\ensuremath{0.080_{-0.049}^{+0.073}}} 
\newcommand{\hatcurextraerrMitwosiglimempiricalxxxxxD}{\ensuremath{}}    
\newcommand{\hatcurextraerrMitwosiglimnoisorestrictempiricalxxxxxD}{\ensuremath{<0.2158}} 
\newcommand{\hatcurextraerrMGempiricalxxxxxD}{\ensuremath{}}             
\newcommand{\hatcurextraerrMGnoisorestrictempiricalxxxxxD}{\ensuremath{0.0230_{-0.0091}^{+0.0146}}} 
\newcommand{\hatcurextraerrMGtwosiglimempiricalxxxxxD}{\ensuremath{}}    
\newcommand{\hatcurextraerrMGtwosiglimnoisorestrictempiricalxxxxxD}{\ensuremath{<0.0499}} 
\newcommand{\hatcurextraerrMBPtwoempiricalxxxxxD}{\ensuremath{}}         
\newcommand{\hatcurextraerrMBPtwonoisorestrictempiricalxxxxxD}{\ensuremath{0.00083_{-0.00081}^{+0.00775}}} 
\newcommand{\hatcurextraerrMBPtwotwosiglimempiricalxxxxxD}{\ensuremath{}} 
\newcommand{\hatcurextraerrMBPtwotwosiglimnoisorestrictempiricalxxxxxD}{\ensuremath{<0.0187}} 
\newcommand{\hatcurextraerrMRPempiricalxxxxxD}{\ensuremath{}}            
\newcommand{\hatcurextraerrMRPnoisorestrictempiricalxxxxxD}{\ensuremath{0.00012_{-0.00011}^{+0.00108}}} 
\newcommand{\hatcurextraerrMRPtwosiglimempiricalxxxxxD}{\ensuremath{}}   
\newcommand{\hatcurextraerrMRPtwosiglimnoisorestrictempiricalxxxxxD}{\ensuremath{<0.0026}} 
\newcommand{\hatcurextraerrMJempiricalxxxxxD}{\ensuremath{}}             
\newcommand{\hatcurextraerrMJnoisorestrictempiricalxxxxxD}{\ensuremath{0.00035_{-0.00035}^{+0.00701}}} 
\newcommand{\hatcurextraerrMJtwosiglimempiricalxxxxxD}{\ensuremath{}}    
\newcommand{\hatcurextraerrMJtwosiglimnoisorestrictempiricalxxxxxD}{\ensuremath{<0.0197}} 
\newcommand{\hatcurextraerrMHempiricalxxxxxD}{\ensuremath{}}             
\newcommand{\hatcurextraerrMHnoisorestrictempiricalxxxxxD}{\ensuremath{0.030\pm0.017}} 
\newcommand{\hatcurextraerrMHtwosiglimempiricalxxxxxD}{\ensuremath{}}    
\newcommand{\hatcurextraerrMHtwosiglimnoisorestrictempiricalxxxxxD}{\ensuremath{<0.0627}} 
\newcommand{\hatcurextraerrMKsempiricalxxxxxD}{\ensuremath{}}            
\newcommand{\hatcurextraerrMKsnoisorestrictempiricalxxxxxD}{\ensuremath{0.00087_{-0.00086}^{+0.00770}}} 
\newcommand{\hatcurextraerrMKstwosiglimempiricalxxxxxD}{\ensuremath{}}   
\newcommand{\hatcurextraerrMKstwosiglimnoisorestrictempiricalxxxxxD}{\ensuremath{<0.0171}} 
\newcommand{\hatcurextraerrMWoneempiricalxxxxxD}{\ensuremath{}}          
\newcommand{\hatcurextraerrMWonenoisorestrictempiricalxxxxxD}{\ensuremath{0.0011_{-0.0011}^{+0.0165}}} 
\newcommand{\hatcurextraerrMWonetwosiglimempiricalxxxxxD}{\ensuremath{}} 
\newcommand{\hatcurextraerrMWonetwosiglimnoisorestrictempiricalxxxxxD}{\ensuremath{<0.0312}} 
\newcommand{\hatcurextraerrMWtwoempiricalxxxxxD}{\ensuremath{}}          
\newcommand{\hatcurextraerrMWtwonoisorestrictempiricalxxxxxD}{\ensuremath{0.00015_{-0.00015}^{+0.00140}}} 
\newcommand{\hatcurextraerrMWtwotwosiglimempiricalxxxxxD}{\ensuremath{}} 
\newcommand{\hatcurextraerrMWtwotwosiglimnoisorestrictempiricalxxxxxD}{\ensuremath{<0.0032}} 
\newcommand{\hatcurextraerrMWthreeempiricalxxxxxD}{\ensuremath{}}        
\newcommand{\hatcurextraerrMWthreenoisorestrictempiricalxxxxxD}{\ensuremath{0.011_{-0.011}^{+0.055}}} 
\newcommand{\hatcurextraerrMWthreetwosiglimempiricalxxxxxD}{\ensuremath{}} 
\newcommand{\hatcurextraerrMWthreetwosiglimnoisorestrictempiricalxxxxxD}{\ensuremath{<0.1254}} 
\newcommand{\hatcurLBiBempiricalxxxxxD}{\ensuremath{0.9774}}             
\newcommand{\hatcurLBiiBempiricalxxxxxD}{\ensuremath{-0.1216}}           
\newcommand{\hatcurLBiVempiricalxxxxxD}{\ensuremath{0.7619}}             
\newcommand{\hatcurLBiiVempiricalxxxxxD}{\ensuremath{0.0363}}            
\newcommand{\hatcurLBiRempiricalxxxxxD}{\ensuremath{0.6278}}             
\newcommand{\hatcurLBiiRempiricalxxxxxD}{\ensuremath{0.0963}}            
\newcommand{\hatcurLBiIempiricalxxxxxD}{\ensuremath{}}                   
\newcommand{\hatcurLBiInoisorestrictempiricalxxxxxD}{\ensuremath{0.220\pm0.078}} 
\newcommand{\hatcurLBiiIempiricalxxxxxD}{\ensuremath{}}                  
\newcommand{\hatcurLBiiInoisorestrictempiricalxxxxxD}{\ensuremath{0.24\pm0.11}} 
\newcommand{\hatcurLBiuempiricalxxxxxD}{\ensuremath{1.1854}}             
\newcommand{\hatcurLBiiuempiricalxxxxxD}{\ensuremath{-0.3416}}           
\newcommand{\hatcurLBigempiricalxxxxxD}{\ensuremath{0.8805}}             
\newcommand{\hatcurLBiigempiricalxxxxxD}{\ensuremath{-0.0557}}           
\newcommand{\hatcurLBirempiricalxxxxxD}{\ensuremath{}}                   
\newcommand{\hatcurLBirnoisorestrictempiricalxxxxxD}{\ensuremath{0.41\pm0.12}} 
\newcommand{\hatcurLBiirempiricalxxxxxD}{\ensuremath{}}                  
\newcommand{\hatcurLBiirnoisorestrictempiricalxxxxxD}{\ensuremath{0.32\pm0.12}} 
\newcommand{\hatcurLBiiempiricalxxxxxD}{\ensuremath{}}                   
\newcommand{\hatcurLBiinoisorestrictempiricalxxxxxD}{\ensuremath{0.342\pm0.086}} 
\newcommand{\hatcurLBiiiempiricalxxxxxD}{\ensuremath{}}                  
\newcommand{\hatcurLBiiinoisorestrictempiricalxxxxxD}{\ensuremath{0.31\pm0.14}} 
\newcommand{\hatcurLBizempiricalxxxxxD}{\ensuremath{0.4296}}             
\newcommand{\hatcurLBiizempiricalxxxxxD}{\ensuremath{0.1493}}            
\newcommand{\hatcurLBiJempiricalxxxxxD}{\ensuremath{0.2991}}             
\newcommand{\hatcurLBiiJempiricalxxxxxD}{\ensuremath{0.2072}}            
\newcommand{\hatcurLBiHempiricalxxxxxD}{\ensuremath{0.1534}}             
\newcommand{\hatcurLBiiHempiricalxxxxxD}{\ensuremath{0.3100}}            
\newcommand{\hatcurLBiKempiricalxxxxxD}{\ensuremath{0.1318}}             
\newcommand{\hatcurLBiiKempiricalxxxxxD}{\ensuremath{0.2539}}            
\newcommand{\hatcurLBiTempiricalxxxxxD}{\ensuremath{}}                   
\newcommand{\hatcurLBiTnoisorestrictempiricalxxxxxD}{\ensuremath{0.373\pm0.079}} 
\newcommand{\hatcurLBiiTempiricalxxxxxD}{\ensuremath{}}                  
\newcommand{\hatcurLBiiTnoisorestrictempiricalxxxxxD}{\ensuremath{0.478\pm0.100}} 
\newcommand{\hatcurLBikepempiricalxxxxxD}{\ensuremath{}}                 
\newcommand{\hatcurLBikepnoisorestrictempiricalxxxxxD}{\ensuremath{0.362_{-0.094}^{+0.128}}} 
\newcommand{\hatcurLBiikepempiricalxxxxxD}{\ensuremath{}}                
\newcommand{\hatcurLBiikepnoisorestrictempiricalxxxxxD}{\ensuremath{0.54\pm0.12}} 
\newcommand{\hatcurLBiCempiricalxxxxxD}{\ensuremath{0.6027}}             
\newcommand{\hatcurLBiiCempiricalxxxxxD}{\ensuremath{0.1141}}            
\newcommand{\hatcurLBiMempiricalxxxxxD}{\ensuremath{0.7303}}             
\newcommand{\hatcurLBiiMempiricalxxxxxD}{\ensuremath{0.0552}}            
\newcommand{\hatcurLBiSoneempiricalxxxxxD}{\ensuremath{0.1161}}            
\newcommand{\hatcurLBiiSoneempiricalxxxxxD}{\ensuremath{0.1617}}           
\newcommand{\hatcurLBiStwoempiricalxxxxxD}{\ensuremath{0.1020}}            
\newcommand{\hatcurLBiiStwoempiricalxxxxxD}{\ensuremath{0.1270}}           
\newcommand{\hatcurLBiSthreeempiricalxxxxxD}{\ensuremath{0.0831}}            
\newcommand{\hatcurLBiiSthreeempiricalxxxxxD}{\ensuremath{0.1042}}           
\newcommand{\hatcurLBiSfourempiricalxxxxxD}{\ensuremath{0.0692}}            
\newcommand{\hatcurLBiiSfourempiricalxxxxxD}{\ensuremath{0.0886}}           
\newcommand{\hatcurISOmempiricalxxxxxD}{\ensuremath{}}                   
\newcommand{\hatcurISOmnoisorestrictempiricalxxxxxD}{\ensuremath{0.688\pm0.029}} 
\newcommand{\hatcurISOmshortempiricalxxxxxD}{\ensuremath{}}              
\newcommand{\hatcurISOmshortnoisorestrictempiricalxxxxxD}{\ensuremath{0.69}} 
\newcommand{\hatcurISOmlongempiricalxxxxxD}{\ensuremath{}}               
\newcommand{\hatcurISOmlongnoisorestrictempiricalxxxxxD}{\ensuremath{0.688\pm0.029}} 
\newcommand{\hatcurISOrempiricalxxxxxD}{\ensuremath{}}                   
\newcommand{\hatcurISOrnoisorestrictempiricalxxxxxD}{\ensuremath{0.7347\pm0.0069}} 
\newcommand{\hatcurISOrshortempiricalxxxxxD}{\ensuremath{}}              
\newcommand{\hatcurISOrshortnoisorestrictempiricalxxxxxD}{\ensuremath{0.73}} 
\newcommand{\hatcurISOrlongempiricalxxxxxD}{\ensuremath{}}               
\newcommand{\hatcurISOrlongnoisorestrictempiricalxxxxxD}{\ensuremath{0.7347\pm0.0069}} 
\newcommand{\hatcurISOloggempiricalxxxxxD}{\ensuremath{}}                
\newcommand{\hatcurISOloggnoisorestrictempiricalxxxxxD}{\ensuremath{4.543\pm0.013}} 
\newcommand{\hatcurISOlumempiricalxxxxxD}{\ensuremath{}}                 
\newcommand{\hatcurISOlumnoisorestrictempiricalxxxxxD}{\ensuremath{0.2237\pm0.0025}} 
\newcommand{\hatcurISOlumshortempiricalxxxxxD}{\ensuremath{}}            
\newcommand{\hatcurISOlumshortnoisorestrictempiricalxxxxxD}{\ensuremath{0.22}} 
\newcommand{\hatcurISOfehempiricalxxxxxD}{\ensuremath{}}                 
\newcommand{\hatcurISOfehnoisorestrictempiricalxxxxxD}{\ensuremath{-0.088\pm0.029}} 
\newcommand{\hatcurISOteffempiricalxxxxxD}{\ensuremath{}}                
\newcommand{\hatcurISOteffnoisorestrictempiricalxxxxxD}{\ensuremath{4631\pm18}} 
\newcommand{\hatcurISOageempiricalxxxxxD}{\ensuremath{}}                 
\newcommand{\hatcurISOagenoisorestrictempiricalxxxxxD}{\ensuremath{0\pm0}} 
\newcommand{\hatcurISOspecempiricalxxxxxD}{K}                            
\newcommand{\hatcurRVKempiricalxxxxxD}{\ensuremath{}}                    
\newcommand{\hatcurRVKnoisorestrictempiricalxxxxxD}{\ensuremath{17.6\pm1.0}} 
\newcommand{\hatcurRVrkempiricalxxxxxD}{\ensuremath{}}                   
\newcommand{\hatcurRVrknoisorestrictempiricalxxxxxD}{\ensuremath{0\pm0}} 
\newcommand{\hatcurRVrhempiricalxxxxxD}{\ensuremath{}}                   
\newcommand{\hatcurRVrhnoisorestrictempiricalxxxxxD}{\ensuremath{0\pm0}} 
\newcommand{\hatcurRVkempiricalxxxxxD}{\ensuremath{}}                    
\newcommand{\hatcurRVknoisorestrictempiricalxxxxxD}{\ensuremath{0\pm0}}  
\newcommand{\hatcurRVhempiricalxxxxxD}{\ensuremath{}}                    
\newcommand{\hatcurRVhnoisorestrictempiricalxxxxxD}{\ensuremath{0\pm0}}  
\newcommand{\hatcurRVtroneempiricalxxxxxD}{\ensuremath{}}                
\newcommand{\hatcurRVtronenoisorestrictempiricalxxxxxD}{\ensuremath{0\pm0}} 
\newcommand{\hatcurRVtrtwoempiricalxxxxxD}{\ensuremath{}}                
\newcommand{\hatcurRVtrtwonoisorestrictempiricalxxxxxD}{\ensuremath{0\pm0}} 
\newcommand{\hatcurRVgammaAempiricalxxxxxD}{\ensuremath{}}               
\newcommand{\hatcurRVgammaAnoisorestrictempiricalxxxxxD}{\ensuremath{15957.0\pm4.0}} 
\newcommand{\hatcurRVjitterAempiricalxxxxxD}{\ensuremath{}}              
\newcommand{\hatcurRVjitterAnoisorestrictempiricalxxxxxD}{\ensuremath{24.6\pm8.5}} 
\newcommand{\hatcurRVjittertwosiglimAempiricalxxxxxD}{\ensuremath{}}     
\newcommand{\hatcurRVjittertwosiglimAnoisorestrictempiricalxxxxxD}{\ensuremath{<40.0}} 
\newcommand{\hatcurRVfitrmsAempiricalxxxxxD}{\ensuremath{0.0}}           
\newcommand{\hatcurRVgammaBempiricalxxxxxD}{\ensuremath{}}               
\newcommand{\hatcurRVgammaBnoisorestrictempiricalxxxxxD}{\ensuremath{15954.0\pm2.7}} 
\newcommand{\hatcurRVjitterBempiricalxxxxxD}{\ensuremath{}}              
\newcommand{\hatcurRVjitterBnoisorestrictempiricalxxxxxD}{\ensuremath{0.60\pm0.47}} 
\newcommand{\hatcurRVjittertwosiglimBempiricalxxxxxD}{\ensuremath{}}     
\newcommand{\hatcurRVjittertwosiglimBnoisorestrictempiricalxxxxxD}{\ensuremath{<1.4}} 
\newcommand{\hatcurRVfitrmsBempiricalxxxxxD}{\ensuremath{0.0}}           
\newcommand{\hatcurRVgammaCempiricalxxxxxD}{\ensuremath{}}               
\newcommand{\hatcurRVgammaCnoisorestrictempiricalxxxxxD}{\ensuremath{2.44\pm0.86}} 
\newcommand{\hatcurRVjitterCempiricalxxxxxD}{\ensuremath{}}              
\newcommand{\hatcurCCbbHmagempirical}[1]{\ifnum#1=47 %
\hatcurCCbbHmagempiricalxxxxxA
\else
\ifnum#1=48 %
\hatcurCCbbHmagempiricalxxxxxB
\else
\ifnum#1=49 %
\hatcurCCbbHmagempiricalxxxxxC
\else
\ifnum#1=72 %
\hatcurCCbbHmagempiricalxxxxxD
\else
??????\fi
\fi
\fi
\fi
}
\newcommand{\hatcurCCbbJmagempirical}[1]{\ifnum#1=47 %
\hatcurCCbbJmagempiricalxxxxxA
\else
\ifnum#1=48 %
\hatcurCCbbJmagempiricalxxxxxB
\else
\ifnum#1=49 %
\hatcurCCbbJmagempiricalxxxxxC
\else
\ifnum#1=72 %
\hatcurCCbbJmagempiricalxxxxxD
\else
??????\fi
\fi
\fi
\fi
}
\newcommand{\hatcurCCbbKmagempirical}[1]{\ifnum#1=47 %
\hatcurCCbbKmagempiricalxxxxxA
\else
\ifnum#1=48 %
\hatcurCCbbKmagempiricalxxxxxB
\else
\ifnum#1=49 %
\hatcurCCbbKmagempiricalxxxxxC
\else
\ifnum#1=72 %
\hatcurCCbbKmagempiricalxxxxxD
\else
??????\fi
\fi
\fi
\fi
}
\newcommand{\hatcurCCcitHmagempirical}[1]{\ifnum#1=47 %
\hatcurCCcitHmagempiricalxxxxxA
\else
\ifnum#1=48 %
\hatcurCCcitHmagempiricalxxxxxB
\else
\ifnum#1=49 %
\hatcurCCcitHmagempiricalxxxxxC
\else
\ifnum#1=72 %
\hatcurCCcitHmagempiricalxxxxxD
\else
??????\fi
\fi
\fi
\fi
}
\newcommand{\hatcurCCcitJmagempirical}[1]{\ifnum#1=47 %
\hatcurCCcitJmagempiricalxxxxxA
\else
\ifnum#1=48 %
\hatcurCCcitJmagempiricalxxxxxB
\else
\ifnum#1=49 %
\hatcurCCcitJmagempiricalxxxxxC
\else
\ifnum#1=72 %
\hatcurCCcitJmagempiricalxxxxxD
\else
??????\fi
\fi
\fi
\fi
}
\newcommand{\hatcurCCcitKmagempirical}[1]{\ifnum#1=47 %
\hatcurCCcitKmagempiricalxxxxxA
\else
\ifnum#1=48 %
\hatcurCCcitKmagempiricalxxxxxB
\else
\ifnum#1=49 %
\hatcurCCcitKmagempiricalxxxxxC
\else
\ifnum#1=72 %
\hatcurCCcitKmagempiricalxxxxxD
\else
??????\fi
\fi
\fi
\fi
}
\newcommand{\hatcurCCdecempirical}[1]{\ifnum#1=47 %
\hatcurCCdecempiricalxxxxxA
\else
\ifnum#1=48 %
\hatcurCCdecempiricalxxxxxB
\else
\ifnum#1=49 %
\hatcurCCdecempiricalxxxxxC
\else
\ifnum#1=72 %
\hatcurCCdecempiricalxxxxxD
\else
??????\fi
\fi
\fi
\fi
}
\newcommand{\hatcurCCesoHKmagempirical}[1]{\ifnum#1=47 %
\hatcurCCesoHKmagempiricalxxxxxA
\else
\ifnum#1=48 %
\hatcurCCesoHKmagempiricalxxxxxB
\else
\ifnum#1=49 %
\hatcurCCesoHKmagempiricalxxxxxC
\else
\ifnum#1=72 %
\hatcurCCesoHKmagempiricalxxxxxD
\else
??????\fi
\fi
\fi
\fi
}
\newcommand{\hatcurCCesoHmagempirical}[1]{\ifnum#1=47 %
\hatcurCCesoHmagempiricalxxxxxA
\else
\ifnum#1=48 %
\hatcurCCesoHmagempiricalxxxxxB
\else
\ifnum#1=49 %
\hatcurCCesoHmagempiricalxxxxxC
\else
\ifnum#1=72 %
\hatcurCCesoHmagempiricalxxxxxD
\else
??????\fi
\fi
\fi
\fi
}
\newcommand{\hatcurCCesoJHmagempirical}[1]{\ifnum#1=47 %
\hatcurCCesoJHmagempiricalxxxxxA
\else
\ifnum#1=48 %
\hatcurCCesoJHmagempiricalxxxxxB
\else
\ifnum#1=49 %
\hatcurCCesoJHmagempiricalxxxxxC
\else
\ifnum#1=72 %
\hatcurCCesoJHmagempiricalxxxxxD
\else
??????\fi
\fi
\fi
\fi
}
\newcommand{\hatcurCCesoJKmagempirical}[1]{\ifnum#1=47 %
\hatcurCCesoJKmagempiricalxxxxxA
\else
\ifnum#1=48 %
\hatcurCCesoJKmagempiricalxxxxxB
\else
\ifnum#1=49 %
\hatcurCCesoJKmagempiricalxxxxxC
\else
\ifnum#1=72 %
\hatcurCCesoJKmagempiricalxxxxxD
\else
??????\fi
\fi
\fi
\fi
}
\newcommand{\hatcurCCesoJmagempirical}[1]{\ifnum#1=47 %
\hatcurCCesoJmagempiricalxxxxxA
\else
\ifnum#1=48 %
\hatcurCCesoJmagempiricalxxxxxB
\else
\ifnum#1=49 %
\hatcurCCesoJmagempiricalxxxxxC
\else
\ifnum#1=72 %
\hatcurCCesoJmagempiricalxxxxxD
\else
??????\fi
\fi
\fi
\fi
}
\newcommand{\hatcurCCesoKmagempirical}[1]{\ifnum#1=47 %
\hatcurCCesoKmagempiricalxxxxxA
\else
\ifnum#1=48 %
\hatcurCCesoKmagempiricalxxxxxB
\else
\ifnum#1=49 %
\hatcurCCesoKmagempiricalxxxxxC
\else
\ifnum#1=72 %
\hatcurCCesoKmagempiricalxxxxxD
\else
??????\fi
\fi
\fi
\fi
}
\newcommand{\hatcurCCgaiadrtwoempirical}[1]{\ifnum#1=47 %
\hatcurCCgaiadrtwoempiricalxxxxxA
\else
\ifnum#1=48 %
\hatcurCCgaiadrtwoempiricalxxxxxB
\else
\ifnum#1=49 %
\hatcurCCgaiadrtwoempiricalxxxxxC
\else
\ifnum#1=72 %
\hatcurCCgaiadrtwoempiricalxxxxxD
\else
??????\fi
\fi
\fi
\fi
}
\newcommand{\hatcurCCgaiaempirical}[1]{\ifnum#1=47 %
\hatcurCCgaiaempiricalxxxxxA
\else
\ifnum#1=48 %
\hatcurCCgaiaempiricalxxxxxB
\else
\ifnum#1=49 %
\hatcurCCgaiaempiricalxxxxxC
\else
\ifnum#1=72 %
\hatcurCCgaiaempiricalxxxxxD
\else
??????\fi
\fi
\fi
\fi
}
\newcommand{\hatcurCCgaiamBPempirical}[1]{\ifnum#1=47 %
\hatcurCCgaiamBPempiricalxxxxxA
\else
\ifnum#1=48 %
\hatcurCCgaiamBPempiricalxxxxxB
\else
\ifnum#1=49 %
\hatcurCCgaiamBPempiricalxxxxxC
\else
\ifnum#1=72 %
\hatcurCCgaiamBPempiricalxxxxxD
\else
??????\fi
\fi
\fi
\fi
}
\newcommand{\hatcurCCgaiamGempirical}[1]{\ifnum#1=47 %
\hatcurCCgaiamGempiricalxxxxxA
\else
\ifnum#1=48 %
\hatcurCCgaiamGempiricalxxxxxB
\else
\ifnum#1=49 %
\hatcurCCgaiamGempiricalxxxxxC
\else
\ifnum#1=72 %
\hatcurCCgaiamGempiricalxxxxxD
\else
??????\fi
\fi
\fi
\fi
}
\newcommand{\hatcurCCgaiamRPempirical}[1]{\ifnum#1=47 %
\hatcurCCgaiamRPempiricalxxxxxA
\else
\ifnum#1=48 %
\hatcurCCgaiamRPempiricalxxxxxB
\else
\ifnum#1=49 %
\hatcurCCgaiamRPempiricalxxxxxC
\else
\ifnum#1=72 %
\hatcurCCgaiamRPempiricalxxxxxD
\else
??????\fi
\fi
\fi
\fi
}
\newcommand{\hatcurCCgscempirical}[1]{\ifnum#1=47 %
\hatcurCCgscempiricalxxxxxA
\else
\ifnum#1=48 %
\hatcurCCgscempiricalxxxxxB
\else
\ifnum#1=49 %
\hatcurCCgscempiricalxxxxxC
\else
\ifnum#1=72 %
\hatcurCCgscempiricalxxxxxD
\else
??????\fi
\fi
\fi
\fi
}
\newcommand{\hatcurCCmagempirical}[1]{\ifnum#1=47 %
\hatcurCCmagempiricalxxxxxA
\else
\ifnum#1=48 %
\hatcurCCmagempiricalxxxxxB
\else
\ifnum#1=49 %
\hatcurCCmagempiricalxxxxxC
\else
\ifnum#1=72 %
\hatcurCCmagempiricalxxxxxD
\else
??????\fi
\fi
\fi
\fi
}
\newcommand{\hatcurCCparallaxempirical}[1]{\ifnum#1=47 %
\hatcurCCparallaxempiricalxxxxxA
\else
\ifnum#1=48 %
\hatcurCCparallaxempiricalxxxxxB
\else
\ifnum#1=49 %
\hatcurCCparallaxempiricalxxxxxC
\else
\ifnum#1=72 %
\hatcurCCparallaxempiricalxxxxxD
\else
??????\fi
\fi
\fi
\fi
}
\newcommand{\hatcurCCpmdecempirical}[1]{\ifnum#1=47 %
\hatcurCCpmdecempiricalxxxxxA
\else
\ifnum#1=48 %
\hatcurCCpmdecempiricalxxxxxB
\else
\ifnum#1=49 %
\hatcurCCpmdecempiricalxxxxxC
\else
??????\fi
\fi
\fi
}
\newcommand{\hatcurCCpmempirical}[1]{\ifnum#1=47 %
\hatcurCCpmempiricalxxxxxA
\else
\ifnum#1=48 %
\hatcurCCpmempiricalxxxxxB
\else
\ifnum#1=49 %
\hatcurCCpmempiricalxxxxxC
\else
??????\fi
\fi
\fi
}
\newcommand{\hatcurCCpmraempirical}[1]{\ifnum#1=47 %
\hatcurCCpmraempiricalxxxxxA
\else
\ifnum#1=48 %
\hatcurCCpmraempiricalxxxxxB
\else
\ifnum#1=49 %
\hatcurCCpmraempiricalxxxxxC
\else
??????\fi
\fi
\fi
}
\newcommand{\hatcurCCraempirical}[1]{\ifnum#1=47 %
\hatcurCCraempiricalxxxxxA
\else
\ifnum#1=48 %
\hatcurCCraempiricalxxxxxB
\else
\ifnum#1=49 %
\hatcurCCraempiricalxxxxxC
\else
\ifnum#1=72 %
\hatcurCCraempiricalxxxxxD
\else
??????\fi
\fi
\fi
\fi
}
\newcommand{\hatcurCCtassmBempirical}[1]{\ifnum#1=47 %
\hatcurCCtassmBempiricalxxxxxA
\else
\ifnum#1=48 %
\hatcurCCtassmBempiricalxxxxxB
\else
\ifnum#1=49 %
\hatcurCCtassmBempiricalxxxxxC
\else
\ifnum#1=72 %
\hatcurCCtassmBempiricalxxxxxD
\else
??????\fi
\fi
\fi
\fi
}
\newcommand{\hatcurCCtassmBshortempirical}[1]{\ifnum#1=47 %
\hatcurCCtassmBshortempiricalxxxxxA
\else
\ifnum#1=48 %
\hatcurCCtassmBshortempiricalxxxxxB
\else
\ifnum#1=49 %
\hatcurCCtassmBshortempiricalxxxxxC
\else
\ifnum#1=72 %
\hatcurCCtassmBshortempiricalxxxxxD
\else
??????\fi
\fi
\fi
\fi
}
\newcommand{\hatcurCCtassmgempirical}[1]{\ifnum#1=47 %
\hatcurCCtassmgempiricalxxxxxA
\else
\ifnum#1=48 %
\hatcurCCtassmgempiricalxxxxxB
\else
\ifnum#1=49 %
\hatcurCCtassmgempiricalxxxxxC
\else
\ifnum#1=72 %
\hatcurCCtassmgempiricalxxxxxD
\else
??????\fi
\fi
\fi
\fi
}
\newcommand{\hatcurCCtassmgshortempirical}[1]{\ifnum#1=47 %
\hatcurCCtassmgshortempiricalxxxxxA
\else
\ifnum#1=48 %
\hatcurCCtassmgshortempiricalxxxxxB
\else
\ifnum#1=49 %
\hatcurCCtassmgshortempiricalxxxxxC
\else
\ifnum#1=72 %
\hatcurCCtassmgshortempiricalxxxxxD
\else
??????\fi
\fi
\fi
\fi
}
\newcommand{\hatcurCCtassmiempirical}[1]{\ifnum#1=47 %
\hatcurCCtassmiempiricalxxxxxA
\else
\ifnum#1=48 %
\hatcurCCtassmiempiricalxxxxxB
\else
\ifnum#1=49 %
\hatcurCCtassmiempiricalxxxxxC
\else
\ifnum#1=72 %
\hatcurCCtassmiempiricalxxxxxD
\else
??????\fi
\fi
\fi
\fi
}
\newcommand{\hatcurCCtassmIempirical}[1]{\ifnum#1=47 %
\hatcurCCtassmIempiricalxxxxxA
\else
\ifnum#1=48 %
\hatcurCCtassmIempiricalxxxxxB
\else
\ifnum#1=49 %
\hatcurCCtassmIempiricalxxxxxC
\else
\ifnum#1=72 %
\hatcurCCtassmIempiricalxxxxxD
\else
??????\fi
\fi
\fi
\fi
}
\newcommand{\hatcurCCtassmishortempirical}[1]{\ifnum#1=47 %
\hatcurCCtassmishortempiricalxxxxxA
\else
\ifnum#1=48 %
\hatcurCCtassmishortempiricalxxxxxB
\else
\ifnum#1=49 %
\hatcurCCtassmishortempiricalxxxxxC
\else
\ifnum#1=72 %
\hatcurCCtassmishortempiricalxxxxxD
\else
??????\fi
\fi
\fi
\fi
}
\newcommand{\hatcurCCtassmIshortempirical}[1]{\ifnum#1=47 %
\hatcurCCtassmIshortempiricalxxxxxA
\else
\ifnum#1=48 %
\hatcurCCtassmIshortempiricalxxxxxB
\else
\ifnum#1=49 %
\hatcurCCtassmIshortempiricalxxxxxC
\else
\ifnum#1=72 %
\hatcurCCtassmIshortempiricalxxxxxD
\else
??????\fi
\fi
\fi
\fi
}
\newcommand{\hatcurCCtassmrempirical}[1]{\ifnum#1=47 %
\hatcurCCtassmrempiricalxxxxxA
\else
\ifnum#1=48 %
\hatcurCCtassmrempiricalxxxxxB
\else
\ifnum#1=49 %
\hatcurCCtassmrempiricalxxxxxC
\else
\ifnum#1=72 %
\hatcurCCtassmrempiricalxxxxxD
\else
??????\fi
\fi
\fi
\fi
}
\newcommand{\hatcurCCtassmrshortempirical}[1]{\ifnum#1=47 %
\hatcurCCtassmrshortempiricalxxxxxA
\else
\ifnum#1=48 %
\hatcurCCtassmrshortempiricalxxxxxB
\else
\ifnum#1=49 %
\hatcurCCtassmrshortempiricalxxxxxC
\else
\ifnum#1=72 %
\hatcurCCtassmrshortempiricalxxxxxD
\else
??????\fi
\fi
\fi
\fi
}
\newcommand{\hatcurCCtassmvempirical}[1]{\ifnum#1=47 %
\hatcurCCtassmvempiricalxxxxxA
\else
\ifnum#1=48 %
\hatcurCCtassmvempiricalxxxxxB
\else
\ifnum#1=49 %
\hatcurCCtassmvempiricalxxxxxC
\else
\ifnum#1=72 %
\hatcurCCtassmvempiricalxxxxxD
\else
??????\fi
\fi
\fi
\fi
}
\newcommand{\hatcurCCtassmvshortempirical}[1]{\ifnum#1=47 %
\hatcurCCtassmvshortempiricalxxxxxA
\else
\ifnum#1=48 %
\hatcurCCtassmvshortempiricalxxxxxB
\else
\ifnum#1=49 %
\hatcurCCtassmvshortempiricalxxxxxC
\else
\ifnum#1=72 %
\hatcurCCtassmvshortempiricalxxxxxD
\else
??????\fi
\fi
\fi
\fi
}
\newcommand{\hatcurCCtwomassempirical}[1]{\ifnum#1=47 %
\hatcurCCtwomassempiricalxxxxxA
\else
\ifnum#1=48 %
\hatcurCCtwomassempiricalxxxxxB
\else
\ifnum#1=49 %
\hatcurCCtwomassempiricalxxxxxC
\else
\ifnum#1=72 %
\hatcurCCtwomassempiricalxxxxxD
\else
??????\fi
\fi
\fi
\fi
}
\newcommand{\hatcurCCtwomassHmagempirical}[1]{\ifnum#1=47 %
\hatcurCCtwomassHmagempiricalxxxxxA
\else
\ifnum#1=48 %
\hatcurCCtwomassHmagempiricalxxxxxB
\else
\ifnum#1=49 %
\hatcurCCtwomassHmagempiricalxxxxxC
\else
\ifnum#1=72 %
\hatcurCCtwomassHmagempiricalxxxxxD
\else
??????\fi
\fi
\fi
\fi
}
\newcommand{\hatcurCCtwomassJmagempirical}[1]{\ifnum#1=47 %
\hatcurCCtwomassJmagempiricalxxxxxA
\else
\ifnum#1=48 %
\hatcurCCtwomassJmagempiricalxxxxxB
\else
\ifnum#1=49 %
\hatcurCCtwomassJmagempiricalxxxxxC
\else
\ifnum#1=72 %
\hatcurCCtwomassJmagempiricalxxxxxD
\else
??????\fi
\fi
\fi
\fi
}
\newcommand{\hatcurCCtwomassKmagempirical}[1]{\ifnum#1=47 %
\hatcurCCtwomassKmagempiricalxxxxxA
\else
\ifnum#1=48 %
\hatcurCCtwomassKmagempiricalxxxxxB
\else
\ifnum#1=49 %
\hatcurCCtwomassKmagempiricalxxxxxC
\else
\ifnum#1=72 %
\hatcurCCtwomassKmagempiricalxxxxxD
\else
??????\fi
\fi
\fi
\fi
}
\newcommand{\hatcurextraerrMBempirical}[1]{\ifnum#1=72 %
\hatcurextraerrMBempiricalxxxxxD
\else
??????\fi
}
\newcommand{\hatcurextraerrMBnoisorestrictempirical}[1]{\ifnum#1=72 %
\hatcurextraerrMBnoisorestrictempiricalxxxxxD
\else
??????\fi
}
\newcommand{\hatcurextraerrMBPtwoempirical}[1]{\ifnum#1=47 %
\hatcurextraerrMBPtwoempiricalxxxxxA
\else
\ifnum#1=48 %
\hatcurextraerrMBPtwoempiricalxxxxxB
\else
\ifnum#1=49 %
\hatcurextraerrMBPtwoempiricalxxxxxC
\else
\ifnum#1=72 %
\hatcurextraerrMBPtwoempiricalxxxxxD
\else
??????\fi
\fi
\fi
\fi
}
\newcommand{\hatcurextraerrMBPtwonoisorestrictempirical}[1]{\ifnum#1=47 %
\hatcurextraerrMBPtwonoisorestrictempiricalxxxxxA
\else
\ifnum#1=48 %
\hatcurextraerrMBPtwonoisorestrictempiricalxxxxxB
\else
\ifnum#1=49 %
\hatcurextraerrMBPtwonoisorestrictempiricalxxxxxC
\else
\ifnum#1=72 %
\hatcurextraerrMBPtwonoisorestrictempiricalxxxxxD
\else
??????\fi
\fi
\fi
\fi
}
\newcommand{\hatcurextraerrMBPtwotwosiglimempirical}[1]{\ifnum#1=47 %
\hatcurextraerrMBPtwotwosiglimempiricalxxxxxA
\else
\ifnum#1=48 %
\hatcurextraerrMBPtwotwosiglimempiricalxxxxxB
\else
\ifnum#1=49 %
\hatcurextraerrMBPtwotwosiglimempiricalxxxxxC
\else
\ifnum#1=72 %
\hatcurextraerrMBPtwotwosiglimempiricalxxxxxD
\else
??????\fi
\fi
\fi
\fi
}
\newcommand{\hatcurextraerrMBPtwotwosiglimnoisorestrictempirical}[1]{\ifnum#1=47 %
\hatcurextraerrMBPtwotwosiglimnoisorestrictempiricalxxxxxA
\else
\ifnum#1=48 %
\hatcurextraerrMBPtwotwosiglimnoisorestrictempiricalxxxxxB
\else
\ifnum#1=49 %
\hatcurextraerrMBPtwotwosiglimnoisorestrictempiricalxxxxxC
\else
\ifnum#1=72 %
\hatcurextraerrMBPtwotwosiglimnoisorestrictempiricalxxxxxD
\else
??????\fi
\fi
\fi
\fi
}
\newcommand{\hatcurextraerrMBtwosiglimempirical}[1]{\ifnum#1=72 %
\hatcurextraerrMBtwosiglimempiricalxxxxxD
\else
??????\fi
}
\newcommand{\hatcurextraerrMBtwosiglimnoisorestrictempirical}[1]{\ifnum#1=72 %
\hatcurextraerrMBtwosiglimnoisorestrictempiricalxxxxxD
\else
??????\fi
}
\newcommand{\hatcurextraerrMgempirical}[1]{\ifnum#1=47 %
\hatcurextraerrMgempiricalxxxxxA
\else
\ifnum#1=48 %
\hatcurextraerrMgempiricalxxxxxB
\else
\ifnum#1=49 %
\hatcurextraerrMgempiricalxxxxxC
\else
\ifnum#1=72 %
\hatcurextraerrMgempiricalxxxxxD
\else
??????\fi
\fi
\fi
\fi
}
\newcommand{\hatcurextraerrMGempirical}[1]{\ifnum#1=47 %
\hatcurextraerrMGempiricalxxxxxA
\else
\ifnum#1=48 %
\hatcurextraerrMGempiricalxxxxxB
\else
\ifnum#1=49 %
\hatcurextraerrMGempiricalxxxxxC
\else
\ifnum#1=72 %
\hatcurextraerrMGempiricalxxxxxD
\else
??????\fi
\fi
\fi
\fi
}
\newcommand{\hatcurextraerrMgnoisorestrictempirical}[1]{\ifnum#1=47 %
\hatcurextraerrMgnoisorestrictempiricalxxxxxA
\else
\ifnum#1=48 %
\hatcurextraerrMgnoisorestrictempiricalxxxxxB
\else
\ifnum#1=49 %
\hatcurextraerrMgnoisorestrictempiricalxxxxxC
\else
\ifnum#1=72 %
\hatcurextraerrMgnoisorestrictempiricalxxxxxD
\else
??????\fi
\fi
\fi
\fi
}
\newcommand{\hatcurextraerrMGnoisorestrictempirical}[1]{\ifnum#1=47 %
\hatcurextraerrMGnoisorestrictempiricalxxxxxA
\else
\ifnum#1=48 %
\hatcurextraerrMGnoisorestrictempiricalxxxxxB
\else
\ifnum#1=49 %
\hatcurextraerrMGnoisorestrictempiricalxxxxxC
\else
\ifnum#1=72 %
\hatcurextraerrMGnoisorestrictempiricalxxxxxD
\else
??????\fi
\fi
\fi
\fi
}
\newcommand{\hatcurextraerrMgtwosiglimempirical}[1]{\ifnum#1=47 %
\hatcurextraerrMgtwosiglimempiricalxxxxxA
\else
\ifnum#1=48 %
\hatcurextraerrMgtwosiglimempiricalxxxxxB
\else
\ifnum#1=49 %
\hatcurextraerrMgtwosiglimempiricalxxxxxC
\else
\ifnum#1=72 %
\hatcurextraerrMgtwosiglimempiricalxxxxxD
\else
??????\fi
\fi
\fi
\fi
}
\newcommand{\hatcurextraerrMGtwosiglimempirical}[1]{\ifnum#1=47 %
\hatcurextraerrMGtwosiglimempiricalxxxxxA
\else
\ifnum#1=48 %
\hatcurextraerrMGtwosiglimempiricalxxxxxB
\else
\ifnum#1=49 %
\hatcurextraerrMGtwosiglimempiricalxxxxxC
\else
\ifnum#1=72 %
\hatcurextraerrMGtwosiglimempiricalxxxxxD
\else
??????\fi
\fi
\fi
\fi
}
\newcommand{\hatcurextraerrMgtwosiglimnoisorestrictempirical}[1]{\ifnum#1=47 %
\hatcurextraerrMgtwosiglimnoisorestrictempiricalxxxxxA
\else
\ifnum#1=48 %
\hatcurextraerrMgtwosiglimnoisorestrictempiricalxxxxxB
\else
\ifnum#1=49 %
\hatcurextraerrMgtwosiglimnoisorestrictempiricalxxxxxC
\else
\ifnum#1=72 %
\hatcurextraerrMgtwosiglimnoisorestrictempiricalxxxxxD
\else
??????\fi
\fi
\fi
\fi
}
\newcommand{\hatcurextraerrMGtwosiglimnoisorestrictempirical}[1]{\ifnum#1=47 %
\hatcurextraerrMGtwosiglimnoisorestrictempiricalxxxxxA
\else
\ifnum#1=48 %
\hatcurextraerrMGtwosiglimnoisorestrictempiricalxxxxxB
\else
\ifnum#1=49 %
\hatcurextraerrMGtwosiglimnoisorestrictempiricalxxxxxC
\else
\ifnum#1=72 %
\hatcurextraerrMGtwosiglimnoisorestrictempiricalxxxxxD
\else
??????\fi
\fi
\fi
\fi
}
\newcommand{\hatcurextraerrMHempirical}[1]{\ifnum#1=47 %
\hatcurextraerrMHempiricalxxxxxA
\else
\ifnum#1=48 %
\hatcurextraerrMHempiricalxxxxxB
\else
\ifnum#1=49 %
\hatcurextraerrMHempiricalxxxxxC
\else
\ifnum#1=72 %
\hatcurextraerrMHempiricalxxxxxD
\else
??????\fi
\fi
\fi
\fi
}
\newcommand{\hatcurextraerrMHnoisorestrictempirical}[1]{\ifnum#1=47 %
\hatcurextraerrMHnoisorestrictempiricalxxxxxA
\else
\ifnum#1=48 %
\hatcurextraerrMHnoisorestrictempiricalxxxxxB
\else
\ifnum#1=49 %
\hatcurextraerrMHnoisorestrictempiricalxxxxxC
\else
\ifnum#1=72 %
\hatcurextraerrMHnoisorestrictempiricalxxxxxD
\else
??????\fi
\fi
\fi
\fi
}
\newcommand{\hatcurextraerrMHtwosiglimempirical}[1]{\ifnum#1=47 %
\hatcurextraerrMHtwosiglimempiricalxxxxxA
\else
\ifnum#1=48 %
\hatcurextraerrMHtwosiglimempiricalxxxxxB
\else
\ifnum#1=49 %
\hatcurextraerrMHtwosiglimempiricalxxxxxC
\else
\ifnum#1=72 %
\hatcurextraerrMHtwosiglimempiricalxxxxxD
\else
??????\fi
\fi
\fi
\fi
}
\newcommand{\hatcurextraerrMHtwosiglimnoisorestrictempirical}[1]{\ifnum#1=47 %
\hatcurextraerrMHtwosiglimnoisorestrictempiricalxxxxxA
\else
\ifnum#1=48 %
\hatcurextraerrMHtwosiglimnoisorestrictempiricalxxxxxB
\else
\ifnum#1=49 %
\hatcurextraerrMHtwosiglimnoisorestrictempiricalxxxxxC
\else
\ifnum#1=72 %
\hatcurextraerrMHtwosiglimnoisorestrictempiricalxxxxxD
\else
??????\fi
\fi
\fi
\fi
}
\newcommand{\hatcurextraerrMiempirical}[1]{\ifnum#1=47 %
\hatcurextraerrMiempiricalxxxxxA
\else
\ifnum#1=48 %
\hatcurextraerrMiempiricalxxxxxB
\else
\ifnum#1=49 %
\hatcurextraerrMiempiricalxxxxxC
\else
\ifnum#1=72 %
\hatcurextraerrMiempiricalxxxxxD
\else
??????\fi
\fi
\fi
\fi
}
\newcommand{\hatcurextraerrMinoisorestrictempirical}[1]{\ifnum#1=47 %
\hatcurextraerrMinoisorestrictempiricalxxxxxA
\else
\ifnum#1=48 %
\hatcurextraerrMinoisorestrictempiricalxxxxxB
\else
\ifnum#1=49 %
\hatcurextraerrMinoisorestrictempiricalxxxxxC
\else
\ifnum#1=72 %
\hatcurextraerrMinoisorestrictempiricalxxxxxD
\else
??????\fi
\fi
\fi
\fi
}
\newcommand{\hatcurextraerrMitwosiglimempirical}[1]{\ifnum#1=47 %
\hatcurextraerrMitwosiglimempiricalxxxxxA
\else
\ifnum#1=48 %
\hatcurextraerrMitwosiglimempiricalxxxxxB
\else
\ifnum#1=49 %
\hatcurextraerrMitwosiglimempiricalxxxxxC
\else
\ifnum#1=72 %
\hatcurextraerrMitwosiglimempiricalxxxxxD
\else
??????\fi
\fi
\fi
\fi
}
\newcommand{\hatcurextraerrMitwosiglimnoisorestrictempirical}[1]{\ifnum#1=47 %
\hatcurextraerrMitwosiglimnoisorestrictempiricalxxxxxA
\else
\ifnum#1=48 %
\hatcurextraerrMitwosiglimnoisorestrictempiricalxxxxxB
\else
\ifnum#1=49 %
\hatcurextraerrMitwosiglimnoisorestrictempiricalxxxxxC
\else
\ifnum#1=72 %
\hatcurextraerrMitwosiglimnoisorestrictempiricalxxxxxD
\else
??????\fi
\fi
\fi
\fi
}
\newcommand{\hatcurextraerrMJempirical}[1]{\ifnum#1=47 %
\hatcurextraerrMJempiricalxxxxxA
\else
\ifnum#1=48 %
\hatcurextraerrMJempiricalxxxxxB
\else
\ifnum#1=49 %
\hatcurextraerrMJempiricalxxxxxC
\else
\ifnum#1=72 %
\hatcurextraerrMJempiricalxxxxxD
\else
??????\fi
\fi
\fi
\fi
}
\newcommand{\hatcurextraerrMJnoisorestrictempirical}[1]{\ifnum#1=47 %
\hatcurextraerrMJnoisorestrictempiricalxxxxxA
\else
\ifnum#1=48 %
\hatcurextraerrMJnoisorestrictempiricalxxxxxB
\else
\ifnum#1=49 %
\hatcurextraerrMJnoisorestrictempiricalxxxxxC
\else
\ifnum#1=72 %
\hatcurextraerrMJnoisorestrictempiricalxxxxxD
\else
??????\fi
\fi
\fi
\fi
}
\newcommand{\hatcurextraerrMJtwosiglimempirical}[1]{\ifnum#1=47 %
\hatcurextraerrMJtwosiglimempiricalxxxxxA
\else
\ifnum#1=48 %
\hatcurextraerrMJtwosiglimempiricalxxxxxB
\else
\ifnum#1=49 %
\hatcurextraerrMJtwosiglimempiricalxxxxxC
\else
\ifnum#1=72 %
\hatcurextraerrMJtwosiglimempiricalxxxxxD
\else
??????\fi
\fi
\fi
\fi
}
\newcommand{\hatcurextraerrMJtwosiglimnoisorestrictempirical}[1]{\ifnum#1=47 %
\hatcurextraerrMJtwosiglimnoisorestrictempiricalxxxxxA
\else
\ifnum#1=48 %
\hatcurextraerrMJtwosiglimnoisorestrictempiricalxxxxxB
\else
\ifnum#1=49 %
\hatcurextraerrMJtwosiglimnoisorestrictempiricalxxxxxC
\else
\ifnum#1=72 %
\hatcurextraerrMJtwosiglimnoisorestrictempiricalxxxxxD
\else
??????\fi
\fi
\fi
\fi
}
\newcommand{\hatcurextraerrMKsempirical}[1]{\ifnum#1=47 %
\hatcurextraerrMKsempiricalxxxxxA
\else
\ifnum#1=48 %
\hatcurextraerrMKsempiricalxxxxxB
\else
\ifnum#1=49 %
\hatcurextraerrMKsempiricalxxxxxC
\else
\ifnum#1=72 %
\hatcurextraerrMKsempiricalxxxxxD
\else
??????\fi
\fi
\fi
\fi
}
\newcommand{\hatcurextraerrMKsnoisorestrictempirical}[1]{\ifnum#1=47 %
\hatcurextraerrMKsnoisorestrictempiricalxxxxxA
\else
\ifnum#1=48 %
\hatcurextraerrMKsnoisorestrictempiricalxxxxxB
\else
\ifnum#1=49 %
\hatcurextraerrMKsnoisorestrictempiricalxxxxxC
\else
\ifnum#1=72 %
\hatcurextraerrMKsnoisorestrictempiricalxxxxxD
\else
??????\fi
\fi
\fi
\fi
}
\newcommand{\hatcurextraerrMKstwosiglimempirical}[1]{\ifnum#1=47 %
\hatcurextraerrMKstwosiglimempiricalxxxxxA
\else
\ifnum#1=48 %
\hatcurextraerrMKstwosiglimempiricalxxxxxB
\else
\ifnum#1=49 %
\hatcurextraerrMKstwosiglimempiricalxxxxxC
\else
\ifnum#1=72 %
\hatcurextraerrMKstwosiglimempiricalxxxxxD
\else
??????\fi
\fi
\fi
\fi
}
\newcommand{\hatcurextraerrMKstwosiglimnoisorestrictempirical}[1]{\ifnum#1=47 %
\hatcurextraerrMKstwosiglimnoisorestrictempiricalxxxxxA
\else
\ifnum#1=48 %
\hatcurextraerrMKstwosiglimnoisorestrictempiricalxxxxxB
\else
\ifnum#1=49 %
\hatcurextraerrMKstwosiglimnoisorestrictempiricalxxxxxC
\else
\ifnum#1=72 %
\hatcurextraerrMKstwosiglimnoisorestrictempiricalxxxxxD
\else
??????\fi
\fi
\fi
\fi
}
\newcommand{\hatcurextraerrMrempirical}[1]{\ifnum#1=47 %
\hatcurextraerrMrempiricalxxxxxA
\else
\ifnum#1=48 %
\hatcurextraerrMrempiricalxxxxxB
\else
\ifnum#1=49 %
\hatcurextraerrMrempiricalxxxxxC
\else
\ifnum#1=72 %
\hatcurextraerrMrempiricalxxxxxD
\else
??????\fi
\fi
\fi
\fi
}
\newcommand{\hatcurextraerrMrnoisorestrictempirical}[1]{\ifnum#1=47 %
\hatcurextraerrMrnoisorestrictempiricalxxxxxA
\else
\ifnum#1=48 %
\hatcurextraerrMrnoisorestrictempiricalxxxxxB
\else
\ifnum#1=49 %
\hatcurextraerrMrnoisorestrictempiricalxxxxxC
\else
\ifnum#1=72 %
\hatcurextraerrMrnoisorestrictempiricalxxxxxD
\else
??????\fi
\fi
\fi
\fi
}
\newcommand{\hatcurextraerrMRPempirical}[1]{\ifnum#1=47 %
\hatcurextraerrMRPempiricalxxxxxA
\else
\ifnum#1=48 %
\hatcurextraerrMRPempiricalxxxxxB
\else
\ifnum#1=49 %
\hatcurextraerrMRPempiricalxxxxxC
\else
\ifnum#1=72 %
\hatcurextraerrMRPempiricalxxxxxD
\else
??????\fi
\fi
\fi
\fi
}
\newcommand{\hatcurextraerrMRPnoisorestrictempirical}[1]{\ifnum#1=47 %
\hatcurextraerrMRPnoisorestrictempiricalxxxxxA
\else
\ifnum#1=48 %
\hatcurextraerrMRPnoisorestrictempiricalxxxxxB
\else
\ifnum#1=49 %
\hatcurextraerrMRPnoisorestrictempiricalxxxxxC
\else
\ifnum#1=72 %
\hatcurextraerrMRPnoisorestrictempiricalxxxxxD
\else
??????\fi
\fi
\fi
\fi
}
\newcommand{\hatcurextraerrMRPtwosiglimempirical}[1]{\ifnum#1=47 %
\hatcurextraerrMRPtwosiglimempiricalxxxxxA
\else
\ifnum#1=48 %
\hatcurextraerrMRPtwosiglimempiricalxxxxxB
\else
\ifnum#1=49 %
\hatcurextraerrMRPtwosiglimempiricalxxxxxC
\else
\ifnum#1=72 %
\hatcurextraerrMRPtwosiglimempiricalxxxxxD
\else
??????\fi
\fi
\fi
\fi
}
\newcommand{\hatcurextraerrMRPtwosiglimnoisorestrictempirical}[1]{\ifnum#1=47 %
\hatcurextraerrMRPtwosiglimnoisorestrictempiricalxxxxxA
\else
\ifnum#1=48 %
\hatcurextraerrMRPtwosiglimnoisorestrictempiricalxxxxxB
\else
\ifnum#1=49 %
\hatcurextraerrMRPtwosiglimnoisorestrictempiricalxxxxxC
\else
\ifnum#1=72 %
\hatcurextraerrMRPtwosiglimnoisorestrictempiricalxxxxxD
\else
??????\fi
\fi
\fi
\fi
}
\newcommand{\hatcurextraerrMrtwosiglimempirical}[1]{\ifnum#1=47 %
\hatcurextraerrMrtwosiglimempiricalxxxxxA
\else
\ifnum#1=48 %
\hatcurextraerrMrtwosiglimempiricalxxxxxB
\else
\ifnum#1=49 %
\hatcurextraerrMrtwosiglimempiricalxxxxxC
\else
\ifnum#1=72 %
\hatcurextraerrMrtwosiglimempiricalxxxxxD
\else
??????\fi
\fi
\fi
\fi
}
\newcommand{\hatcurextraerrMrtwosiglimnoisorestrictempirical}[1]{\ifnum#1=47 %
\hatcurextraerrMrtwosiglimnoisorestrictempiricalxxxxxA
\else
\ifnum#1=48 %
\hatcurextraerrMrtwosiglimnoisorestrictempiricalxxxxxB
\else
\ifnum#1=49 %
\hatcurextraerrMrtwosiglimnoisorestrictempiricalxxxxxC
\else
\ifnum#1=72 %
\hatcurextraerrMrtwosiglimnoisorestrictempiricalxxxxxD
\else
??????\fi
\fi
\fi
\fi
}
\newcommand{\hatcurextraerrMVempirical}[1]{\ifnum#1=72 %
\hatcurextraerrMVempiricalxxxxxD
\else
??????\fi
}
\newcommand{\hatcurextraerrMVnoisorestrictempirical}[1]{\ifnum#1=72 %
\hatcurextraerrMVnoisorestrictempiricalxxxxxD
\else
??????\fi
}
\newcommand{\hatcurextraerrMVtwosiglimempirical}[1]{\ifnum#1=72 %
\hatcurextraerrMVtwosiglimempiricalxxxxxD
\else
??????\fi
}
\newcommand{\hatcurextraerrMVtwosiglimnoisorestrictempirical}[1]{\ifnum#1=72 %
\hatcurextraerrMVtwosiglimnoisorestrictempiricalxxxxxD
\else
??????\fi
}
\newcommand{\hatcurextraerrMWoneempirical}[1]{\ifnum#1=47 %
\hatcurextraerrMWoneempiricalxxxxxA
\else
\ifnum#1=48 %
\hatcurextraerrMWoneempiricalxxxxxB
\else
\ifnum#1=49 %
\hatcurextraerrMWoneempiricalxxxxxC
\else
\ifnum#1=72 %
\hatcurextraerrMWoneempiricalxxxxxD
\else
??????\fi
\fi
\fi
\fi
}
\newcommand{\hatcurextraerrMWonenoisorestrictempirical}[1]{\ifnum#1=47 %
\hatcurextraerrMWonenoisorestrictempiricalxxxxxA
\else
\ifnum#1=48 %
\hatcurextraerrMWonenoisorestrictempiricalxxxxxB
\else
\ifnum#1=49 %
\hatcurextraerrMWonenoisorestrictempiricalxxxxxC
\else
\ifnum#1=72 %
\hatcurextraerrMWonenoisorestrictempiricalxxxxxD
\else
??????\fi
\fi
\fi
\fi
}
\newcommand{\hatcurextraerrMWonetwosiglimempirical}[1]{\ifnum#1=47 %
\hatcurextraerrMWonetwosiglimempiricalxxxxxA
\else
\ifnum#1=48 %
\hatcurextraerrMWonetwosiglimempiricalxxxxxB
\else
\ifnum#1=49 %
\hatcurextraerrMWonetwosiglimempiricalxxxxxC
\else
\ifnum#1=72 %
\hatcurextraerrMWonetwosiglimempiricalxxxxxD
\else
??????\fi
\fi
\fi
\fi
}
\newcommand{\hatcurextraerrMWonetwosiglimnoisorestrictempirical}[1]{\ifnum#1=47 %
\hatcurextraerrMWonetwosiglimnoisorestrictempiricalxxxxxA
\else
\ifnum#1=48 %
\hatcurextraerrMWonetwosiglimnoisorestrictempiricalxxxxxB
\else
\ifnum#1=49 %
\hatcurextraerrMWonetwosiglimnoisorestrictempiricalxxxxxC
\else
\ifnum#1=72 %
\hatcurextraerrMWonetwosiglimnoisorestrictempiricalxxxxxD
\else
??????\fi
\fi
\fi
\fi
}
\newcommand{\hatcurextraerrMWthreeempirical}[1]{\ifnum#1=72 %
\hatcurextraerrMWthreeempiricalxxxxxD
\else
??????\fi
}
\newcommand{\hatcurextraerrMWthreenoisorestrictempirical}[1]{\ifnum#1=72 %
\hatcurextraerrMWthreenoisorestrictempiricalxxxxxD
\else
??????\fi
}
\newcommand{\hatcurextraerrMWthreetwosiglimempirical}[1]{\ifnum#1=72 %
\hatcurextraerrMWthreetwosiglimempiricalxxxxxD
\else
??????\fi
}
\newcommand{\hatcurextraerrMWthreetwosiglimnoisorestrictempirical}[1]{\ifnum#1=72 %
\hatcurextraerrMWthreetwosiglimnoisorestrictempiricalxxxxxD
\else
??????\fi
}
\newcommand{\hatcurextraerrMWtwoempirical}[1]{\ifnum#1=47 %
\hatcurextraerrMWtwoempiricalxxxxxA
\else
\ifnum#1=48 %
\hatcurextraerrMWtwoempiricalxxxxxB
\else
\ifnum#1=49 %
\hatcurextraerrMWtwoempiricalxxxxxC
\else
\ifnum#1=72 %
\hatcurextraerrMWtwoempiricalxxxxxD
\else
??????\fi
\fi
\fi
\fi
}
\newcommand{\hatcurextraerrMWtwonoisorestrictempirical}[1]{\ifnum#1=47 %
\hatcurextraerrMWtwonoisorestrictempiricalxxxxxA
\else
\ifnum#1=48 %
\hatcurextraerrMWtwonoisorestrictempiricalxxxxxB
\else
\ifnum#1=49 %
\hatcurextraerrMWtwonoisorestrictempiricalxxxxxC
\else
\ifnum#1=72 %
\hatcurextraerrMWtwonoisorestrictempiricalxxxxxD
\else
??????\fi
\fi
\fi
\fi
}
\newcommand{\hatcurextraerrMWtwotwosiglimempirical}[1]{\ifnum#1=47 %
\hatcurextraerrMWtwotwosiglimempiricalxxxxxA
\else
\ifnum#1=48 %
\hatcurextraerrMWtwotwosiglimempiricalxxxxxB
\else
\ifnum#1=49 %
\hatcurextraerrMWtwotwosiglimempiricalxxxxxC
\else
\ifnum#1=72 %
\hatcurextraerrMWtwotwosiglimempiricalxxxxxD
\else
??????\fi
\fi
\fi
\fi
}
\newcommand{\hatcurextraerrMWtwotwosiglimnoisorestrictempirical}[1]{\ifnum#1=47 %
\hatcurextraerrMWtwotwosiglimnoisorestrictempiricalxxxxxA
\else
\ifnum#1=48 %
\hatcurextraerrMWtwotwosiglimnoisorestrictempiricalxxxxxB
\else
\ifnum#1=49 %
\hatcurextraerrMWtwotwosiglimnoisorestrictempiricalxxxxxC
\else
\ifnum#1=72 %
\hatcurextraerrMWtwotwosiglimnoisorestrictempiricalxxxxxD
\else
??????\fi
\fi
\fi
\fi
}
\newcommand{\hatcurfieldempirical}[1]{\ifnum#1=47 %
\hatcurfieldempiricalxxxxxA
\else
\ifnum#1=48 %
\hatcurfieldempiricalxxxxxB
\else
\ifnum#1=49 %
\hatcurfieldempiricalxxxxxC
\else
\ifnum#1=72 %
\hatcurfieldempiricalxxxxxD
\else
??????\fi
\fi
\fi
\fi
}
\newcommand{\hatcurhtrempirical}[1]{\ifnum#1=47 %
\hatcurhtrempiricalxxxxxA
\else
\ifnum#1=48 %
\hatcurhtrempiricalxxxxxB
\else
\ifnum#1=49 %
\hatcurhtrempiricalxxxxxC
\else
\ifnum#1=72 %
\hatcurhtrempiricalxxxxxD
\else
??????\fi
\fi
\fi
\fi
}
\newcommand{\hatcurISOageempirical}[1]{\ifnum#1=47 %
\hatcurISOageempiricalxxxxxA
\else
\ifnum#1=48 %
\hatcurISOageempiricalxxxxxB
\else
\ifnum#1=49 %
\hatcurISOageempiricalxxxxxC
\else
\ifnum#1=72 %
\hatcurISOageempiricalxxxxxD
\else
??????\fi
\fi
\fi
\fi
}
\newcommand{\hatcurISOagenoisorestrictempirical}[1]{\ifnum#1=47 %
\hatcurISOagenoisorestrictempiricalxxxxxA
\else
\ifnum#1=48 %
\hatcurISOagenoisorestrictempiricalxxxxxB
\else
\ifnum#1=49 %
\hatcurISOagenoisorestrictempiricalxxxxxC
\else
\ifnum#1=72 %
\hatcurISOagenoisorestrictempiricalxxxxxD
\else
??????\fi
\fi
\fi
\fi
}
\newcommand{\hatcurISOfehempirical}[1]{\ifnum#1=47 %
\hatcurISOfehempiricalxxxxxA
\else
\ifnum#1=48 %
\hatcurISOfehempiricalxxxxxB
\else
\ifnum#1=49 %
\hatcurISOfehempiricalxxxxxC
\else
\ifnum#1=72 %
\hatcurISOfehempiricalxxxxxD
\else
??????\fi
\fi
\fi
\fi
}
\newcommand{\hatcurISOfehnoisorestrictempirical}[1]{\ifnum#1=47 %
\hatcurISOfehnoisorestrictempiricalxxxxxA
\else
\ifnum#1=48 %
\hatcurISOfehnoisorestrictempiricalxxxxxB
\else
\ifnum#1=49 %
\hatcurISOfehnoisorestrictempiricalxxxxxC
\else
\ifnum#1=72 %
\hatcurISOfehnoisorestrictempiricalxxxxxD
\else
??????\fi
\fi
\fi
\fi
}
\newcommand{\hatcurISOloggempirical}[1]{\ifnum#1=47 %
\hatcurISOloggempiricalxxxxxA
\else
\ifnum#1=48 %
\hatcurISOloggempiricalxxxxxB
\else
\ifnum#1=49 %
\hatcurISOloggempiricalxxxxxC
\else
\ifnum#1=72 %
\hatcurISOloggempiricalxxxxxD
\else
??????\fi
\fi
\fi
\fi
}
\newcommand{\hatcurISOloggnoisorestrictempirical}[1]{\ifnum#1=47 %
\hatcurISOloggnoisorestrictempiricalxxxxxA
\else
\ifnum#1=48 %
\hatcurISOloggnoisorestrictempiricalxxxxxB
\else
\ifnum#1=49 %
\hatcurISOloggnoisorestrictempiricalxxxxxC
\else
\ifnum#1=72 %
\hatcurISOloggnoisorestrictempiricalxxxxxD
\else
??????\fi
\fi
\fi
\fi
}
\newcommand{\hatcurISOlumempirical}[1]{\ifnum#1=47 %
\hatcurISOlumempiricalxxxxxA
\else
\ifnum#1=48 %
\hatcurISOlumempiricalxxxxxB
\else
\ifnum#1=49 %
\hatcurISOlumempiricalxxxxxC
\else
\ifnum#1=72 %
\hatcurISOlumempiricalxxxxxD
\else
??????\fi
\fi
\fi
\fi
}
\newcommand{\hatcurISOlumnoisorestrictempirical}[1]{\ifnum#1=47 %
\hatcurISOlumnoisorestrictempiricalxxxxxA
\else
\ifnum#1=48 %
\hatcurISOlumnoisorestrictempiricalxxxxxB
\else
\ifnum#1=49 %
\hatcurISOlumnoisorestrictempiricalxxxxxC
\else
\ifnum#1=72 %
\hatcurISOlumnoisorestrictempiricalxxxxxD
\else
??????\fi
\fi
\fi
\fi
}
\newcommand{\hatcurISOlumshortempirical}[1]{\ifnum#1=47 %
\hatcurISOlumshortempiricalxxxxxA
\else
\ifnum#1=48 %
\hatcurISOlumshortempiricalxxxxxB
\else
\ifnum#1=49 %
\hatcurISOlumshortempiricalxxxxxC
\else
\ifnum#1=72 %
\hatcurISOlumshortempiricalxxxxxD
\else
??????\fi
\fi
\fi
\fi
}
\newcommand{\hatcurISOlumshortnoisorestrictempirical}[1]{\ifnum#1=47 %
\hatcurISOlumshortnoisorestrictempiricalxxxxxA
\else
\ifnum#1=48 %
\hatcurISOlumshortnoisorestrictempiricalxxxxxB
\else
\ifnum#1=49 %
\hatcurISOlumshortnoisorestrictempiricalxxxxxC
\else
\ifnum#1=72 %
\hatcurISOlumshortnoisorestrictempiricalxxxxxD
\else
??????\fi
\fi
\fi
\fi
}
\newcommand{\hatcurISOmempirical}[1]{\ifnum#1=47 %
\hatcurISOmempiricalxxxxxA
\else
\ifnum#1=48 %
\hatcurISOmempiricalxxxxxB
\else
\ifnum#1=49 %
\hatcurISOmempiricalxxxxxC
\else
\ifnum#1=72 %
\hatcurISOmempiricalxxxxxD
\else
??????\fi
\fi
\fi
\fi
}
\newcommand{\hatcurISOmlongempirical}[1]{\ifnum#1=47 %
\hatcurISOmlongempiricalxxxxxA
\else
\ifnum#1=48 %
\hatcurISOmlongempiricalxxxxxB
\else
\ifnum#1=49 %
\hatcurISOmlongempiricalxxxxxC
\else
\ifnum#1=72 %
\hatcurISOmlongempiricalxxxxxD
\else
??????\fi
\fi
\fi
\fi
}
\newcommand{\hatcurISOmlongnoisorestrictempirical}[1]{\ifnum#1=47 %
\hatcurISOmlongnoisorestrictempiricalxxxxxA
\else
\ifnum#1=48 %
\hatcurISOmlongnoisorestrictempiricalxxxxxB
\else
\ifnum#1=49 %
\hatcurISOmlongnoisorestrictempiricalxxxxxC
\else
\ifnum#1=72 %
\hatcurISOmlongnoisorestrictempiricalxxxxxD
\else
??????\fi
\fi
\fi
\fi
}
\newcommand{\hatcurISOmnoisorestrictempirical}[1]{\ifnum#1=47 %
\hatcurISOmnoisorestrictempiricalxxxxxA
\else
\ifnum#1=48 %
\hatcurISOmnoisorestrictempiricalxxxxxB
\else
\ifnum#1=49 %
\hatcurISOmnoisorestrictempiricalxxxxxC
\else
\ifnum#1=72 %
\hatcurISOmnoisorestrictempiricalxxxxxD
\else
??????\fi
\fi
\fi
\fi
}
\newcommand{\hatcurISOmshortempirical}[1]{\ifnum#1=47 %
\hatcurISOmshortempiricalxxxxxA
\else
\ifnum#1=48 %
\hatcurISOmshortempiricalxxxxxB
\else
\ifnum#1=49 %
\hatcurISOmshortempiricalxxxxxC
\else
\ifnum#1=72 %
\hatcurISOmshortempiricalxxxxxD
\else
??????\fi
\fi
\fi
\fi
}
\newcommand{\hatcurISOmshortnoisorestrictempirical}[1]{\ifnum#1=47 %
\hatcurISOmshortnoisorestrictempiricalxxxxxA
\else
\ifnum#1=48 %
\hatcurISOmshortnoisorestrictempiricalxxxxxB
\else
\ifnum#1=49 %
\hatcurISOmshortnoisorestrictempiricalxxxxxC
\else
\ifnum#1=72 %
\hatcurISOmshortnoisorestrictempiricalxxxxxD
\else
??????\fi
\fi
\fi
\fi
}
\newcommand{\hatcurISOrempirical}[1]{\ifnum#1=47 %
\hatcurISOrempiricalxxxxxA
\else
\ifnum#1=48 %
\hatcurISOrempiricalxxxxxB
\else
\ifnum#1=49 %
\hatcurISOrempiricalxxxxxC
\else
\ifnum#1=72 %
\hatcurISOrempiricalxxxxxD
\else
??????\fi
\fi
\fi
\fi
}
\newcommand{\hatcurISOrlongempirical}[1]{\ifnum#1=47 %
\hatcurISOrlongempiricalxxxxxA
\else
\ifnum#1=48 %
\hatcurISOrlongempiricalxxxxxB
\else
\ifnum#1=49 %
\hatcurISOrlongempiricalxxxxxC
\else
\ifnum#1=72 %
\hatcurISOrlongempiricalxxxxxD
\else
??????\fi
\fi
\fi
\fi
}
\newcommand{\hatcurISOrlongnoisorestrictempirical}[1]{\ifnum#1=47 %
\hatcurISOrlongnoisorestrictempiricalxxxxxA
\else
\ifnum#1=48 %
\hatcurISOrlongnoisorestrictempiricalxxxxxB
\else
\ifnum#1=49 %
\hatcurISOrlongnoisorestrictempiricalxxxxxC
\else
\ifnum#1=72 %
\hatcurISOrlongnoisorestrictempiricalxxxxxD
\else
??????\fi
\fi
\fi
\fi
}
\newcommand{\hatcurISOrnoisorestrictempirical}[1]{\ifnum#1=47 %
\hatcurISOrnoisorestrictempiricalxxxxxA
\else
\ifnum#1=48 %
\hatcurISOrnoisorestrictempiricalxxxxxB
\else
\ifnum#1=49 %
\hatcurISOrnoisorestrictempiricalxxxxxC
\else
\ifnum#1=72 %
\hatcurISOrnoisorestrictempiricalxxxxxD
\else
??????\fi
\fi
\fi
\fi
}
\newcommand{\hatcurISOrshortempirical}[1]{\ifnum#1=47 %
\hatcurISOrshortempiricalxxxxxA
\else
\ifnum#1=48 %
\hatcurISOrshortempiricalxxxxxB
\else
\ifnum#1=49 %
\hatcurISOrshortempiricalxxxxxC
\else
\ifnum#1=72 %
\hatcurISOrshortempiricalxxxxxD
\else
??????\fi
\fi
\fi
\fi
}
\newcommand{\hatcurISOrshortnoisorestrictempirical}[1]{\ifnum#1=47 %
\hatcurISOrshortnoisorestrictempiricalxxxxxA
\else
\ifnum#1=48 %
\hatcurISOrshortnoisorestrictempiricalxxxxxB
\else
\ifnum#1=49 %
\hatcurISOrshortnoisorestrictempiricalxxxxxC
\else
\ifnum#1=72 %
\hatcurISOrshortnoisorestrictempiricalxxxxxD
\else
??????\fi
\fi
\fi
\fi
}
\newcommand{\hatcurISOspecempirical}[1]{\ifnum#1=47 %
\hatcurISOspecempiricalxxxxxA
\else
\ifnum#1=48 %
\hatcurISOspecempiricalxxxxxB
\else
\ifnum#1=49 %
\hatcurISOspecempiricalxxxxxC
\else
\ifnum#1=72 %
\hatcurISOspecempiricalxxxxxD
\else
??????\fi
\fi
\fi
\fi
}
\newcommand{\hatcurISOteffempirical}[1]{\ifnum#1=47 %
\hatcurISOteffempiricalxxxxxA
\else
\ifnum#1=48 %
\hatcurISOteffempiricalxxxxxB
\else
\ifnum#1=49 %
\hatcurISOteffempiricalxxxxxC
\else
\ifnum#1=72 %
\hatcurISOteffempiricalxxxxxD
\else
??????\fi
\fi
\fi
\fi
}
\newcommand{\hatcurISOteffnoisorestrictempirical}[1]{\ifnum#1=47 %
\hatcurISOteffnoisorestrictempiricalxxxxxA
\else
\ifnum#1=48 %
\hatcurISOteffnoisorestrictempiricalxxxxxB
\else
\ifnum#1=49 %
\hatcurISOteffnoisorestrictempiricalxxxxxC
\else
\ifnum#1=72 %
\hatcurISOteffnoisorestrictempiricalxxxxxD
\else
??????\fi
\fi
\fi
\fi
}
\newcommand{\hatcurLBiBempirical}[1]{\ifnum#1=47 %
\hatcurLBiBempiricalxxxxxA
\else
\ifnum#1=48 %
\hatcurLBiBempiricalxxxxxB
\else
\ifnum#1=49 %
\hatcurLBiBempiricalxxxxxC
\else
\ifnum#1=72 %
\hatcurLBiBempiricalxxxxxD
\else
??????\fi
\fi
\fi
\fi
}
\newcommand{\hatcurLBiCempirical}[1]{\ifnum#1=47 %
\hatcurLBiCempiricalxxxxxA
\else
\ifnum#1=48 %
\hatcurLBiCempiricalxxxxxB
\else
\ifnum#1=49 %
\hatcurLBiCempiricalxxxxxC
\else
\ifnum#1=72 %
\hatcurLBiCempiricalxxxxxD
\else
??????\fi
\fi
\fi
\fi
}
\newcommand{\hatcurLBigempirical}[1]{\ifnum#1=47 %
\hatcurLBigempiricalxxxxxA
\else
\ifnum#1=48 %
\hatcurLBigempiricalxxxxxB
\else
\ifnum#1=49 %
\hatcurLBigempiricalxxxxxC
\else
\ifnum#1=72 %
\hatcurLBigempiricalxxxxxD
\else
??????\fi
\fi
\fi
\fi
}
\newcommand{\hatcurLBiHempirical}[1]{\ifnum#1=47 %
\hatcurLBiHempiricalxxxxxA
\else
\ifnum#1=48 %
\hatcurLBiHempiricalxxxxxB
\else
\ifnum#1=49 %
\hatcurLBiHempiricalxxxxxC
\else
\ifnum#1=72 %
\hatcurLBiHempiricalxxxxxD
\else
??????\fi
\fi
\fi
\fi
}
\newcommand{\hatcurLBiiBempirical}[1]{\ifnum#1=47 %
\hatcurLBiiBempiricalxxxxxA
\else
\ifnum#1=48 %
\hatcurLBiiBempiricalxxxxxB
\else
\ifnum#1=49 %
\hatcurLBiiBempiricalxxxxxC
\else
\ifnum#1=72 %
\hatcurLBiiBempiricalxxxxxD
\else
??????\fi
\fi
\fi
\fi
}
\newcommand{\hatcurLBiiCempirical}[1]{\ifnum#1=47 %
\hatcurLBiiCempiricalxxxxxA
\else
\ifnum#1=48 %
\hatcurLBiiCempiricalxxxxxB
\else
\ifnum#1=49 %
\hatcurLBiiCempiricalxxxxxC
\else
\ifnum#1=72 %
\hatcurLBiiCempiricalxxxxxD
\else
??????\fi
\fi
\fi
\fi
}
\newcommand{\hatcurLBiiempirical}[1]{\ifnum#1=47 %
\hatcurLBiiempiricalxxxxxA
\else
\ifnum#1=48 %
\hatcurLBiiempiricalxxxxxB
\else
\ifnum#1=49 %
\hatcurLBiiempiricalxxxxxC
\else
\ifnum#1=72 %
\hatcurLBiiempiricalxxxxxD
\else
??????\fi
\fi
\fi
\fi
}
\newcommand{\hatcurLBiIempirical}[1]{\ifnum#1=47 %
\hatcurLBiIempiricalxxxxxA
\else
\ifnum#1=48 %
\hatcurLBiIempiricalxxxxxB
\else
\ifnum#1=49 %
\hatcurLBiIempiricalxxxxxC
\else
\ifnum#1=72 %
\hatcurLBiIempiricalxxxxxD
\else
??????\fi
\fi
\fi
\fi
}
\newcommand{\hatcurLBiigempirical}[1]{\ifnum#1=47 %
\hatcurLBiigempiricalxxxxxA
\else
\ifnum#1=48 %
\hatcurLBiigempiricalxxxxxB
\else
\ifnum#1=49 %
\hatcurLBiigempiricalxxxxxC
\else
\ifnum#1=72 %
\hatcurLBiigempiricalxxxxxD
\else
??????\fi
\fi
\fi
\fi
}
\newcommand{\hatcurLBiiHempirical}[1]{\ifnum#1=47 %
\hatcurLBiiHempiricalxxxxxA
\else
\ifnum#1=48 %
\hatcurLBiiHempiricalxxxxxB
\else
\ifnum#1=49 %
\hatcurLBiiHempiricalxxxxxC
\else
\ifnum#1=72 %
\hatcurLBiiHempiricalxxxxxD
\else
??????\fi
\fi
\fi
\fi
}
\newcommand{\hatcurLBiiiempirical}[1]{\ifnum#1=47 %
\hatcurLBiiiempiricalxxxxxA
\else
\ifnum#1=48 %
\hatcurLBiiiempiricalxxxxxB
\else
\ifnum#1=49 %
\hatcurLBiiiempiricalxxxxxC
\else
\ifnum#1=72 %
\hatcurLBiiiempiricalxxxxxD
\else
??????\fi
\fi
\fi
\fi
}
\newcommand{\hatcurLBiiIempirical}[1]{\ifnum#1=47 %
\hatcurLBiiIempiricalxxxxxA
\else
\ifnum#1=48 %
\hatcurLBiiIempiricalxxxxxB
\else
\ifnum#1=49 %
\hatcurLBiiIempiricalxxxxxC
\else
\ifnum#1=72 %
\hatcurLBiiIempiricalxxxxxD
\else
??????\fi
\fi
\fi
\fi
}
\newcommand{\hatcurLBiiinoisorestrictempirical}[1]{\ifnum#1=47 %
\hatcurLBiiinoisorestrictempiricalxxxxxA
\else
\ifnum#1=48 %
\hatcurLBiiinoisorestrictempiricalxxxxxB
\else
\ifnum#1=49 %
\hatcurLBiiinoisorestrictempiricalxxxxxC
\else
\ifnum#1=72 %
\hatcurLBiiinoisorestrictempiricalxxxxxD
\else
??????\fi
\fi
\fi
\fi
}
\newcommand{\hatcurLBiiInoisorestrictempirical}[1]{\ifnum#1=72 %
\hatcurLBiiInoisorestrictempiricalxxxxxD
\else
??????\fi
}
\newcommand{\hatcurLBiiJempirical}[1]{\ifnum#1=47 %
\hatcurLBiiJempiricalxxxxxA
\else
\ifnum#1=48 %
\hatcurLBiiJempiricalxxxxxB
\else
\ifnum#1=49 %
\hatcurLBiiJempiricalxxxxxC
\else
\ifnum#1=72 %
\hatcurLBiiJempiricalxxxxxD
\else
??????\fi
\fi
\fi
\fi
}
\newcommand{\hatcurLBiiKempirical}[1]{\ifnum#1=47 %
\hatcurLBiiKempiricalxxxxxA
\else
\ifnum#1=48 %
\hatcurLBiiKempiricalxxxxxB
\else
\ifnum#1=49 %
\hatcurLBiiKempiricalxxxxxC
\else
\ifnum#1=72 %
\hatcurLBiiKempiricalxxxxxD
\else
??????\fi
\fi
\fi
\fi
}
\newcommand{\hatcurLBiikepempirical}[1]{\ifnum#1=47 %
\hatcurLBiikepempiricalxxxxxA
\else
\ifnum#1=48 %
\hatcurLBiikepempiricalxxxxxB
\else
\ifnum#1=49 %
\hatcurLBiikepempiricalxxxxxC
\else
\ifnum#1=72 %
\hatcurLBiikepempiricalxxxxxD
\else
??????\fi
\fi
\fi
\fi
}
\newcommand{\hatcurLBiikepnoisorestrictempirical}[1]{\ifnum#1=72 %
\hatcurLBiikepnoisorestrictempiricalxxxxxD
\else
??????\fi
}
\newcommand{\hatcurLBiiMempirical}[1]{\ifnum#1=47 %
\hatcurLBiiMempiricalxxxxxA
\else
\ifnum#1=48 %
\hatcurLBiiMempiricalxxxxxB
\else
\ifnum#1=49 %
\hatcurLBiiMempiricalxxxxxC
\else
\ifnum#1=72 %
\hatcurLBiiMempiricalxxxxxD
\else
??????\fi
\fi
\fi
\fi
}
\newcommand{\hatcurLBiinoisorestrictempirical}[1]{\ifnum#1=47 %
\hatcurLBiinoisorestrictempiricalxxxxxA
\else
\ifnum#1=48 %
\hatcurLBiinoisorestrictempiricalxxxxxB
\else
\ifnum#1=49 %
\hatcurLBiinoisorestrictempiricalxxxxxC
\else
\ifnum#1=72 %
\hatcurLBiinoisorestrictempiricalxxxxxD
\else
??????\fi
\fi
\fi
\fi
}
\newcommand{\hatcurLBiInoisorestrictempirical}[1]{\ifnum#1=72 %
\hatcurLBiInoisorestrictempiricalxxxxxD
\else
??????\fi
}
\newcommand{\hatcurLBiirempirical}[1]{\ifnum#1=47 %
\hatcurLBiirempiricalxxxxxA
\else
\ifnum#1=48 %
\hatcurLBiirempiricalxxxxxB
\else
\ifnum#1=49 %
\hatcurLBiirempiricalxxxxxC
\else
\ifnum#1=72 %
\hatcurLBiirempiricalxxxxxD
\else
??????\fi
\fi
\fi
\fi
}
\newcommand{\hatcurLBiiRempirical}[1]{\ifnum#1=47 %
\hatcurLBiiRempiricalxxxxxA
\else
\ifnum#1=48 %
\hatcurLBiiRempiricalxxxxxB
\else
\ifnum#1=49 %
\hatcurLBiiRempiricalxxxxxC
\else
\ifnum#1=72 %
\hatcurLBiiRempiricalxxxxxD
\else
??????\fi
\fi
\fi
\fi
}
\newcommand{\hatcurLBiirnoisorestrictempirical}[1]{\ifnum#1=47 %
\hatcurLBiirnoisorestrictempiricalxxxxxA
\else
\ifnum#1=48 %
\hatcurLBiirnoisorestrictempiricalxxxxxB
\else
\ifnum#1=49 %
\hatcurLBiirnoisorestrictempiricalxxxxxC
\else
\ifnum#1=72 %
\hatcurLBiirnoisorestrictempiricalxxxxxD
\else
??????\fi
\fi
\fi
\fi
}
\newcommand{\hatcurLBiiSfourempirical}[1]{\ifnum#1=47 %
\hatcurLBiiSfourempiricalxxxxxA
\else
\ifnum#1=48 %
\hatcurLBiiSfourempiricalxxxxxB
\else
\ifnum#1=49 %
\hatcurLBiiSfourempiricalxxxxxC
\else
\ifnum#1=72 %
\hatcurLBiiSfourempiricalxxxxxD
\else
??????\fi
\fi
\fi
\fi
}
\newcommand{\hatcurLBiiSoneempirical}[1]{\ifnum#1=47 %
\hatcurLBiiSoneempiricalxxxxxA
\else
\ifnum#1=48 %
\hatcurLBiiSoneempiricalxxxxxB
\else
\ifnum#1=49 %
\hatcurLBiiSoneempiricalxxxxxC
\else
\ifnum#1=72 %
\hatcurLBiiSoneempiricalxxxxxD
\else
??????\fi
\fi
\fi
\fi
}
\newcommand{\hatcurLBiiSthreeempirical}[1]{\ifnum#1=47 %
\hatcurLBiiSthreeempiricalxxxxxA
\else
\ifnum#1=48 %
\hatcurLBiiSthreeempiricalxxxxxB
\else
\ifnum#1=49 %
\hatcurLBiiSthreeempiricalxxxxxC
\else
\ifnum#1=72 %
\hatcurLBiiSthreeempiricalxxxxxD
\else
??????\fi
\fi
\fi
\fi
}
\newcommand{\hatcurLBiiStwoempirical}[1]{\ifnum#1=47 %
\hatcurLBiiStwoempiricalxxxxxA
\else
\ifnum#1=48 %
\hatcurLBiiStwoempiricalxxxxxB
\else
\ifnum#1=49 %
\hatcurLBiiStwoempiricalxxxxxC
\else
\ifnum#1=72 %
\hatcurLBiiStwoempiricalxxxxxD
\else
??????\fi
\fi
\fi
\fi
}
\newcommand{\hatcurLBiiTempirical}[1]{\ifnum#1=47 %
\hatcurLBiiTempiricalxxxxxA
\else
\ifnum#1=48 %
\hatcurLBiiTempiricalxxxxxB
\else
\ifnum#1=49 %
\hatcurLBiiTempiricalxxxxxC
\else
\ifnum#1=72 %
\hatcurLBiiTempiricalxxxxxD
\else
??????\fi
\fi
\fi
\fi
}
\newcommand{\hatcurLBiiTnoisorestrictempirical}[1]{\ifnum#1=47 %
\hatcurLBiiTnoisorestrictempiricalxxxxxA
\else
\ifnum#1=48 %
\hatcurLBiiTnoisorestrictempiricalxxxxxB
\else
\ifnum#1=49 %
\hatcurLBiiTnoisorestrictempiricalxxxxxC
\else
\ifnum#1=72 %
\hatcurLBiiTnoisorestrictempiricalxxxxxD
\else
??????\fi
\fi
\fi
\fi
}
\newcommand{\hatcurLBiiuempirical}[1]{\ifnum#1=47 %
\hatcurLBiiuempiricalxxxxxA
\else
\ifnum#1=48 %
\hatcurLBiiuempiricalxxxxxB
\else
\ifnum#1=49 %
\hatcurLBiiuempiricalxxxxxC
\else
\ifnum#1=72 %
\hatcurLBiiuempiricalxxxxxD
\else
??????\fi
\fi
\fi
\fi
}
\newcommand{\hatcurLBiiVempirical}[1]{\ifnum#1=47 %
\hatcurLBiiVempiricalxxxxxA
\else
\ifnum#1=48 %
\hatcurLBiiVempiricalxxxxxB
\else
\ifnum#1=49 %
\hatcurLBiiVempiricalxxxxxC
\else
\ifnum#1=72 %
\hatcurLBiiVempiricalxxxxxD
\else
??????\fi
\fi
\fi
\fi
}
\newcommand{\hatcurLBiizempirical}[1]{\ifnum#1=47 %
\hatcurLBiizempiricalxxxxxA
\else
\ifnum#1=48 %
\hatcurLBiizempiricalxxxxxB
\else
\ifnum#1=49 %
\hatcurLBiizempiricalxxxxxC
\else
\ifnum#1=72 %
\hatcurLBiizempiricalxxxxxD
\else
??????\fi
\fi
\fi
\fi
}
\newcommand{\hatcurLBiJempirical}[1]{\ifnum#1=47 %
\hatcurLBiJempiricalxxxxxA
\else
\ifnum#1=48 %
\hatcurLBiJempiricalxxxxxB
\else
\ifnum#1=49 %
\hatcurLBiJempiricalxxxxxC
\else
\ifnum#1=72 %
\hatcurLBiJempiricalxxxxxD
\else
??????\fi
\fi
\fi
\fi
}
\newcommand{\hatcurLBiKempirical}[1]{\ifnum#1=47 %
\hatcurLBiKempiricalxxxxxA
\else
\ifnum#1=48 %
\hatcurLBiKempiricalxxxxxB
\else
\ifnum#1=49 %
\hatcurLBiKempiricalxxxxxC
\else
\ifnum#1=72 %
\hatcurLBiKempiricalxxxxxD
\else
??????\fi
\fi
\fi
\fi
}
\newcommand{\hatcurLBikepempirical}[1]{\ifnum#1=47 %
\hatcurLBikepempiricalxxxxxA
\else
\ifnum#1=48 %
\hatcurLBikepempiricalxxxxxB
\else
\ifnum#1=49 %
\hatcurLBikepempiricalxxxxxC
\else
\ifnum#1=72 %
\hatcurLBikepempiricalxxxxxD
\else
??????\fi
\fi
\fi
\fi
}
\newcommand{\hatcurLBikepnoisorestrictempirical}[1]{\ifnum#1=72 %
\hatcurLBikepnoisorestrictempiricalxxxxxD
\else
??????\fi
}
\newcommand{\hatcurLBiMempirical}[1]{\ifnum#1=47 %
\hatcurLBiMempiricalxxxxxA
\else
\ifnum#1=48 %
\hatcurLBiMempiricalxxxxxB
\else
\ifnum#1=49 %
\hatcurLBiMempiricalxxxxxC
\else
\ifnum#1=72 %
\hatcurLBiMempiricalxxxxxD
\else
??????\fi
\fi
\fi
\fi
}
\newcommand{\hatcurLBirempirical}[1]{\ifnum#1=47 %
\hatcurLBirempiricalxxxxxA
\else
\ifnum#1=48 %
\hatcurLBirempiricalxxxxxB
\else
\ifnum#1=49 %
\hatcurLBirempiricalxxxxxC
\else
\ifnum#1=72 %
\hatcurLBirempiricalxxxxxD
\else
??????\fi
\fi
\fi
\fi
}
\newcommand{\hatcurLBiRempirical}[1]{\ifnum#1=47 %
\hatcurLBiRempiricalxxxxxA
\else
\ifnum#1=48 %
\hatcurLBiRempiricalxxxxxB
\else
\ifnum#1=49 %
\hatcurLBiRempiricalxxxxxC
\else
\ifnum#1=72 %
\hatcurLBiRempiricalxxxxxD
\else
??????\fi
\fi
\fi
\fi
}
\newcommand{\hatcurLBirnoisorestrictempirical}[1]{\ifnum#1=47 %
\hatcurLBirnoisorestrictempiricalxxxxxA
\else
\ifnum#1=48 %
\hatcurLBirnoisorestrictempiricalxxxxxB
\else
\ifnum#1=49 %
\hatcurLBirnoisorestrictempiricalxxxxxC
\else
\ifnum#1=72 %
\hatcurLBirnoisorestrictempiricalxxxxxD
\else
??????\fi
\fi
\fi
\fi
}
\newcommand{\hatcurLBiSfourempirical}[1]{\ifnum#1=47 %
\hatcurLBiSfourempiricalxxxxxA
\else
\ifnum#1=48 %
\hatcurLBiSfourempiricalxxxxxB
\else
\ifnum#1=49 %
\hatcurLBiSfourempiricalxxxxxC
\else
\ifnum#1=72 %
\hatcurLBiSfourempiricalxxxxxD
\else
??????\fi
\fi
\fi
\fi
}
\newcommand{\hatcurLBiSoneempirical}[1]{\ifnum#1=47 %
\hatcurLBiSoneempiricalxxxxxA
\else
\ifnum#1=48 %
\hatcurLBiSoneempiricalxxxxxB
\else
\ifnum#1=49 %
\hatcurLBiSoneempiricalxxxxxC
\else
\ifnum#1=72 %
\hatcurLBiSoneempiricalxxxxxD
\else
??????\fi
\fi
\fi
\fi
}
\newcommand{\hatcurLBiSthreeempirical}[1]{\ifnum#1=47 %
\hatcurLBiSthreeempiricalxxxxxA
\else
\ifnum#1=48 %
\hatcurLBiSthreeempiricalxxxxxB
\else
\ifnum#1=49 %
\hatcurLBiSthreeempiricalxxxxxC
\else
\ifnum#1=72 %
\hatcurLBiSthreeempiricalxxxxxD
\else
??????\fi
\fi
\fi
\fi
}
\newcommand{\hatcurLBiStwoempirical}[1]{\ifnum#1=47 %
\hatcurLBiStwoempiricalxxxxxA
\else
\ifnum#1=48 %
\hatcurLBiStwoempiricalxxxxxB
\else
\ifnum#1=49 %
\hatcurLBiStwoempiricalxxxxxC
\else
\ifnum#1=72 %
\hatcurLBiStwoempiricalxxxxxD
\else
??????\fi
\fi
\fi
\fi
}
\newcommand{\hatcurLBiTempirical}[1]{\ifnum#1=47 %
\hatcurLBiTempiricalxxxxxA
\else
\ifnum#1=48 %
\hatcurLBiTempiricalxxxxxB
\else
\ifnum#1=49 %
\hatcurLBiTempiricalxxxxxC
\else
\ifnum#1=72 %
\hatcurLBiTempiricalxxxxxD
\else
??????\fi
\fi
\fi
\fi
}
\newcommand{\hatcurLBiTnoisorestrictempirical}[1]{\ifnum#1=47 %
\hatcurLBiTnoisorestrictempiricalxxxxxA
\else
\ifnum#1=48 %
\hatcurLBiTnoisorestrictempiricalxxxxxB
\else
\ifnum#1=49 %
\hatcurLBiTnoisorestrictempiricalxxxxxC
\else
\ifnum#1=72 %
\hatcurLBiTnoisorestrictempiricalxxxxxD
\else
??????\fi
\fi
\fi
\fi
}
\newcommand{\hatcurLBiuempirical}[1]{\ifnum#1=47 %
\hatcurLBiuempiricalxxxxxA
\else
\ifnum#1=48 %
\hatcurLBiuempiricalxxxxxB
\else
\ifnum#1=49 %
\hatcurLBiuempiricalxxxxxC
\else
\ifnum#1=72 %
\hatcurLBiuempiricalxxxxxD
\else
??????\fi
\fi
\fi
\fi
}
\newcommand{\hatcurLBiVempirical}[1]{\ifnum#1=47 %
\hatcurLBiVempiricalxxxxxA
\else
\ifnum#1=48 %
\hatcurLBiVempiricalxxxxxB
\else
\ifnum#1=49 %
\hatcurLBiVempiricalxxxxxC
\else
\ifnum#1=72 %
\hatcurLBiVempiricalxxxxxD
\else
??????\fi
\fi
\fi
\fi
}
\newcommand{\hatcurLBizempirical}[1]{\ifnum#1=47 %
\hatcurLBizempiricalxxxxxA
\else
\ifnum#1=48 %
\hatcurLBizempiricalxxxxxB
\else
\ifnum#1=49 %
\hatcurLBizempiricalxxxxxC
\else
\ifnum#1=72 %
\hatcurLBizempiricalxxxxxD
\else
??????\fi
\fi
\fi
\fi
}
\newcommand{\hatcurLCbsqempirical}[1]{\ifnum#1=47 %
\hatcurLCbsqempiricalxxxxxA
\else
\ifnum#1=48 %
\hatcurLCbsqempiricalxxxxxB
\else
\ifnum#1=49 %
\hatcurLCbsqempiricalxxxxxC
\else
\ifnum#1=72 %
\hatcurLCbsqempiricalxxxxxD
\else
??????\fi
\fi
\fi
\fi
}
\newcommand{\hatcurLCbsqnoisorestrictempirical}[1]{\ifnum#1=47 %
\hatcurLCbsqnoisorestrictempiricalxxxxxA
\else
\ifnum#1=48 %
\hatcurLCbsqnoisorestrictempiricalxxxxxB
\else
\ifnum#1=49 %
\hatcurLCbsqnoisorestrictempiricalxxxxxC
\else
\ifnum#1=72 %
\hatcurLCbsqnoisorestrictempiricalxxxxxD
\else
??????\fi
\fi
\fi
\fi
}
\newcommand{\hatcurLCdipempirical}[1]{\ifnum#1=47 %
\hatcurLCdipempiricalxxxxxA
\else
\ifnum#1=48 %
\hatcurLCdipempiricalxxxxxB
\else
\ifnum#1=49 %
\hatcurLCdipempiricalxxxxxC
\else
\ifnum#1=72 %
\hatcurLCdipempiricalxxxxxD
\else
??????\fi
\fi
\fi
\fi
}
\newcommand{\hatcurLCdurempirical}[1]{\ifnum#1=47 %
\hatcurLCdurempiricalxxxxxA
\else
\ifnum#1=48 %
\hatcurLCdurempiricalxxxxxB
\else
\ifnum#1=49 %
\hatcurLCdurempiricalxxxxxC
\else
\ifnum#1=72 %
\hatcurLCdurempiricalxxxxxD
\else
??????\fi
\fi
\fi
\fi
}
\newcommand{\hatcurLCdurhrempirical}[1]{\ifnum#1=47 %
\hatcurLCdurhrempiricalxxxxxA
\else
\ifnum#1=48 %
\hatcurLCdurhrempiricalxxxxxB
\else
\ifnum#1=49 %
\hatcurLCdurhrempiricalxxxxxC
\else
\ifnum#1=72 %
\hatcurLCdurhrempiricalxxxxxD
\else
??????\fi
\fi
\fi
\fi
}
\newcommand{\hatcurLCdurhrnoisorestrictempirical}[1]{\ifnum#1=47 %
\hatcurLCdurhrnoisorestrictempiricalxxxxxA
\else
\ifnum#1=48 %
\hatcurLCdurhrnoisorestrictempiricalxxxxxB
\else
\ifnum#1=49 %
\hatcurLCdurhrnoisorestrictempiricalxxxxxC
\else
\ifnum#1=72 %
\hatcurLCdurhrnoisorestrictempiricalxxxxxD
\else
??????\fi
\fi
\fi
\fi
}
\newcommand{\hatcurLCdurhrshortempirical}[1]{\ifnum#1=47 %
\hatcurLCdurhrshortempiricalxxxxxA
\else
\ifnum#1=48 %
\hatcurLCdurhrshortempiricalxxxxxB
\else
\ifnum#1=49 %
\hatcurLCdurhrshortempiricalxxxxxC
\else
\ifnum#1=72 %
\hatcurLCdurhrshortempiricalxxxxxD
\else
??????\fi
\fi
\fi
\fi
}
\newcommand{\hatcurLCdurhrshortnoisorestrictempirical}[1]{\ifnum#1=47 %
\hatcurLCdurhrshortnoisorestrictempiricalxxxxxA
\else
\ifnum#1=48 %
\hatcurLCdurhrshortnoisorestrictempiricalxxxxxB
\else
\ifnum#1=49 %
\hatcurLCdurhrshortnoisorestrictempiricalxxxxxC
\else
\ifnum#1=72 %
\hatcurLCdurhrshortnoisorestrictempiricalxxxxxD
\else
??????\fi
\fi
\fi
\fi
}
\newcommand{\hatcurLCdurnoisorestrictempirical}[1]{\ifnum#1=47 %
\hatcurLCdurnoisorestrictempiricalxxxxxA
\else
\ifnum#1=48 %
\hatcurLCdurnoisorestrictempiricalxxxxxB
\else
\ifnum#1=49 %
\hatcurLCdurnoisorestrictempiricalxxxxxC
\else
\ifnum#1=72 %
\hatcurLCdurnoisorestrictempiricalxxxxxD
\else
??????\fi
\fi
\fi
\fi
}
\newcommand{\hatcurLCdurshortempirical}[1]{\ifnum#1=47 %
\hatcurLCdurshortempiricalxxxxxA
\else
\ifnum#1=48 %
\hatcurLCdurshortempiricalxxxxxB
\else
\ifnum#1=49 %
\hatcurLCdurshortempiricalxxxxxC
\else
\ifnum#1=72 %
\hatcurLCdurshortempiricalxxxxxD
\else
??????\fi
\fi
\fi
\fi
}
\newcommand{\hatcurLCdurshortnoisorestrictempirical}[1]{\ifnum#1=47 %
\hatcurLCdurshortnoisorestrictempiricalxxxxxA
\else
\ifnum#1=48 %
\hatcurLCdurshortnoisorestrictempiricalxxxxxB
\else
\ifnum#1=49 %
\hatcurLCdurshortnoisorestrictempiricalxxxxxC
\else
\ifnum#1=72 %
\hatcurLCdurshortnoisorestrictempiricalxxxxxD
\else
??????\fi
\fi
\fi
\fi
}
\newcommand{\hatcurLChatnetmAempirical}[1]{\ifnum#1=47 %
\hatcurLChatnetmAempiricalxxxxxA
\else
\ifnum#1=48 %
\hatcurLChatnetmAempiricalxxxxxB
\else
\ifnum#1=49 %
\hatcurLChatnetmAempiricalxxxxxC
\else
\ifnum#1=72 %
\hatcurLChatnetmAempiricalxxxxxD
\else
??????\fi
\fi
\fi
\fi
}
\newcommand{\hatcurLChatnetmAnoisorestrictempirical}[1]{\ifnum#1=47 %
\hatcurLChatnetmAnoisorestrictempiricalxxxxxA
\else
\ifnum#1=48 %
\hatcurLChatnetmAnoisorestrictempiricalxxxxxB
\else
\ifnum#1=49 %
\hatcurLChatnetmAnoisorestrictempiricalxxxxxC
\else
\ifnum#1=72 %
\hatcurLChatnetmAnoisorestrictempiricalxxxxxD
\else
??????\fi
\fi
\fi
\fi
}
\newcommand{\hatcurLChatnetmBempirical}[1]{\ifnum#1=47 %
\hatcurLChatnetmBempiricalxxxxxA
\else
\ifnum#1=48 %
\hatcurLChatnetmBempiricalxxxxxB
\else
\ifnum#1=49 %
\hatcurLChatnetmBempiricalxxxxxC
\else
\ifnum#1=72 %
\hatcurLChatnetmBempiricalxxxxxD
\else
??????\fi
\fi
\fi
\fi
}
\newcommand{\hatcurLChatnetmBnoisorestrictempirical}[1]{\ifnum#1=47 %
\hatcurLChatnetmBnoisorestrictempiricalxxxxxA
\else
\ifnum#1=48 %
\hatcurLChatnetmBnoisorestrictempiricalxxxxxB
\else
\ifnum#1=49 %
\hatcurLChatnetmBnoisorestrictempiricalxxxxxC
\else
\ifnum#1=72 %
\hatcurLChatnetmBnoisorestrictempiricalxxxxxD
\else
??????\fi
\fi
\fi
\fi
}
\newcommand{\hatcurLChatnetmCempirical}[1]{\ifnum#1=49 %
\hatcurLChatnetmCempiricalxxxxxC
\else
\ifnum#1=72 %
\hatcurLChatnetmCempiricalxxxxxD
\else
??????\fi
\fi
}
\newcommand{\hatcurLChatnetmCnoisorestrictempirical}[1]{\ifnum#1=49 %
\hatcurLChatnetmCnoisorestrictempiricalxxxxxC
\else
\ifnum#1=72 %
\hatcurLChatnetmCnoisorestrictempiricalxxxxxD
\else
??????\fi
\fi
}
\newcommand{\hatcurLCiblendAempirical}[1]{\ifnum#1=47 %
\hatcurLCiblendAempiricalxxxxxA
\else
\ifnum#1=48 %
\hatcurLCiblendAempiricalxxxxxB
\else
\ifnum#1=49 %
\hatcurLCiblendAempiricalxxxxxC
\else
\ifnum#1=72 %
\hatcurLCiblendAempiricalxxxxxD
\else
??????\fi
\fi
\fi
\fi
}
\newcommand{\hatcurLCiblendAnoisorestrictempirical}[1]{\ifnum#1=47 %
\hatcurLCiblendAnoisorestrictempiricalxxxxxA
\else
\ifnum#1=48 %
\hatcurLCiblendAnoisorestrictempiricalxxxxxB
\else
\ifnum#1=49 %
\hatcurLCiblendAnoisorestrictempiricalxxxxxC
\else
\ifnum#1=72 %
\hatcurLCiblendAnoisorestrictempiricalxxxxxD
\else
??????\fi
\fi
\fi
\fi
}
\newcommand{\hatcurLCiblendBempirical}[1]{\ifnum#1=47 %
\hatcurLCiblendBempiricalxxxxxA
\else
\ifnum#1=48 %
\hatcurLCiblendBempiricalxxxxxB
\else
\ifnum#1=49 %
\hatcurLCiblendBempiricalxxxxxC
\else
\ifnum#1=72 %
\hatcurLCiblendBempiricalxxxxxD
\else
??????\fi
\fi
\fi
\fi
}
\newcommand{\hatcurLCiblendBnoisorestrictempirical}[1]{\ifnum#1=47 %
\hatcurLCiblendBnoisorestrictempiricalxxxxxA
\else
\ifnum#1=48 %
\hatcurLCiblendBnoisorestrictempiricalxxxxxB
\else
\ifnum#1=49 %
\hatcurLCiblendBnoisorestrictempiricalxxxxxC
\else
\ifnum#1=72 %
\hatcurLCiblendBnoisorestrictempiricalxxxxxD
\else
??????\fi
\fi
\fi
\fi
}
\newcommand{\hatcurLCiblendCempirical}[1]{\ifnum#1=49 %
\hatcurLCiblendCempiricalxxxxxC
\else
\ifnum#1=72 %
\hatcurLCiblendCempiricalxxxxxD
\else
??????\fi
\fi
}
\newcommand{\hatcurLCiblendCnoisorestrictempirical}[1]{\ifnum#1=49 %
\hatcurLCiblendCnoisorestrictempiricalxxxxxC
\else
\ifnum#1=72 %
\hatcurLCiblendCnoisorestrictempiricalxxxxxD
\else
??????\fi
\fi
}
\newcommand{\hatcurLCimpempirical}[1]{\ifnum#1=47 %
\hatcurLCimpempiricalxxxxxA
\else
\ifnum#1=48 %
\hatcurLCimpempiricalxxxxxB
\else
\ifnum#1=49 %
\hatcurLCimpempiricalxxxxxC
\else
\ifnum#1=72 %
\hatcurLCimpempiricalxxxxxD
\else
??????\fi
\fi
\fi
\fi
}
\newcommand{\hatcurLCimpnoisorestrictempirical}[1]{\ifnum#1=47 %
\hatcurLCimpnoisorestrictempiricalxxxxxA
\else
\ifnum#1=48 %
\hatcurLCimpnoisorestrictempiricalxxxxxB
\else
\ifnum#1=49 %
\hatcurLCimpnoisorestrictempiricalxxxxxC
\else
\ifnum#1=72 %
\hatcurLCimpnoisorestrictempiricalxxxxxD
\else
??????\fi
\fi
\fi
\fi
}
\newcommand{\hatcurLCingdurempirical}[1]{\ifnum#1=47 %
\hatcurLCingdurempiricalxxxxxA
\else
\ifnum#1=48 %
\hatcurLCingdurempiricalxxxxxB
\else
\ifnum#1=49 %
\hatcurLCingdurempiricalxxxxxC
\else
\ifnum#1=72 %
\hatcurLCingdurempiricalxxxxxD
\else
??????\fi
\fi
\fi
\fi
}
\newcommand{\hatcurLCingdurnoisorestrictempirical}[1]{\ifnum#1=47 %
\hatcurLCingdurnoisorestrictempiricalxxxxxA
\else
\ifnum#1=48 %
\hatcurLCingdurnoisorestrictempiricalxxxxxB
\else
\ifnum#1=49 %
\hatcurLCingdurnoisorestrictempiricalxxxxxC
\else
\ifnum#1=72 %
\hatcurLCingdurnoisorestrictempiricalxxxxxD
\else
??????\fi
\fi
\fi
\fi
}
\newcommand{\hatcurLCPempirical}[1]{\ifnum#1=47 %
\hatcurLCPempiricalxxxxxA
\else
\ifnum#1=48 %
\hatcurLCPempiricalxxxxxB
\else
\ifnum#1=49 %
\hatcurLCPempiricalxxxxxC
\else
\ifnum#1=72 %
\hatcurLCPempiricalxxxxxD
\else
??????\fi
\fi
\fi
\fi
}
\newcommand{\hatcurLCPnoisorestrictempirical}[1]{\ifnum#1=47 %
\hatcurLCPnoisorestrictempiricalxxxxxA
\else
\ifnum#1=48 %
\hatcurLCPnoisorestrictempiricalxxxxxB
\else
\ifnum#1=49 %
\hatcurLCPnoisorestrictempiricalxxxxxC
\else
\ifnum#1=72 %
\hatcurLCPnoisorestrictempiricalxxxxxD
\else
??????\fi
\fi
\fi
\fi
}
\newcommand{\hatcurLCPprecempirical}[1]{\ifnum#1=47 %
\hatcurLCPprecempiricalxxxxxA
\else
\ifnum#1=48 %
\hatcurLCPprecempiricalxxxxxB
\else
\ifnum#1=49 %
\hatcurLCPprecempiricalxxxxxC
\else
\ifnum#1=72 %
\hatcurLCPprecempiricalxxxxxD
\else
??????\fi
\fi
\fi
\fi
}
\newcommand{\hatcurLCPprecnoisorestrictempirical}[1]{\ifnum#1=47 %
\hatcurLCPprecnoisorestrictempiricalxxxxxA
\else
\ifnum#1=48 %
\hatcurLCPprecnoisorestrictempiricalxxxxxB
\else
\ifnum#1=49 %
\hatcurLCPprecnoisorestrictempiricalxxxxxC
\else
\ifnum#1=72 %
\hatcurLCPprecnoisorestrictempiricalxxxxxD
\else
??????\fi
\fi
\fi
\fi
}
\newcommand{\hatcurLCPshortempirical}[1]{\ifnum#1=47 %
\hatcurLCPshortempiricalxxxxxA
\else
\ifnum#1=48 %
\hatcurLCPshortempiricalxxxxxB
\else
\ifnum#1=49 %
\hatcurLCPshortempiricalxxxxxC
\else
\ifnum#1=72 %
\hatcurLCPshortempiricalxxxxxD
\else
??????\fi
\fi
\fi
\fi
}
\newcommand{\hatcurLCPshortnoisorestrictempirical}[1]{\ifnum#1=47 %
\hatcurLCPshortnoisorestrictempiricalxxxxxA
\else
\ifnum#1=48 %
\hatcurLCPshortnoisorestrictempiricalxxxxxB
\else
\ifnum#1=49 %
\hatcurLCPshortnoisorestrictempiricalxxxxxC
\else
\ifnum#1=72 %
\hatcurLCPshortnoisorestrictempiricalxxxxxD
\else
??????\fi
\fi
\fi
\fi
}
\newcommand{\hatcurLCqempirical}[1]{\ifnum#1=47 %
\hatcurLCqempiricalxxxxxA
\else
\ifnum#1=48 %
\hatcurLCqempiricalxxxxxB
\else
\ifnum#1=49 %
\hatcurLCqempiricalxxxxxC
\else
\ifnum#1=72 %
\hatcurLCqempiricalxxxxxD
\else
??????\fi
\fi
\fi
\fi
}
\newcommand{\hatcurLCqnoisorestrictempirical}[1]{\ifnum#1=47 %
\hatcurLCqnoisorestrictempiricalxxxxxA
\else
\ifnum#1=48 %
\hatcurLCqnoisorestrictempiricalxxxxxB
\else
\ifnum#1=49 %
\hatcurLCqnoisorestrictempiricalxxxxxC
\else
\ifnum#1=72 %
\hatcurLCqnoisorestrictempiricalxxxxxD
\else
??????\fi
\fi
\fi
\fi
}
\newcommand{\hatcurLCqshortempirical}[1]{\ifnum#1=47 %
\hatcurLCqshortempiricalxxxxxA
\else
\ifnum#1=48 %
\hatcurLCqshortempiricalxxxxxB
\else
\ifnum#1=49 %
\hatcurLCqshortempiricalxxxxxC
\else
\ifnum#1=72 %
\hatcurLCqshortempiricalxxxxxD
\else
??????\fi
\fi
\fi
\fi
}
\newcommand{\hatcurLCqshortnoisorestrictempirical}[1]{\ifnum#1=47 %
\hatcurLCqshortnoisorestrictempiricalxxxxxA
\else
\ifnum#1=48 %
\hatcurLCqshortnoisorestrictempiricalxxxxxB
\else
\ifnum#1=49 %
\hatcurLCqshortnoisorestrictempiricalxxxxxC
\else
\ifnum#1=72 %
\hatcurLCqshortnoisorestrictempiricalxxxxxD
\else
??????\fi
\fi
\fi
\fi
}
\newcommand{\hatcurLCrhoempirical}[1]{\ifnum#1=47 %
\hatcurLCrhoempiricalxxxxxA
\else
\ifnum#1=48 %
\hatcurLCrhoempiricalxxxxxB
\else
\ifnum#1=49 %
\hatcurLCrhoempiricalxxxxxC
\else
\ifnum#1=72 %
\hatcurLCrhoempiricalxxxxxD
\else
??????\fi
\fi
\fi
\fi
}
\newcommand{\hatcurLCrhonoisorestrictempirical}[1]{\ifnum#1=47 %
\hatcurLCrhonoisorestrictempiricalxxxxxA
\else
\ifnum#1=48 %
\hatcurLCrhonoisorestrictempiricalxxxxxB
\else
\ifnum#1=49 %
\hatcurLCrhonoisorestrictempiricalxxxxxC
\else
\ifnum#1=72 %
\hatcurLCrhonoisorestrictempiricalxxxxxD
\else
??????\fi
\fi
\fi
\fi
}
\newcommand{\hatcurLCrprstarempirical}[1]{\ifnum#1=47 %
\hatcurLCrprstarempiricalxxxxxA
\else
\ifnum#1=48 %
\hatcurLCrprstarempiricalxxxxxB
\else
\ifnum#1=49 %
\hatcurLCrprstarempiricalxxxxxC
\else
\ifnum#1=72 %
\hatcurLCrprstarempiricalxxxxxD
\else
??????\fi
\fi
\fi
\fi
}
\newcommand{\hatcurLCrprstarnoisorestrictempirical}[1]{\ifnum#1=47 %
\hatcurLCrprstarnoisorestrictempiricalxxxxxA
\else
\ifnum#1=48 %
\hatcurLCrprstarnoisorestrictempiricalxxxxxB
\else
\ifnum#1=49 %
\hatcurLCrprstarnoisorestrictempiricalxxxxxC
\else
\ifnum#1=72 %
\hatcurLCrprstarnoisorestrictempiricalxxxxxD
\else
??????\fi
\fi
\fi
\fi
}
\newcommand{\hatcurLCTAempirical}[1]{\ifnum#1=47 %
\hatcurLCTAempiricalxxxxxA
\else
\ifnum#1=48 %
\hatcurLCTAempiricalxxxxxB
\else
\ifnum#1=49 %
\hatcurLCTAempiricalxxxxxC
\else
\ifnum#1=72 %
\hatcurLCTAempiricalxxxxxD
\else
??????\fi
\fi
\fi
\fi
}
\newcommand{\hatcurLCTAnoisorestrictempirical}[1]{\ifnum#1=47 %
\hatcurLCTAnoisorestrictempiricalxxxxxA
\else
\ifnum#1=48 %
\hatcurLCTAnoisorestrictempiricalxxxxxB
\else
\ifnum#1=49 %
\hatcurLCTAnoisorestrictempiricalxxxxxC
\else
\ifnum#1=72 %
\hatcurLCTAnoisorestrictempiricalxxxxxD
\else
??????\fi
\fi
\fi
\fi
}
\newcommand{\hatcurLCTBempirical}[1]{\ifnum#1=47 %
\hatcurLCTBempiricalxxxxxA
\else
\ifnum#1=48 %
\hatcurLCTBempiricalxxxxxB
\else
\ifnum#1=49 %
\hatcurLCTBempiricalxxxxxC
\else
\ifnum#1=72 %
\hatcurLCTBempiricalxxxxxD
\else
??????\fi
\fi
\fi
\fi
}
\newcommand{\hatcurLCTBnoisorestrictempirical}[1]{\ifnum#1=47 %
\hatcurLCTBnoisorestrictempiricalxxxxxA
\else
\ifnum#1=48 %
\hatcurLCTBnoisorestrictempiricalxxxxxB
\else
\ifnum#1=49 %
\hatcurLCTBnoisorestrictempiricalxxxxxC
\else
\ifnum#1=72 %
\hatcurLCTBnoisorestrictempiricalxxxxxD
\else
??????\fi
\fi
\fi
\fi
}
\newcommand{\hatcurLCTempirical}[1]{\ifnum#1=47 %
\hatcurLCTempiricalxxxxxA
\else
\ifnum#1=48 %
\hatcurLCTempiricalxxxxxB
\else
\ifnum#1=49 %
\hatcurLCTempiricalxxxxxC
\else
\ifnum#1=72 %
\hatcurLCTempiricalxxxxxD
\else
??????\fi
\fi
\fi
\fi
}
\newcommand{\hatcurLCTnoisorestrictempirical}[1]{\ifnum#1=47 %
\hatcurLCTnoisorestrictempiricalxxxxxA
\else
\ifnum#1=48 %
\hatcurLCTnoisorestrictempiricalxxxxxB
\else
\ifnum#1=49 %
\hatcurLCTnoisorestrictempiricalxxxxxC
\else
\ifnum#1=72 %
\hatcurLCTnoisorestrictempiricalxxxxxD
\else
??????\fi
\fi
\fi
\fi
}
\newcommand{\hatcurLCzetaempirical}[1]{\ifnum#1=47 %
\hatcurLCzetaempiricalxxxxxA
\else
\ifnum#1=48 %
\hatcurLCzetaempiricalxxxxxB
\else
\ifnum#1=49 %
\hatcurLCzetaempiricalxxxxxC
\else
\ifnum#1=72 %
\hatcurLCzetaempiricalxxxxxD
\else
??????\fi
\fi
\fi
\fi
}
\newcommand{\hatcurLCzetanoisorestrictempirical}[1]{\ifnum#1=47 %
\hatcurLCzetanoisorestrictempiricalxxxxxA
\else
\ifnum#1=48 %
\hatcurLCzetanoisorestrictempiricalxxxxxB
\else
\ifnum#1=49 %
\hatcurLCzetanoisorestrictempiricalxxxxxC
\else
\ifnum#1=72 %
\hatcurLCzetanoisorestrictempiricalxxxxxD
\else
??????\fi
\fi
\fi
\fi
}
\newcommand{\hatcurPPaequivempirical}[1]{\ifnum#1=47 %
\hatcurPPaequivempiricalxxxxxA
\else
\ifnum#1=48 %
\hatcurPPaequivempiricalxxxxxB
\else
\ifnum#1=49 %
\hatcurPPaequivempiricalxxxxxC
\else
??????\fi
\fi
\fi
}
\newcommand{\hatcurPPaequivnoisorestrictempirical}[1]{\ifnum#1=47 %
\hatcurPPaequivnoisorestrictempiricalxxxxxA
\else
\ifnum#1=48 %
\hatcurPPaequivnoisorestrictempiricalxxxxxB
\else
\ifnum#1=49 %
\hatcurPPaequivnoisorestrictempiricalxxxxxC
\else
??????\fi
\fi
\fi
}
\newcommand{\hatcurPParelempirical}[1]{\ifnum#1=47 %
\hatcurPParelempiricalxxxxxA
\else
\ifnum#1=48 %
\hatcurPParelempiricalxxxxxB
\else
\ifnum#1=49 %
\hatcurPParelempiricalxxxxxC
\else
??????\fi
\fi
\fi
}
\newcommand{\hatcurPParelnoisorestrictempirical}[1]{\ifnum#1=47 %
\hatcurPParelnoisorestrictempiricalxxxxxA
\else
\ifnum#1=48 %
\hatcurPParelnoisorestrictempiricalxxxxxB
\else
\ifnum#1=49 %
\hatcurPParelnoisorestrictempiricalxxxxxC
\else
??????\fi
\fi
\fi
}
\newcommand{\hatcurPParempirical}[1]{\ifnum#1=47 %
\hatcurPParempiricalxxxxxA
\else
\ifnum#1=48 %
\hatcurPParempiricalxxxxxB
\else
\ifnum#1=49 %
\hatcurPParempiricalxxxxxC
\else
??????\fi
\fi
\fi
}
\newcommand{\hatcurPParnoisorestrictempirical}[1]{\ifnum#1=47 %
\hatcurPParnoisorestrictempiricalxxxxxA
\else
\ifnum#1=48 %
\hatcurPParnoisorestrictempiricalxxxxxB
\else
\ifnum#1=49 %
\hatcurPParnoisorestrictempiricalxxxxxC
\else
??????\fi
\fi
\fi
}
\newcommand{\hatcurPPfluxapdimempirical}[1]{\ifnum#1=47 %
\hatcurPPfluxapdimempiricalxxxxxA
\else
\ifnum#1=48 %
\hatcurPPfluxapdimempiricalxxxxxB
\else
\ifnum#1=49 %
\hatcurPPfluxapdimempiricalxxxxxC
\else
??????\fi
\fi
\fi
}
\newcommand{\hatcurPPfluxapempirical}[1]{\ifnum#1=47 %
\hatcurPPfluxapempiricalxxxxxA
\else
\ifnum#1=48 %
\hatcurPPfluxapempiricalxxxxxB
\else
\ifnum#1=49 %
\hatcurPPfluxapempiricalxxxxxC
\else
??????\fi
\fi
\fi
}
\newcommand{\hatcurPPfluxapnoisorestrictempirical}[1]{\ifnum#1=47 %
\hatcurPPfluxapnoisorestrictempiricalxxxxxA
\else
\ifnum#1=48 %
\hatcurPPfluxapnoisorestrictempiricalxxxxxB
\else
\ifnum#1=49 %
\hatcurPPfluxapnoisorestrictempiricalxxxxxC
\else
??????\fi
\fi
\fi
}
\newcommand{\hatcurPPfluxavgdimempirical}[1]{\ifnum#1=47 %
\hatcurPPfluxavgdimempiricalxxxxxA
\else
\ifnum#1=48 %
\hatcurPPfluxavgdimempiricalxxxxxB
\else
\ifnum#1=49 %
\hatcurPPfluxavgdimempiricalxxxxxC
\else
??????\fi
\fi
\fi
}
\newcommand{\hatcurPPfluxavgempirical}[1]{\ifnum#1=47 %
\hatcurPPfluxavgempiricalxxxxxA
\else
\ifnum#1=48 %
\hatcurPPfluxavgempiricalxxxxxB
\else
\ifnum#1=49 %
\hatcurPPfluxavgempiricalxxxxxC
\else
??????\fi
\fi
\fi
}
\newcommand{\hatcurPPfluxavglogempirical}[1]{\ifnum#1=47 %
\hatcurPPfluxavglogempiricalxxxxxA
\else
\ifnum#1=48 %
\hatcurPPfluxavglogempiricalxxxxxB
\else
\ifnum#1=49 %
\hatcurPPfluxavglogempiricalxxxxxC
\else
??????\fi
\fi
\fi
}
\newcommand{\hatcurPPfluxavglognoisorestrictempirical}[1]{\ifnum#1=47 %
\hatcurPPfluxavglognoisorestrictempiricalxxxxxA
\else
\ifnum#1=48 %
\hatcurPPfluxavglognoisorestrictempiricalxxxxxB
\else
\ifnum#1=49 %
\hatcurPPfluxavglognoisorestrictempiricalxxxxxC
\else
??????\fi
\fi
\fi
}
\newcommand{\hatcurPPfluxavgnoisorestrictempirical}[1]{\ifnum#1=47 %
\hatcurPPfluxavgnoisorestrictempiricalxxxxxA
\else
\ifnum#1=48 %
\hatcurPPfluxavgnoisorestrictempiricalxxxxxB
\else
\ifnum#1=49 %
\hatcurPPfluxavgnoisorestrictempiricalxxxxxC
\else
??????\fi
\fi
\fi
}
\newcommand{\hatcurPPfluxperidimempirical}[1]{\ifnum#1=47 %
\hatcurPPfluxperidimempiricalxxxxxA
\else
\ifnum#1=48 %
\hatcurPPfluxperidimempiricalxxxxxB
\else
\ifnum#1=49 %
\hatcurPPfluxperidimempiricalxxxxxC
\else
??????\fi
\fi
\fi
}
\newcommand{\hatcurPPfluxperiempirical}[1]{\ifnum#1=47 %
\hatcurPPfluxperiempiricalxxxxxA
\else
\ifnum#1=48 %
\hatcurPPfluxperiempiricalxxxxxB
\else
\ifnum#1=49 %
\hatcurPPfluxperiempiricalxxxxxC
\else
??????\fi
\fi
\fi
}
\newcommand{\hatcurPPfluxperinoisorestrictempirical}[1]{\ifnum#1=47 %
\hatcurPPfluxperinoisorestrictempiricalxxxxxA
\else
\ifnum#1=48 %
\hatcurPPfluxperinoisorestrictempiricalxxxxxB
\else
\ifnum#1=49 %
\hatcurPPfluxperinoisorestrictempiricalxxxxxC
\else
??????\fi
\fi
\fi
}
\newcommand{\hatcurPPgempirical}[1]{\ifnum#1=47 %
\hatcurPPgempiricalxxxxxA
\else
\ifnum#1=48 %
\hatcurPPgempiricalxxxxxB
\else
\ifnum#1=49 %
\hatcurPPgempiricalxxxxxC
\else
??????\fi
\fi
\fi
}
\newcommand{\hatcurPPgnoisorestrictempirical}[1]{\ifnum#1=47 %
\hatcurPPgnoisorestrictempiricalxxxxxA
\else
\ifnum#1=48 %
\hatcurPPgnoisorestrictempiricalxxxxxB
\else
\ifnum#1=49 %
\hatcurPPgnoisorestrictempiricalxxxxxC
\else
??????\fi
\fi
\fi
}
\newcommand{\hatcurPPiempirical}[1]{\ifnum#1=47 %
\hatcurPPiempiricalxxxxxA
\else
\ifnum#1=48 %
\hatcurPPiempiricalxxxxxB
\else
\ifnum#1=49 %
\hatcurPPiempiricalxxxxxC
\else
??????\fi
\fi
\fi
}
\newcommand{\hatcurPPinoisorestrictempirical}[1]{\ifnum#1=47 %
\hatcurPPinoisorestrictempiricalxxxxxA
\else
\ifnum#1=48 %
\hatcurPPinoisorestrictempiricalxxxxxB
\else
\ifnum#1=49 %
\hatcurPPinoisorestrictempiricalxxxxxC
\else
??????\fi
\fi
\fi
}
\newcommand{\hatcurPPloggempirical}[1]{\ifnum#1=47 %
\hatcurPPloggempiricalxxxxxA
\else
\ifnum#1=48 %
\hatcurPPloggempiricalxxxxxB
\else
\ifnum#1=49 %
\hatcurPPloggempiricalxxxxxC
\else
??????\fi
\fi
\fi
}
\newcommand{\hatcurPPloggnoisorestrictempirical}[1]{\ifnum#1=47 %
\hatcurPPloggnoisorestrictempiricalxxxxxA
\else
\ifnum#1=48 %
\hatcurPPloggnoisorestrictempiricalxxxxxB
\else
\ifnum#1=49 %
\hatcurPPloggnoisorestrictempiricalxxxxxC
\else
??????\fi
\fi
\fi
}
\newcommand{\hatcurPPmeempirical}[1]{\ifnum#1=47 %
\hatcurPPmeempiricalxxxxxA
\else
\ifnum#1=48 %
\hatcurPPmeempiricalxxxxxB
\else
\ifnum#1=49 %
\hatcurPPmeempiricalxxxxxC
\else
??????\fi
\fi
\fi
}
\newcommand{\hatcurPPmelongempirical}[1]{\ifnum#1=47 %
\hatcurPPmelongempiricalxxxxxA
\else
\ifnum#1=48 %
\hatcurPPmelongempiricalxxxxxB
\else
\ifnum#1=49 %
\hatcurPPmelongempiricalxxxxxC
\else
??????\fi
\fi
\fi
}
\newcommand{\hatcurPPmelongnoisorestrictempirical}[1]{\ifnum#1=47 %
\hatcurPPmelongnoisorestrictempiricalxxxxxA
\else
\ifnum#1=48 %
\hatcurPPmelongnoisorestrictempiricalxxxxxB
\else
\ifnum#1=49 %
\hatcurPPmelongnoisorestrictempiricalxxxxxC
\else
??????\fi
\fi
\fi
}
\newcommand{\hatcurPPmempirical}[1]{\ifnum#1=47 %
\hatcurPPmempiricalxxxxxA
\else
\ifnum#1=48 %
\hatcurPPmempiricalxxxxxB
\else
\ifnum#1=49 %
\hatcurPPmempiricalxxxxxC
\else
??????\fi
\fi
\fi
}
\newcommand{\hatcurPPmenoisorestrictempirical}[1]{\ifnum#1=47 %
\hatcurPPmenoisorestrictempiricalxxxxxA
\else
\ifnum#1=48 %
\hatcurPPmenoisorestrictempiricalxxxxxB
\else
\ifnum#1=49 %
\hatcurPPmenoisorestrictempiricalxxxxxC
\else
??????\fi
\fi
\fi
}
\newcommand{\hatcurPPmeshortempirical}[1]{\ifnum#1=47 %
\hatcurPPmeshortempiricalxxxxxA
\else
\ifnum#1=48 %
\hatcurPPmeshortempiricalxxxxxB
\else
\ifnum#1=49 %
\hatcurPPmeshortempiricalxxxxxC
\else
??????\fi
\fi
\fi
}
\newcommand{\hatcurPPmeshortnoisorestrictempirical}[1]{\ifnum#1=47 %
\hatcurPPmeshortnoisorestrictempiricalxxxxxA
\else
\ifnum#1=48 %
\hatcurPPmeshortnoisorestrictempiricalxxxxxB
\else
\ifnum#1=49 %
\hatcurPPmeshortnoisorestrictempiricalxxxxxC
\else
??????\fi
\fi
\fi
}
\newcommand{\hatcurPPmlongempirical}[1]{\ifnum#1=47 %
\hatcurPPmlongempiricalxxxxxA
\else
\ifnum#1=48 %
\hatcurPPmlongempiricalxxxxxB
\else
\ifnum#1=49 %
\hatcurPPmlongempiricalxxxxxC
\else
??????\fi
\fi
\fi
}
\newcommand{\hatcurPPmlongnoisorestrictempirical}[1]{\ifnum#1=47 %
\hatcurPPmlongnoisorestrictempiricalxxxxxA
\else
\ifnum#1=48 %
\hatcurPPmlongnoisorestrictempiricalxxxxxB
\else
\ifnum#1=49 %
\hatcurPPmlongnoisorestrictempiricalxxxxxC
\else
??????\fi
\fi
\fi
}
\newcommand{\hatcurPPmnoisorestrictempirical}[1]{\ifnum#1=47 %
\hatcurPPmnoisorestrictempiricalxxxxxA
\else
\ifnum#1=48 %
\hatcurPPmnoisorestrictempiricalxxxxxB
\else
\ifnum#1=49 %
\hatcurPPmnoisorestrictempiricalxxxxxC
\else
??????\fi
\fi
\fi
}
\newcommand{\hatcurPPmrcorrempirical}[1]{\ifnum#1=47 %
\hatcurPPmrcorrempiricalxxxxxA
\else
\ifnum#1=48 %
\hatcurPPmrcorrempiricalxxxxxB
\else
\ifnum#1=49 %
\hatcurPPmrcorrempiricalxxxxxC
\else
??????\fi
\fi
\fi
}
\newcommand{\hatcurPPmshortempirical}[1]{\ifnum#1=47 %
\hatcurPPmshortempiricalxxxxxA
\else
\ifnum#1=48 %
\hatcurPPmshortempiricalxxxxxB
\else
\ifnum#1=49 %
\hatcurPPmshortempiricalxxxxxC
\else
??????\fi
\fi
\fi
}
\newcommand{\hatcurPPmshortnoisorestrictempirical}[1]{\ifnum#1=47 %
\hatcurPPmshortnoisorestrictempiricalxxxxxA
\else
\ifnum#1=48 %
\hatcurPPmshortnoisorestrictempiricalxxxxxB
\else
\ifnum#1=49 %
\hatcurPPmshortnoisorestrictempiricalxxxxxC
\else
??????\fi
\fi
\fi
}
\newcommand{\hatcurPPperiempirical}[1]{\ifnum#1=47 %
\hatcurPPperiempiricalxxxxxA
\else
\ifnum#1=48 %
\hatcurPPperiempiricalxxxxxB
\else
\ifnum#1=49 %
\hatcurPPperiempiricalxxxxxC
\else
??????\fi
\fi
\fi
}
\newcommand{\hatcurPPperinoisorestrictempirical}[1]{\ifnum#1=47 %
\hatcurPPperinoisorestrictempiricalxxxxxA
\else
\ifnum#1=48 %
\hatcurPPperinoisorestrictempiricalxxxxxB
\else
\ifnum#1=49 %
\hatcurPPperinoisorestrictempiricalxxxxxC
\else
??????\fi
\fi
\fi
}
\newcommand{\hatcurPPphiconjempirical}[1]{\ifnum#1=47 %
\hatcurPPphiconjempiricalxxxxxA
\else
\ifnum#1=48 %
\hatcurPPphiconjempiricalxxxxxB
\else
\ifnum#1=49 %
\hatcurPPphiconjempiricalxxxxxC
\else
??????\fi
\fi
\fi
}
\newcommand{\hatcurPPphiconjnoisorestrictempirical}[1]{\ifnum#1=47 %
\hatcurPPphiconjnoisorestrictempiricalxxxxxA
\else
\ifnum#1=48 %
\hatcurPPphiconjnoisorestrictempiricalxxxxxB
\else
\ifnum#1=49 %
\hatcurPPphiconjnoisorestrictempiricalxxxxxC
\else
??????\fi
\fi
\fi
}
\newcommand{\hatcurPPreempirical}[1]{\ifnum#1=47 %
\hatcurPPreempiricalxxxxxA
\else
\ifnum#1=48 %
\hatcurPPreempiricalxxxxxB
\else
\ifnum#1=49 %
\hatcurPPreempiricalxxxxxC
\else
??????\fi
\fi
\fi
}
\newcommand{\hatcurPPrelongempirical}[1]{\ifnum#1=47 %
\hatcurPPrelongempiricalxxxxxA
\else
\ifnum#1=48 %
\hatcurPPrelongempiricalxxxxxB
\else
\ifnum#1=49 %
\hatcurPPrelongempiricalxxxxxC
\else
??????\fi
\fi
\fi
}
\newcommand{\hatcurPPrelongnoisorestrictempirical}[1]{\ifnum#1=47 %
\hatcurPPrelongnoisorestrictempiricalxxxxxA
\else
\ifnum#1=48 %
\hatcurPPrelongnoisorestrictempiricalxxxxxB
\else
\ifnum#1=49 %
\hatcurPPrelongnoisorestrictempiricalxxxxxC
\else
??????\fi
\fi
\fi
}
\newcommand{\hatcurPPrempirical}[1]{\ifnum#1=47 %
\hatcurPPrempiricalxxxxxA
\else
\ifnum#1=48 %
\hatcurPPrempiricalxxxxxB
\else
\ifnum#1=49 %
\hatcurPPrempiricalxxxxxC
\else
??????\fi
\fi
\fi
}
\newcommand{\hatcurPPrenoisorestrictempirical}[1]{\ifnum#1=47 %
\hatcurPPrenoisorestrictempiricalxxxxxA
\else
\ifnum#1=48 %
\hatcurPPrenoisorestrictempiricalxxxxxB
\else
\ifnum#1=49 %
\hatcurPPrenoisorestrictempiricalxxxxxC
\else
??????\fi
\fi
\fi
}
\newcommand{\hatcurPPreshortempirical}[1]{\ifnum#1=47 %
\hatcurPPreshortempiricalxxxxxA
\else
\ifnum#1=48 %
\hatcurPPreshortempiricalxxxxxB
\else
\ifnum#1=49 %
\hatcurPPreshortempiricalxxxxxC
\else
??????\fi
\fi
\fi
}
\newcommand{\hatcurPPreshortnoisorestrictempirical}[1]{\ifnum#1=47 %
\hatcurPPreshortnoisorestrictempiricalxxxxxA
\else
\ifnum#1=48 %
\hatcurPPreshortnoisorestrictempiricalxxxxxB
\else
\ifnum#1=49 %
\hatcurPPreshortnoisorestrictempiricalxxxxxC
\else
??????\fi
\fi
\fi
}
\newcommand{\hatcurPPrhoempirical}[1]{\ifnum#1=47 %
\hatcurPPrhoempiricalxxxxxA
\else
\ifnum#1=48 %
\hatcurPPrhoempiricalxxxxxB
\else
\ifnum#1=49 %
\hatcurPPrhoempiricalxxxxxC
\else
??????\fi
\fi
\fi
}
\newcommand{\hatcurPPrhonoisorestrictempirical}[1]{\ifnum#1=47 %
\hatcurPPrhonoisorestrictempiricalxxxxxA
\else
\ifnum#1=48 %
\hatcurPPrhonoisorestrictempiricalxxxxxB
\else
\ifnum#1=49 %
\hatcurPPrhonoisorestrictempiricalxxxxxC
\else
??????\fi
\fi
\fi
}
\newcommand{\hatcurPPrlongempirical}[1]{\ifnum#1=47 %
\hatcurPPrlongempiricalxxxxxA
\else
\ifnum#1=48 %
\hatcurPPrlongempiricalxxxxxB
\else
\ifnum#1=49 %
\hatcurPPrlongempiricalxxxxxC
\else
??????\fi
\fi
\fi
}
\newcommand{\hatcurPPrlongnoisorestrictempirical}[1]{\ifnum#1=47 %
\hatcurPPrlongnoisorestrictempiricalxxxxxA
\else
\ifnum#1=48 %
\hatcurPPrlongnoisorestrictempiricalxxxxxB
\else
\ifnum#1=49 %
\hatcurPPrlongnoisorestrictempiricalxxxxxC
\else
??????\fi
\fi
\fi
}
\newcommand{\hatcurPPrnoisorestrictempirical}[1]{\ifnum#1=47 %
\hatcurPPrnoisorestrictempiricalxxxxxA
\else
\ifnum#1=48 %
\hatcurPPrnoisorestrictempiricalxxxxxB
\else
\ifnum#1=49 %
\hatcurPPrnoisorestrictempiricalxxxxxC
\else
??????\fi
\fi
\fi
}
\newcommand{\hatcurPPrshortempirical}[1]{\ifnum#1=47 %
\hatcurPPrshortempiricalxxxxxA
\else
\ifnum#1=48 %
\hatcurPPrshortempiricalxxxxxB
\else
\ifnum#1=49 %
\hatcurPPrshortempiricalxxxxxC
\else
??????\fi
\fi
\fi
}
\newcommand{\hatcurPPrshortnoisorestrictempirical}[1]{\ifnum#1=47 %
\hatcurPPrshortnoisorestrictempiricalxxxxxA
\else
\ifnum#1=48 %
\hatcurPPrshortnoisorestrictempiricalxxxxxB
\else
\ifnum#1=49 %
\hatcurPPrshortnoisorestrictempiricalxxxxxC
\else
??????\fi
\fi
\fi
}
\newcommand{\hatcurPPtcircempirical}[1]{\ifnum#1=47 %
\hatcurPPtcircempiricalxxxxxA
\else
\ifnum#1=48 %
\hatcurPPtcircempiricalxxxxxB
\else
\ifnum#1=49 %
\hatcurPPtcircempiricalxxxxxC
\else
??????\fi
\fi
\fi
}
\newcommand{\hatcurPPtcircnoisorestrictempirical}[1]{\ifnum#1=47 %
\hatcurPPtcircnoisorestrictempiricalxxxxxA
\else
\ifnum#1=48 %
\hatcurPPtcircnoisorestrictempiricalxxxxxB
\else
\ifnum#1=49 %
\hatcurPPtcircnoisorestrictempiricalxxxxxC
\else
??????\fi
\fi
\fi
}
\newcommand{\hatcurPPteffempirical}[1]{\ifnum#1=47 %
\hatcurPPteffempiricalxxxxxA
\else
\ifnum#1=48 %
\hatcurPPteffempiricalxxxxxB
\else
\ifnum#1=49 %
\hatcurPPteffempiricalxxxxxC
\else
??????\fi
\fi
\fi
}
\newcommand{\hatcurPPteffnoisorestrictempirical}[1]{\ifnum#1=47 %
\hatcurPPteffnoisorestrictempiricalxxxxxA
\else
\ifnum#1=48 %
\hatcurPPteffnoisorestrictempiricalxxxxxB
\else
\ifnum#1=49 %
\hatcurPPteffnoisorestrictempiricalxxxxxC
\else
??????\fi
\fi
\fi
}
\newcommand{\hatcurPPthetaempirical}[1]{\ifnum#1=47 %
\hatcurPPthetaempiricalxxxxxA
\else
\ifnum#1=48 %
\hatcurPPthetaempiricalxxxxxB
\else
\ifnum#1=49 %
\hatcurPPthetaempiricalxxxxxC
\else
??????\fi
\fi
\fi
}
\newcommand{\hatcurPPthetanoisorestrictempirical}[1]{\ifnum#1=47 %
\hatcurPPthetanoisorestrictempiricalxxxxxA
\else
\ifnum#1=48 %
\hatcurPPthetanoisorestrictempiricalxxxxxB
\else
\ifnum#1=49 %
\hatcurPPthetanoisorestrictempiricalxxxxxC
\else
??????\fi
\fi
\fi
}
\newcommand{\hatcurPPtinfallempirical}[1]{\ifnum#1=47 %
\hatcurPPtinfallempiricalxxxxxA
\else
\ifnum#1=48 %
\hatcurPPtinfallempiricalxxxxxB
\else
\ifnum#1=49 %
\hatcurPPtinfallempiricalxxxxxC
\else
??????\fi
\fi
\fi
}
\newcommand{\hatcurPPtinfallnoisorestrictempirical}[1]{\ifnum#1=47 %
\hatcurPPtinfallnoisorestrictempiricalxxxxxA
\else
\ifnum#1=48 %
\hatcurPPtinfallnoisorestrictempiricalxxxxxB
\else
\ifnum#1=49 %
\hatcurPPtinfallnoisorestrictempiricalxxxxxC
\else
??????\fi
\fi
\fi
}
\newcommand{\hatcurRVeccenempirical}[1]{\ifnum#1=47 %
\hatcurRVeccenempiricalxxxxxA
\else
\ifnum#1=48 %
\hatcurRVeccenempiricalxxxxxB
\else
\ifnum#1=49 %
\hatcurRVeccenempiricalxxxxxC
\else
??????\fi
\fi
\fi
}
\newcommand{\hatcurRVeccennoisorestrictempirical}[1]{\ifnum#1=47 %
\hatcurRVeccennoisorestrictempiricalxxxxxA
\else
\ifnum#1=48 %
\hatcurRVeccennoisorestrictempiricalxxxxxB
\else
\ifnum#1=49 %
\hatcurRVeccennoisorestrictempiricalxxxxxC
\else
??????\fi
\fi
\fi
}
\newcommand{\hatcurRVeccentwosiglimempirical}[1]{\ifnum#1=47 %
\hatcurRVeccentwosiglimempiricalxxxxxA
\else
\ifnum#1=48 %
\hatcurRVeccentwosiglimempiricalxxxxxB
\else
\ifnum#1=49 %
\hatcurRVeccentwosiglimempiricalxxxxxC
\else
??????\fi
\fi
\fi
}
\newcommand{\hatcurRVeccentwosiglimnoisorestrictempirical}[1]{\ifnum#1=47 %
\hatcurRVeccentwosiglimnoisorestrictempiricalxxxxxA
\else
\ifnum#1=48 %
\hatcurRVeccentwosiglimnoisorestrictempiricalxxxxxB
\else
\ifnum#1=49 %
\hatcurRVeccentwosiglimnoisorestrictempiricalxxxxxC
\else
??????\fi
\fi
\fi
}
\newcommand{\hatcurRVfitrmsAempirical}[1]{\ifnum#1=72 %
\hatcurRVfitrmsAempiricalxxxxxD
\else
??????\fi
}
\newcommand{\hatcurRVfitrmsBempirical}[1]{\ifnum#1=72 %
\hatcurRVfitrmsBempiricalxxxxxD
\else
??????\fi
}
\newcommand{\hatcurRVfitrmsempirical}[1]{\ifnum#1=47 %
\hatcurRVfitrmsempiricalxxxxxA
\else
\ifnum#1=48 %
\hatcurRVfitrmsempiricalxxxxxB
\else
\ifnum#1=49 %
\hatcurRVfitrmsempiricalxxxxxC
\else
??????\fi
\fi
\fi
}
\newcommand{\hatcurRVgammaAempirical}[1]{\ifnum#1=72 %
\hatcurRVgammaAempiricalxxxxxD
\else
??????\fi
}
\newcommand{\hatcurRVgammaAnoisorestrictempirical}[1]{\ifnum#1=72 %
\hatcurRVgammaAnoisorestrictempiricalxxxxxD
\else
??????\fi
}
\newcommand{\hatcurRVgammaBempirical}[1]{\ifnum#1=72 %
\hatcurRVgammaBempiricalxxxxxD
\else
??????\fi
}
\newcommand{\hatcurRVgammaBnoisorestrictempirical}[1]{\ifnum#1=72 %
\hatcurRVgammaBnoisorestrictempiricalxxxxxD
\else
??????\fi
}
\newcommand{\hatcurRVgammaCempirical}[1]{\ifnum#1=72 %
\hatcurRVgammaCempiricalxxxxxD
\else
??????\fi
}
\newcommand{\hatcurRVgammaCnoisorestrictempirical}[1]{\ifnum#1=72 %
\hatcurRVgammaCnoisorestrictempiricalxxxxxD
\else
??????\fi
}
\newcommand{\hatcurRVgammaempirical}[1]{\ifnum#1=47 %
\hatcurRVgammaempiricalxxxxxA
\else
\ifnum#1=48 %
\hatcurRVgammaempiricalxxxxxB
\else
\ifnum#1=49 %
\hatcurRVgammaempiricalxxxxxC
\else
??????\fi
\fi
\fi
}
\newcommand{\hatcurRVgammanoisorestrictempirical}[1]{\ifnum#1=47 %
\hatcurRVgammanoisorestrictempiricalxxxxxA
\else
\ifnum#1=48 %
\hatcurRVgammanoisorestrictempiricalxxxxxB
\else
\ifnum#1=49 %
\hatcurRVgammanoisorestrictempiricalxxxxxC
\else
??????\fi
\fi
\fi
}
\newcommand{\hatcurRVhempirical}[1]{\ifnum#1=47 %
\hatcurRVhempiricalxxxxxA
\else
\ifnum#1=48 %
\hatcurRVhempiricalxxxxxB
\else
\ifnum#1=49 %
\hatcurRVhempiricalxxxxxC
\else
\ifnum#1=72 %
\hatcurRVhempiricalxxxxxD
\else
??????\fi
\fi
\fi
\fi
}
\newcommand{\hatcurRVhnoisorestrictempirical}[1]{\ifnum#1=47 %
\hatcurRVhnoisorestrictempiricalxxxxxA
\else
\ifnum#1=48 %
\hatcurRVhnoisorestrictempiricalxxxxxB
\else
\ifnum#1=49 %
\hatcurRVhnoisorestrictempiricalxxxxxC
\else
\ifnum#1=72 %
\hatcurRVhnoisorestrictempiricalxxxxxD
\else
??????\fi
\fi
\fi
\fi
}
\newcommand{\hatcurRVjitterAempirical}[1]{\ifnum#1=72 %
\hatcurRVjitterAempiricalxxxxxD
\else
??????\fi
}
\newcommand{\hatcurRVjitterAnoisorestrictempirical}[1]{\ifnum#1=72 %
\hatcurRVjitterAnoisorestrictempiricalxxxxxD
\else
??????\fi
}
\newcommand{\hatcurRVjitterBempirical}[1]{\ifnum#1=72 %
\hatcurRVjitterBempiricalxxxxxD
\else
??????\fi
}
\newcommand{\hatcurRVjitterBnoisorestrictempirical}[1]{\ifnum#1=72 %
\hatcurRVjitterBnoisorestrictempiricalxxxxxD
\else
??????\fi
}
\newcommand{\hatcurRVjitterCempirical}[1]{\ifnum#1=72 %
\hatcurRVjitterCempiricalxxxxxD
\else
??????\fi
}
\newcommand{\hatcurRVjitterempirical}[1]{\ifnum#1=47 %
\hatcurRVjitterempiricalxxxxxA
\else
\ifnum#1=48 %
\hatcurRVjitterempiricalxxxxxB
\else
\ifnum#1=49 %
\hatcurRVjitterempiricalxxxxxC
\else
??????\fi
\fi
\fi
}
\newcommand{\hatcurRVjitternoisorestrictempirical}[1]{\ifnum#1=47 %
\hatcurRVjitternoisorestrictempiricalxxxxxA
\else
\ifnum#1=48 %
\hatcurRVjitternoisorestrictempiricalxxxxxB
\else
\ifnum#1=49 %
\hatcurRVjitternoisorestrictempiricalxxxxxC
\else
??????\fi
\fi
\fi
}
\newcommand{\hatcurRVjittertwosiglimAempirical}[1]{\ifnum#1=72 %
\hatcurRVjittertwosiglimAempiricalxxxxxD
\else
??????\fi
}
\newcommand{\hatcurRVjittertwosiglimAnoisorestrictempirical}[1]{\ifnum#1=72 %
\hatcurRVjittertwosiglimAnoisorestrictempiricalxxxxxD
\else
??????\fi
}
\newcommand{\hatcurRVjittertwosiglimBempirical}[1]{\ifnum#1=72 %
\hatcurRVjittertwosiglimBempiricalxxxxxD
\else
??????\fi
}
\newcommand{\hatcurRVjittertwosiglimBnoisorestrictempirical}[1]{\ifnum#1=72 %
\hatcurRVjittertwosiglimBnoisorestrictempiricalxxxxxD
\else
??????\fi
}
\newcommand{\hatcurRVjittertwosiglimempirical}[1]{\ifnum#1=47 %
\hatcurRVjittertwosiglimempiricalxxxxxA
\else
\ifnum#1=48 %
\hatcurRVjittertwosiglimempiricalxxxxxB
\else
\ifnum#1=49 %
\hatcurRVjittertwosiglimempiricalxxxxxC
\else
??????\fi
\fi
\fi
}
\newcommand{\hatcurRVjittertwosiglimnoisorestrictempirical}[1]{\ifnum#1=47 %
\hatcurRVjittertwosiglimnoisorestrictempiricalxxxxxA
\else
\ifnum#1=48 %
\hatcurRVjittertwosiglimnoisorestrictempiricalxxxxxB
\else
\ifnum#1=49 %
\hatcurRVjittertwosiglimnoisorestrictempiricalxxxxxC
\else
??????\fi
\fi
\fi
}
\newcommand{\hatcurRVkempirical}[1]{\ifnum#1=47 %
\hatcurRVkempiricalxxxxxA
\else
\ifnum#1=48 %
\hatcurRVkempiricalxxxxxB
\else
\ifnum#1=49 %
\hatcurRVkempiricalxxxxxC
\else
\ifnum#1=72 %
\hatcurRVkempiricalxxxxxD
\else
??????\fi
\fi
\fi
\fi
}
\newcommand{\hatcurRVKempirical}[1]{\ifnum#1=47 %
\hatcurRVKempiricalxxxxxA
\else
\ifnum#1=48 %
\hatcurRVKempiricalxxxxxB
\else
\ifnum#1=49 %
\hatcurRVKempiricalxxxxxC
\else
\ifnum#1=72 %
\hatcurRVKempiricalxxxxxD
\else
??????\fi
\fi
\fi
\fi
}
\newcommand{\hatcurRVknoisorestrictempirical}[1]{\ifnum#1=47 %
\hatcurRVknoisorestrictempiricalxxxxxA
\else
\ifnum#1=48 %
\hatcurRVknoisorestrictempiricalxxxxxB
\else
\ifnum#1=49 %
\hatcurRVknoisorestrictempiricalxxxxxC
\else
\ifnum#1=72 %
\hatcurRVknoisorestrictempiricalxxxxxD
\else
??????\fi
\fi
\fi
\fi
}
\newcommand{\hatcurRVKnoisorestrictempirical}[1]{\ifnum#1=47 %
\hatcurRVKnoisorestrictempiricalxxxxxA
\else
\ifnum#1=48 %
\hatcurRVKnoisorestrictempiricalxxxxxB
\else
\ifnum#1=49 %
\hatcurRVKnoisorestrictempiricalxxxxxC
\else
\ifnum#1=72 %
\hatcurRVKnoisorestrictempiricalxxxxxD
\else
??????\fi
\fi
\fi
\fi
}
\newcommand{\hatcurRVomegaempirical}[1]{\ifnum#1=47 %
\hatcurRVomegaempiricalxxxxxA
\else
\ifnum#1=48 %
\hatcurRVomegaempiricalxxxxxB
\else
\ifnum#1=49 %
\hatcurRVomegaempiricalxxxxxC
\else
??????\fi
\fi
\fi
}
\newcommand{\hatcurRVomeganoisorestrictempirical}[1]{\ifnum#1=47 %
\hatcurRVomeganoisorestrictempiricalxxxxxA
\else
\ifnum#1=48 %
\hatcurRVomeganoisorestrictempiricalxxxxxB
\else
\ifnum#1=49 %
\hatcurRVomeganoisorestrictempiricalxxxxxC
\else
??????\fi
\fi
\fi
}
\newcommand{\hatcurRVrhempirical}[1]{\ifnum#1=47 %
\hatcurRVrhempiricalxxxxxA
\else
\ifnum#1=48 %
\hatcurRVrhempiricalxxxxxB
\else
\ifnum#1=49 %
\hatcurRVrhempiricalxxxxxC
\else
\ifnum#1=72 %
\hatcurRVrhempiricalxxxxxD
\else
??????\fi
\fi
\fi
\fi
}
\newcommand{\hatcurRVrhnoisorestrictempirical}[1]{\ifnum#1=47 %
\hatcurRVrhnoisorestrictempiricalxxxxxA
\else
\ifnum#1=48 %
\hatcurRVrhnoisorestrictempiricalxxxxxB
\else
\ifnum#1=49 %
\hatcurRVrhnoisorestrictempiricalxxxxxC
\else
\ifnum#1=72 %
\hatcurRVrhnoisorestrictempiricalxxxxxD
\else
??????\fi
\fi
\fi
\fi
}
\newcommand{\hatcurRVrkempirical}[1]{\ifnum#1=47 %
\hatcurRVrkempiricalxxxxxA
\else
\ifnum#1=48 %
\hatcurRVrkempiricalxxxxxB
\else
\ifnum#1=49 %
\hatcurRVrkempiricalxxxxxC
\else
\ifnum#1=72 %
\hatcurRVrkempiricalxxxxxD
\else
??????\fi
\fi
\fi
\fi
}
\newcommand{\hatcurRVrknoisorestrictempirical}[1]{\ifnum#1=47 %
\hatcurRVrknoisorestrictempiricalxxxxxA
\else
\ifnum#1=48 %
\hatcurRVrknoisorestrictempiricalxxxxxB
\else
\ifnum#1=49 %
\hatcurRVrknoisorestrictempiricalxxxxxC
\else
\ifnum#1=72 %
\hatcurRVrknoisorestrictempiricalxxxxxD
\else
??????\fi
\fi
\fi
\fi
}
\newcommand{\hatcurRVtroneempirical}[1]{\ifnum#1=47 %
\hatcurRVtroneempiricalxxxxxA
\else
\ifnum#1=48 %
\hatcurRVtroneempiricalxxxxxB
\else
\ifnum#1=49 %
\hatcurRVtroneempiricalxxxxxC
\else
\ifnum#1=72 %
\hatcurRVtroneempiricalxxxxxD
\else
??????\fi
\fi
\fi
\fi
}
\newcommand{\hatcurRVtronenoisorestrictempirical}[1]{\ifnum#1=47 %
\hatcurRVtronenoisorestrictempiricalxxxxxA
\else
\ifnum#1=48 %
\hatcurRVtronenoisorestrictempiricalxxxxxB
\else
\ifnum#1=49 %
\hatcurRVtronenoisorestrictempiricalxxxxxC
\else
\ifnum#1=72 %
\hatcurRVtronenoisorestrictempiricalxxxxxD
\else
??????\fi
\fi
\fi
\fi
}
\newcommand{\hatcurRVtrtwoempirical}[1]{\ifnum#1=47 %
\hatcurRVtrtwoempiricalxxxxxA
\else
\ifnum#1=48 %
\hatcurRVtrtwoempiricalxxxxxB
\else
\ifnum#1=49 %
\hatcurRVtrtwoempiricalxxxxxC
\else
\ifnum#1=72 %
\hatcurRVtrtwoempiricalxxxxxD
\else
??????\fi
\fi
\fi
\fi
}
\newcommand{\hatcurRVtrtwonoisorestrictempirical}[1]{\ifnum#1=47 %
\hatcurRVtrtwonoisorestrictempiricalxxxxxA
\else
\ifnum#1=48 %
\hatcurRVtrtwonoisorestrictempiricalxxxxxB
\else
\ifnum#1=49 %
\hatcurRVtrtwonoisorestrictempiricalxxxxxC
\else
\ifnum#1=72 %
\hatcurRVtrtwonoisorestrictempiricalxxxxxD
\else
??????\fi
\fi
\fi
\fi
}
\newcommand{\hatcurSMEiiloggempirical}[1]{\ifnum#1=47 %
\hatcurSMEiiloggempiricalxxxxxA
\else
??????\fi
}
\newcommand{\hatcurSMEiiteffempirical}[1]{\ifnum#1=47 %
\hatcurSMEiiteffempiricalxxxxxA
\else
??????\fi
}
\newcommand{\hatcurSMEiivsinempirical}[1]{\ifnum#1=47 %
\hatcurSMEiivsinempiricalxxxxxA
\else
??????\fi
}
\newcommand{\hatcurSMEiizfehempirical}[1]{\ifnum#1=47 %
\hatcurSMEiizfehempiricalxxxxxA
\else
??????\fi
}
\newcommand{\hatcurSMEiizfehshortempirical}[1]{\ifnum#1=47 %
\hatcurSMEiizfehshortempiricalxxxxxA
\else
??????\fi
}
\newcommand{\hatcurSMEiloggempirical}[1]{\ifnum#1=47 %
\hatcurSMEiloggempiricalxxxxxA
\else
\ifnum#1=48 %
\hatcurSMEiloggempiricalxxxxxB
\else
\ifnum#1=49 %
\hatcurSMEiloggempiricalxxxxxC
\else
\ifnum#1=72 %
\hatcurSMEiloggempiricalxxxxxD
\else
??????\fi
\fi
\fi
\fi
}
\newcommand{\hatcurSMEiteffempirical}[1]{\ifnum#1=47 %
\hatcurSMEiteffempiricalxxxxxA
\else
\ifnum#1=48 %
\hatcurSMEiteffempiricalxxxxxB
\else
\ifnum#1=49 %
\hatcurSMEiteffempiricalxxxxxC
\else
\ifnum#1=72 %
\hatcurSMEiteffempiricalxxxxxD
\else
??????\fi
\fi
\fi
\fi
}
\newcommand{\hatcurSMEivmacempirical}[1]{\ifnum#1=47 %
\hatcurSMEivmacempiricalxxxxxA
\else
\ifnum#1=48 %
\hatcurSMEivmacempiricalxxxxxB
\else
\ifnum#1=49 %
\hatcurSMEivmacempiricalxxxxxC
\else
\ifnum#1=72 %
\hatcurSMEivmacempiricalxxxxxD
\else
??????\fi
\fi
\fi
\fi
}
\newcommand{\hatcurSMEivmicempirical}[1]{\ifnum#1=47 %
\hatcurSMEivmicempiricalxxxxxA
\else
\ifnum#1=48 %
\hatcurSMEivmicempiricalxxxxxB
\else
\ifnum#1=49 %
\hatcurSMEivmicempiricalxxxxxC
\else
\ifnum#1=72 %
\hatcurSMEivmicempiricalxxxxxD
\else
??????\fi
\fi
\fi
\fi
}
\newcommand{\hatcurSMEivsinempirical}[1]{\ifnum#1=47 %
\hatcurSMEivsinempiricalxxxxxA
\else
\ifnum#1=48 %
\hatcurSMEivsinempiricalxxxxxB
\else
\ifnum#1=49 %
\hatcurSMEivsinempiricalxxxxxC
\else
\ifnum#1=72 %
\hatcurSMEivsinempiricalxxxxxD
\else
??????\fi
\fi
\fi
\fi
}
\newcommand{\hatcurSMEizfehempirical}[1]{\ifnum#1=47 %
\hatcurSMEizfehempiricalxxxxxA
\else
\ifnum#1=48 %
\hatcurSMEizfehempiricalxxxxxB
\else
\ifnum#1=49 %
\hatcurSMEizfehempiricalxxxxxC
\else
\ifnum#1=72 %
\hatcurSMEizfehempiricalxxxxxD
\else
??????\fi
\fi
\fi
\fi
}
\newcommand{\hatcurSMEizfehshortempirical}[1]{\ifnum#1=47 %
\hatcurSMEizfehshortempiricalxxxxxA
\else
\ifnum#1=48 %
\hatcurSMEizfehshortempiricalxxxxxB
\else
\ifnum#1=49 %
\hatcurSMEizfehshortempiricalxxxxxC
\else
\ifnum#1=72 %
\hatcurSMEizfehshortempiricalxxxxxD
\else
??????\fi
\fi
\fi
\fi
}
\newcommand{\hatcurXAvempirical}[1]{\ifnum#1=47 %
\hatcurXAvempiricalxxxxxA
\else
\ifnum#1=48 %
\hatcurXAvempiricalxxxxxB
\else
\ifnum#1=49 %
\hatcurXAvempiricalxxxxxC
\else
??????\fi
\fi
\fi
}
\newcommand{\hatcurXAvnoisorestrictempirical}[1]{\ifnum#1=47 %
\hatcurXAvnoisorestrictempiricalxxxxxA
\else
\ifnum#1=48 %
\hatcurXAvnoisorestrictempiricalxxxxxB
\else
\ifnum#1=49 %
\hatcurXAvnoisorestrictempiricalxxxxxC
\else
??????\fi
\fi
\fi
}
\newcommand{\hatcurXdistempirical}[1]{\ifnum#1=47 %
\hatcurXdistempiricalxxxxxA
\else
\ifnum#1=48 %
\hatcurXdistempiricalxxxxxB
\else
\ifnum#1=49 %
\hatcurXdistempiricalxxxxxC
\else
??????\fi
\fi
\fi
}
\newcommand{\hatcurXdistnoisorestrictempirical}[1]{\ifnum#1=47 %
\hatcurXdistnoisorestrictempiricalxxxxxA
\else
\ifnum#1=48 %
\hatcurXdistnoisorestrictempiricalxxxxxB
\else
\ifnum#1=49 %
\hatcurXdistnoisorestrictempiricalxxxxxC
\else
??????\fi
\fi
\fi
}
\newcommand{\hatcurXdistredempirical}[1]{\ifnum#1=47 %
\hatcurXdistredempiricalxxxxxA
\else
\ifnum#1=48 %
\hatcurXdistredempiricalxxxxxB
\else
\ifnum#1=49 %
\hatcurXdistredempiricalxxxxxC
\else
??????\fi
\fi
\fi
}
\newcommand{\hatcurXdistrednoisorestrictempirical}[1]{\ifnum#1=47 %
\hatcurXdistrednoisorestrictempiricalxxxxxA
\else
\ifnum#1=48 %
\hatcurXdistrednoisorestrictempiricalxxxxxB
\else
\ifnum#1=49 %
\hatcurXdistrednoisorestrictempiricalxxxxxC
\else
??????\fi
\fi
\fi
}
\newcommand{\hatcurXEBVempirical}[1]{\ifnum#1=47 %
\hatcurXEBVempiricalxxxxxA
\else
\ifnum#1=48 %
\hatcurXEBVempiricalxxxxxB
\else
\ifnum#1=49 %
\hatcurXEBVempiricalxxxxxC
\else
??????\fi
\fi
\fi
}
\newcommand{\hatcurXEBVnoisorestrictempirical}[1]{\ifnum#1=47 %
\hatcurXEBVnoisorestrictempiricalxxxxxA
\else
\ifnum#1=48 %
\hatcurXEBVnoisorestrictempiricalxxxxxB
\else
\ifnum#1=49 %
\hatcurXEBVnoisorestrictempiricalxxxxxC
\else
??????\fi
\fi
\fi
}
\newcommand{\hatcurXsecdurempirical}[1]{\ifnum#1=47 %
\hatcurXsecdurempiricalxxxxxA
\else
\ifnum#1=48 %
\hatcurXsecdurempiricalxxxxxB
\else
\ifnum#1=49 %
\hatcurXsecdurempiricalxxxxxC
\else
??????\fi
\fi
\fi
}
\newcommand{\hatcurXsecdurnoisorestrictempirical}[1]{\ifnum#1=47 %
\hatcurXsecdurnoisorestrictempiricalxxxxxA
\else
\ifnum#1=48 %
\hatcurXsecdurnoisorestrictempiricalxxxxxB
\else
\ifnum#1=49 %
\hatcurXsecdurnoisorestrictempiricalxxxxxC
\else
??????\fi
\fi
\fi
}
\newcommand{\hatcurXsecingdurempirical}[1]{\ifnum#1=47 %
\hatcurXsecingdurempiricalxxxxxA
\else
\ifnum#1=48 %
\hatcurXsecingdurempiricalxxxxxB
\else
\ifnum#1=49 %
\hatcurXsecingdurempiricalxxxxxC
\else
??????\fi
\fi
\fi
}
\newcommand{\hatcurXsecingdurnoisorestrictempirical}[1]{\ifnum#1=47 %
\hatcurXsecingdurnoisorestrictempiricalxxxxxA
\else
\ifnum#1=48 %
\hatcurXsecingdurnoisorestrictempiricalxxxxxB
\else
\ifnum#1=49 %
\hatcurXsecingdurnoisorestrictempiricalxxxxxC
\else
??????\fi
\fi
\fi
}
\newcommand{\hatcurXsecondaryempirical}[1]{\ifnum#1=47 %
\hatcurXsecondaryempiricalxxxxxA
\else
\ifnum#1=48 %
\hatcurXsecondaryempiricalxxxxxB
\else
\ifnum#1=49 %
\hatcurXsecondaryempiricalxxxxxC
\else
??????\fi
\fi
\fi
}
\newcommand{\hatcurXsecondarynoisorestrictempirical}[1]{\ifnum#1=47 %
\hatcurXsecondarynoisorestrictempiricalxxxxxA
\else
\ifnum#1=48 %
\hatcurXsecondarynoisorestrictempiricalxxxxxB
\else
\ifnum#1=49 %
\hatcurXsecondarynoisorestrictempiricalxxxxxC
\else
??????\fi
\fi
\fi
}
\newcommand{\hatcurXsecphaseempirical}[1]{\ifnum#1=47 %
\hatcurXsecphaseempiricalxxxxxA
\else
\ifnum#1=48 %
\hatcurXsecphaseempiricalxxxxxB
\else
\ifnum#1=49 %
\hatcurXsecphaseempiricalxxxxxC
\else
??????\fi
\fi
\fi
}
\newcommand{\hatcurXsecphasenoisorestrictempirical}[1]{\ifnum#1=47 %
\hatcurXsecphasenoisorestrictempiricalxxxxxA
\else
\ifnum#1=48 %
\hatcurXsecphasenoisorestrictempiricalxxxxxB
\else
\ifnum#1=49 %
\hatcurXsecphasenoisorestrictempiricalxxxxxC
\else
??????\fi
\fi
\fi
}
\newcommand{\hatcurxxxxxA}{HATS-47}
\newcommand{\hatcurbxxxxxA}{HATS-47b}
\newcommand{\hatcurcxxxxxA}{HATS-47c}

\newcommand{\hatcurplanetnumxxxxxA}{47}

\newcommand{\hatcurCCtwomassshortxxxxxA}{19095625-4939538}

\newcommand{\hatcurCCticxxxxxA}{158297421}
\newcommand{\hatcurCCtoixxxxxA}{1073.01}
\newcommand{\hatcurCCwaspxxxxxA}{}

\newcommand{\hatcurRVgammaabsxxxxxA}{\ensuremath{3197\pm31}} 

\newcommand{\hatcurRVgammarelxxxxxA}{\hatcurRVgamma{\hatcurplanetnumxxxxxA}}                           

\newcommand{\hatcurCCtassvixxxxxA}{\ensuremath{NULL\pm NULL}}                  

\newcommand{\hatcurSMEversionxxxxxA}{i}                                       

\newcommand{\hatcurisoshortxxxxxA}{PARSEC}
\newcommand{\hatcurisofullxxxxxA}{PARSEC}
\newcommand{\hatcurisocitexxxxxA}{marigo:2017}

\newcommand{\hatcurlumindxxxxxA}{\rstar}

\newcommand{\hatcurjhkfilsetxxxxxA}{MASS}

\newcommand{\hatcurSMEteffxxxxxA}{\ifthenelse{\equal{\hatcurSMEversionxxxxxA}{i}}{\hatcurSMEiteff{\hatcurplanetnumxxxxxA}}{\hatcurSMEiiteff{\hatcurplanetnumxxxxxA}}}
\newcommand{\hatcurSMEzfehxxxxxA}{\ifthenelse{\equal{\hatcurSMEversionxxxxxA}{i}}{\hatcurSMEizfeh{\hatcurplanetnumxxxxxA}}{\hatcurSMEiizfeh{\hatcurplanetnumxxxxxA}}}
\newcommand{\hatcurSMEzfehshortxxxxxA}{\ifthenelse{\equal{\hatcurSMEversionxxxxxA}{i}}{\hatcurSMEizfehshort{\hatcurplanetnumxxxxxA}}{\hatcurSMEiizfehshort{\hatcurplanetnumxxxxxA}}}
\newcommand{\hatcurSMEloggxxxxxA}{\ifthenelse{\equal{\hatcurSMEversionxxxxxA}{i}}{\hatcurSMEilogg{\hatcurplanetnumxxxxxA}}{\hatcurSMEiilogg{\hatcurplanetnumxxxxxA}}}
\newcommand{\hatcurSMEvsinxxxxxA}{\ifthenelse{\equal{\hatcurSMEversionxxxxxA}{i}}{\hatcurSMEivsin{\hatcurplanetnumxxxxxA}}{\hatcurSMEiivsin{\hatcurplanetnumxxxxxA}}}
\newcommand{\hatcurSMEvmacxxxxxA}{\ifthenelse{\equal{\hatcurSMEversionxxxxxA}{i}}{\hatcurSMEivmac{\hatcurplanetnumxxxxxA}}{\hatcurSMEiivmac{\hatcurplanetnumxxxxxA}}}
\newcommand{\hatcurSMEvmicxxxxxA}{\ifthenelse{\equal{\hatcurSMEversionxxxxxA}{i}}{\hatcurSMEivmic{\hatcurplanetnumxxxxxA}}{\hatcurSMEiivmic{\hatcurplanetnumxxxxxA}}}

\newcommand{\hatcurxxxxxB}{HATS-48A}
\newcommand{\hatcurbxxxxxB}{HATS-48Ab}
\newcommand{\hatcurcxxxxxB}{HATS-48Ac}

\newcommand{\hatcurplanetnumxxxxxB}{48}

\newcommand{\hatcurCCtwomassshortxxxxxB}{19144126-5934458}

\newcommand{\hatcurCCticxxxxxB}{201642601}
\newcommand{\hatcurCCtoixxxxxB}{}
\newcommand{\hatcurCCwaspxxxxxB}{}

\newcommand{\hatcurRVgammaabsxxxxxB}{\ensuremath{-22460\pm21}} 

\newcommand{\hatcurRVgammarelxxxxxB}{\hatcurRVgamma{\hatcurplanetnumxxxxxB}}                           

\newcommand{\hatcurCCtassvixxxxxB}{\ensuremath{NULL\pm NULL}}                  

\newcommand{\hatcurSMEversionxxxxxB}{i}                                       

\newcommand{\hatcurisoshortxxxxxB}{PARSEC}
\newcommand{\hatcurisofullxxxxxB}{PARSEC}
\newcommand{\hatcurisocitexxxxxB}{marigo:2017}

\newcommand{\hatcurlumindxxxxxB}{\rstar}

\newcommand{\hatcurjhkfilsetxxxxxB}{MASS}

\newcommand{\hatcurSMEteffxxxxxB}{\ifthenelse{\equal{\hatcurSMEversionxxxxxB}{i}}{\hatcurSMEiteff{\hatcurplanetnumxxxxxB}}{\hatcurSMEiiteff{\hatcurplanetnumxxxxxB}}}
\newcommand{\hatcurSMEzfehxxxxxB}{\ifthenelse{\equal{\hatcurSMEversionxxxxxB}{i}}{\hatcurSMEizfeh{\hatcurplanetnumxxxxxB}}{\hatcurSMEiizfeh{\hatcurplanetnumxxxxxB}}}
\newcommand{\hatcurSMEzfehshortxxxxxB}{\ifthenelse{\equal{\hatcurSMEversionxxxxxB}{i}}{\hatcurSMEizfehshort{\hatcurplanetnumxxxxxB}}{\hatcurSMEiizfehshort{\hatcurplanetnumxxxxxB}}}
\newcommand{\hatcurSMEloggxxxxxB}{\ifthenelse{\equal{\hatcurSMEversionxxxxxB}{i}}{\hatcurSMEilogg{\hatcurplanetnumxxxxxB}}{\hatcurSMEiilogg{\hatcurplanetnumxxxxxB}}}
\newcommand{\hatcurSMEvsinxxxxxB}{\ifthenelse{\equal{\hatcurSMEversionxxxxxB}{i}}{\hatcurSMEivsin{\hatcurplanetnumxxxxxB}}{\hatcurSMEiivsin{\hatcurplanetnumxxxxxB}}}
\newcommand{\hatcurSMEvmacxxxxxB}{\ifthenelse{\equal{\hatcurSMEversionxxxxxB}{i}}{\hatcurSMEivmac{\hatcurplanetnumxxxxxB}}{\hatcurSMEiivmac{\hatcurplanetnumxxxxxB}}}
\newcommand{\hatcurSMEvmicxxxxxB}{\ifthenelse{\equal{\hatcurSMEversionxxxxxB}{i}}{\hatcurSMEivmic{\hatcurplanetnumxxxxxB}}{\hatcurSMEiivmic{\hatcurplanetnumxxxxxB}}}

\newcommand{\hatcurxxxxxC}{HATS-49}
\newcommand{\hatcurbxxxxxC}{HATS-49b}
\newcommand{\hatcurcxxxxxC}{HATS-49c}

\newcommand{\hatcurplanetnumxxxxxC}{49}

\newcommand{\hatcurCCtwomassshortxxxxxC}{00262717-5620395}

\newcommand{\hatcurCCticxxxxxC}{281541545}
\newcommand{\hatcurCCtoixxxxxC}{}
\newcommand{\hatcurCCwaspxxxxxC}{}

\newcommand{\hatcurRVgammaabsxxxxxC}{\ensuremath{8100\pm1500}} 

\newcommand{\hatcurRVgammarelxxxxxC}{\hatcurRVgamma{\hatcurplanetnumxxxxxC}}                           

\newcommand{\hatcurCCtassvixxxxxC}{\ensuremath{NULL\pm NULL}}                  

\newcommand{\hatcurSMEversionxxxxxC}{i}                                       

\newcommand{\hatcurisoshortxxxxxC}{PARSEC}
\newcommand{\hatcurisofullxxxxxC}{PARSEC}
\newcommand{\hatcurisocitexxxxxC}{marigo:2017}

\newcommand{\hatcurlumindxxxxxC}{\rstar}

\newcommand{\hatcurjhkfilsetxxxxxC}{MASS}

\newcommand{\hatcurSMEteffxxxxxC}{\ifthenelse{\equal{\hatcurSMEversionxxxxxC}{i}}{\hatcurSMEiteff{\hatcurplanetnumxxxxxC}}{\hatcurSMEiiteff{\hatcurplanetnumxxxxxC}}}
\newcommand{\hatcurSMEzfehxxxxxC}{\ifthenelse{\equal{\hatcurSMEversionxxxxxC}{i}}{\hatcurSMEizfeh{\hatcurplanetnumxxxxxC}}{\hatcurSMEiizfeh{\hatcurplanetnumxxxxxC}}}
\newcommand{\hatcurSMEzfehshortxxxxxC}{\ifthenelse{\equal{\hatcurSMEversionxxxxxC}{i}}{\hatcurSMEizfehshort{\hatcurplanetnumxxxxxC}}{\hatcurSMEiizfehshort{\hatcurplanetnumxxxxxC}}}
\newcommand{\hatcurSMEloggxxxxxC}{\ifthenelse{\equal{\hatcurSMEversionxxxxxC}{i}}{\hatcurSMEilogg{\hatcurplanetnumxxxxxC}}{\hatcurSMEiilogg{\hatcurplanetnumxxxxxC}}}
\newcommand{\hatcurSMEvsinxxxxxC}{\ifthenelse{\equal{\hatcurSMEversionxxxxxC}{i}}{\hatcurSMEivsin{\hatcurplanetnumxxxxxC}}{\hatcurSMEiivsin{\hatcurplanetnumxxxxxC}}}
\newcommand{\hatcurSMEvmacxxxxxC}{\ifthenelse{\equal{\hatcurSMEversionxxxxxC}{i}}{\hatcurSMEivmac{\hatcurplanetnumxxxxxC}}{\hatcurSMEiivmac{\hatcurplanetnumxxxxxC}}}
\newcommand{\hatcurSMEvmicxxxxxC}{\ifthenelse{\equal{\hatcurSMEversionxxxxxC}{i}}{\hatcurSMEivmic{\hatcurplanetnumxxxxxC}}{\hatcurSMEiivmic{\hatcurplanetnumxxxxxC}}}

\newcommand{\hatcurxxxxxD}{HATS-72}
\newcommand{\hatcurbxxxxxD}{HATS-72b}
\newcommand{\hatcurcxxxxxD}{HATS-72c}

\newcommand{\hatcurplanetnumxxxxxD}{72}

\newcommand{\hatcurCCtwomassshortxxxxxD}{22360631-1659597}

\newcommand{\hatcurCCticxxxxxD}{188570092}
\newcommand{\hatcurCCtoixxxxxD}{294.01}
\newcommand{\hatcurCCwaspxxxxxD}{WASP-191b}

\newcommand{\hatcurRVgammaabsxxxxxD}{\hatcurRVgammaB{\hatcurplanetnumxxxxxD}} 

\newcommand{\hatcurRVgammarelxxxxxD}{\hatcurRVgammaB{\hatcurplanetnumxxxxxD}}                           

\newcommand{\hatcurCCtassvixxxxxD}{\ensuremath{NULL\pm NULL}}                  

\newcommand{\hatcurSMEversionxxxxxD}{i}                                       

\newcommand{\hatcurisoshortxxxxxD}{PARSEC}
\newcommand{\hatcurisofullxxxxxD}{PARSEC}
\newcommand{\hatcurisocitexxxxxD}{marigo:2017}

\newcommand{\hatcurlumindxxxxxD}{\rstar}

\newcommand{\hatcurjhkfilsetxxxxxD}{MASS}

\newcommand{\hatcurSMEteffxxxxxD}{\ifthenelse{\equal{\hatcurSMEversionxxxxxD}{i}}{\hatcurSMEiteff{\hatcurplanetnumxxxxxD}}{\hatcurSMEiiteff{\hatcurplanetnumxxxxxD}}}
\newcommand{\hatcurSMEzfehxxxxxD}{\ifthenelse{\equal{\hatcurSMEversionxxxxxD}{i}}{\hatcurSMEizfeh{\hatcurplanetnumxxxxxD}}{\hatcurSMEiizfeh{\hatcurplanetnumxxxxxD}}}
\newcommand{\hatcurSMEzfehshortxxxxxD}{\ifthenelse{\equal{\hatcurSMEversionxxxxxD}{i}}{\hatcurSMEizfehshort{\hatcurplanetnumxxxxxD}}{\hatcurSMEiizfehshort{\hatcurplanetnumxxxxxD}}}
\newcommand{\hatcurSMEloggxxxxxD}{\ifthenelse{\equal{\hatcurSMEversionxxxxxD}{i}}{\hatcurSMEilogg{\hatcurplanetnumxxxxxD}}{\hatcurSMEiilogg{\hatcurplanetnumxxxxxD}}}
\newcommand{\hatcurSMEvsinxxxxxD}{\ifthenelse{\equal{\hatcurSMEversionxxxxxD}{i}}{\hatcurSMEivsin{\hatcurplanetnumxxxxxD}}{\hatcurSMEiivsin{\hatcurplanetnumxxxxxD}}}
\newcommand{\hatcurSMEvmacxxxxxD}{\ifthenelse{\equal{\hatcurSMEversionxxxxxD}{i}}{\hatcurSMEivmac{\hatcurplanetnumxxxxxD}}{\hatcurSMEiivmac{\hatcurplanetnumxxxxxD}}}
\newcommand{\hatcurSMEvmicxxxxxD}{\ifthenelse{\equal{\hatcurSMEversionxxxxxD}{i}}{\hatcurSMEivmic{\hatcurplanetnumxxxxxD}}{\hatcurSMEiivmic{\hatcurplanetnumxxxxxD}}}
\newcommand{\hatcur}[1]{\ifnum#1=47 %
\hatcurxxxxxA
\else
\ifnum#1=48 %
\hatcurxxxxxB
\else
\ifnum#1=49 %
\hatcurxxxxxC
\else
\ifnum#1=72 %
\hatcurxxxxxD
\else
??????\fi
\fi
\fi
\fi
}
\newcommand{\hatcurb}[1]{\ifnum#1=47 %
\hatcurbxxxxxA
\else
\ifnum#1=48 %
\hatcurbxxxxxB
\else
\ifnum#1=49 %
\hatcurbxxxxxC
\else
\ifnum#1=72 %
\hatcurbxxxxxD
\else
??????\fi
\fi
\fi
\fi
}
\newcommand{\hatcurc}[1]{\ifnum#1=47 %
\hatcurcxxxxxA
\else
\ifnum#1=48 %
\hatcurcxxxxxB
\else
\ifnum#1=49 %
\hatcurcxxxxxC
\else
\ifnum#1=72 %
\hatcurcxxxxxD
\else
??????\fi
\fi
\fi
\fi
}
\newcommand{\hatcurCCtassvi}[1]{\ifnum#1=47 %
\hatcurCCtassvixxxxxA
\else
\ifnum#1=48 %
\hatcurCCtassvixxxxxB
\else
\ifnum#1=49 %
\hatcurCCtassvixxxxxC
\else
\ifnum#1=72 %
\hatcurCCtassvixxxxxD
\else
??????\fi
\fi
\fi
\fi
}
\newcommand{\hatcurCCtic}[1]{\ifnum#1=47 %
\hatcurCCticxxxxxA
\else
\ifnum#1=48 %
\hatcurCCticxxxxxB
\else
\ifnum#1=49 %
\hatcurCCticxxxxxC
\else
\ifnum#1=72 %
\hatcurCCticxxxxxD
\else
??????\fi
\fi
\fi
\fi
}
\newcommand{\hatcurCCtoi}[1]{\ifnum#1=47 %
\hatcurCCtoixxxxxA
\else
\ifnum#1=48 %
\hatcurCCtoixxxxxB
\else
\ifnum#1=49 %
\hatcurCCtoixxxxxC
\else
\ifnum#1=72 %
\hatcurCCtoixxxxxD
\else
??????\fi
\fi
\fi
\fi
}
\newcommand{\hatcurCCtwomassshort}[1]{\ifnum#1=47 %
\hatcurCCtwomassshortxxxxxA
\else
\ifnum#1=48 %
\hatcurCCtwomassshortxxxxxB
\else
\ifnum#1=49 %
\hatcurCCtwomassshortxxxxxC
\else
\ifnum#1=72 %
\hatcurCCtwomassshortxxxxxD
\else
??????\fi
\fi
\fi
\fi
}
\newcommand{\hatcurCCwasp}[1]{\ifnum#1=47 %
\hatcurCCwaspxxxxxA
\else
\ifnum#1=48 %
\hatcurCCwaspxxxxxB
\else
\ifnum#1=49 %
\hatcurCCwaspxxxxxC
\else
\ifnum#1=72 %
\hatcurCCwaspxxxxxD
\else
??????\fi
\fi
\fi
\fi
}
\newcommand{\hatcurisocite}[1]{\ifnum#1=47 %
\hatcurisocitexxxxxA
\else
\ifnum#1=48 %
\hatcurisocitexxxxxB
\else
\ifnum#1=49 %
\hatcurisocitexxxxxC
\else
\ifnum#1=72 %
\hatcurisocitexxxxxD
\else
??????\fi
\fi
\fi
\fi
}
\newcommand{\hatcurisofull}[1]{\ifnum#1=47 %
\hatcurisofullxxxxxA
\else
\ifnum#1=48 %
\hatcurisofullxxxxxB
\else
\ifnum#1=49 %
\hatcurisofullxxxxxC
\else
\ifnum#1=72 %
\hatcurisofullxxxxxD
\else
??????\fi
\fi
\fi
\fi
}
\newcommand{\hatcurisoshort}[1]{\ifnum#1=47 %
\hatcurisoshortxxxxxA
\else
\ifnum#1=48 %
\hatcurisoshortxxxxxB
\else
\ifnum#1=49 %
\hatcurisoshortxxxxxC
\else
\ifnum#1=72 %
\hatcurisoshortxxxxxD
\else
??????\fi
\fi
\fi
\fi
}
\newcommand{\hatcurjhkfilset}[1]{\ifnum#1=47 %
\hatcurjhkfilsetxxxxxA
\else
\ifnum#1=48 %
\hatcurjhkfilsetxxxxxB
\else
\ifnum#1=49 %
\hatcurjhkfilsetxxxxxC
\else
\ifnum#1=72 %
\hatcurjhkfilsetxxxxxD
\else
??????\fi
\fi
\fi
\fi
}
\newcommand{\hatcurlumind}[1]{\ifnum#1=47 %
\hatcurlumindxxxxxA
\else
\ifnum#1=48 %
\hatcurlumindxxxxxB
\else
\ifnum#1=49 %
\hatcurlumindxxxxxC
\else
\ifnum#1=72 %
\hatcurlumindxxxxxD
\else
??????\fi
\fi
\fi
\fi
}
\newcommand{\hatcurplanetnum}[1]{\ifnum#1=47 %
\hatcurplanetnumxxxxxA
\else
\ifnum#1=48 %
\hatcurplanetnumxxxxxB
\else
\ifnum#1=49 %
\hatcurplanetnumxxxxxC
\else
\ifnum#1=72 %
\hatcurplanetnumxxxxxD
\else
??????\fi
\fi
\fi
\fi
}
\newcommand{\hatcurRVgammaabs}[1]{\ifnum#1=47 %
\hatcurRVgammaabsxxxxxA
\else
\ifnum#1=48 %
\hatcurRVgammaabsxxxxxB
\else
\ifnum#1=49 %
\hatcurRVgammaabsxxxxxC
\else
\ifnum#1=72 %
\hatcurRVgammaabsxxxxxD
\else
??????\fi
\fi
\fi
\fi
}
\newcommand{\hatcurRVgammarel}[1]{\ifnum#1=47 %
\hatcurRVgammarelxxxxxA
\else
\ifnum#1=48 %
\hatcurRVgammarelxxxxxB
\else
\ifnum#1=49 %
\hatcurRVgammarelxxxxxC
\else
\ifnum#1=72 %
\hatcurRVgammarelxxxxxD
\else
??????\fi
\fi
\fi
\fi
}
\newcommand{\hatcurSMElogg}[1]{\ifnum#1=47 %
\hatcurSMEloggxxxxxA
\else
\ifnum#1=48 %
\hatcurSMEloggxxxxxB
\else
\ifnum#1=49 %
\hatcurSMEloggxxxxxC
\else
\ifnum#1=72 %
\hatcurSMEloggxxxxxD
\else
??????\fi
\fi
\fi
\fi
}
\newcommand{\hatcurSMEteff}[1]{\ifnum#1=47 %
\hatcurSMEteffxxxxxA
\else
\ifnum#1=48 %
\hatcurSMEteffxxxxxB
\else
\ifnum#1=49 %
\hatcurSMEteffxxxxxC
\else
\ifnum#1=72 %
\hatcurSMEteffxxxxxD
\else
??????\fi
\fi
\fi
\fi
}
\newcommand{\hatcurSMEversion}[1]{\ifnum#1=47 %
\hatcurSMEversionxxxxxA
\else
\ifnum#1=48 %
\hatcurSMEversionxxxxxB
\else
\ifnum#1=49 %
\hatcurSMEversionxxxxxC
\else
\ifnum#1=72 %
\hatcurSMEversionxxxxxD
\else
??????\fi
\fi
\fi
\fi
}
\newcommand{\hatcurSMEvmac}[1]{\ifnum#1=47 %
\hatcurSMEvmacxxxxxA
\else
\ifnum#1=48 %
\hatcurSMEvmacxxxxxB
\else
\ifnum#1=49 %
\hatcurSMEvmacxxxxxC
\else
\ifnum#1=72 %
\hatcurSMEvmacxxxxxD
\else
??????\fi
\fi
\fi
\fi
}
\newcommand{\hatcurSMEvmic}[1]{\ifnum#1=47 %
\hatcurSMEvmicxxxxxA
\else
\ifnum#1=48 %
\hatcurSMEvmicxxxxxB
\else
\ifnum#1=49 %
\hatcurSMEvmicxxxxxC
\else
\ifnum#1=72 %
\hatcurSMEvmicxxxxxD
\else
??????\fi
\fi
\fi
\fi
}
\newcommand{\hatcurSMEvsin}[1]{\ifnum#1=47 %
\hatcurSMEvsinxxxxxA
\else
\ifnum#1=48 %
\hatcurSMEvsinxxxxxB
\else
\ifnum#1=49 %
\hatcurSMEvsinxxxxxC
\else
\ifnum#1=72 %
\hatcurSMEvsinxxxxxD
\else
??????\fi
\fi
\fi
\fi
}
\newcommand{\hatcurSMEzfeh}[1]{\ifnum#1=47 %
\hatcurSMEzfehxxxxxA
\else
\ifnum#1=48 %
\hatcurSMEzfehxxxxxB
\else
\ifnum#1=49 %
\hatcurSMEzfehxxxxxC
\else
\ifnum#1=72 %
\hatcurSMEzfehxxxxxD
\else
??????\fi
\fi
\fi
\fi
}
\newcommand{\hatcurSMEzfehshort}[1]{\ifnum#1=47 %
\hatcurSMEzfehshortxxxxxA
\else
\ifnum#1=48 %
\hatcurSMEzfehshortxxxxxB
\else
\ifnum#1=49 %
\hatcurSMEzfehshortxxxxxC
\else
\ifnum#1=72 %
\hatcurSMEzfehshortxxxxxD
\else
??????\fi
\fi
\fi
\fi
}

\newcounter{planetcounter}

\newboolean{emulateapj}
\setboolean{emulateapj}{true}

\newboolean{rvtablelong}
\setboolean{rvtablelong}{true}

\newboolean{astroph}
\setboolean{astroph}{true}

\shortauthors{Hartman et al.}
\shorttitle{\hatcur{47}\lowercase{b}, \hatcur{48}\lowercase{b}, \hatcur{49}\lowercase{b} and \hatcur{72}\lowercase{b}}
\ifthenelse{\boolean{emulateapj}}{
    
}{
    
}

\begin{document}

\title{
\hatcur{47}\lowercase{b}, \hatcur{48}\lowercase{b}, \hatcur{49}\lowercase{b} and \hatcur{72}\lowercase{b}: Four Warm Giant Planets Transiting K Dwarfs\footnote{The HATSouth network is operated by a collaboration consisting of
Princeton University (PU), the Max Planck Institute f\"ur Astronomie
(MPIA), the Australian National University (ANU), and the Pontificia
Universidad Cat\'olica de Chile (PUC).  The station at Las Campanas
Observatory (LCO) of the Carnegie Institute is operated by PU in
conjunction with PUC, the station at the High Energy Spectroscopic
Survey (H.E.S.S.) site is operated in conjunction with MPIA, and the
station at Siding Spring Observatory (SSO) is operated jointly with
ANU.
 Based in
 part on observations made with the MPG~2.2\,m Telescope at the ESO
 Observatory in La Silla.
Based on observations collected at the European Southern Observatory.
 This paper includes data gathered with the 6.5 meter Magellan Telescopes at Las Campanas Observatory, Chile.
}
}

\correspondingauthor{Joel Hartman}
\email{jhartman@astro.princeton.edu}

\author[0000-0001-8732-6166]{J. D. Hartman}
\affil{Department of Astrophysical Sciences, Princeton University, NJ 08544, USA}

\author[0000-0002-5389-3944]{Andr\'es Jord\'an}
\affiliation{Facultad de Ingenier\'ia y Ciencias, Universidad Adolfo Ib\'a\~nez, Av.\ Diagonal las Torres 2640, Pe\~nalol\'en, Santiago, Chile}
\affiliation{Millennium Institute for Astrophysics, Chile}

\author[0000-0001-6023-1335]{D. Bayliss}
\affil{Department of Physics, University of Warwick, Gibbet Hill Road, Coventry CV4 7AL, UK}

\author[0000-0001-7204-6727]{G. \'A. Bakos}
\altaffiliation{Packard Fellow}
\affil{Department of Astrophysical Sciences, Princeton University, NJ 08544, USA}
\affil{MTA Distinguished Guest Fellow, Konkoly Observatory, Hungary}

\author[0000-0002-9832-9271]{J. Bento}
\affil{Research School of Astronomy and Astrophysics, Australian National University, Canberra, ACT 2611, Australia}

\author[0000-0002-0628-0088]{W. Bhatti}
\affil{Department of Astrophysical Sciences, Princeton University, NJ 08544, USA}

\author[0000-0002-9158-7315]{R. Brahm}
\affil{Center of Astro-Engineering UC, Pontificia Universidad Cat\'olica de Chile, Av. Vicu\~{n}a Mackenna 4860, 7820436 Macul, Santiago, Chile}
\affil{Instituto de Astrof{\'{i}}sica, Pontificia Universidad Cat{\'{o}}lica de Chile, Av. Vicu{\~{n}}a Mackenna 4860, 7820436 Macul, Santiago, Chile}
\affiliation{Millennium Institute for Astrophysics, Chile}

\author{Z. Csubry}
\affil{Department of Astrophysical Sciences, Princeton University, NJ 08544, USA}

\author[0000-0001-9513-1449]{N. Espinoza}
\affil{Space Telescope Science Institute, 3700 San Martin Drive, Baltimore, MD 21218, USA}

\author{Th. Henning}
\affil{Max Planck Institute for Astronomy, K{\"{o}}nigstuhl 17, 69117 - Heidelberg, Germany}

\author[0000-0002-9428-8732]{L. Mancini}
\affil{Department of Physics, University of Rome Tor Vergata, Via della
Ricerca Scientifica 1, I-00133 - Roma, Italy}
\affil{Max Planck Institute for Astronomy, K{\"{o}}nigstuhl 17, 69117 - Heidelberg, Germany}
\affil{INAF - Astrophysical Observatory of Turin, Via Osservatorio 20, I-10025 - Pino Torinese, Italy}

\author[0000-0003-4464-1371]{K. Penev}
\affil{Department of Physics, University of Texas at Dallas, Richardson, TX 75080, USA}

\author[0000-0003-2935-7196]{M. Rabus}
\affil{Las Cumbres Observatory Global Telescope Network, 6740 Cortona Dr. Suite 102, Goleta, CA 93117}
\affil{Department of Physics, University of California, Santa Barbara, CA 93106-9530, USA}

\author[0000-0001-8128-3126]{P. Sarkis}
\affil{Max Planck Institute for Astronomy, K{\"{o}}nigstuhl 17, 69117 - Heidelberg, Germany}

\author[0000-0001-7070-3842]{V. Suc}
\affil{Instituto de Astrof{\'{i}}sica, Pontificia Universidad Cat{\'{o}}lica de Chile, Av. Vicu{\~{n}}a Mackenna 4860, 7820436 Macul, Santiago, Chile}

\author[0000-0002-0455-9384]{M. de Val-Borro}
\affil{Astrochemistry Laboratory, Goddard Space Flight Center, NASA, 8800 Greenbelt Rd, Greenbelt, MD 20771, USA}

\author[0000-0002-4891-3517]{G. Zhou}
\affil{Harvard-Smithsonian Center for Astrophysics, 60 Garden St., Cambridge, MA 02138, USA}

\author[0000-0002-5226-787X]{J. D. Crane}
\affil{The Observatories of the Carnegie Institution for Science, 813 Santa Barbara St, Pasadena, CA 91101, USA}

\author{S. Shectman}
\affil{The Observatories of the Carnegie Institution for Science, 813 Santa Barbara St, Pasadena, CA 91101, USA}

\author{J. K. Teske}
\altaffiliation{NASA Hubble Fellow}
\affil{The Observatories of the Carnegie Institution for Science, 813 Santa Barbara St, Pasadena, CA 91101, USA}

\author{S. X. Wang}
\affil{The Observatories of the Carnegie Institution for Science, 813 Santa Barbara St, Pasadena, CA 91101, USA}

\author[0000-0003-1305-3761]{R. P. Butler}
\affil{Department of Terrestrial Magnetism, Carnegie Institution for Science, Washington, DC 20015, USA}

\author{J. L\'az\'ar}
\affil{Hungarian Astronomical Association, 1451 Budapest, Hungary}

\author{I. Papp}
\affil{Hungarian Astronomical Association, 1451 Budapest, Hungary}

\author{P. S\'ari}
\affil{Hungarian Astronomical Association, 1451 Budapest, Hungary}

\author{D. R. Anderson}
\affil{Astrophysics Group, Keele University, Staffordshire, ST5 5BG, UK}
\affil{Department of Physics, University of Warwick, Gibbet Hill Road, Coventry CV4 7AL, UK}

\author{C. Hellier}
\affil{Astrophysics Group, Keele University, Staffordshire, ST5 5BG, UK}

\author{R. G. West}
\affil{Department of Physics, University of Warwick, Gibbet Hill Road, Coventry CV4 7AL, UK}

\author[0000-0003-1464-9276]{K. Barkaoui}
\affil{Astrobiology Research Unit, University of Liege, All\'{e}e du 6 Ao\^{u}t 19C, 4000 Li\`{e}ge, Belgium}
\affil{LPHEA Laboratory, Oukaimeden Observatory, Cadi Ayyad University/FSSM, BP 2390, Marrakesh, Morocco}

\author[0000-0003-1572-7707]{F.~J. Pozuelos}
\affil{Astrobiology Research Unit, University of Liege, All\'{e}e du 6 Ao\^{u}t 19C, 4000 Li\`{e}ge, Belgium}
\affil{STAR Research Unit, University of Liege, All\'{e}e du 6 Ao\^{u}t 19C, 4000 Li\`{e}ge, Belgium}

\author{E. Jehin}
\affil{STAR Research Unit, University of Liege, All\'{e}e du 6 Ao\^{u}t 19C, 4000 Li\`{e}ge, Belgium}

\author[0000-0003-1462-7739]{M. Gillon}
\affil{Astrobiology Research Unit, University of Liege, All\'{e}e du 6 Ao\^{u}t 19C, 4000 Li\`{e}ge, Belgium}

\author{L. Nielsen}
\affil{Observatoire de Gen\`eve, Universit\'e de Gen\`eve, 51 Ch. des Maillettes, 1290 Sauverny, Switzerland}

\author[0000-0001-9699-1459]{M. Lendl}
\affil{Observatoire de Gen\`eve, Universit\'e de Gen\`eve, 51 Ch. des Maillettes, 1290 Sauverny, Switzerland}
\affil{Space Research Institute, Austrian Academy of Sciences, Schmiedlstr. 6, 8042 Graz, Austria}

\author{S. Udry}
\affil{Observatoire de Gen\`eve, Universit\'e de Gen\`eve, 51 Ch. des Maillettes, 1290 Sauverny, Switzerland}

\author[0000-0003-2058-6662]{George~R.~Ricker}
\affiliation{Department of Physics and Kavli Institute for Astrophysics and Space Research, Massachusetts Institute of Technology, Cambridge, MA 02139, USA}

\author[0000-0001-6763-6562]{Roland~Vanderspek}
\affiliation{Department of Physics and Kavli Institute for Astrophysics and Space Research, Massachusetts Institute of Technology, Cambridge, MA 02139, USA}

\author[0000-0001-9911-7388]{David~W.~Latham}
\affiliation{Harvard-Smithsonian Center for Astrophysics, 60 Garden St, Cambridge, MA 02138, USA}

\author[0000-0002-6892-6948]{S.~Seager}
\affiliation{Department of Physics and Kavli Institute for Astrophysics and Space Research, Massachusetts Institute of Technology, Cambridge, MA 02139, USA}
\affiliation{Department of Earth, Atmospheric and Planetary Sciences, Massachusetts Institute of Technology, Cambridge, MA 02139, USA}
\affiliation{Department of Aeronautics and Astronautics, MIT, 77 Massachusetts Avenue, Cambridge, MA 02139, USA}

\author[0000-0002-4265-047X]{Joshua~N.~Winn}
\affil{Department of Astrophysical Sciences, Princeton University, NJ 08544, USA}

\author[0000-0002-8035-4778]{Jessie~Christiansen}
\affiliation{Caltech/IPAC-NASA Exoplanet Science Institute, 770 S. Wilson Avenue, Pasadena, CA 91106, USA}

\author{Ian~J.~M.~Crossfield}
\affiliation{Department of Physics and Astronomy, University of Kansas,
1251 Wescoe Hall Dr., Lawrence, KS 66045, USA}
\affiliation{Department of Physics and Kavli Institute for Astrophysics and Space Research, Massachusetts Institute of Technology, Cambridge, MA 02139, USA}

\author{Christopher~E.~Henze}
\affiliation{NASA Ames Research Center, Moffett Field, CA, 94035, USA}

\author[0000-0002-4715-9460]{Jon~M.~Jenkins}
\affiliation{NASA Ames Research Center, Moffett Field, CA, 94035, USA}

\author{Jeffrey~C.~Smith}
\affiliation{SETI Institute, Mountain View, CA 94043, USA}
\affiliation{NASA Ames Research Center, Moffett Field, CA, 94035, USA}

\author[0000-0002-8219-9505]{Eric~B.~Ting}
\affiliation{NASA Ames Research Center, Moffett Field, CA, 94035, USA}


\begin{abstract}

\setcounter{footnote}{10}
We report the discovery of four transiting giant planets around K dwarfs. The planets \hatcurb{47}, \hatcurb{48}, \hatcurb{49}, and \hatcurb{72} have masses of \hatcurPPm{47}\,\mjup, \hatcurPPm{48}\,\mjup, \hatcurPPm{49}\,\mjup\ and \hatcurPPm{72}\,\mjup, respectively, and radii of \hatcurPPr{47}\,\rjup, \hatcurPPr{48}\,\rjup, \hatcurPPr{49}\,\rjup, and \hatcurPPr{72}\,\rjup, respectively. The planets orbit close to their host stars with orbital periods of \hatcurLCPshort{47}\,d, \hatcurLCPshort{48}\,d, \hatcurLCPshort{49}\,d and \hatcurLCPshort{72}\,d, respectively. The hosts are main sequence K dwarfs with masses of \hatcurISOm{47}\,\msun, \hatcurISOm{48}\,\msun, \hatcurISOm{49}\,\msun, and \hatcurISOm{72}, and with $V$-band magnitudes of $V = \hatcurCCtassmv{47}$, \hatcurCCtassmv{48}, \hatcurCCtassmv{49} and \hatcurCCtassmv{72}. The Super-Neptune \hatcurb{72} (a.k.a.\ \hatcurCCwasp{72} and TOI~\hatcurCCtoi{72}) was independently identified as a transiting planet candidate by the HATSouth, WASP and {\em TESS} surveys, and we present a combined analysis of all of the data gathered by each of these projects (and their follow-up programs). An exceptionally precise mass is measured for \hatcurb{72} thanks to high-precision radial velocity (RV) measurements obtained with VLT/ESPRESSO, FEROS, HARPS and Magellan/PFS.  We also incorporate {\em TESS} observations of the warm Saturn-hosting systems \hatcur{47} (a.k.a.\ TOI~\hatcurCCtoi{47}), \hatcur{48} and \hatcur{49}. \hatcur{47} was independently identified as a candidate by the {\em TESS} team, while the other two systems were not previously identified from the {\em TESS} data. The RV orbital variations are measured for these systems using Magellan/PFS. \hatcur{48} has a resolved $5\farcs4$ neighbor in Gaia~DR2, which is a common-proper-motion binary star companion to \hatcur{48} with a mass of $0.22$\,\msun\ and
a current projected physical separation of $\sim$1{,}400\,au.
\setcounter{footnote}{0}
\end{abstract}

\keywords{
    planetary systems ---
    stars: individual (
\setcounter{planetcounter}{1}
\hatcur{47},
TOI~\hatcurCCtoi{47},
TIC~\hatcurCCtic{47}\loopcommanoperiod
\setcounter{planetcounter}{2}
\hatcur{48},
TIC~\hatcurCCtic{48}\loopcommanoperiod
\setcounter{planetcounter}{3}
\hatcur{49},
TIC~\hatcurCCtic{49}\loopcommanoperiod
\setcounter{planetcounter}{4}
\hatcur{72},
TIC~\hatcurCCtic{72},
TOI~\hatcurCCtoi{72}\loopcommanoperiod
\setcounter{planetcounter}{5}
) 
    techniques: spectroscopic, photometric
}


\section{Introduction}
\label{sec:introduction}

Much of our empirical knowledge about the physical properties of
planets beyond the Solar System (exoplanets) comes from observing
planets with orbits that are fortuitously oriented such that the
planets transit in front of their host stars from our vantage
point. 

Of particular importance are transiting planets with masses
that have been measured either via high-precision radial velocity (RV)
observations
\citep[e.g.,][]{henry:2000,charbonneau:2000,konacki:2003}, or by
observing deviations from strict periodicity in the transit times of
other planets in the planetary system
\citep[e.g.,][]{holman:2010}. Measuring both the planetary mass and
radius (the latter being measurable from the transits once the physical
properties of the host star are determined), together with the
incident stellar flux (determined from the period once the stellar
luminosity and mass are known) and system age (determined as one of
the host star properties), allows constraints to be set on the
composition of the planet \citep[e.g.,][]{guillot:2006}. The necessary stellar
properties can be determined by comparing photometric, astrometric,
and spectroscopic observations of the star to empirical or theoretical
relations between stellar physical and observable parameters. 

To date, more than 3{,}000 transiting planets have been confirmed or validated.\footnote{The NASA Exoplanet Archive,
  \url{https://exoplanetarchive.ipac.caltech.edu}, accessed 2019 Sep.\ 9}. About 90\% of these were found by the NASA {\em Kepler}
mission \citep{borucki:2010}, or its successor {\em K2}
\citep{howell:2014}. However, the planetary masses have not yet been
measured for the majority of these. Of the $\sim 600$ transiting
planets with measured masses, only about one-third were
discovered by {\em Kepler} or {\em K2}. The masses
of typical {\em Kepler} and
{\em K2} transiting planets are difficult to measure
because the planets are too small, have periods that
are too long, or are orbiting stars that are too faint for effective RV monitoring.

In order to increase the
number of exoplanets --- particularly planets
smaller than Neptune --- with measured masses,
NASA launched the {\em TESS} space-based photometer \citep{ricker:2015}. For its primary mission, {\em TESS} is carrying out a two-year survey of approximately three-quarters of the sky to
find transiting planets around bright stars.
As of the time of writing, the mission has been
operational for a year and a half, and has led to the identification of about
1{,}000 transiting planet candidates, of which 17 new planets have so
far been confirmed and have had their masses measured. The follow-up
observations needed to confirm and characterize the transiting planet
candidates identified by {\em TESS} are being carried out by the
community, with organization provided by the {\em TESS} Follow-up
Program \citep[TFOP;][]{collins:2018}. 

Approximately half of the transiting planets with measured masses were
discovered by wide-field ground-based transit surveys, especially the
WASP \citep{pollacco:2006}, HAT \citep{bakos:2004:hatnet,bakos:2013:hatsouth} and KELT \citep{pepper:2007,pepper:2012} projects. These
projects have primarily been sensitive to short-period gas-giant
planets. They have discovered the majority of exoplanets that have been
the subject of observational studies of the planetary atmospheres, and
measurements of the stellar-spin--planetary-orbit alignment. These
projects, some of which have been in operation for more than a decade,
are now contributing to the follow-up and confirmation
of transiting planets from the {\em TESS} mission. This includes both
providing photometric observations carried out with the ground-based
transit survey instruments, and follow-up spectroscopic and
photometric observations using facilities, procedures and teams that
were originally brought together to confirm candidates from the
ground-based surveys. 

Many of the transiting giant planet candidates detected by {\em TESS}
had already been identified by the ground-based projects. Some are
confirmed and published planets, some have
already been ruled out as
false positives, some have been confirmed but not yet published
(including two cases presented in this paper), some have survived initial
follow-up vetting observations but are not yet confirmed, and some have simply
not been followed-up yet. There are also cases in which objects
that were selected as candidates by the ground-based surveys have not
been identified publicly by the {\em TESS} team. Many of these latter objects are
false alarms. However, in some cases the transiting planets have been
confirmed, and the reason they were not
identified by {\it TESS} is that the signals are weak due to crowding,
substantial scattered light, or other issues. In other cases, the
candidates are around stars that are fainter than the magnitude
thresholds being applied by the {\em TESS} team in searching for
transit signals. Two of the planets presented
in this paper have stars that are fainter than the magnitude
limits currently being searched with the
standard {\em TESS} procedures.

In this paper we present the discovery and characterization --- including
precise mass measurements --- of four giant planets transiting K dwarf stars.
The four planets \hatcurb{47}, \hatcurb{48}, \hatcurb{49} and
\hatcurb{72} were identified by the HATSouth transit survey, and one
was also independently identified by the WASP survey. All four of
these objects also have transits that can be detected using {\em TESS}
data, but only two of them have been selected as candidate transiting
planet systems by the {\em TESS} team. The planets have masses between
\hatcurPPmlong{72}\,\mjup\ (\hatcurb{72}) and
\hatcurPPmlong{47}\,\mjup\ (\hatcurb{47}), radii between
\hatcurPPrlong{72}\,\rjup\ (\hatcurb{72}) and \hatcurPPrlong{47}\,\rjup\ (\hatcurb{47}), and orbit stars with masses
between \hatcurISOmlong{47}\,\msun\ (\hatcur{47}) and
\hatcurISOmlong{72}\,\msun\ (\hatcur{72}). Thanks to the relatively
low masses and luminosities of these host stars, the planets can all be classified
as ``warm'' giant planets rather than ``hot'' giant planets: in
all four cases, the expected equilibrium temperature computed under the assumptions of
zero albedo and isotropic reradiation,
is below 1{,}000\,K. While hot giant planets exhibit a radius inflation anomaly
\citep[e.g.,][among many other examples]{hartman:2011:hat3233}, warm
giant planets are observed to have radii that are consistent
with theoretical expectations \citep[e.g.,][]{kovacs:2010:hat15}. More recently, \citet{sestovic:2018} have argued that planets with masses $\mpl < 0.37$\,\mjup, which includes all the planets here, do not exhibit anomalous radius inflation, even when highly irradiated. As a result, it is reasonable to suppose that meaningful inferences about the bulk planet metal content
can be derived for these warm giant planets \citep[e.g.,][]{thorngren:2016}. This makes warm low-mass giant planets, like those presented here,
particularly useful for testing theories of giant planet formation and
structure.

In the following section we discuss the observations used to detect, confirm and characterize the transiting planet systems \hatcur{47}, \hatcur{48}, \hatcur{49}, and \hatcur{72}. The analysis of these data is described in Section~\ref{sec:analysis}. We discuss the results in Section~\ref{sec:discussion}.

\section{Observations}
\label{sec:obs}

Figures~\ref{fig:hats47}, \ref{fig:hats48}, \ref{fig:hats49} and
\ref{fig:hats72} show some of the observations collected for
\hatcur{47}, \hatcur{48}, \hatcur{49} and \hatcur{72},
respectively. Each figure shows the HATSouth light curve used to
detect the transits, the ground-based follow-up transit light curves,
the high-precision RVs and spectral line bisector spans (BSs), and the
catalog broad-band photometry, including parallax corrections from
Gaia~DR2 \citep{gaiadr2}, used in characterizing the host stars. We also show the TESS
light curves for each system in
Figures~\ref{fig:hats47tess}, \ref{fig:hats48tess},
\ref{fig:hats49tess} and \ref{fig:hats72tess}, and the WASP light
curve for \hatcur{72} in Figure~\ref{fig:hats72wasp}. Below we
describe the observations of these objects that were collected and analyzed here.

\subsection{Photometric detection}
\label{sec:detection}

All four of the systems presented here were detected as transiting
planet candidates by the HATSouth ground-based transiting planet
survey \citep{bakos:2013:hatsouth} as we discuss in
Section~\ref{sec:hatsouth}.  The transits of all four objects are also
detected in time-series observations collected by the {\em TESS}
mission (Section~\ref{sec:tess}), and \hatcurb{47} and \hatcurb{72}
were independently identified as transiting planet candidates by the
{\em TESS} team based on these data. The transits of \hatcurb{72} were
also independently identified by the WASP project, as discussed in
Section~\ref{sec:wasp}.

\subsubsection{HATSouth}
\label{sec:hatsouth}

HATSouth uses a network of 24 telescopes, each 0.18\,m in aperture,
and 4K$\times$4K front-side-illuminated CCD cameras. These are
attached to a total of six fully-automated mounts, each with an
associated enclosure, which are in turn located at three sites around
the Southern hemisphere. The three sites are Las Campanas Observatory (LCO) in
Chile, the site of the H.E.S.S.\ gamma-ray observatory in Namibia, and
Siding Spring Observatory (SSO) in Australia. The operations and
observing procedures of the network were described by
\citet{bakos:2013:hatsouth}, while the method for reducing the images
to trend-filtered light curves and searching for candidate transiting
planets were described by \citet{penev:2013:hats1}. We note that the
trend-filtering makes use of the Trend-Filtering Algorithm (TFA) of
\citet{kovacs:2005:TFA}, while transit signals are detected using the
Box-fitting Least Squares (BLS) method of \citet{kovacs:2002:BLS}. The
HATSouth observations of each system are summarized in
\reftabl{photobs}, and displayed in Figures~\ref{fig:hats47},
\ref{fig:hats48}, \ref{fig:hats49}, and \ref{fig:hats72}, while the
light curve data are made available in \reftabl{phfu}.

We also searched the HATSouth light curves for other periodic signals
using the Generalized Lomb-Scargle method
\citep[GLS;][]{zechmeister:2009}, and for additional transit signals
by applying a second iteration of BLS. Both of these searches were
performed on the residual light curves after subtracting the best-fit
transit models. 

\hatcur{47} shows quasi-sinusoidal periodic variability at a period of $6.621581\pm0.000078$\,days and semi-amplitude of $1.96\pm0.18$\,mmag in the $r^{\prime}$ band (Fig.~\ref{fig:gls}). The GLS false alarm probability, calibrated using a bootstrap procedure, is $10^{-34}$, indicating a strong detection. We interpret this signal as the photometric rotation period of the star. BLS also picks up on this modulation as the strongest ``transit-like'' signal in the residuals, though the duration of the ``transit'' feature is much too long for this to be due to the transit of a planet or star. No other notable transit signals are seen in the light curve.

No periodic signals are detected in the light curves of \hatcur{48} or \hatcur{49}. 

For \hatcur{72}, GLS identifies a periodic signal at a period of $P = 48.725\pm0.015$\,days and with a semi-amplitude of $0.52\pm0.12$\,mmag (Fig.~\ref{fig:gls}). We estimate a bootstrap-calibrated false alarm probability of $10^{-4}$, indicating that this is a marginal detection, though we list this as our best estimate for the photometric rotation period of the star. We do not find evidence with BLS for any additional transit signals in our light curve of this object.

%
%
\ifthenelse{\boolean{emulateapj}}{
    \begin{figure*}[!ht]
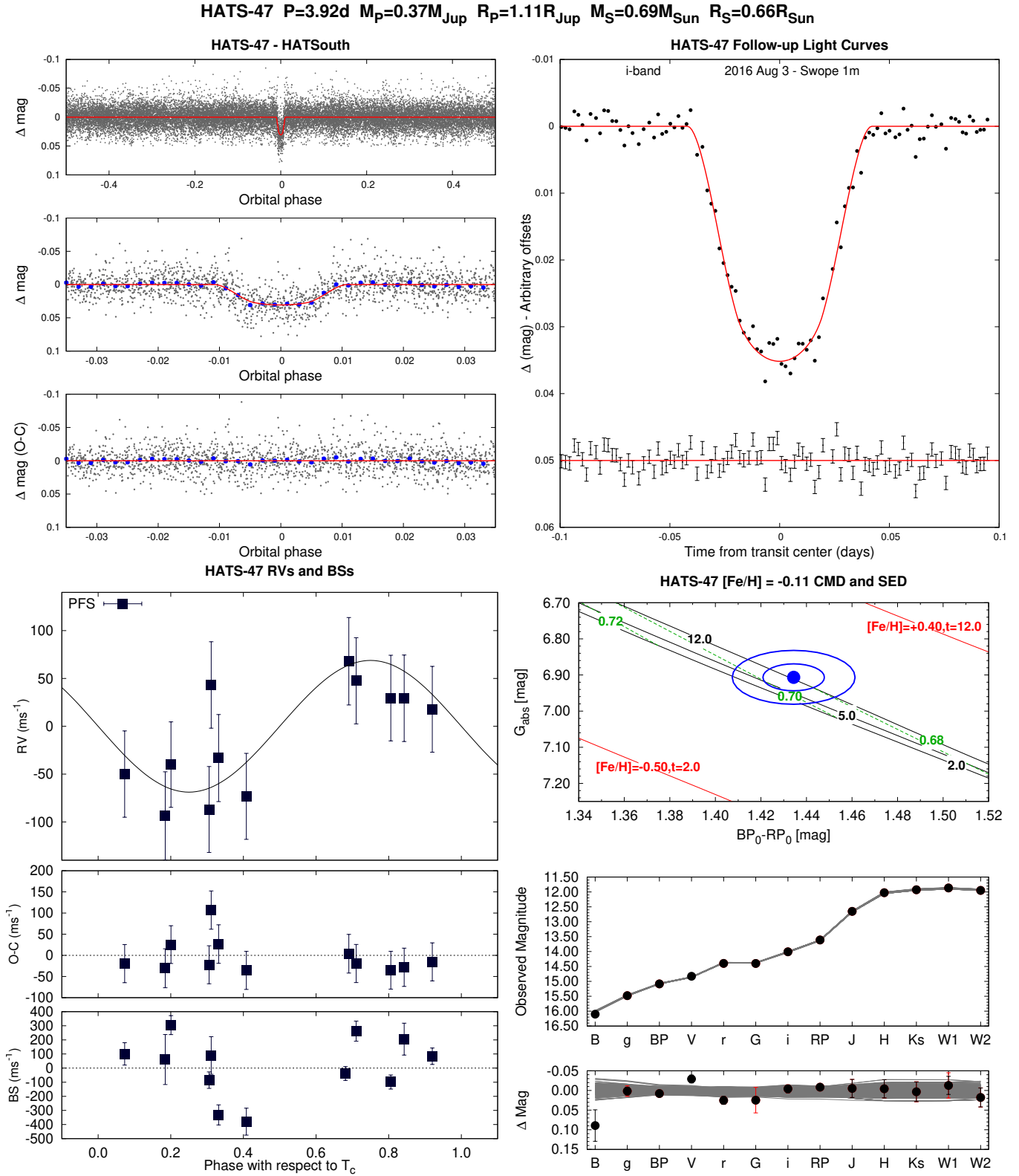

}{
    \begin{figure}[!ht]
}
 {
 \centering
 \leavevmode
 \includegraphics[width={1.0\linewidth}]{\hatcurhtr{47}-banner}
}
 {
 \centering
 \leavevmode
 \includegraphics[width={0.5\linewidth}]{\hatcurhtr{47}-hs}%
 \hfil
 \includegraphics[width={0.5\linewidth}]{\hatcurhtr{47}-lc}%
 }
 {
 \centering
 \leavevmode
 \includegraphics[width={0.5\linewidth}]{\hatcurhtr{47}-rv}%
 \hfil
 \includegraphics[width={0.5\linewidth}]{\hatcurhtr{47}-iso-bprp-gabs-isofeh-SED}%
 }                        
\caption{
    Observations used to confirm the transiting planet system \hatcur{47}, excluding data from the NASA {\em TESS} mission which are shown in Figure~\ref{fig:hats47tess}. {\em Top Left:} Phase-folded unbinned HATSouth light curve. The
    top panel shows the full light curve, the middle panel shows
    the light curve zoomed-in on the transit, and the bottom panel shows the residuals from the best-fit model zoomed-in on the transit. The solid lines show the
    model fits to the light curves. The dark filled circles show the light curves binned in phase with a bin
    size of 0.002. (Caption continued on next page.) 
\label{fig:hats47}
}
\ifthenelse{\boolean{emulateapj}}{
    \end{figure*}
}{
    \end{figure}
}

%
%
\addtocounter{figure}{-1}
\ifthenelse{\boolean{emulateapj}}{
    \begin{figure*}[!ht]
}{
    \begin{figure}[!ht]
}
\caption{
    (Caption continued from previous page.)
{\em Top Right:} Unbinned follow-up transit light curves
    corrected for instrumental trends fitted
    simultaneously with the transit model, which is overplotted.
    The dates, filters and instruments used are indicated.  In this
    figure the residuals are shown below the light curve. In
    Figures~\ref{fig:hats48},~\ref{fig:hats49} and~\ref{fig:hats72},
    the residuals are shown on the right-hand-side in the same order
    as the original light curves.  The error bars represent the photon
    and background shot noise, plus the readout noise. Note that these
    uncertainties are scaled up in the fitting procedure to achieve a
    reduced $\chi^2$ of unity, but the uncertainties shown in the plot
    have not been scaled.
{\em Bottom Left:}
High-precision RVs phased with respect to the mid-transit time. The instruments used are labelled in the plot. 
The top panel shows the phased measurements together with the best-fit model.
The center-of-mass velocity has been subtracted. The second panel shows the velocity $O\!-\!C$ residuals.
The error bars include the estimated jitter.
The third panel shows the bisector spans. 
{\em Bottom Right:} Color-magnitude diagram (CMD) and spectral energy distribution (SED). The top panel shows the absolute $G$ magnitude vs.\ the de-reddened $BP - RP$ color compared to
  theoretical isochrones (black lines) and stellar evolution tracks
  (green lines) from the PARSEC models interpolated at
  the best-estimate value for the metallicity of the host. The age
  of each isochrone is listed in black in Gyr, while the mass of each
  evolution track is listed in green in solar masses. The solid red lines show isochrones at higher and lower metallicities than the best-estimate value, with the metallicity and age in Gyr of each isochrone labelled on the plot. The filled
  blue circles show the measured reddening- and distance-corrected
  values from Gaia DR2, while the blue lines indicate
  the $1\sigma$ and $2\sigma$ confidence regions, including the
  estimated systematic errors in the photometry. The middle panel shows the SED as measured via broadband photometry through the listed filters. Here we plot the observed magnitudes without correcting for distance or extinction. Overplotted are 200 model SEDs randomly selected from the MCMC posterior distribution produced through the global analysis (gray lines). 
The model makes use of the predicted absolute magnitudes in each bandpass from the PARSEC isochrones, the distance to the system (constrained largely via Gaia DR2) and extinction (constrained from the SED with a prior coming from the {\sc mwdust} 3D Galactic extinction model).  
The bottom panel shows the $O\!-\!C$ residuals from the best-fit model SED.
\label{fig:hats47:labcontinue}}
\ifthenelse{\boolean{emulateapj}}{
    \end{figure*}
}{
    \end{figure}
}

%
%
\ifthenelse{\boolean{emulateapj}}{
    \begin{figure*}[!ht]
}{
    \begin{figure}[!ht]
}
 {
 \centering
 \leavevmode
 \includegraphics[width={1.0\linewidth}]{\hatcurhtr{48}-banner}
}
 {
 \centering
 \leavevmode
 \includegraphics[width={0.5\linewidth}]{\hatcurhtr{48}-hs}%
 \hfil
 \includegraphics[width={0.5\linewidth}]{\hatcurhtr{48}-lc}%
 }
 {
 \centering
 \leavevmode
 \includegraphics[width={0.5\linewidth}]{\hatcurhtr{48}-rv}%
 \hfil
 \includegraphics[width={0.5\linewidth}]{\hatcurhtr{48}-iso-bprp-gabs-isofeh-SED}%
 }                        
\caption{
    Same as Figure~\ref{fig:hats47}, here we show the observations of \hatcur{48} together with our best-fit model. The {\em TESS} light curve for this system is shown in Figure~\ref{fig:hats48tess}.
\label{fig:hats48}
}
\ifthenelse{\boolean{emulateapj}}{
    \end{figure*}
}{
    \end{figure}
}

%
%
\ifthenelse{\boolean{emulateapj}}{
    \begin{figure*}[!ht]
}{
    \begin{figure}[!ht]
}
 {
 \centering
 \leavevmode
 \includegraphics[width={1.0\linewidth}]{\hatcurhtr{49}-banner}
}
 {
 \centering
 \leavevmode
 \includegraphics[width={0.5\linewidth}]{\hatcurhtr{49}-hs}%
 \hfil
 \includegraphics[width={0.5\linewidth}]{\hatcurhtr{49}-lc}%
 }
 {
 \centering
 \leavevmode
 \includegraphics[width={0.5\linewidth}]{\hatcurhtr{49}-rv}%
 \hfil
 \includegraphics[width={0.5\linewidth}]{\hatcurhtr{49}-iso-bprp-gabs-isofeh-SED}%
 }                        
\caption{
    Same as Figure~\ref{fig:hats47}, here we show the observations of \hatcur{49} together with our best-fit model. The {\em TESS} light curve for this system is shown in Figure~\ref{fig:hats49tess}.
\label{fig:hats49}
}
\ifthenelse{\boolean{emulateapj}}{
    \end{figure*}
}{
    \end{figure}
}

%
%
\ifthenelse{\boolean{emulateapj}}{
    \begin{figure*}[!ht]
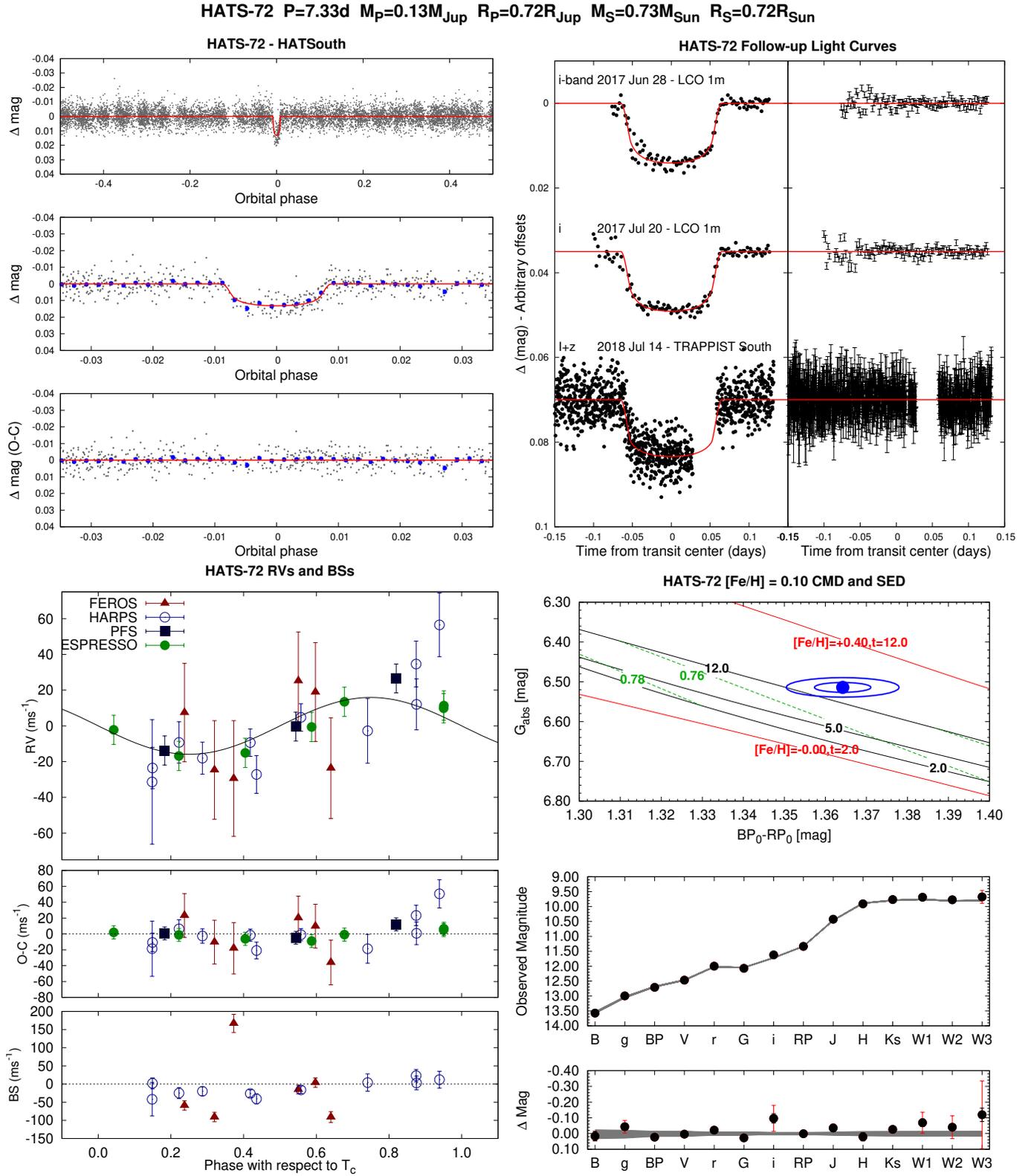

}{
    \begin{figure}[!ht]
}
 {
 \centering
 \leavevmode
 \includegraphics[width={1.0\linewidth}]{\hatcurhtr{72}-banner}
}
 {
 \centering
 \leavevmode
 \includegraphics[width={0.5\linewidth}]{\hatcurhtr{72}-hs}%
 \hfil
 \includegraphics[width={0.5\linewidth}]{\hatcurhtr{72}-lc}%
 }
 {
 \centering
 \leavevmode
 \includegraphics[width={0.5\linewidth}]{\hatcurhtr{72}-rv}%
 \hfil
 \includegraphics[width={0.5\linewidth}]{\hatcurhtr{72}-iso-bprp-gabs-isofeh-SED}%
 }                        
\caption{
    Same as Figure~\ref{fig:hats47}, here we show the observations of \hatcur{72} together with our best-fit model. The {\em TESS} light curve for this system is shown in Figure~\ref{fig:hats72tess}, while the WASP light curve is shown in Figure~\ref{fig:hats72wasp}.
\label{fig:hats72}
}
\ifthenelse{\boolean{emulateapj}}{
    \end{figure*}
}{
    \end{figure}
}

\ifthenelse{\boolean{emulateapj}}{
    \begin{figure*}[!ht]
}{
    \begin{figure}[!ht]
}
 {
 \centering
 \leavevmode
 \includegraphics[width={1.0\linewidth}]{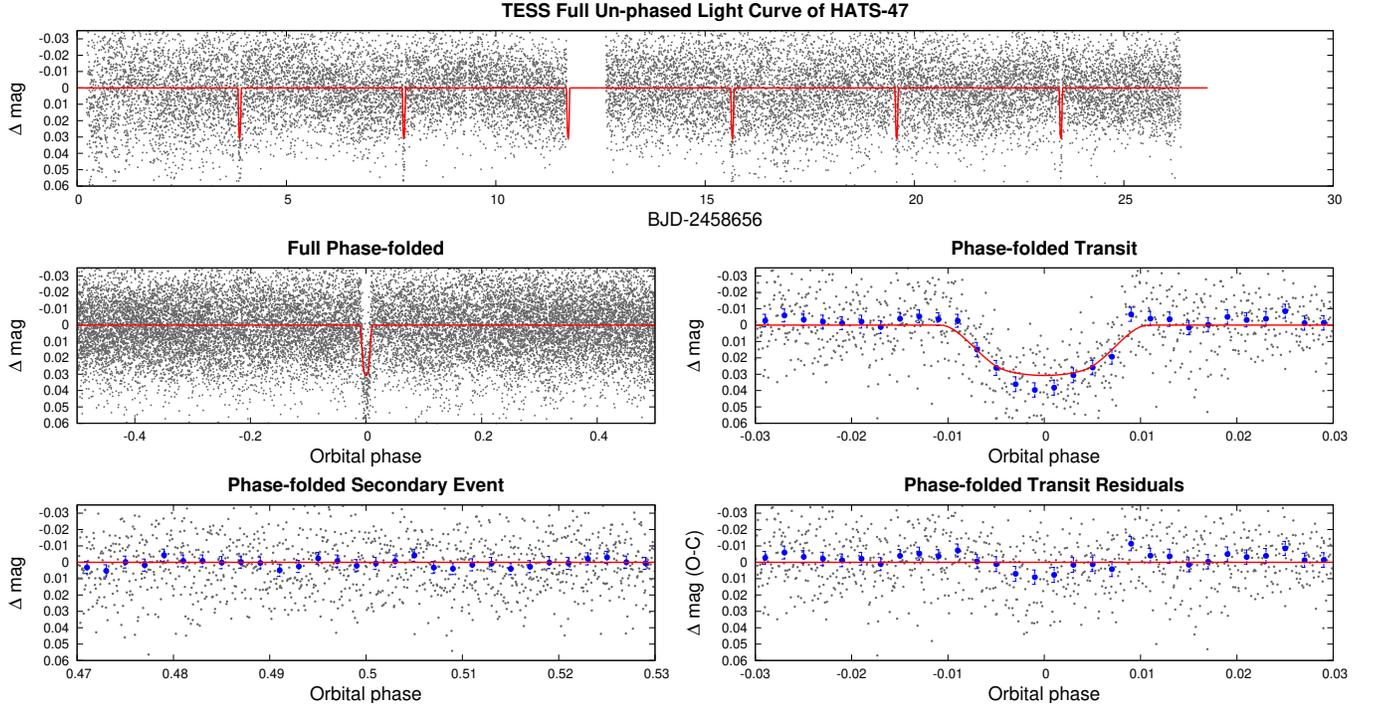}
}
\caption{
    {\em TESS} short-cadence light curve for \hatcur{47}. We show the full un-phased light curve as a function of time ({\em top}), the full phase-folded light curve ({\em middle left}), the phase-folded light curve zoomed-in on the planetary transit ({\em middle right}), the phase-folded light curve zoomed-in on the secondary eclipse ({\em bottom left}), and the residuals from the best-fit model, phase-folded and zoomed-in on the planetary transit ({\em bottom right}). The solid line in each panel shows the model fit to the light curve. The dark filled circles show the light curve binned in phase with a bin size of 0.002. Other observations included in our analysis of this system are shown in Figure~\ref{fig:hats47}. The {\em TESS} light curve has been corrected for dilution from neighbors as part of the PDC process, but in this case, where two bright neighbors are blended with the target in the {\em TESS} images, the correction appears to have been somewhat overestimated. This leads to a slightly deeper apparent transit in the {\em TESS} PDC light curve compared to the model. The model, however, fits the HATSouth and Swope~1\,m transit observations (Fig.~\ref{fig:hats47}), which were obtained at sufficiently high spatial resolution to resolve the neighbors. 
\label{fig:hats47tess}
}
\ifthenelse{\boolean{emulateapj}}{
    \end{figure*}
}{
    \end{figure}
}

\ifthenelse{\boolean{emulateapj}}{
    \begin{figure*}[!ht]
}{
    \begin{figure}[!ht]
}
 {
 \centering
 \leavevmode
 \includegraphics[width={1.0\linewidth}]{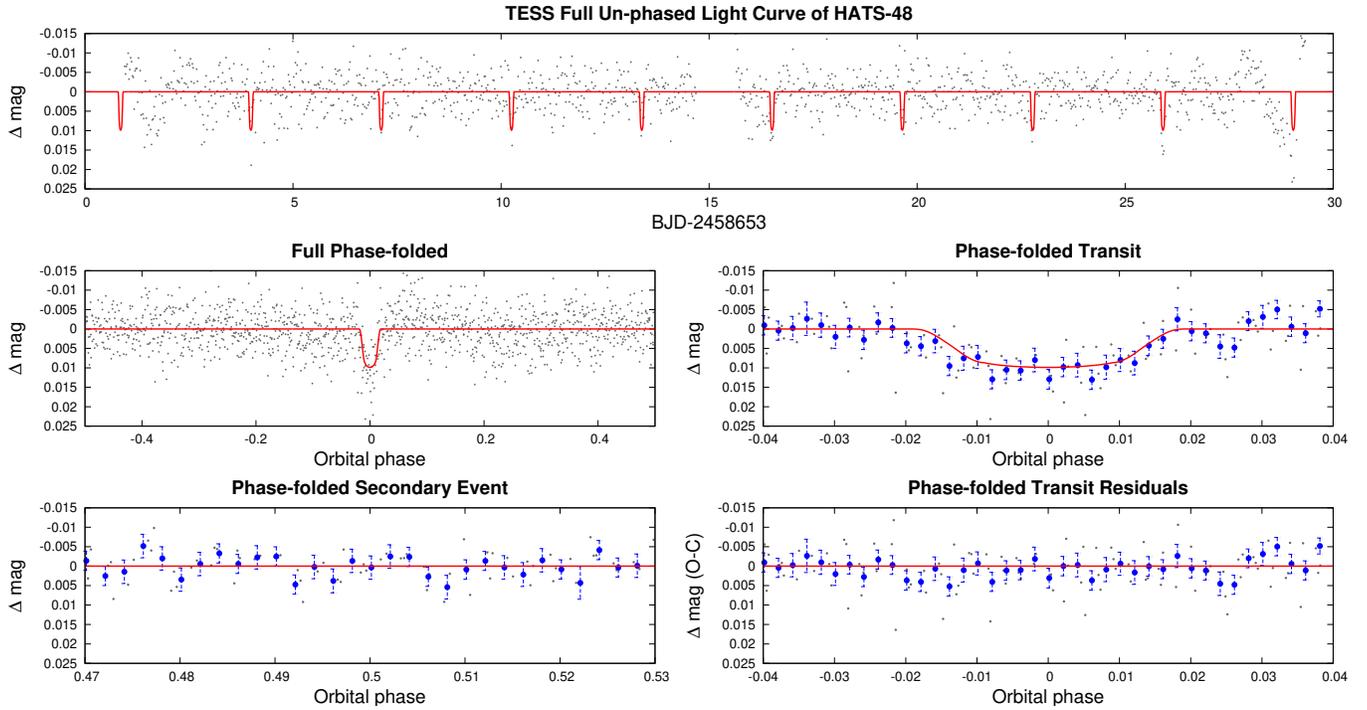}
}
\caption{
    Similar to Figure~\ref{fig:hats47tess}, here we show the {\em TESS} long-cadence light curve for \hatcur{48}. The model lines account for the 30\,min integrations. Other observations included in our analysis of this system are shown in Figure~\ref{fig:hats48}.
\label{fig:hats48tess}
}
\ifthenelse{\boolean{emulateapj}}{
    \end{figure*}
}{
    \end{figure}
}

\ifthenelse{\boolean{emulateapj}}{
    \begin{figure*}[!ht]
}{
    \begin{figure}[!ht]
}
 {
 \centering
 \leavevmode
 \includegraphics[width={1.0\linewidth}]{\hatcurhtr{49}-TESS}
}
\caption{
    Similar to Figure~\ref{fig:hats47tess}, here we show the {\em TESS} long-cadence light curve for \hatcur{49}. Other observations included in our analysis of this system are shown in Figure~\ref{fig:hats49}.
\label{fig:hats49tess}
}
\ifthenelse{\boolean{emulateapj}}{
    \end{figure*}
}{
    \end{figure}
}

\ifthenelse{\boolean{emulateapj}}{
    \begin{figure*}[!ht]
}{
    \begin{figure}[!ht]
}
 {
 \centering
 \leavevmode
 \includegraphics[width={1.0\linewidth}]{\hatcurhtr{72}-TESS}
}
\caption{
    Similar to Figure~\ref{fig:hats72tess}, here we show the {\em TESS} long-cadence light curve for \hatcur{72}. Other observations included in our analysis of this system are shown in Figures~\ref{fig:hats72}, and~\ref{fig:hats72wasp}.
\label{fig:hats72tess}
}
\ifthenelse{\boolean{emulateapj}}{
    \end{figure*}
}{
    \end{figure}
}

\begin{figure}[!ht]
 {
 \centering
 \leavevmode
 \includegraphics[width={1.0\linewidth}]{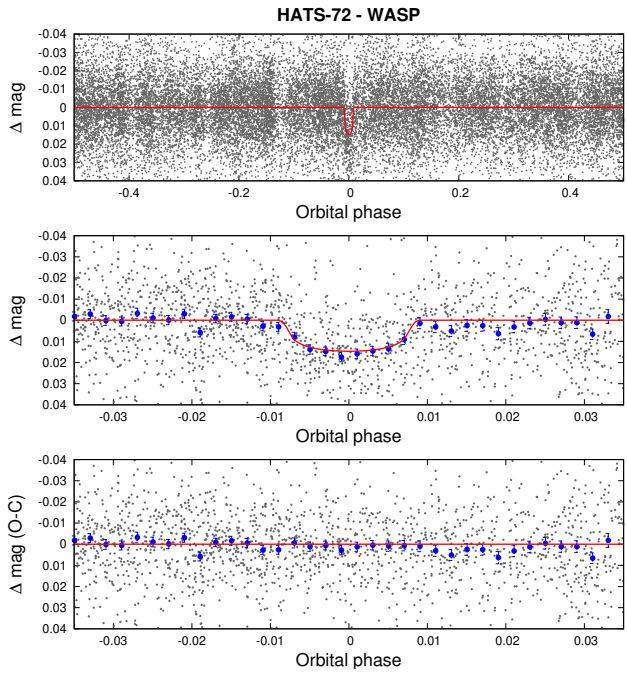}
}
\caption{
    WASP light curve for \hatcur{72}, displayed in a similar fashion to the HATSouth light curve shown in Figure~\ref{fig:hats72} (see the caption in Figure~\ref{fig:hats47}). Other observations included in our analysis of this system are shown in Figure~\ref{fig:hats72} and~\ref{fig:hats72tess}.
\label{fig:hats72wasp}
}
\end{figure}

\ifthenelse{\boolean{emulateapj}}{
    \begin{figure*}[!ht]
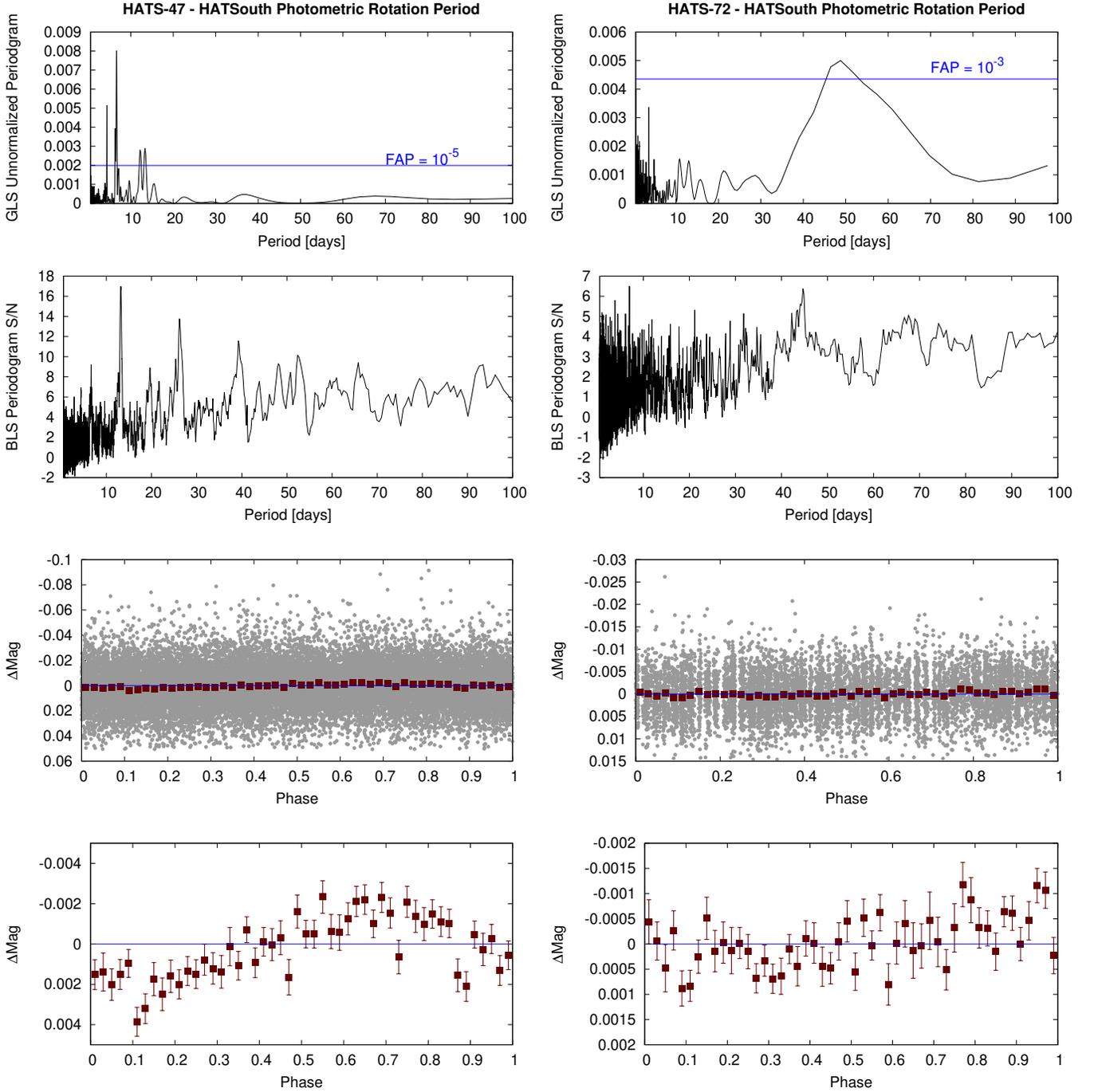

}{
    \begin{figure}[!ht]
}
 {
 \centering
 \leavevmode
 \includegraphics[width={0.5\linewidth}]{\hatcurhtr{47}-GLS}
 \hfil
 \includegraphics[width={0.5\linewidth}]{\hatcurhtr{72}-GLS}
 }
\caption{
Detection of a strong $P = 6.621581\pm0.000078$\,days photometric rotation period signal in the HATSouth light curve of \hatcur{47} ({\em left}) and a tentative $P = 48.725\pm0.015$\,days signal in the HATSouth light curve of \hatcur{72} ({\em right}). In each case we show the following panels. {\em Top:} the Generalized Lomb-Scargle (GLS) periodogram of the combined HATSouth light curve after subtracting the best-fit transit model. The horizontal blue line shows the bootstrap-calibrated $10^{-5}$ false alarm probability level for \hatcur{47} and the $10^{-3}$ false alarm probability level for \hatcur{72}. {\em Second from top:} The Box-fitting Least Squares (BLS) periodogram of the same light curve. For \hatcur{47} there is a peak in the BLS periodogram at twice the period of the strongest peak in the GLS periodogram. For \hatcur{72} no significant peak is present in the BLS periodogram. {\em Second from bottom:} The HATSouth light curve phase-folded at the peak GLS period. The gray points show the individual photometric measurements, while the dark red filled squares show the observations binned in phase with a bin size of 0.02. {\em Bottom:} Same as the second from bottom, here we restrict the vertical range of the plot to better show the variation seen in the phase-binned measurements.\label{fig:gls}
}
\ifthenelse{\boolean{emulateapj}}{
    \end{figure*}
}{
    \end{figure}
}

\clearpage

\subsubsection{TESS}
\label{sec:tess}

All four systems were observed by the NASA {\em TESS} mission as
summarized in Table~\ref{tab:photobs}.

\hatcur{47} was observed in
short-cadence mode through an approved {\em TESS} Guest-Investigator
program (G011214; PI Bakos) to observe HATSouth transiting planet
candidates with {\em TESS}. The short-cadence observations were
reduced to light curves by the NASA Science Processing Operations
Center (SPOC) Pipeline at NASA Ames Research Center \citep{jenkinsSPOC2016,jenkins:2010}. Two threshold crossing events were identified by this pipeline for this target, and the object was selected as a transiting planet candidate and assigned the {\em TESS} Object of Interest (TOI)
identifier TOI~\hatcurCCtoi{47} by the {\em TESS} Science Office team after inspecting the data validation report produced by the SPOC pipeline. The target passed all of the data validation tests conducted by the pipeline, including no discernable difference between odd and even transits, no evidence for a weak secondary event, no evidence for stronger transits in a halo aperture compared to the optimal aperture used to extract the light curve, strong evidence that the target is not a false alarm due to correlated noise, and no evidence for variations in the difference image centroid. We obtained
the SPOC PDC light curve \citep{2014PASP..126..100S,2012PASP..124.1000S} for \hatcur{47} from the Barbara A.\ Mikulski
Archive for Space Telescopes (MAST).

\hatcur{47} is blended in the {\em TESS} images with two other comparably bright stars (the two neighbors are separated from \hatcur{47} by $33$\arcsec and $42$\arcsec). These neighbors are both well-resolved by HATSouth and the facility used for photometric follow-up observations (Section~\ref{sec:phot}). A correction for dilution is applied in the PDC process, but may have been overestimated in this case leading to a slightly deeper transit seen in the {\em TESS} light curve (Fig.~\ref{fig:hats47tess}), compared to the other light curves (Fig.~\ref{fig:hats47}).

The {\em TESS} light curve of \hatcur{47} shows a clear
quasi-sinusoidal out-of-transit variability with a period of $6.22158
\pm 0.00011$\,days and a semiamplitude of $5.79\pm0.33$\,mmag. The
period is close to, but different from, the rotation period of
$6.621581\pm0.000078$\,days estimated independently from the HATSouth
light curve.  The difference is presumably a result of starspot
evolution and/or differential rotation. We take the average of these
two measurements as our estimate for the photometric rotation period
of the star ($6.42\pm0.28$\,days). We filtered the quasi-sinusoidal
variation out of the {\em TESS} light curve of \hatcur{47} by fitting
and subtracting a harmonic series to the data. The harmonic-filtered
light curve is then used in the analysis (Section~\ref{sec:analysis}).

The other three targets were not included in the set of stars observed
in short-cadence mode by the mission, so only $\sim 30$\,min
integrations from the Full Frame Images (FFIs) are available for these
objects. Of these, \hatcur{72} was bright enough to have a light curve
produced from the FFI observations by the {\em TESS} Quick Look
Pipeline \citep[QLP;][]{huang:2018} at MIT. We made use of the
detrended light curve for \hatcur{72} produced by this pipeline. For
\hatcur{48} and \hatcur{49}, which were not processed by SPOC, and
were too faint to be processed through the QLP, we extracted light
curves from the {\em TESS} FFIs using Lightkurve \citep{lightkurve:2018} and
TESSCut \citep{brasseur:2019}, using a B-spline to remove trends on
time-scales longer than the transits, and made use of these data in
our analysis.

We note that \hatcur{47} and \hatcur{48} suffer significant blending
from nearby stars in the low spatial-resolution {\em TESS} images, and
the Sector 13 observations also have significant scattered light.

\subsubsection{WASP}
\label{sec:wasp}

\hatcur{72} was observed by WASP-South over the period 2006 May to 2012 June, accumulating 24{,}200 data points.  WASP-South is an array of 8 cameras combining 200-mm f/1.8 lenses with 2k$\times$2k CCDs and observing with a broad, 400--700\,nm bandpass \citep{pollacco:2006}. It observed visible
fields with a typical cadence of 15 min on each clear night. After identification of a candidate 7.33\,d transit periodicity (see \citealt{colliercameron:2007}), \hatcur{72} was placed on the WASP-South follow-up program in 2013 January. This led to nine RVs being observed with the Euler/CORALIE spectrograph (e.g.~\citealt{triaud:2013}) over 2013--2018, and the observation in 2018 July of a transit with TRAPPIST-South (e.g.~\citealt{gillon:2013}), using an {\it I+z} filter. The data were compatible with the transiting object being a planet, leading to a provisional designation as \hatcurCCwasp{72}, but the low-amplitude meant that detection of orbital motion was not secure.  Plans to acquire HARPS observations were then superseded by confirmation of the planet by the HATSouth team.

\subsection{Spectroscopic Observations}
\label{sec:obsspec}

The spectroscopic observations carried out to confirm and characterize
the four transiting planet systems presented here are summarized in
\reftabl{specobs}. The facilities are: WiFeS on the ANU~2.3\,m
\citep{dopita:2007}, PFS on the Magellan~6.5\,m \citep{crane:2006,crane:2008,crane:2010},
FEROS on the MPG~2.2\,m \citep{kaufer:1998}, HARPS on the ESO~3.6\,m
\citep{mayor:2003}, Coralie on the Euler~1.2\,m \citep{queloz:2001},
and ESPRESSO on the VLT~8.2\,m \citep{megevand:2014}.

The WiFeS observations were obtained for \hatcur{47}, \hatcur{48} and
\hatcur{49}, and were used for reconnaissance to rule out common false
positives such as transiting very low mass stars, or eclipsing stellar
binaries blended with a brighter giant star. The data were reduced
following \citet{bayliss:2013:hats3}. For each object we obtained
spectra at resolving power $R \equiv \Delta\,\lambda\,/\,\lambda \approx
3000$ to estimate the effective temperature, \logg\ and \feh\ of the
star. Additional observations at $R \approx 7000$ were also obtained
to search for any large amplitude radial velocity variations at the
$\sim 4$\,\kms\ level, which would indicate a stellar mass
companion. All three systems were confirmed to be dwarf stars with RV
variations below $4$\,\kms.

We obtained PFS observations of all four systems. For each system we
obtained observations with an I$_{2}$ cell, and observations without
the cell. The I$_{2}$-free observations were used to construct a
template for measuring high-precision RVs from observations made with
the cell following the method of \citet{butler:1996}. The PFS RV
observations were included in the modelling that we performed for all
four systems (Section~\ref{sec:transitmodel}). Spectral line Bisector Span
(BS) measurements and their uncertainties were measured as described
by \citet{jordan:2014:hats4} and \citet{brahm:2017:ceres}.

We obtained FEROS observations for \hatcur{47}, \hatcur{48} and
\hatcur{72}. These were reduced to wavelength-calibrated spectra, and
high-precision RV and BS measurements using the CERES software package
\citep{brahm:2017:ceres}. Due to the faintness of \hatcur{47} and
\hatcur{48}, we found that the scatter in the FEROS RV measurements for
these two systems was too high to be useful in constraining the RV
orbital variation of the host star. For the much brighter host
\hatcur{72}, however, we did incorporate FEROS data into the analysis.

The HARPS observations of \hatcur{72} were also reduced using
CERES. The RVs were of high enough precision to be included in our
analysis of this system. The HARPS observations reported here were
obtained by the HATSouth team. We note that a single HARPS observation
of this system was also independently obtained by the WASP team, but
we do not include it in the analysis since it was gathered and reduced
in a different manner from the observations obtained by the HATSouth
team.

Coralie observations of \hatcur{72} were carried out by the WASP team
independently of the other reported spectroscopic observations of this
system, which were gathered by the HATSouth team. The observations
constrain the orbital variation of the host star due to the planet to
have a semi-amplitude less than $\sim 50$\,\ms.  Additionally, no
correlation is seen between the Coralie RV and BS measurements
supporting a planetary interpretation of the observations.

Finally, we obtained ESPRESSO observations of \hatcur{72} after the
HARPS, PFS and FEROS observations indicated a likely Super-Neptune
mass for the planet. 
The observations were
carried out through the queue service mode between 2019 May 11 and
2019 June 4. We obtained seven exposures of 1800\,s, all made at an
airmass below $1.2$, achieving a mean S/N per resolution element of
$50$ at 550\,nm. The ESPRESSO observations were reduced using version
1.3.2 of the ESPRESSO Data Analysis System \citep{cupani:2018} in the
ESO Reflex environment \citep{freudling:2013}, with the spectra
cross-correlated against a K5 spectral mask to produce high-precision
RVs. A more complete description of the observational setup and reduction procedure for the ESPRESSO observations obtained through our program will be provided in a forthcoming publication on the HATS-73 system (Bayliss et al.~2020, in preparation). 

We also used the FEROS and I$_{2}$-free PFS observations to determine
high-precision stellar atmospheric parameters for the host stars using
the ZASPE package \citep{brahm:2017:zaspe}. The parameters that we
measured include the effective temperature \teffstar, surface gravity
\logg, metallicity \feh, and \vsini. The method involves
cross-correlating the observed spectra against synthetic model
spectra, and then obtaining error estimates for the parameters by
performing a bootstrap analysis where the regions in the spectra that
are most sensitive to changes in the atmospheric parameters are
randomly adjusted based on the observed distribution of systematic
mismatches between the observations and the best-matching model. This
method allows for realistic parameter uncertainties that account, in a
principled fashion, for systematic errors in the theoretical
models. We performed this analysis on the PFS template spectra for
\hatcur{47} and \hatcur{49}, and on the FEROS spectra for \hatcur{48}
and \hatcur{72}. The resulting parameters are listed in Table~\ref{tab:stellarobserved}.

The high-precision RV and BS measurements that were used in the
analysis are given in \reftabl{rvs} for all four systems.

\startlongtable
\ifthenelse{\boolean{emulateapj}}{
    \begin{deluxetable*}{llrrrr}
}{
    \begin{deluxetable}{llrrrr}
}
\tablewidth{0pc}
\tabletypesize{\scriptsize}
\tablecaption{
    Summary of photometric observations
    \label{tab:photobs}
}
\tablehead{
    \multicolumn{1}{c}{Instrument/Field\tablenotemark{a}} &
    \multicolumn{1}{c}{Date(s)} &
    \multicolumn{1}{c}{\# Images\tablenotemark{b}} &
    \multicolumn{1}{c}{Cadence\tablenotemark{c}} &
    \multicolumn{1}{c}{Filter} &
    \multicolumn{1}{c}{Precision\tablenotemark{d}} \\
    \multicolumn{1}{c}{} &
    \multicolumn{1}{c}{} &
    \multicolumn{1}{c}{} &
    \multicolumn{1}{c}{(sec)} &
    \multicolumn{1}{c}{} &
    \multicolumn{1}{c}{(mmag)}
}
\startdata
\sidehead{\textbf{\hatcur{47}}}
~~~~HS-1/G747 & 2013 Mar--Oct & 4231 & 287 & $r$ & 15.9 \\
~~~~HS-2/G747 & 2013 Sep--Oct & 646 & 287 & $r$ & 15.7 \\
~~~~HS-3/G747 & 2013 Apr--Nov & 9045 & 297 & $r$ & 16.6 \\
~~~~HS-4/G747 & 2013 Sep--Nov & 1467 & 297 & $r$ & 19.1 \\
~~~~HS-5/G747 & 2013 Mar--Nov & 6022 & 297 & $r$ & 15.7 \\
~~~~HS-6/G747 & 2013 Sep--Nov & 1568 & 290 & $r$ & 14.9 \\
~~~~TESS/Sector~13 & 2019 Jun--Jul & 1159 & 1798 & $T$ & 3.3 \\
~~~~Swope~1\,m & 2016 Aug 03 & 151 & 160 & $i$ & 1.9 \\
\sidehead{\textbf{\hatcur{48}}}
~~~~HS-2/G778 & 2011 May--2012 Nov & 2982 & 287 & $r$ & 14.6 \\
~~~~HS-4/G778 & 2011 Jul--2012 Nov & 3726 & 298 & $r$ & 13.5 \\
~~~~HS-6/G778 & 2011 Apr--2012 Oct & 2215 & 298 & $r$ & 14.3 \\
~~~~TESS/Sector~13 & 2019 Jun--Jul & 1301 & 1798 & $T$ & 5.2 \\
~~~~LCO~1\,m/SBIG & 2015 Jul 12 & 61 & 201 & $i$ & 2.3 \\
~~~~Swope~1\,m & 2015 Jul 15 & 62 & 139 & $i$ & 2.7 \\
\sidehead{\textbf{\hatcur{49}}}
~~~~HS-2/G754 & 2012 Sep--Dec      & 3875 & 282 & $r$ & 16.1 \\
~~~~HS-4/G754 & 2012 Sep--2013 Jan & 3197 & 292 & $r$ & 17.7 \\
~~~~HS-6/G754 & 2012 Sep--Dec      & 3002 & 285 & $r$ & 17.0 \\
~~~~HS-1/G755 & 2011 Jul--2012 Oct & 5249 & 292 & $r$ & 18.1 \\
~~~~HS-3/G755 & 2011 Jul--2012 Oct & 4828 & 287 & $r$ & 19.0 \\
~~~~HS-5/G755 & 2011 Jul--2012 Oct & 6024 & 296 & $r$ & 16.0 \\
~~~~TESS/Sector~1 & 2018 Jul--Aug & 1078 & 1798 & $T$ & 4.8 \\
~~~~TESS/Sector~2 & 2018 Aug--Sep & 1219 & 1798 & $T$ & 3.7 \\
~~~~LCO~1\,m/Sinistro & 2014 Nov 19 & 34 & 288 & $i$ & 1.5 \\
~~~~LCO~1\,m/Sinistro & 2015 Aug 31 & 37 & 223 & $i$ & 1.8 \\
~~~~LCO~1\,m/SBIG & 2015 Sep 13 & 73 & 201 & $i$ & 3.9 \\
~~~~LCO~1\,m/Sinistro & 2015 Sep 17 & 30 & 223 & $i$ & 3.0 \\
\sidehead{\textbf{\hatcur{72}}}
~~~~HS-1/G537 & 2016 Jun--2016 Dec & 4125 & 333 & $r$ & 4.8 \\
~~~~HS-3/G537 & 2016 Oct--2016 Dec & 901 & 346 & $r$ & 4.9 \\
~~~~HS-5/G537 & 2016 Jun--2016 Dec & 3354 & 365 & $r$ & 4.8 \\
~~~~WASP-South  & 2006 May--2011 Nov & 24431 & 169 & 400--700\,nm & 21.8 \\
~~~~TESS/Sector~3 & 2018 Aug--Sep & 1036 & 1798 & $T$ & 0.66 \\
~~~~LCO~1\,m/Sinistro & 2017 Jun 28 & 107 & 163 & $i$ & 1.3 \\
~~~~LCO~1\,m/Sinistro & 2017 Jul 20 & 113 & 163 & $i$ & 1.3 \\
~~~~TRAPPIST-South & 2018 Jul 14 & 1195 & 21 & $I+z$ & 4.1 \\
\enddata \tablenotetext{a}{ For HATSouth data we list the HATSouth
  unit, CCD and field name from which the observations are taken. HS-1
  and -2 are located at Las Campanas Observatory in Chile, HS-3 and -4
  are located at the H.E.S.S. site in Namibia, and HS-5 and -6 are
  located at Siding Spring Observatory in Australia. Each unit has 4
  CCDs. Each field corresponds to one of 838 fixed pointings used to
  cover the full 4$\pi$ celestial sphere. All data from a given
  HATSouth field and CCD number are reduced together, while detrending
  through External Parameter Decorrelation (EPD) is done independently
  for each unique unit+CCD+field combination.
}
\tablenotetext{b}{ Excluding any outliers or other data not included in the modelling. }
\tablenotetext{c}{ The median time between consecutive images rounded
  to the nearest second. Due to factors such as weather, the
  day--night cycle, guiding and focus corrections the cadence is only
  approximately uniform over short timescales.  } 
\tablenotetext{d}{
  The RMS of the residuals from the best-fit model. Note that in the case of HATSouth and {\em TESS} observations the transit may appear artificially shallower due to over-filtering and/or blending from unresolved neighbors. As a result the S/N of the transit may be less than what would be calculated from $\rpl/\rstar$ and the RMS estimates given here. }
\ifthenelse{\boolean{emulateapj}}{
    \end{deluxetable*}
}{
    \end{deluxetable}
}

%
%
\ifthenelse{\boolean{emulateapj}}{
    \begin{deluxetable*}{llrrrrl}
}{
    \begin{deluxetable}{llrrrrl}
}
\tablewidth{0pc}
\tablecaption{
    Light curve data for \hatcur{47}, \hatcur{48}, \hatcur{49}, and \hatcur{72}\label{tab:phfu}.
}
\tablehead{
    \colhead{Object\tablenotemark{a}} &
    \colhead{BJD\tablenotemark{b}} & 
    \colhead{Mag\tablenotemark{c}} & 
    \colhead{\ensuremath{\sigma_{\rm Mag}}} &
    \colhead{Mag(orig)\tablenotemark{d}} & 
    \colhead{Filter} &
    \colhead{Instrument} \\
    \colhead{} &
    \colhead{\hbox{~~~~(2,400,000$+$)~~~~}} & 
    \colhead{} & 
    \colhead{} &
    \colhead{} & 
    \colhead{} &
    \colhead{}
}
\startdata
HATS-47 & $ 2456492.53516 $ & $  -0.01555 $ & $   0.01509 $ & $ \cdots $ & $ r$ &         HS\\
HATS-47 & $ 2456570.99127 $ & $  -0.01117 $ & $   0.01041 $ & $ \cdots $ & $ r$ &         HS\\
HATS-47 & $ 2456555.30015 $ & $  -0.05056 $ & $   0.01484 $ & $ \cdots $ & $ r$ &         HS\\
HATS-47 & $ 2456551.37754 $ & $   0.03031 $ & $   0.01653 $ & $ \cdots $ & $ r$ &         HS\\
HATS-47 & $ 2456500.38125 $ & $  -0.01951 $ & $   0.01123 $ & $ \cdots $ & $ r$ &         HS\\
HATS-47 & $ 2456433.69366 $ & $  -0.00988 $ & $   0.01094 $ & $ \cdots $ & $ r$ &         HS\\
HATS-47 & $ 2456574.91475 $ & $  -0.00262 $ & $   0.00940 $ & $ \cdots $ & $ r$ &         HS\\
HATS-47 & $ 2456606.29725 $ & $  -0.00778 $ & $   0.01329 $ & $ \cdots $ & $ r$ &         HS\\
HATS-47 & $ 2456516.07293 $ & $  -0.00670 $ & $   0.01008 $ & $ \cdots $ & $ r$ &         HS\\
HATS-47 & $ 2456465.07680 $ & $  -0.02574 $ & $   0.01287 $ & $ \cdots $ & $ r$ &         HS\\
\enddata
\tablenotetext{a}{
    Either \hatcur{47}, \hatcur{48}, \hatcur{49}, or \hatcur{72}.
}
\tablenotetext{b}{
    Barycentric Julian Dates in this paper are reported on the
    Barycentric Dynamical Time (TDB) system.  
} \tablenotetext{c}{
    The out-of-transit level has been subtracted. For observations
    made with the HATSouth instruments (identified by ``HS'' in the
    ``Instrument'' column) these magnitudes have been corrected for
    trends using the EPD and TFA procedures applied {\em prior} to
    fitting the transit model. This procedure may lead to an
    artificial dilution in the transit depths. For several of these
    systems neighboring stars are blended into the TESS observations
    as well. The blend factors for the HATSouth and TESS light curves
    are listed in Table~\ref{tab:planetparam}. For observations made
    with follow-up instruments (anything other than ``HS'', ``TESS''
    and ``WASP'' in the ``Instrument'' column), the magnitudes have
    been corrected for a quadratic trend in time, and for variations
    correlated with up to three PSF shape parameters, fit
    simultaneously with the transit. For the Swope~1\,m observations of \hatcur{47}, these observations have been further detrended against a set of 20 light curves for other stars observed in the field.
}
\tablenotetext{d}{
    Raw magnitude values without correction for the quadratic trend in
    time, or for trends correlated with the seeing. These are only
    reported for the follow-up observations.
}
\tablecomments{
    This table is available in a machine-readable form in the online
    journal.  A portion is shown here for guidance regarding its form
    and content.
}
\ifthenelse{\boolean{emulateapj}}{
    \end{deluxetable*}
}{
    \end{deluxetable}
}

\ifthenelse{\boolean{emulateapj}}{
    \begin{deluxetable*}{llrrrrr}
}{
    \begin{deluxetable}{llrrrrrrrr}
}
\tablewidth{0pc}
\tabletypesize{\scriptsize}
\tablecaption{
    Summary of spectroscopy observations.
    \label{tab:specobs}
}
\tablehead{
    \multicolumn{1}{c}{Instrument}          &
    \multicolumn{1}{c}{UT Date(s)}             &
    \multicolumn{1}{c}{\# Spec.}   &
    \multicolumn{1}{c}{Res.}          &
    \multicolumn{1}{c}{S/N Range\tablenotemark{a}}           &
    \multicolumn{1}{c}{$\gamma_{\rm RV}$\tablenotemark{b}} &
    \multicolumn{1}{c}{RV Precision\tablenotemark{c}} \\
    &
    &
    &
    \multicolumn{1}{c}{$\Delta \lambda$/$\lambda$/1000} &
    &
    \multicolumn{1}{c}{(\kms)}              &
    \multicolumn{1}{c}{(\ms)}
}
\startdata
\noalign{\vskip -3pt}
\sidehead{\textbf{\hatcur{47}}}\\
\noalign{\vskip -13pt}
ANU~2.3\,m/WiFeS & 2015 Jun 1 & 1 & 3 & 28 & $\cdots$ & $\cdots$ \\
ANU~2.3\,m/WiFeS & 2015 Jul 27--30 & 4 & 7 & 17--40 & 2.0 & 4000 \\
Magellan~6.5\,m/PFS+I$_{2}$ & 2016 Mar--Aug & 12 & 76 & $\cdots$ & $\cdots$ & 37 \\
Magellan~6.5\,m/PFS & 2016 Mar 30 & 1 & 76 & $\cdots$ & $\cdots$ & $\cdots$ \\
MPG~2.2\,m/FEROS & 2016 Jul 1--26 & 5 & 48 & 14--33 & 3.179 & 55 \\
\noalign{\vskip -3pt}
\sidehead{\textbf{\hatcur{48}}}\\
\noalign{\vskip -13pt}
ANU~2.3\,m/WiFeS & 2014 Oct 4 & 1 & 3 & 49 & $\cdots$ & $\cdots$ \\
ANU~2.3\,m/WiFeS & 2014 Oct 6--8 & 2 & 7 & 52--63 & -22.5 & 4000 \\
MPG~2.2\,m/FEROS & 2015 Jun--Oct & 10 & 48 & 17--36 & -22.457 & 75 \\
Magellan~6.5\,m/PFS+I$_{2}$ & 2015 Jun--Oct & 12 & 76 & $\cdots$ & $\cdots$ & 30 \\
Magellan~6.5\,m/PFS & 2015 Jul 1 & 1 & 76 & $\cdots$ & $\cdots$ & $\cdots$ \\
\noalign{\vskip -3pt}
\sidehead{\textbf{\hatcur{49}}}\\
\noalign{\vskip -13pt}
ANU~2.3\,m/WiFeS & 2014 Oct 4--5 & 3 & 3 & 24--45 & $\cdots$ & $\cdots$ \\
ANU~2.3\,m/WiFeS & 2014 Oct 5--7 & 2 & 7 & 24--48 & 8.1 & 4000 \\
Magellan~6.5\,m/PFS+I$_{2}$ & 2015 Jan--2016 Jun & 8 & 76 & $\cdots$ & $\cdots$ & 53 \\
Magellan~6.5\,m/PFS & 2015 Jan 8 & 1 & 76 & $\cdots$ & $\cdots$ & $\cdots$ \\
\noalign{\vskip -3pt}
\sidehead{\textbf{\hatcur{72}}}\\
\noalign{\vskip -13pt}
ESO~3.6\,m/HARPS & 2017 Apr--2018 Aug & 11 & 115 & 9--31 & 15.955 & 12.0 \\
MPG~2.2\,m/FEROS & 2017 Jun--Aug & 6 & 48 & 24--68 & 15.946 & 12.7 \\
Euler~1.2\,m/Coralie & 2013 Jul--2018 Jul & 9 & 60 & 42--46 & $15.934$ & $15$ \\
Magellan~6.5\,m/PFS+I$_{2}$ & 2018 May--Aug & 3 & 76 & $\cdots$ & $\cdots$ & 14.7 \\
Magellan~6.5\,m/PFS & 2018 Jun 23 & 1 & 76 & $\cdots$ & $\cdots$ & $\cdots$ \\
VLT~8.2\,m/ESPRESSO & 2019 May--Jun & 7 & 140 & 50 & 15.997 & 10.5 \\
\enddata 
\tablenotetext{a}{
    S/N per resolution element near 5180\,\AA. This was not measured for all of the instruments.
}
\tablenotetext{b}{
    For high-precision RV observations included in the orbit determination this is the zero-point RV from the best-fit orbit. For other instruments it is the mean value. We only provide this quantity when applicable.
}
\tablenotetext{c}{
    For high-precision RV observations included in the orbit
    determination this is the scatter in the RV residuals from the
    best-fit orbit (which may include astrophysical jitter), for other
    instruments this is either an estimate of the precision (not
    including jitter), or the measured standard deviation.  We only provide this quantity when applicable.
}
\ifthenelse{\boolean{emulateapj}}{
    \end{deluxetable*}
}{
    \end{deluxetable}
}


\subsection{Photometric follow-up observations}
\label{sec:phot}

Follow-up higher-precision ground-based photometric transit
observations were obtained for all four systems, as summarized in
Table~\ref{tab:photobs}. The facilities used for this purpose are:
the Swope 1\,m telescope at Las Campanas Observatory in Chile, 1\,m
telescopes from the Las Cumbres Observatory (LCOGT) network
\citep{brown:2013:lcogt}, and the 0.6\,m TRAPPIST-South telescope at
La Silla Observatory \citep{gillon:2013}. The follow-up observations using the
Swope 1\,m and the LCOGT 1\,m network were performed by the HATSouth
team, while the TRAPPIST-South observations of \hatcur{72} were
performed by the WASP team. The exposure time for the TRAPPIST-South observations was 10\,s, though with read-out, the median cadence was 21\,s as listed in Table~\ref{tab:photobs}.

\ifthenelse{\boolean{emulateapj}}{
    \begin{figure*}[!ht]
}{
    \begin{figure}[!ht]
}
 {
 \centering
 \leavevmode
 \includegraphics[width={0.5\linewidth}]{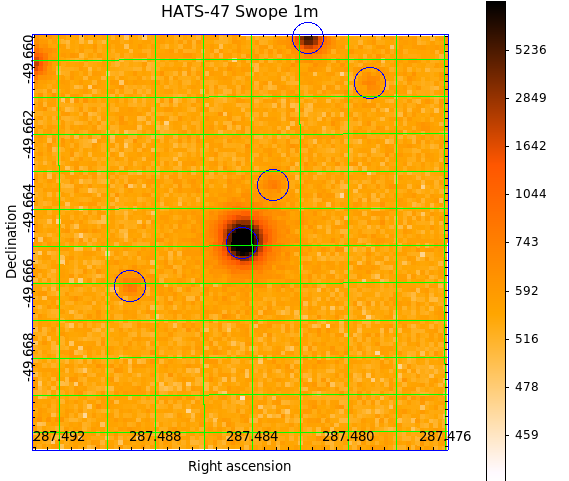}
 \hfil
 \includegraphics[width={0.5\linewidth}]{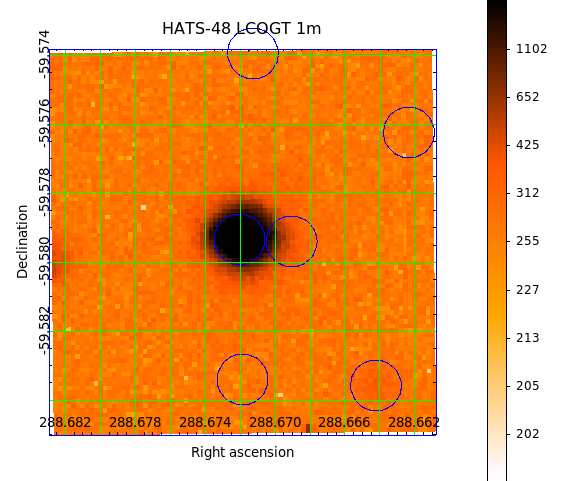}
 }
 {
 \centering
 \leavevmode
 \includegraphics[width={0.5\linewidth}]{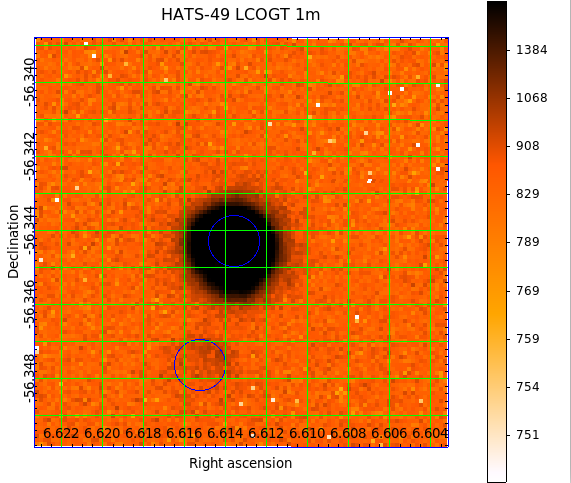}
 \hfil
 \includegraphics[width={0.5\linewidth}]{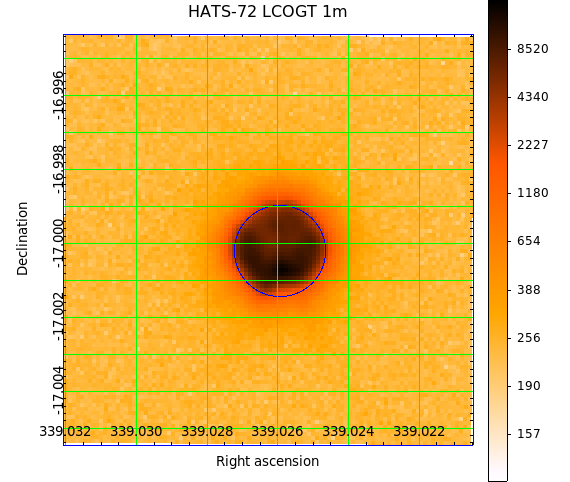}
 }
\caption{
Snapshot $40\arcsec \times 40\arcsec$ images of \hatcur{47}, \hatcur{48}, \hatcur{49} and \hatcur{72} selected from the ground-based time-series photometric follow-up observations obtained for each object. Each image is centered on the transiting planet host star. The blue circles indicate the positions of sources in the Gaia~DR2 catalog, which is based on higher spatial resolution and deeper observations than the images shown here. The radius of each circle is equal to the approximate HWHM of the image PSF. Note that the LCOGT~1\,m observations were carried out with defocusing to improve the photometric precision, while the Swope~1\,m observations were carried out in-focus with the resolution limited by atmospheric seeing. The color scale indicates the number of counts (in ADU) in each calibrated image pixel, and is shown on an inverted logarithmic scale. All known neighbors are resolved in the ground-based photometric follow-up observations carried out for each system.\label{fig:phfusnapshots}
}
\ifthenelse{\boolean{emulateapj}}{
    \end{figure*}
}{
    \end{figure}
}

Figure~\ref{fig:phfusnapshots} shows example $40\arcsec \times
  40\arcsec$ images, centered on each target, selected from our photometric
  follow-up observations. In each case we overlay sources from the
  Gaia~DR2 catalog, which is based on higher spatial resolution and
  deeper imaging than the photometric follow-up observations
  themselves. For all four objects all known neighbors have been
  resolved by the ground-based photometric follow-up observations.

Observations with the Swope 1\,m and the reduction of the data to
light curves were performed as described by \citet{penev:2013:hats1}. The LCOGT
1\,m observations were carried out in a similar manner to that
described by \citet{hartman:2015:hats6}, but were reduced using the
methods applied by \citet{penev:2013:hats1} to data from the Faulkes
Telescope South (FTS) 2\,m, with some updates for automation to be
described by Espinoza et al.\ (2020; in preparation). The
TRAPPIST-South observations were carried out and reduced as described
in \citet{gillon:2013}.

The time-series photometry data are available in Table~\ref{tab:phfu},
and are plotted Figures~\ref{fig:hats47}, \ref{fig:hats48},
\ref{fig:hats49}, and \ref{fig:hats72}.

\subsection{Search for Resolved Stellar Companions}
\label{sec:luckyimaging}

For \hatcur{47} and \hatcur{48}, the highest spatial resolution optical
imaging available is from the Gaia mission \citep{gaiadr2}. Gaia~DR2 is sensitive to
neighbors with $G \la 20$\,mag down to a limiting resolution of $\sim
1\arcsec$ \citep[e.g.,][]{ziegler:2018}.

There is a faint neighboring source to \hatcur{47} listed in the
Gaia~DR2 catalog at a projected separation of $6\farcs4$ with $\Delta
G = 5.8$\,mag. This object is fully resolved by the Swope~1\,m photometric follow-up observations for which the seeing-limited resolution ranged between $1\farcs4$ and $2\farcs5$ FWHM. These observations show that the neighbor is not responsible for the transits, nor does it
impact the transit depths measured in the follow-up observations.
Based on the Gaia~DR2 parallax measurements ($0.3 \pm 1.3$\,mas, compared to \hatcurCCparallax{47}\,mas for \hatcur{47}), the neighboring source is
in the background of \hatcur{47} and is not physically associated with
it.

Similarly, there is a faint neighbor to \hatcur{48} in the Gaia~DR2
catalog at a projected separation of $5\farcs4$ and with $\Delta G =
5.4$\,mag. The neighbor has a parallax of $3.64 \pm 0.44$\,mas
compared to $\hatcurCCparallax{48}$\,mas for \hatcur{48}, and a proper
motion of $\mu_{\rm R.A.} = 2.95 \pm 0.47$\,\masy\ and $\mu_{\rm Dec.}
= 5.20 \pm 0.45$\,\masy, compared to \hatcurCCpmra{48}\,\masy\ and
\hatcurCCpmdec{48}\,\masy\ for \hatcur{48}. If we assume that this
source is a physical binary companion to \hatcur{48}, and that it is a
single star, then adopting the age, mass and metallicity for
\hatcur{48} determined in Section~\ref{sec:transitmodel} and listed in
Table~\ref{tab:stellarderived}, and using the PARSEC stellar evolution
models \citep{marigo:2017}, we find that $\Delta G = 5.442 \pm
0.004$\,mag implies a mass of $0.224 \pm 0.001$\,\msun\ for the
companion. In that case the predicted $BP$ and $RP$ magnitude
differences between the companion and \hatcur{48} are $\Delta BP =
6.74$\,mag and $\Delta RP = 4.88$\,mag, which are comparable to the
observed differences of $\Delta BP = 6.11 \pm 0.12$\,mag and $\Delta
RP = 4.80 \pm 0.02$\,mag. Note that theoretical isochrones are known
to have errors in matching the optical photometry of late M
dwarf stars, particularly in blue filters, so although the observed
magnitude differences are off by more than the formal uncertainties,
the results are close enough for us to conclude that the faint
neighbor is most likely a bound physical companion to
\hatcur{48}. Given the distance measured to \hatcur{48}, the neighbor
is currently at a projected physical separation of $\sim$1{,}400\,au from
\hatcur{48}.

\ifthenelse{\boolean{emulateapj}}{
    \begin{figure*}[!ht]
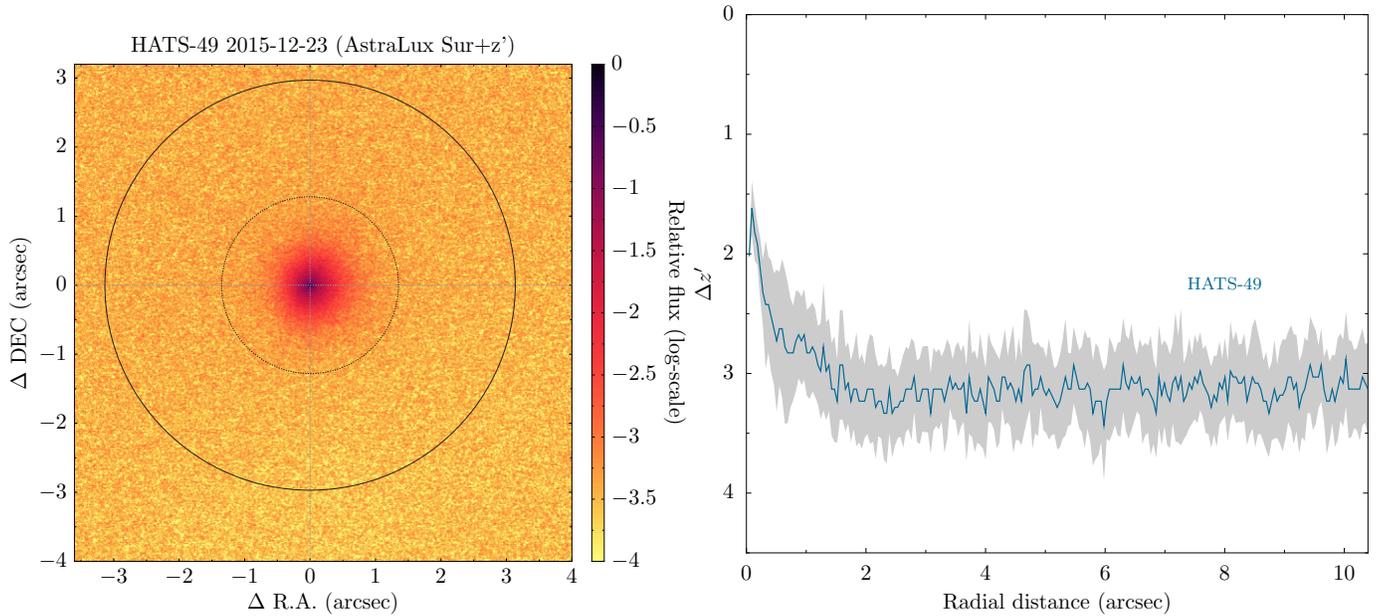

}{
    \begin{figure}[!ht]
}
 {
 \centering
 \leavevmode
 \includegraphics[width={0.5\linewidth}]{\hatcurhtr{49}-astralux}
 \hfil
 \includegraphics[width={0.5\linewidth}]{\hatcurhtr{49}-astralux-contrast-curve}
 }
\caption{
    ({\em left)} High-resolution image of \hatcur{49} obtained with AstraLux Sur on the NTT through the $z^{\prime}$ filter. No neighboring sources are detected. ({\em right}) 5$\sigma$ upper limit on the magnitude contrast of any resolved neighbor to \hatcur{49} based on the Astralux Sur image. The gray band shows the variation in the limit with position angle. \label{fig:hats49astralux}
}
\ifthenelse{\boolean{emulateapj}}{
    \end{figure*}
}{
    \end{figure}
}

We obtained high-angular resolution imaging of \hatcur{49} using the
Astralux Sur imager \citep{hippler:2009} on the New Technology
Telescope (NTT). Observations were carried out in $z^{\prime}$-band on
the night of 2015 December 23, and were reduced as in
\citet{espinoza:2016:hats25hats30}. No neighbor was detected, and we
place limits of $\Delta z^{\prime} > 1.8\pm0.2$\,mag on neighbors down to $0\farcs138$, and $\Delta z^{\prime} > 3.1\pm0.4$\,mag at $\sim 1\farcs5$. The image and contrast curve are shown in Figure~\ref{fig:hats49astralux}. We also note that there is no
neighbor within 10\arcsec\ of \hatcur{49} listed in Gaia~DR2.

High-angular resolution imaging of \hatcur{72} was reported by
\citet{ziegler:2019} who carried out speckle imaging with SOAR to
search for resolved stellar companions to 542 TESS planet candidate
hosts. They report that no companion to \hatcur{72} was detected, and
place magnitude contrast limits of $\Delta m > 2.11$\,mag and $\Delta
m > 3.60$\,mag at separations of $0\farcs 15$ and 1\arcsec,
respectively. We also note that there is no neighbor within
10\arcsec\ of \hatcur{72} listed in Gaia~DR2.

\section{Analysis}
\label{sec:analysis}

\subsection{Transiting Planet Modelling}
\label{sec:transitmodel}

We analyzed the photometric, spectroscopic and astrometric
observations of each system to determine the stellar and planetary
parameters following the methods described by
\citet{hartman:2019:hats6069}, with modifications as summarized most
recently by \citet{bakos:2018:hats71}. 

We perform a global fit to the light curves, RV curves,
spectroscopically measured stellar atmospheric parameters, catalog
broad-band photometry, and astrometric parallax from Gaia~DR2. The fit is carried out using a modified version of the {\sc lfit} program which is included in the {\sc fitsh} software package \citep{pal:2012}. The light curves are modelled using the \citet{mandel:2002} semi-analytic transit model with quadratic limb-darkening. The limb darkening coefficients are allowed to vary in the fit, using the tabulations from \citet{claret:2012,claret:2013} and \citet{claret:2018} to place Gaussian prior constraints on their values, assuming a prior uncertainty of $0.2$ for each coefficient. The RV curves are modelled using the appropriate relations for Keplerian orbits.

We include in the model several parameters for the physical and observed properties of the host star, including the effective photospheric temperature, the metallicity, the distance modulus, and the $V$-band extinction $A_{V}$. These parameters are, in turn, constrained by the observed spectroscopic stellar atmospheric parameters (as measured in Section~\ref{sec:obsspec}), the catalog photometry, and the parallax. Together with the parameters used to describe the transit and RV observations, these parameters are sufficient to determine the bulk physical properties of the stars and their transiting planets. We fit the data using two different methods for relating the stellar mass to the stellar radius, metallicity and luminosity: (1) an empirical method which uses the stellar bulk density measured from the transit and RV observations to determine the stellar mass from the stellar radius, which is itself inferred from the effective temperature and luminosity \citep[this method is similar to that of, e.g.,][]{stassun:2017}, and (2) using the PARSEC theoretical stellar evolution models \citep{marigo:2017} to impose an additional constraint on the stellar relations that is typically tighter than the observed constraint on the stellar bulk density.

In each case, we model the data assuming the orbital
eccentricity is zero, and we
separately try allowing the eccentricity to be a free parameter.

A Differential
Evolution Markov Chain Monte Carlo (DEMCMC) method is used to explore
the parameter space and estimate the uncertainties based on the
posterior parameter distribution. 
See \citet{hartman:2019:hats6069} for a full list of the parameters that we vary, and their assumed priors.

We include in the fit the
optical broad-band photometry from Gaia~DR2 and APASS, NIR photometry
from 2MASS, and IR photometry from WISE. For WISE we exclude the W4
band for all systems as none of the objects were detected in that
bandpass, while for \hatcur{47}, \hatcur{48}, and \hatcur{49} we
also exclude the W3 band as the photometric uncertainty exceeds
0.1\,mag in this bandpass for these three objects.  These observations, together with the stellar atmospheric parameters, the parallax, and the reddening, constrain the luminosity of the star. To model the reddening, we assume a $R_{V} = 3.1$ \citet{cardelli:1989} dust law parameterized by $A_{V}$, and use the {\sc mwdust} 3D Galactic extinction model \citep{bovy:2016} to place a prior constraint on its value.

We find that for all four transiting planet systems the orbits are consistent with being
circular when the eccentricities are varied, and that the stellar
parameters are more robustly constrained when imposing the theoretical
stellar evolution model constraints. We therefore choose to adopt
the parameters that stem from fixing the orbit to be circular, and
imposing the stellar evolution models as a constraint on the stellar
physical parameters.

The best-fit models are compared to the various observational data for
the four transiting planet systems in
Figures~\ref{fig:hats47}--\ref{fig:hats72wasp}. The adopted stellar
parameters derived from the analysis are listed in
Table~\ref{tab:stellarderived}, while the adopted planetary parameters
are listed in Table~\ref{tab:planetparam}. We also list in
Table~\ref{tab:planetparam} the 95\% confidence upper limit on the
eccentricity that comes from allowing the eccentricity to vary in the
fit.

\subsection{Stellar Blend Modelling}
\label{sec:blendmodel}

We also performed a blend modelling of each system following
\citet{hartman:2019:hats6069}, where we attempt to fit all of the
observations (except the RV data) using various combinations of stars,
with parameters constrained by the PARSEC models. We find that for all
four objects a model consisting of a single star with a transiting
planet provides a better fit (a greater likelihood and a lower $\chi^2$)
to the light curves, spectroscopic stellar atmospheric parameters, broad-band
catalog photometry, and astrometric parallax measurements than the
best-fit blended stellar eclipsing binary models. The blended stellar
eclipsing binary models involve more free parameters than the
transiting planet model, and thus can be rejected on the grounds that
they are both poorer-fitting and more complicated models.
Moreover, the fact that Keplerian orbital variations
consistent with transiting planets are observed for all four objects,
and that large spectral line bisector span variations are not observed
is further evidence in favor of the transiting planet interpretation
of the observations.

We also attempted to fit the systems as unresolved stellar binaries
with a planet transiting the brighter stellar component. We find that
for \hatcur{47} and \hatcur{48} there is no significant improvement in
$\chi^2$ when adding an unresolved stellar binary companion compared
to the model of a single star with a transiting planet. For
\hatcur{49} and \hatcur{72} adding an unresolved companion does
improve the fit, with $\Delta \chi^2 = -17.4$ for \hatcur{49}, and
$\Delta \chi^2 = -39.8$ for \hatcur{72}. At face value this may be
taken as evidence for an unresolved stellar companion to both of these
objects. For \hatcur{49} this modelling yields a mass of
$0.245\pm0.039$\,\msun\ for the unresolved stellar companion, while
for \hatcur{72} we find $0.317\pm0.025$\,\msun. However, as there is
no other independent evidence for a stellar companion (such as a
long-term trend in the RVs) for either of these objects, and as some low-mass stars (including late K dwarfs) have been observed to have larger radii
than predicted by theoretical isochrones \citep[e.g.,][in this case invoking an additional star would appear to reconcile the observations to the model leading to a better fit, but an erroneous conclusion]{torres:2013}, we do
not consider this to be a clear detection of a stellar companion for
either \hatcur{49} or \hatcur{72}. Instead we present in
Table~\ref{tab:stellarderived} the 95\% confidence upper limits on the
mass of an unresolved companion for all four objects. High angular
resolution imaging, long-term RV observations, and/or Gaia astrometry
could potentially detect the companions if they are present.



\section{Discussion}
\label{sec:discussion}

We have presented the discovery of four transiting giant planets on
close-in orbits around K dwarf
stars. Three of the planets presented in this paper,
\hatcurb{47}, \hatcurb{48}, and \hatcurb{49}, are comparable in mass to Saturn.
The masses are 
\hatcurPPmlong{47}\,\mjup, \hatcurPPmlong{48}\,\mjup, and
\hatcurPPmlong{49}\,\mjup, respectively. Despite their relatively
short orbital periods of \hatcurLCPshort{47}\,days,
\hatcurLCPshort{48}\,days, and \hatcurLCPshort{49}\,days, all three of
these planets can be considered warm giants with predicted
equilibrium temperatures of \hatcurPPteff{47}\,K, \hatcurPPteff{48}\,K,
and \hatcurPPteff{49}\,K (estimated assuming zero-albedo and full
redistribution of heat; note that the small uncertainties listed here do not account for the possibility that these assumptions are wrong, or for systematic errors in the stellar evolution models). This is due to the planets orbiting cool K
dwarf stars with respective masses of \hatcurISOmlong{47}\,\msun,
\hatcurISOmlong{48}\,\msun, and \hatcurISOmlong{49}\,\msun. The fourth
planet, \hatcurb{72}, is a Super-Neptune with a mass of
\hatcurPPmlong{72}\,\mjup, and with a somewhat longer orbital period of
\hatcurLCPshort{72}\,days. This planet also orbits a cool K dwarf star
of mass \hatcurISOmlong{72}\,\msun, and has a modest predicted
equilibrium temperature of \hatcurPPteff{72}\,K.

Figures~\ref{fig:planetmassradiusfluxhostmass}
and~\ref{fig:planetmassfluxsemimajorhostmass} compare the planet
masses, planet radii, average incident fluxes, host star masses, and
orbital semi-major axes of the four systems presented in this paper to
other published giant transiting planets listed in the NASA Exoplanet
Archive with $\rpl > 0.5$\,\rjup\ and with measured masses. We show
the comparison to all planets that satisfy these restrictions, and to
only those found around stars with $M < 0.8$\,\msun. The four objects
presented here are consistent with established trends. Notably, all
four objects have relatively small radii (\hatcurPPrlong{47}\,\rjup,
\hatcurPPrlong{48}\,\rjup, \hatcurPPrlong{49}\,\rjup, and
\hatcurPPrlong{72}\,\rjup, for \hatcurb{47}--\hatcurb{49}, and
\hatcurb{72}, respectively) as expected for their low masses and
modest irradiation. This makes the planets potentially useful objects
for comparing to theoretical models of giant planet structure to infer
their bulk heavy element contents. The planets all have semi-major
axes that are beyond the empirical minimum semi-major axis as a
function of planet mass, as seen in the top-right panel of
Figure~\ref{fig:planetmassfluxsemimajorhostmass}. As seen in the
bottom panels of Figures~\ref{fig:planetmassradiusfluxhostmass}
and~\ref{fig:planetmassfluxsemimajorhostmass}, the four planets
discovered here are among a still fairly small sample of giant planets
known around stars with $M < 0.8$\,\msun, and may be useful in that
sense for studying the formation and properties of close-in giant
planets around low mass stars.

\hatcurb{47} and \hatcurb{72} are also notable for their relatively
deep transits. With a transit depth of 3\%, (e.g.,
Fig.~\ref{fig:hats47}), \hatcur{47} is among the deepest known
transiting planet systems. Only HATS-71b \citep{bakos:2018:hats71},
WTS-2b \citep{birkby:2014}, HATS-6b \citep{hartman:2015:hats6}, and
Kepler-45b \citep{johnson:2012} are known to have deeper transits.
Nearly as deep are the transits of
WASP-80b \citep{triaud:2013}, POTS-1b \citep{koppenhoefer:2013},
Qatar-2b \citep{bryan:2012}, and CoRoT-2b \citep{alonso:2008}. The large transit depth makes \hatcurb{47} a potentially attractive target
for follow-up observations, such as transmission spectroscopy, for which
the signal strength scales with the transit depth. With a transit
depth of 1.1\% (e.g., Fig.~\ref{fig:hats72}), \hatcurb{72} stands out
as having the deepest transits among all known planets with $\mpl <
0.15$\,\mjup, making it a valuable target as well for transmission
spectroscopy, in this case to study the atmosphere of a
Super-Neptune. With an optical magnitude of $V =
\hatcurCCtassmv{72}$\,mag and NIR magnitude of $J =
\hatcurCCtwomassJmag{72}$\,mag, \hatcur{72} is only slightly fainter
than HAT-P-26 ($V = 11.74$\,mag, $J = 10.08$\,mag) which hosts a
Neptune, and is significantly brighter than the super-Earth-hosting
K2-18 in the visual band ($V = 13.5$) and only somewhat fainter at near-infrared wavelenghts ($J =
9.76$\,mag). Both of these planets produce shallower transits than
\hatcurb{72}, and have had molecules detected in their
atmospheres via transmission spectroscopy
\citep{tsiaras:2019,benneke:2019,wakeford:2017}.

\hatcur{72} and HATS-73 (Bayliss et al.~2020, in preparation) are the first two systems confirmed by our team using ESPRESSO. This facility has been vital in confirming a relatively low amplitude signal (\hatcurRVK{72}\,\ms) around cool stars. The benefits of this facility are derived not only from the increased mirror size and spectrograph efficiency, but also from the redder wavelength coverage of ESPRESSO which is important as we push to cooler host stars. 

The combination of transit survey and follow-up data from three
separate projects (HATSouth, {\em TESS} and WASP)
also demonstrates the benefits of collaboration between surveys going
forward.  This is particularly so for
\hatcurb{72}, which was independently detected by all three surveys.
As {\em TESS} continues its survey of the sky for transiting
planets around bright stars, most if not all systems that have
previously been identified by ground-based surveys will be observed by
{\em TESS}. Through the coordination of TFOP, redundant follow-up
observations can be avoided for transiting planet systems that have
already been identified and confirmed by ground-based surveys, but
have not yet been published. Coordination by TFOP also ensures greater
efficiency in the analysis and publication of transiting planets such
as these, by enabling data independently collected by different groups
to be combined and analyzed in a single work.

Finally, this paper also illustrates a useful science contribution of ground-based transit surveys that is complimentary to the primary {\em TESS} mission. Neither of the planets \hatcurb{48} or \hatcurb{49} were
identified as transiting planet candidates by the {\em TESS}
team. Both of these objects are relatively faint with $V =
\hatcurCCtassmv{48}$\,mag for \hatcur{48} and $V =
\hatcurCCtassmv{49}$\,mag for \hatcur{49}, and are not among the
targets that have been searched for transits by the {\em TESS}
team, which is focused on searching bright stars around which small planets may be detectable. However, to discover transiting giant planets around low-mass
stars it is necessary to search a large number of M dwarfs and late K
dwarfs, and thus to consider stars that are faint in the optical
band-passes. Ground-based surveys like HATSouth, combined with independent analyses of the {\em TESS} FFIs being released to the public, are providing valuable contributions addressing this science topic.

\ifthenelse{\boolean{emulateapj}}{
    \begin{figure*}[!ht]
}{
    \begin{figure}[!ht]
}
 {
 \centering
 \leavevmode
 \includegraphics[width={0.5\linewidth}]{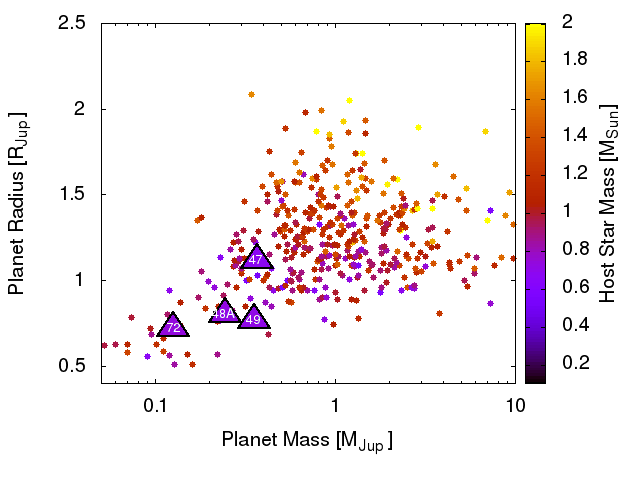}
 \hfil
 \includegraphics[width={0.5\linewidth}]{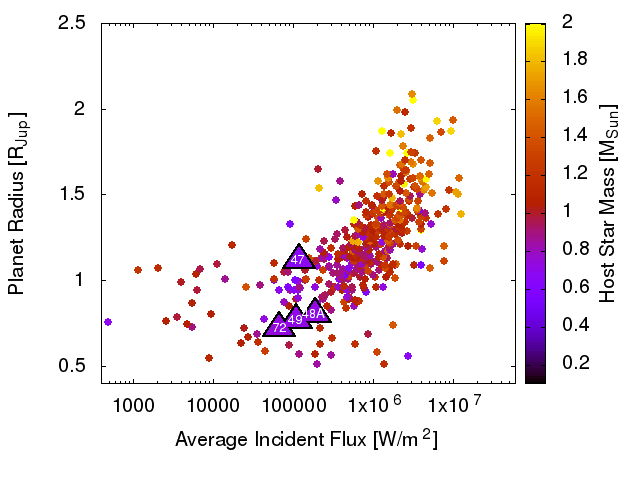}
 }
 {
 \centering
 \leavevmode
 \includegraphics[width={0.5\linewidth}]{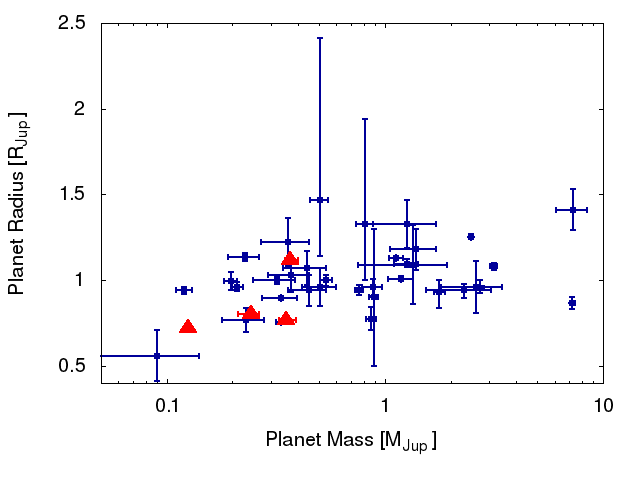}
 \hfil
 \includegraphics[width={0.5\linewidth}]{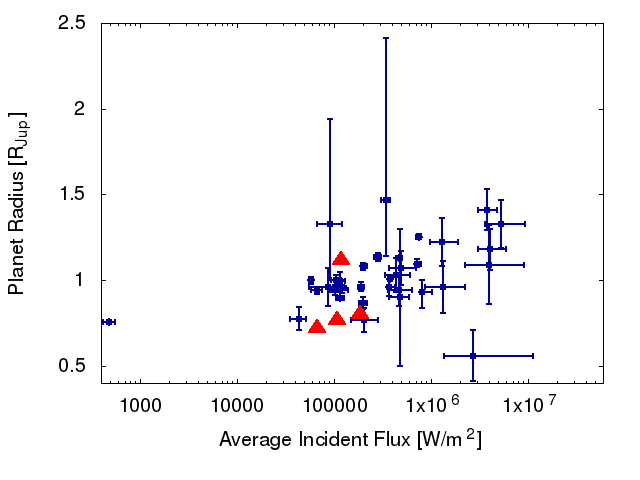}
 }
\caption{
    {\em Top:} Giant transiting planet radius vs.\ planet mass ({\em left}) and average incident flux ({\em right}). The four transiting planets presented in this paper are shown with large filled triangles, and are labelled by the HATS object number. The smaller filled circles show confirmed transiting giant planets with $R_{p} > 0.5$\,\rjup\ and with measured masses (excluding objects where only an upper limit on the mass has been set) taken from the NASA Exoplanet Archive accessed 2019 Sep.\ 9. The color of each symbol indicates the mass of the transiting planet host star. {\em Bottom:} Similar to the top, but here we show only transiting planets around stars with $M < 0.8$\,\msun, we do not show the host star mass, but we do include the $1\sigma$ errorbars. The four planets presented in this paper are shown by the red filled triangles. Errorbars on the planet radius and incident flux are smaller than the symbols for these four planets.\label{fig:planetmassradiusfluxhostmass}
}
\ifthenelse{\boolean{emulateapj}}{
    \end{figure*}
}{
    \end{figure}
}

\ifthenelse{\boolean{emulateapj}}{
    \begin{figure*}[!ht]
}{
    \begin{figure}[!ht]
}
 {
 \centering
 \leavevmode
 \includegraphics[width={0.5\linewidth}]{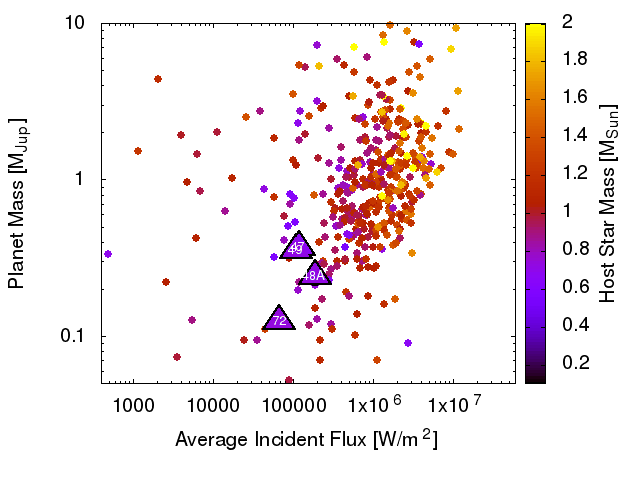}
 \hfil
 \includegraphics[width={0.5\linewidth}]{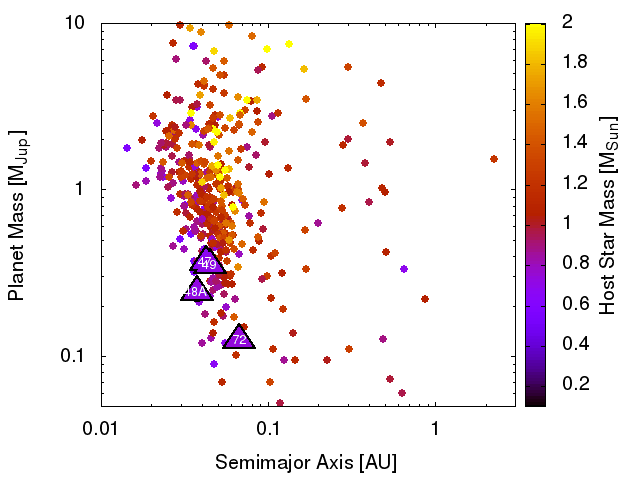}
 }
 {
 \centering
 \leavevmode
 \includegraphics[width={0.5\linewidth}]{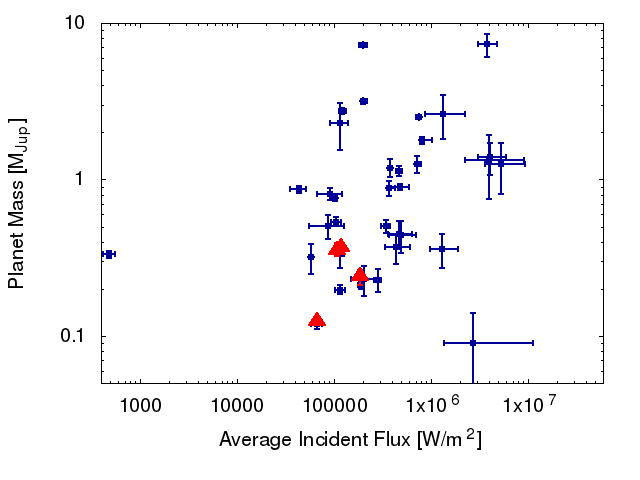}
 \hfil
 \includegraphics[width={0.5\linewidth}]{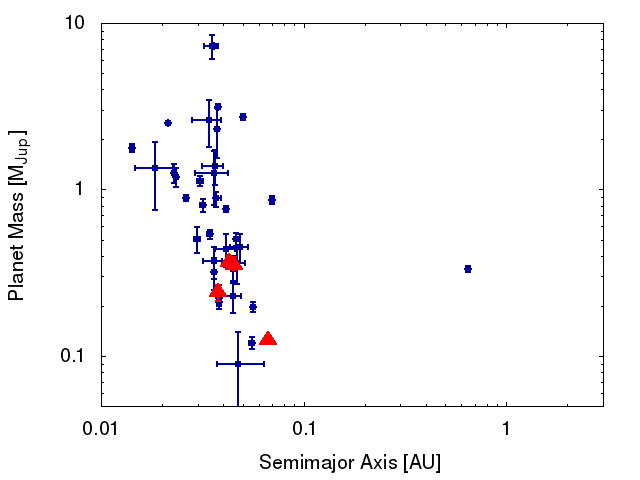}
 }
\caption{
    Similar to Fig.~\ref{fig:planetmassradiusfluxhostmass}, here we show giant transiting planet mass vs.\ average incident flux ({\em left}) and semimajor axis ({\em right}). \hatcurb{47} and \hatcurb{49} overlap on these plots. As in Fig.~\ref{fig:planetmassradiusfluxhostmass}, only planets with $\rpl > 0.5$\,\rjup, and with measured masses are shown. \label{fig:planetmassfluxsemimajorhostmass}
}
\ifthenelse{\boolean{emulateapj}}{
    \end{figure*}
}{
    \end{figure}
}



\acknowledgements 

We thank the anonymous referee for valuable feedback which has improved the quality of this paper. Development of the HATSouth
project was funded by NSF MRI grant NSF/AST-0723074, operations have
been supported by NASA grants NNX09AB29G, NNX12AH91H, and NNX17AB61G, and follow-up
observations have received partial support from grant NSF/AST-1108686.
A.J.\ acknowledges support from FONDECYT project 1171208 and by the Ministry for the Economy, Development, and Tourism's Programa Iniciativa Cient\'{i}fica Milenio through grant IC\,120009, awarded to the Millennium Institute of Astrophysics (MAS).
L.M.\ acknowledges support from the Italian Minister of Instruction, University and Research (MIUR) through FFABR 2017 fund. L.M.\ acknowledges support from the University of Rome Tor Vergata through ``Mission: Sustainability 2016'' fund.
K.P.\ acknowledges support from NASA ATP grant 80NSSC18K1009.
V.S.\ acknowledges support from BASAL CATA PFB-06.  
J.N.W.\ thanks the Heising-Simons foundation for support.
I.J.M.C. acknowledges support from the NSF through grant AST-1824644,
and from NASA through Caltech/JPL grant RSA-1610091.  
Support for this work was provided to J.K.T. by NASA through Hubble Fellowship grant HST-HF2-51399.001 awarded by the Space Telescope Science Institute, which is operated by the Association of Universities for Research in Astronomy, Inc., for NASA, under contract NAS5-26555.
This work is based on observations made with ESO Telescopes at the La
Silla Observatory.
This paper also makes use of observations from the LCOGT network. Some of this time was awarded by NOAO.
We acknowledge the use of the AAVSO Photometric All-Sky Survey (APASS),
funded by the Robert Martin Ayers Sciences Fund, and the SIMBAD
database, operated at CDS, Strasbourg, France.
Operations at the MPG~2.2\,m Telescope are jointly performed by the
Max Planck Gesellschaft and the European Southern Observatory.  
We thank the MPG 2.2m telescope support team for their technical
assistance during observations.
TRAPPIST-South is a project funded by the Belgian F.R.S.-FNRS under
grant FRFC 2.5.594.09.F, with the participation of the Swiss FNS.  The
research leading to these results has received funding from the ARC
grant for Concerted Research Actions, financed by the
Wallonia-Brussels Federation. EJ and MG are F.R.S.-FNRS Senior
Research Associates.
Contributions at the University of Geneva by LN, ML, and SU were carried out within the framework of the National Centre for Competence in Research "PlanetS" supported by the Swiss National Science Foundation (SNSF).
ML acknowledges support from the Austrian Research Promotion Agency (FFG) under project 859724 ``GRAPPA''.
This work has made use of data from the European Space Agency (ESA)
mission {\it Gaia} (\url{https://www.cosmos.esa.int/gaia}), processed by
the {\it Gaia} Data Processing and Analysis Consortium (DPAC,
\url{https://www.cosmos.esa.int/web/gaia/dpac/consortium}). Funding
for the DPAC has been provided by national institutions, in particular
the institutions participating in the {\it Gaia} Multilateral Agreement.
This research has made use of the NASA Exoplanet Archive, which is
operated by the California Institute of Technology, under contract
with the National Aeronautics and Space Administration under the
Exoplanet Exploration Program.
This research has made use NASA's Astrophysics Data System Bibliographic Services.

\facilities{HATSouth, TESS, SuperWASP, Swope, LCOGT, TRAPPIST, Max
  Planck:2.2m (FEROS), ESO:3.6m (HARPS), Euler1.2m (Coralie), ATT
  (WiFeS), Magellan:Clay (PFS), VLT (ESPRESSO), NTT (Astralux Sur),
  SOAR, Gaia, Exoplanet Archive}

\software{FITSH \citep{pal:2012}, BLS \citep{kovacs:2002:BLS},
  VARTOOLS \citep{hartman:2016:vartools}, CERES
  \citep{brahm:2017:ceres}, ZASPE \citep{brahm:2017:zaspe}, SPEX-tool
  \citep{cushing:2004,vacca:2004}, SExtractor \citep{bertin:1996},
  Astrometry.net \citep{lang:2010}, MWDUST \citep{bovy:2016}, TESSCut
  \citep{brasseur:2019}, Lightkurve \citep{lightkurve:2018}, Astropy \citep{astropy:2013,astropy:2018}}


\bibliographystyle{aasjournal}

\clearpage

%
%
\ifthenelse{\boolean{emulateapj}}{
    \begin{deluxetable*}{lccccl}
}{
    \begin{deluxetable}{lccccl}
}
\tablewidth{0pc}
\tabletypesize{\tiny}
\tablecaption{
    Astrometric, Spectroscopic and Photometric parameters for \hatcur{47}, \hatcur{48}, \hatcur{49} and \hatcur{72}
    \label{tab:stellarobserved}
}
\tablehead{
    \multicolumn{1}{c}{} &
    \multicolumn{1}{c}{\bf HATS-47} &
    \multicolumn{1}{c}{\bf HATS-48A} &
    \multicolumn{1}{c}{\bf HATS-49} &
    \multicolumn{1}{c}{\bf HATS-72} &
    \multicolumn{1}{c}{} \\
    \multicolumn{1}{c}{~~~~~~~~Parameter~~~~~~~~} &
    \multicolumn{1}{c}{Value}                     &
    \multicolumn{1}{c}{Value}                     &
    \multicolumn{1}{c}{Value}                     &
    \multicolumn{1}{c}{Value}                     &
    \multicolumn{1}{c}{Source}
}
\startdata
\noalign{\vskip -3pt}
\sidehead{Astrometric properties and cross-identifications}
~~~~2MASS-ID\dotfill               & \hatcurCCtwomassshort{47}  & \hatcurCCtwomassshort{48} & \hatcurCCtwomassshort{49} & \hatcurCCtwomassshort{72} & \\
~~~~TIC-ID\dotfill                 & \hatcurCCtic{47} & \hatcurCCtic{48} & \hatcurCCtic{49} & \hatcurCCtic{72} & \\
~~~~TOI-ID\dotfill                 & \hatcurCCtoi{47} & $\cdots$ & $\cdots$ & \hatcurCCtoi{72} & \\
~~~~GAIA~DR2-ID\dotfill                 & \hatcurCCgaiadrtwo{47}      & \hatcurCCgaiadrtwo{48} & \hatcurCCgaiadrtwo{49} & \hatcurCCgaiadrtwo{72} & \\
~~~~R.A. (J2000)\dotfill            & \hatcurCCra{47}       & \hatcurCCra{48}    & \hatcurCCra{49}    & \hatcurCCra{72}    & GAIA DR2\\
~~~~Dec. (J2000)\dotfill            & \hatcurCCdec{47}      & \hatcurCCdec{48}   & \hatcurCCdec{49}   & \hatcurCCdec{72}   & GAIA DR2\\
~~~~$\mu_{\rm R.A.}$ (\masy)              & \hatcurCCpmra{47}     & \hatcurCCpmra{48} & \hatcurCCpmra{49} & \hatcurCCpmra{72} & GAIA DR2\\
~~~~$\mu_{\rm Dec.}$ (\masy)              & \hatcurCCpmdec{47}    & \hatcurCCpmdec{48} & \hatcurCCpmdec{49} & \hatcurCCpmdec{72} & GAIA DR2\\
~~~~parallax (mas)              & \hatcurCCparallax{47}    & \hatcurCCparallax{48} & \hatcurCCparallax{49} & \hatcurCCparallax{72} & GAIA DR2\\
\sidehead{Spectroscopic properties}
~~~~$\teffstar$ (K)\dotfill         &  \hatcurSMEteff{47}   & \hatcurSMEteff{48} & \hatcurSMEteff{49} & \hatcurSMEteff{72} & ZASPE\tablenotemark{a}\\
~~~~$\feh$\dotfill                  &  \hatcurSMEzfeh{47}   & \hatcurSMEzfeh{48} & \hatcurSMEzfeh{49} & \hatcurSMEzfeh{72} & ZASPE               \\
~~~~$\vsini$ (\kms)\dotfill         &  \hatcurSMEvsin{47}   & \hatcurSMEvsin{48} & \hatcurSMEvsin{49} & \hatcurSMEvsin{72} & ZASPE                \\
~~~~$\vmac$ (\kms)\dotfill          &  \hatcurSMEvmac{47}   & \hatcurSMEvmac{48} & \hatcurSMEvmac{49} & \hatcurSMEvmac{72} & Assumed \\
~~~~$\vmic$ (\kms)\dotfill          &  \hatcurSMEvmic{47}   & \hatcurSMEvmic{48} & \hatcurSMEvmic{49} & \hatcurSMEvmic{72} & Assumed              \\
~~~~$\gamma_{\rm RV}$ (\ms)\dotfill&  $\hatcurRVgammaabs{47}$  & $\hatcurRVgammaabs{48}$ & $\hatcurRVgammaabs{49}$ & $\hatcurRVgammaabs{72}$ & FEROS or WiFeS\tablenotemark{b}  \\
\sidehead{Photometric properties}
~~~~$P_{\rm rot}$ (d)\tablenotemark{c}   & $6.42\pm0.28$ & $\cdots$ & $\cdots$ & $48.725\pm0.015$ & HATSouth \\
~~~~$G$ (mag)\tablenotemark{d}\dotfill               &  \hatcurCCgaiamG{47}  & \hatcurCCgaiamG{48} & \hatcurCCgaiamG{49} & \hatcurCCgaiamG{72} & GAIA DR2 \\
~~~~$BP$ (mag)\tablenotemark{d}\dotfill               &  \hatcurCCgaiamBP{47}  & \hatcurCCgaiamBP{48} & \hatcurCCgaiamBP{49} & \hatcurCCgaiamBP{72} & GAIA DR2 \\
~~~~$RP$ (mag)\tablenotemark{d}\dotfill               &  \hatcurCCgaiamRP{47}  & \hatcurCCgaiamRP{48} & \hatcurCCgaiamRP{49} & \hatcurCCgaiamRP{72} & GAIA DR2 \\
~~~~$B$ (mag)\dotfill               &  \hatcurCCtassmB{47}  & \hatcurCCtassmB{48} & \hatcurCCtassmB{49} & \hatcurCCtassmB{72} & APASS\tablenotemark{e} \\
~~~~$V$ (mag)\dotfill               &  \hatcurCCtassmv{47}  & \hatcurCCtassmv{48} & \hatcurCCtassmv{49} & \hatcurCCtassmv{72} & APASS\tablenotemark{e} \\
~~~~$g$ (mag)\dotfill               &  \hatcurCCtassmg{47}  & \hatcurCCtassmg{48} & \hatcurCCtassmg{49} & \hatcurCCtassmg{72} & APASS\tablenotemark{e} \\
~~~~$r$ (mag)\dotfill               &  \hatcurCCtassmr{47}  & \hatcurCCtassmr{48} & \hatcurCCtassmr{49} & \hatcurCCtassmr{72} & APASS\tablenotemark{e} \\
~~~~$i$ (mag)\dotfill               &  \hatcurCCtassmi{47}  & \hatcurCCtassmi{48} & \hatcurCCtassmi{49} & \hatcurCCtassmi{72} & APASS\tablenotemark{e} \\
~~~~$J$ (mag)\dotfill               &  \hatcurCCtwomassJmag{47} & \hatcurCCtwomassJmag{48} & \hatcurCCtwomassJmag{49} & \hatcurCCtwomassJmag{72} & 2MASS           \\
~~~~$H$ (mag)\dotfill               &  \hatcurCCtwomassHmag{47} & \hatcurCCtwomassHmag{48} & \hatcurCCtwomassHmag{49} & \hatcurCCtwomassHmag{72} & 2MASS           \\
~~~~$K_s$ (mag)\dotfill             &  \hatcurCCtwomassKmag{47} & \hatcurCCtwomassKmag{48} & \hatcurCCtwomassKmag{49} & \hatcurCCtwomassKmag{72} & 2MASS           \\
~~~~$W1$ (mag)\dotfill             &  \hatcurCCWonemag{47} & \hatcurCCWonemag{48} & \hatcurCCWonemag{49} & \hatcurCCWonemag{72} & WISE           \\
~~~~$W2$ (mag)\dotfill             &  \hatcurCCWtwomag{47} & \hatcurCCWtwomag{48} & \hatcurCCWtwomag{49} & \hatcurCCWtwomag{72} & WISE           \\
~~~~$W3$ (mag)\dotfill             &  $11.707\pm0.192$ & $11.433\pm0.135$ & $12.212\pm0.256$ & \hatcurCCWthreemag{72} & WISE           \\
\enddata
\tablenotetext{a}{
    ZASPE = Zonal Atmospherical Stellar Parameter Estimator routine
    for the analysis of high-resolution spectra
    \citep{brahm:2017:zaspe}, applied to the PFS spectra of
    \hatcur{47} and \hatcur{49}, and to the FEROS spectra of
    \hatcur{48} and \hatcur{72}.
}
\tablenotetext{b}{
    The error on $\gamma_{\rm RV}$ is determined from the
    orbital fit to the RV measurements, and does not include the
    systematic uncertainty in transforming the velocities to the IAU
    standard system. The velocities have not been corrected for
    gravitational redshifts. We report the value from FEROS for \hatcur{47}, \hatcur{48} and \hatcur{72}. For \hatcur{49} we report the value from WiFeS.
} 
\tablenotetext{c}{
    Photometric rotation period.
}
\tablenotetext{d}{
    The listed uncertainties for the Gaia DR2 photometry are taken from the catalog. For the analysis we assume additional systematic uncertainties of 0.002\,mag, 0.005\,mag and 0.003\,mag for the G, BP and RP bands, respectively.
}
\tablenotetext{e}{
    From APASS DR6 as
    listed in the UCAC 4 catalog \citep{zacharias:2013:ucac4}.  
}
\ifthenelse{\boolean{emulateapj}}{
    \end{deluxetable*}
}{
    \end{deluxetable}
}

\tabletypesize{\scriptsize}
\ifthenelse{\boolean{emulateapj}}{
    \begin{deluxetable*}{lrrrrrrl}
}{
    \begin{deluxetable}{lrrrrrrl}
}
\tablewidth{0pc}
\tablecaption{
    Relative radial velocities and bisector spans for \hatcur{47}, \hatcur{48}, \hatcur{49} and \hatcur{72}.
    \label{tab:rvs}
}
\tablehead{
    \colhead{System} &
    \colhead{BJD} &
    \colhead{RV\tablenotemark{a}} &
    \colhead{\ensuremath{\sigma_{\rm RV}}\tablenotemark{b}} &
    \colhead{BS} &
    \colhead{\ensuremath{\sigma_{\rm BS}}} &
    \colhead{Phase} &
    \colhead{Instrument}\\
    \colhead{} &
    \colhead{\hbox{(2,450,000$+$)}} &
    \colhead{(\ms)} &
    \colhead{(\ms)} &
    \colhead{(\ms)} &
    \colhead{(\ms)} &
    \colhead{} &
    \colhead{}
}
\startdata
HATS-47 & $ 7472.87421 $ & $   -73.25 $ & $     9.58 $ & $ -379.5 $ & $   95.6 $ & $   0.408 $ & PFS \\
HATS-47 & $ 7474.88339 $ & $    17.75 $ & $     9.26 $ & $   84.1 $ & $   58.3 $ & $   0.920 $ & PFS \\
HATS-47 & $ 7477.86812 $ & \nodata      & \nodata      & $  -38.4 $ & $   48.8 $ & $   0.681 $ & PFS \\
HATS-47 & $ 7477.90415 $ & $    68.00 $ & $    12.60 $ & \nodata      & \nodata      & $   0.690 $ & PFS \\
HATS-47 & $ 7507.88072 $ & $   -33.23 $ & $    11.94 $ & $ -331.8 $ & $   71.1 $ & $   0.332 $ & PFS \\
HATS-47 & $ 7530.84061 $ & $   -93.65 $ & $    13.98 $ & $   60.7 $ & $  177.1 $ & $   0.185 $ & PFS \\
HATS-47 & $ 7534.82504 $ & $   -40.07 $ & $     8.01 $ & $  304.7 $ & $   67.2 $ & $   0.200 $ & PFS \\
HATS-47 & $ 7536.82554 $ & $    47.53 $ & $     9.89 $ & $  262.1 $ & $   71.0 $ & $   0.710 $ & PFS \\
HATS-47 & $ 7558.79646 $ & $    43.22 $ & $    10.65 $ & $   89.5 $ & $  133.0 $ & $   0.311 $ & PFS \\
HATS-47 & $ 7615.65549 $ & $    29.50 $ & $     8.55 $ & $  -98.7 $ & $   49.9 $ & $   0.806 $ & PFS \\
\enddata
\tablenotetext{a}{
    The zero-point of these velocities is arbitrary. An overall offset
    $\gamma_{\rm rel}$ fitted independently to the velocities from
    each instrument has been subtracted.
}
\tablenotetext{b}{
    Internal errors excluding the component of astrophysical jitter
    allowed to vary in the fit.
}
\tablecomments{
    This table is available in a machine-readable form in the online
    journal.  A portion is shown here for guidance regarding its form
    and content.
}
\ifthenelse{\boolean{emulateapj}}{
    \end{deluxetable*}
}{
    \end{deluxetable}
}

%
%
\ifthenelse{\boolean{emulateapj}}{
    \begin{deluxetable*}{lcccc}
}{
    \begin{deluxetable}{lcccc}
}
\tablewidth{0pc}
\tabletypesize{\footnotesize}
\tablecaption{
    Adopted derived stellar parameters for \hatcur{47}, \hatcur{48}, \hatcur{49} and \hatcur{72}.
    \label{tab:stellarderived}
}
\tablehead{
    \multicolumn{1}{c}{} &
    \multicolumn{1}{c}{\bf HATS-47} &
    \multicolumn{1}{c}{\bf HATS-48A} &
    \multicolumn{1}{c}{\bf HATS-49} &
    \multicolumn{1}{c}{\bf HATS-72} \\
    \multicolumn{1}{c}{~~~~~~~~Parameter~~~~~~~~} &
    \multicolumn{1}{c}{Value}                     &
    \multicolumn{1}{c}{Value}                     &
    \multicolumn{1}{c}{Value}                     &
    \multicolumn{1}{c}{Value}                     
}
\startdata
~~~~$\mstar$ ($\msun$)\dotfill      &  \hatcurISOmlong{47}   & \hatcurISOmlong{48} & \hatcurISOmlong{49} & \hatcurISOmlong{72} \\
~~~~$\rstar$ ($\rsun$)\dotfill      &  \hatcurISOrlong{47}   & \hatcurISOrlong{48} & \hatcurISOrlong{49} & \hatcurISOrlong{72} \\
~~~~$\loggstar$ (cgs)\dotfill       &  \hatcurISOlogg{47}    & \hatcurISOlogg{48} & \hatcurISOlogg{49} & \hatcurISOlogg{72} \\
~~~~$\rhostar$ (\gcmc)\dotfill       &  \hatcurLCrho{47}    & \hatcurLCrho{48} & \hatcurLCrho{49} & \hatcurLCrho{72} \\
~~~~$\lstar$ ($\lsun$)\dotfill      &  \hatcurISOlum{47}     & \hatcurISOlum{48} & \hatcurISOlum{49} & \hatcurISOlum{72} \\
~~~~$\teffstar$ (K)\dotfill      &  \hatcurISOteff{47} &  \hatcurISOteff{48} &  \hatcurISOteff{49} &  \hatcurISOteff{72} \\
~~~~$\feh$\dotfill      &  \hatcurISOzfeh{47} &  \hatcurISOzfeh{48} &  \hatcurISOzfeh{49} &  \hatcurISOzfeh{72} \\
~~~~Age (Gyr)\dotfill               &  \hatcurISOage{47}     & \hatcurISOage{48} & \hatcurISOage{49} & \hatcurISOage{72} \\
~~~~$A_{V}$ (mag)\dotfill               &  \hatcurXAv{47}     & \hatcurXAv{48} & \hatcurXAv{49} & \hatcurXAv{72} \\
~~~~Distance (pc)\dotfill           &  \hatcurXdistred{47}\phn  & \hatcurXdistred{48} & \hatcurXdistred{49} & \hatcurXdistred{72} \\
~~~~M$_{\rm B}$ (\msun)\tablenotemark{a} & $< 0.23$ & $0.22$ & $< 0.31$ & $< 0.36$ \\
\enddata
\tablenotetext{a}{
    For \hatcur{47}, \hatcur{49} and \hatcur{72} we list the 95\% confidence upper limit on the mass of any unresolved stellar companion based on modelling the system as a blend between a transiting planet system and an unresolved wide stellar binary companion (Section~\ref{sec:blendmodel}). For \hatcur{48} we list the estimated mass for the 5\farcs4 neighbor in Gaia~DR2 which we determined to be a common-proper-motion and common-parallax companion to \hatcur{48} (Section~\ref{sec:luckyimaging}).
}
\tablecomments{
The listed parameters are those determined through the joint differential evolution Markov Chain analysis described in Section~\ref{sec:transitmodel}. For all four systems the RV observations are consistent with a circular orbit, and we assume a fixed circular orbit in generating the parameters listed here. Systematic errors in the bolometric correction tables or stellar evolution models are not included, and may dominate the error budget for some of these parameters. 
}
\ifthenelse{\boolean{emulateapj}}{
    \end{deluxetable*}
}{
    \end{deluxetable}
}

%
\ifthenelse{\boolean{emulateapj}}{
    \begin{deluxetable*}{lcccc}
}{
    \begin{deluxetable}{lcccc}
}
\tabletypesize{\tiny}
\tablecaption{Adopted orbital and planetary parameters for \hatcurb{47}, \hatcurb{48}, \hatcurb{49} and \hatcurb{72}\label{tab:planetparam}}
\tablehead{
    \multicolumn{1}{c}{} &
    \multicolumn{1}{c}{\bf HATS-47b} &
    \multicolumn{1}{c}{\bf HATS-48Ab} &
    \multicolumn{1}{c}{\bf HATS-49b} &
    \multicolumn{1}{c}{\bf HATS-72b} \\
    \multicolumn{1}{c}{~~~~~~~~~~~~~~~Parameter~~~~~~~~~~~~~~~} &
    \multicolumn{1}{c}{Value} &
    \multicolumn{1}{c}{Value} &
    \multicolumn{1}{c}{Value} &
    \multicolumn{1}{c}{Value}
}
\startdata
\noalign{\vskip -3pt}
\sidehead{\Lc{} parameters}
~~~$P$ (days)             \dotfill    & $\hatcurLCP{47}$ & $\hatcurLCP{48}$ & $\hatcurLCP{49}$ & $\hatcurLCP{72}$ \\
~~~$T_c$ (${\rm BJD\_{}TDB}$)    
      \tablenotemark{a}   \dotfill    & $\hatcurLCT{47}$ & $\hatcurLCT{48}$ & $\hatcurLCT{49}$ & $\hatcurLCT{72}$ \\
~~~$T_{14}$ (days)
      \tablenotemark{a}   \dotfill    & $\hatcurLCdur{47}$ & $\hatcurLCdur{48}$ & $\hatcurLCdur{49}$ & $\hatcurLCdur{72}$ \\
~~~$T_{12} = T_{34}$ (days)
      \tablenotemark{a}   \dotfill    & $\hatcurLCingdur{47}$ & $\hatcurLCingdur{48}$ & $\hatcurLCingdur{49}$ & $\hatcurLCingdur{72}$ \\
~~~$\arstar$              \dotfill    & $\hatcurPPar{47}$ & $\hatcurPPar{48}$ & $\hatcurPPar{49}$ & $\hatcurPPar{72}$ \\
~~~$\zrstar$ \tablenotemark{b}             \dotfill    & $\hatcurLCzeta{47}$\phn & $\hatcurLCzeta{48}$\phn& $\hatcurLCzeta{49}$\phn& $\hatcurLCzeta{72}$\phn\\
~~~$\rpl/\rstar$          \dotfill    & $\hatcurLCrprstar{47}$ & $\hatcurLCrprstar{48}$& $\hatcurLCrprstar{49}$& $\hatcurLCrprstar{72}$\\
~~~$b^2$                  \dotfill    & $\hatcurLCbsq{47}$ & $\hatcurLCbsq{48}$& $\hatcurLCbsq{49}$& $\hatcurLCbsq{72}$\\
~~~$b \equiv a \cos i/\rstar$
                          \dotfill    & $\hatcurLCimp{47}$ & $\hatcurLCimp{48}$& $\hatcurLCimp{49}$& $\hatcurLCimp{72}$\\
~~~$i$ (deg)              \dotfill    & $\hatcurPPi{47}$\phn & $\hatcurPPi{48}$\phn& $\hatcurPPi{49}$\phn& $\hatcurPPi{72}$\phn\\
\sidehead{Dilution factors \tablenotemark{c}}
~~~HATSouth 1\dotfill & \hatcurLCiblendA{47} & \hatcurLCiblendA{48}& \hatcurLCiblendA{49}& \hatcurLCiblendA{72}\\
~~~HATSouth 2\dotfill & $\cdots$ & $\cdots$ & \hatcurLCiblendB{49}& $\cdots$ \\
~~~{\em TESS} \dotfill & \hatcurLCiblendB{47} & \hatcurLCiblendB{48}& \hatcurLCiblendC{49}& \hatcurLCiblendB{72}\\
~~~WASP \dotfill & $\cdots$ & $\cdots$ & $\cdots$ & \hatcurLCiblendC{72}\\
\sidehead{Limb-darkening coefficients \tablenotemark{d}}
~~~$c_1,r$                  \dotfill    & $\hatcurLBir{47}$ & $\hatcurLBir{48}$& $\hatcurLBir{49}$& $\hatcurLBir{72}$\\
~~~$c_2,r$                  \dotfill    & $\hatcurLBiir{47}$ & $\hatcurLBiir{48}$& $\hatcurLBiir{49}$& $\hatcurLBiir{72}$\\
~~~$c_1,i$                  \dotfill    & $\hatcurLBii{47}$ & $\hatcurLBii{48}$& $\hatcurLBii{49}$& $\hatcurLBii{72}$\\
~~~$c_2,i$                  \dotfill    & $\hatcurLBiii{47}$ & $\hatcurLBiii{48}$& $\hatcurLBiii{49}$& $\hatcurLBiii{72}$\\
~~~$c_1,T$                  \dotfill    & $\hatcurLBiT{47}$ & $\hatcurLBiT{48}$& $\hatcurLBiT{49}$& $\hatcurLBiT{72}$\\
~~~$c_2,T$                  \dotfill    & $\hatcurLBiiT{47}$ & $\hatcurLBiiT{48}$& $\hatcurLBiiT{49}$& $\hatcurLBiiT{72}$\\
~~~$c_1,I+z$                  \dotfill    & $\cdots$ & $\cdots$& $\cdots$& $\hatcurLBiI{72}$\\
~~~$c_2,I+z$                  \dotfill    & $\cdots$ & $\cdots$& $\cdots$& $\hatcurLBiiI{72}$\\
~~~$c_1,{\rm WASP}$                  \dotfill    & $\cdots$ & $\cdots$& $\cdots$& $\hatcurLBikep{72}$\\
~~~$c_2,{\rm WASP}$                  \dotfill    & $\cdots$ & $\cdots$& $\cdots$& $\hatcurLBiikep{72}$\\
\sidehead{RV parameters}
~~~$K$ (\ms)              \dotfill    & $\hatcurRVK{47}$\phn\phn & $\hatcurRVK{48}$\phn\phn& $\hatcurRVK{49}$\phn\phn& $\hatcurRVK{72}$\phn\phn\\
~~~$e$ \tablenotemark{e}               \dotfill    & $\hatcurRVeccentwosiglimeccen{47}$ & $\hatcurRVeccentwosiglimeccen{48}$ & $\hatcurRVeccentwosiglimeccen{49}$ & $\hatcurRVeccentwosiglimeccen{72}$ \\
~~~RV jitter PFS (\ms)        \dotfill    & $\hatcurRVjitter{47}$ & $\hatcurRVjitter{48}$& $\hatcurRVjitter{49}$& $\hatcurRVjitterC{72}$\\
~~~RV jitter ESPRESSO (\ms)        \dotfill    & $\cdots$ & $\cdots$& $\cdots$& $\hatcurRVjitterD{72}$\\
~~~RV jitter HARPS (\ms)        \dotfill    & $\cdots$ & $\cdots$& $\cdots$& $\hatcurRVjitterB{72}$\\
~~~RV jitter FEROS (\ms)        \dotfill    & $\cdots$ & $\cdots$& $\cdots$& $\hatcurRVjitterA{72}$\\
\sidehead{Planetary parameters}
~~~$\mpl$ ($\mjup$)       \dotfill    & $\hatcurPPmlong{47}$ & $\hatcurPPmlong{48}$& $\hatcurPPmlong{49}$& $\hatcurPPmlong{72}$\\
~~~$\rpl$ ($\rjup$)       \dotfill    & $\hatcurPPrlong{47}$ & $\hatcurPPrlong{48}$& $\hatcurPPrlong{49}$& $\hatcurPPrlong{72}$\\
~~~$C(\mpl,\rpl)$
    \tablenotemark{g}     \dotfill    & $\hatcurPPmrcorr{47}$ & $\hatcurPPmrcorr{48}$& $\hatcurPPmrcorr{49}$& $\hatcurPPmrcorr{72}$\\
~~~$\rhopl$ (\gcmc)       \dotfill    & $\hatcurPPrho{47}$ & $\hatcurPPrho{48}$& $\hatcurPPrho{49}$& $\hatcurPPrho{72}$\\
~~~$\log g_p$ (cgs)       \dotfill    & $\hatcurPPlogg{47}$ & $\hatcurPPlogg{48}$& $\hatcurPPlogg{49}$& $\hatcurPPlogg{72}$\\
~~~$a$ (AU)               \dotfill    & $\hatcurPParel{47}$ & $\hatcurPParel{48}$& $\hatcurPParel{49}$& $\hatcurPParel{72}$\\
~~~$T_{\rm eq}$ (K)        \dotfill   & $\hatcurPPteff{47}$ & $\hatcurPPteff{48}$& $\hatcurPPteff{49}$& $\hatcurPPteff{72}$\\
~~~$\Theta$ \tablenotemark{h} \dotfill & $\hatcurPPtheta{47}$ & $\hatcurPPtheta{48}$& $\hatcurPPtheta{49}$& $\hatcurPPtheta{72}$\\
~~~$\log_{10}\langle F \rangle$ (cgs) \tablenotemark{i}
                          \dotfill    & $\hatcurPPfluxavglog{47}$ & $\hatcurPPfluxavglog{48}$& $\hatcurPPfluxavglog{49}$& $\hatcurPPfluxavglog{72}$\\
\enddata
\tablecomments{
For all systems we adopt a model in which the orbit is assumed to be circular. See the discussion in Section~\ref{sec:transitmodel}.
}
\tablenotetext{a}{
    Times are in Barycentric Julian Date calculated on the Barycentric Dynamical Time (TDB) system.
    \ensuremath{T_c}: Reference epoch of
    mid transit that minimizes the correlation with the orbital
    period.
    \ensuremath{T_{12}}: total transit duration, time
    between first to last contact;
    \ensuremath{T_{12}=T_{34}}: ingress/egress time, time between first
    and second, or third and fourth contact.
}
\tablenotetext{b}{
   Reciprocal of the half duration of the transit used as a jump parameter in our MCMC analysis in place of $\arstar$. It is related to $\arstar$ by the expression $\zrstar = \arstar(2\pi(1+e\sin\omega))/(P\sqrt{1-b^2}\sqrt{1-e^2})$ \citep{bakos:2010:hat11}.
}
\tablenotetext{c}{
    Scaling factor applied to the model transit that is fit to the HATSouth, {\em TESS} and WASP light curves. This factor accounts for dilution of the transit due to blending from neighboring stars and/or over-filtering of the light curve.  These factors are varied in the fit, with independent values adopted for each light curve. \hatcur{49} was observed in two separate HATSouth fields, and we list the two independent dilution factors fitted for the light curves from each of these fields.
}
\tablenotetext{d}{
    Values for a quadratic law. The limb darkening parameters were
    directly varied in the fit, using the tabulations from
    \cite{claret:2012,claret:2013,claret:2018} to place Gaussian prior
    constraints on their values, assuming a prior uncertainty of $0.2$
    for each coefficient.
}
\tablenotetext{e}{
    The 95\% confidence upper limit on the eccentricity determined
    when $\sqrt{e}\cos\omega$ and $\sqrt{e}\sin\omega$ are allowed to
    vary in the fit.
}
\tablenotetext{f}{
    Term added in quadrature to the formal RV uncertainties for each
    instrument. This is treated as a free parameter in the fitting
    routine. 
}
\tablenotetext{g}{
    Correlation coefficient between the planetary mass \mpl\ and radius
    \rpl\ estimated from the posterior parameter distribution.
}
\tablenotetext{h}{
    The Safronov number is given by $\Theta = \frac{1}{2}(V_{\rm
    esc}/V_{\rm orb})^2 = (a/\rpl)(\mpl / \mstar )$
    \citep[see][]{hansen:2007}.
}
\tablenotetext{i}{
    Incoming flux per unit surface area, averaged over the orbit.
}
\ifthenelse{\boolean{emulateapj}}{
    \end{deluxetable*}
}{
    \end{deluxetable}
}

\end{document}